\documentclass[final,3p,times]{elsarticle} %Option 1

\usepackage{lineno}
\usepackage{graphicx}
\usepackage{tikz}
\definecolor{cube}{HTML}{EBDEF0} 
\definecolor{redchannel}{HTML}{CD6155}
\definecolor{greenchannel}{HTML}{82E0AA}
\definecolor{bluechannel}{HTML}{85C1E9}
\usetikzlibrary{shapes.geometric, arrows.meta}
\usetikzlibrary{backgrounds}
\usepackage{tabularx}
\usepackage{pdflscape}
\usepackage{amsmath}
\usepackage{multirow}
\usepackage{adjustbox}
\usepackage{xcolor}
\usepackage{rotating}
\usepackage[hidelinks]{hyperref}
\usepackage{booktabs} 
\usepackage{soul}
\usepackage{subcaption}
\usepackage{float}
\usepackage{pifont}
\usepackage{algorithm}
\usepackage{algpseudocode}

\usepackage{amssymb}
\usepackage{todonotes}
\journal{Image and Vision Computing}

\begin{document}

\begin{frontmatter}



\title{UNIR-Net: A Novel Approach for Restoring Underwater Images with Non-Uniform Illumination Using Synthetic Data}

\author[add1]{Ezequiel Perez-Zarate}
\ead{isaias.perez@alumnos.udg.mx}

\author[add1]{Chunxiao Liu}
\ead{cxliu@zjgsu.edu.cn}

\author[add2]{Oscar Ramos-Soto}
\ead{oscar.ramos9279@alumnos.udg.mx}

\author[add2]{Diego Oliva\corref{cor1}}
\cortext[cor1]{Corresponding author}
\ead{diego.oliva@cucei.udg.mx}

\author[add2]{Marco Perez-Cisneros}
\ead{marco.perez@cucei.udg.mx}

\affiliation[add1]{
    organization={School of Computer Science and Technology, Zhejiang Gongshang University},
    city={Hangzhou},
    postcode={310018},
    state={Zhejiang},
    country={China}
}

\affiliation[add2]{
    organization={Departamento de Ingeniería Electro-fotónica, Universidad de Guadalajara, CUCEI},
    addressline={Av. Revolución 1500},
    city={Guadalajara},
    postcode={44430},
    state={Jal.},
    country={México}
}

\begin{abstract}
Restoring underwater images affected by non-uniform illumination (NUI) is essential to improve visual quality and usability in marine applications. Conventional methods often fall short in handling complex illumination patterns, while learning-based approaches face challenges due to the lack of targeted datasets. To address these limitations, the Underwater Non-uniform Illumination Restoration Network (UNIR-Net) is proposed. UNIR-Net integrates multiple components, including illumination enhancement, attention mechanisms, visual refinement, and contrast correction, to effectively restore underwater images affected by NUI. In addition, the Paired Underwater Non-uniform Illumination (PUNI) dataset is introduced, specifically designed for training and evaluating models under NUI conditions. Experimental results on PUNI and the large-scale real-world Non-Uniform Illumination Dataset (NUID) show that UNIR-Net achieves superior performance in both quantitative metrics and visual outcomes. UNIR-Net also improves downstream tasks such as underwater semantic segmentation, highlighting its practical relevance. The code is available at \href{https://github.com/xingyumex/UNIR-Net}{https://github.com/xingyumex/UNIR-Net}.
\end{abstract}


\end{frontmatter}

%\linenumbers

\section{Introduction}
Underwater images are captured under non-uniform illumination conditions, which present significant challenges due to the absorption and scattering of light in submarine environments. This phenomenon leads to loss of detail, low visibility, and a marked degradation of contrast, affecting visual quality and hindering accurate interpretation of the images \cite{cong2024comprehensive}. As depth increases, natural light diminishes drastically, making the use of artificial illumination essential. However, these light sources often produce uneven illumination, concentrating light in specific areas of the image. This results in overexposed zones at the center and underexposed regions at the periphery, causing the loss of critical details and complicating information recovery for detailed analysis.

Conventional and Learning-Based methods have been developed to enhance underwater images; however, most of them focus on general problems, such as contrast enhancement \cite{zhang2022underwater}, color correction \cite{li2023tctl}, underwater haze removal, or blur reduction \cite{shen2023udaformer}. Nonetheless, few studies have specifically addressed the unique challenges associated with non-uniform illumination in underwater environments. Recent methods \cite{marques2020l2uwe, zhang2023framework, hou2023non, zhao2024psnet,ma2025pqgal} have started to focus on this issue, aiming to simultaneously correct overexposed and underexposed regions without introducing additional distortions.

Building upon these efforts, advancements in applications such as Autonomous Underwater Vehicles (AUVs) \cite{sahoo2019advancements, hasan2024oceanic} and Remotely Operated Vehicles (ROVs) \cite{he2024introduction} have been instrumental in the development of new methods to address non-uniform illumination in underwater images. These innovations are crucial not only to overcome challenges in ocean exploration but also to enhance precision in tasks such as object detection\cite{xu2023systematic} and semantic segmentation\cite{abdullah2024caveseg} in underwater environments.

\subsection{Motivation}
The correction of non-uniform illumination in underwater images poses a persistent challenge due to the complex optical properties of the aquatic environment. Factors such as spatially varying light attenuation, strong scattering, and wavelength-dependent absorption lead to uneven exposure, color distortion, and the loss of important structural details.~\cite{ji2024dual} These challenges are especially pronounced in real-world underwater scenarios, where lighting conditions can change drastically within a single image and standard enhancement techniques fall short.

\begin{figure*}[ht]
\newlength{\Mheight}
\setlength{\Mheight}{2.4cm}
	\centering
	\begin{subfigure}{0.20\linewidth}
		\centering
		\includegraphics[width=\linewidth, height=\Mheight]{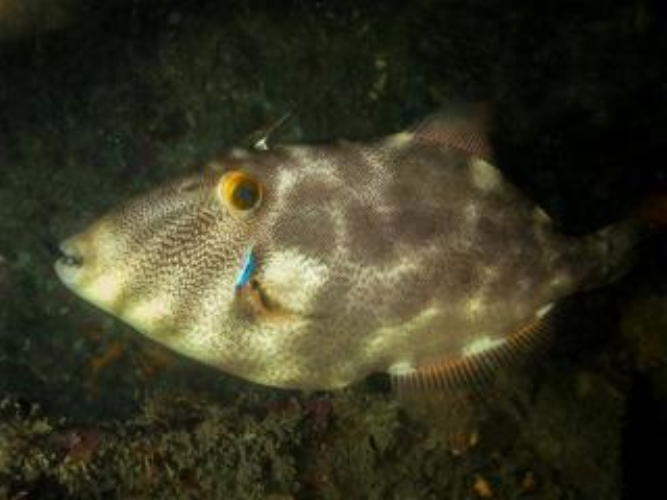} 
        \includegraphics[width=\linewidth, height=\Mheight]{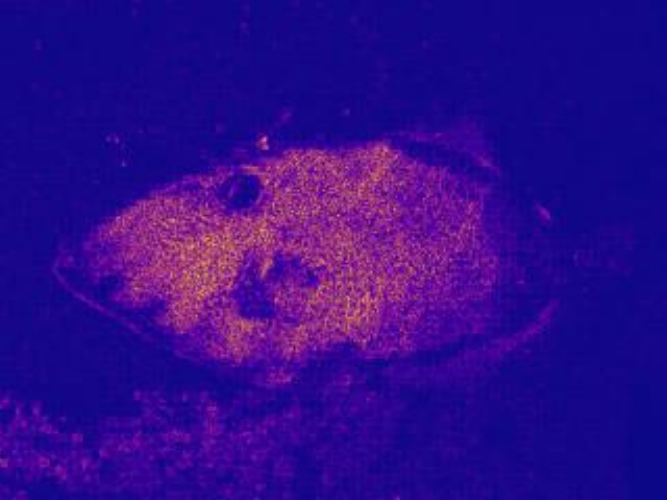} 
        \caption{\footnotesize NUI image} 
	\end{subfigure}
	\begin{subfigure}{0.20\linewidth}
		\centering
		\includegraphics[width=\linewidth, height=\Mheight]{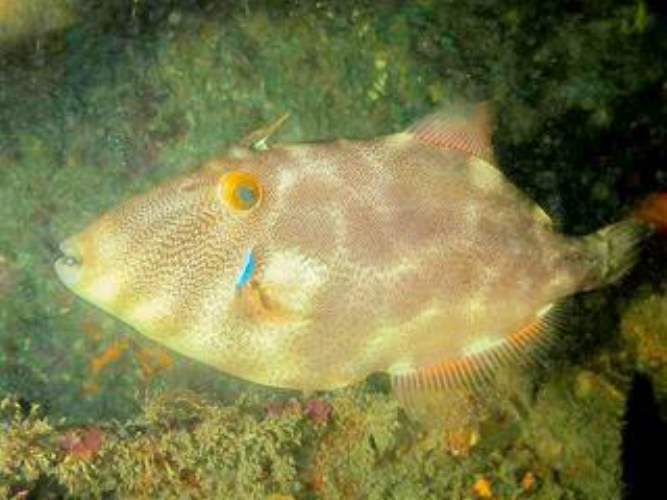}
        \includegraphics[width=\linewidth, height=\Mheight]{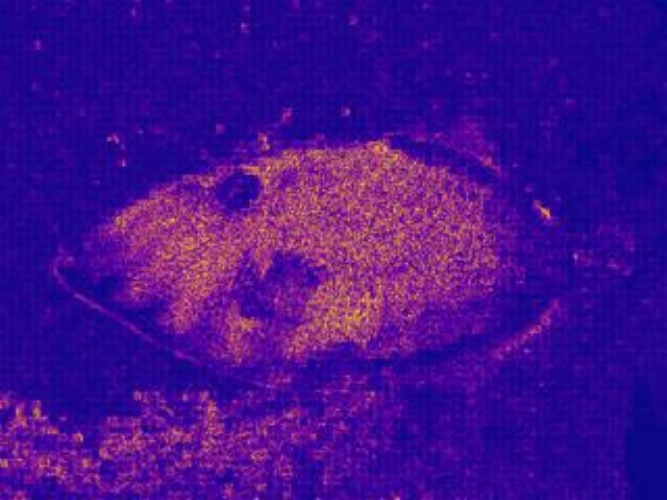} 
        \caption{\footnotesize GCP}
	\end{subfigure}
	\begin{subfigure}{0.20\linewidth}
		\centering
		\includegraphics[width=\linewidth, height=\Mheight]{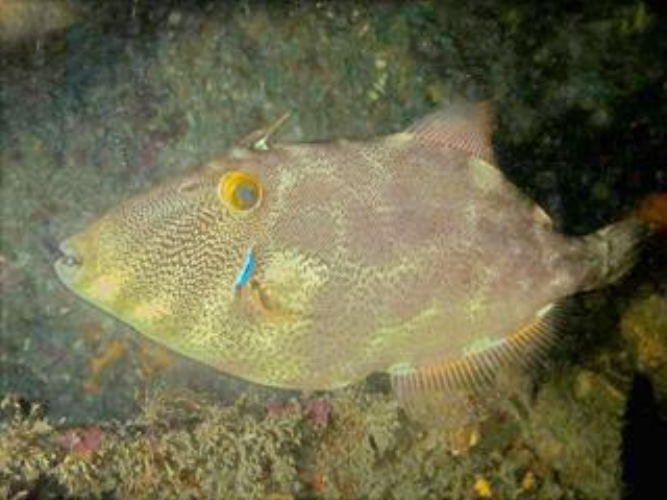}
        \includegraphics[width=\linewidth, height=\Mheight]{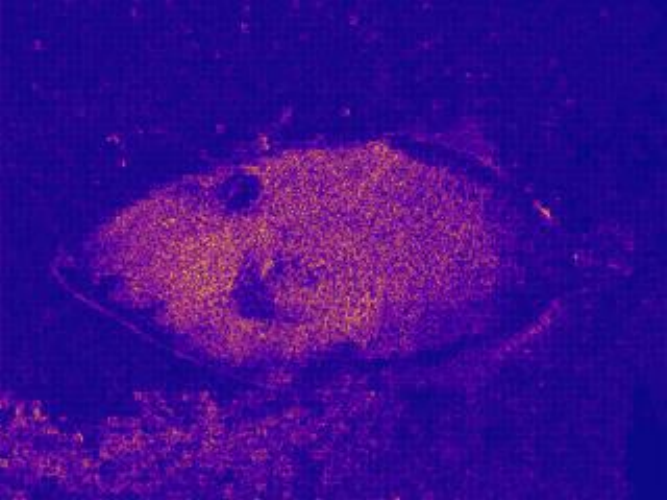} 
        \caption{\footnotesize EIB-FNDL}
	\end{subfigure}

	\begin{subfigure}{0.20\linewidth}
		\centering
		\includegraphics[width=\linewidth, height=\Mheight]{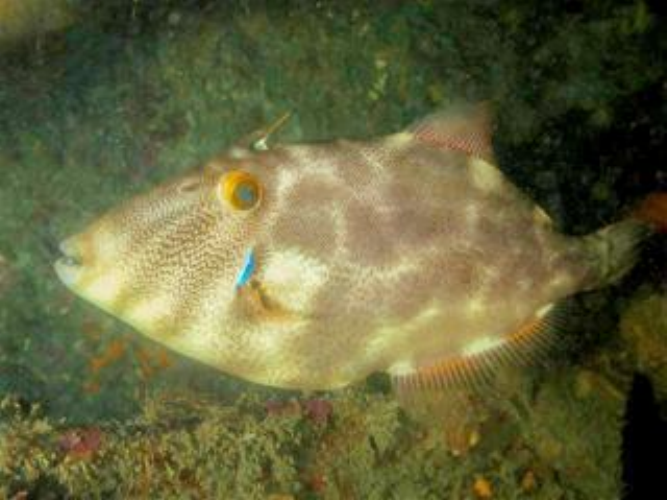} 
        \includegraphics[width=\linewidth, height=\Mheight]{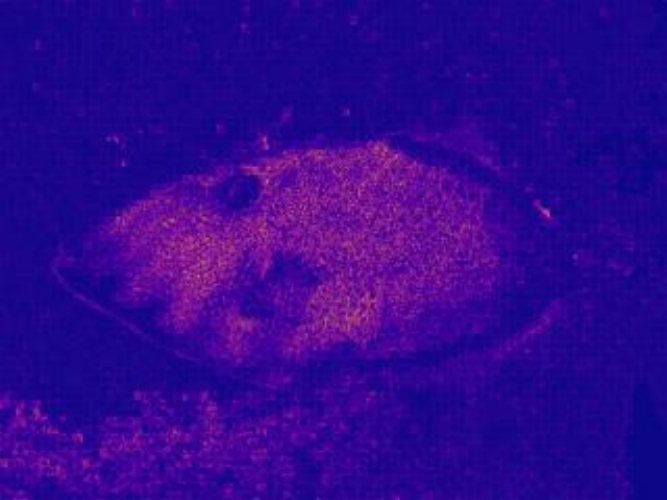} 
		\caption{\footnotesize ALEN}
	\end{subfigure}
    \begin{subfigure}{0.20\linewidth}
		\centering
		\includegraphics[width=\linewidth, height=\Mheight]{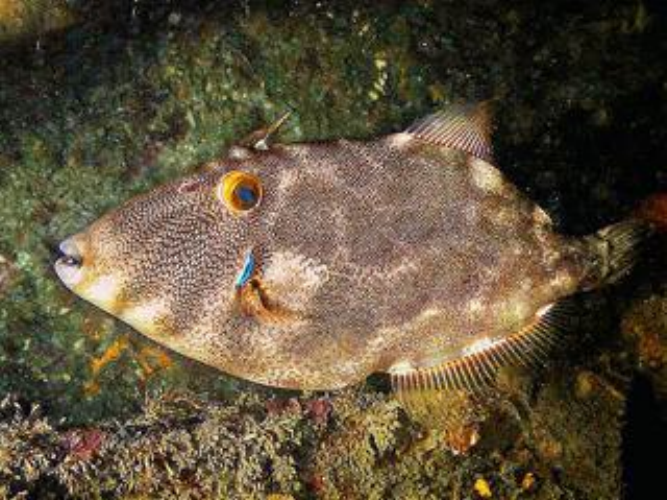}\\
        \includegraphics[width=\linewidth, height=\Mheight]{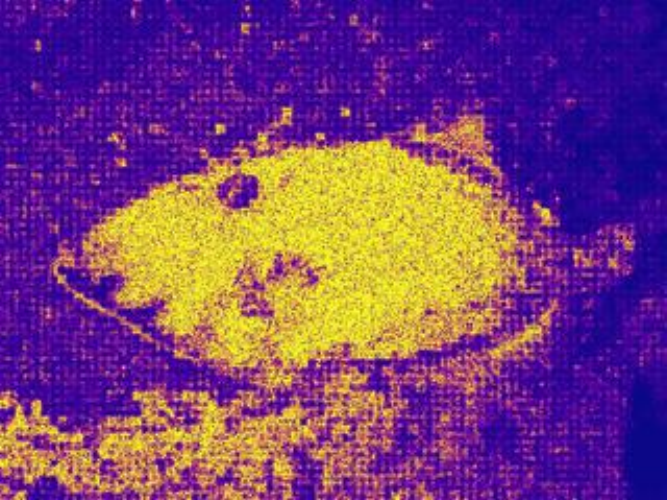} 
		\caption{\footnotesize UNIR-Net}
	\end{subfigure}
    
	\caption{Visual comparison of enhancement methods on an underwater image affected by non-uniform illumination (NUI). The figure includes the input NUI image followed by results from Gamma Correction Prior (GCP), Enhancing Image Brightness - Fast No-reference Deep Learning (EIB-FNDL), Adaptive Light Enhancement Network (ALEN), and the underwater-specific UNIR-Net. For each method, both the enhanced result and its corresponding contrast map are shown. }
	\label{IN_AIR_SOTA}
\end{figure*}

Although significant progress has been made through both conventional and deep learning approaches, existing methods often introduce haze, produce unnatural colors, or fail to preserve fine textures. Techniques originally developed for terrestrial low-light enhancement \cite{perez2023loli, jeon2024low, khajehvandi2025enhancing, perez2025alen} typically assume uniform degradation across the image and do not consider the particularities of underwater environments. When applied to underwater images, these methods frequently generate greenish tones and blurred results due to their inability to handle selective color absorption and complex light scattering, as evidenced in Fig. \ref{IN_AIR_SOTA}.

To overcome these limitations, the present study aims to develop a deep learning-based enhancement method specifically tailored for underwater images affected by non-uniform illumination. UNIR-Net integrates domain-specific knowledge to guide luminance correction and chromatic adaptation, selectively enhancing poorly lit regions while preserving well-exposed areas. Unlike existing methods, it explicitly models the spatial variation in lighting and accounts for wavelength-dependent distortion, enabling more accurate and perceptually consistent results. This design allows the method to better preserve structural detail, correct color bias, and avoid common artifacts such as haze and oversaturation, thus offering a robust and original solution to a persistent problem in underwater image enhancement.

\subsection{Contributions}
This study presents the Underwater Non-uniform Illumination Restoration Network (UNIR-Net), a method specifically designed to enhance underwater images captured under spatially uneven lighting conditions. Unlike previous approaches developed for terrestrial low-light scenarios or uniformly lit underwater scenes, UNIR-Net is tailored to address the complex visual degradations caused by non-uniform illumination in underwater environments. The main contributions of this research are summarized as follows:

\begin{itemize}
\item \textbf{An approach for enhancing underwater images with non-uniform illumination:} UNIR-Net is proposed as a dedicated deep learning model that integrates blocks for illumination enhancement, attention mechanisms, and modules for visual refinement and contrast correction. This architectural design enables the model to adaptively correct uneven lighting while preserving structural details and enhancing visual clarity.

\item \textbf{A novel synthetic paired dataset:}  The Paired Underwater Non-uniform Illumination (PUNI) dataset is introduced to address the lack of paired training data representing underwater scenes with spatially varying illumination. PUNI provides a controlled environment for supervised learning and benchmarking in this specific context.

\item \textbf{Validation in practical underwater applications:}  UNIR-Net is evaluated not only on enhancement quality but also in the context of downstream tasks. Specifically, its application as a preprocessing step in semantic segmentation demonstrates tangible improvements in the accuracy and robustness of vision-based analysis in real underwater scenarios.
\end{itemize}

The remainder of the sections are organized as follows: Section \ref{rewo} presents the related work. Section \ref{prometh} introduces the materials and methods. Section \ref{expdis} presents the experimental setup and analyzes the results. Finally, Section \ref{conlusions} discusses some conclusions.

\section{Related Work}
\label{rewo}

In this section, the most relevant previous approaches in underwater image enhancement are presented, focusing on conventional methods and learning-based methods, highlighting their strengths, limitations, and key contributions.

\subsection{Conventional Methods}
Conventional methods refer to traditional image enhancement techniques that do not rely on learning from data. These include model-based approaches grounded in physical principles of image formation, as well as heuristic methods based on algorithmic rules, illumination priors, or statistical assumptions.

In 2016, Guo et al.~\cite{guo2016lime} proposed a Low-light Image Enhancement Method (LIME) that estimates the illumination of each pixel by using the maximum value of the R, G, and B channels. The illumination map is subsequently refined by applying a structural prior. Three years later, in 2019, Zhang et al.~\cite{zhang2019dual} presented an automatic exposure correction method that uses a dual estimation of illumination to correct underexposed and overexposed regions. The corrected images are then fused using a multiple-exposure fusion technique to obtain a globally well-exposed image. Later in 2020, Marques and Albu introduced L$^2$UWE~\cite{marques2020l2uwe}, an underwater low-light image enhancer that uses contrast models to estimate illumination. Two enhanced images are fused, emphasizing luminance and local contrast. In the same year, Yuan et al.~\cite{yuan2020underwater} proposed an underwater image enhancement method using Contour Bougie (CB) morphology. It enhances scene contours and visibility through morphological operations with two structuring elements. The images are then normalized and stretched to improve white balance in the RGB channels.

Moving to 2021, Xie et al.~\cite{xie2021variational} proposed a variational framework guided by the red channel to enhance underwater images affected by low contrast, fog, and blur, incorporating forward scattering and blur kernel estimation in the refinement process. In 2022, Zhang et al.~\cite{zhang2022underwater} presented a method for enhancing underwater images by correcting colors, improving contrast, and sharpening details using attenuation matrices, histogram-based techniques, and a multi-scale unsharp mask. In the same year, Zhuang et al.~\cite{zhuang2022underwater} introduced a Retinex variational model with hyper-laplacian reflectance priors for underwater image enhancement. It uses $l_{1/2}$ and $l_2$ norms to enhance fine details and structures while estimating illumination. Similarly, Zhang et al.~\cite{zhang2022underwaterMMLE} proposed MLLE, an underwater image enhancement method that adjusts color, details, contrast, and balances color in the CIELAB space, producing vivid colors and improved contrast.

More recently, in 2023, Hou et al.~\cite{hou2023non} proposed a variational framework with an Illumination Channel Sparsity Prior (ICSP) to restore underwater images, enhancing brightness, correcting color distortions, and highlighting details using a Retinex-based model with $\text{L}_0$ norm constraints. In the same year, Zhang et al.~\cite{zhang2023PCDE} introduced a method for underwater image enhancement using piecewise color correction and dual prior-based contrast enhancement, applying gain-based color correction and decomposing the base and detail layers in the V channel of HSV. Another contribution by Zhang et al.~\cite{zhang2023underwater} presented a technique called WWPF for underwater image enhancement, which corrects color distortion using an attenuation map, improves both global and local contrast, and fuses the images through a wavelet visual perception strategy to produce high-quality results. Fusion-based approaches, such as CCMF~\cite{zhang2023underwatermulti}, combine red channel compensation, guided filtering, and adaptive gamma correction to enhance contrast and color through multi-scale feature fusion. While effective, they require careful weight tuning and struggle under extreme low-light conditions. In 2024, An et al.~\cite{an2024hfm} proposed a Hybrid Fusion Method (HFM) to enhance underwater images, addressing issues such as white balance distortion, color shift, low visibility, and contrast. The process incorporates color correction, visibility recovery, and contrast enhancement through fusion techniques. Additionally, Jeon et al.~\cite{jeon2024low} presented a fast low-light image enhancement method based on an atmospheric scattering model. The transmission map is calculated using the average and maximum values of the original image, which are estimated with gamma correction.

Conventional methods offer the advantage of not relying on extensive datasets and provide interpretable results grounded in theory. However, they face significant limitations when dealing with underwater images exhibiting non-uniform illumination. First, many of these techniques rely on global priors or heuristics that fail to capture the local variations in lighting commonly present in underwater scenes, such as shadows, caustics, and depth-induced color distortions. As a result, enhancement may be uneven, leading to overexposure in some regions and under-enhancement in others. Second, physical-model-based methods often assume simplified or idealized imaging conditions, such as homogeneous water properties or uniform ambient light, which rarely hold true in real-world underwater environments. Third, handcrafted fusion strategies or morphological operations may not adapt well to complex textures or intricate scene content, resulting in artifacts or loss of fine details. Finally, while computationally efficient, these methods generally lack the capacity to adaptively adjust to varying degradation patterns, making them less robust than modern learning-based techniques under diverse or unseen scenarios.

\subsection{Learning-Based Methods}

Learning-based methods have gained significant attention in low-level vision tasks due to their ability to learn complex patterns from large datasets. By leveraging deep/machine learning models, these methods effectively address various image distortions and enhance visual quality across a wide range of tasks. Their capacity to learn from diverse data allows them to generalize well, providing efficient and scalable solutions for real-world image enhancement challenges.

In 2020, Guo et al.~\cite{guo2020zero} introduced Zero-DCE, a method for light enhancement using image-specific curve estimation with a lightweight deep network. It adjusts dynamic range without requiring paired data, using non-reference loss functions for efficient image enhancement across diverse lighting conditions. In 2021, Naik et al.~\cite{naik2021shallow} proposed UWNet, a lightweight neural network architecture for underwater image enhancement, designed to balance effectiveness with reduced computational complexity. The next year, in 2022, Xie et al.~\cite{xie2022lighting} presented a deep learning-based underwater image enhancement network tailored for low-light environments, addressing both low-light degradation and scattering effects. Later, in 2023, Wen et al.~\cite{wen2023syreanet} proposed SyreaNet, a framework for underwater image enhancement that combines synthetic and real data. It uses a revised image formation model, domain adaptation strategies, image synthesis module, and disentangled network to predict more explicit images. In the same year, Li et al.~\cite{li2023tctl} introduced a template-free color transfer learning framework for underwater image enhancement, predicting transfer parameters with attention-driven modules for more flexible and robust enhancement. Similarly, Shen et al.~\cite{shen2023udaformer} proposed UDAformer, a dual attention transformer for underwater image enhancement. It combines channel and pixel self-attention transformers, uses a shifted window technique for efficiency, and restores images with residual connections. In the same year, Zhang et al.~\cite{zhang2023framework}introduce NUIENet, a CNN with an encoder-decoder structure and skip connections to enhance HSV components. While effective, it struggles with over-exposed areas and shows limited generalization in extreme underwater conditions.

More recently, in 2024, Yao et al.~\cite{yao2024gaca} proposed the Gradient-Aware and Contrastive-Adaptive (GACA) framework for low-light image enhancement, using accurate gradient estimation as a structural prior and a novel regularization constraint to address color abnormalities and artifacts. In the same year, Zhang et al.~\cite{zhang2024liteenhancenet} introduced LENet, a lightweight underwater image enhancement network using depthwise separable convolution, one-shot aggregation, and a squeeze-and-excitation module to reduce complexity and improve feature extraction. Zhao et al.~\cite{zhao2024psnet} present PSNet, a pseudo-siamese network that separates NUI from well-illuminated content using iterative refinement and multiple loss functions. Although it improves illumination and color fidelity, it fails under extreme color distortion due to its illumination-focused design. In addition, Zhang et al.~\cite{zhang2024synergistic} proposed SMDR-IS, a method for enhancing underwater details using multi-scale refinement. It employs the Adaptive Selective Intrinsic Supervised Feature (ASISF) module for detail propagation and the Bifocal Intrinsic-Context Attention (BICA) module in the encoder-decoder framework to improve spatial context and restoration. Park and Eom~\cite{park2024underwater} proposed a method with an adaptive standardization network to correct distortion and a normalization network using squeeze-and-excitation blocks to enhance contrast, remove haze, and restore brightness. Alternatively, DeMBFF-Net~\cite{qu2024denoising} follows a decomposition-estimation-reconstruction scheme, using back-projection fusion and denoising to restore degraded images. It generalizes well across degradation types but depends on complex training and remains challenged by severe low-light scenarios.

In 2025, Ma et al.~\cite{ma2025pqgal} proposed PQGAL-Net, a GAN-based model for correcting NUI in underwater images. It balances brightness, contrast, and saturation using a quality-aware module and is trained on the NUIPUI dataset. However, it performs poorly in severely distorted scenes and depends heavily on large training datasets.  In contrast, Du et al.~\cite{du2025uiedp} proposed UIEDP, a diffusion-based framework guided by scene priors for underwater image enhancement. By leveraging a pre-trained diffusion model and integrating existing enhancement algorithms, it addresses the limitations of synthetic training data and improves both illumination and clarity. Nevertheless, its reliance on multiple sampling steps hinders real-time use, and results may vary depending on the guiding algorithm.

Moreover, recent works in related domains have demonstrated the potential of structured priors and multi-task learning to improve generalization and robustness. For instance, Chen et al. \cite{yao2024between} proposed a tensor-based method to complete missing views in multi-view clustering, while Zhang et al. \cite{cai2025focus} introduced a guided multi-task training framework for person search. Although these approaches address different tasks, they reflect a growing trend of exploiting multi-view and multi-task information, which could inspire future directions in underwater image enhancement by incorporating complementary structural cues or auxiliary objectives.

Despite the progress achieved by recent learning-based methods, several limitations persist, particularly in dealing with underwater scenes affected by non-uniform illumination. Many models focus predominantly on global enhancement and often fail to capture localized degradation, resulting in suboptimal correction in areas with severe lighting imbalance. Additionally, some methods are heavily dependent on synthetic datasets or paired supervision, which may not fully represent the complexity of real underwater environments. This leads to poor generalization in the presence of domain shift, color bias, and varying visibility conditions. Furthermore, lightweight models often sacrifice restoration accuracy for efficiency, while more powerful architectures tend to require significant computational resources, limiting their practical deployment on edge devices or in real-time applications. Addressing these limitations remains an open challenge, motivating the development of more adaptive and illumination-aware frameworks capable of enhancing images with complex and spatially varying degradation patterns.

\section{Materials and Methods}
\label{prometh}

\subsection{Underwater Image Generation Dataset}

\begin{figure*}[th]
  \centering
  \includegraphics[width=1\textwidth]{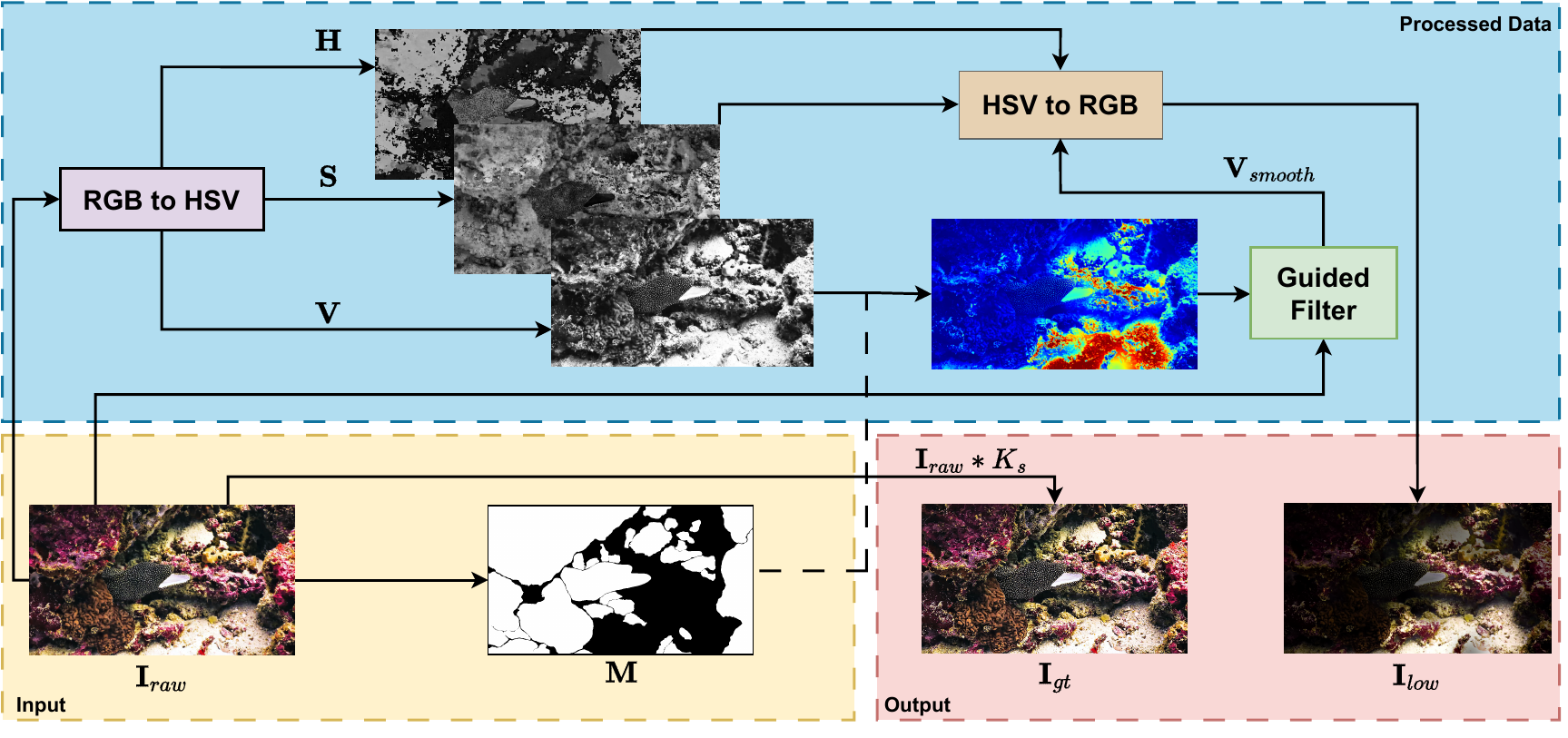}
  \caption{Image synthesis process for the PUNI dataset with non-uniform illumination.}
  \label{PUNIDS}
\end{figure*}

This section introduces the Paired Underwater Non-uniform Illumination (PUNI) dataset, designed for training and evaluating enhancement techniques under non-uniform illumination. The selection of raw underwater images is detailed in section \ref{UIS}, while the synthesis process for generating non-uniform illumination is explained in section \ref{S_NUI}. The diagram in Fig.\ref{PUNIDS} corresponds to the synthesis process described in section \ref{S_NUI}.

\subsubsection{Underwater Image Selection}
\label{UIS}
The enhancement of underwater images has been extensively studied through benchmarks like Enhanced Underwater Visual Perception (EUVP)~\cite{islam2020fast}, Underwater Image Enhancement Benchmark (UIEB)~\cite{li2019underwater}, and Low-light Underwater Image Enhancement (LUIE)~\cite{xie2022lighting}, which aim to address challenges related to contrast, color fidelity, and illumination in low-light scenarios. Despite their contributions, these approaches face notable drawbacks in terms of preserving resolution and generating high-quality outputs. The raw images provided by existing datasets frequently exhibit inherent flaws, including insufficient detail, color distortions, blurring artifacts, and uneven lighting. Additionally, the presence of fog-like effects often obscures essential visual information. Such limitations not only hinder the datasets’ utility for producing synthetic data with non-uniform illumination but may also introduce undesirable biases in training pipelines, ultimately compromising the performance and generalization of enhancement models.

\begin{figure*}[htbp]
	\newlength{\SDheight}
	\setlength{\SDheight}{2.4cm}
	\centering
	\begin{subfigure}{0.24\linewidth}
		\centering
		\includegraphics[width=\linewidth, height=\SDheight]{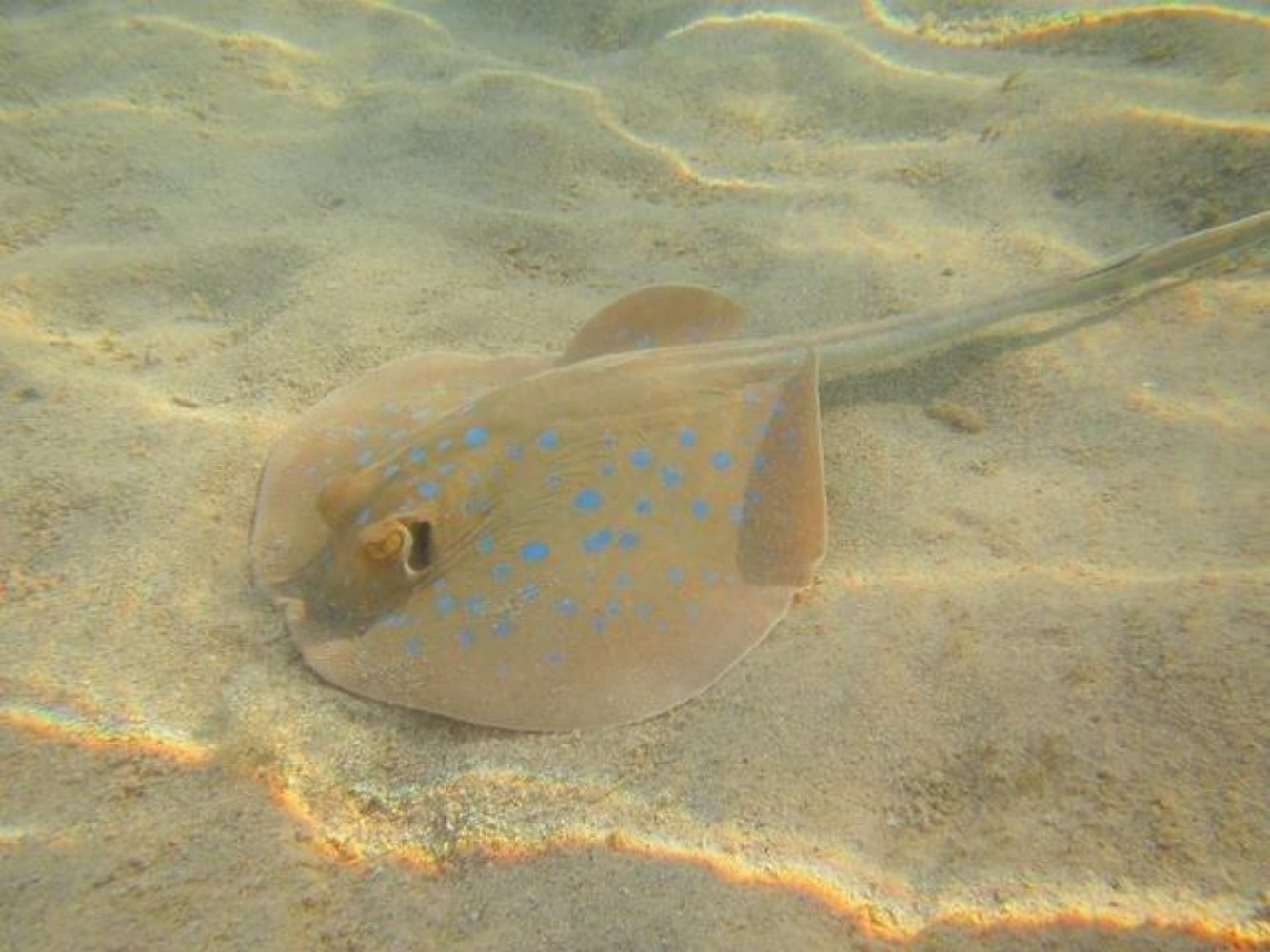} 
        \includegraphics[width=\linewidth]{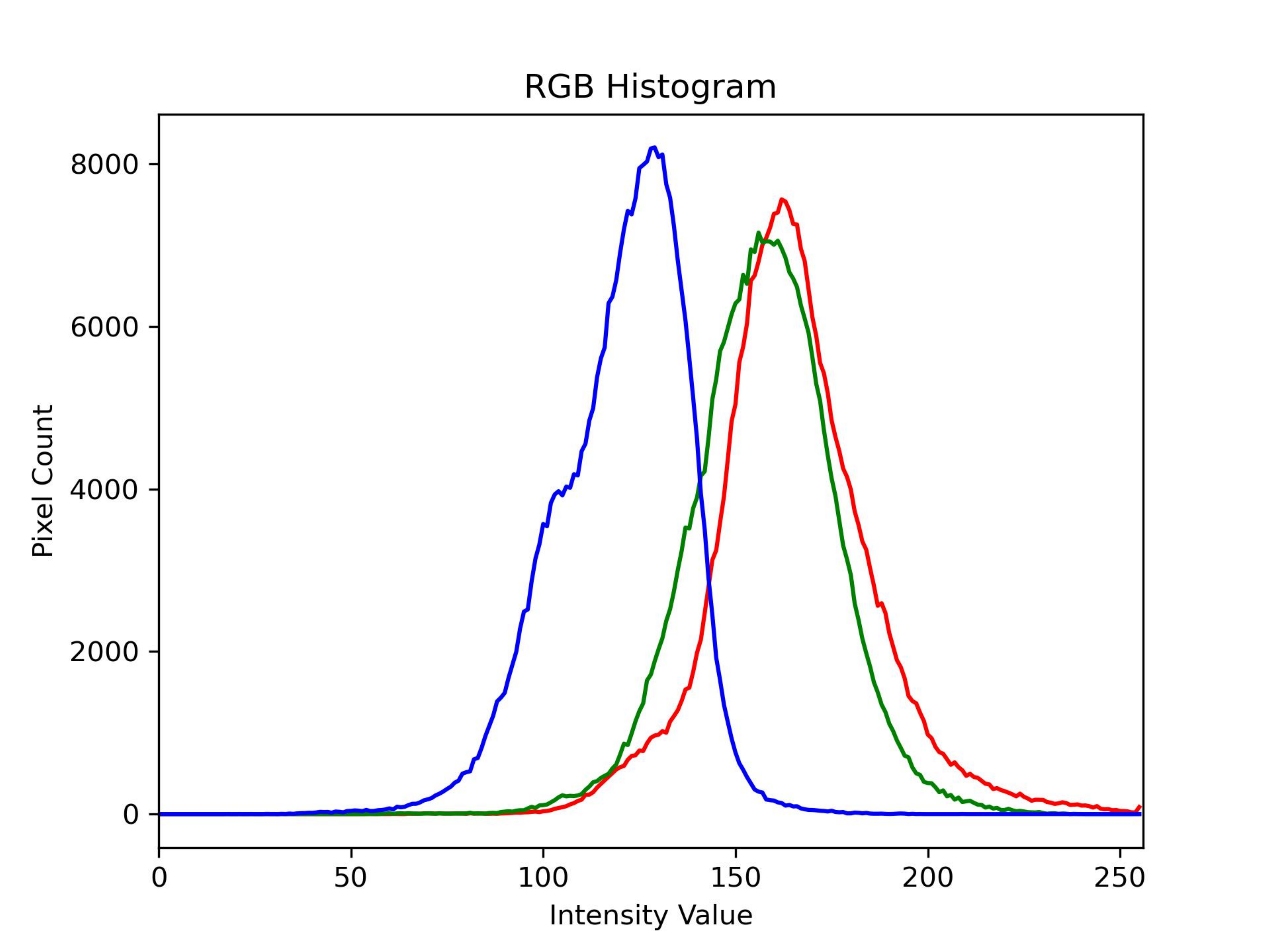} 
        \caption{\footnotesize EUVP raw} 
        \label{EUVP_raw}
	\end{subfigure}
	\begin{subfigure}{0.24\linewidth}
		\centering
		\includegraphics[width=\linewidth, height=\SDheight]{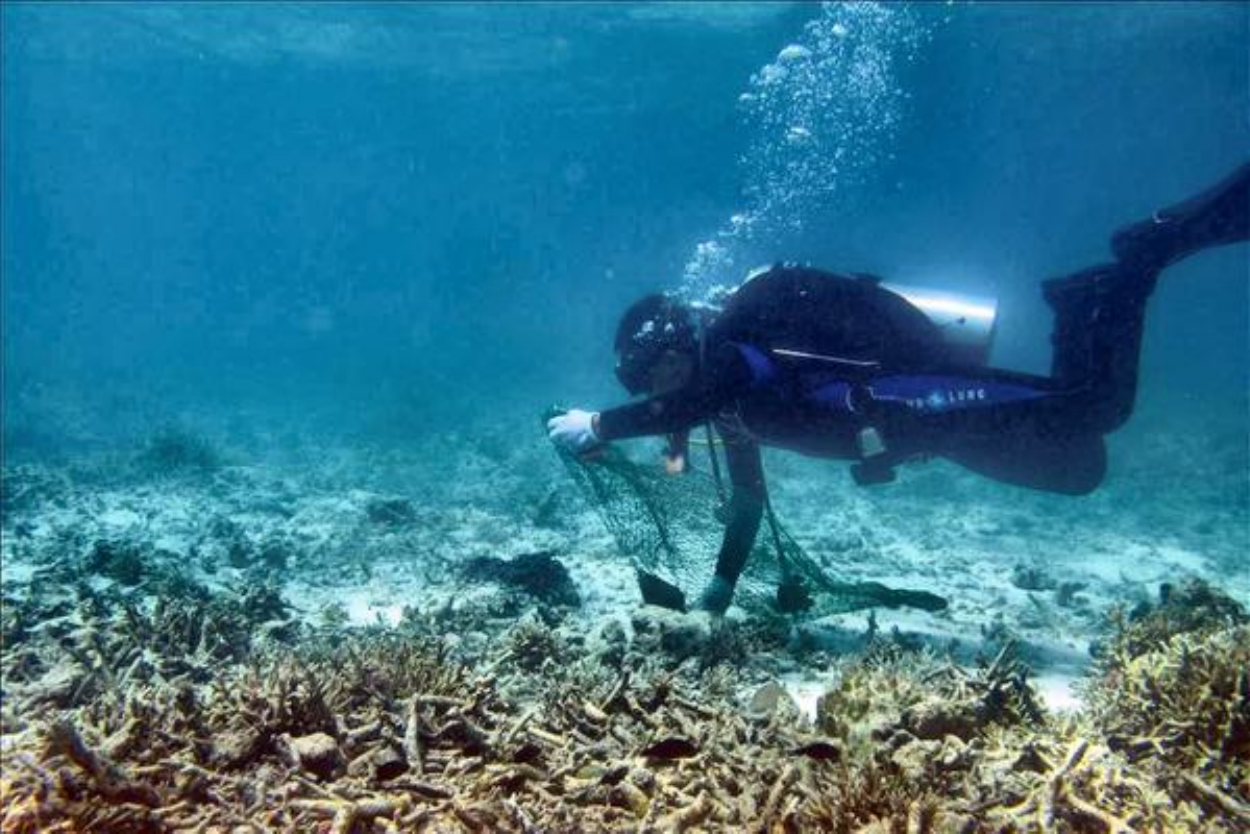}
        \includegraphics[width=\linewidth]{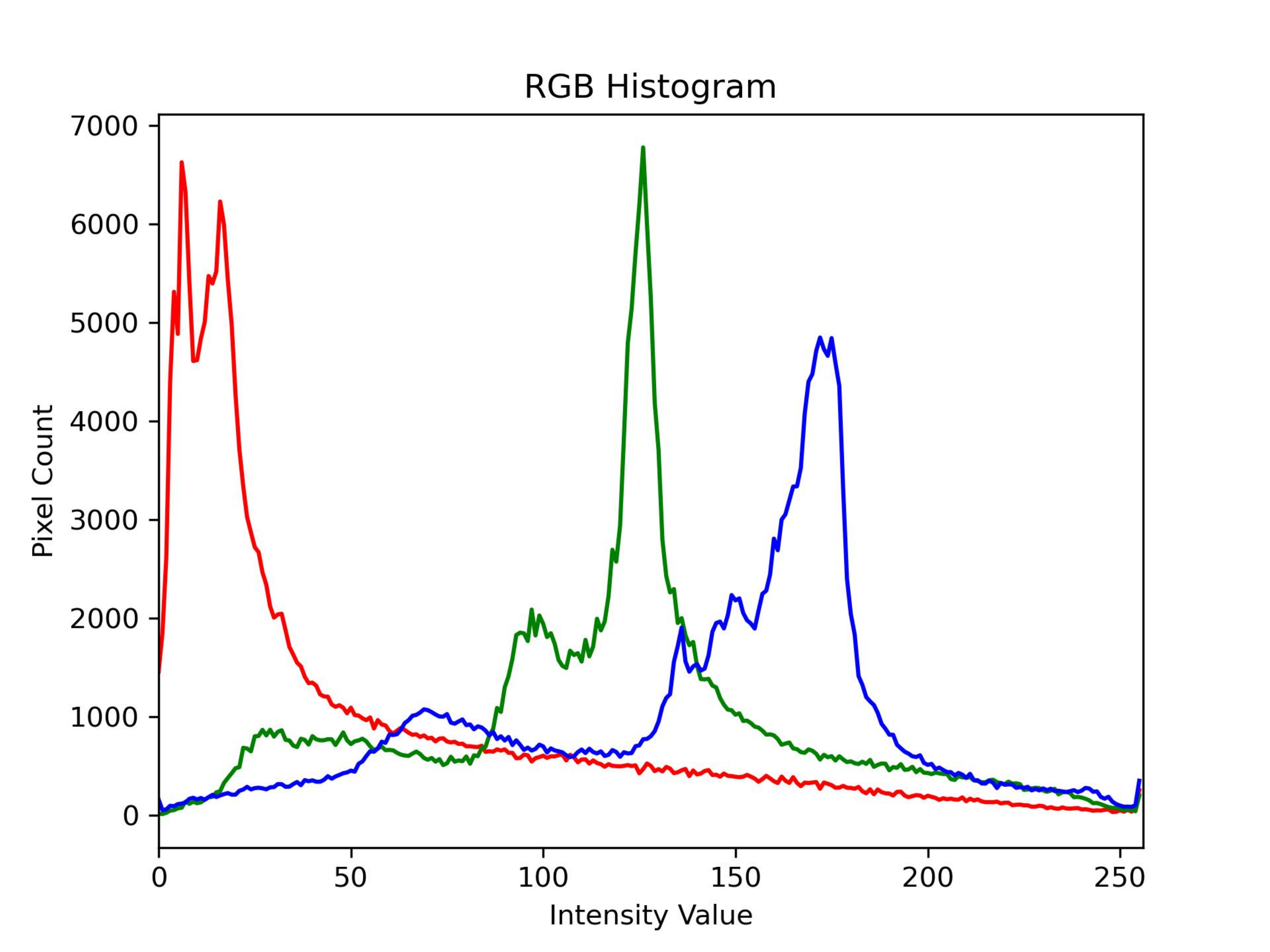} 
        \caption{\footnotesize UIEB raw}
        \label{UIEB_raw}
	\end{subfigure}
	\begin{subfigure}{0.24\linewidth}
		\centering
		\includegraphics[width=\linewidth, height=\SDheight]{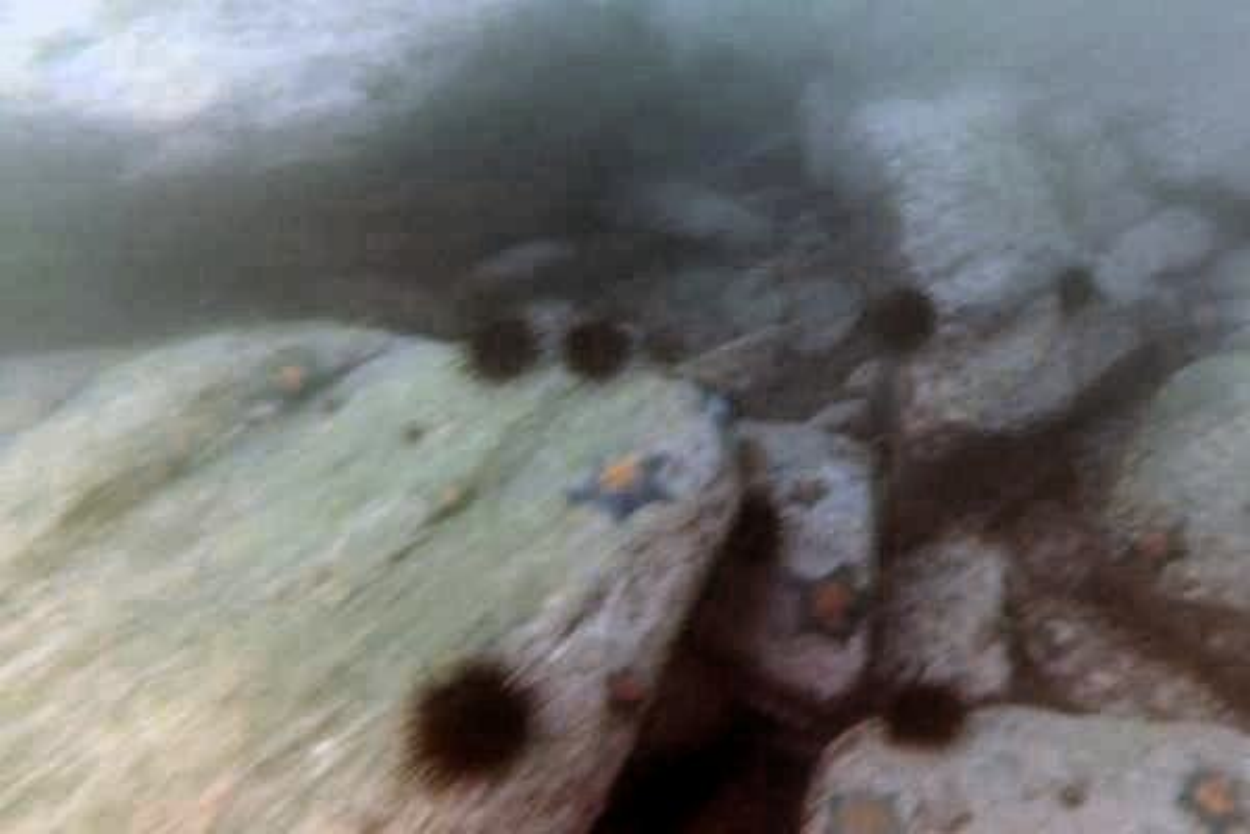}
        \includegraphics[width=\linewidth]{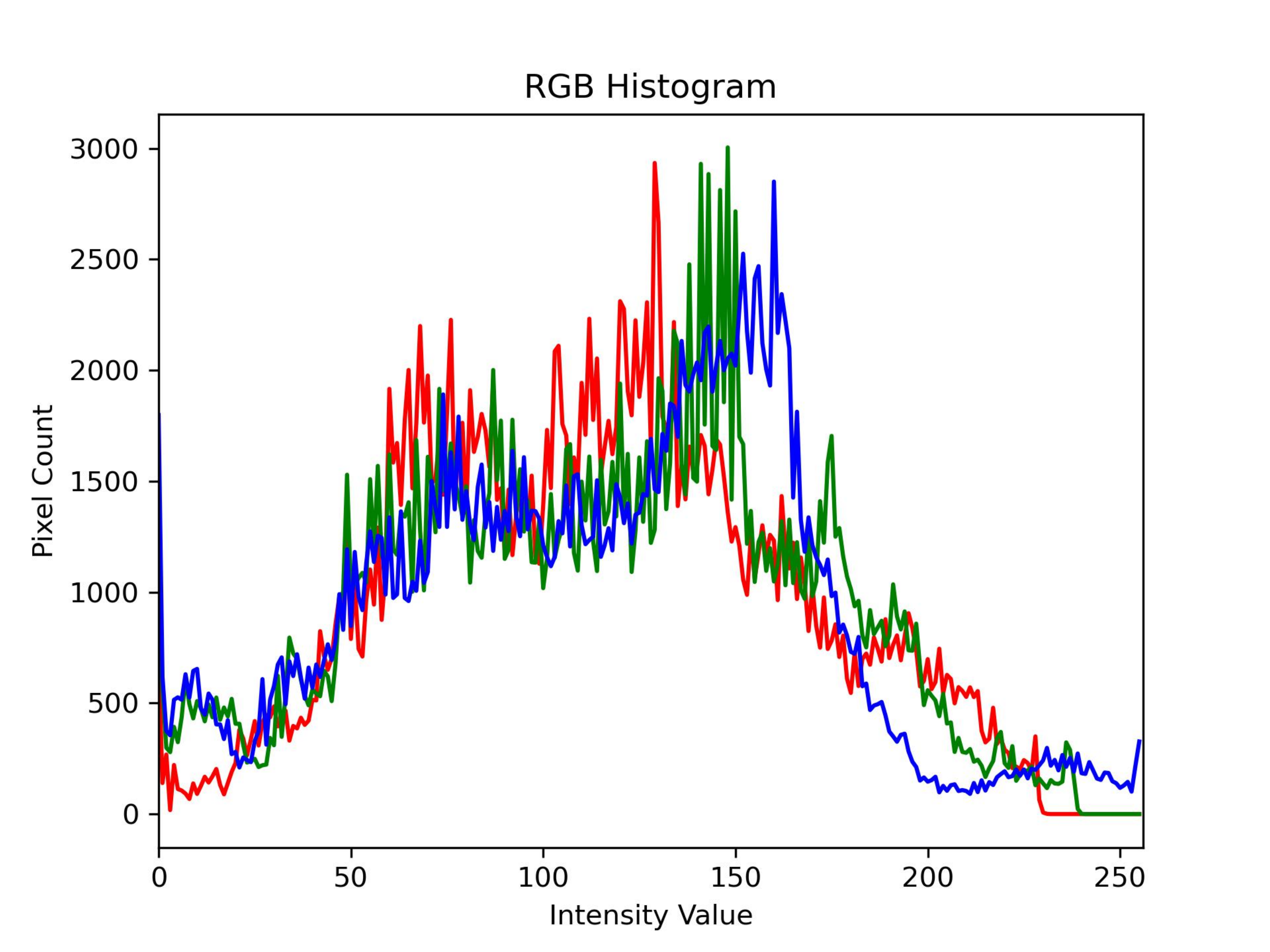} 
        \caption{\footnotesize LUIE raw}
        \label{LUIE_raw}
	\end{subfigure}
	\begin{subfigure}{0.24\linewidth}
		\centering
		\includegraphics[width=\linewidth, height=\SDheight]{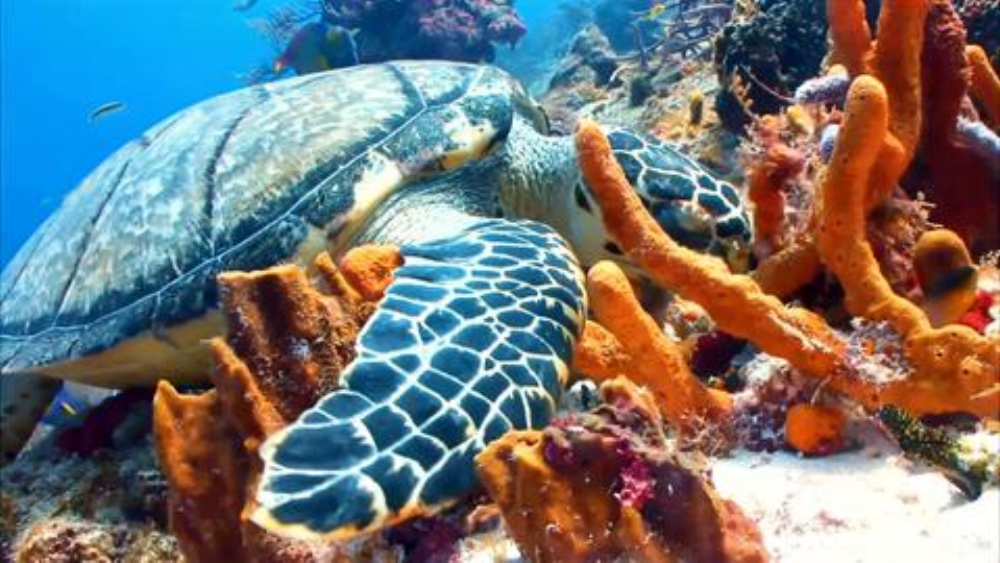} 
        \includegraphics[width=\linewidth]{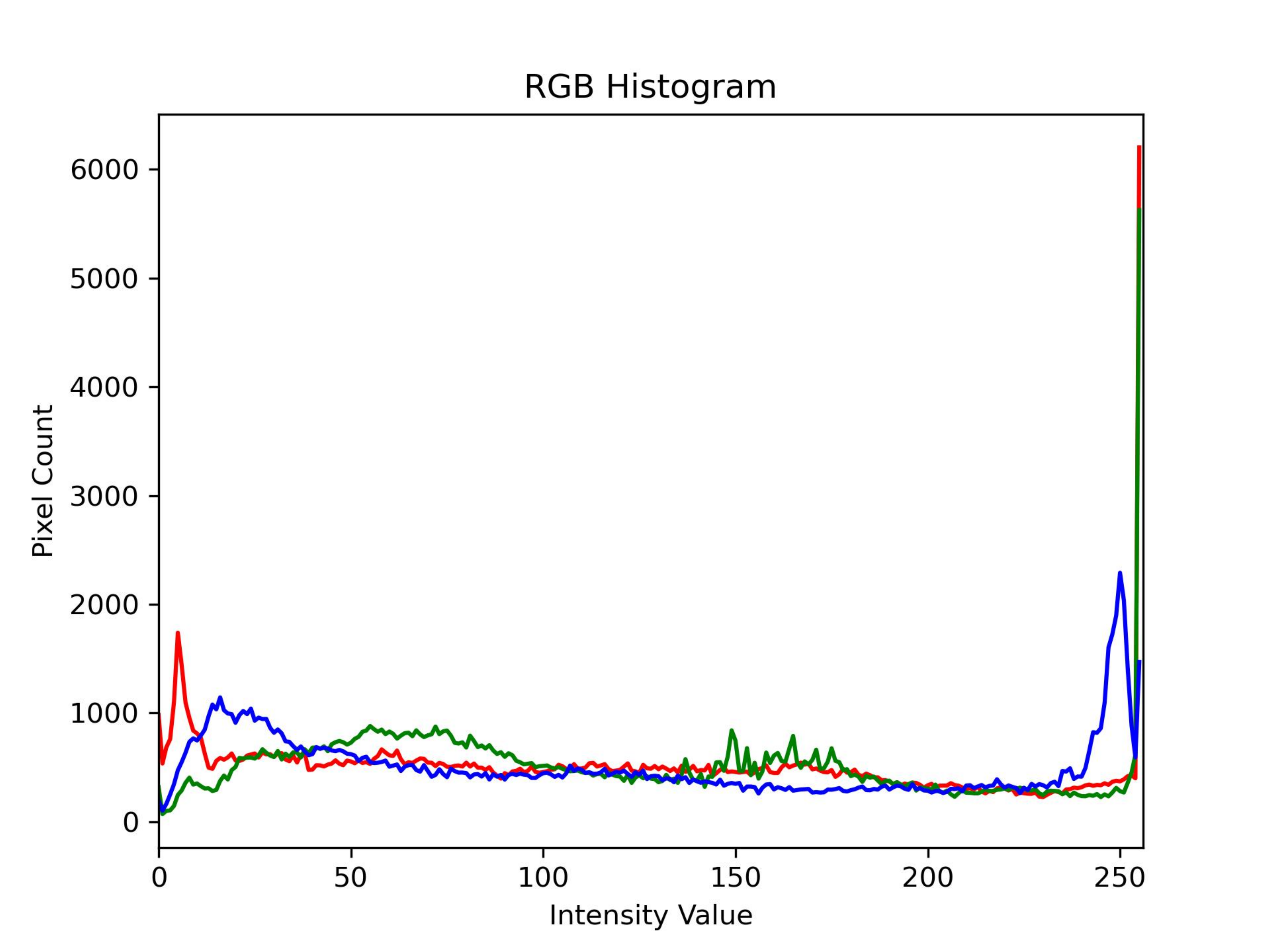} 
		\caption{\footnotesize PUNI raw}
        \label{PUNI_raw}
	\end{subfigure}
\caption{Comparative analysis of various underwater datasets in raw images, including their respective histograms. The comparisons are made using the EUVP, UIEB, LUIE, and PUNI datasets without any synthetic processing.}
\label{DATASETS_SOTA}
\end{figure*}

Figure \ref{DATASETS_SOTA} shows representative examples from these datasets. For instance, Fig. \ref{EUVP_raw}, taken from the EUVP dataset, reveals that most pixel intensities in its RGB histogram are concentrated in the center, resulting in a predominant fog effect in the image. In the case of Fig. \ref{UIEB_raw}, corresponding to the UIEB dataset, most pixel intensities shift towards the left side in the red channel, while in the green and blue channels, they concentrate in the center. Additionally, this image exhibits low-quality textures with noticeable noise in certain areas. Lastly, the image from the LUIE dataset shown in Fig. \ref{LUIE_raw} displays a motion effect that affects its clarity. In its RGB histogram, pixel intensities are primarily distributed in the center, reflecting visual quality issues. In contrast, most of the images collected for creating the Paired Underwater Non-uniform Illumination (PUNI) dataset present a more uniform distribution of pixel intensities in the histogram. This results in better visual quality in terms of texture and lighting, as shown in Fig. \ref{PUNI_raw}.

The limitations present in most raw images from the EUVP, UIEB, and LUIE underwater datasets underscore the need to develop a new, higher-quality raw dataset that can better synthesize the challenges associated with non-uniform illumination in underwater environments. To address this, 3,277 high-resolution, royalty-free images were carefully collected from a human perspective using platforms like Unsplash and Freepik. These images were used solely for academic and research purposes, in compliance with the respective license terms, with no commercial intent or redistribution. Images from Unsplash were used under their license, which allows free use of all images, including for commercial and research purposes, without the need for permission or attribution. Similarly, images from Freepik were accessed via a Premium account, allowing their use in educational contexts without attribution, in accordance with Freepik's licensing terms. The images were selected based on strict criteria, including color accuracy, absence of blur, proper lighting, and no fog. This set of high-quality images serves as the foundation for the development of the new PUNI dataset.

\subsubsection{Synthesis with Non-Uniform Illumination}
\label{S_NUI}

Non-uniform illumination in underwater environments is a distinctive feature caused by the absorption and scattering of water light, preventing it from propagating evenly. In this context, paired terrestrial datasets commonly used for low-light image enhancement~\cite{wei2018deep, hai2023r2rnet, bychkovsky2011learning} are ineffective, as they fail to account for the absorption and scattering phenomena inherent to aquatic environments. Consequently, training models with these datasets leads to undesired effects in the images, such as haze and inaccurate color tones.

Synthesizing a dataset of low-light images with non-uniform illumination has been previously explored to train models for enhancing terrestrial images under poor lighting conditions \cite{liu2022efinet}. However, this approach proves inadequate for underwater images due to the inherent differences between terrestrial and aquatic environments. A notable aspect of the synthesis method in \cite{liu2022efinet} is the application of superpixel-based segmentation \cite{wu2021texture} to divide images into regions and simulate illumination variations within these segments.

In contrast, this work adopts the Segment Anything Model (SAM) \cite{kirillov2023segment}, an advanced method capable of automatically identifying and segmenting objects in images. SAM enables the generation of binary masks from pre-collected raw underwater images, as depicted in Fig. \ref{PUNI_raw}, to construct an initial underwater dataset. These binary masks allow selective attenuation of specific image areas, effectively replicating the non-uniform illumination conditions characteristic of underwater environments. The complete synthesis process for generating underwater images with non-uniform illumination is outlined in \textbf{Algorithm~\ref{PUNI_Synthesis}}.

\begin{algorithm}
\caption{Underwater Image Synthesis with Non-Uniform Illumination}
\label{PUNI_Synthesis}
\begin{algorithmic}[1]
\Require RGB image raw $\mathbf{I}_{raw}$; Binary mask $\mathbf{M}$; Brightness threshold $\mathbf{\alpha}$, Guided Filter Kernel $K_{g}$
\Ensure Synthetic non-uniform illumination image $\mathbf{I}_{low}$
\State Convert $\mathbf{I}_{raw}$ $\rightarrow$ $\mathbf{I}_{HSV}$
\State Extract $\mathbf{V}$ channel from $\mathbf{I}_{HSV}$
\State Average brightness $\mathbf{B}_{average}$ of the masked area $\mathbf{V}_{masked}$
\If{$\mathbf{B}_{average}$ $< \alpha$}
     \State Apply "$\mathbf{Deepen}$" adjustment to the masked area $\mathbf{V}_{masked}$
     \State $\mathbf{V}^\lambda \leftarrow Adjust\;Brightness(\mathbf{V}, "Deepen")$
\Else
     \State Randomly choice apply "$\mathbf{Surface}$", "$\mathbf{Deepen}$", or "$\mathbf{None}$" adjustment to the masked area $\mathbf{V}_{masked}$
     \State $\mathbf{V}^\lambda \leftarrow Adjust\;Brightness(\mathbf{V}, "Random\;Choice")$
\EndIf
\State Compute $\mathbf{V}_{smooth} \leftarrow Guided\;Filter (\mathbf{I}_{raw}, \mathbf{V}^{\lambda}, K_{g})$
\State Update $\mathbf{I}_{HSV} \leftarrow (\mathbf{H}, \mathbf{S}, \mathbf{V}_{smooth})$
\State Convert $\mathbf{I}_{HSV}$ $\rightarrow$  $\mathbf{I}_{low}$ \\
\Return $\mathbf{I}_{low}$
\end{algorithmic}
\end{algorithm}

The process starts with an input image represented as $\mathbf{I}_{raw} \in \mathbb{R}^{H \times W \times 3}$, where $H$ and $W$ are the dimensions of the RGB image. A binary mask $\mathbf{M} \in \{0,1\}^{H \times W}$  is generated using SAM, where masked pixels are assigned a value of 1. Subsequently, the raw RGB image $\mathbf{I}_{raw}$ is converted into the HSV color space, resulting in $\mathbf{I}_{HSV} \in \mathbb{R}^{H \times W \times 3}$. From this, the V channel, denoted as $\mathbf{V} \in \mathbb{R}^{H \times W}$, is extracted. The masked region in the V channel, $\mathbf{V}_{\text{masked}}$, is defined as follows:

\begin{equation}
    \mathbf{V}_{\text{masked}} = \{ \mathbf{V}_{i,j} \mid \mathbf{M}_{i,j} > 0 \},
\end{equation}

\noindent where $\mathbf{V}_{\text{masked}} \in \mathbb{R}^{H \times W}$ represents the masked area in the V channel, $\mathbf{V}_{i,j}$ is the pixel value at position $(i, j)$, and $\mathbf{M}_{i,j}$ is the corresponding mask value. The average brightness of the masked values, $\mathbf{B}_{\text{average}}$, is computed as:

\begin{equation}
    \mathbf{B}_{\text{average}} = \frac{1}{|\mathbf{V}_{\text{masked}}|} \sum_{(i,j) \in \mathbf{V}_{\text{masked}}} \mathbf{V}_{i,j}.
\end{equation}

\begin{figure*}[ht]
	\newlength{\SPheight}
	\setlength{\SPheight}{2.4cm}
	\centering

	\begin{subfigure}{0.24\linewidth}
		\centering
		\includegraphics[width=\linewidth, height=\SPheight]{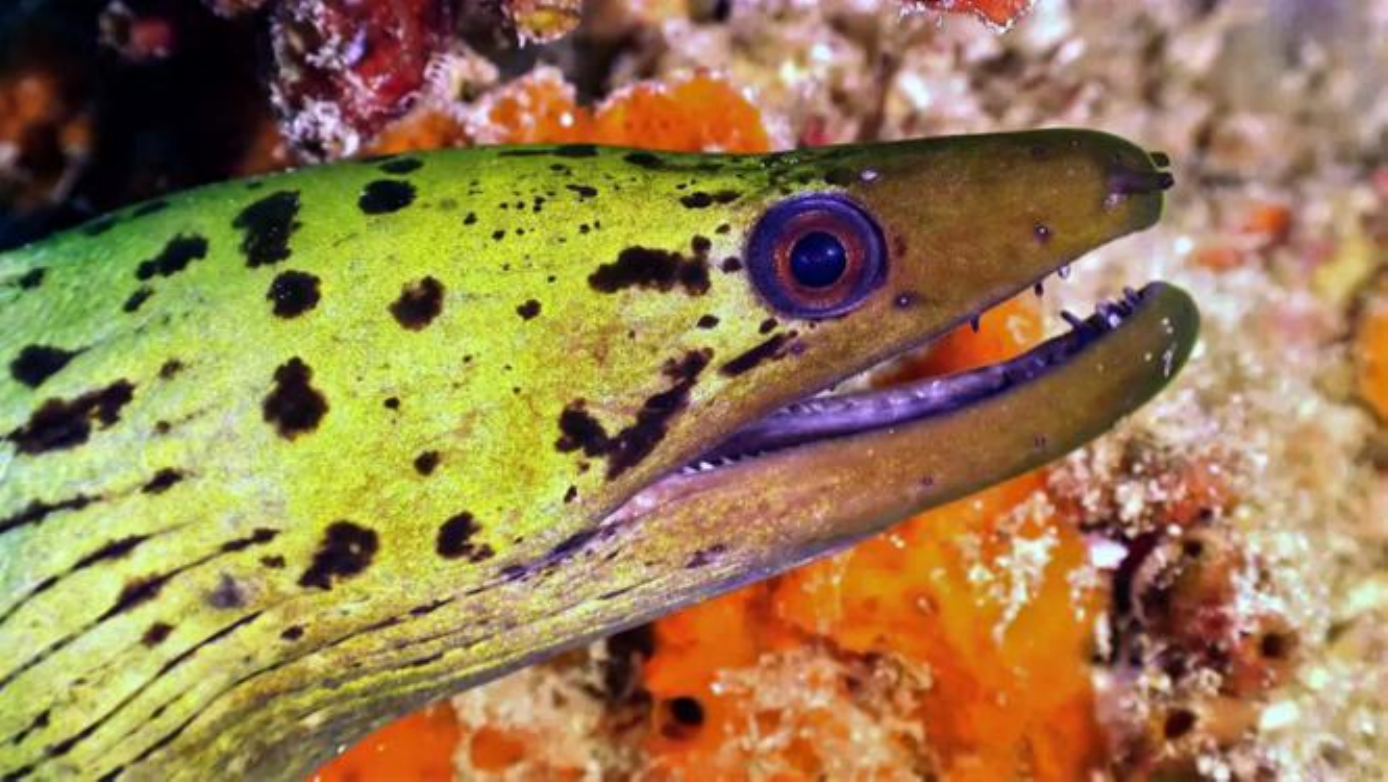} 
        \includegraphics[width=\linewidth]{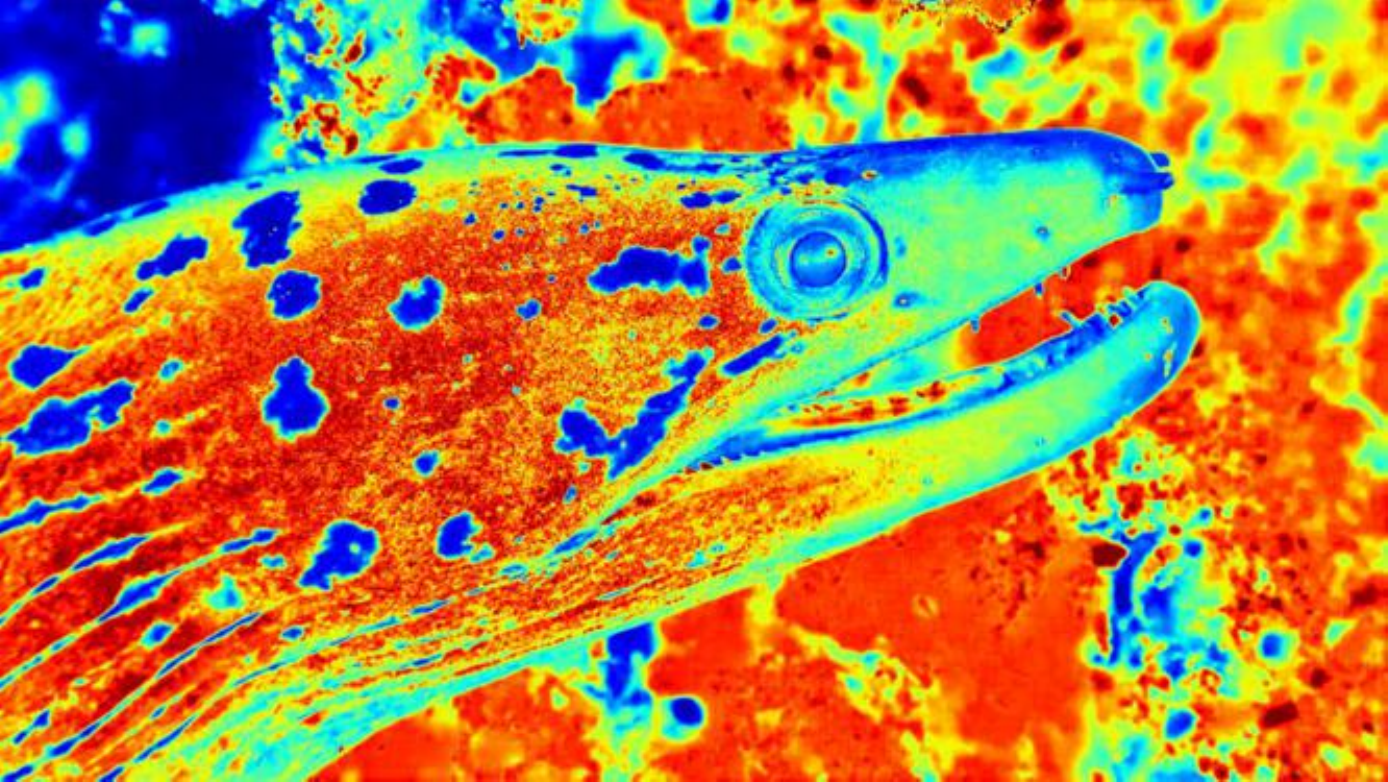} 
        \caption{\footnotesize Raw Image} 
        \label{Original}
	\end{subfigure}
	\begin{subfigure}{0.24\linewidth}
		\centering
		\includegraphics[width=\linewidth, height=\SPheight]{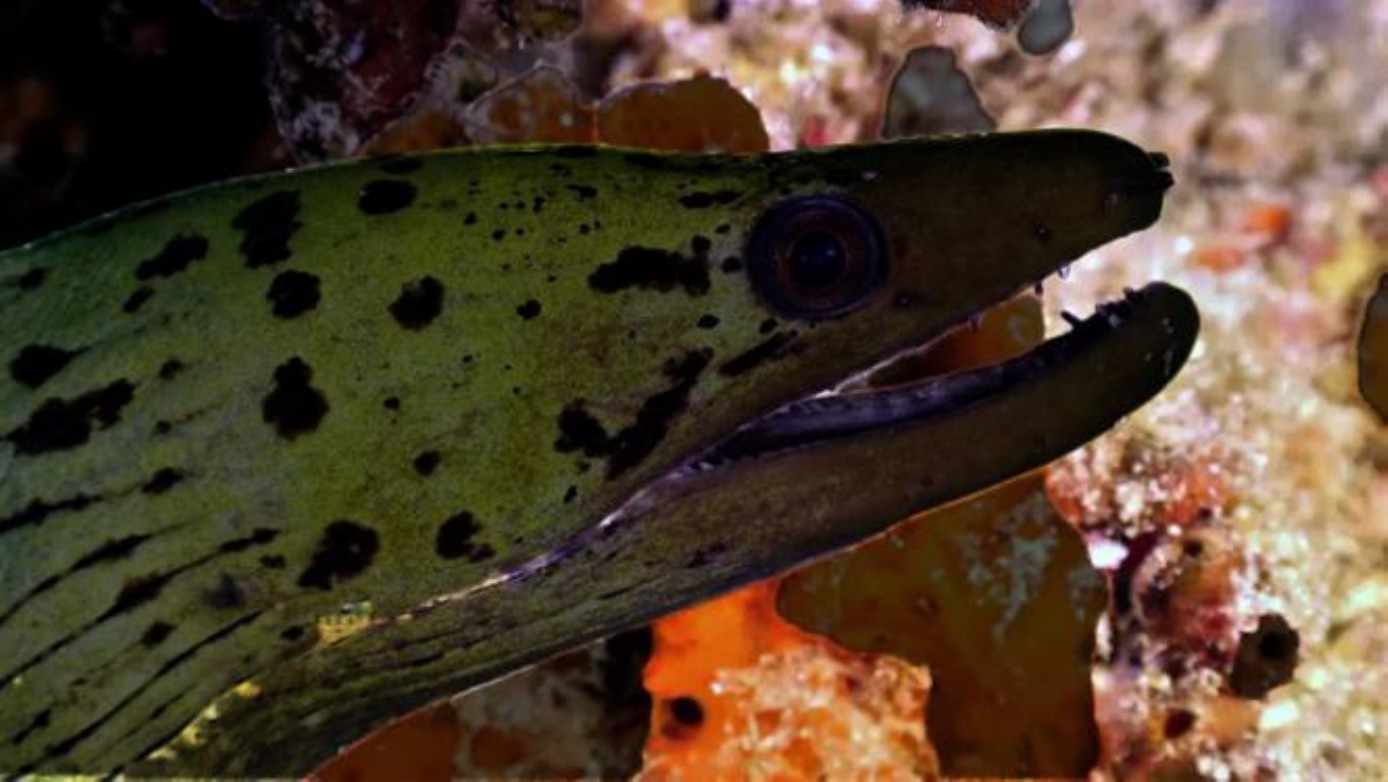}
        \includegraphics[width=\linewidth]{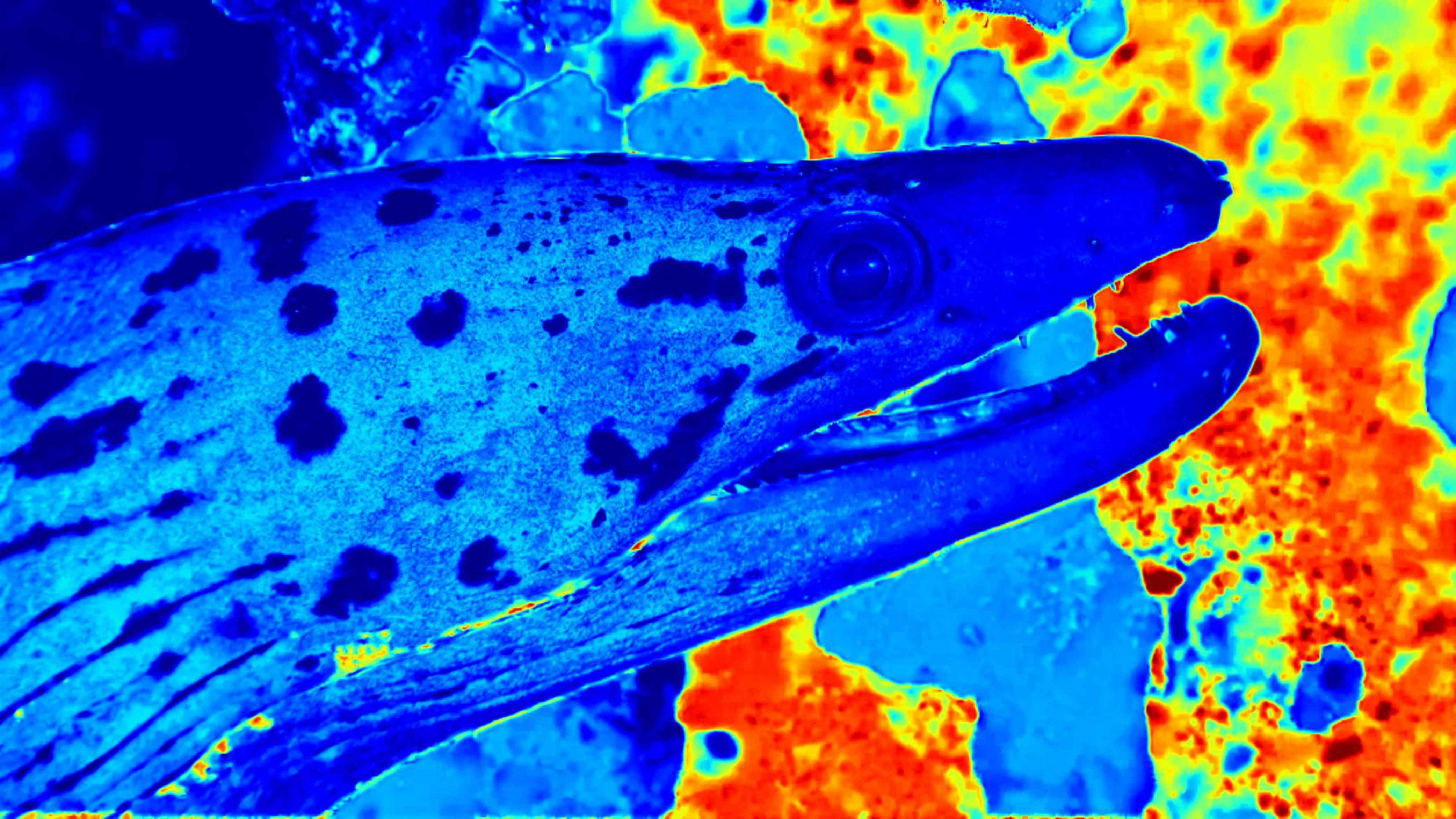} 
        \caption{\footnotesize $K_g=4$}
        \label{r4}
	\end{subfigure}
    \begin{subfigure}{0.24\linewidth}
		\centering
		\includegraphics[width=\linewidth, height=\SPheight]{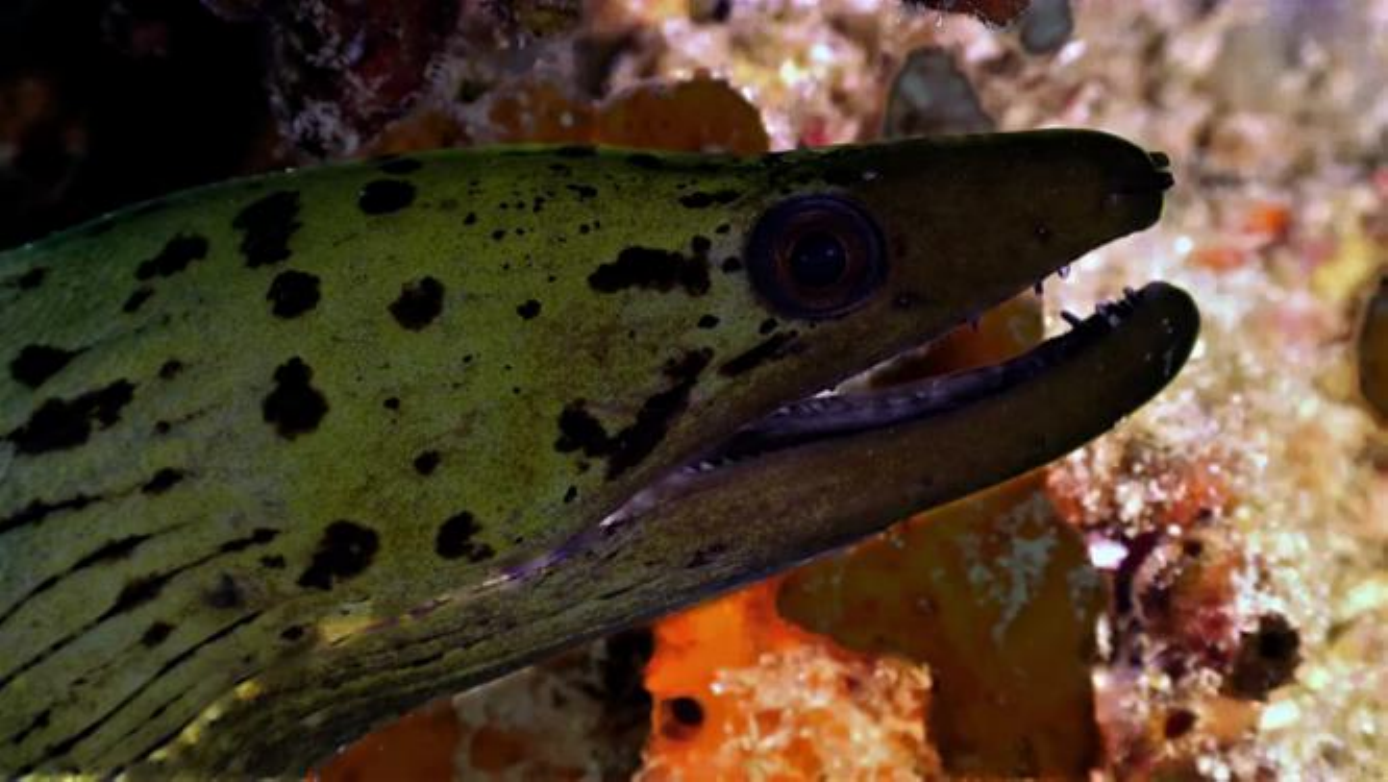}
        \includegraphics[width=\linewidth]{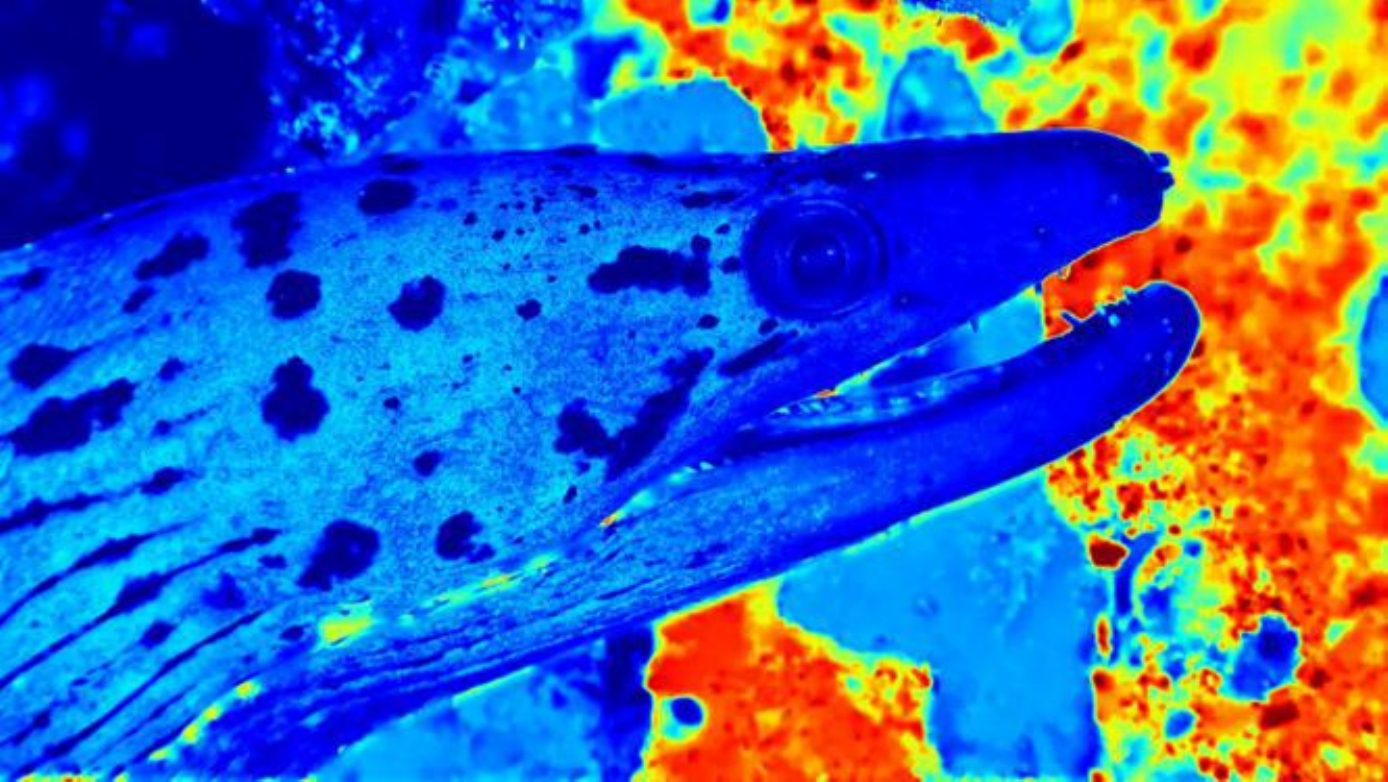} 
        \caption{\footnotesize $K_g=8$}
        \label{r8}
	\end{subfigure}
	\begin{subfigure}{0.24\linewidth}
		\centering
		\includegraphics[width=\linewidth, height=\SPheight]{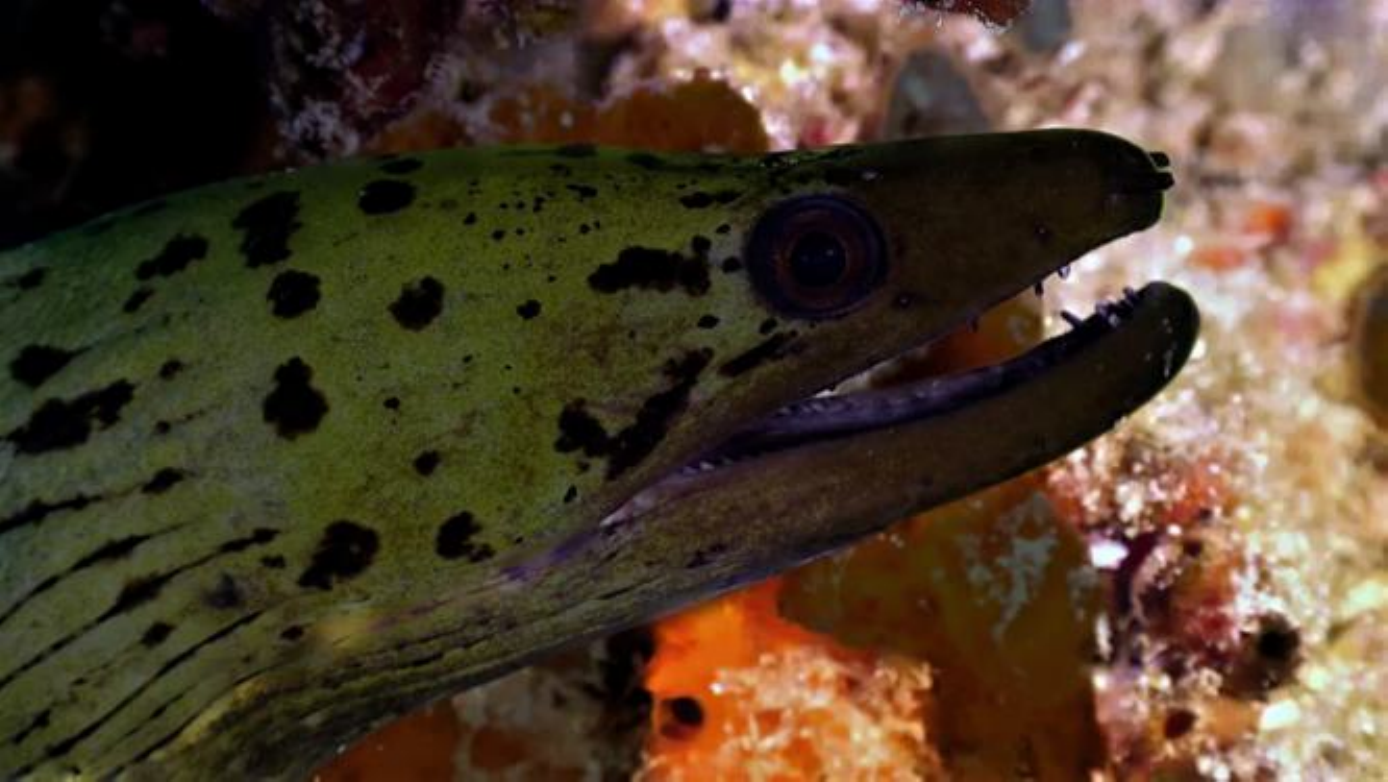} 
        \includegraphics[width=\linewidth]{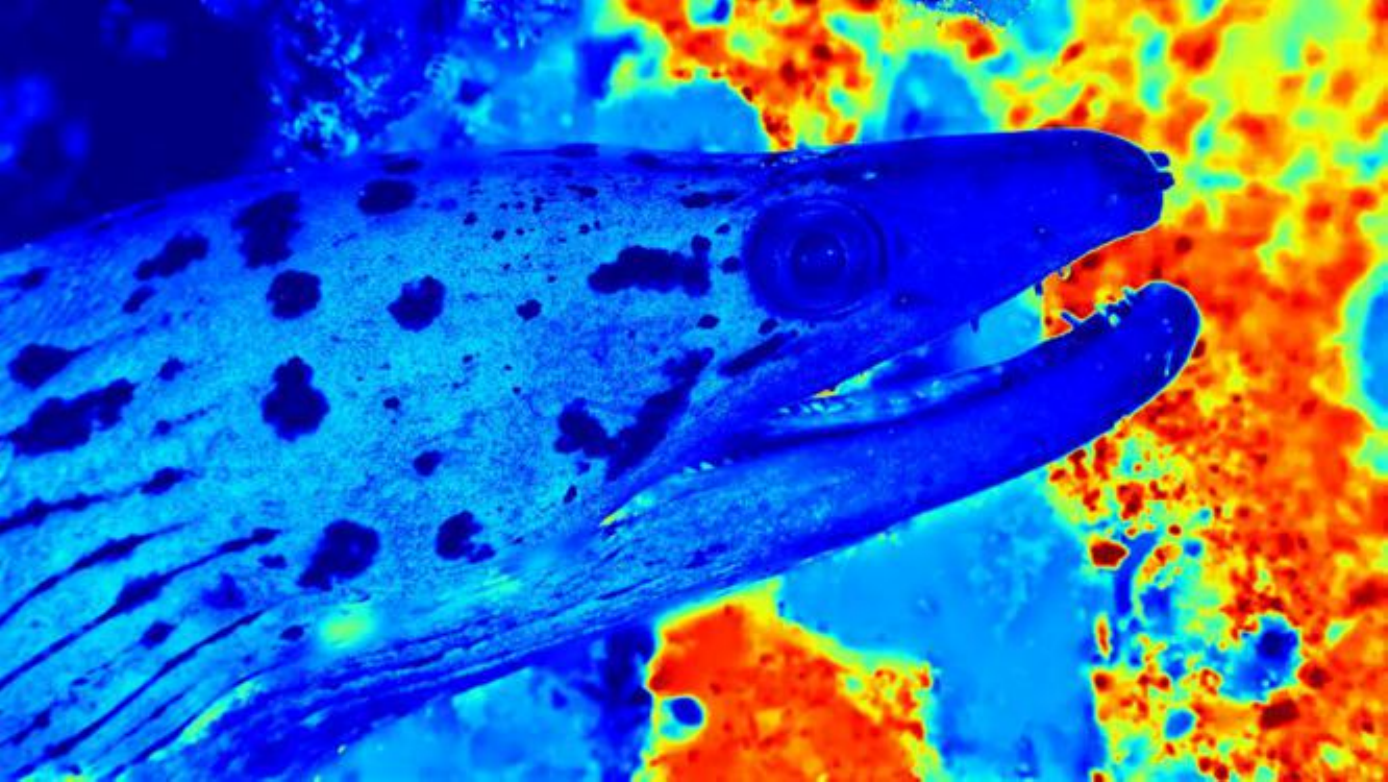} 
		\caption{\footnotesize $K_g=16$}
        \label{r16}
	\end{subfigure}    

	\begin{subfigure}{0.24\linewidth}
		\centering
		\includegraphics[width=\linewidth, height=\SPheight]{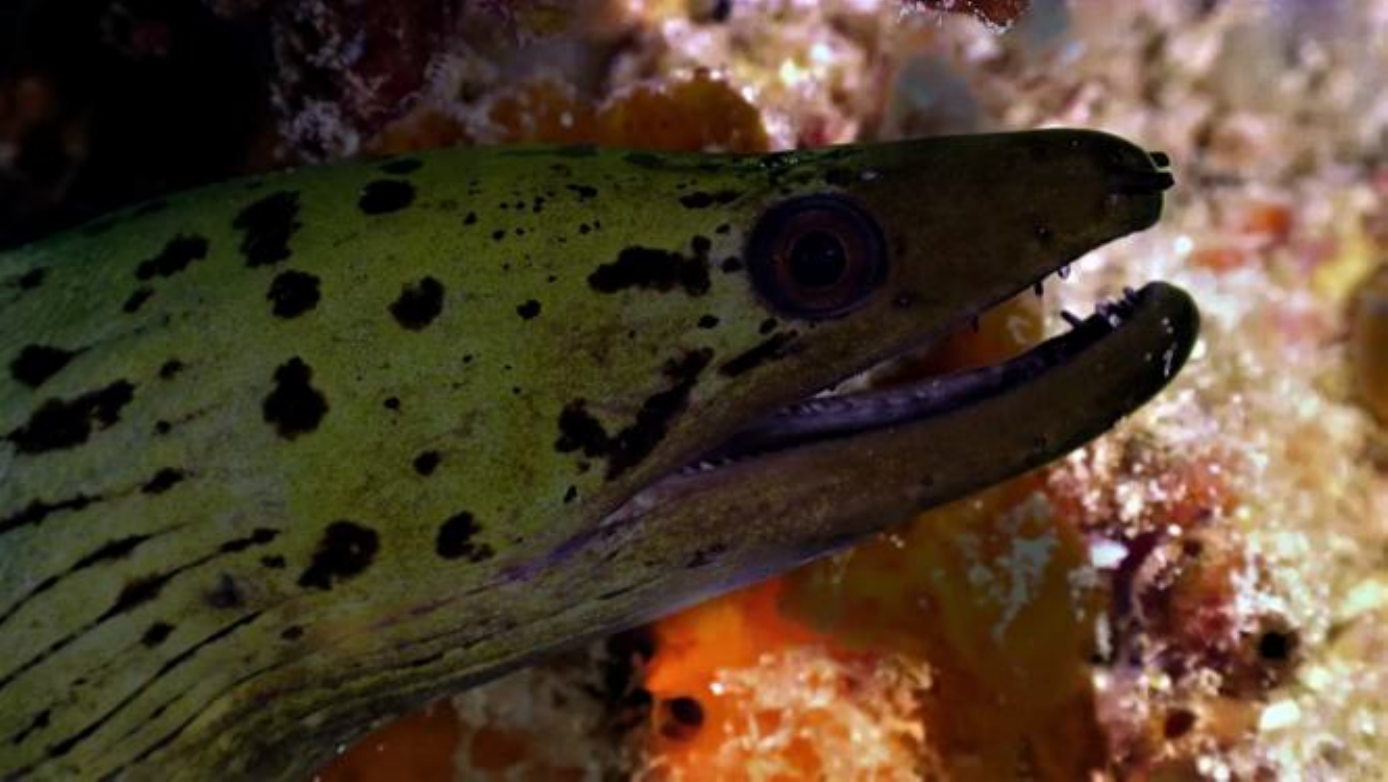} 
        \includegraphics[width=\linewidth]{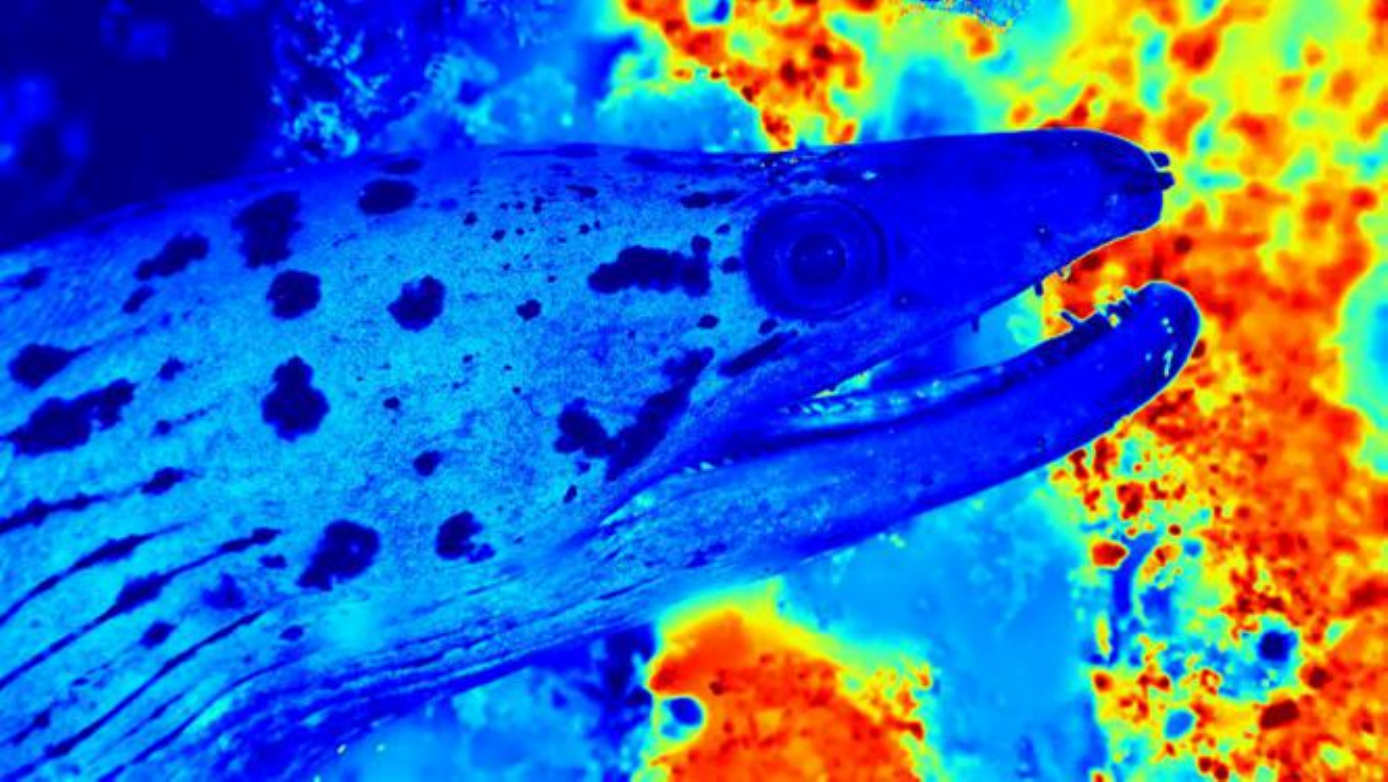} 
        \caption{\footnotesize $K_g=32$} 
        \label{r32}
	\end{subfigure}
	\begin{subfigure}{0.24\linewidth}
		\centering
		\includegraphics[width=\linewidth, height=\SPheight]{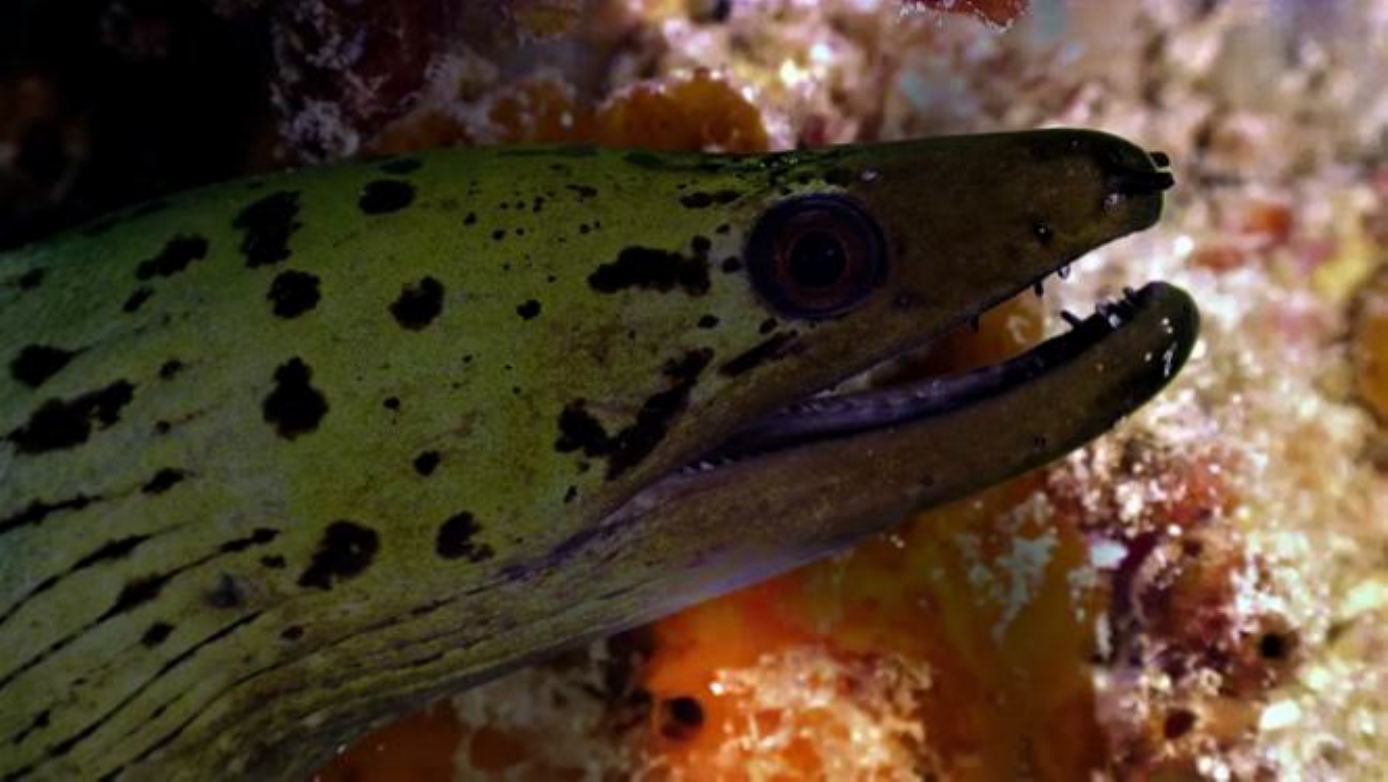}
        \includegraphics[width=\linewidth]{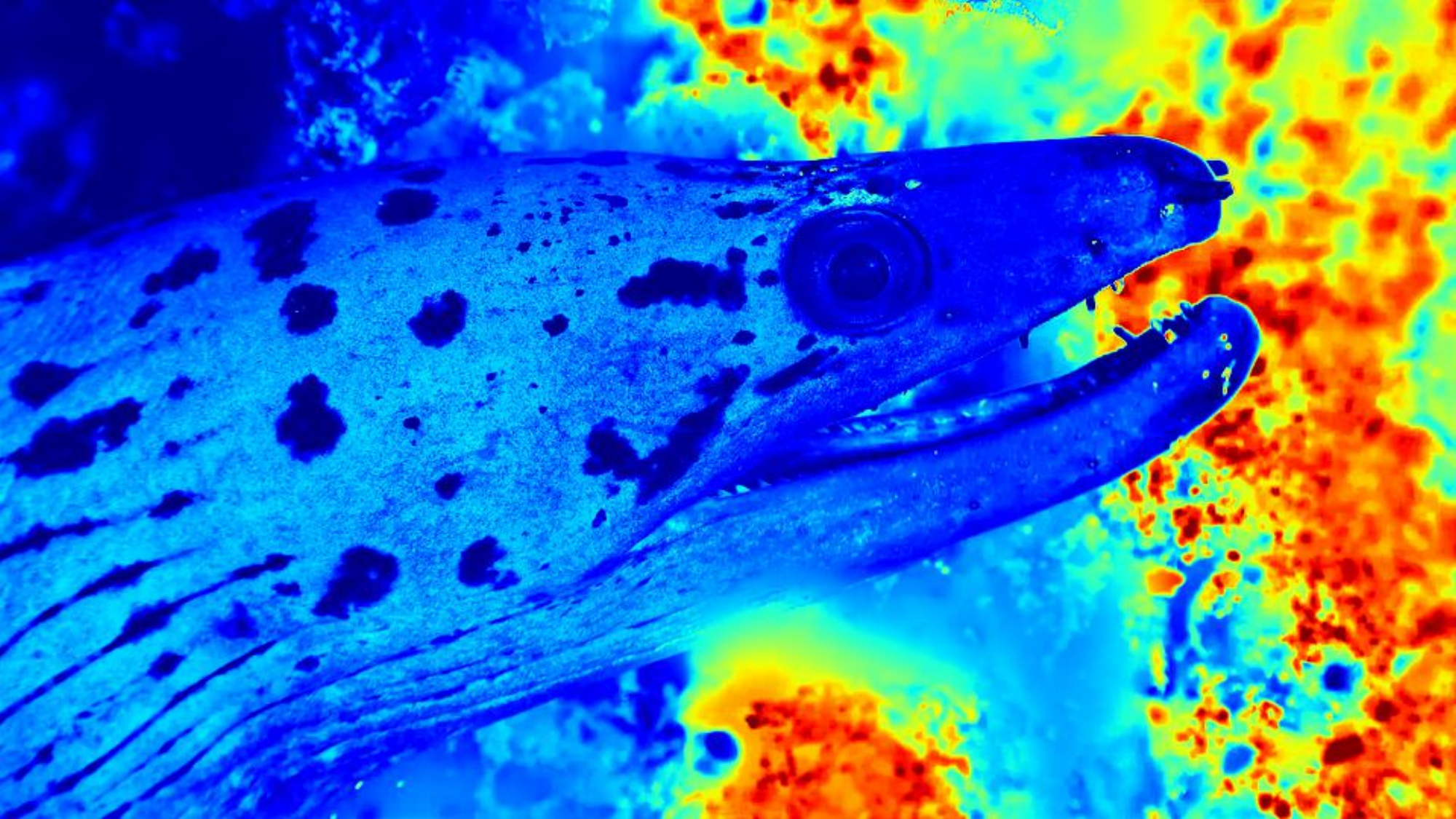} 
        \caption{\footnotesize $K_g=64$}
        \label{r64}
	\end{subfigure}
	\begin{subfigure}{0.24\linewidth}
		\centering
		\includegraphics[width=\linewidth, height=\SPheight]{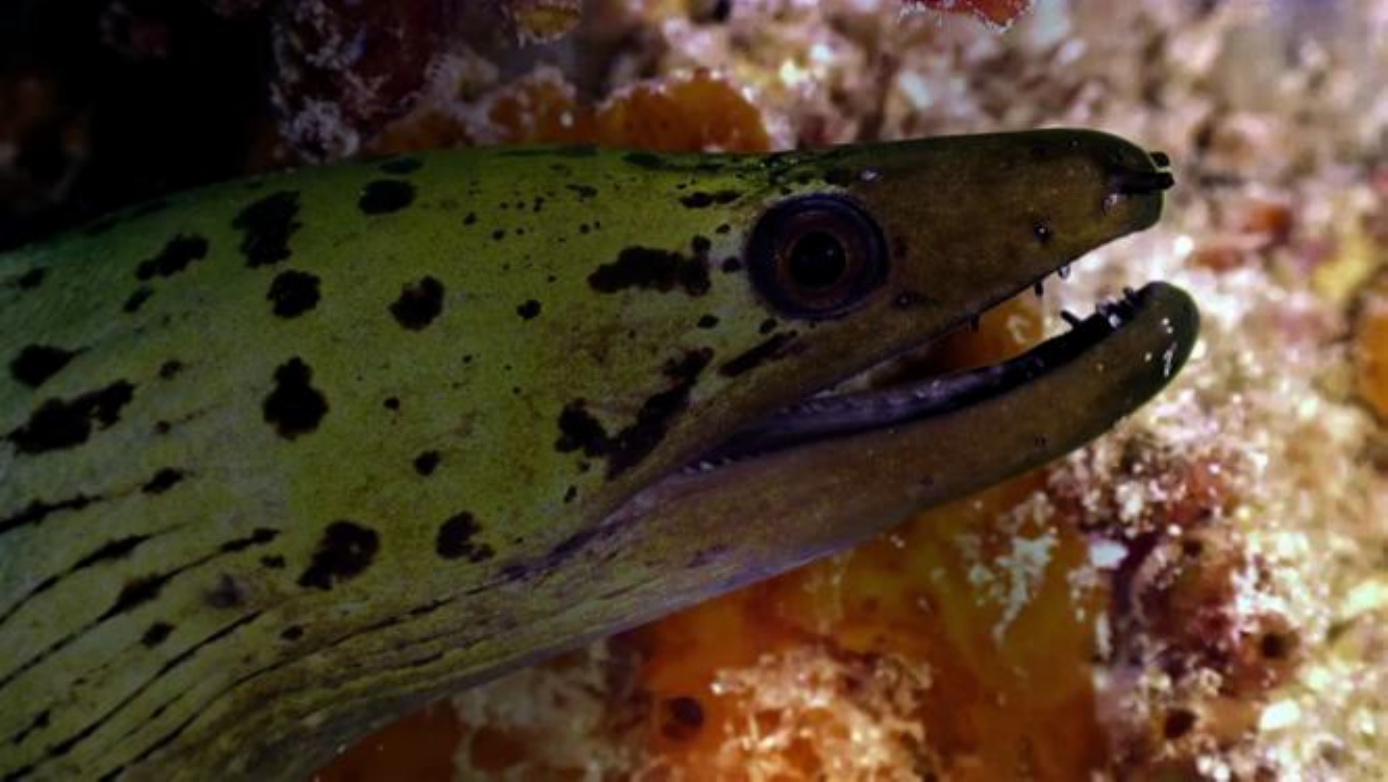}
        \includegraphics[width=\linewidth]{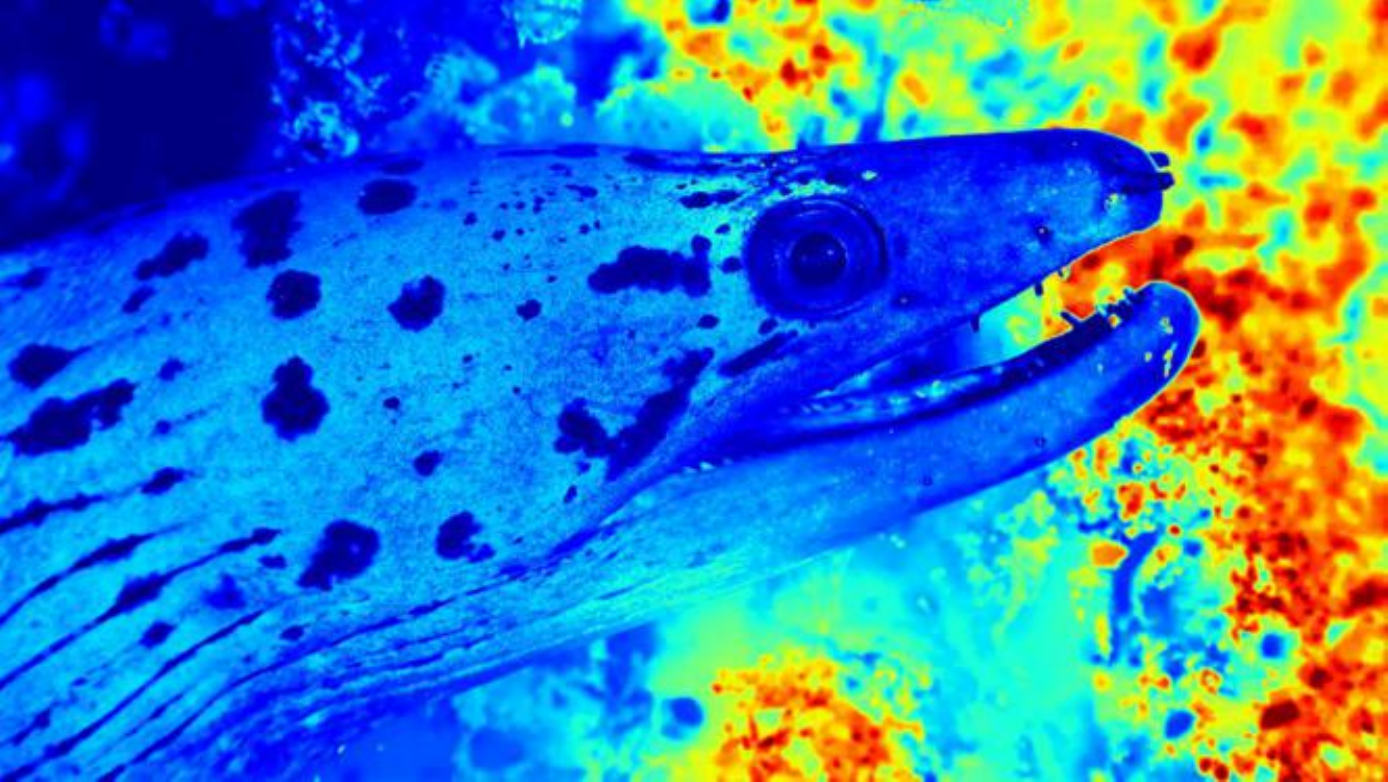} 
        \caption{\footnotesize $K_g=128$}
        \label{r128}
	\end{subfigure}
	\begin{subfigure}{0.24\linewidth}
		\centering
		\includegraphics[width=\linewidth, height=\SPheight]{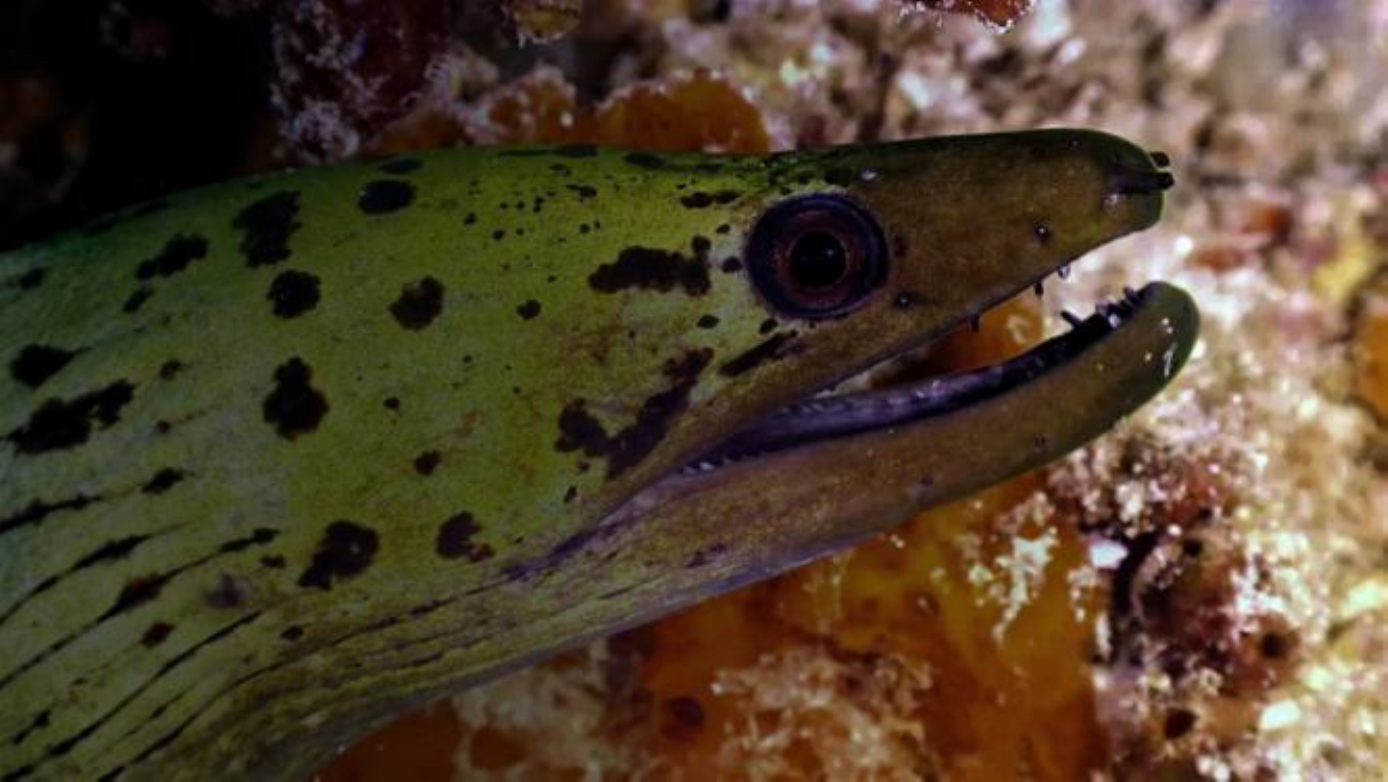} 
        \includegraphics[width=\linewidth]{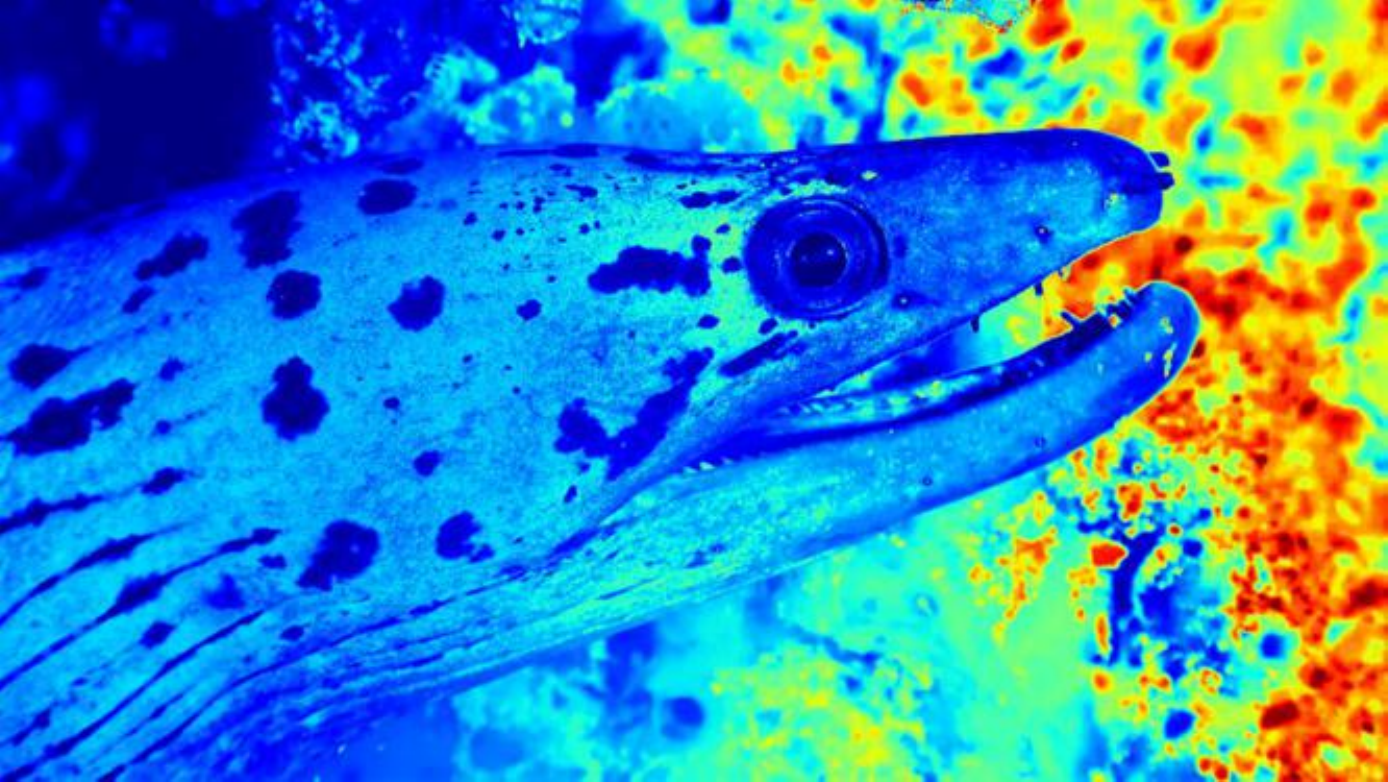} 
		\caption{\footnotesize $K_g=256$}
        \label{r256}
	\end{subfigure}    
    
	\caption{Visual comparison of results using different kernel sizes in the guided filter, each accompanied by its respective illumination map.}
	\label{SAMPLES}
\end{figure*}

To determine the type of brightness adjustment for the segmented masks, a comparison between $\mathbf{B}_{\text{average}}$ and the predefined brightness threshold $\alpha$, which is set to 70, is performed. This value was empirically chosen to suit underwater scenes, where stronger light attenuation justifies a higher threshold than in terrestrial settings~\cite{liu2022efinet}. This decision process is defined as:

\begin{equation}
    \delta = 
    \begin{cases}
        \text{\textit{Deepen}}, & \text{if } \mathbf{B}_{\text{average}} < \alpha, \\
        \text{Random}(\{\text{\textit{Surface}}, \text{\textit{Deepen}}, \text{\textit{None}}\}), & \text{otherwise},
    \end{cases}
\end{equation}

\noindent where ``\textit{Deepen}'' simulates low-light conditions typical of deeper and less illuminated areas, ``\textit{Surface}'' simulates brighter conditions near the surface, and ``\textit{None}'' implies no modification. The local brightness adjustment for the masked V channel, $\mathbf{V}^\lambda$, depending on the selected operation $\delta$, is expressed as follows:

\begin{equation}
    \mathbf{V}^\lambda = 
    \begin{cases} 
    \beta \cdot \mathbf{V}^{\gamma_d}, & \text{if } \delta = \text{Deepen}, \\
    \mathbf{V}^{\gamma_s}, & \text{if } \delta = \text{Surface}, \\
    \mathbf{V}, & \text{otherwise},
    \end{cases}
\end{equation}

\noindent where, $\gamma_d$ and $\gamma_s$ are constant gamma correction values, $\beta$ is a linear adjustment factor, and $\mathbf{V}^\lambda \in \mathbb{R}^{H \times W}$ represents the brightness-modified V channel. The constant values are adopted from the synthesis parameters proposed in \cite{liu2022efinet}. After adjusting the brightness in the segmented regions, a guided filter \cite{he2012guided} is applied to produce more realistic textures under non-uniform illumination conditions. This process is defined as:

\begin{equation}
    \mathbf{V}_{\text{smooth}} = \text{Guided Filter}(\mathbf{I}_{raw}, \mathbf{V}^\lambda, K_g),
\end{equation}

\noindent where $\mathbf{V}_{\text{smooth}} \in \mathbb{R}^{H \times W}$ is the filtered image, and $K_g$ is the kernel size of the filter. A kernel of $64 \times 64$ was chosen as it produced the best results, as shown in Fig. \ref{SAMPLES}. Smaller kernels prominently highlight the mask boundaries, while larger kernels overly blur the image, diminishing the non-uniform illumination effect. Finally, $\mathbf{V}_{\text{smooth}}$ replaces $\mathbf{V}$ as follows:

\begin{equation}
    \mathbf{I}_{HSV} = (\mathbf{H}, \mathbf{S}, \mathbf{V}_{\text{smooth}}).
\end{equation}

The modified $\mathbf{I}_{HSV}$ is then converted back to the RGB color space, resulting in the synthesized non-uniformly illuminated image $\mathbf{I}_{low}$. Once the non-uniformly illuminated input image $\mathbf{I}_{low}$ is generated, the ground truth image $\mathbf{I}_{gt}$ is created by applying a sharpening filter to enhance edges. Let $K_s$ be the sharpening kernel defined as:

\begin{equation}
    K_s = 
    \begin{bmatrix}
        0 & -1 & 0 \\
        -1 & 5 & -1 \\
        0 & -1 & 0
    \end{bmatrix}.
\end{equation}

\noindent To compute $\mathbf{I}_{gt}$, a channel-wise convolution is performed between the raw input image, $\mathbf{I}_{raw}$, and the kernel $K_s$. This operation is expressed as:

\begin{equation} \mathbf{I}_{gt} = \mathbf{I}_{raw} * K_s, \end{equation}

\noindent where, $*$ represents the convolution operator. The sharpened image $\mathbf{I}_{gt}$ serves as the ground truth, providing an enhanced representation of edge information. This procedure finalizes the construction of the PUNI dataset, ensuring that it captures both non-uniform illumination and detailed edge features critical for training enhancement models.

While the PUNI dataset enables supervised training by providing synthetic paired images under controlled non-uniform illumination, it remains a synthetic dataset. Currently, no real-world dataset provides pixel-aligned underwater image pairs capturing both degraded and restored appearances under non-uniform illumination. This lack of real paired data introduces a domain gap between synthetic training and real-world deployment. To account for this, UNIR-Net is evaluated not only on the PUNI test set but also on the unpaired, real-world NUID dataset. The consistent performance observed across both datasets indicates that the model generalizes well to real underwater conditions despite the synthetic origin of the training data.

\subsection{UNIR-Net Design}

\begin{figure*}[ht]
	\centering
	\includegraphics[width=1\textwidth]{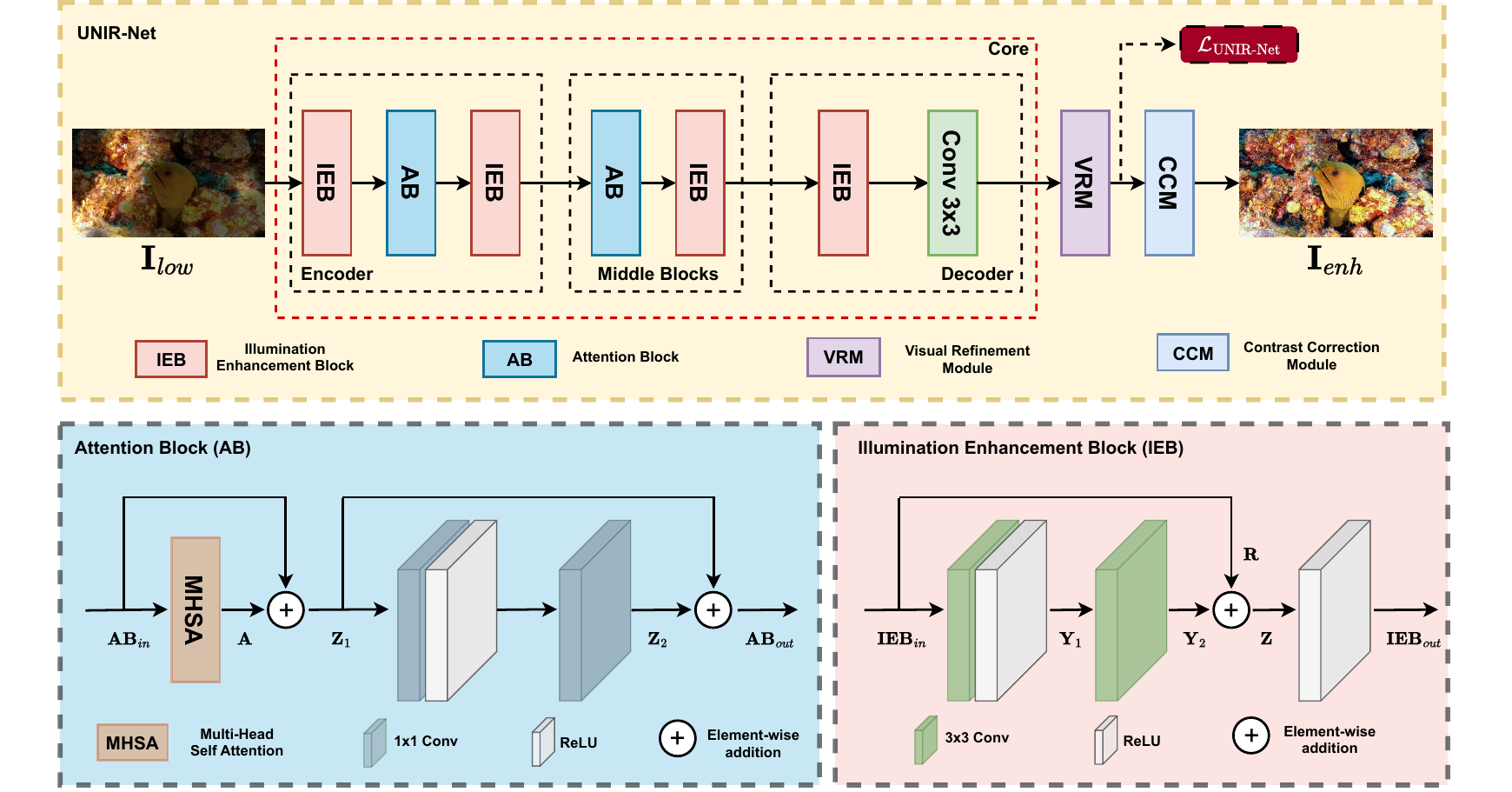}
	\caption{The architecture of UNIR-Net.}
	\label{fig:diagram}
\end{figure*}

This section presents the main study of this article, which consists of developing the Underwater Non-uniform Illumination Restoration Network (UNIR-Net) architecture. The primary objective of this network is to improve the visual perception of images captured in marine environments with non-uniform illumination. The design of UNIR-Net is structured around four essential components that ensure its optimal performance. The first component is the Illumination Enhancement Block (IEB), presented in Section \ref{IEB}. The second is the Attention Block (AB), described in Section \ref{AB}. The third is the Visual Refinement Module (VRM), detailed in Section \ref{VRM}. Finally, the fourth component is the Contrast Correction Module (CCM), explained in Section \ref{CCM}. The modular design of UNIR-Net, which includes convolutional components, allows it to handle input images of arbitrary resolution while maintaining spatial consistency and minimizing artifacts. This structure enables the architecture to effectively address both local and global variations in illumination of the input image $\mathbf{I}_{low} \in \mathbb{R}^{H \times W \times 3}$, leading to significant improvements in the lighting and visual perception of the enhanced image $\mathbf{I}_{enh} \in \mathbb{R}^{H \times W \times 3}$. Figure \ref{fig:diagram} shows the complete architecture, where the encoder part includes two IEBs interspersed with an AB. The middle section combines one AB and one IEB, while the decoder includes an IEB and a convolution with a $3 \times 3$ kernel. Finally, the VRM and CCM modules are placed at the end of the process. In addition, Table~\ref{tab:unirnet_simple} presents the layer-wise configuration of the main modules used in UNIR-Net, detailing the number of convolutional layers, kernel size, stride, padding, and activation function.

\begin{table}[htbp]
\centering
\caption{Layer‑wise configuration of UNIR‑Net.}
\label{tab:unirnet_simple}
\begin{tabular}{|l|c|c|c|c|c|}
\hline
\textbf{Module} & \textbf{\# Conv Layers} & \textbf{Kernel} & \textbf{Stride} & \textbf{Padding} & \textbf{Act.} \\
\hline
IEB     & 2 & $3\times3$ & 1 & 1 & ReLU\\
AB      & 2 & $1\times1$ & 1 & 0 & ReLU \\
Conv 3x3 & 1 & $3\times3$ & 1 & 1 & - \\
VRM & 1 & $3\times3$ & 1 & 1 & Sigmoid \\
\hline
\end{tabular}
\end{table}

\subsubsection{Illumination Enhancement Block (IEB)}
\label{IEB}
The IEB is based on the use of convolutions, ReLU activation functions, and residual connections to enhance lighting features in underwater images. It operates on an input tensor $\mathbf{IEB}_{\text{in}} \in \mathbb{R}^{H \times W \times C_{\text{in}}}$, where $C_{\text{in}}$ denotes the number of input channels, and $H$ and $W$ correspond to the spatial dimensions of the image. The enhancement process begins with a convolution operation performed using a filter $\mathbf{W}_1$, which extracts key features from the input. This is immediately followed by the application of the ReLU activation function to introduce non-linearity, resulting in:

\begin{equation}
    \mathbf{Y}_1 = \text{ReLU}(\mathbf{IEB}_{\text{in}} \ast \mathbf{W}_1).
\end{equation}

Next, the intermediate features $\mathbf{Y}_1$ are refined through a second convolution with a filter $\mathbf{W}_2$, resulting in:

\begin{equation}
    \mathbf{Y}_2 = \mathbf{Y}_1 \ast \mathbf{W}_2.
\end{equation}

If the number of input and output channels does not match ($C_{\text{in}} \neq C_{\text{out}}$), an adjustment filter $\mathbf{W}_m$ is used. The residual $\mathbf{R}$ is then computed as follows:

\begin{equation}
    \mathbf{R} = 
    \begin{cases} 
        \mathbf{IEB}_{\text{in}} \ast \mathbf{W}_m, & \text{if } C_{\text{in}} \neq C_{\text{out}}, \\
        \mathbf{IEB}_{\text{in}}, & \text{otherwise}.
    \end{cases}
\end{equation}

Subsequently, the refined output $\mathbf{Y}_2$ is combined with the adjusted residual $\mathbf{R}$ through an addition operation:

\begin{equation}
    \mathbf{Z} = \mathbf{Y}_2 + \mathbf{R}.
\end{equation}

Finally, $\mathbf{Z}$ passes through an additional ReLU activation to produce the final output of the block:

\begin{equation}
    \mathbf{IEB}_{\text{out}} = \text{ReLU}(\mathbf{Z}),
\end{equation}

\noindent where $\mathbf{IEB}_{\text{out}} \in \mathbb{R}^{H \times W \times C_{\text{out}}}$ contains the enhanced features. This output integrates both the original input information and the transformations performed by the convolutions, optimizing the lighting representation for subsequent stages of the model.

\subsubsection{Attention Block}
\label{AB}
The Attention Block (AB) leverages a Multi-Head Self-Attention (MHSA) mechanism combined with a Feed-Forward Network (FFN) to effectively capture both spatial and contextual relationships within the features processed by a convolutional backbone. This design integrates two key components, an MHSA module and an FFN enhanced by residual connections to retain the original feature representations from the input.

The process begins with an input tensor $\mathbf{AB}_{\text{in}} \in \mathbb{R}^{H \times W \times C}$, where $H$ and $W$ denote the spatial dimensions, and $C$ represents the number of channels. The MHSA module extracts the query ($\mathbf{Q}$), key ($\mathbf{K}$), and value ($\mathbf{V}$) matrices using learned weight matrices through convolution operations, as expressed below:

\begin{equation} \mathbf{Q} = \mathbf{W}_Q * \mathbf{AB}_{in}, \quad \mathbf{K} = \mathbf{W}_K * \mathbf{AB}_{in}, \quad \mathbf{V} = \mathbf{W}_V * \mathbf{AB}_{in}, \end{equation}

\noindent where $\mathbf{W}_Q$, $\mathbf{W}_K$, and $\mathbf{W}_V$ are trainable weight matrices, and $*$ indicates convolution. The attention mechanism is then formulated as:

\begin{equation} \mathbf{A} = \text{softmax}\left(\frac{\mathbf{Q} \cdot \mathbf{K}^\top}{\sqrt{\frac{C}{N_h}}}\right) \cdot \mathbf{V}, \end{equation}

\noindent where $N_h$ is the number of attention heads. This mechanism generates an attention-enhanced feature map $\mathbf{A}$, which is combined with the original input through a residual connection:

\begin{equation} \mathbf{Z}_1 = \mathbf{A} + \mathbf{AB}_{in}. \end{equation}

Subsequently, the FFN further processes $\mathbf{Z}_1$ to refine and enhance the extracted features. The FFN operation is defined as:

\begin{equation} \mathbf{Z}_2 = \mathbf{W}_b * \text{ReLU}(\mathbf{W}_a *\mathbf{Z}_1), \end{equation}

\noindent where $\mathbf{W}_a$ and $\mathbf{W}_b$ are learnable matrices used to project features into higher-dimensional spaces and then reduce them back, respectively. Finally, the output of the FFN is combined residually with $\mathbf{Z}_1$, yielding the final output of the AB:

\begin{equation} \mathbf{AB}_{out} = \mathbf{Z}_1 + \mathbf{Z}_2. \end{equation}

This design ensures that the AB block not only captures intricate spatial and contextual relationships but also preserves critical information from the input features.

\subsubsection{Visual Refinement Module}
\label{VRM}

The Visual Refinement Module (VRM) is designed to refine image features within a defined range of 0 to 1, preventing pixel overflow and enhancing the visual clarity of the processed information. This refinement builds on the results generated by the IEB and AB blocks. The output of the VRM is defined as follows:

\begin{equation} \mathbf{VRM}_{out} = \text{Sigmoid}(\mathbf{W}_s * \mathbf{VRM}_{in}), \end{equation}

\noindent where $\mathbf{W}_s$ is a learnable matrix. The VRM enables additional processing of the features obtained from previous modules, optimizing the visual quality of the resulting images.

\subsubsection{Contrast Correction Module}
\label{CCM}

The Contrast Correction Module (CCM) aims to improve the contrast of images processed earlier by the Visual Refinement Module (VRM). Its primary objective is to enhance contrast's visual quality, ensuring that underwater colors appear more vivid and realistic. The CCM employs the algorithm outlined in \cite{albakri2021rapid}, starting with the input image \(\mathbf{CCM}_{\text{in}}\), which is processed using the Probability Density Function of a Standard Normal Distribution (PDF-SND). This process is mathematically expressed as:

\begin{equation}
\phi = \frac{1}{\sqrt{2 \pi}} \exp\left(-\frac{1}{2} \cdot (\mathbf{CCM}_{\text{in}})^2\right),
\end{equation}

\noindent where \(\phi\) represents the processed image resulting from the PDF-SND transformation. Subsequently, the same input image, \(\mathbf{CCM}_{\text{in}}\), undergoes further enhancement using the Softplus function, denoted as \(\psi\), which is defined as:

\begin{equation}
\psi = \log\left(1 + \exp(\mathbf{CCM}_{\text{in}})\right).
\end{equation}

Once \(\phi\) and \(\psi\) are computed, the module applies a Logarithmic Image Processing (LIP) model to combine these components. The LIP model is described by the following equation:

\begin{equation}
\mathbf{I}^{\prime} = \sqrt{\phi + \psi + \phi \cdot \psi},
\end{equation}

\noindent where \(\mathbf{I}^{\prime}\) denotes the intermediate image generated by this model. The final step in the CCM involves applying a gamma-controlled normalization function to produce the contrast-enhanced image. This step is defined as:

\begin{equation}
\mathbf{CCM}_{\text{out}} = \left(\frac{\mathbf{I}^{\prime} - \mathbf{I}^{\prime}_{\text{min}}}{\mathbf{I}^{\prime}_{\text{max}} - \mathbf{I}^{\prime}_{\text{min}}}\right)^\gamma,
\end{equation}

\noindent where $\mathbf{CCM}_{\text{out}}$ is the final output image with enhanced contrast and $\gamma$ is the gamma correction factor used to control the contrast enhancement intensity, which is set to 1.4 in the experiments. ${\mathbf{I}^{\prime}_{\text{min}}}$ is the minimum intensity value of the intermediate image $\mathbf{I}^{\prime}$, and ${\mathbf{I}^{\prime}_{\text{max}}}$ is the maximum intensity value of $\mathbf{I}^{\prime}$. This value provided a balanced enhancement across various underwater scenes and is consistent with prior studies suggesting gamma values in the 1.x–1.y range for low-light correction \cite{lv2021attention,liu2025dark}. Such gamma values are commonly adopted in image enhancement, including underwater contexts where non-linear illumination compensation is required \cite{liu2025underwater}.

The CCM effectively enhances underwater images captured under non-uniform illumination conditions by systematically applying these steps. The result is a more balanced color distribution and improved visibility of fine details, significantly contributing to the visual enhancement of submarine imagery. 

\subsection{Objective Function}

To improve the visual quality of underwater images with non-uniform illumination, the total loss function, denoted as $\mathcal{L}_{\text{UNIR-Net}}$, is formulated as a combination of three complementary components: contrast loss ($\mathcal{L}_c$), structural loss ($\mathcal{L}_s$), and perceptual loss ($\mathcal{L}p$). This design ensures that the enhanced images exhibit improved contrast, retain essential structural details, and maintain perceptual consistency with reference images. The total loss function $\mathcal{L}_{\text{UNIR-Net}}$ is defined as follows:

\begin{equation} \mathcal{L}_{\text{UNIR-Net}} = \lambda_c\mathcal{L}_c + \lambda_s\mathcal{L}_s + \lambda_p\mathcal{L}_p. \end{equation}

Here, the weights $\lambda_c$, $\lambda_s$, and $\lambda_p$ are all set to 1, ensuring equal contribution from each component to the total loss and simplifying the optimization process.
  
\textbf{Contrast Loss.}  
The contrast loss $\mathcal{L}_c$ evaluates the consistency of brightness and contrast between the enhanced image $\hat{y}$ and the reference image $y$. This term employs the $L_1$ distance, which measures the average absolute difference between the two sets of pixels over a total of $N$ pixels. Its primary objective is to ensure that the brightness and contrast characteristics of the generated image align with those of the reference. The loss is calculated using the following expression:

\begin{equation}
\mathcal{L}_{c} = \frac{1}{N} \sum_{i=1}^N \left\| y - \hat{y} \right\|_{1}.
\label{contrast}
\end{equation}

\textbf{Structural Loss.}  
The structural loss evaluates the Structural Similarity Index (SSIM)~\cite{wang2004image} between the enhanced image $\hat{y}$ and the reference image $y$. This measure penalizes structural discrepancies, aiming to preserve the spatial integrity of the image. It is calculated using the following formula:

\begin{equation}
	\mathcal{L}_{s} = 1 - \frac{(2\mu_{\hat{y}}\mu_{y} + c_1)(2\sigma_{\hat{y} {y}} + c_2)}{(\mu_{\hat{y}}^2 + \mu_{y}^2 + c_1)(\sigma_{\hat{y}}^2 + \sigma_{y}^2 + c_2)}.
	\label{structural}
\end{equation}

\noindent In this expression, $\mu_{\hat{y}}$ and $\mu_{y}$ represent the mean intensities of the enhanced and reference images, respectively. $\sigma_{\hat{y}}$ and $\sigma_{y}$ are the variances of each image, and $\sigma_{\hat{y} y}$ is the covariance between both images. The constants $c_1$ and $c_2$ are used to stabilize the division calculation.

\textbf{Perceptual Loss.}  
The perceptual loss guarantees that the enhanced image $\hat{y}$ closely resembles the reference image $y$ in terms of visual appearance by comparing high-level features extracted using a pre-trained network, specifically the VGG network\cite{johnson2016perceptual,simonyan2014very}. The perceptual loss is defined as:

\begin{equation} 
\mathcal{L}_{p} = \frac{1}{N}\sum_{i=1}^{N}\frac{1}{C_{j}H_{j}W_{j}}\left\|\phi_j(\hat{y})-\phi_j(y)\right\|^2_2 
\label{perceptual}, 
\end{equation}

\noindent where, \(\phi_j\) represents the feature map of the \(j\)-th layer of the pre-trained network, and \(C_j\), \(H_j\), and \(W_j\) are the channel, height, and width dimensions, respectively.

\section{Experiments and Analysis}
\label{expdis}
This section delves into the datasets utilized for performance evaluation, detailing their relevance and specific challenges related to underwater images with non-uniform illumination. It outlines the evaluation framework, including the metrics selected to benchmark UNIR-Net and the comparative approaches employed rigorously. Additionally, the training process of UNIR-Net is thoroughly described, emphasizing the steps taken to optimize its performance. A comprehensive analysis of the results is provided, highlighting both quantitative and qualitative aspects and comparisons across multiple datasets tailored to address the complexities of non-uniform lighting. The section also examines ablation studies to uncover the contributions of individual components, evaluates computational efficiency, and explores the broader implications of using the method as a preprocessing step for downstream tasks in underwater image analysis.

\subsection{Experiment settings}

\subsubsection{Implementation Details}
The training of UNIR-Net leverages the PyTorch framework~\cite{paszke2019pytorch}, a widely adopted open-source library designed for building and optimizing complex neural network architectures. The experiments are conducted on a system equipped with an NVIDIA RTX 3060 GPU, an Intel Core i5-12400F CPU running at 2.50 GHz, and 16 GB of RAM, ensuring a robust computational setup. For the training phase, a total of 2,786 images from the PUNI dataset are utilized, while an additional 491 images are allocated exclusively for testing purposes. To facilitate learning, the training process uses 128x128 image patches and processes mini-batches of 8 images. The model is trained over 100 epochs using the Adam optimizer~\cite{kingma2014adam}, with an initial learning rate set to $1e-4$. This configuration ensures efficient convergence and optimization tailored to the unique challenges posed by the PUNI dataset.

\subsubsection{Benchmark Datasets}
Two datasets are utilized to comprehensively evaluate UNIR-Net: the PUNI test set and a real-world unpaired dataset named Non-Uniform Illumination Dataset (NUID)~\cite{hou2023non}. The PUNI test set, comprising 491 images, enables performance assessments using both full-reference and non-reference metrics, ensuring a detailed evaluation. However, recognizing the limitations posed by the unavailability of training codes for some state-of-the-art methods, a secondary evaluation is conducted using the NUID dataset to ensure a more equitable comparison. NUID, a large-scale dataset tailored for enhancing underwater images with non-uniform illumination, includes 925 real-world images sourced from a variety of collections: UIEB~\cite{li2019underwater} (32 images), EUVP~\cite{islam2020fast} (256 images), OceanDark~\cite{porto2019contrast} (21 images), Google Image (96 images), and Nature Footage (520 images). By incorporating this diverse dataset, the evaluation framework ensures fairness and better reflects real-world conditions, offering a robust comparison of UNIR-Net’s capabilities against other state-of-the-art approaches.

\subsubsection{Evaluation Metrics}
The evaluation process incorporates a combination of Full-reference Image Quality Assessment (FIQA) and No-reference Image Quality Assessment (NIQA) metrics to analyze the performance of UNIR-Net comprehensively. FIQA metrics, such as Peak Signal-to-Noise Ratio (PSNR)~\cite{wang2004image}, Structural Similarity Index Measure (SSIM)~\cite{wang2004image}, Universal Quality Index (UQI)~\cite{wang2002universal}, Learned Perceptual Image Patch Similarity (LPIPS)~\cite{zhang2018unreasonable}, and DeltaE~\cite{xu2024degraded}, are employed to quantify image quality in terms of pixel accuracy, structural fidelity, perceptual similarity, and color consistency. These metrics evaluate reconstructed images against their ground truth.

Complementing this, NIQA metrics are utilized to assess image quality without requiring reference images, focusing on aspects critical to underwater scenarios. This includes Underwater Color Image Quality Evaluation (UCIQE)~\cite{yang2015underwater}, tailored specifically for underwater environments; Fog Aware Density Evaluator (FADE)~\cite{choi2015referenceless}, for analyzing fog-related image degradation; No-reference Image Quality Metric for Contrast (NIQMC)~\cite{gu2016no}, which measures contrast performance; and Multi-Scale Image Quality Transformer (MUSIQ)~\cite{ke2021musiq}, leveraging multi-scale features for a robust assessment. By integrating these diverse metrics, the evaluation framework ensures a thorough comparison of UNIR-Net with state-of-the-art approaches, capturing both objective quality and perceptual improvements.

\subsubsection{Comparison Methods}
To assess the performance of UNIR-Net, a comprehensive comparison was conducted with multiple state-of-the-art methods. Specifically, 13 techniques focusing on the enhancement of underwater images were considered: UNTV~\cite{xie2021variational}, UWNet~\cite{naik2021shallow}, ACDC~\cite{zhang2022underwater}, MMLE~\cite{zhang2022underwaterMMLE}, TCTL-Net~\cite{li2023tctl}, ICSP~\cite{hou2023non}, PCDE~\cite{zhang2023PCDE}, UDAformer~\cite{shen2023udaformer}, HFM~\cite{an2024hfm}, LENet~\cite{zhang2024liteenhancenet}, SMDR-IS~\cite{zhang2024synergistic}, MACT~\cite{zhang2025mact} and UDnet~\cite{saleh2025adaptive}. Additionally, 3 terrestrial methods aimed at enhancing low-light images were included in the evaluation: GCP~\cite{jeon2024low}, EIB-FNDL~\cite{khajehvandi2025enhancing} and ALEN~\cite{perez2025alen}. These methods enable a comprehensive comparison of UNIR-Net with state-of-the-art techniques, ensuring a robust evaluation in scenarios with non-uniform illumination.

\subsection{Evaluation Results and Discussion}

\subsubsection{Qualitative Results}
\begin{figure*}[!ht]
	\newlength{\puniheight}
	\setlength{\puniheight}{1.83cm}
	\centering
	\begin{subfigure}{0.105\linewidth}
		\centering
        \includegraphics[width=\linewidth,  height=\puniheight]{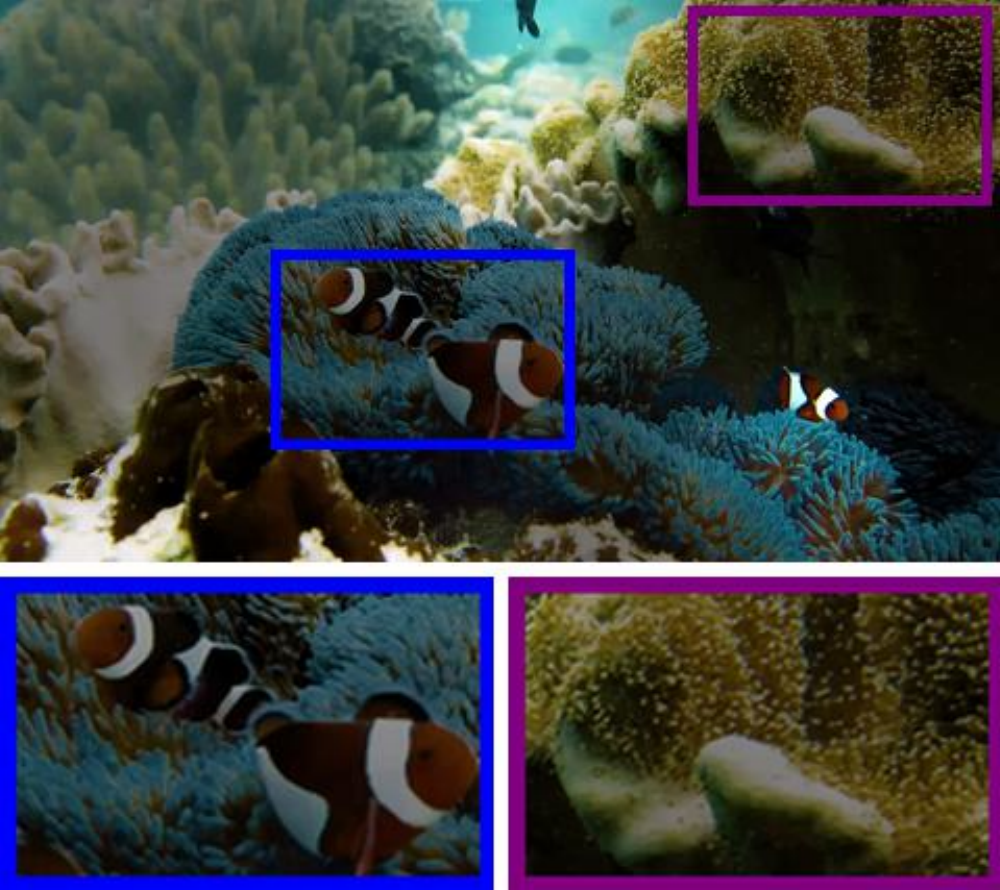} 
		\includegraphics[width=\linewidth,  height=\puniheight]{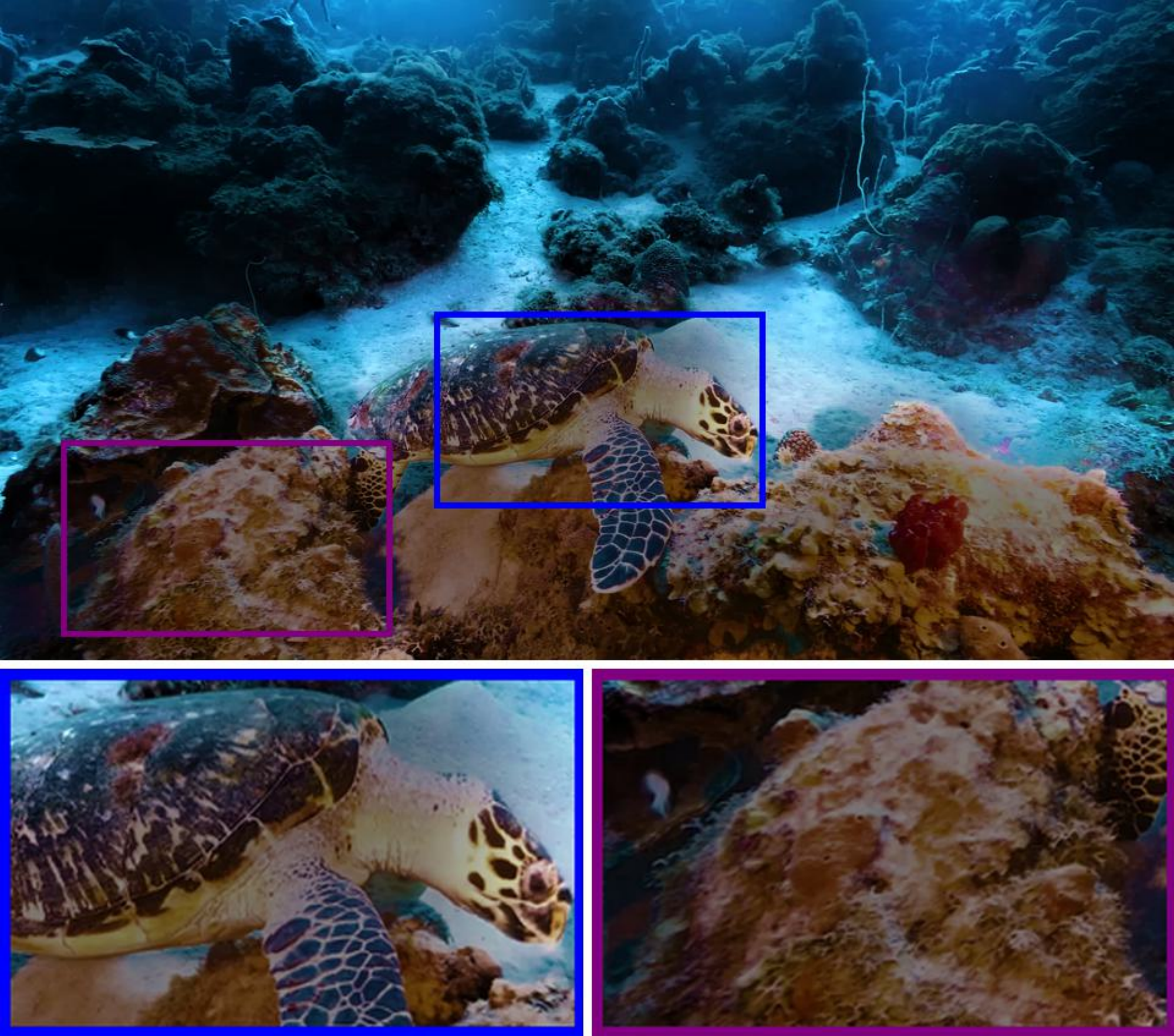} 
        \includegraphics[width=\linewidth,  height=\puniheight]{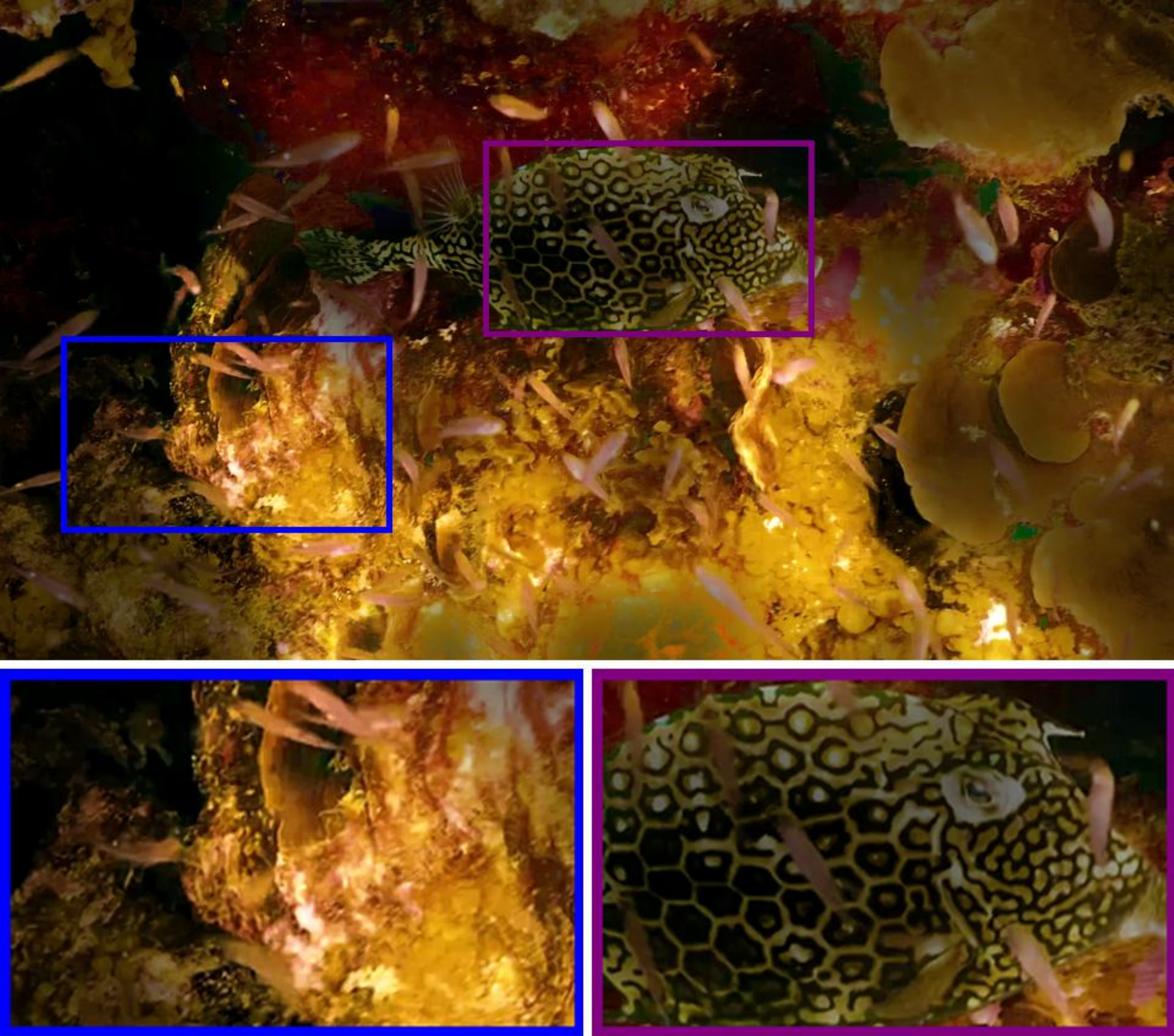} 
        \includegraphics[width=\linewidth,  height=\puniheight]{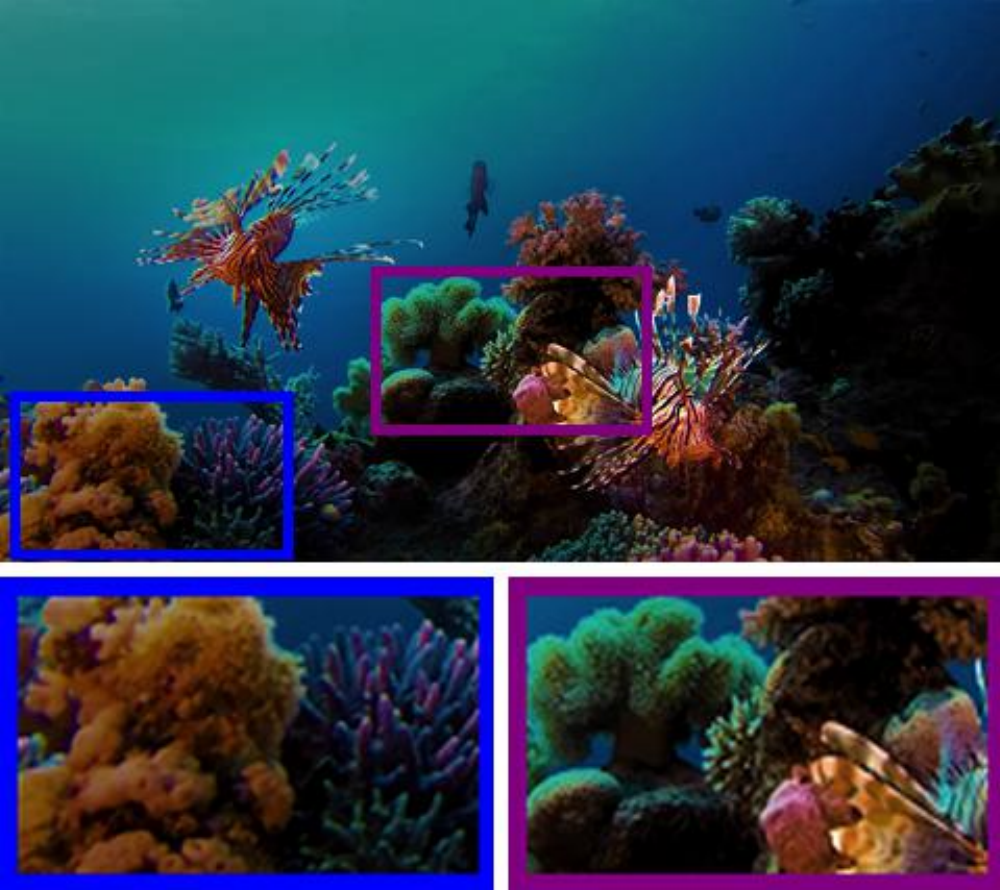} 
        \includegraphics[width=\linewidth,  height=\puniheight]{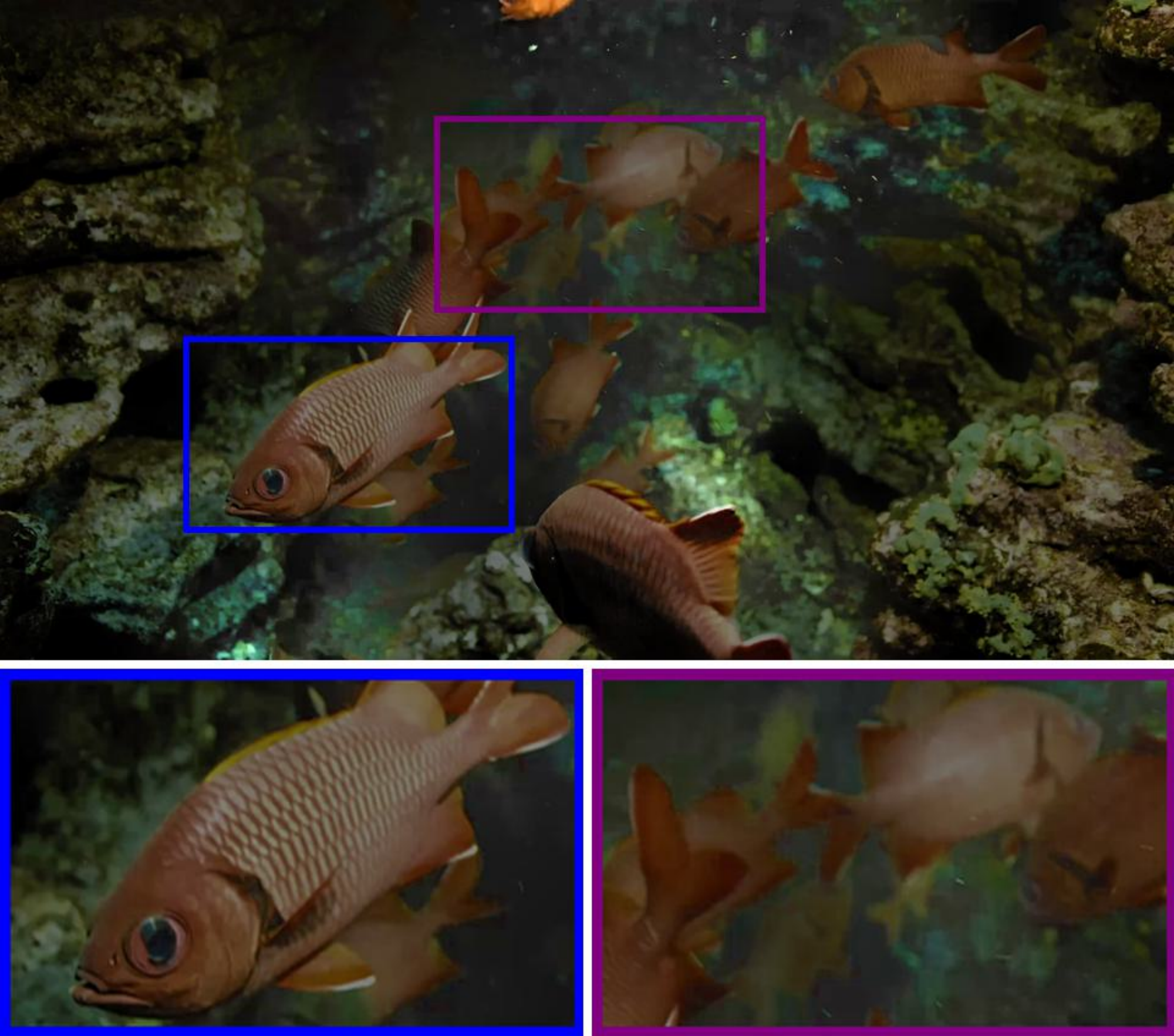} 
        \caption{\footnotesize NUI Image}
	\end{subfigure}
	\begin{subfigure}{0.105\linewidth}
		\centering
		\includegraphics[width=\linewidth,  height=\puniheight]{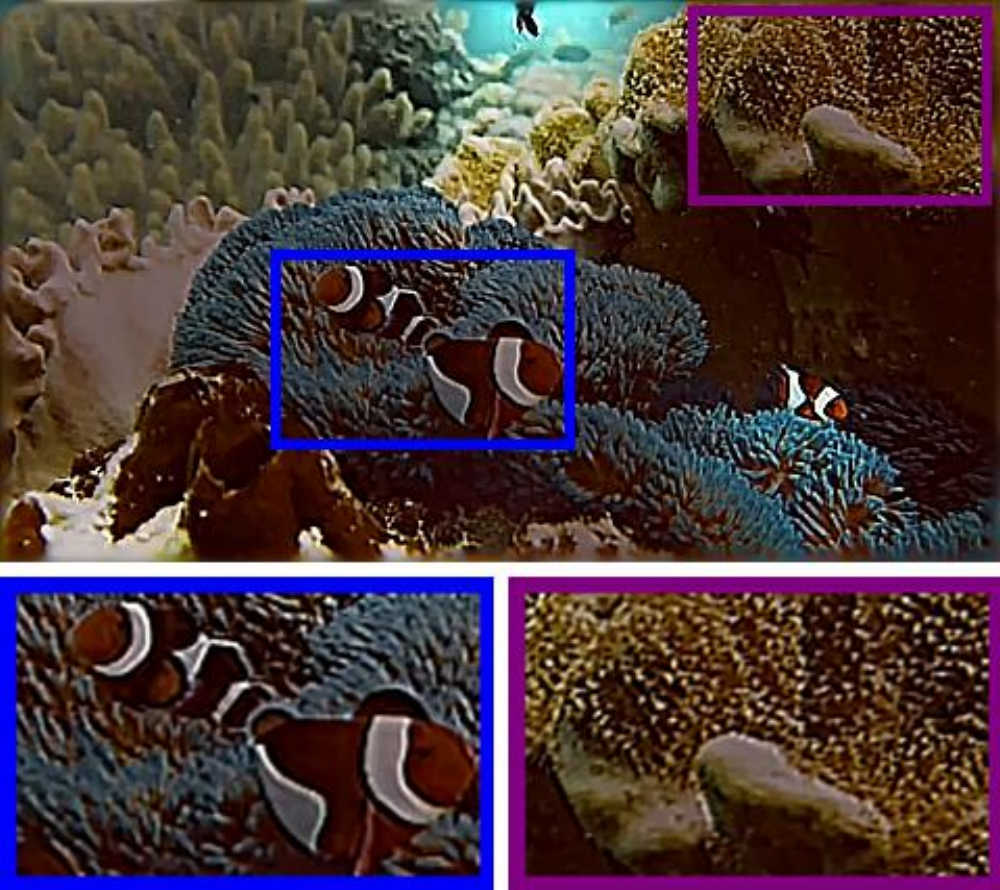} 
        \includegraphics[width=\linewidth,  height=\puniheight]{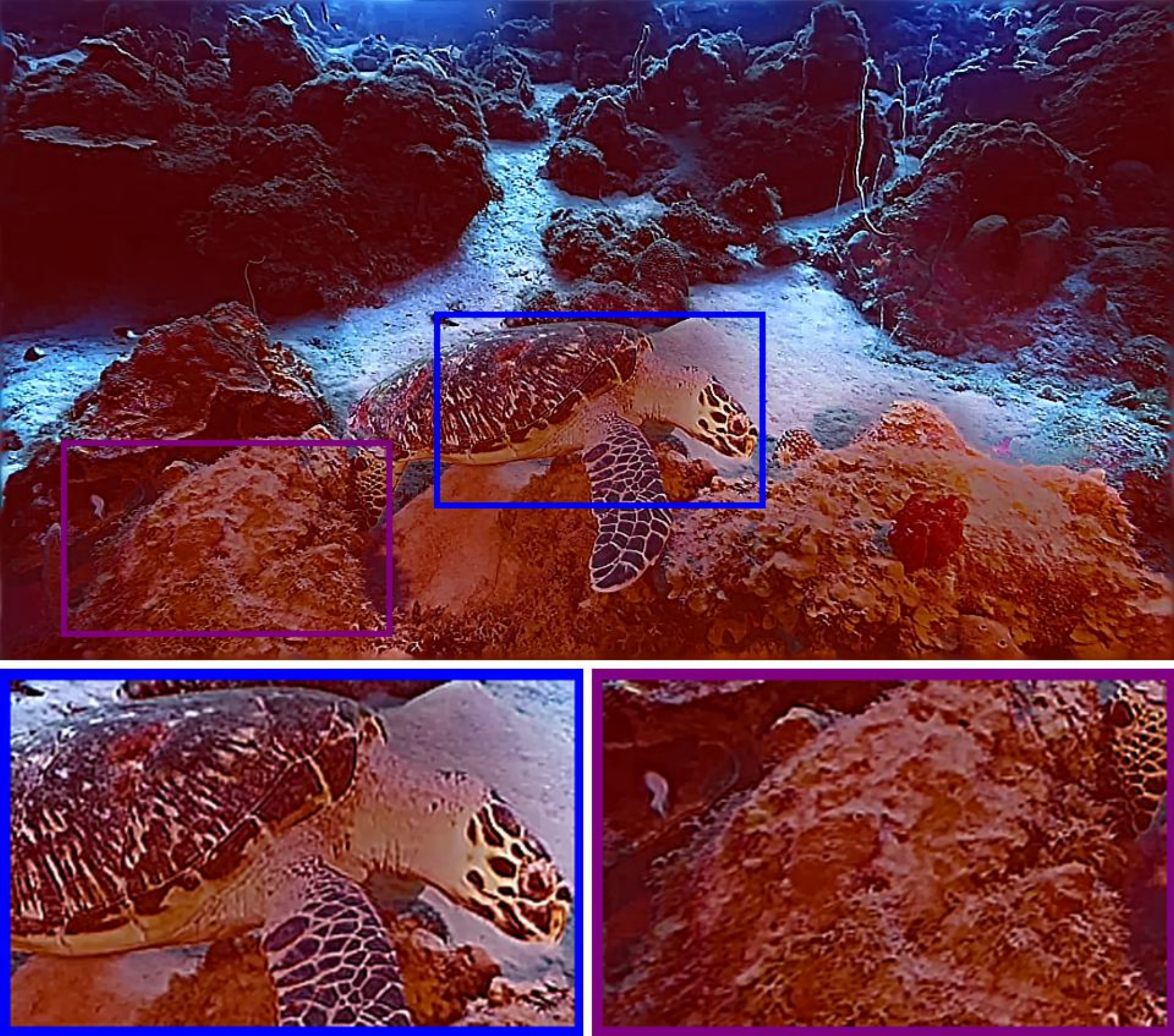}
        \includegraphics[width=\linewidth,  height=\puniheight]{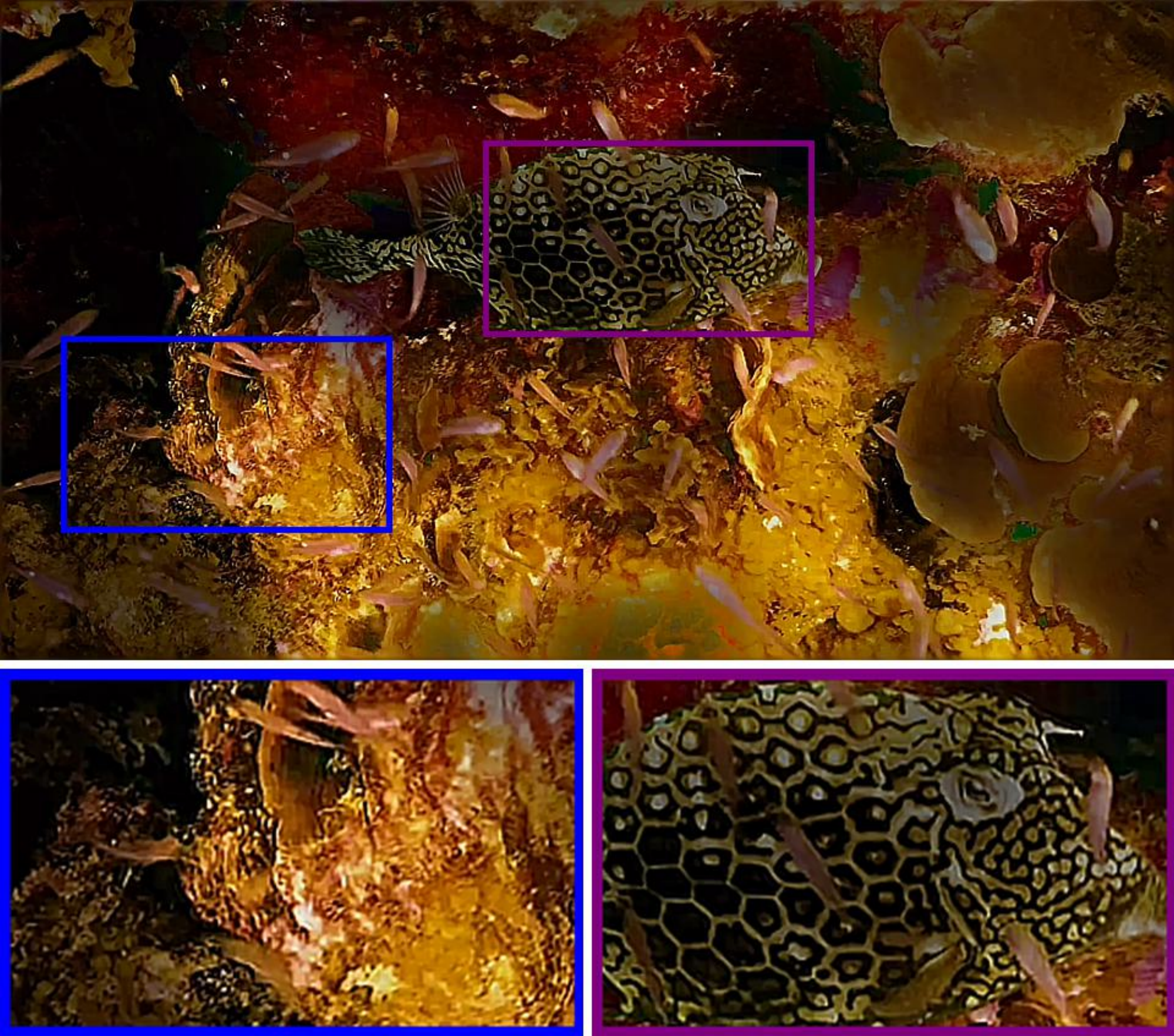}
        \includegraphics[width=\linewidth,  height=\puniheight]{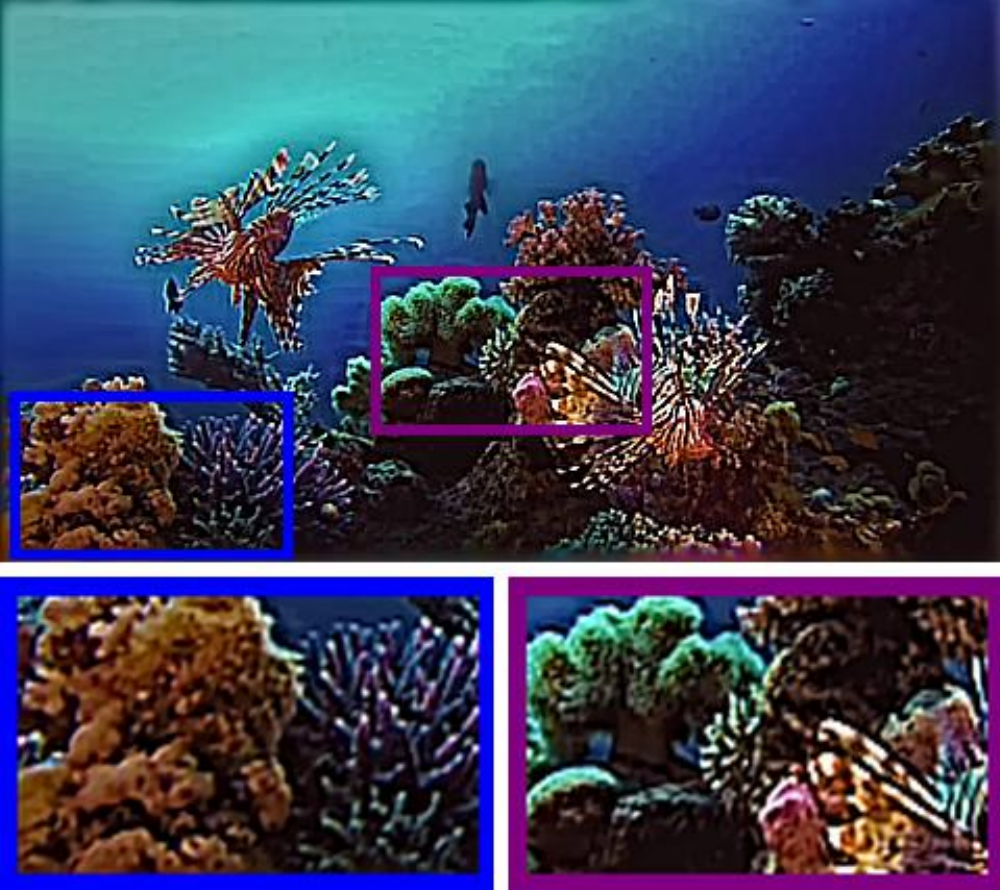}
        \includegraphics[width=\linewidth,  height=\puniheight]{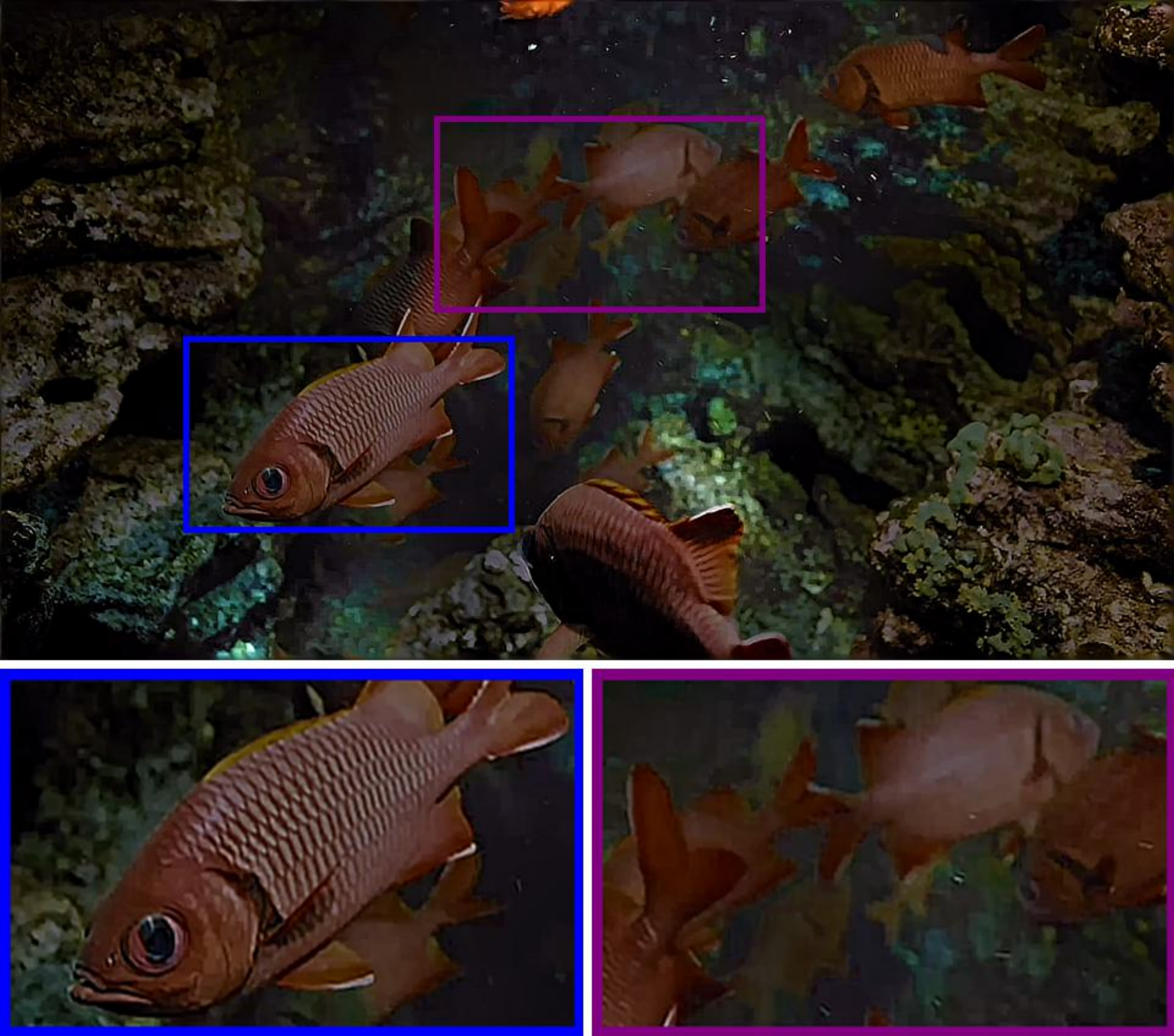}
		\caption{\footnotesize UNTV}
	\end{subfigure}
	\begin{subfigure}{0.105\linewidth}
		\centering
		\includegraphics[width=\linewidth,  height=\puniheight]{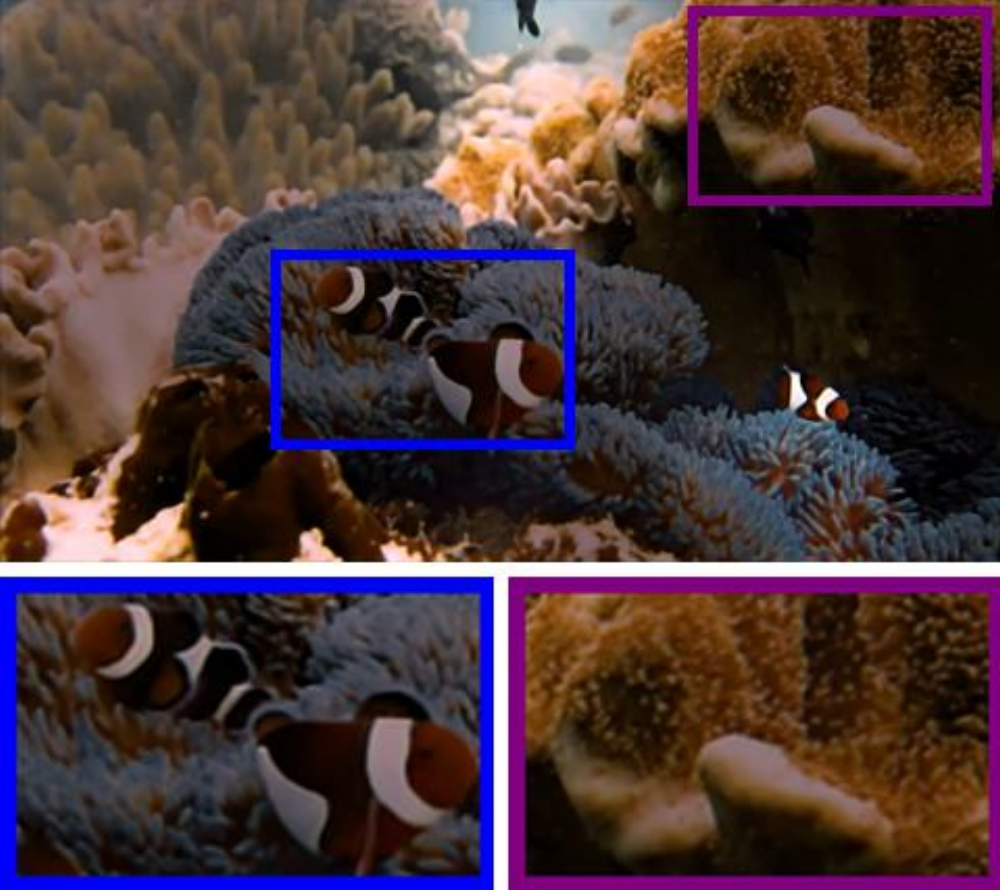}
        \includegraphics[width=\linewidth,  height=\puniheight]{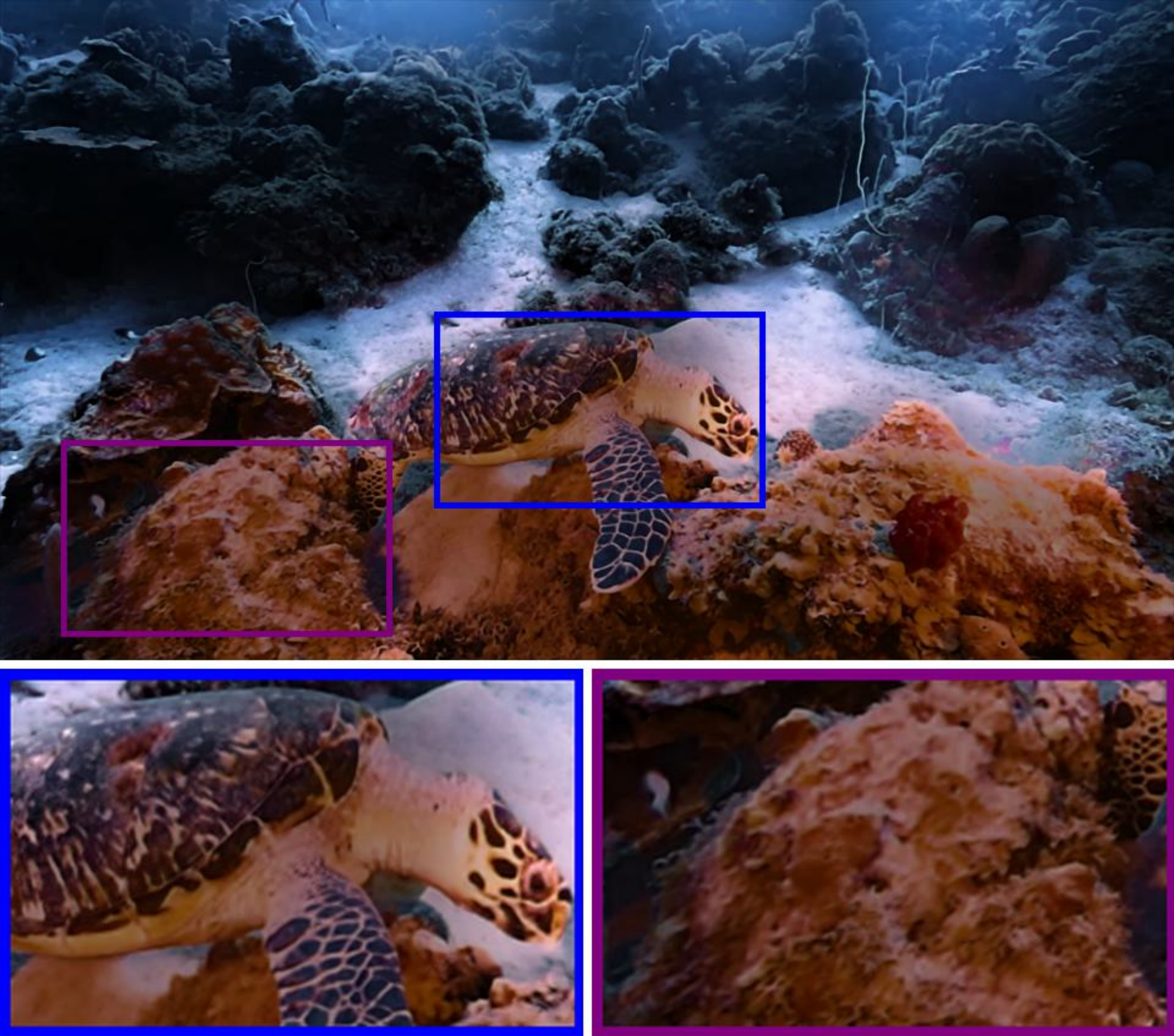}
        \includegraphics[width=\linewidth,  height=\puniheight]{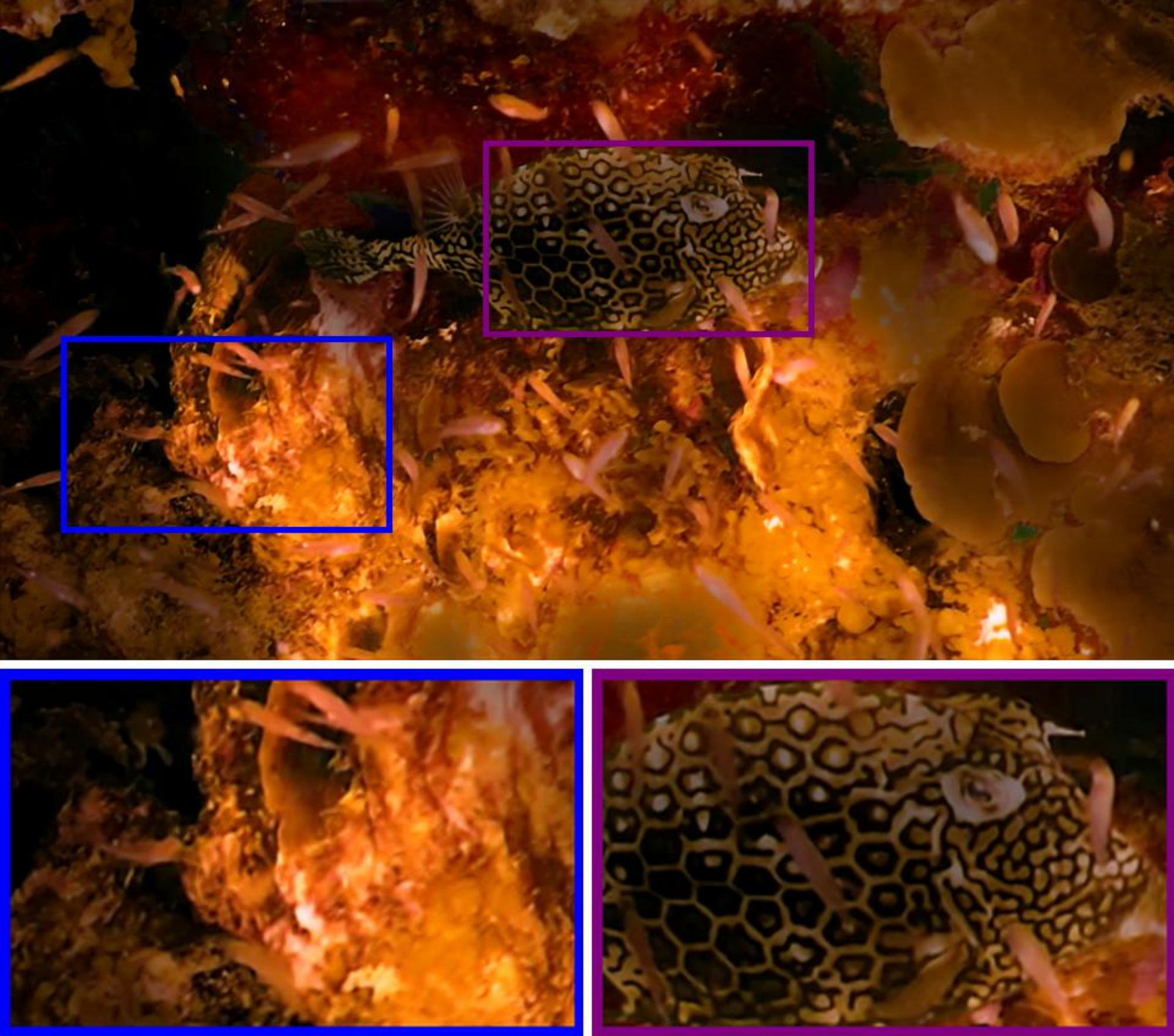}
        \includegraphics[width=\linewidth,  height=\puniheight]{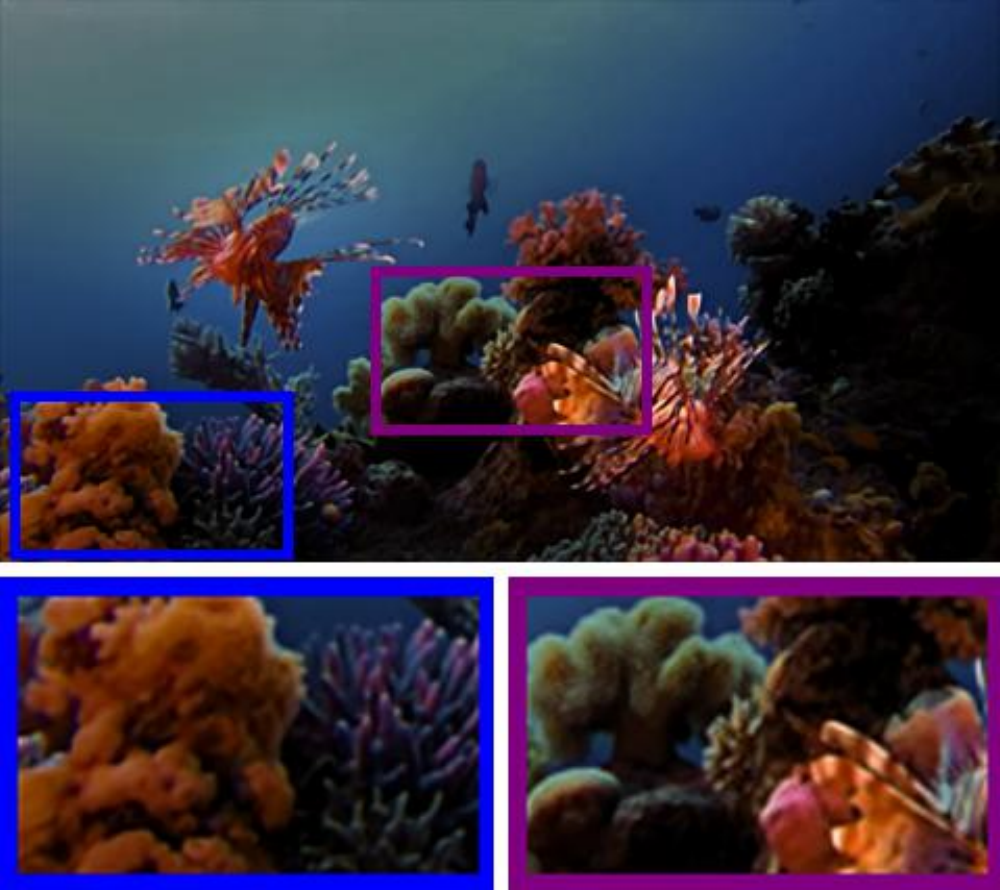}
        \includegraphics[width=\linewidth,  height=\puniheight]{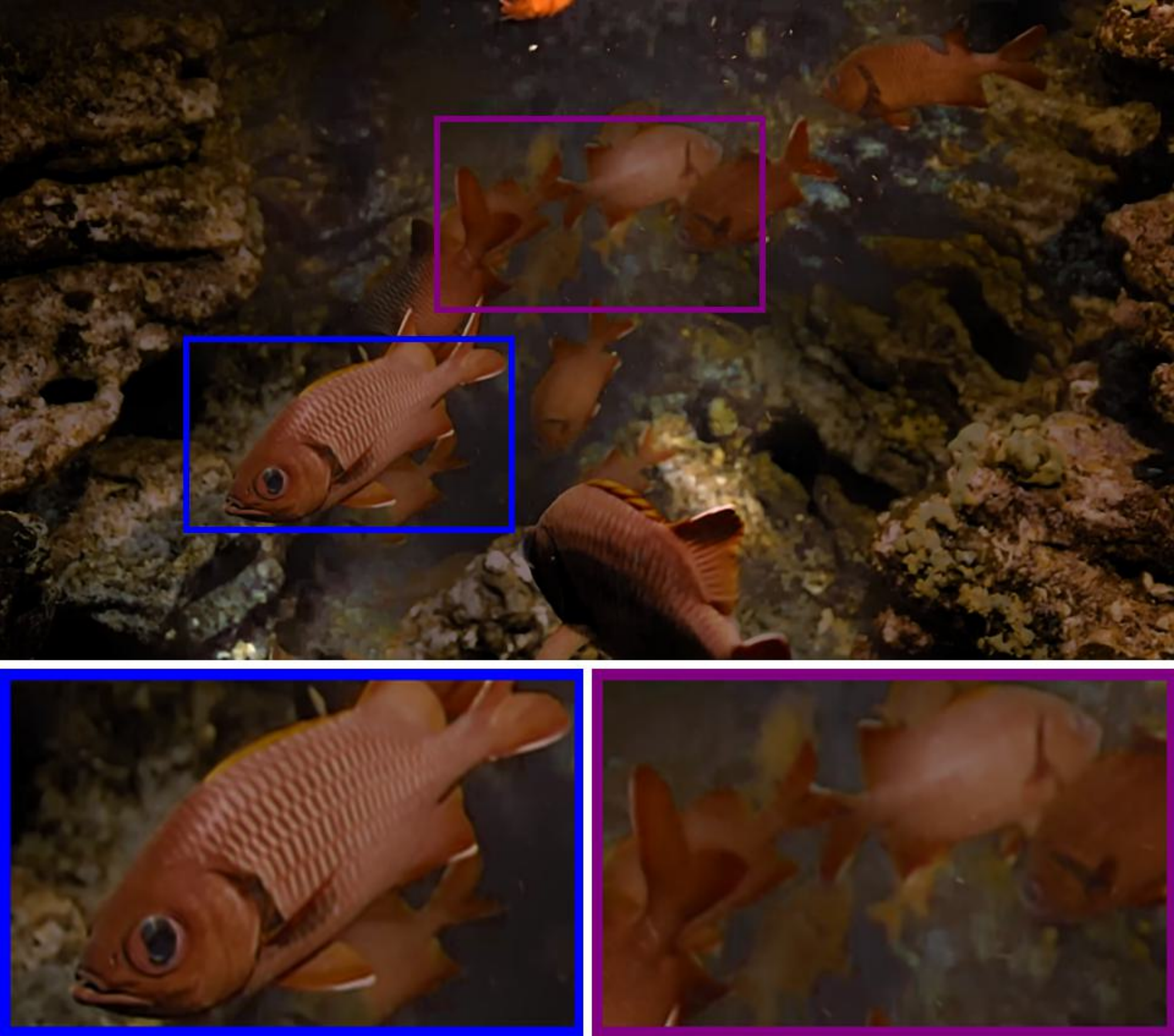}
		\caption{\footnotesize UWNet}
	\end{subfigure}
    \begin{subfigure}{0.105\linewidth}
		\centering
		\includegraphics[width=\linewidth,  height=\puniheight]{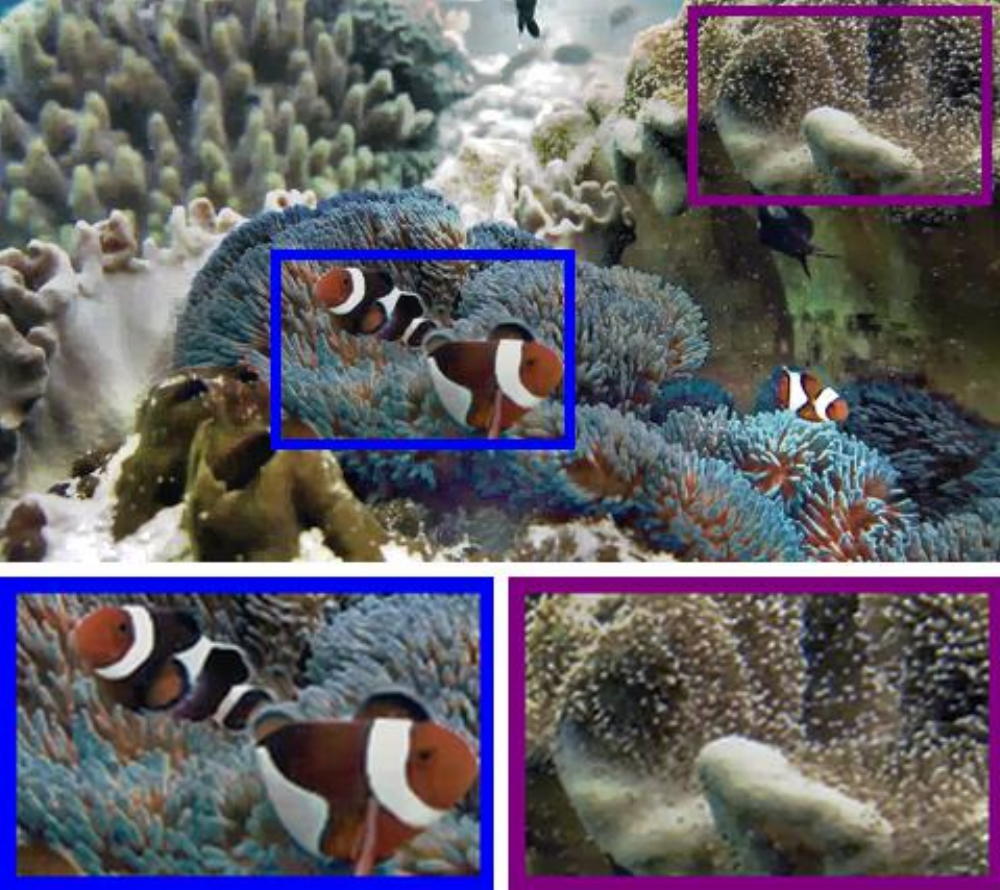} 
        \includegraphics[width=\linewidth,  height=\puniheight]{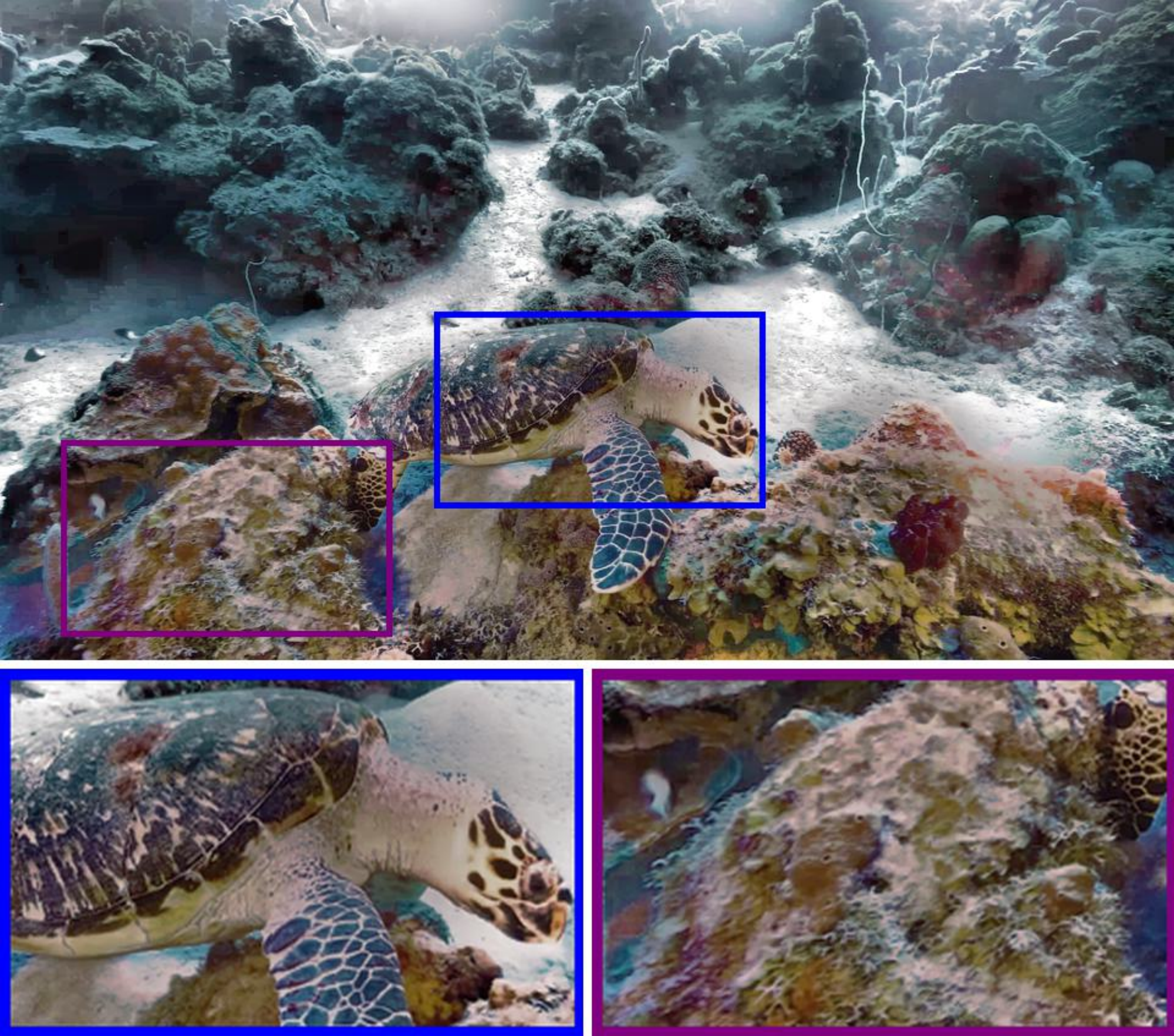} 
        \includegraphics[width=\linewidth,  height=\puniheight]{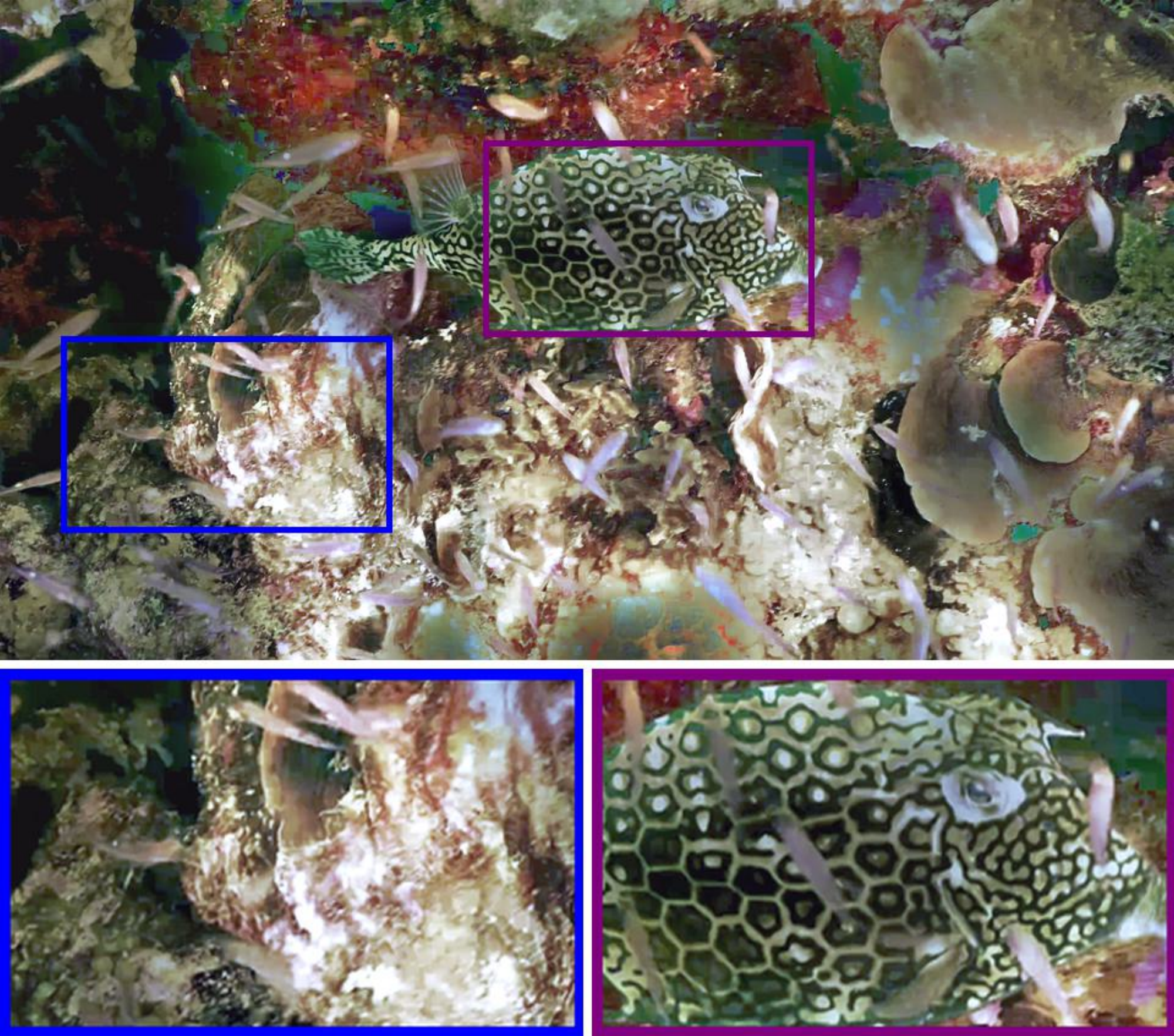} 
        \includegraphics[width=\linewidth,  height=\puniheight]{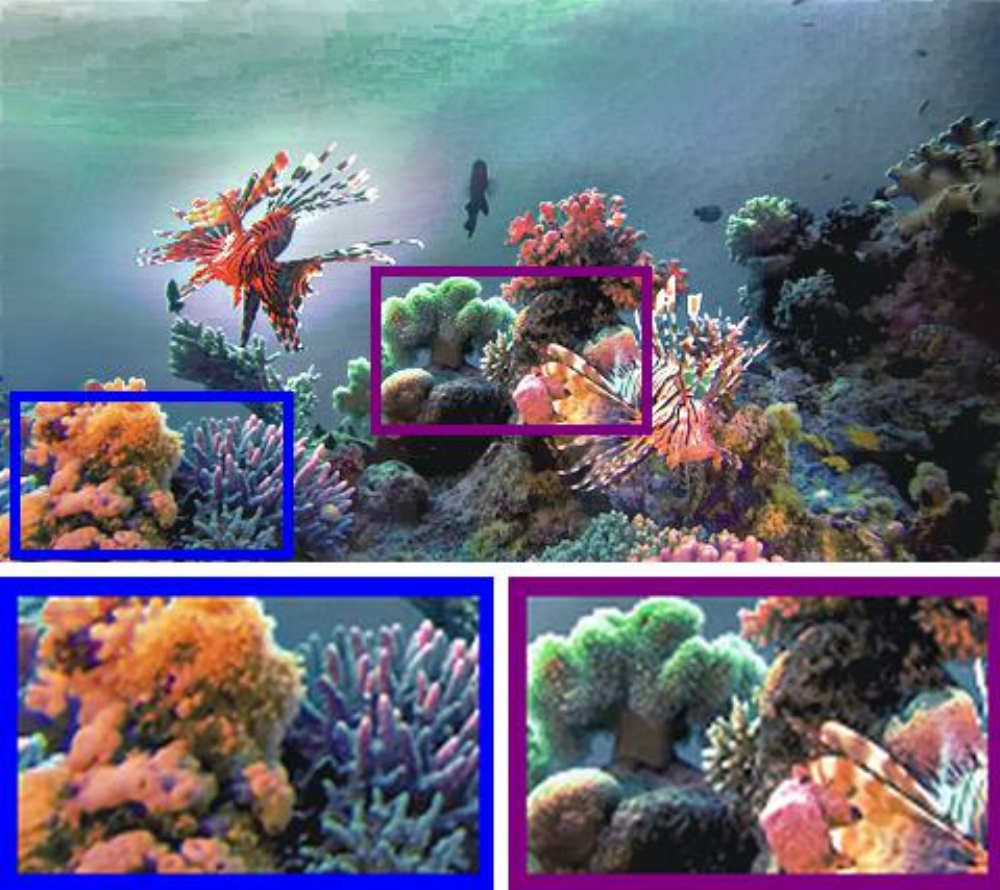} 
        \includegraphics[width=\linewidth,  height=\puniheight]{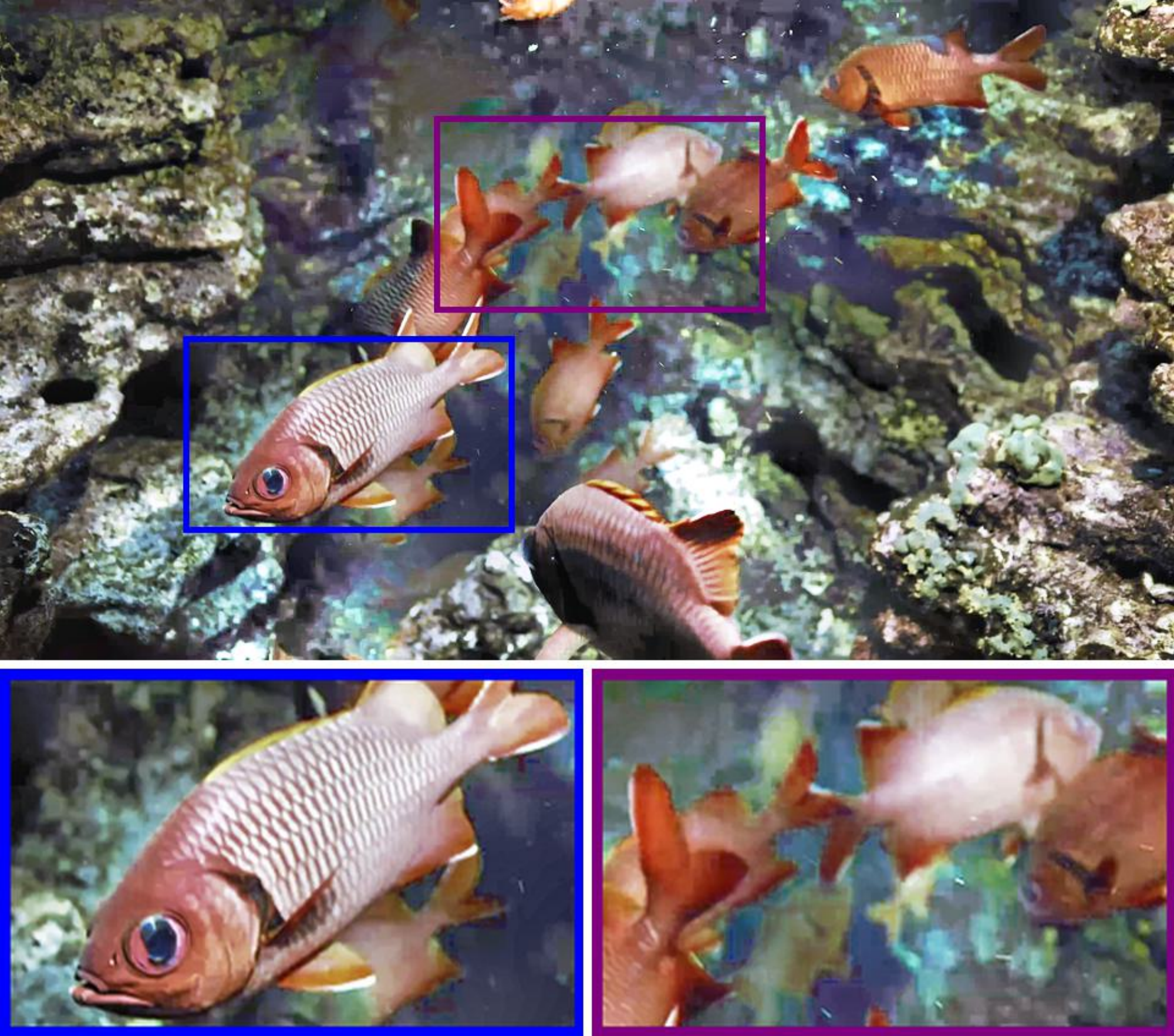}
		\caption{\footnotesize ACDC}
	\end{subfigure}
	\begin{subfigure}{0.105\linewidth}
		\centering
		\includegraphics[width=\linewidth,  height=\puniheight]{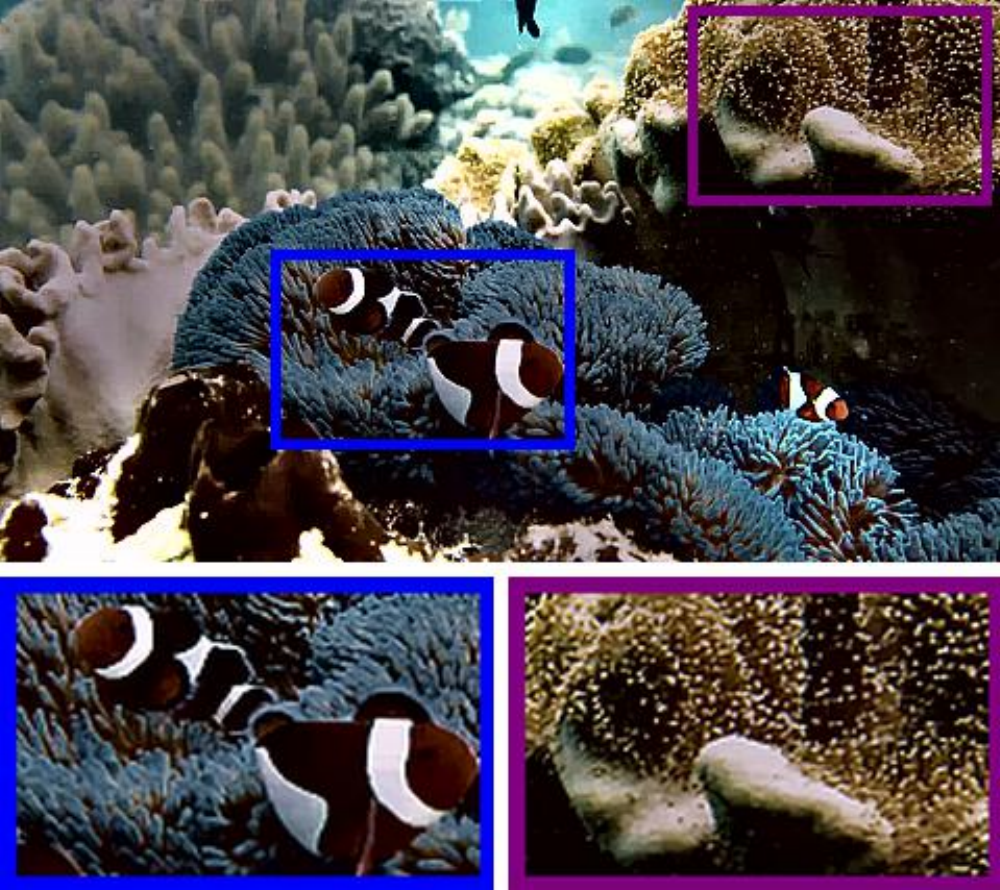} 
        \includegraphics[width=\linewidth,  height=\puniheight]{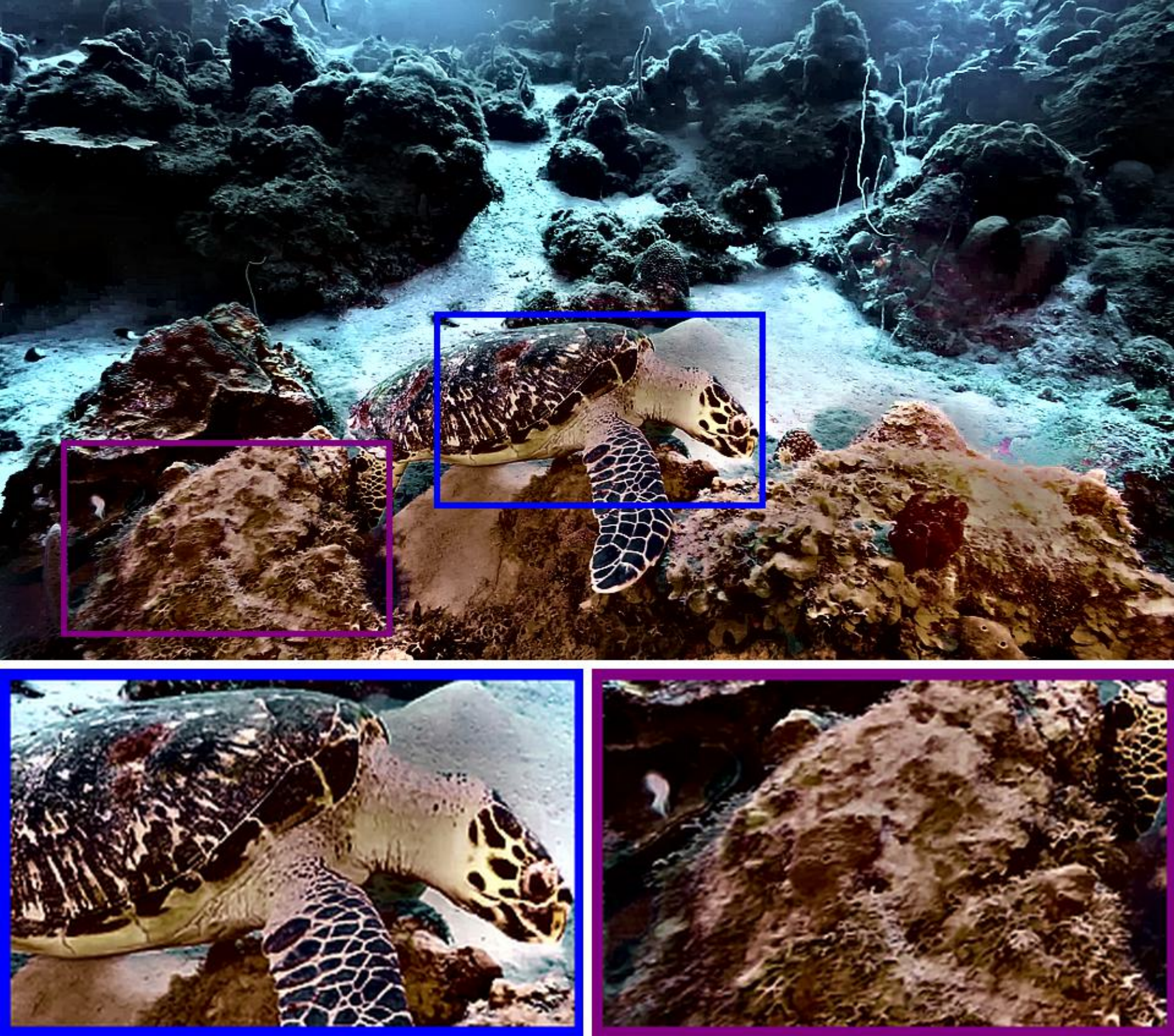} 
        \includegraphics[width=\linewidth,  height=\puniheight]{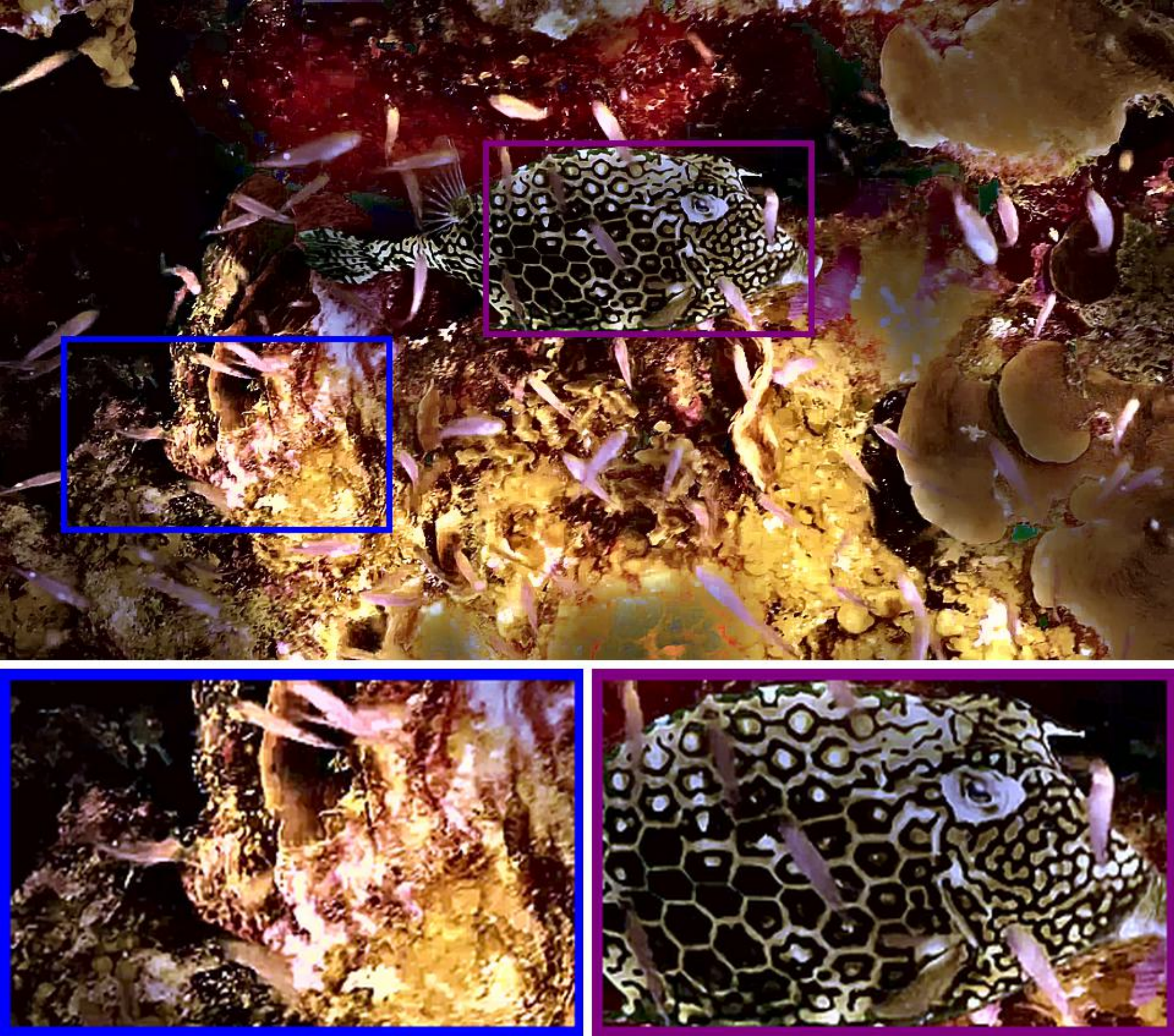} 
        \includegraphics[width=\linewidth,  height=\puniheight]{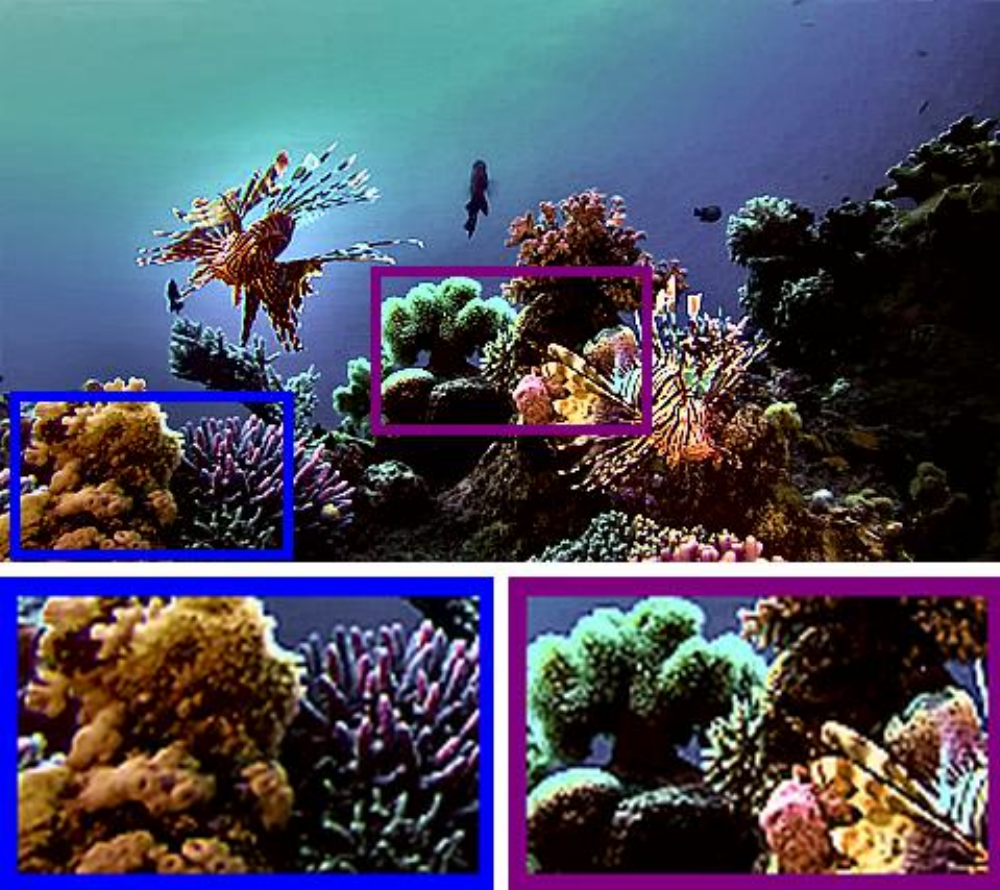} 
        \includegraphics[width=\linewidth,  height=\puniheight]{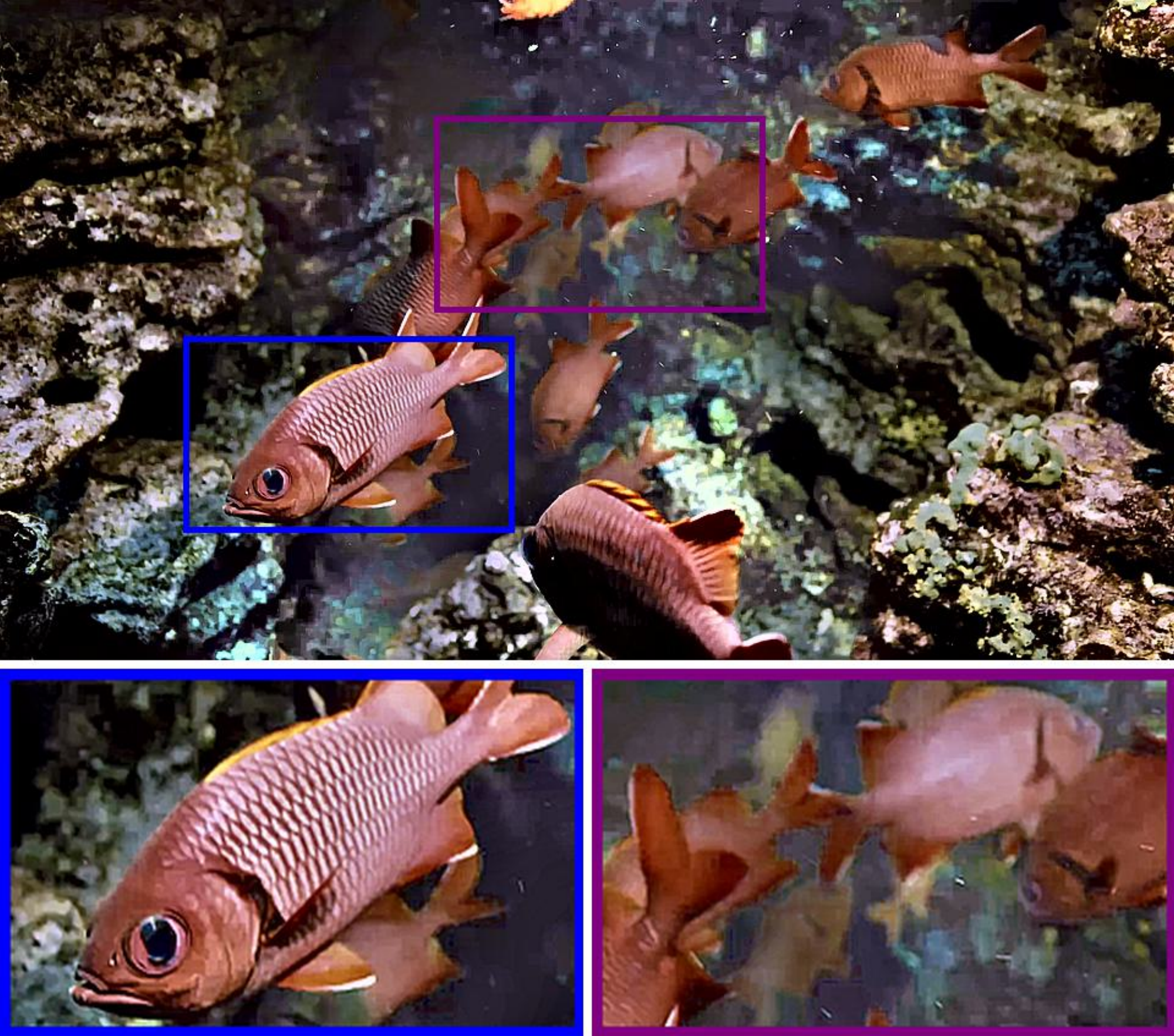} 
		\caption{\footnotesize MMLE}
	\end{subfigure}
    \begin{subfigure}{0.105\linewidth}
		\centering
		\includegraphics[width=\linewidth,  height=\puniheight]{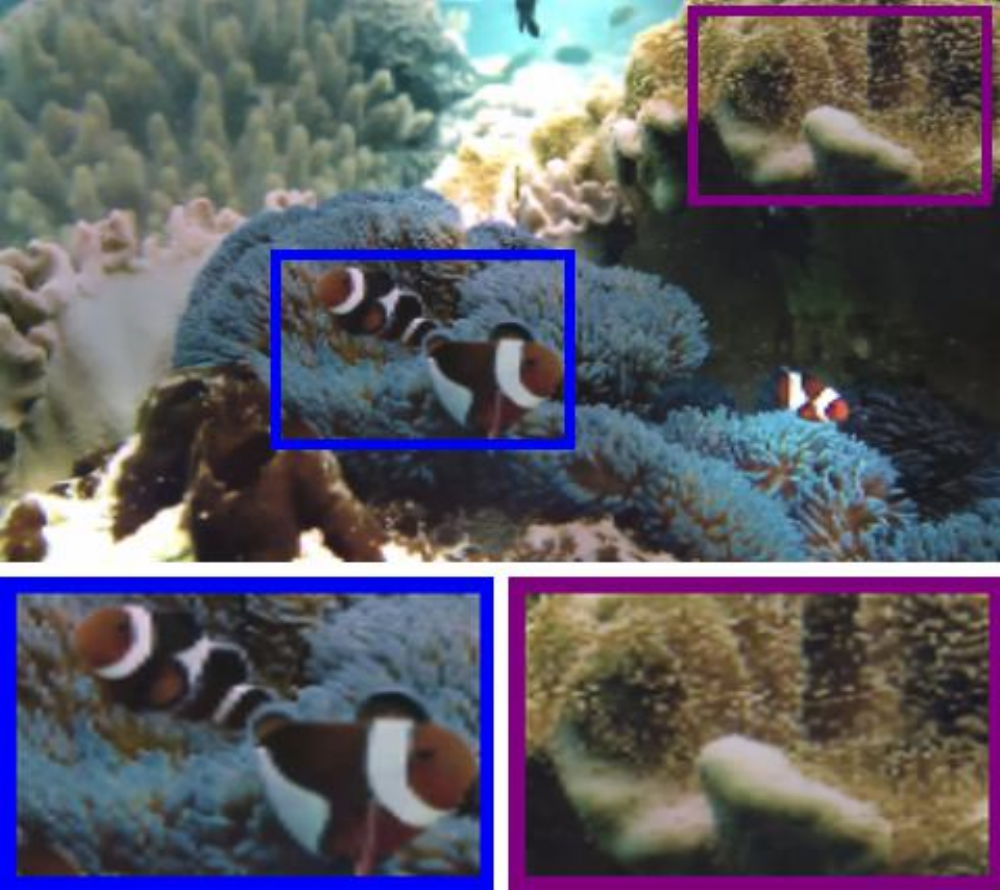} 
        \includegraphics[width=\linewidth,  height=\puniheight]{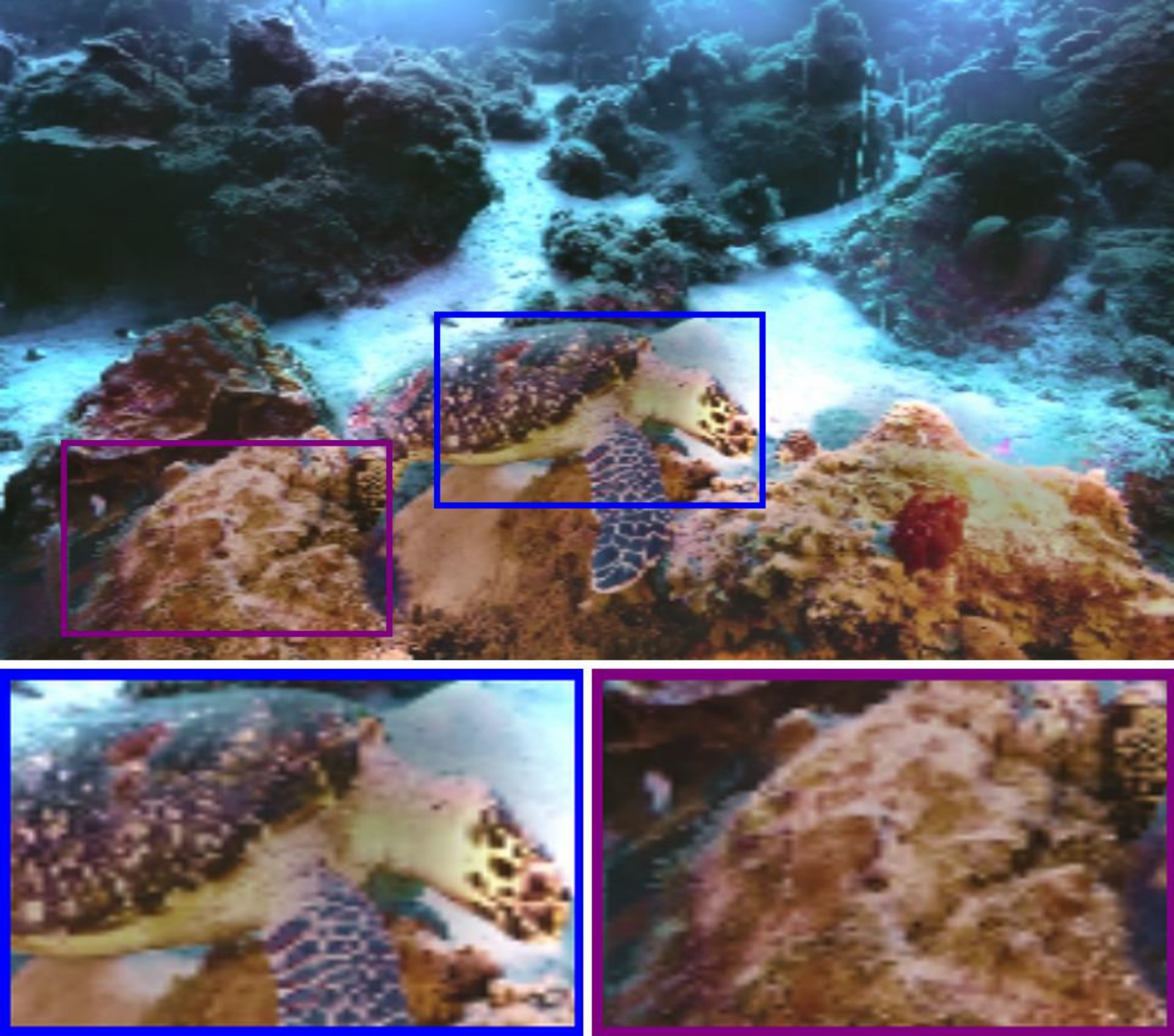} 
        \includegraphics[width=\linewidth,  height=\puniheight]{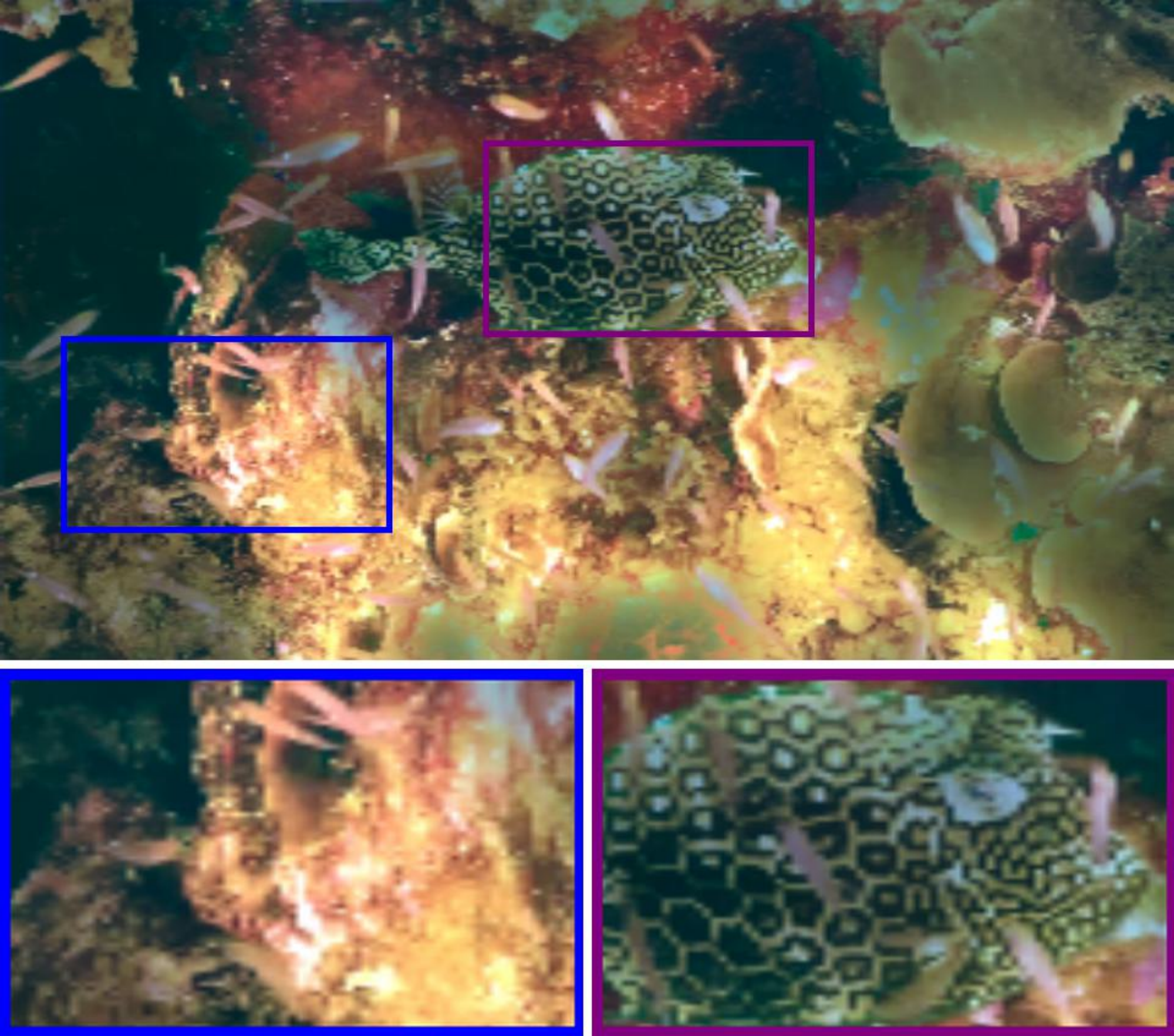} 
        \includegraphics[width=\linewidth,  height=\puniheight]{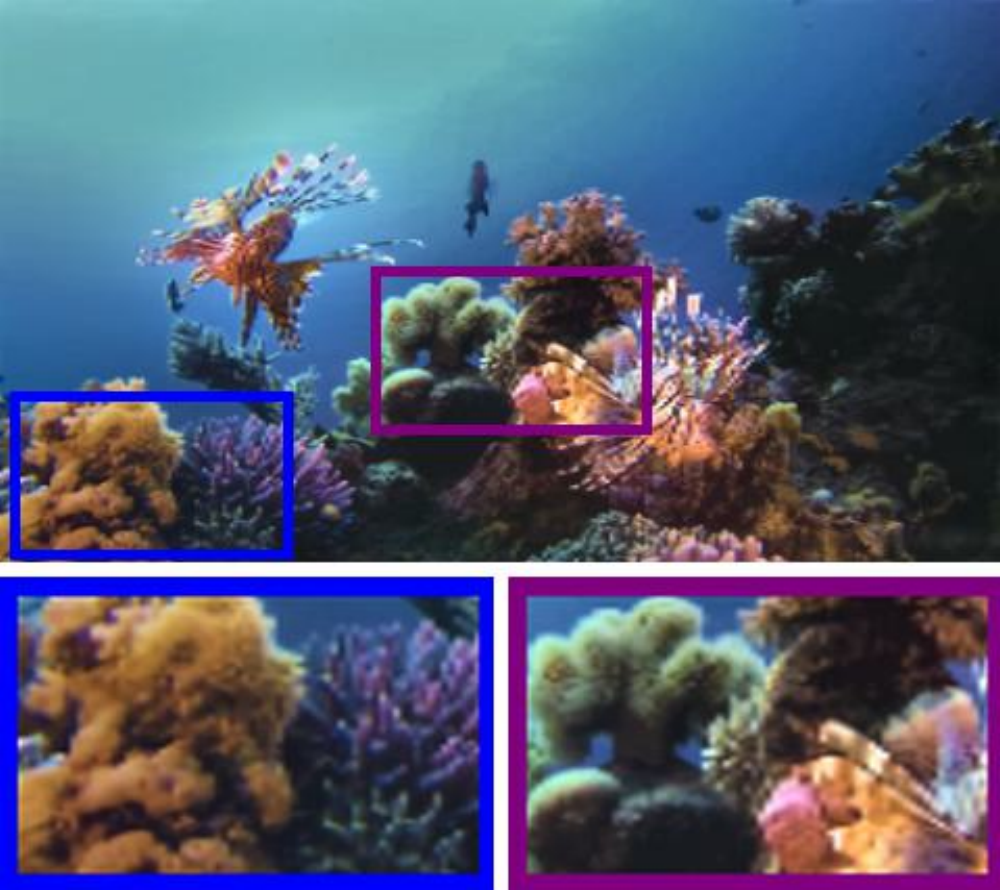}
        \includegraphics[width=\linewidth,  height=\puniheight]{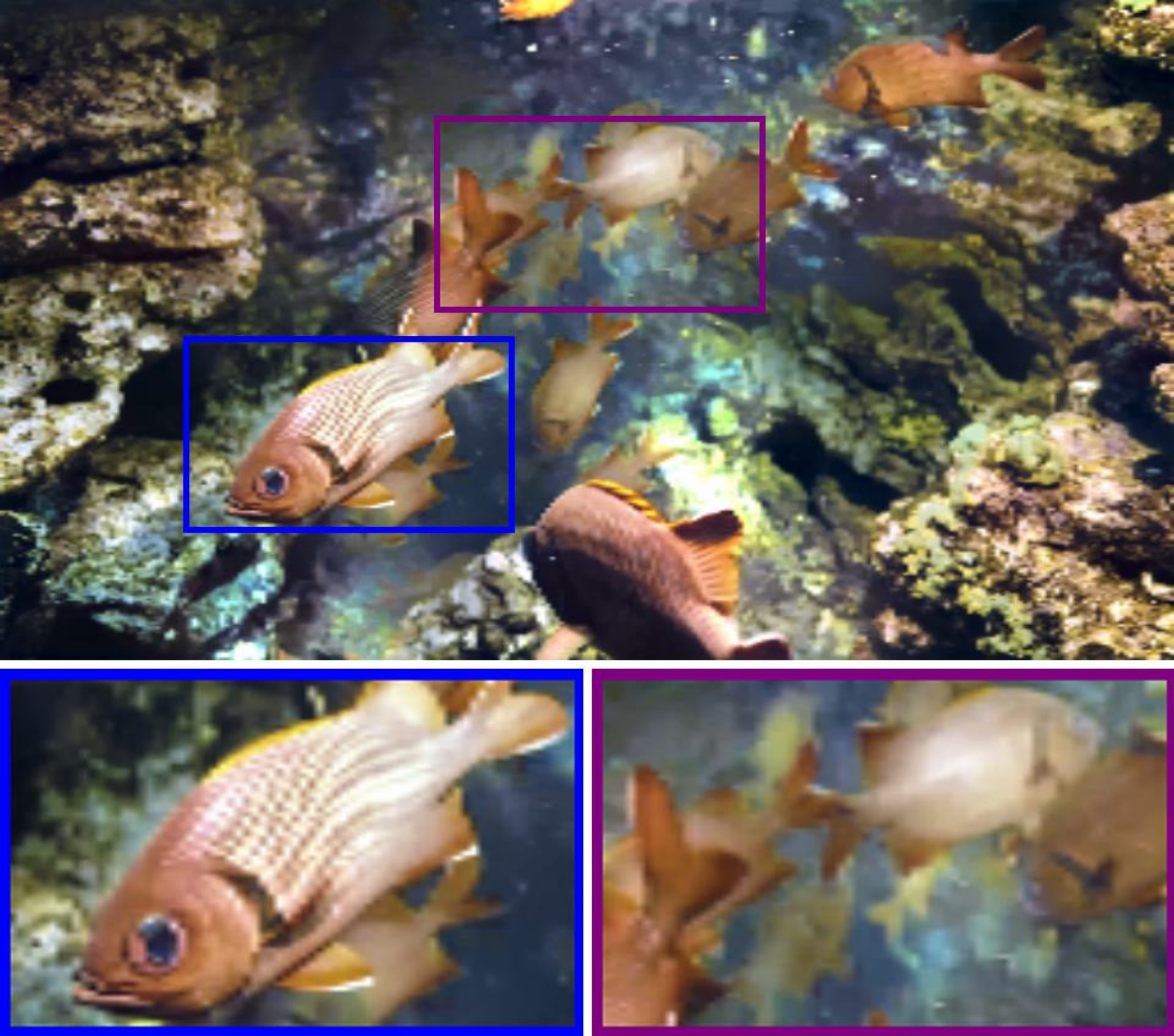} 
		\caption{\footnotesize TCTL-Net}
	\end{subfigure}
    \begin{subfigure}{0.105\linewidth}
		\centering
		\includegraphics[width=\linewidth,  height=\puniheight]{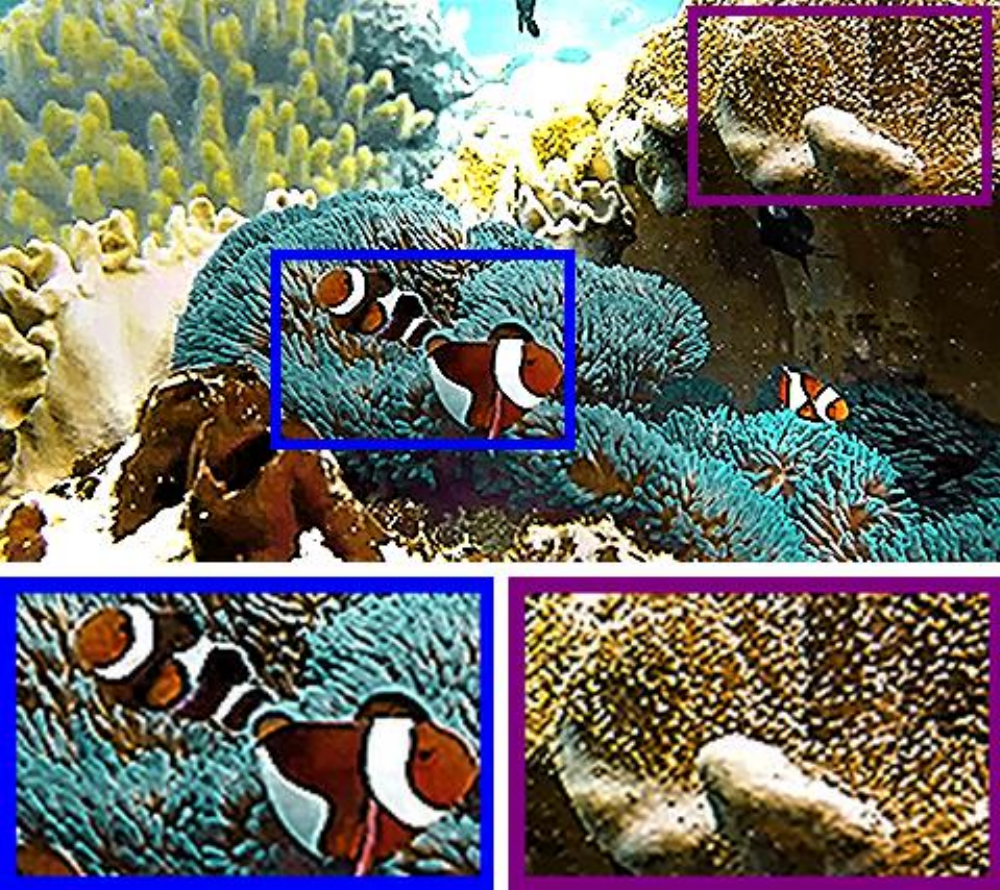} 
        \includegraphics[width=\linewidth,  height=\puniheight]{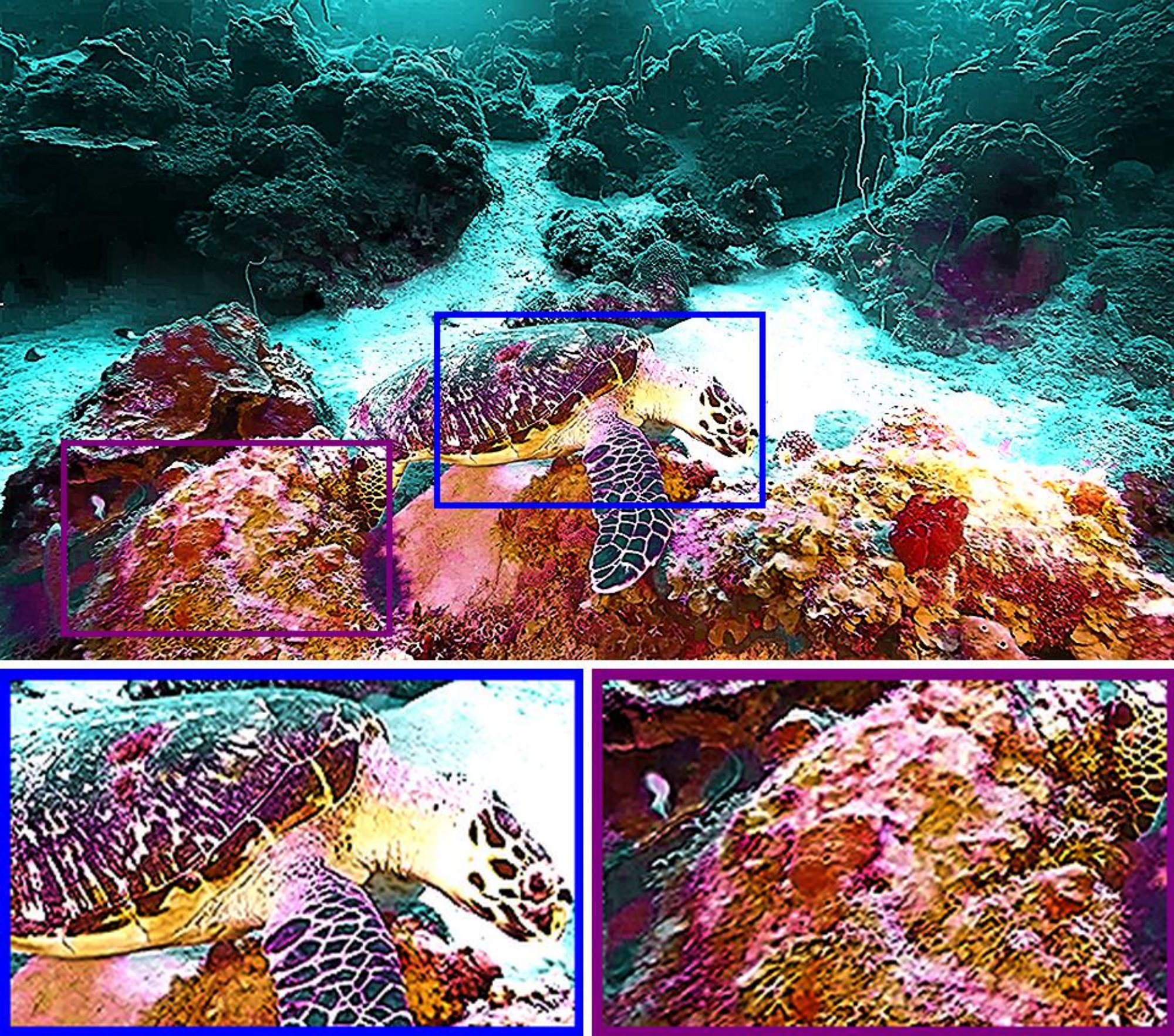} 
        \includegraphics[width=\linewidth,  height=\puniheight]{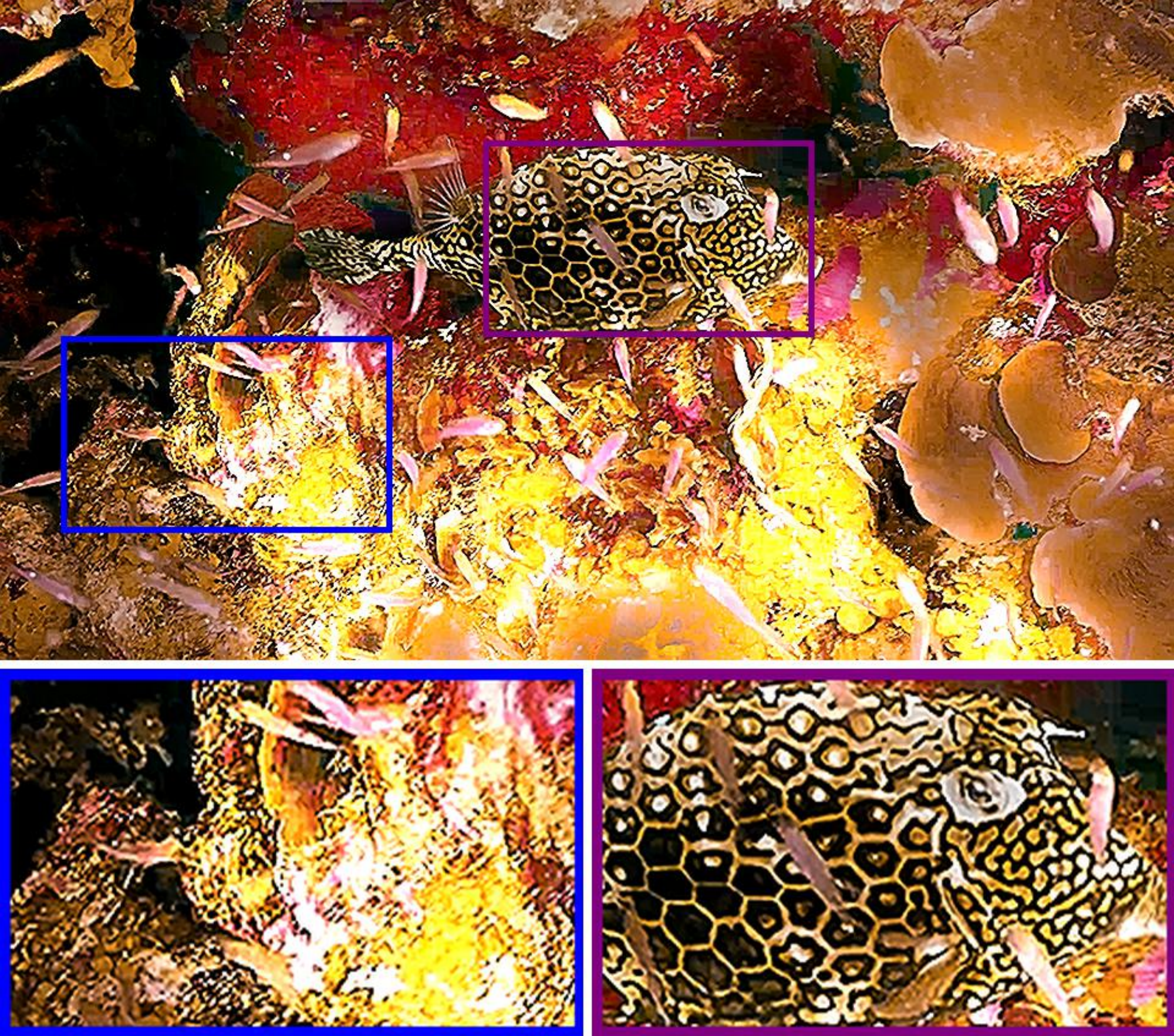}
        \includegraphics[width=\linewidth,  height=\puniheight]{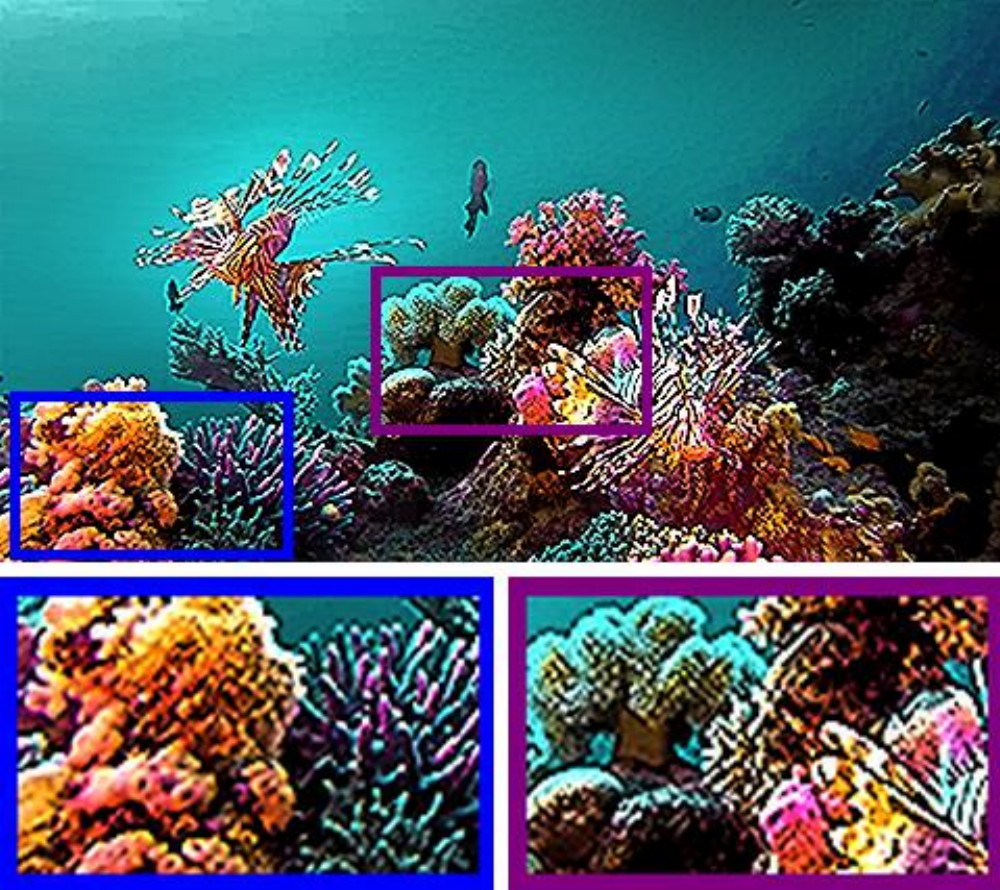}
        \includegraphics[width=\linewidth,  height=\puniheight]{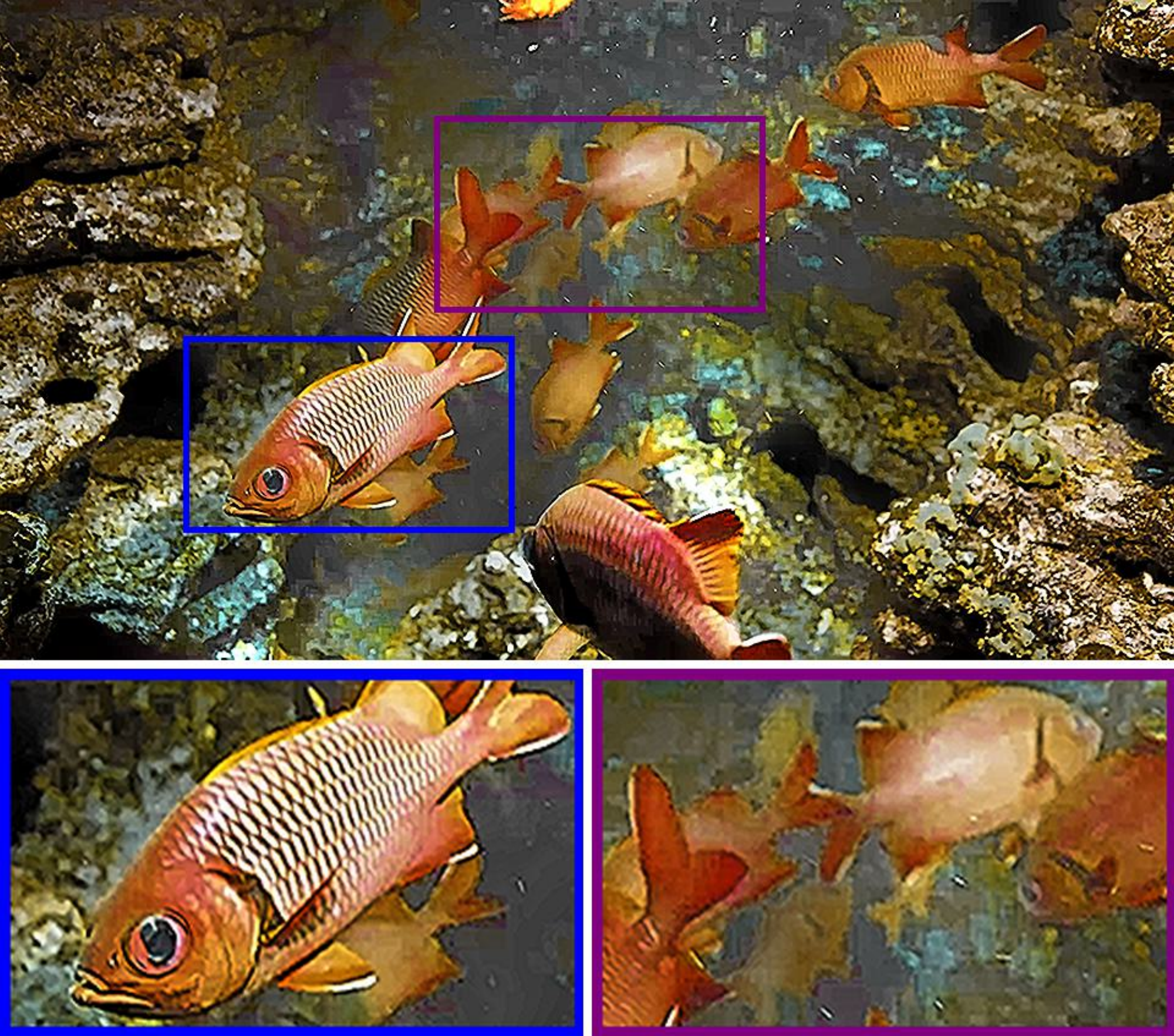}
        \caption{\footnotesize ICSP}
	\end{subfigure}
	\begin{subfigure}{0.105\linewidth}
		\centering
		\includegraphics[width=\linewidth,  height=\puniheight]{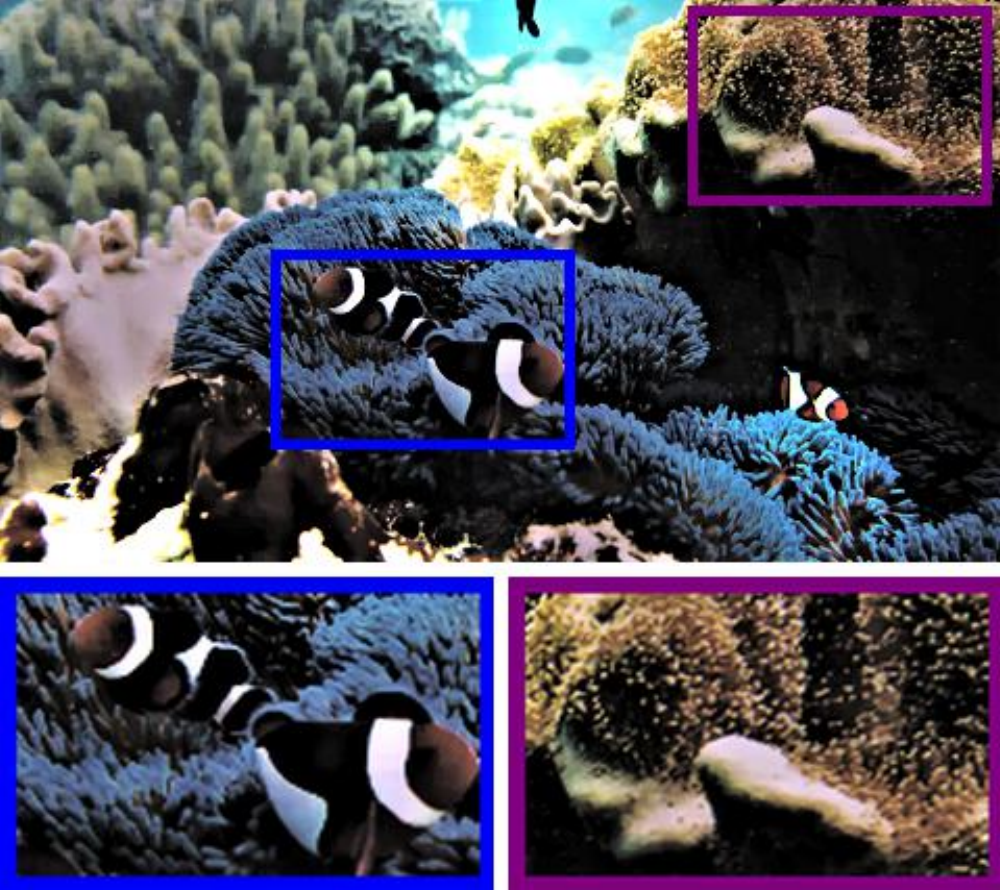} 
        \includegraphics[width=\linewidth,  height=\puniheight]{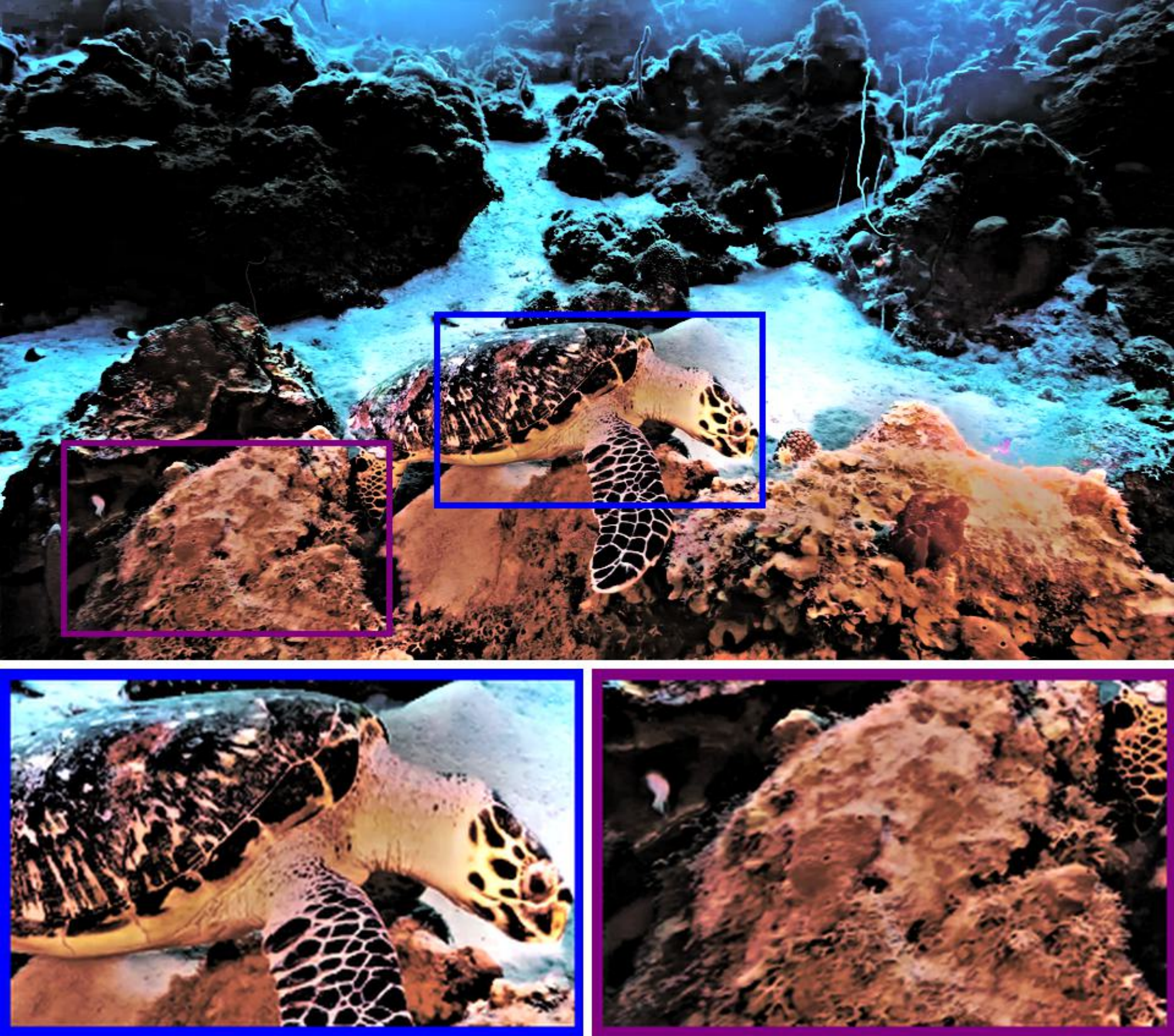} 
        \includegraphics[width=\linewidth,  height=\puniheight]{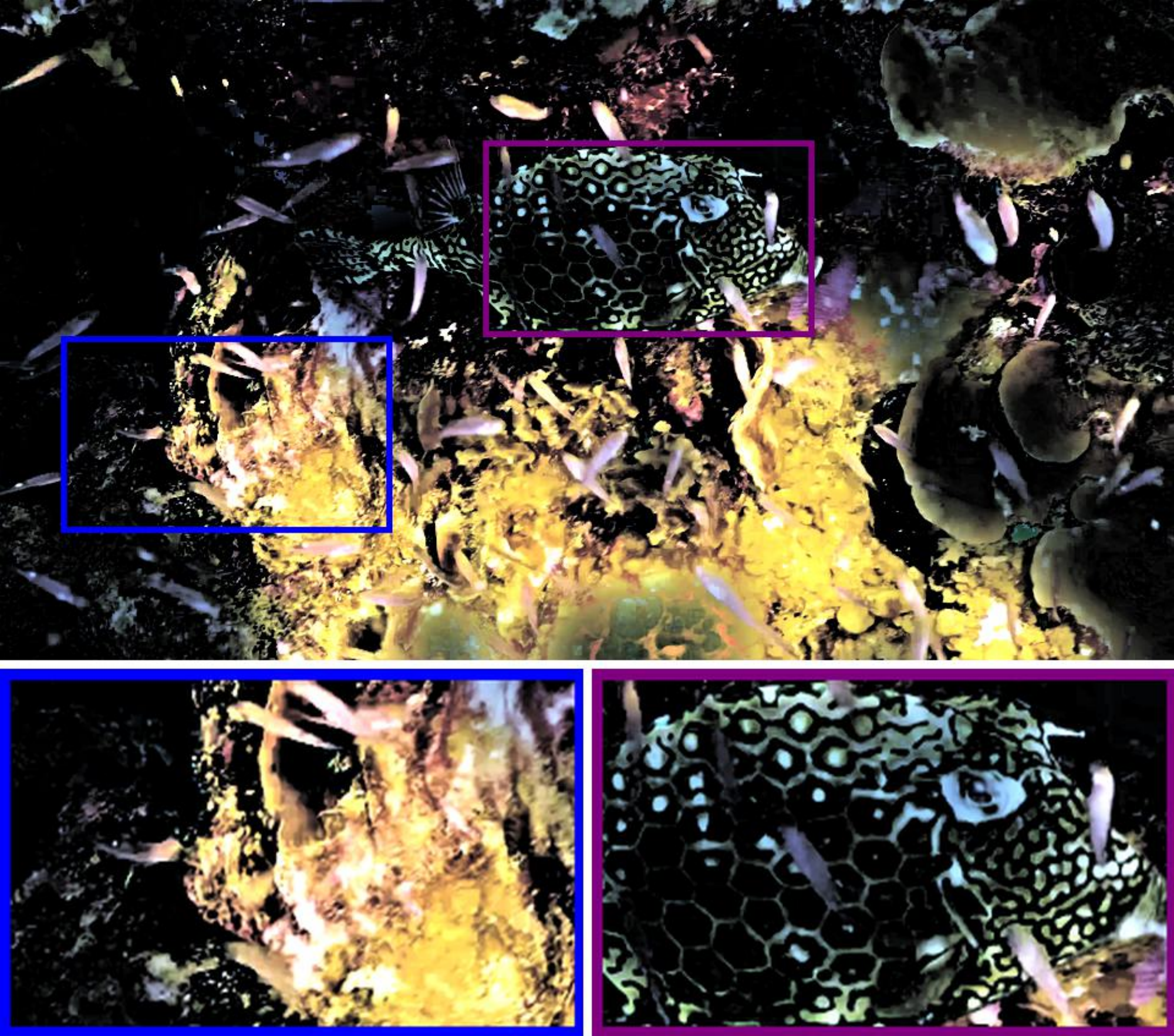}
        \includegraphics[width=\linewidth,  height=\puniheight]{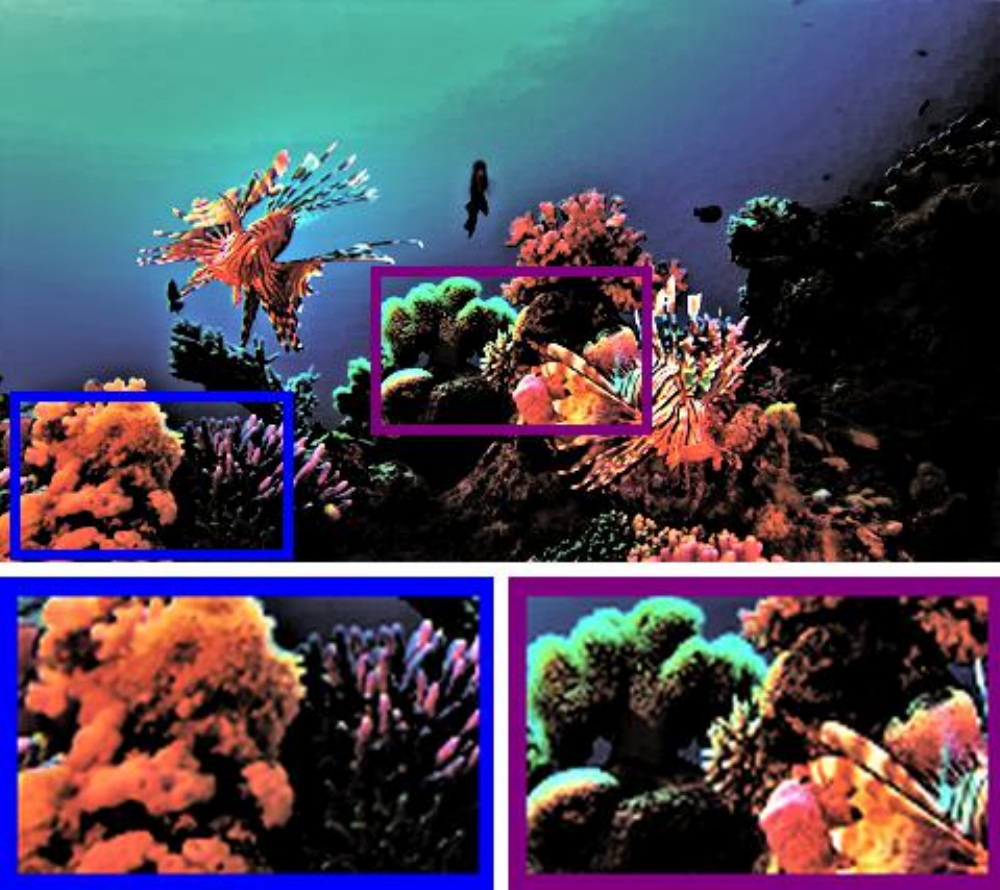} 
         \includegraphics[width=\linewidth,  height=\puniheight]{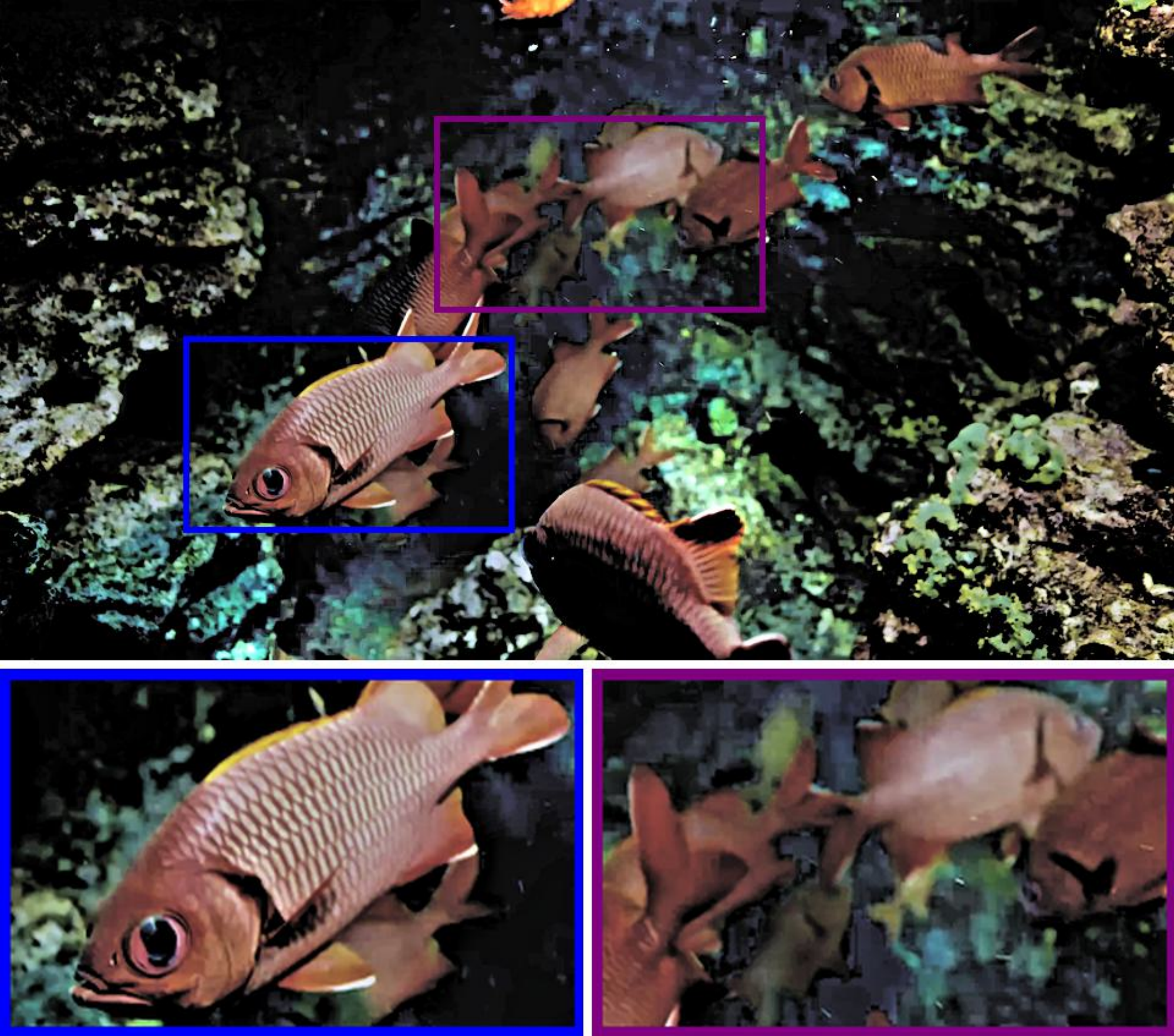} 
        \caption{\footnotesize PCDE}
	\end{subfigure}
	\begin{subfigure}{0.105\linewidth}
		\centering
		\includegraphics[width=\linewidth,  height=\puniheight]{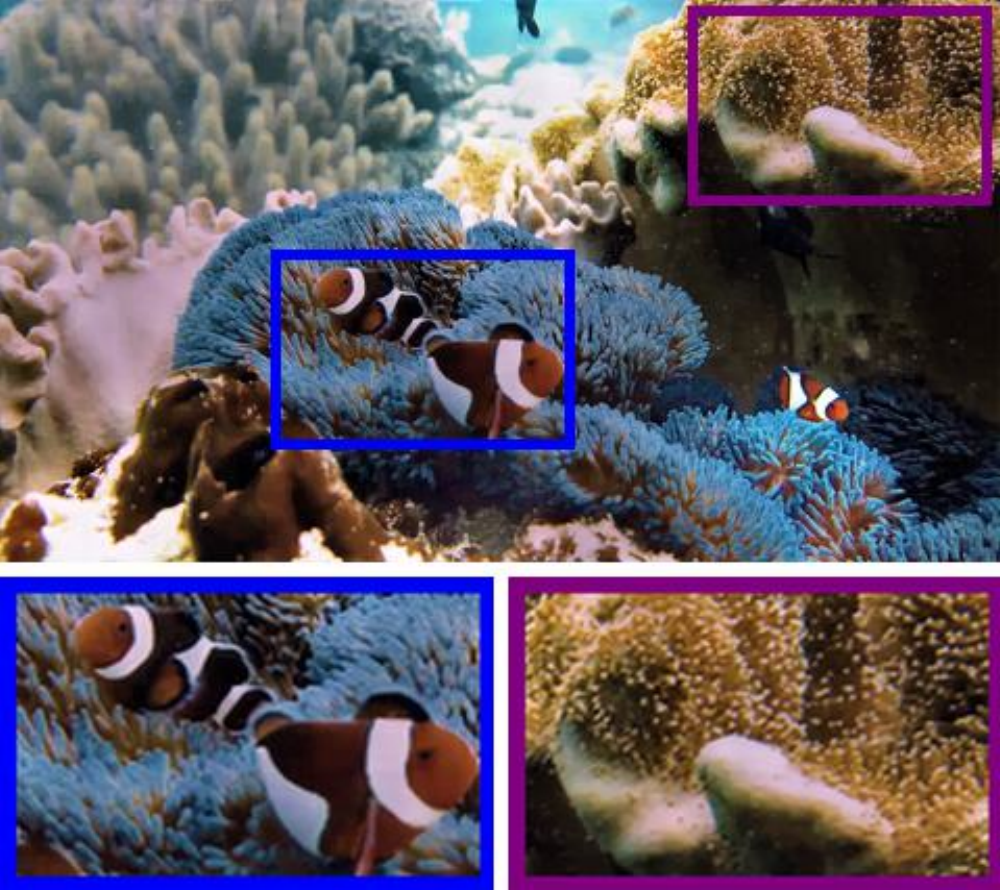}
        \includegraphics[width=\linewidth,  height=\puniheight]{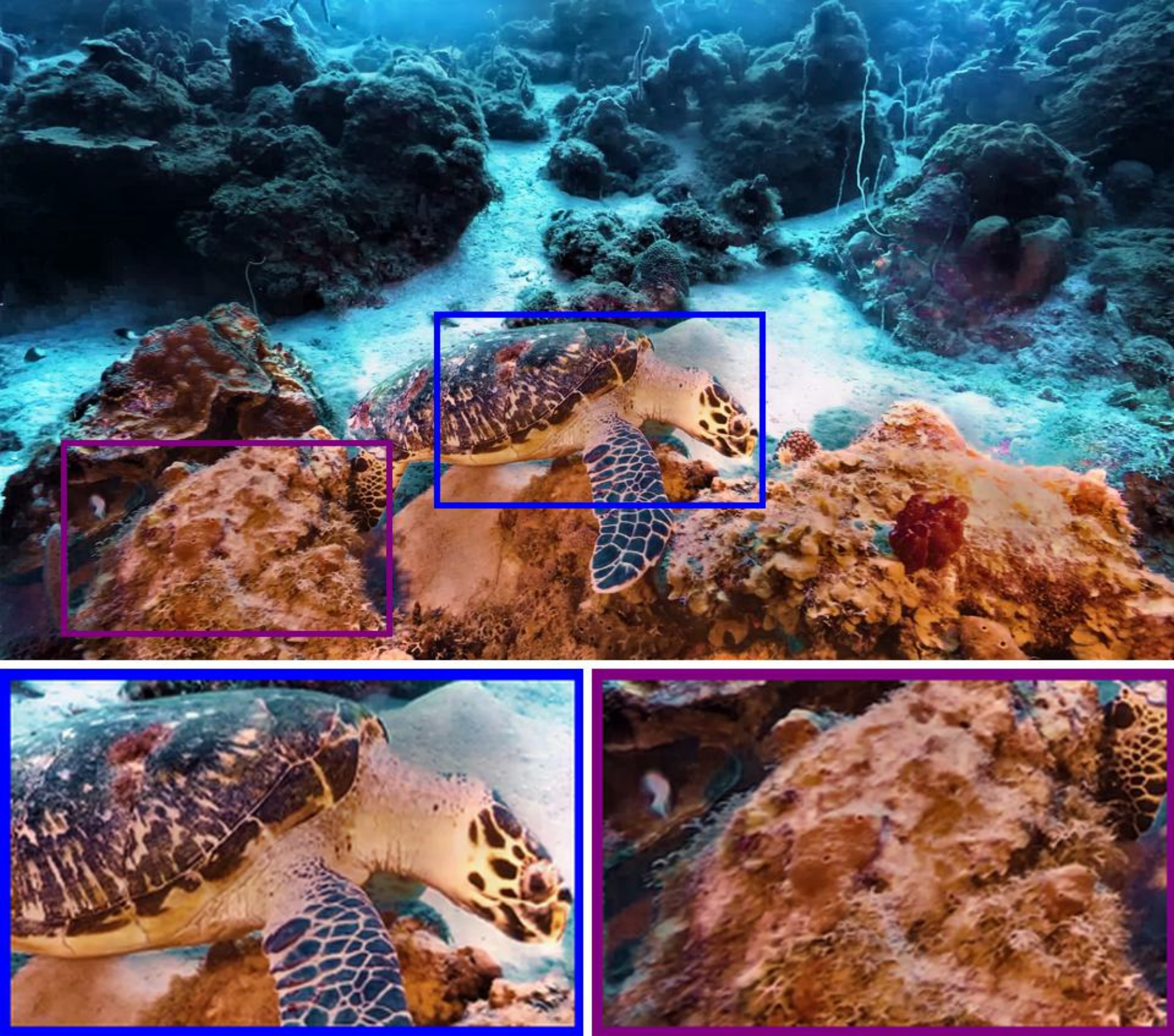}
        \includegraphics[width=\linewidth,  height=\puniheight]{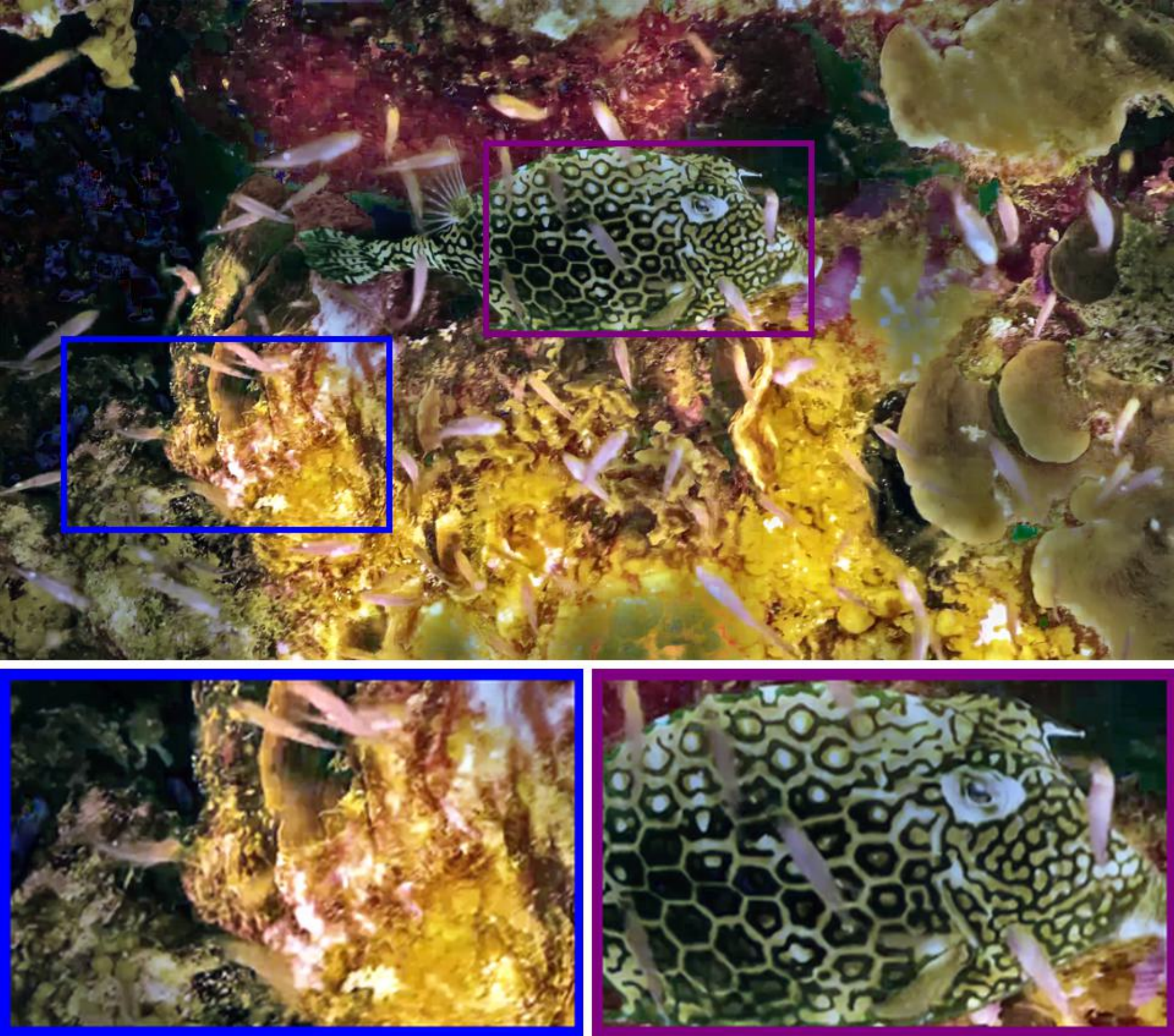}
        \includegraphics[width=\linewidth,  height=\puniheight]{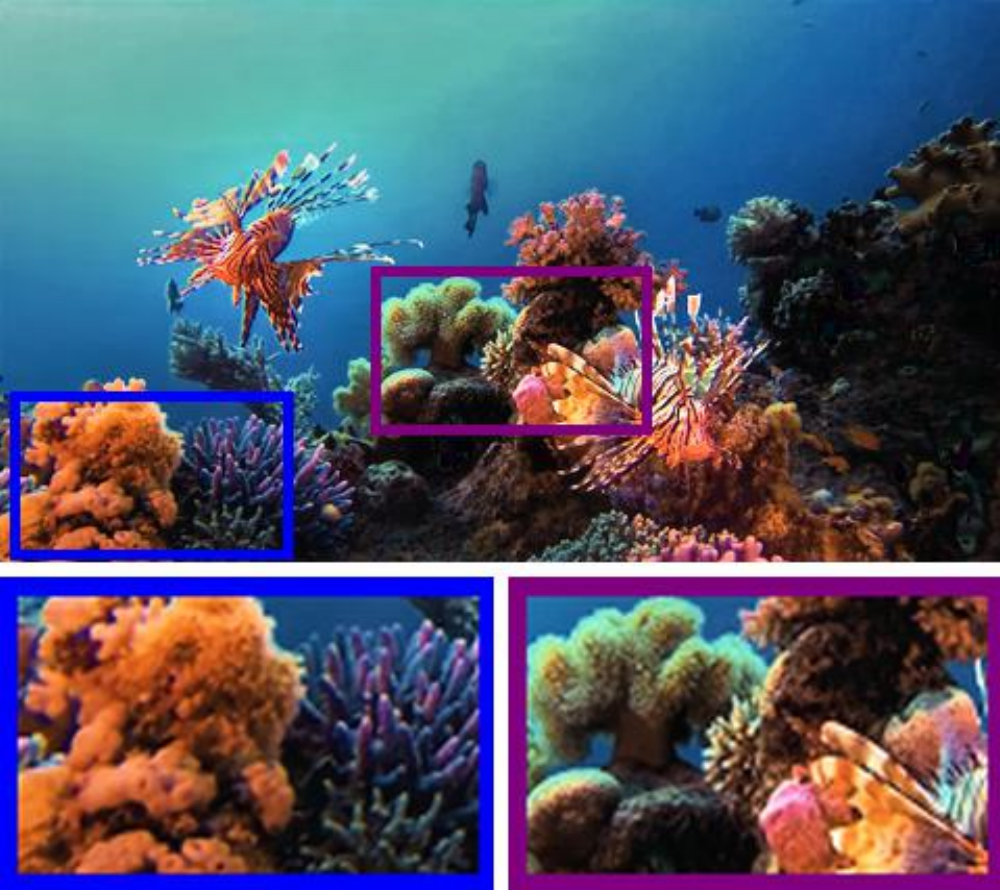}
        \includegraphics[width=\linewidth,  height=\puniheight]{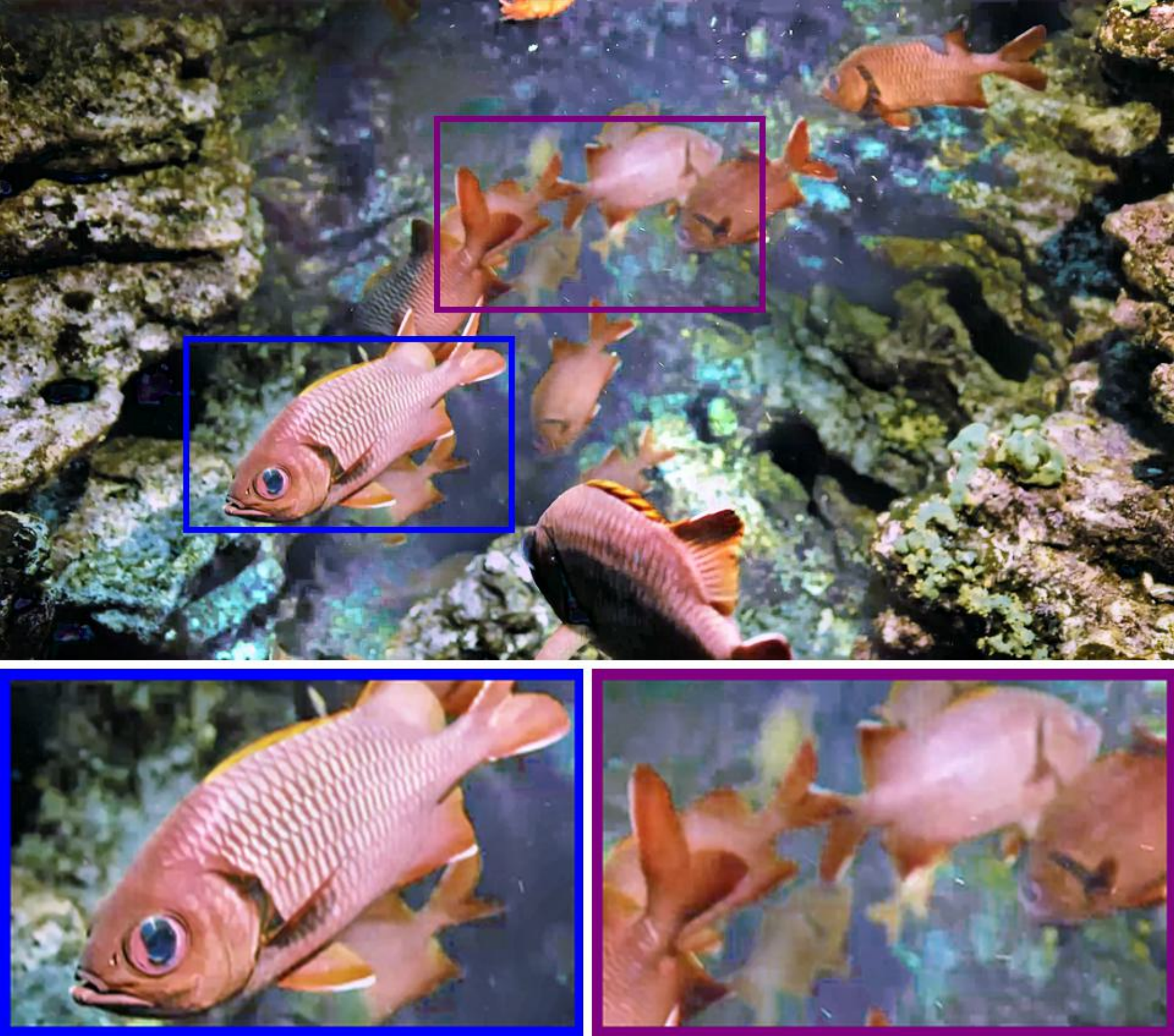}
        \caption{\footnotesize UDAformer}
	\end{subfigure}

    %%%%%%%%%%%%%%%%%%%%%%%%%%%%%

	\begin{subfigure}{0.105\linewidth}
		\centering
		\includegraphics[width=\linewidth,  height=\puniheight]{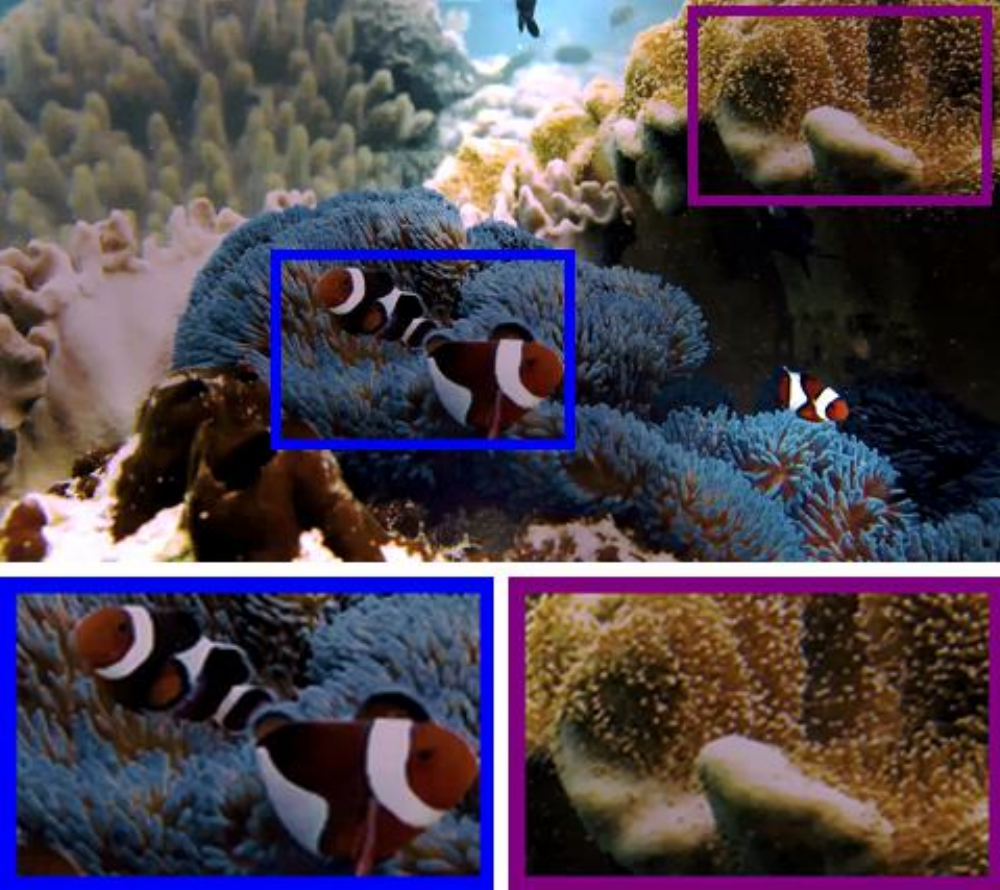} 
        \includegraphics[width=\linewidth,  height=\puniheight]{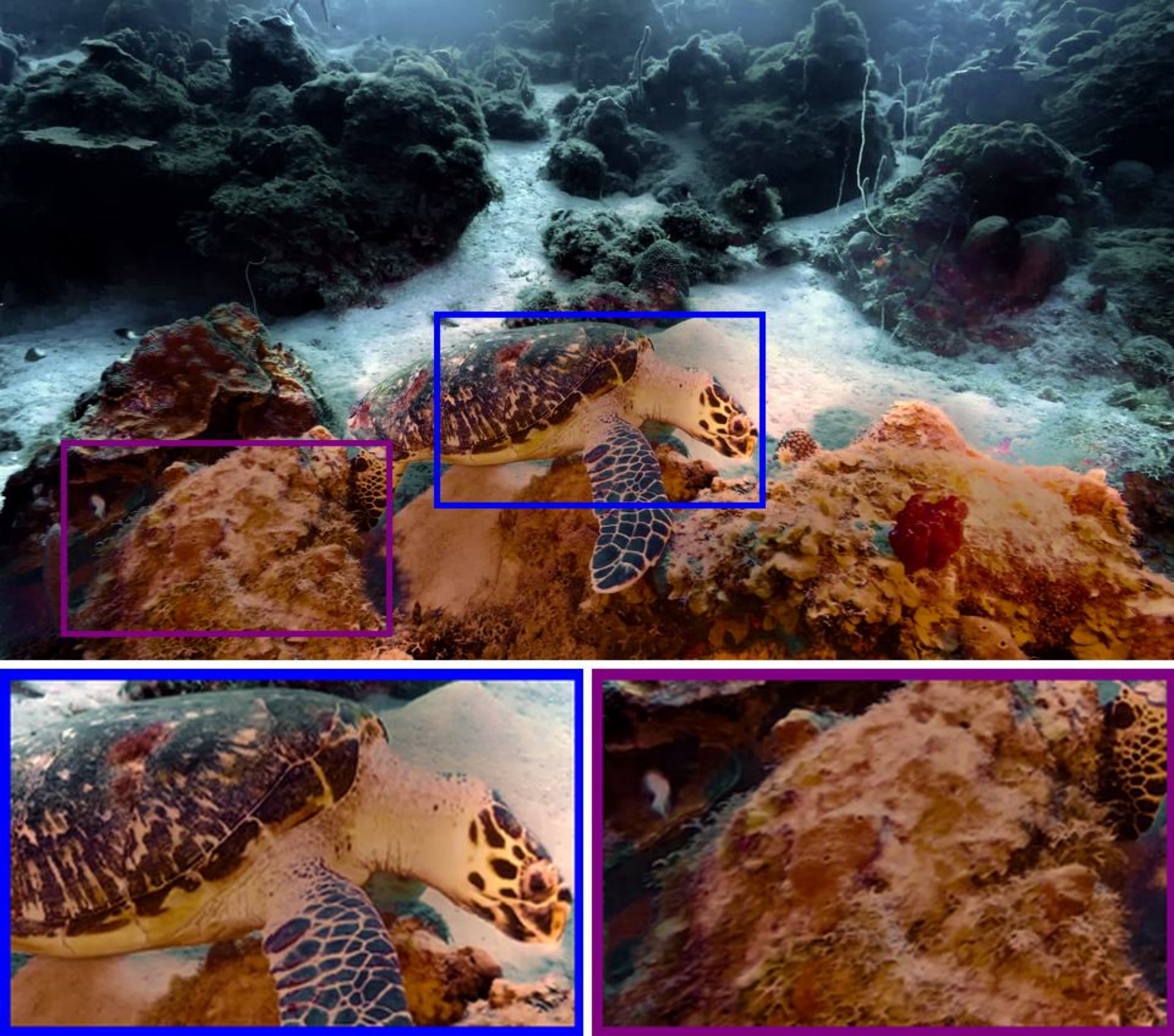} 
        \includegraphics[width=\linewidth,  height=\puniheight]{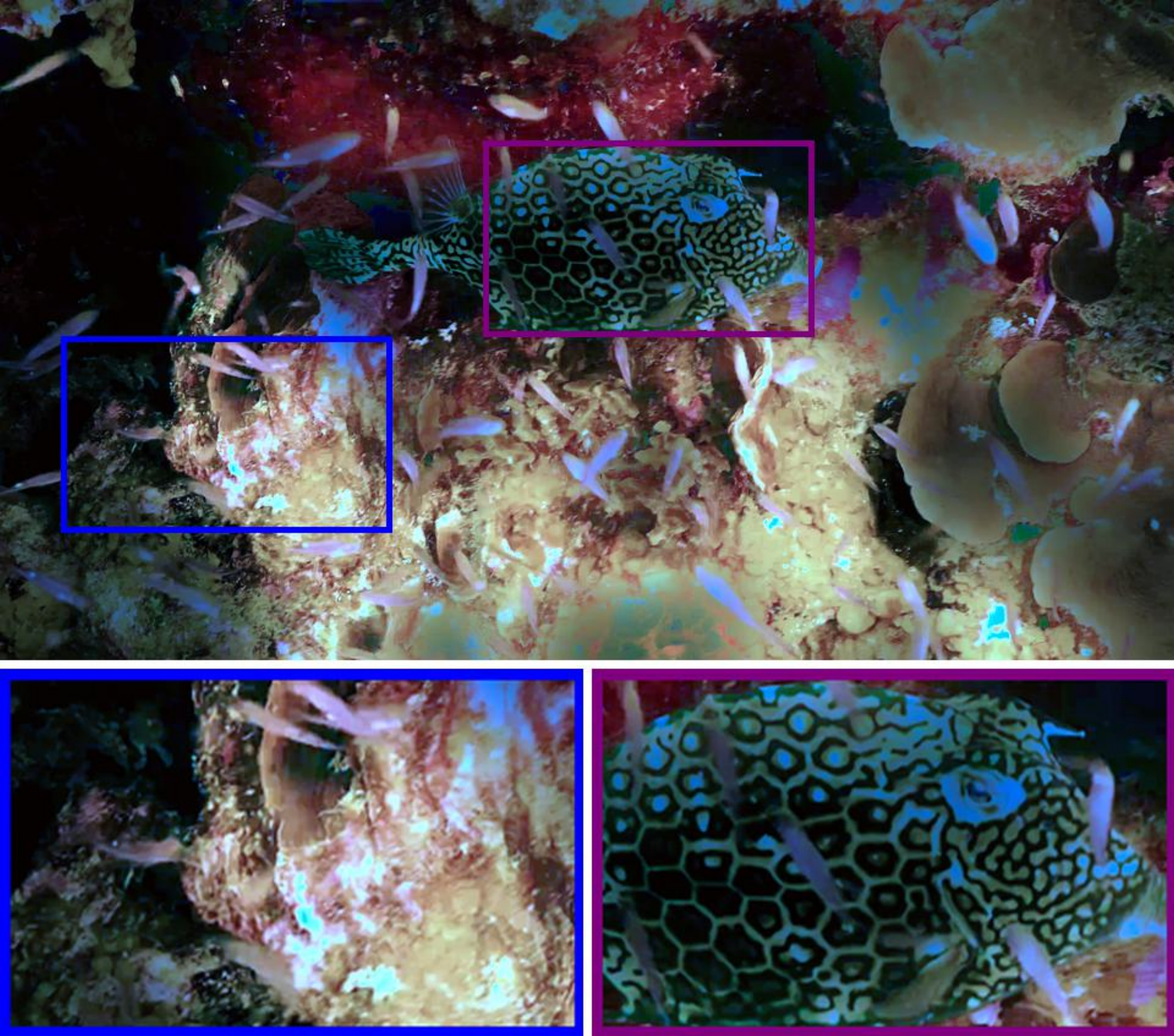} 
        \includegraphics[width=\linewidth,  height=\puniheight]{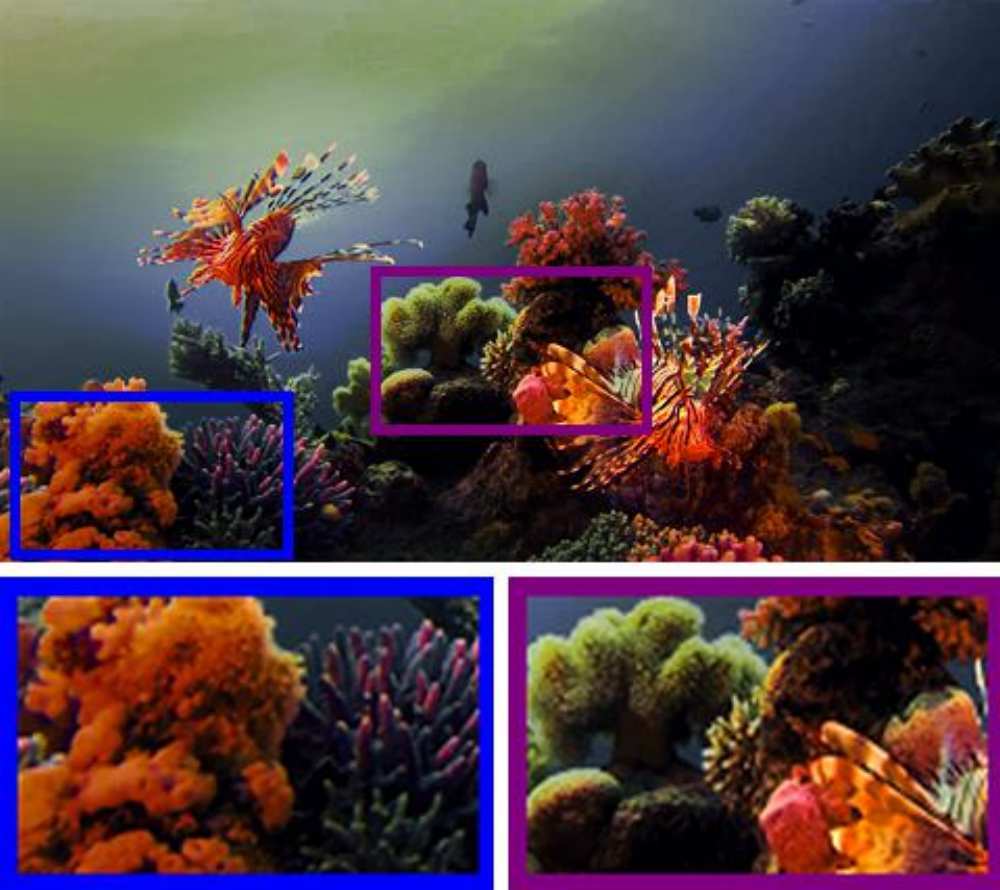} 
         \includegraphics[width=\linewidth,  height=\puniheight]{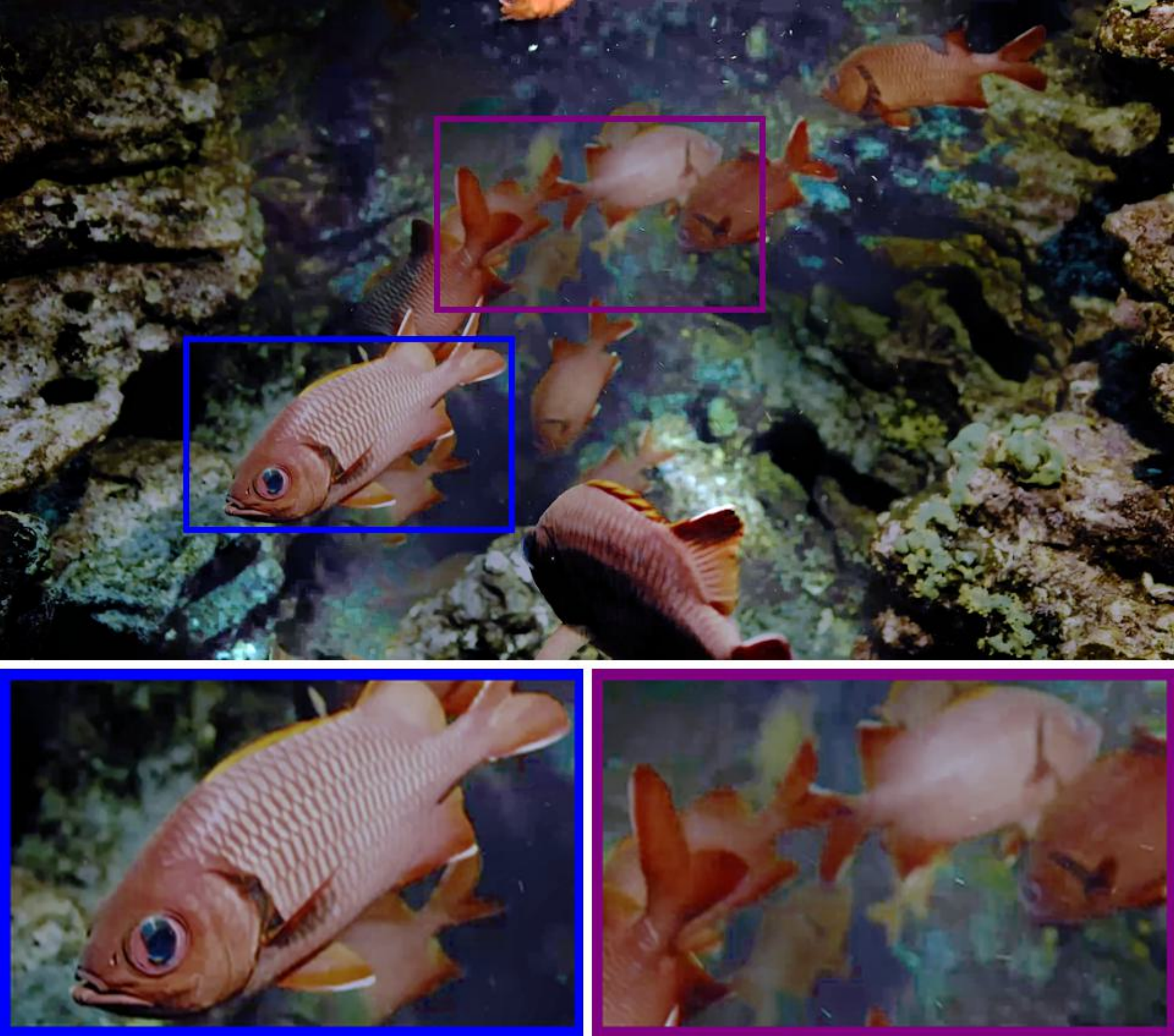} 
		\caption{\footnotesize HFM}
	\end{subfigure}
	\begin{subfigure}{0.105\linewidth}
		\centering
		\includegraphics[width=\linewidth,  height=\puniheight]{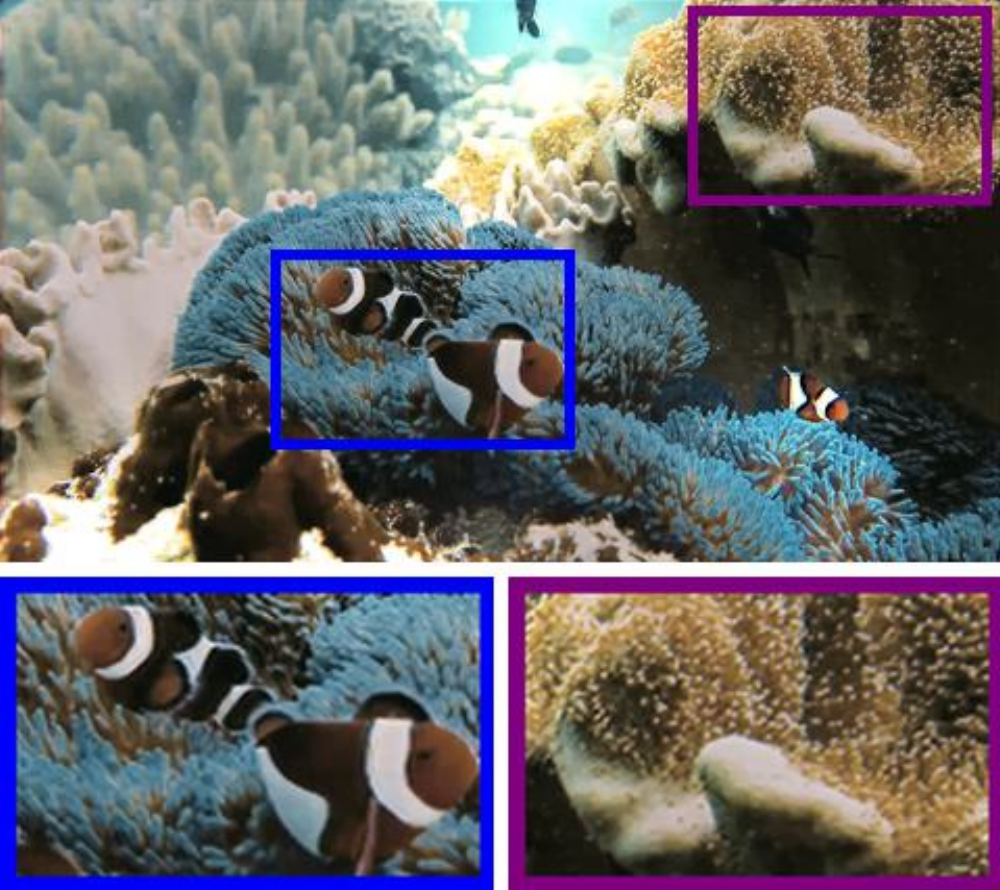} 
        \includegraphics[width=\linewidth,  height=\puniheight]{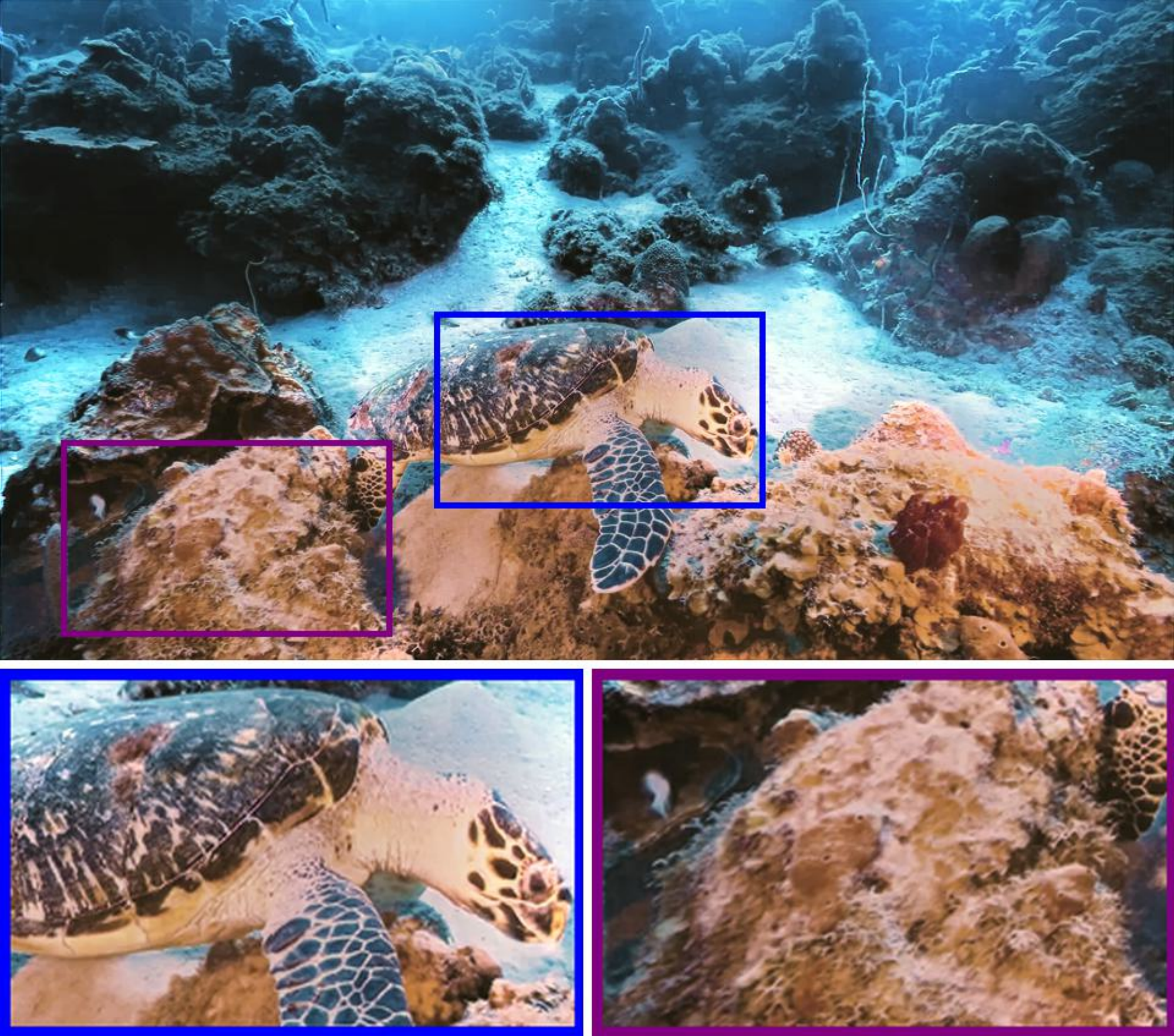} 
        \includegraphics[width=\linewidth,  height=\puniheight]{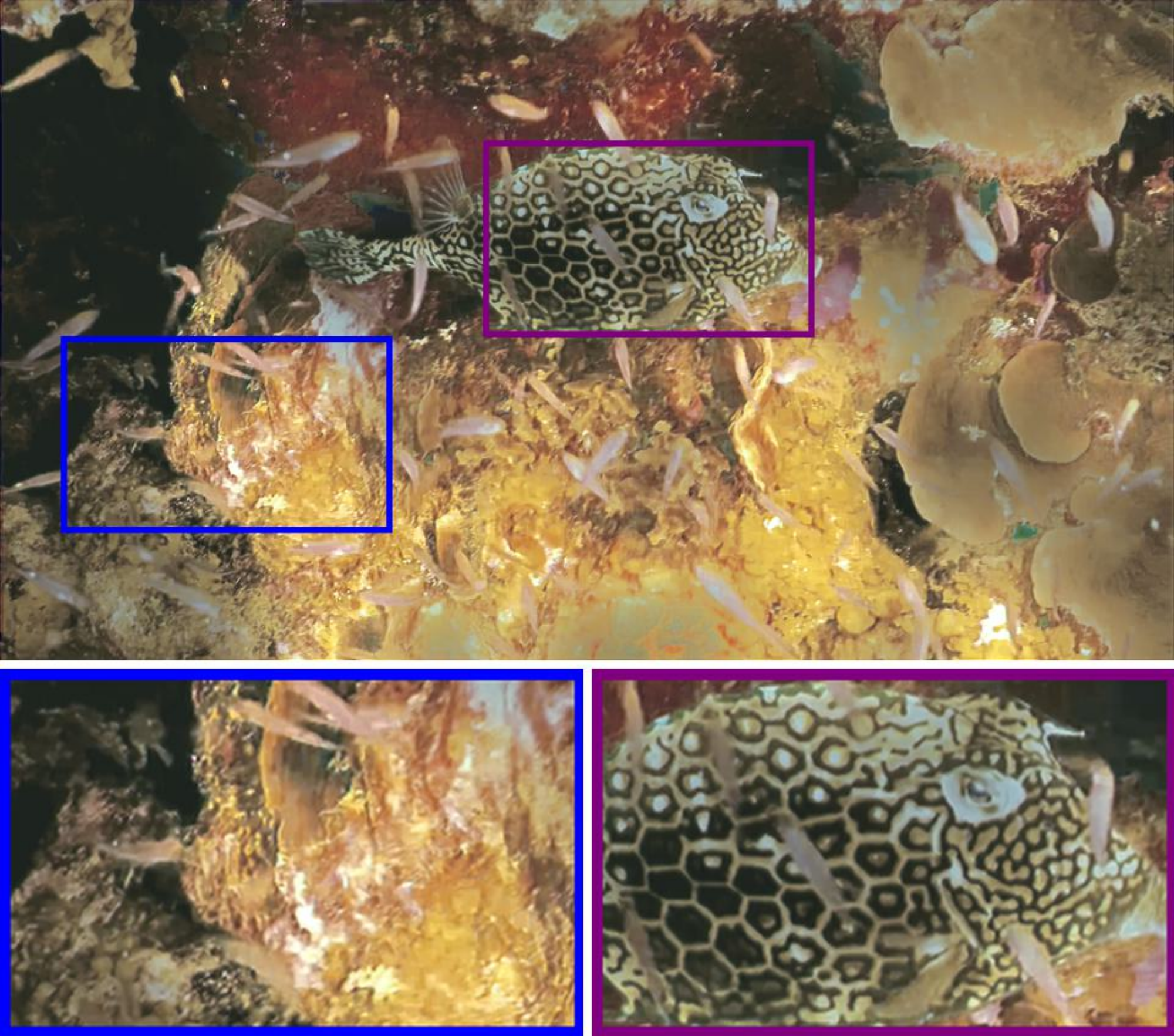}
        \includegraphics[width=\linewidth,  height=\puniheight]{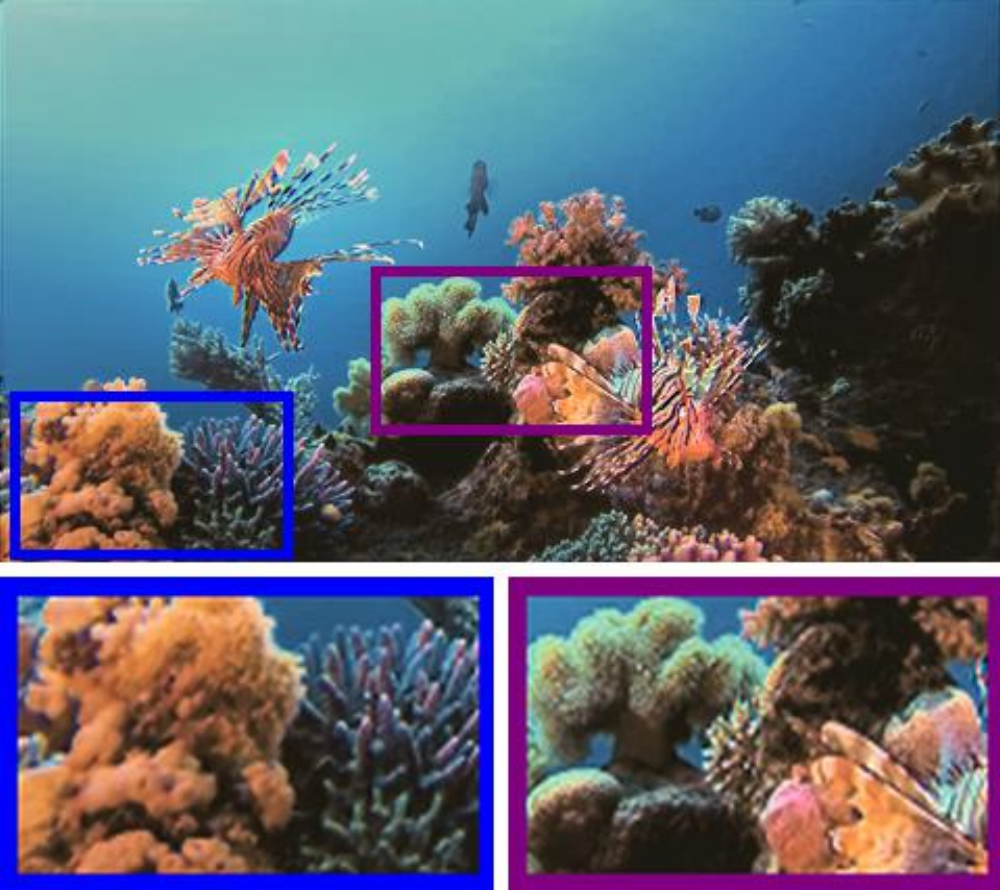}
          \includegraphics[width=\linewidth,  height=\puniheight]{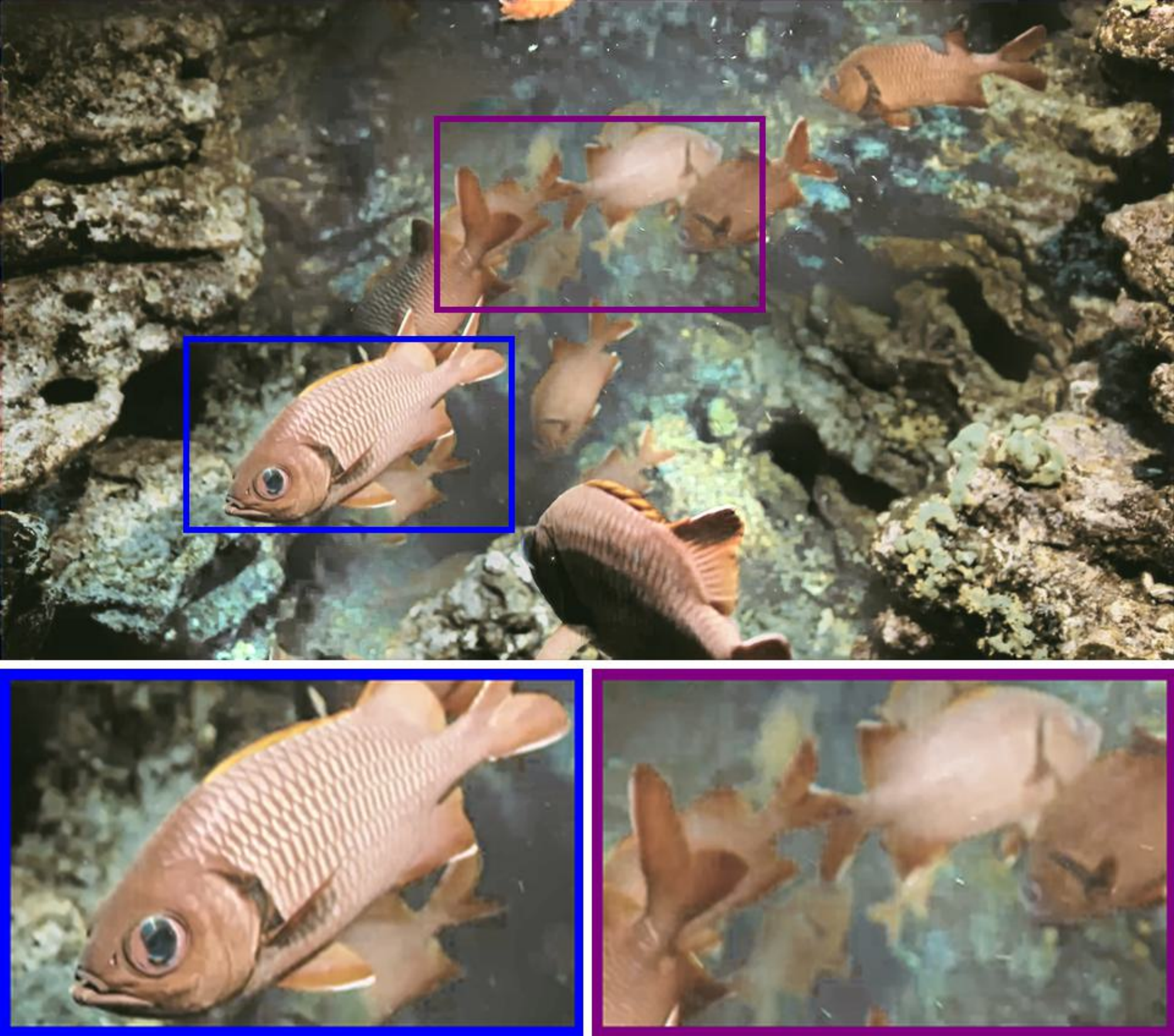} 
		\caption{\footnotesize LENet}
	\end{subfigure}
	\begin{subfigure}{0.105\linewidth}
		\centering
		\includegraphics[width=\linewidth,  height=\puniheight]{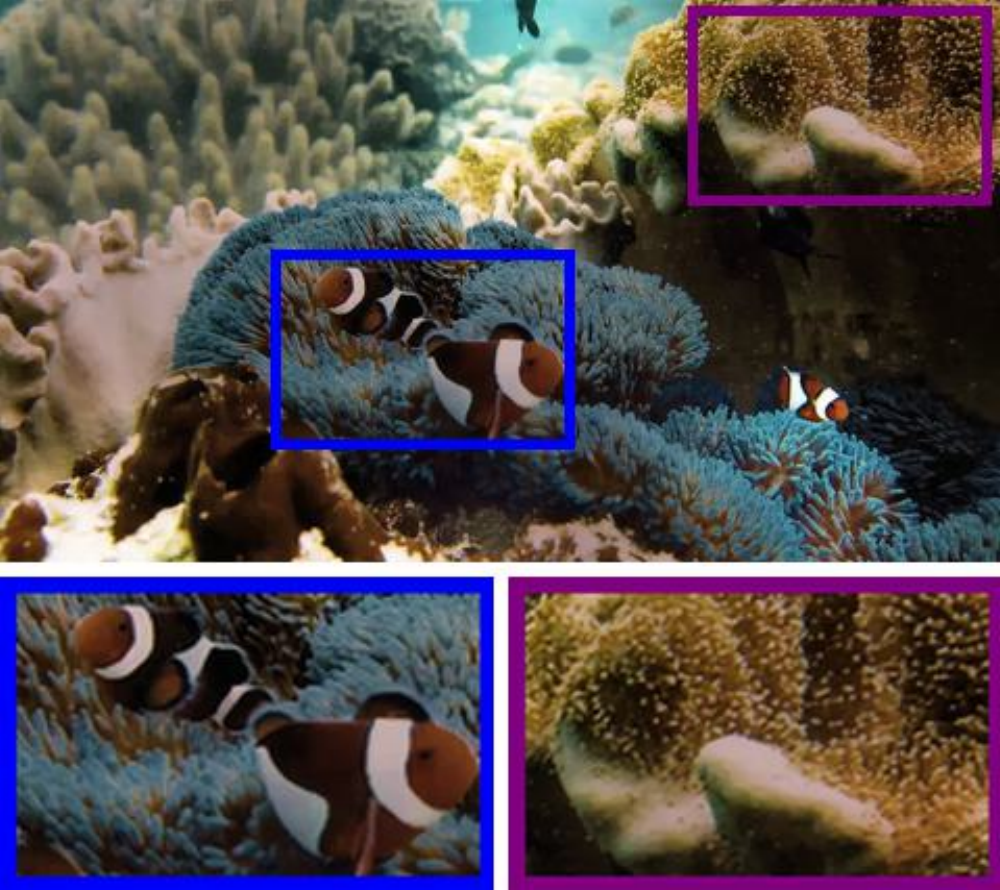}
        \includegraphics[width=\linewidth,  height=\puniheight]{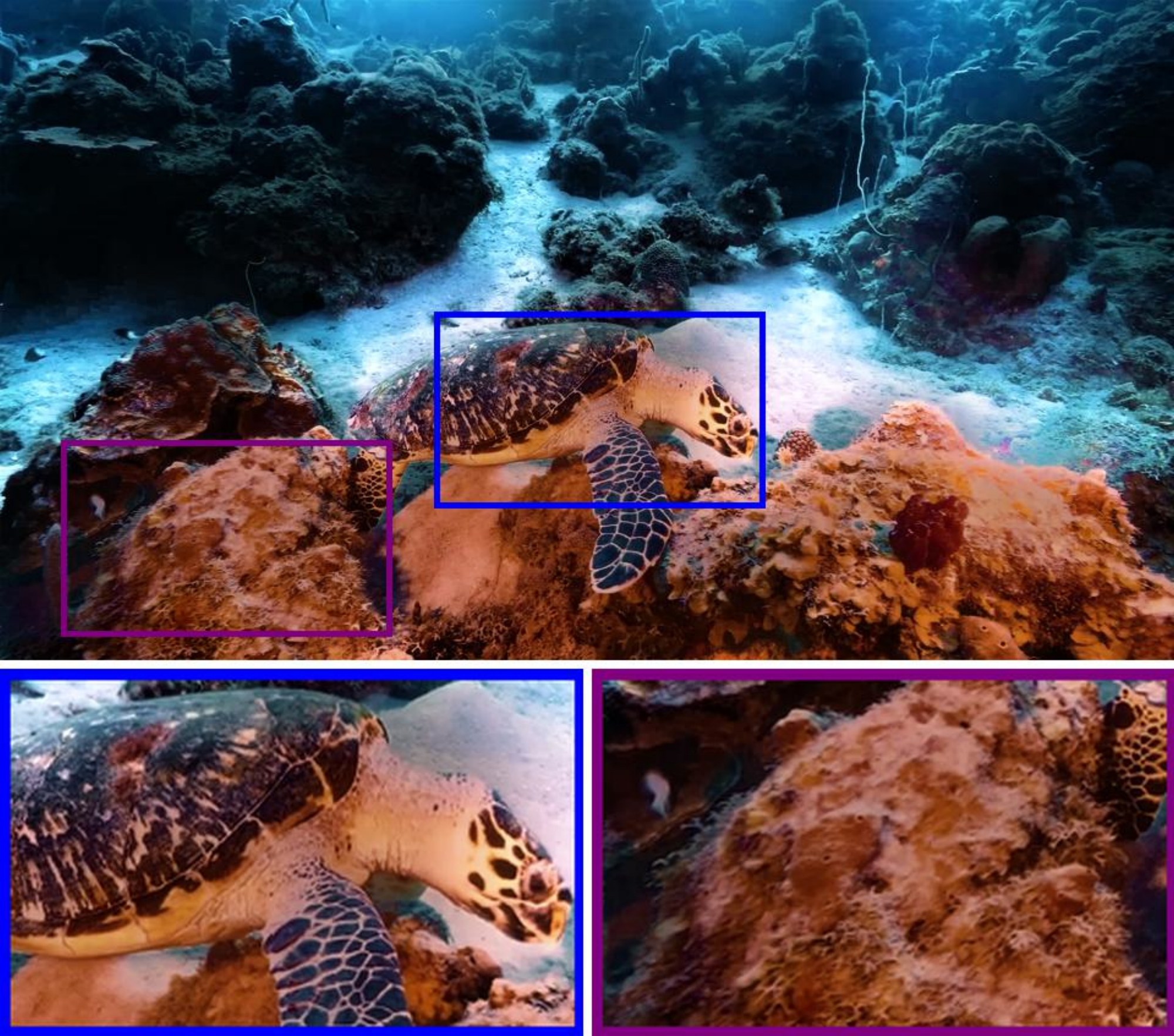}
        \includegraphics[width=\linewidth,  height=\puniheight]{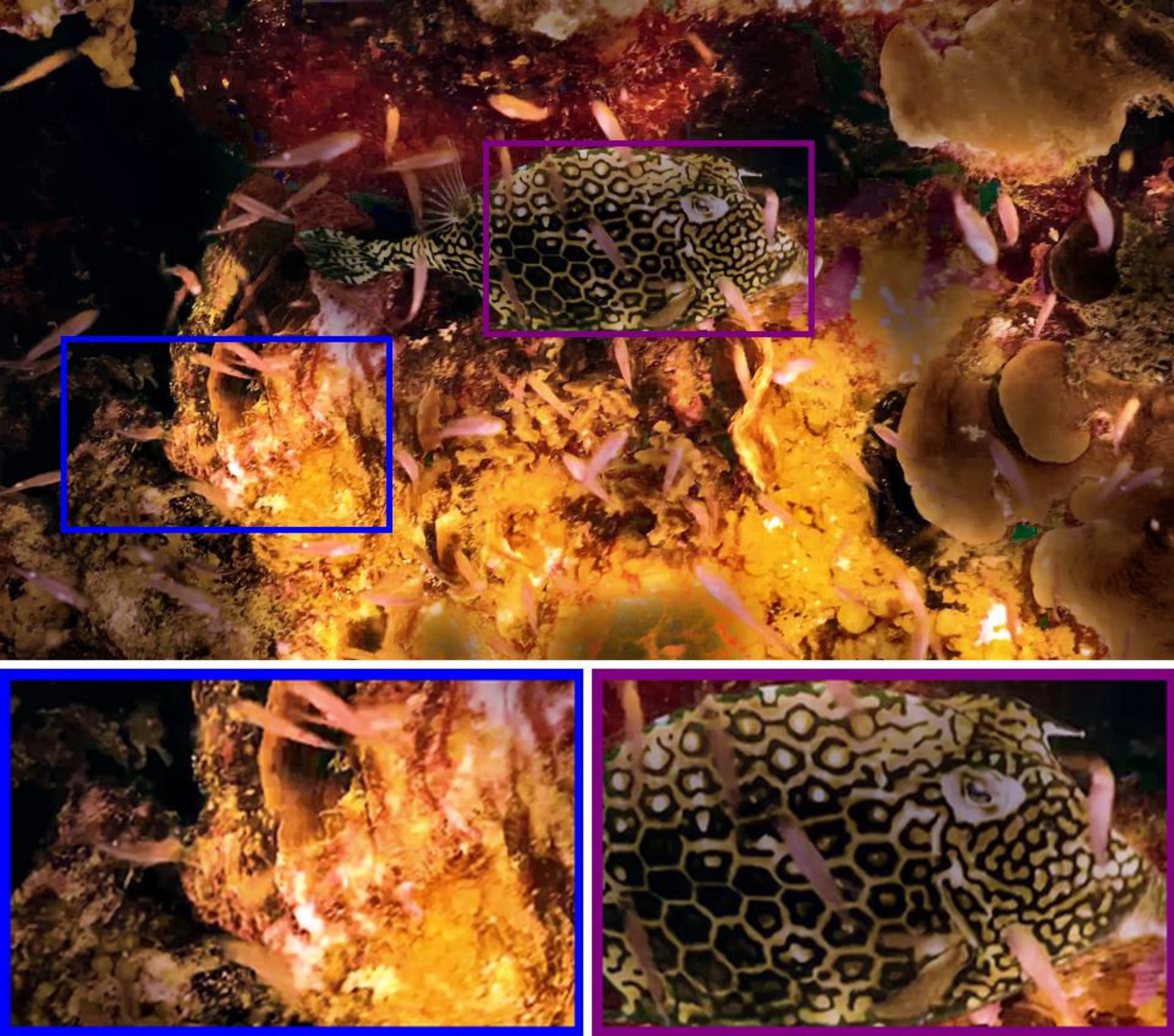}
        \includegraphics[width=\linewidth,  height=\puniheight]{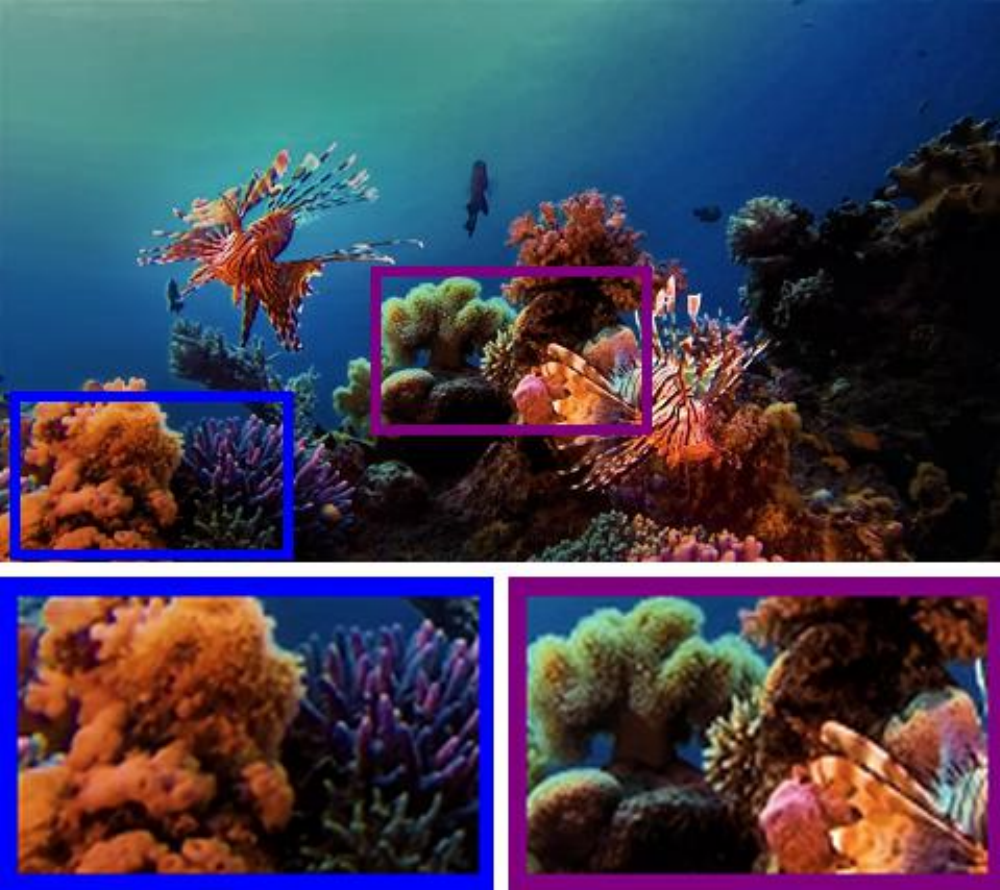}
        \includegraphics[width=\linewidth,  height=\puniheight]{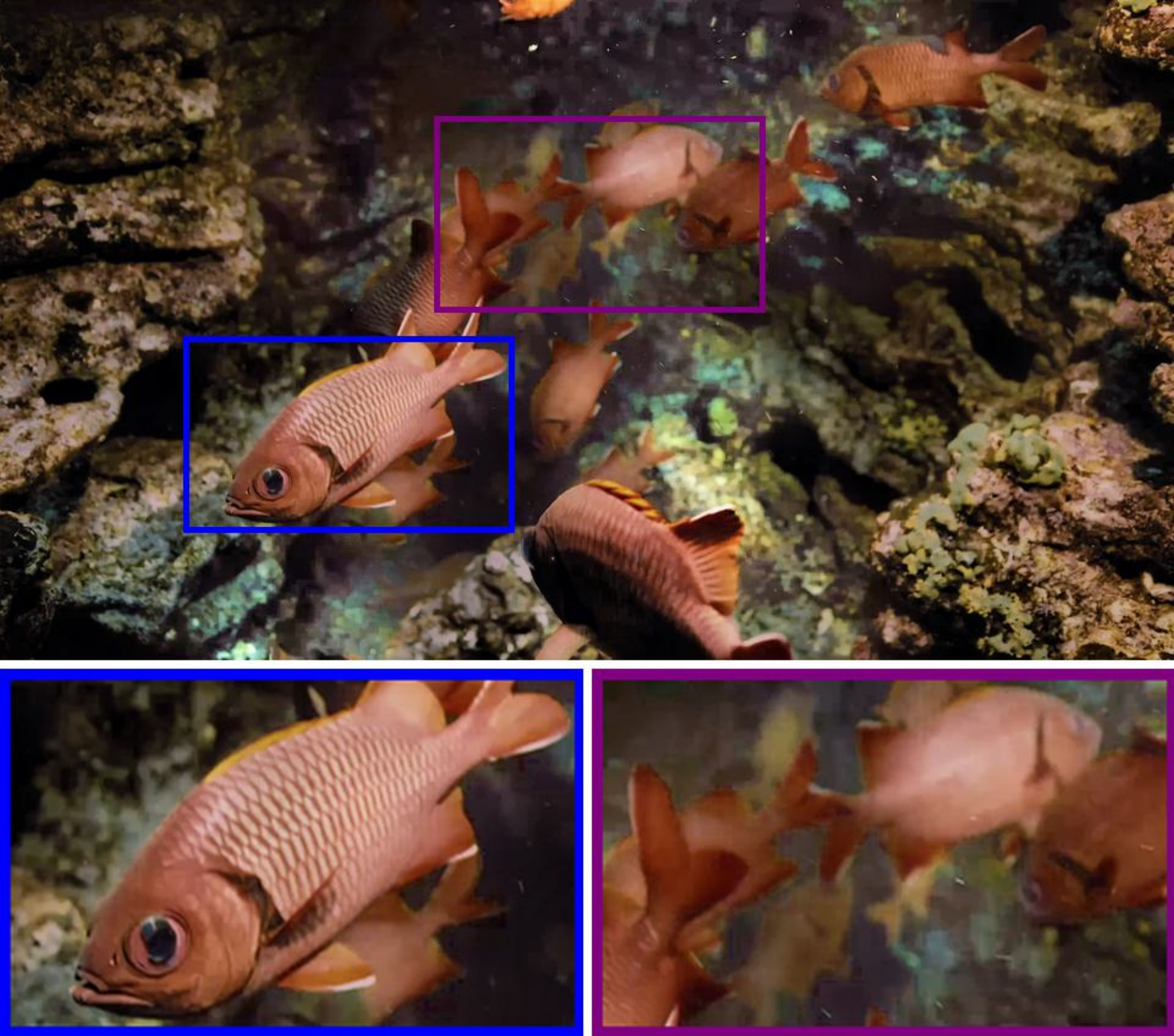}
		\caption{\footnotesize SMDR-IS}
	\end{subfigure}
	\begin{subfigure}{0.105\linewidth}
		\centering
		\includegraphics[width=\linewidth,  height=\puniheight]{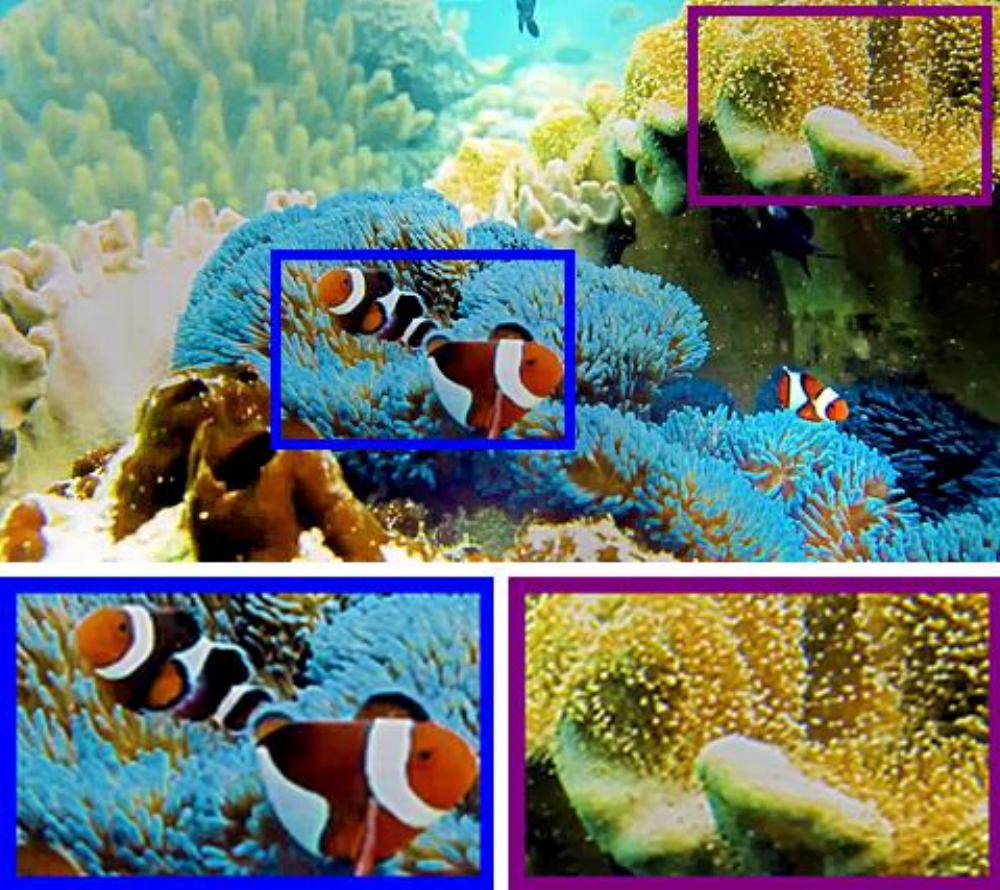} 
        \includegraphics[width=\linewidth,  height=\puniheight]{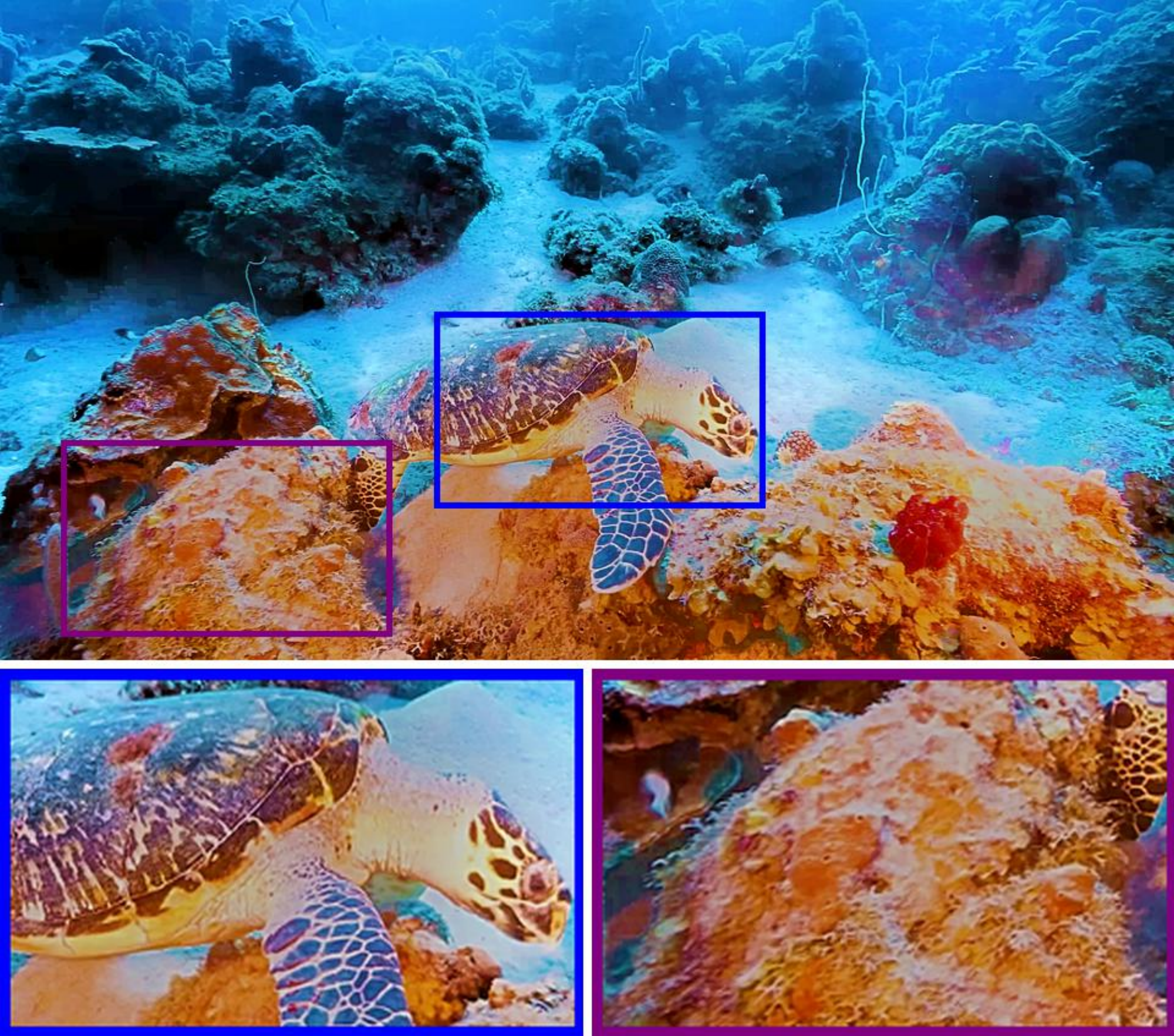} 
        \includegraphics[width=\linewidth,  height=\puniheight]{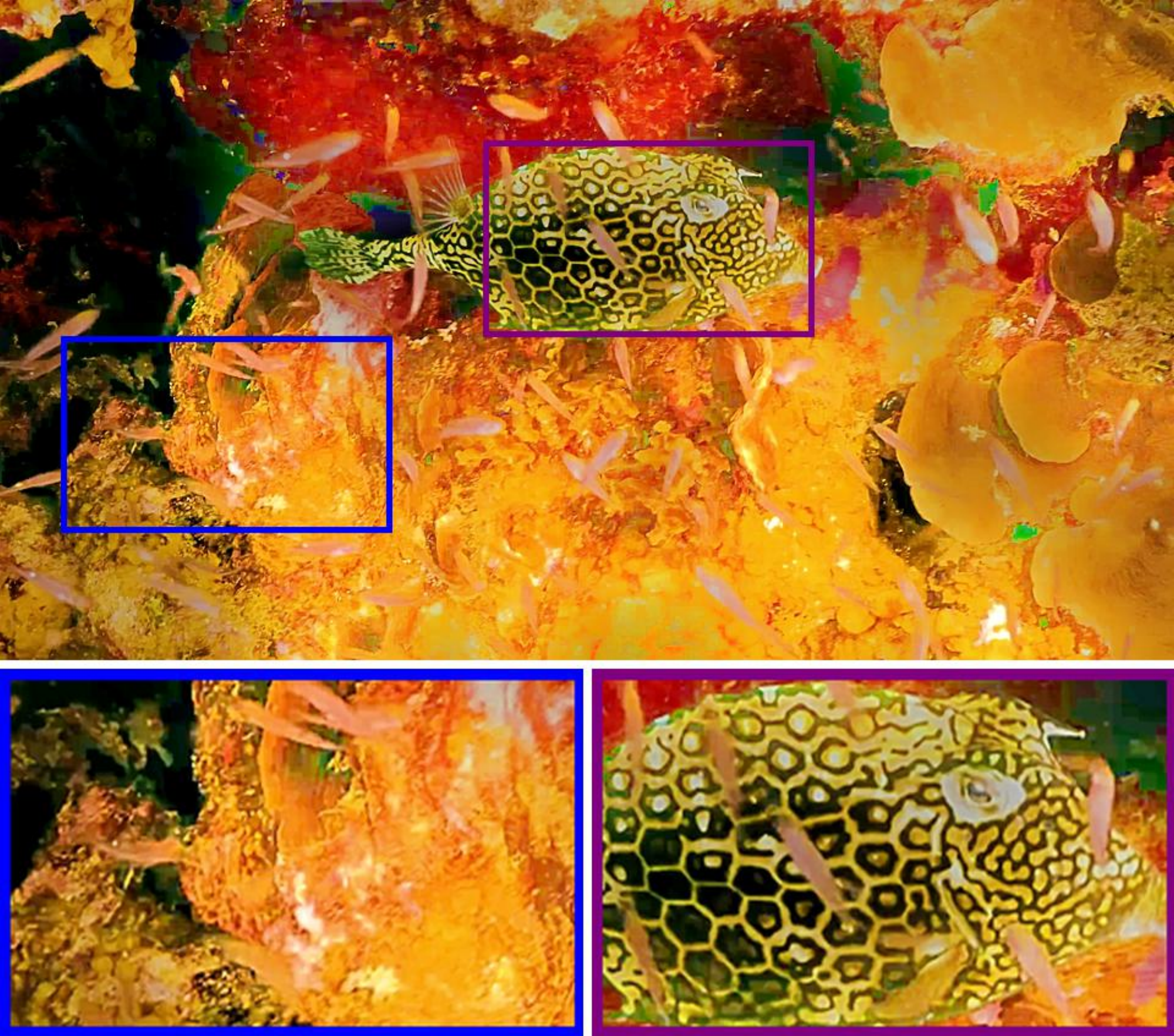} 
        \includegraphics[width=\linewidth,  height=\puniheight]{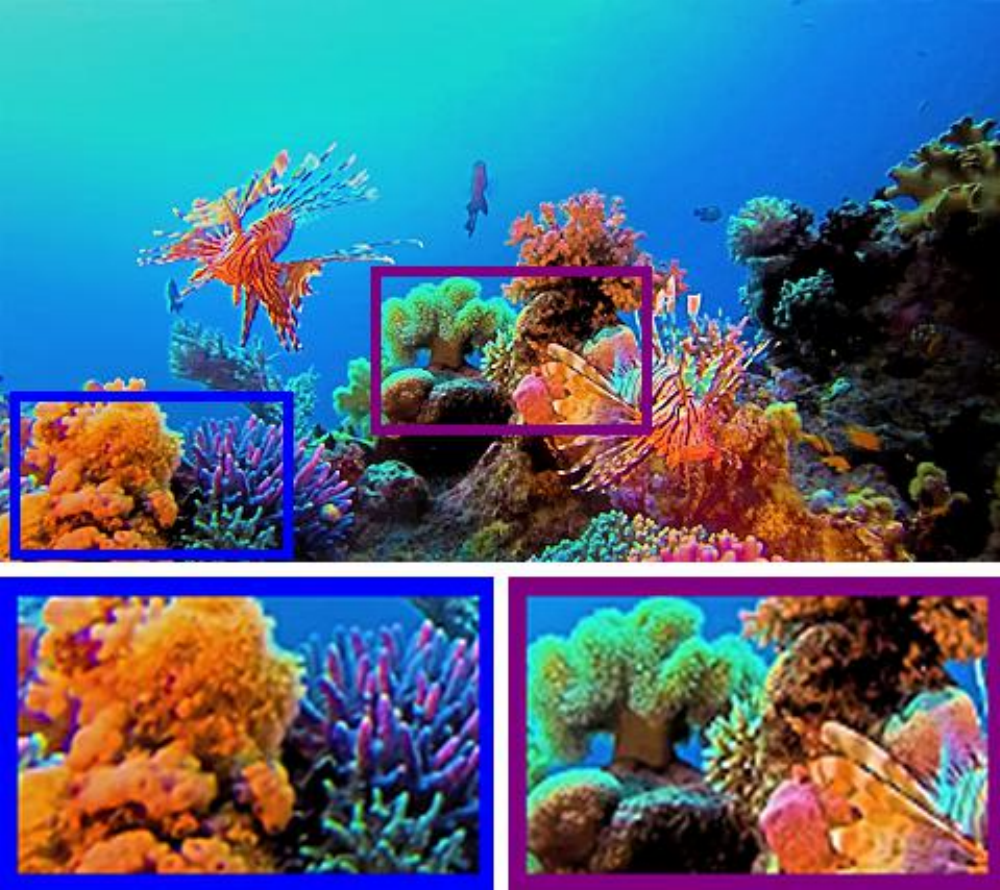}
        \includegraphics[width=\linewidth,  height=\puniheight]{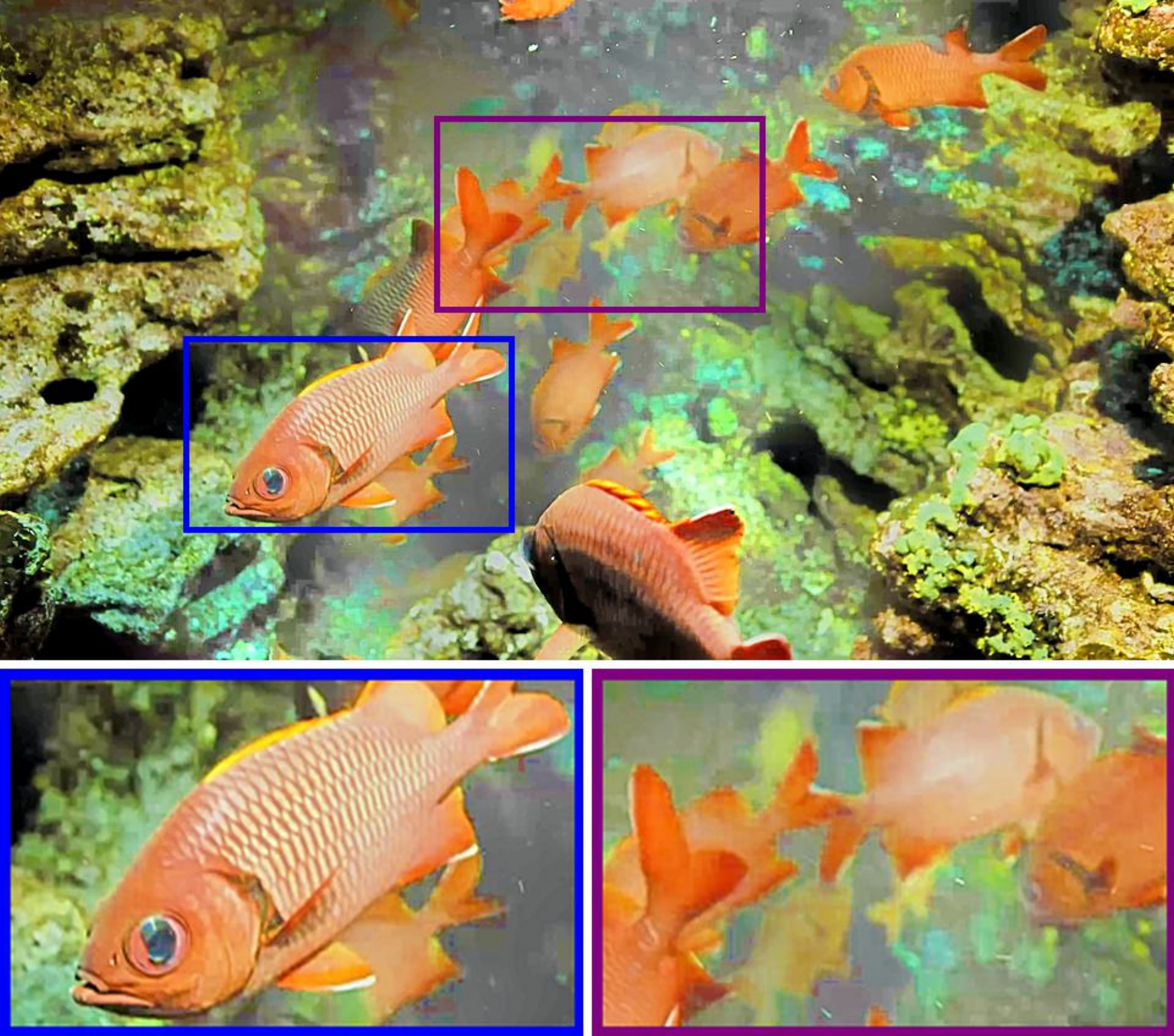}
		\caption{\footnotesize GCP}
	\end{subfigure}
    \begin{subfigure}{0.105\linewidth}
		\centering
		\includegraphics[width=\linewidth,  height=\puniheight]{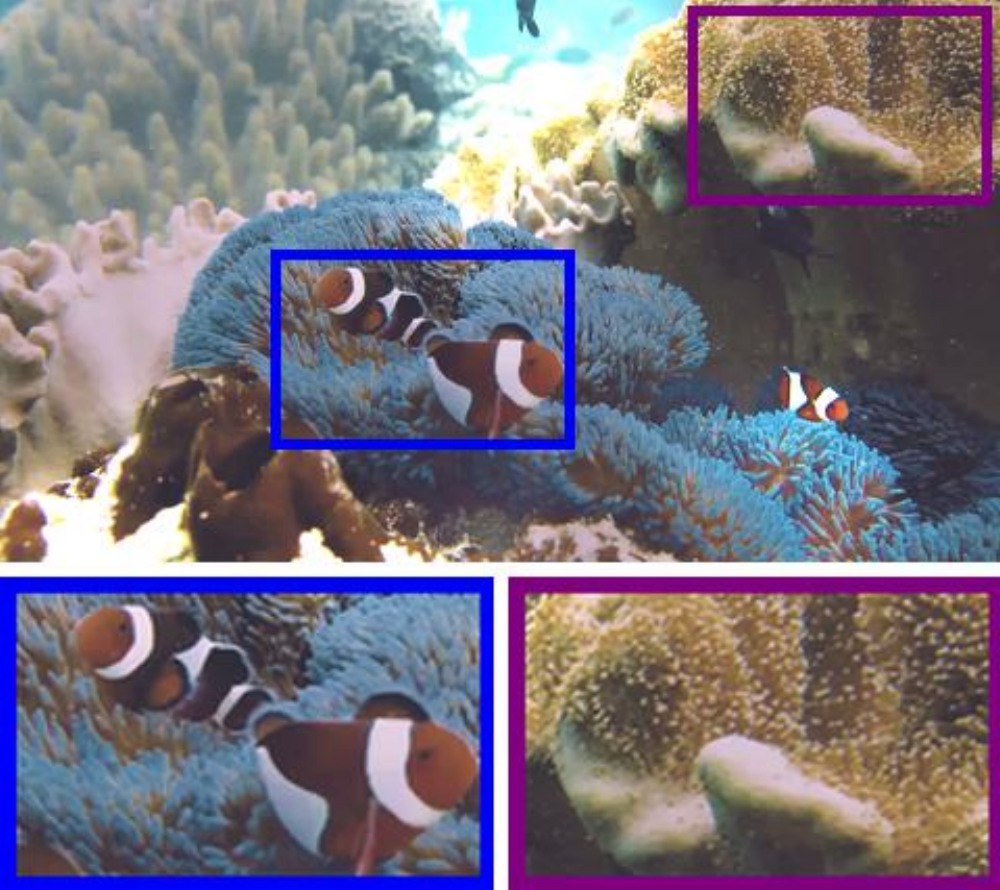} 
        \includegraphics[width=\linewidth,  height=\puniheight]{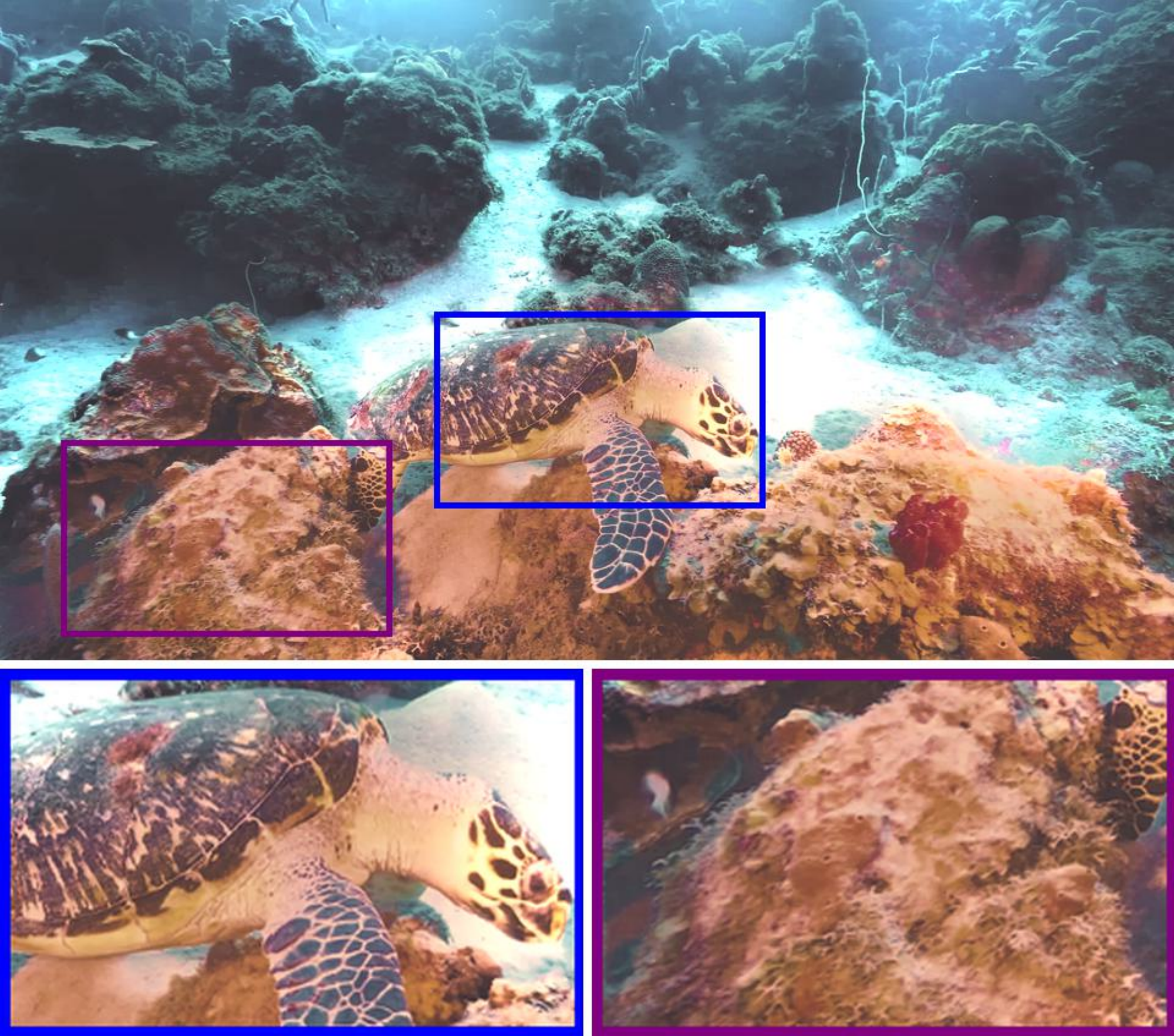} 
        \includegraphics[width=\linewidth,  height=\puniheight]{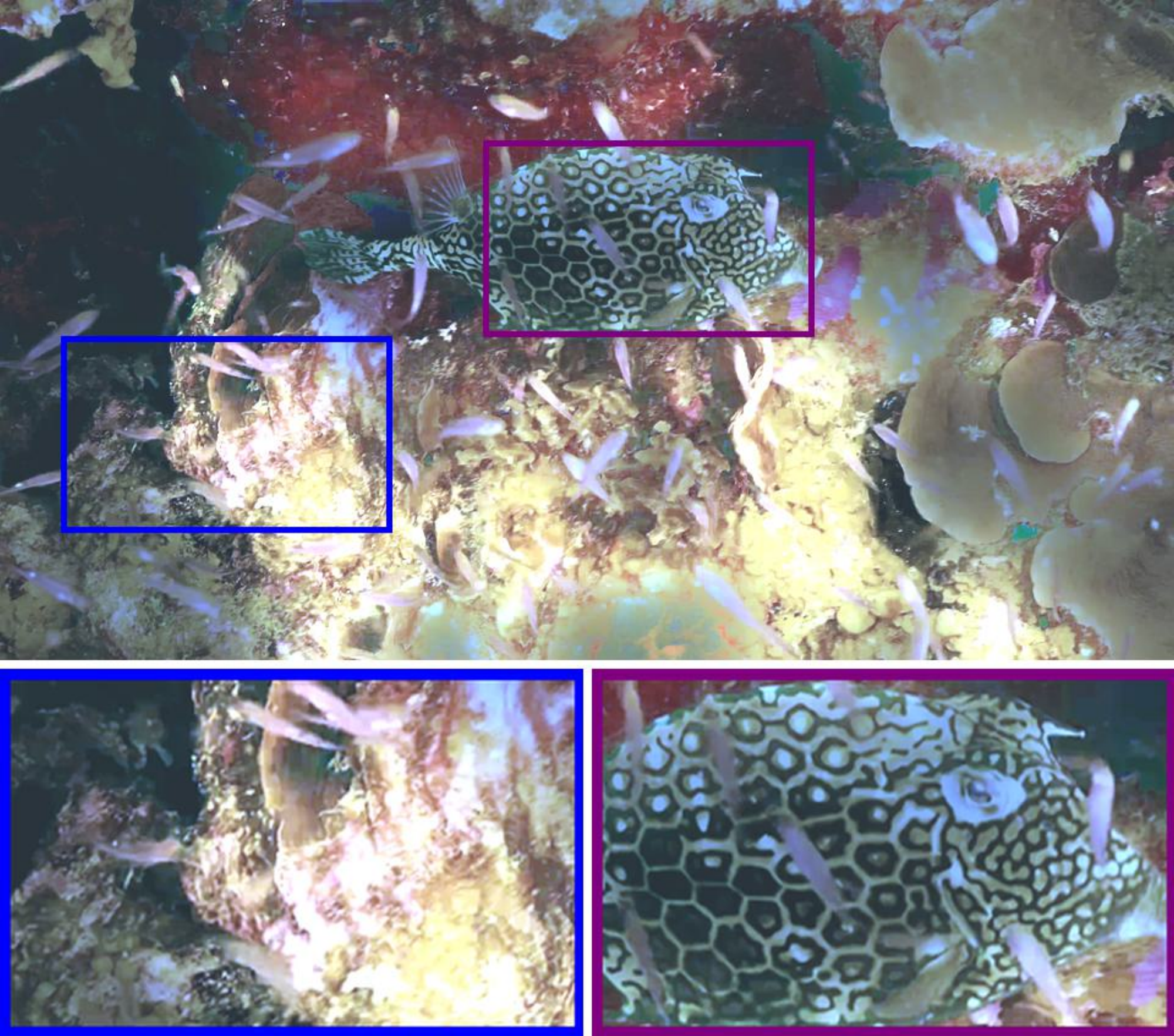}
        \includegraphics[width=\linewidth,  height=\puniheight]{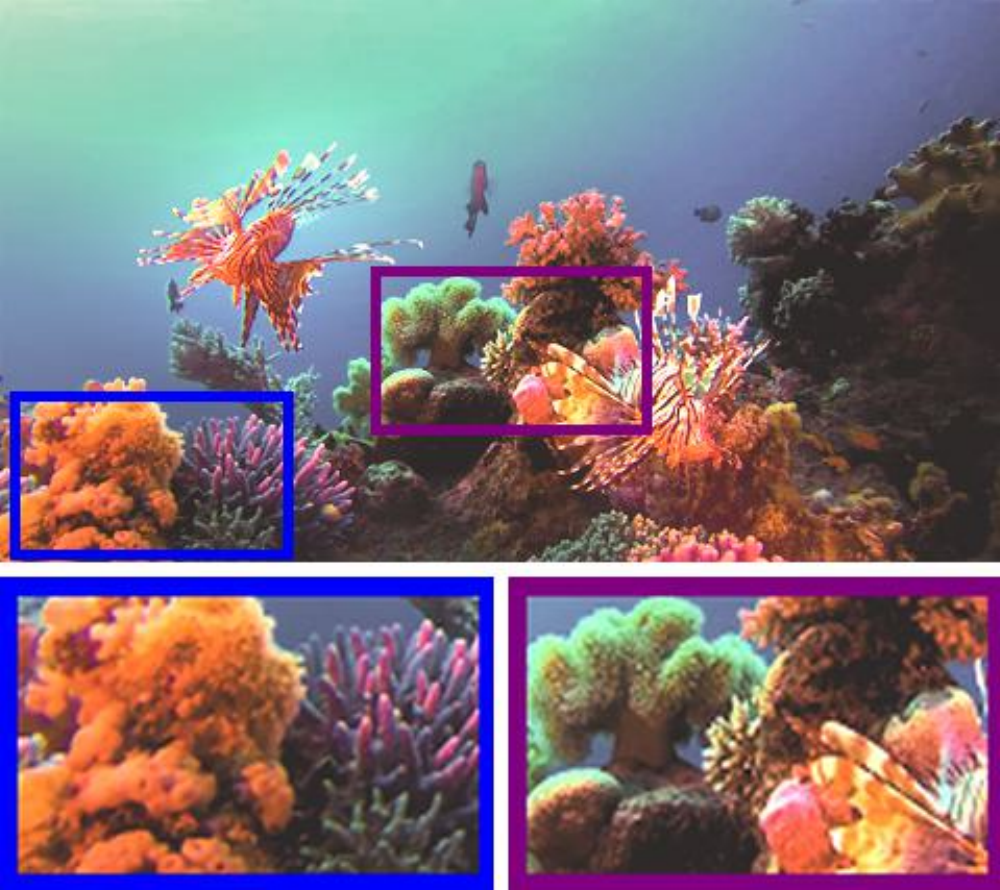}
        \includegraphics[width=\linewidth,  height=\puniheight]{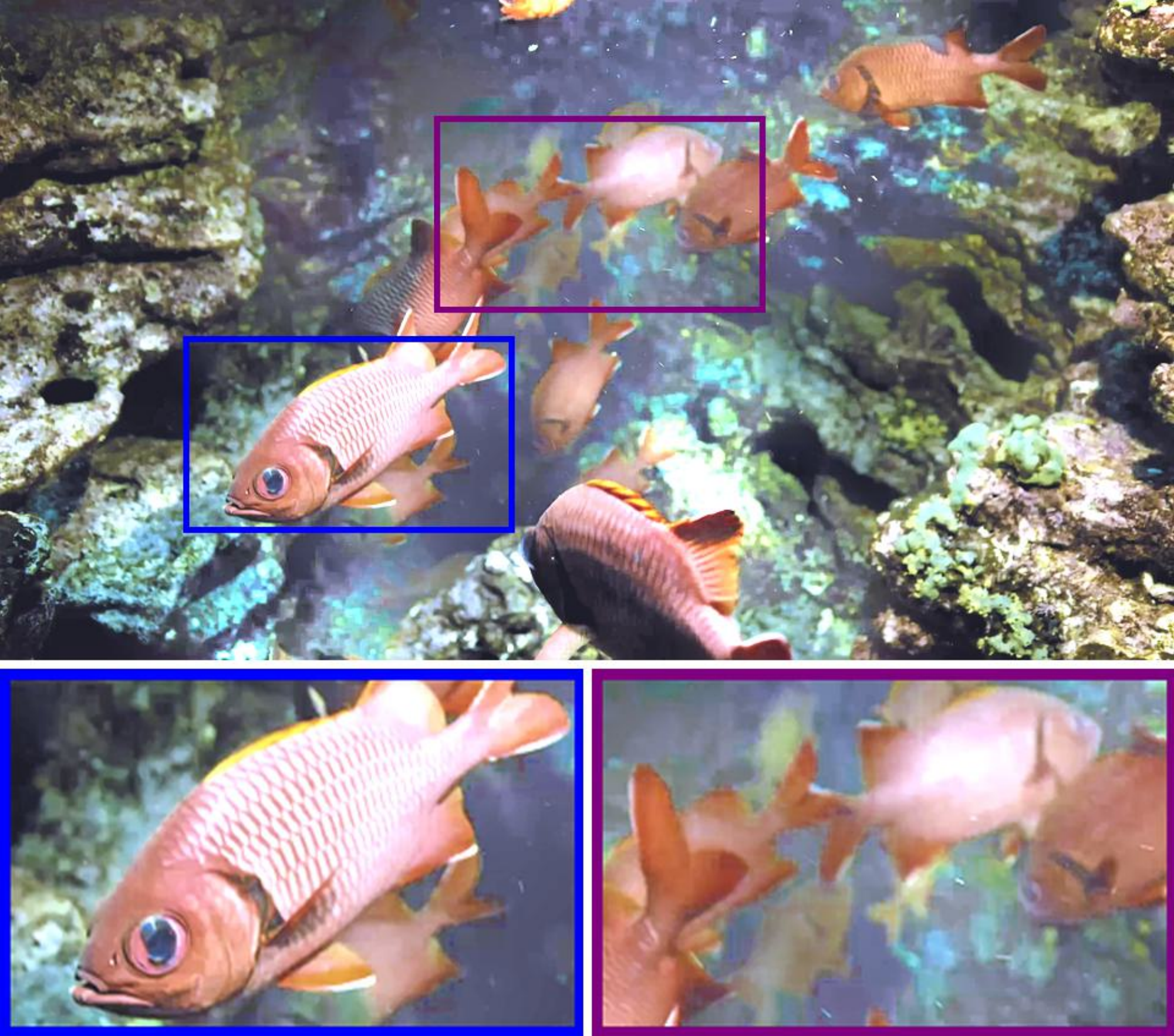}
		\caption{\footnotesize MACT}
	\end{subfigure}
        \begin{subfigure}{0.105\linewidth}
		\centering
		\includegraphics[width=\linewidth,  height=\puniheight]{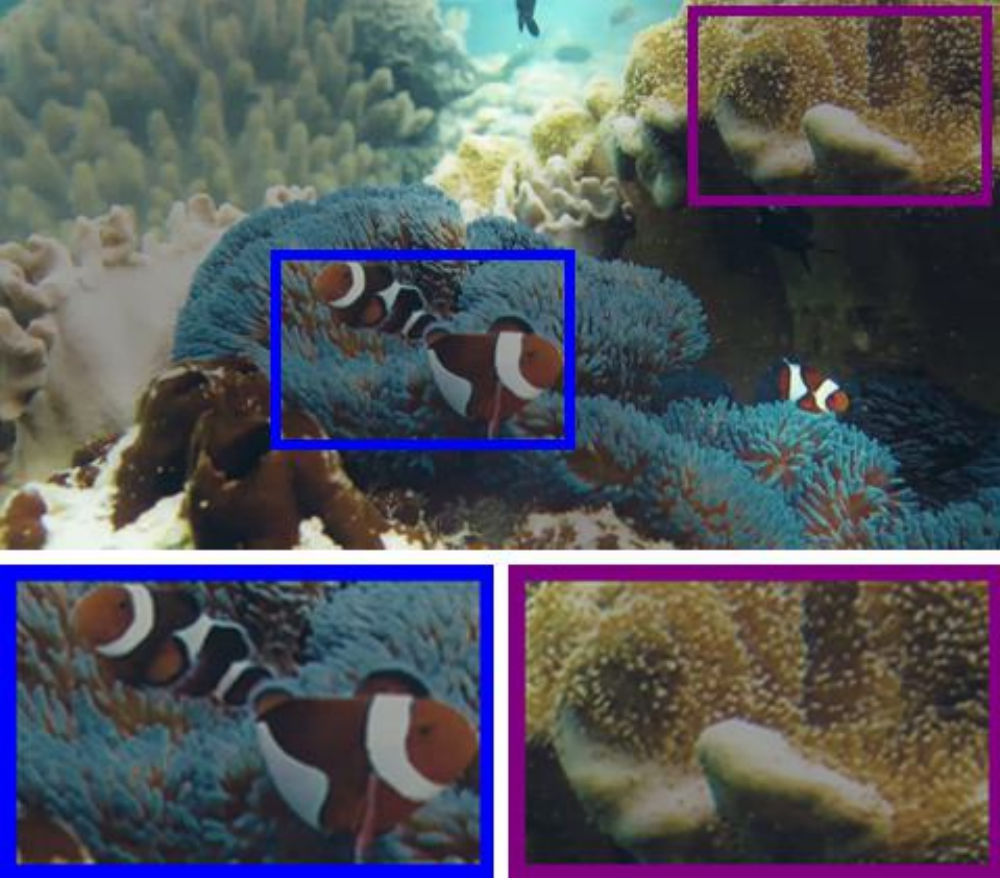} 
        \includegraphics[width=\linewidth,  height=\puniheight]{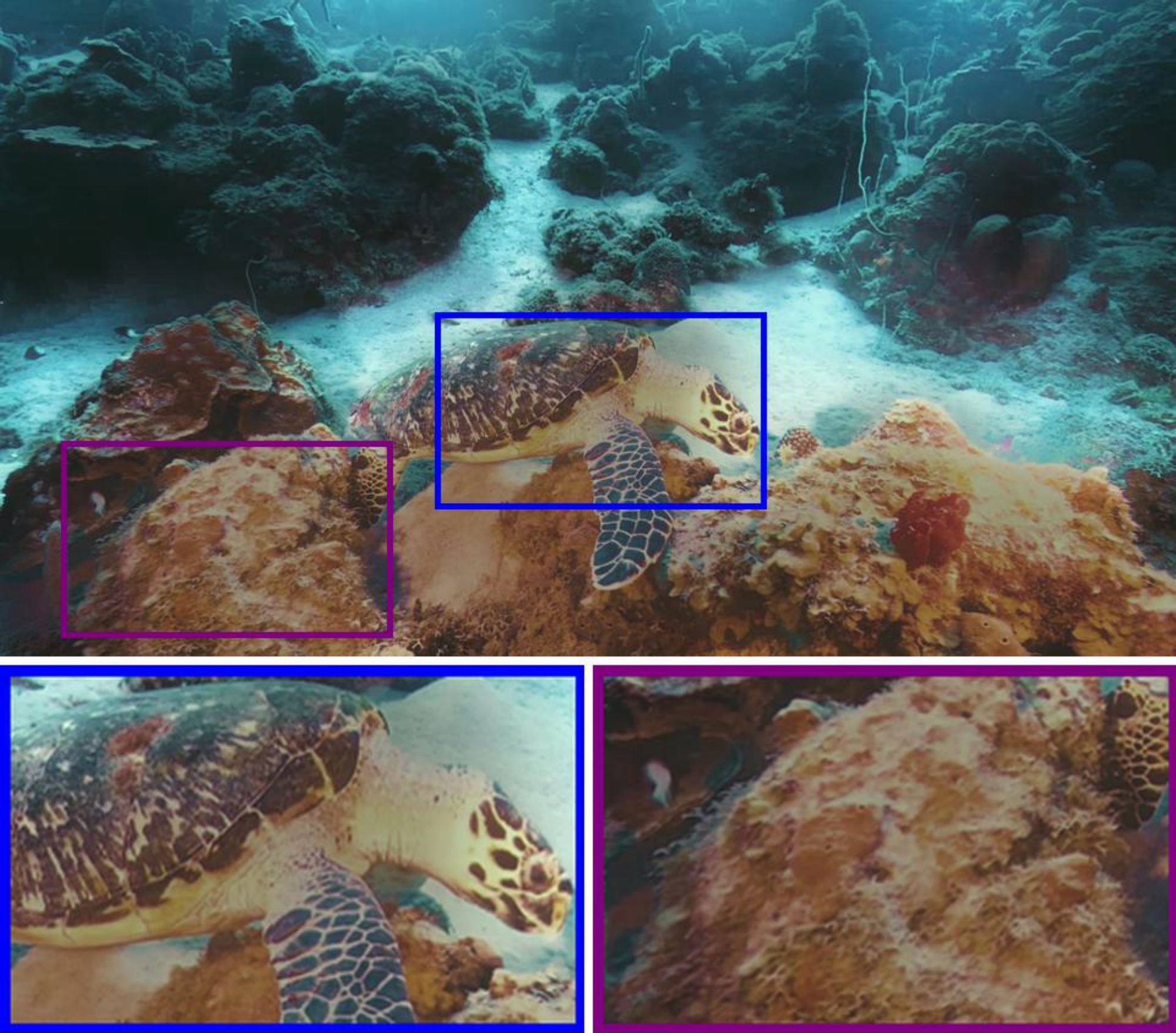} 
        \includegraphics[width=\linewidth,  height=\puniheight]{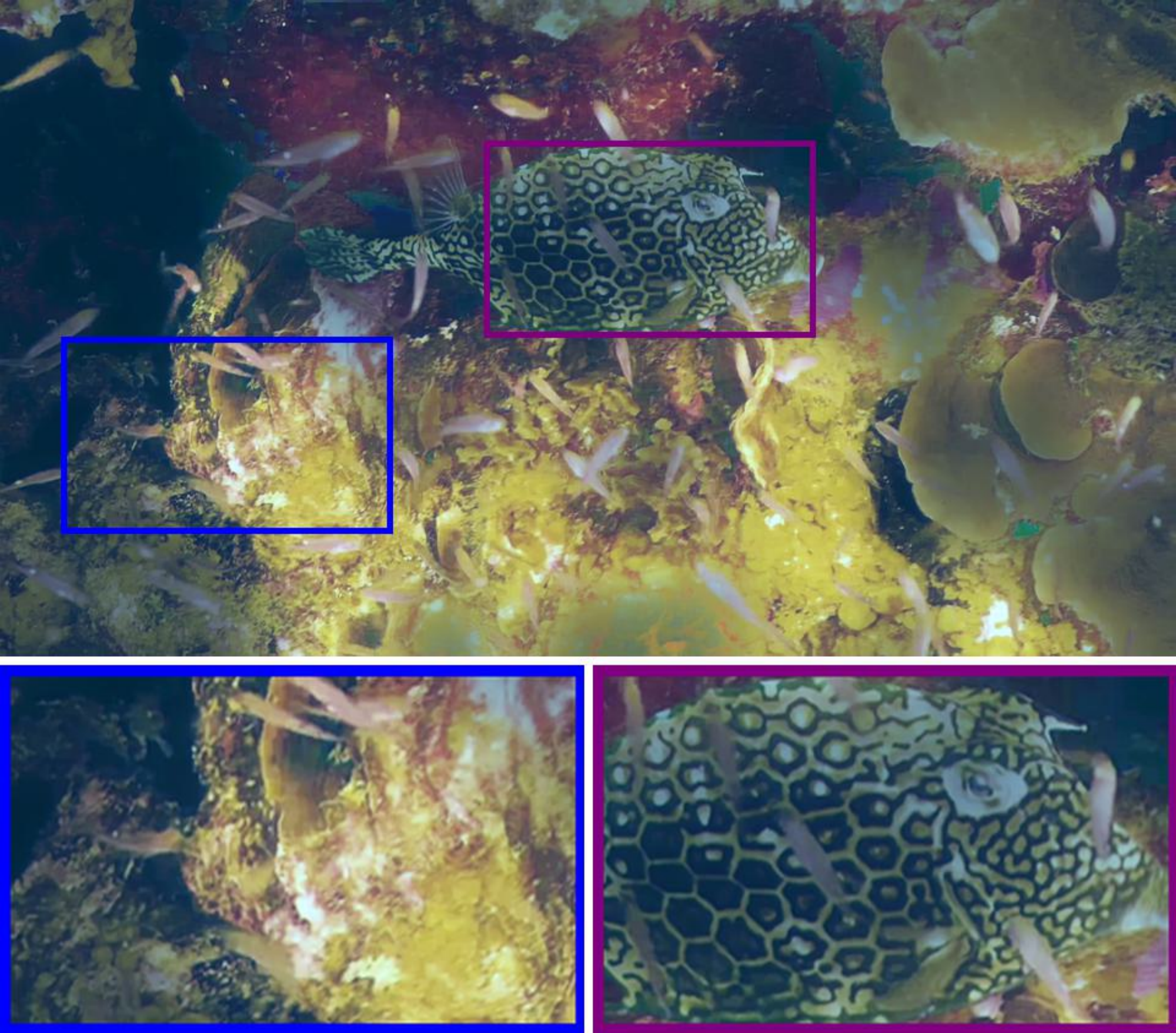}
        \includegraphics[width=\linewidth,  height=\puniheight]{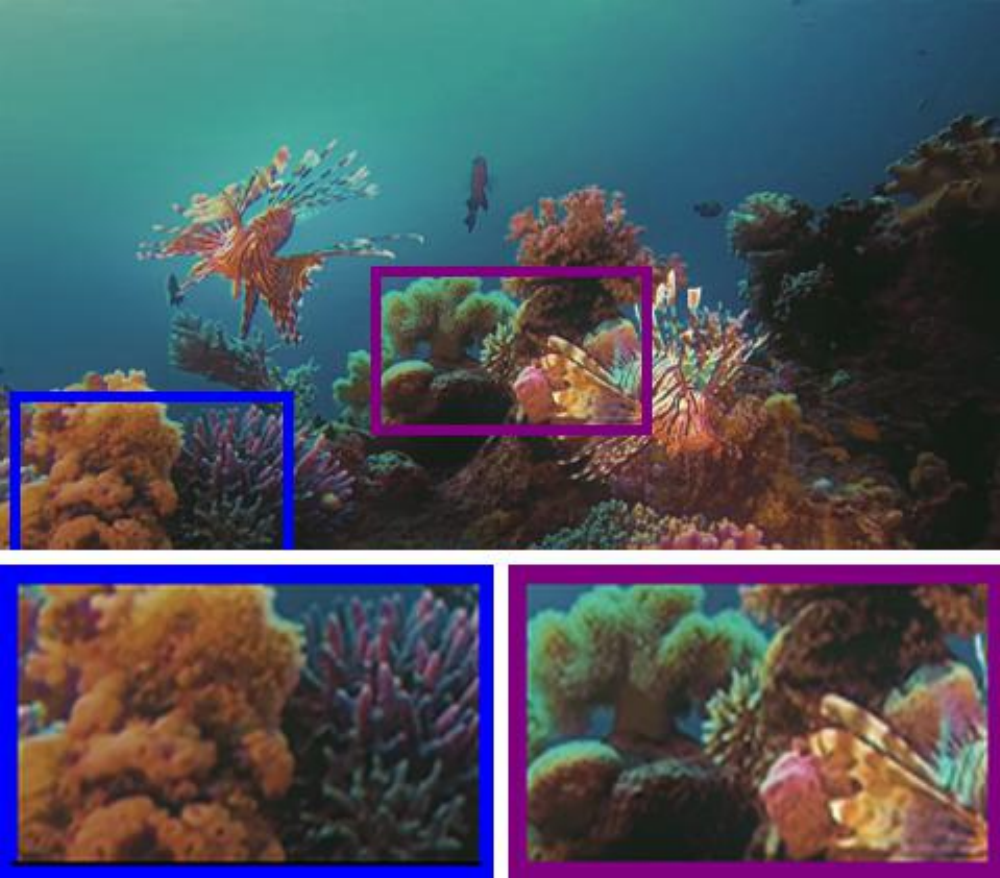}
        \includegraphics[width=\linewidth,  height=\puniheight]{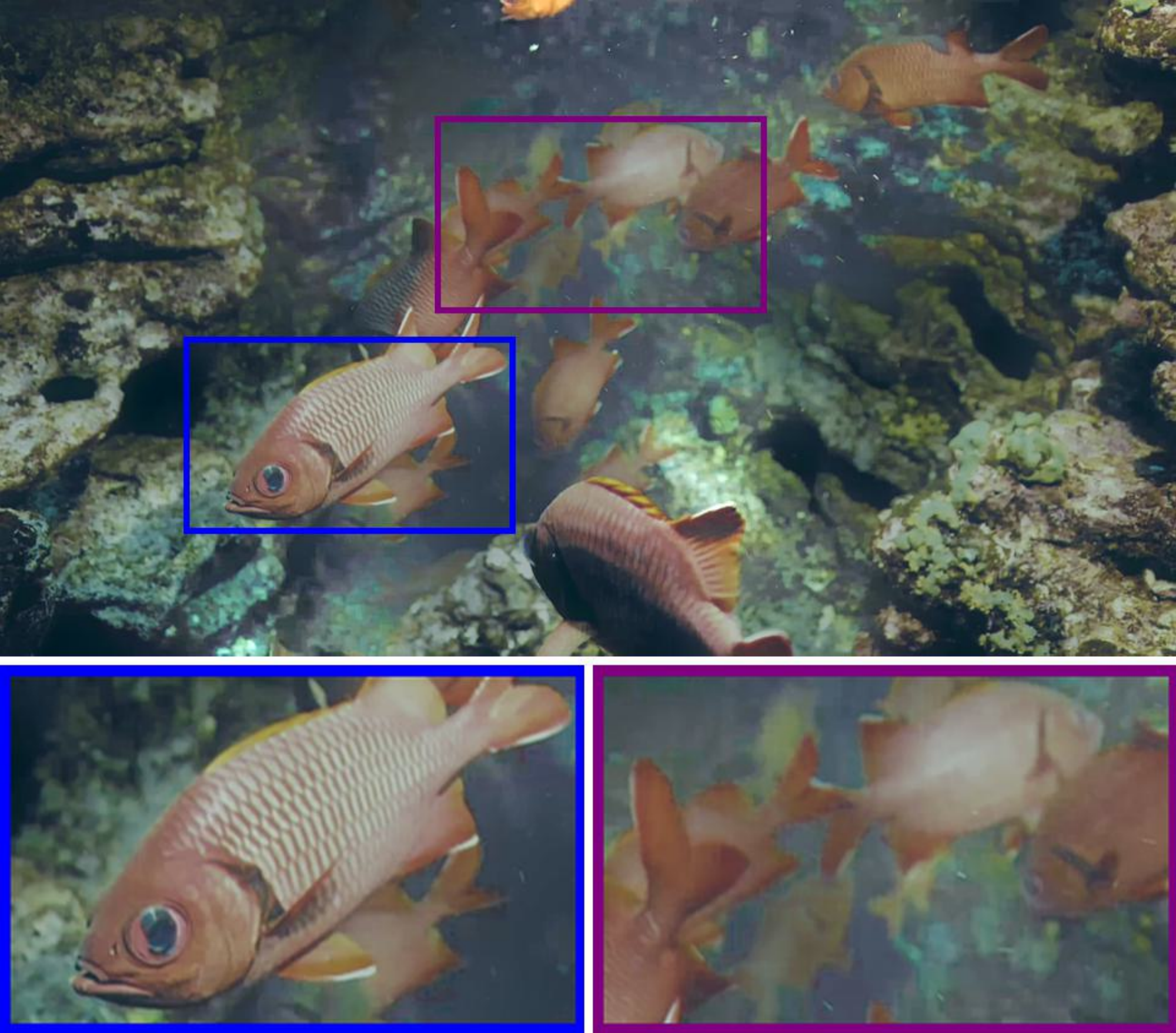}
		\caption{\footnotesize UDnet}
	\end{subfigure}
        \begin{subfigure}{0.105\linewidth}
		\centering
		\includegraphics[width=\linewidth,  height=\puniheight]{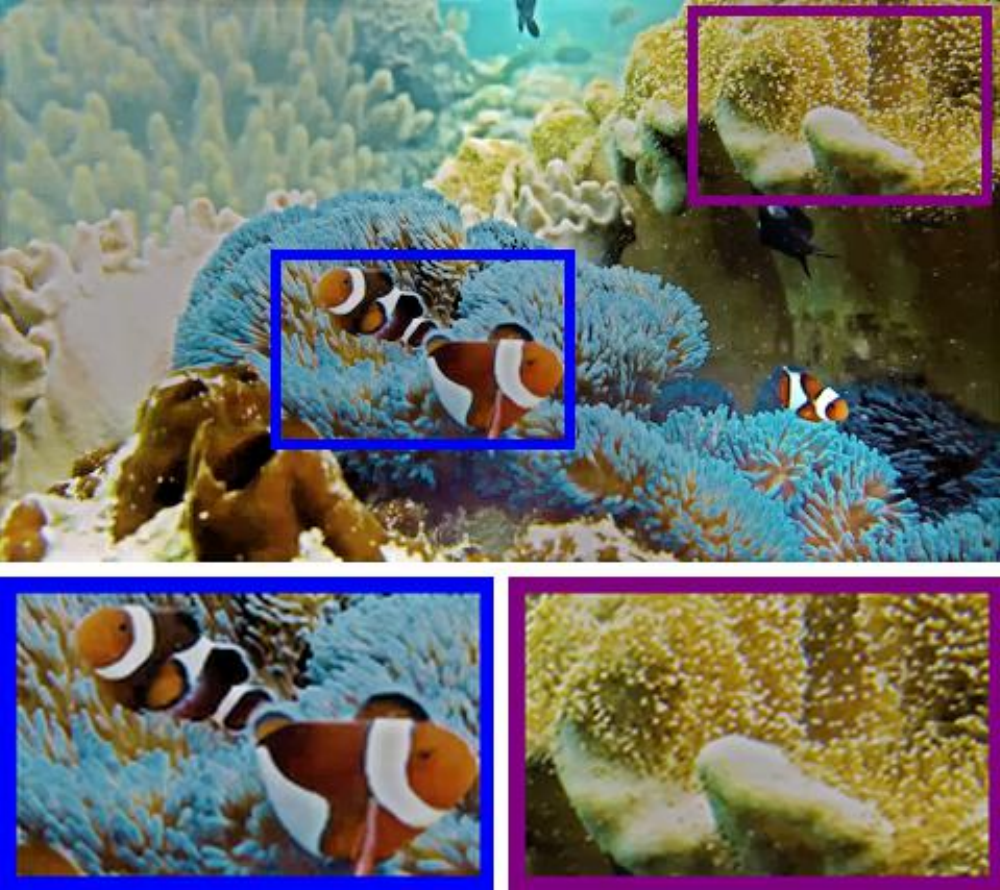} 
        \includegraphics[width=\linewidth,  height=\puniheight]{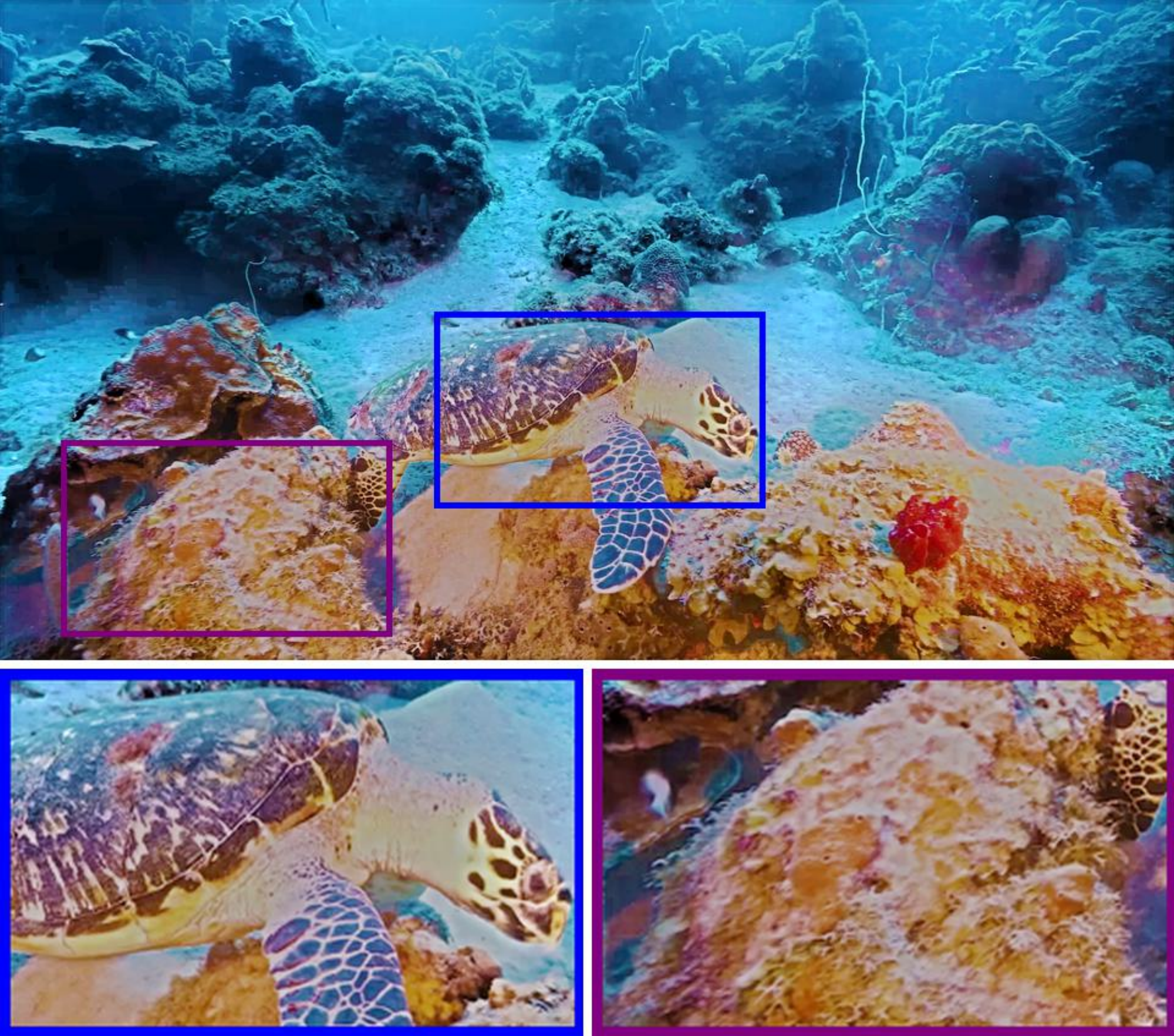} 
        \includegraphics[width=\linewidth,  height=\puniheight]{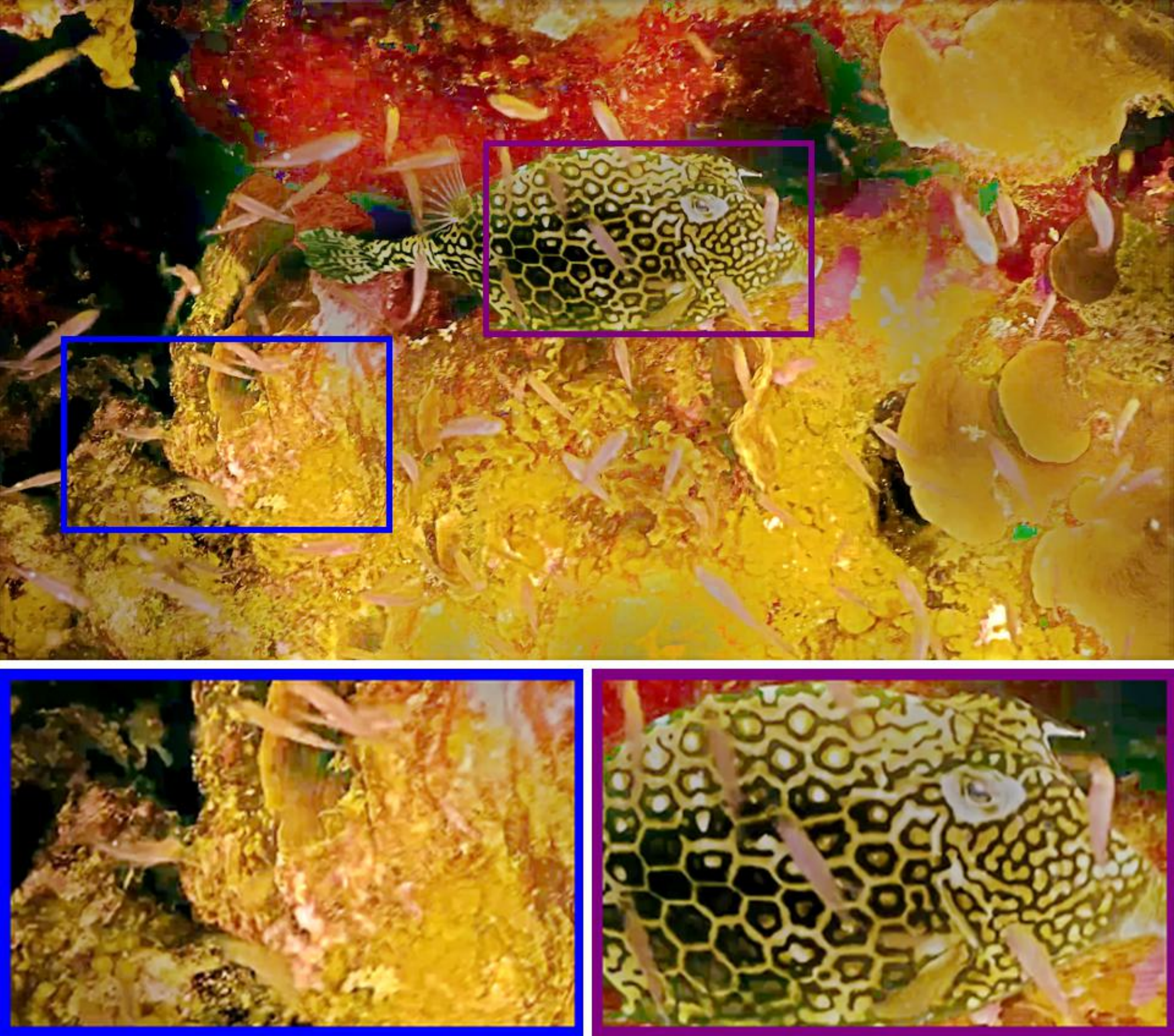}
        \includegraphics[width=\linewidth,  height=\puniheight]{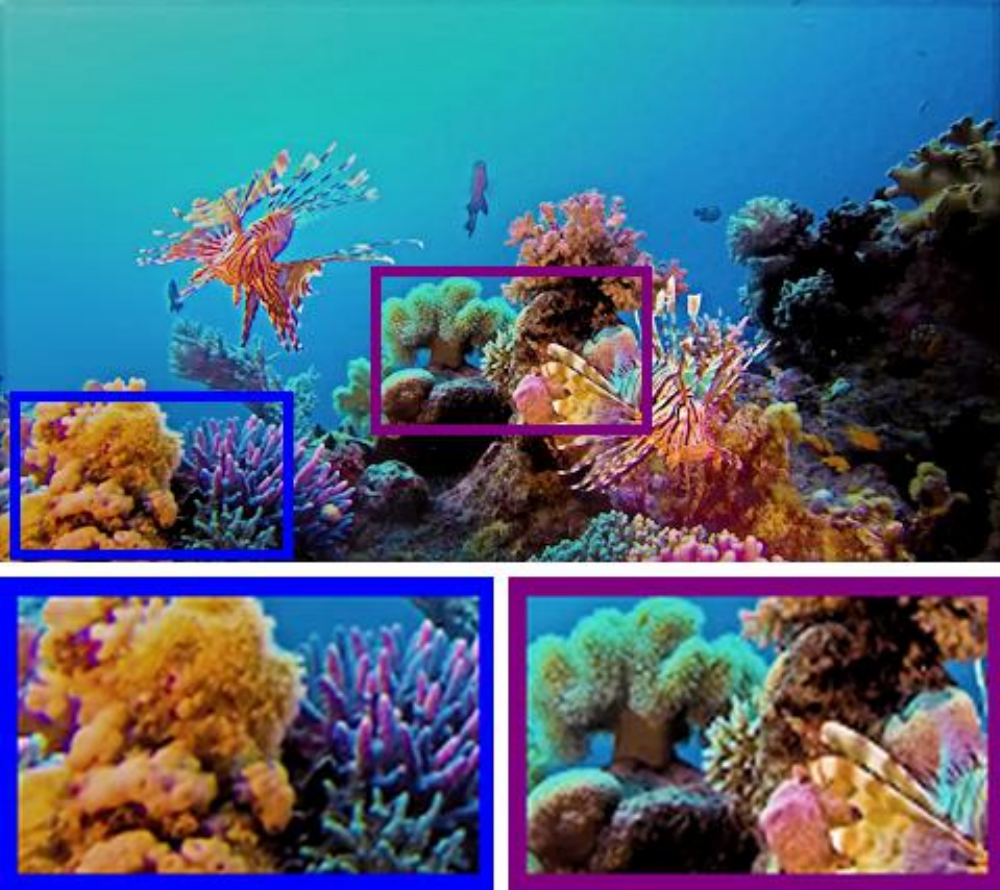}
        \includegraphics[width=\linewidth,  height=\puniheight]{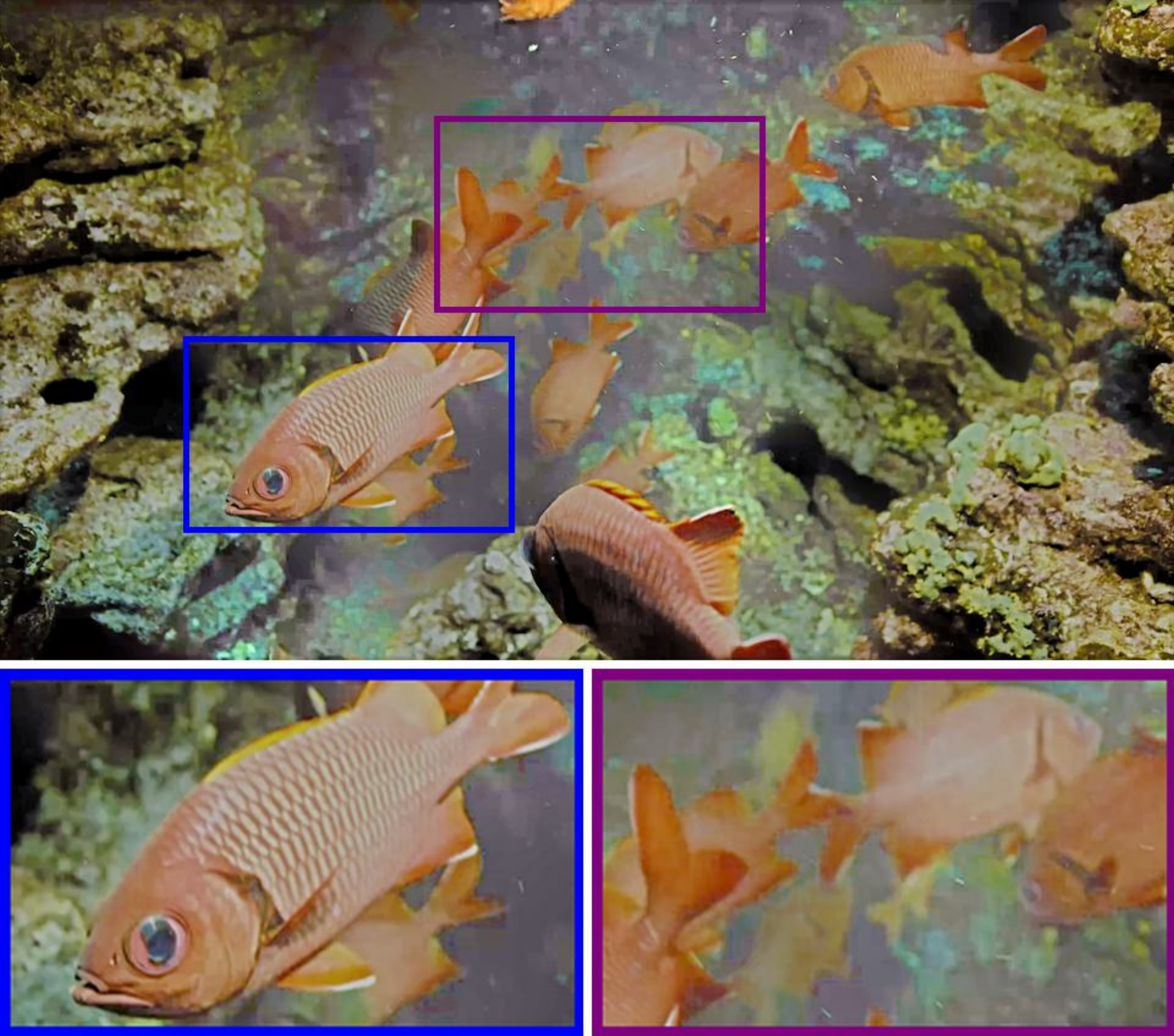}
		\caption{\footnotesize EIB-FNDL}
	\end{subfigure}
    \begin{subfigure}{0.105\linewidth}
		\centering
		\includegraphics[width=\linewidth,  height=\puniheight]{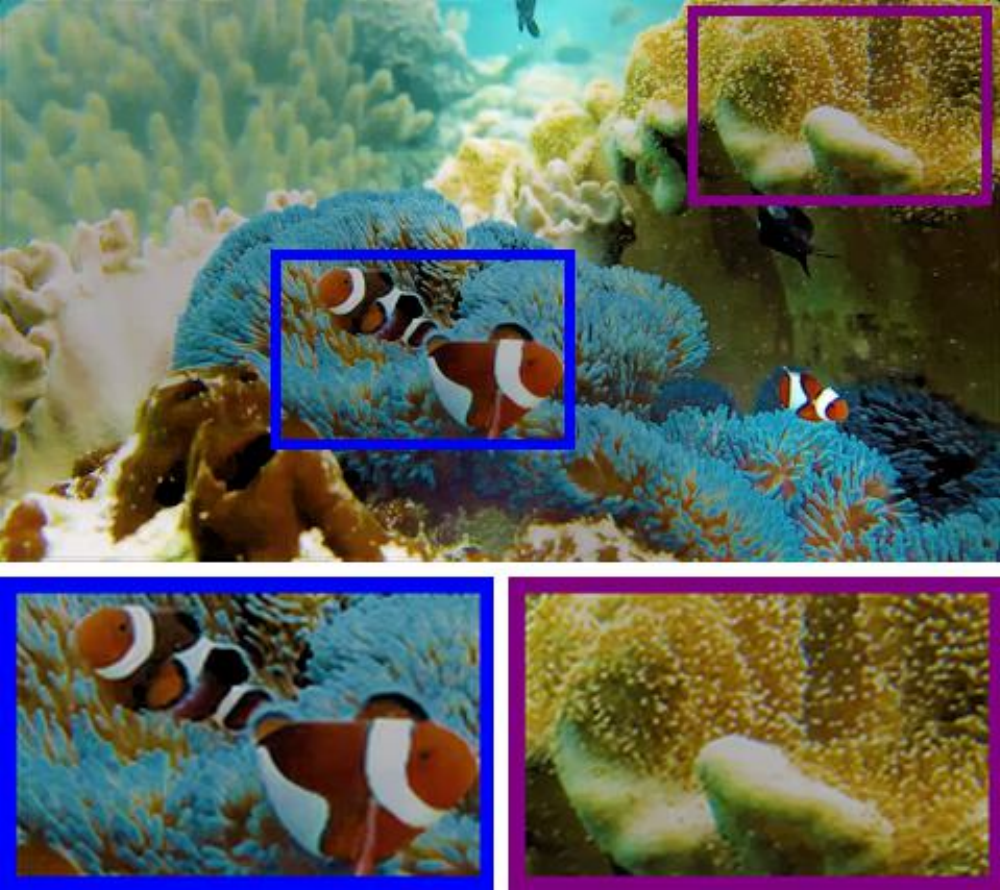} 
        \includegraphics[width=\linewidth,  height=\puniheight]{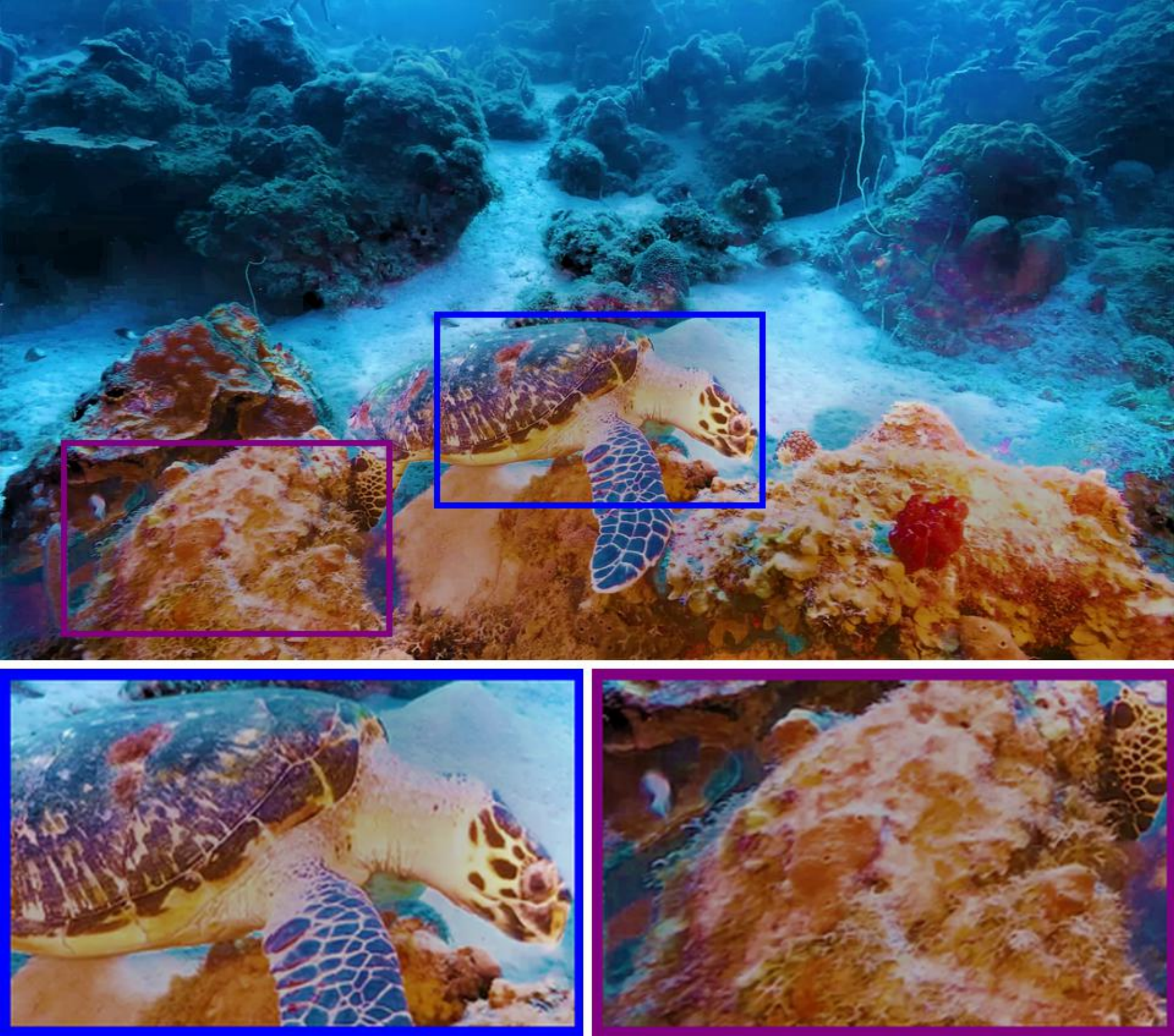} 
        \includegraphics[width=\linewidth,  height=\puniheight]{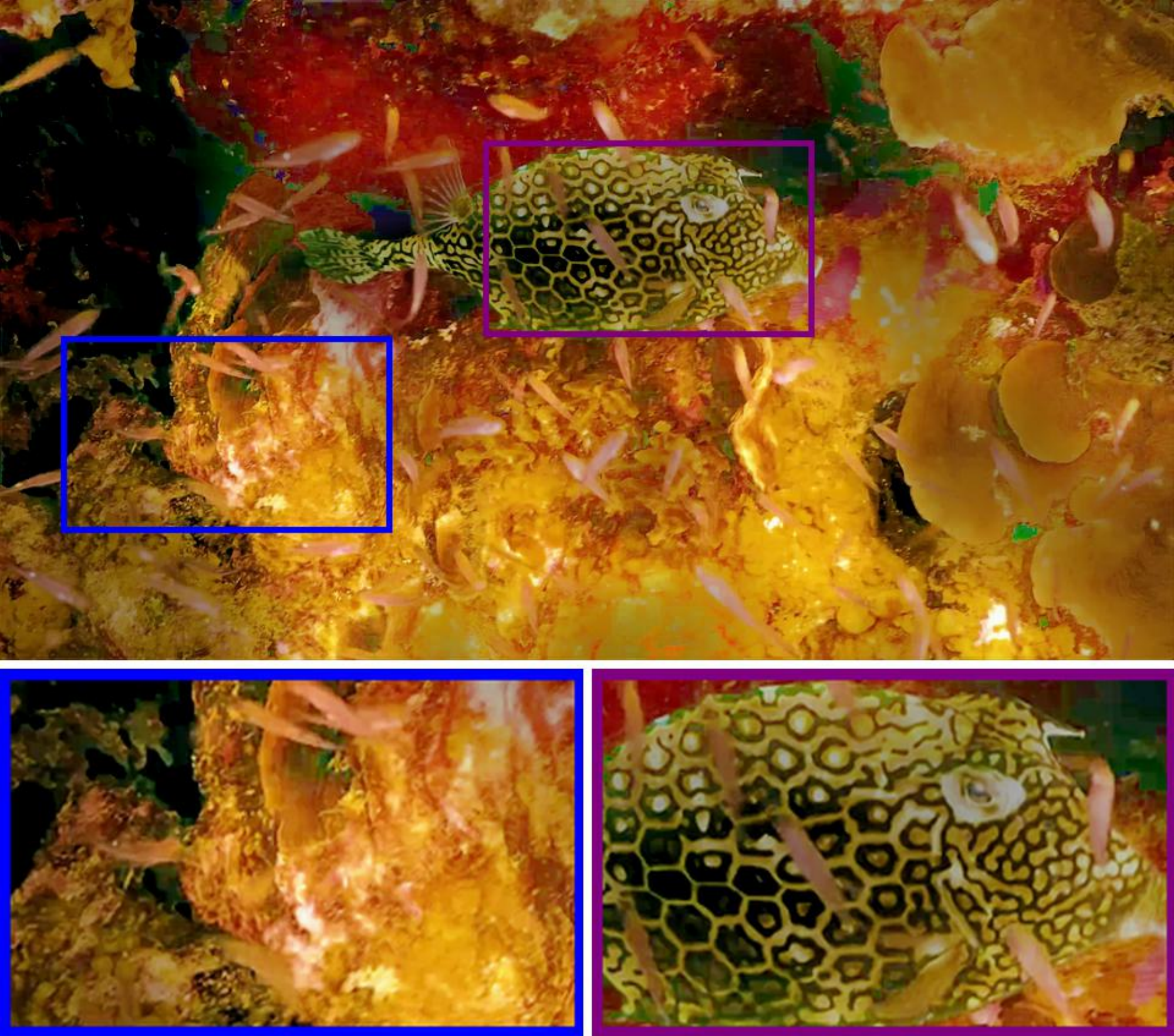}
        \includegraphics[width=\linewidth,  height=\puniheight]{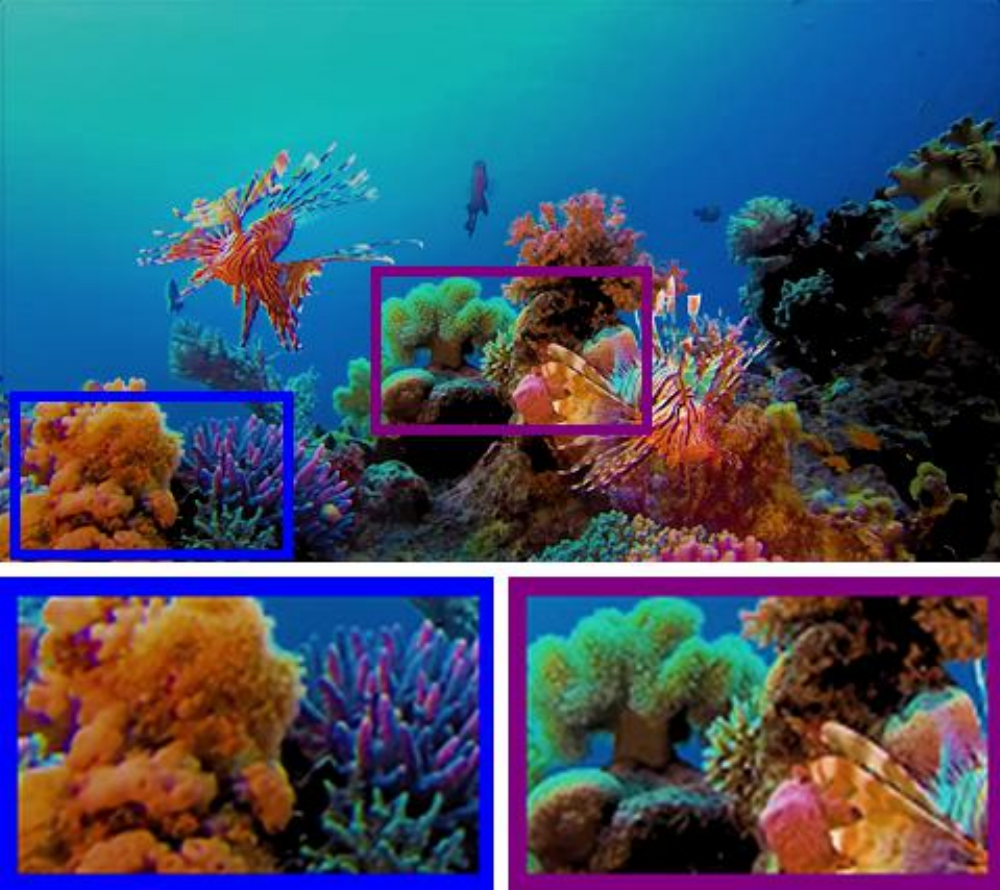}
        \includegraphics[width=\linewidth,  height=\puniheight]{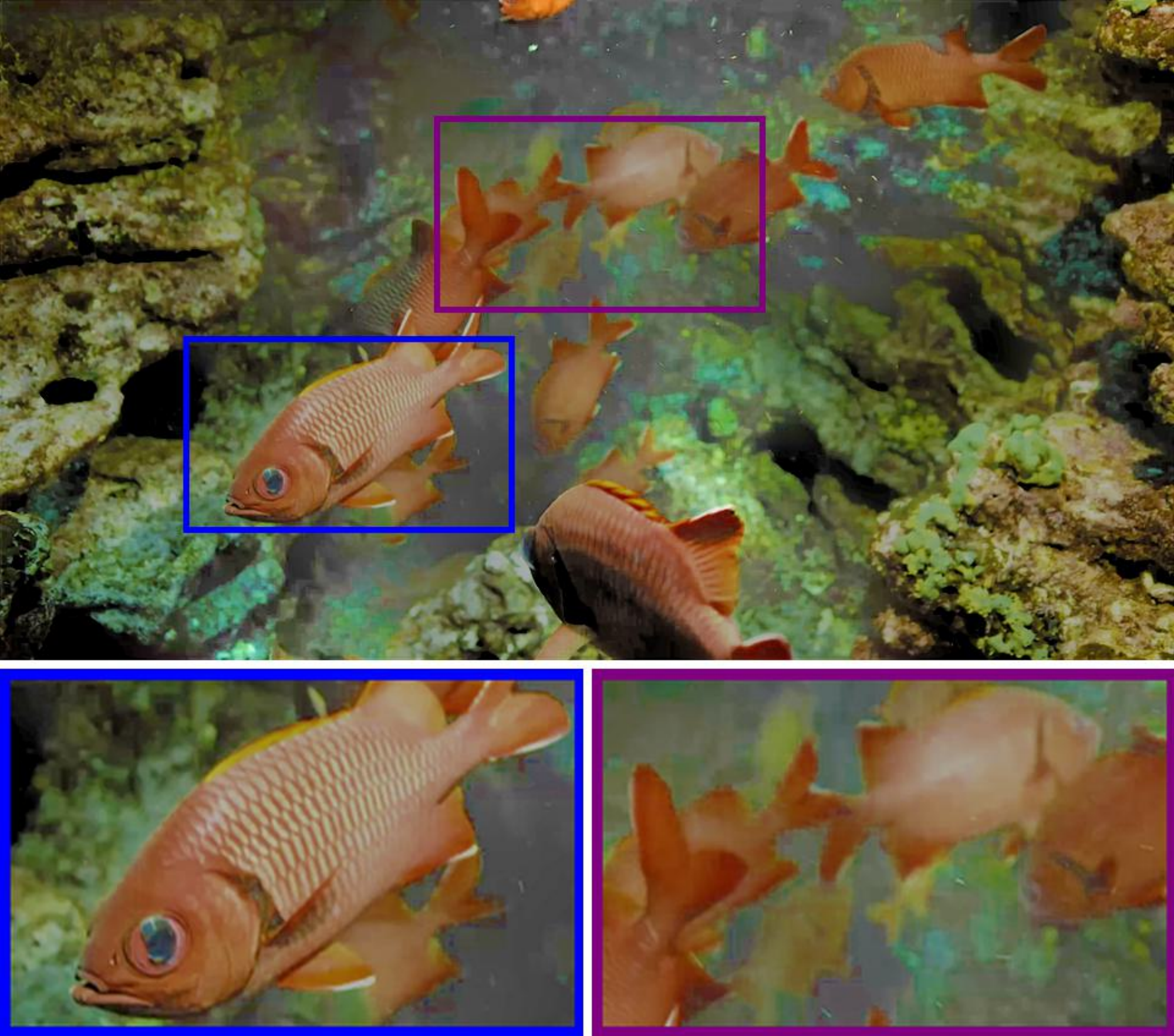}
		\caption{\footnotesize ALEN}
	\end{subfigure}
    \begin{subfigure}{0.105\linewidth}
		\centering
		\includegraphics[width=\linewidth,  height=\puniheight]{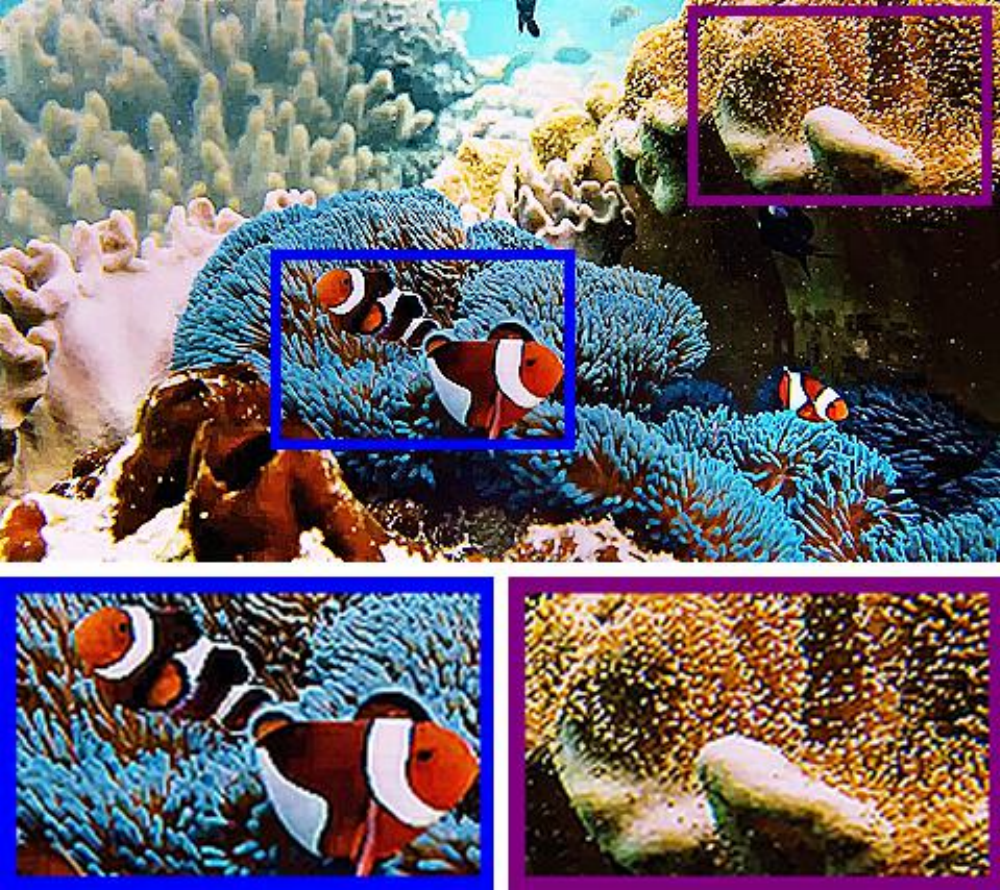} 
        \includegraphics[width=\linewidth,  height=\puniheight]{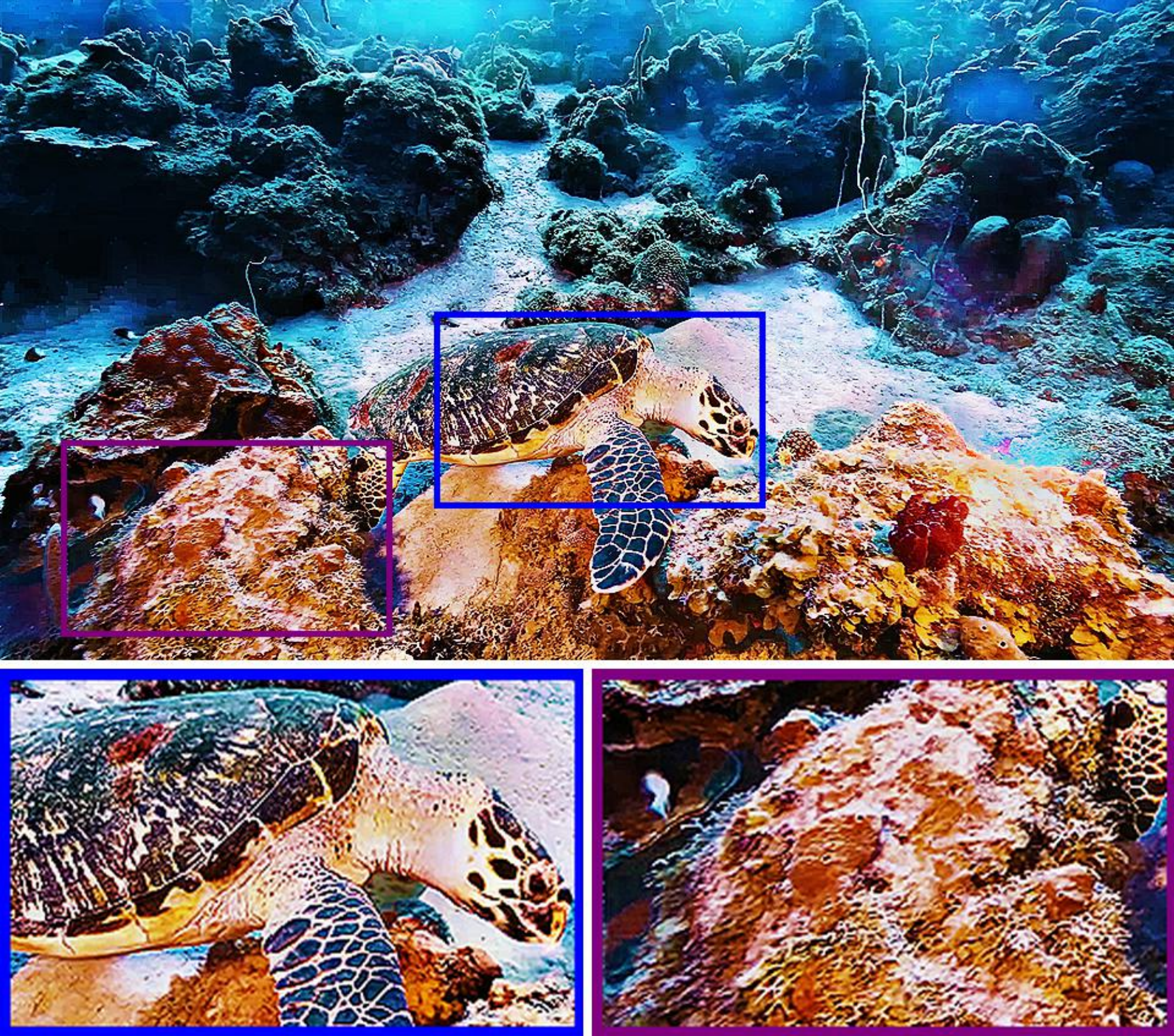} 
        \includegraphics[width=\linewidth,  height=\puniheight]{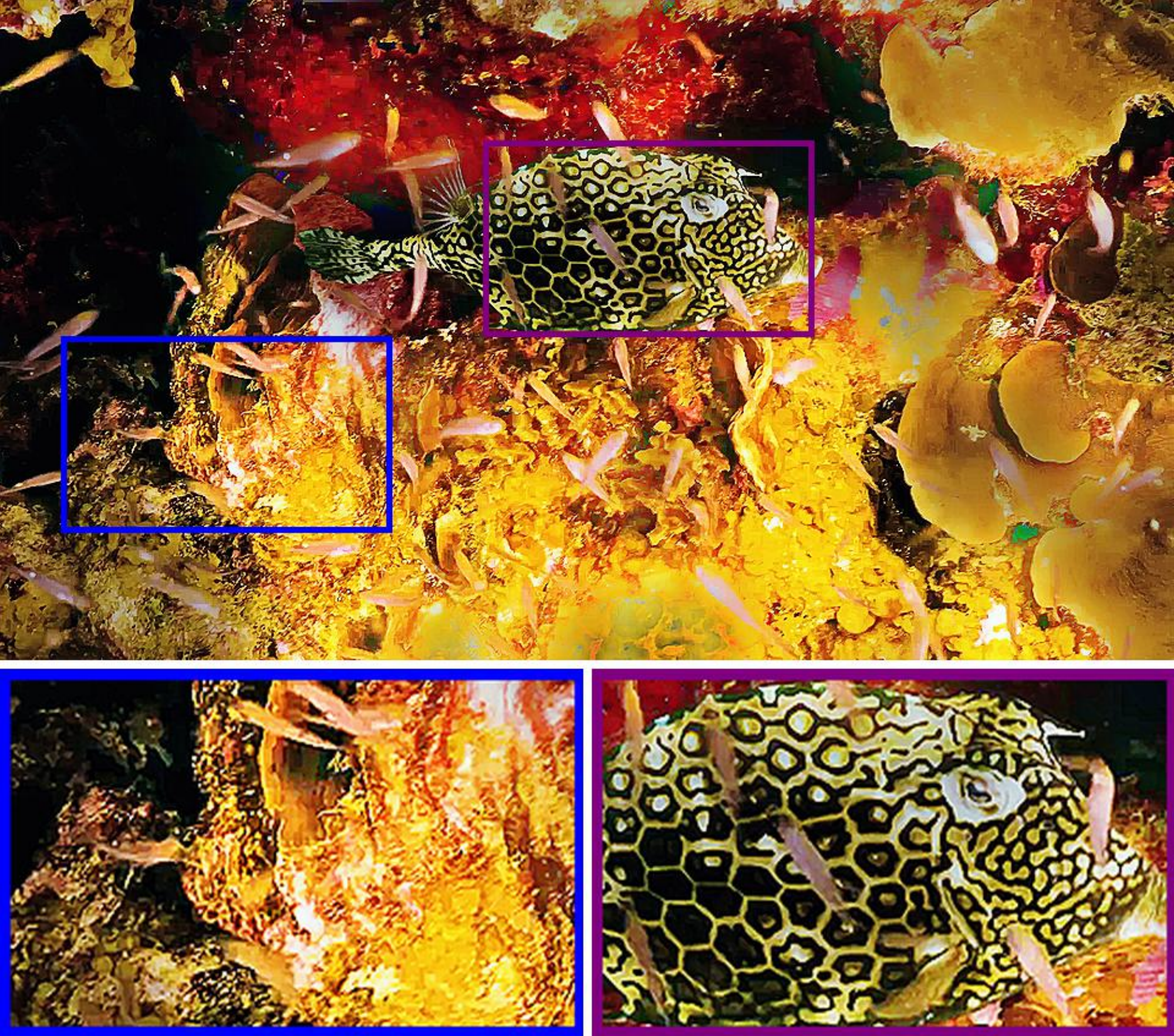}
        \includegraphics[width=\linewidth,  height=\puniheight]{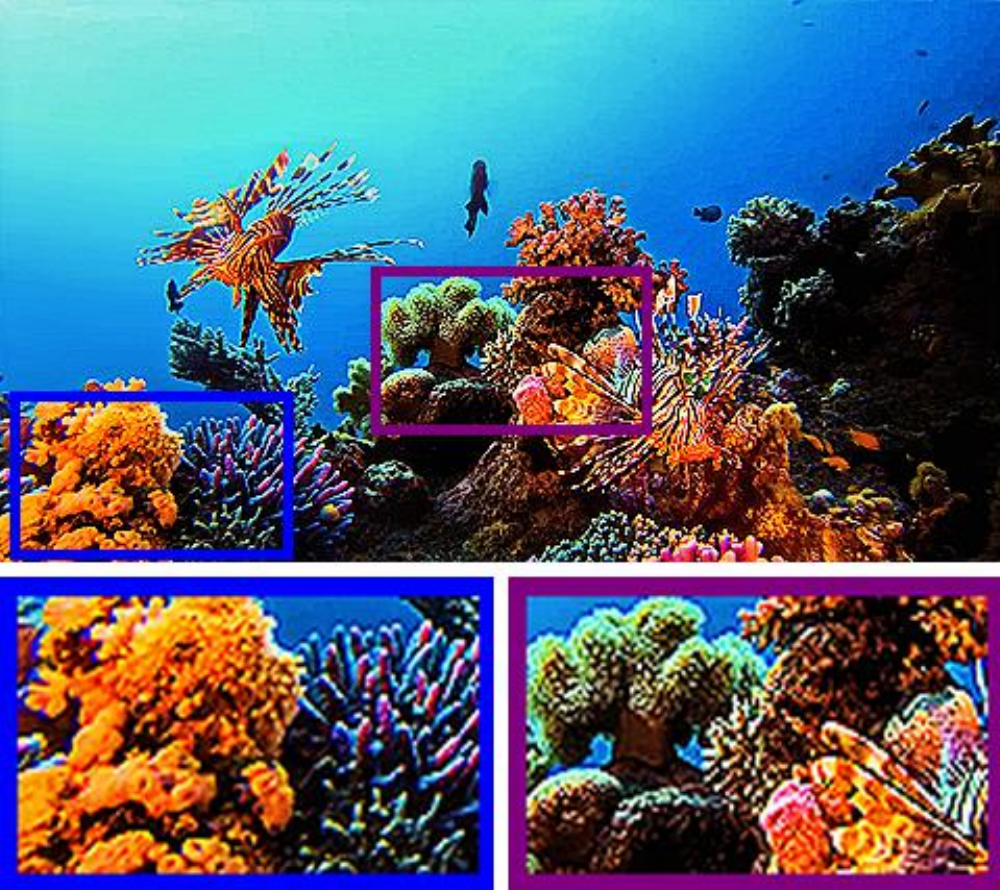} 
        \includegraphics[width=\linewidth,  height=\puniheight]{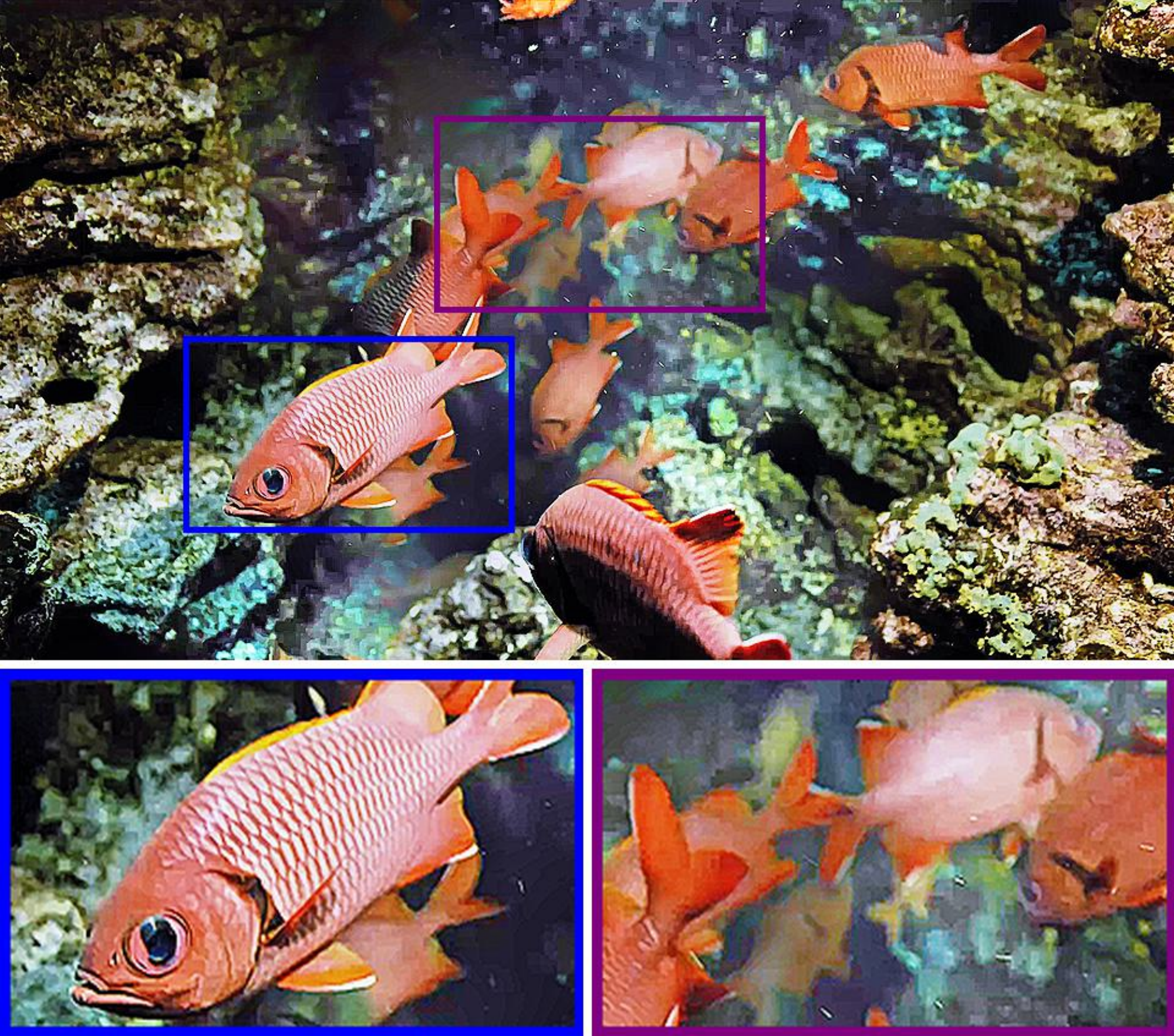} 
		\caption{\footnotesize UNIR-Net}
	\end{subfigure}
 
	\caption{Visual comparison of enhancement results on the synthetic PUNI dataset. Subfigure (a) shows the original underwater image affected by non-uniform illumination (NUI). Subfigures (b) to (r) display the enhancement results produced by different methods, all applied to the same NUI input in (a). The figure highlights visual differences in color correction, contrast, and illumination consistency across the evaluated approaches.}
	\label{Q1}
\end{figure*}

The qualitative evaluation examines images from the PUNI and NUID datasets, highlighting the performance of various methods under non-uniform illumination conditions. Figure \ref{Q1}, which shows results from the PUNI dataset, reveals that methods such as UNTV and UWNet exhibit noticeable color distortions. In contrast, approaches like GCP, EIB-FNDL, and ALEN improve illumination; however, a hazy effect is present in some of the enhancements produced by these methods. Techniques such as MMLE, PCDE, HFM, SMDR-IS, and UDAformer fail to eliminate dark regions, leading to uneven tonal distribution. Although TCTL-Net, ACDC, LENet, MACT, and UDnet enhance illumination, the resulting images exhibit muted colors and reduced detail. Among the evaluated methods, ICSP and UNIR-Net stand out in the PUNI dataset for their effective improvements in illumination and edge definition. UNIR-Net appears to outperform ICSP by offering a more balanced color representation and sharper edge rendering, as observed in Fig.\ref{Q1}.

\begin{figure*}[!ht]
	\newlength{\nuidheight}
	\setlength{\nuidheight}{1.85cm}
	\centering
	\begin{subfigure}{0.105\linewidth}
		\centering
        \includegraphics[width=\linewidth,  height=\nuidheight]{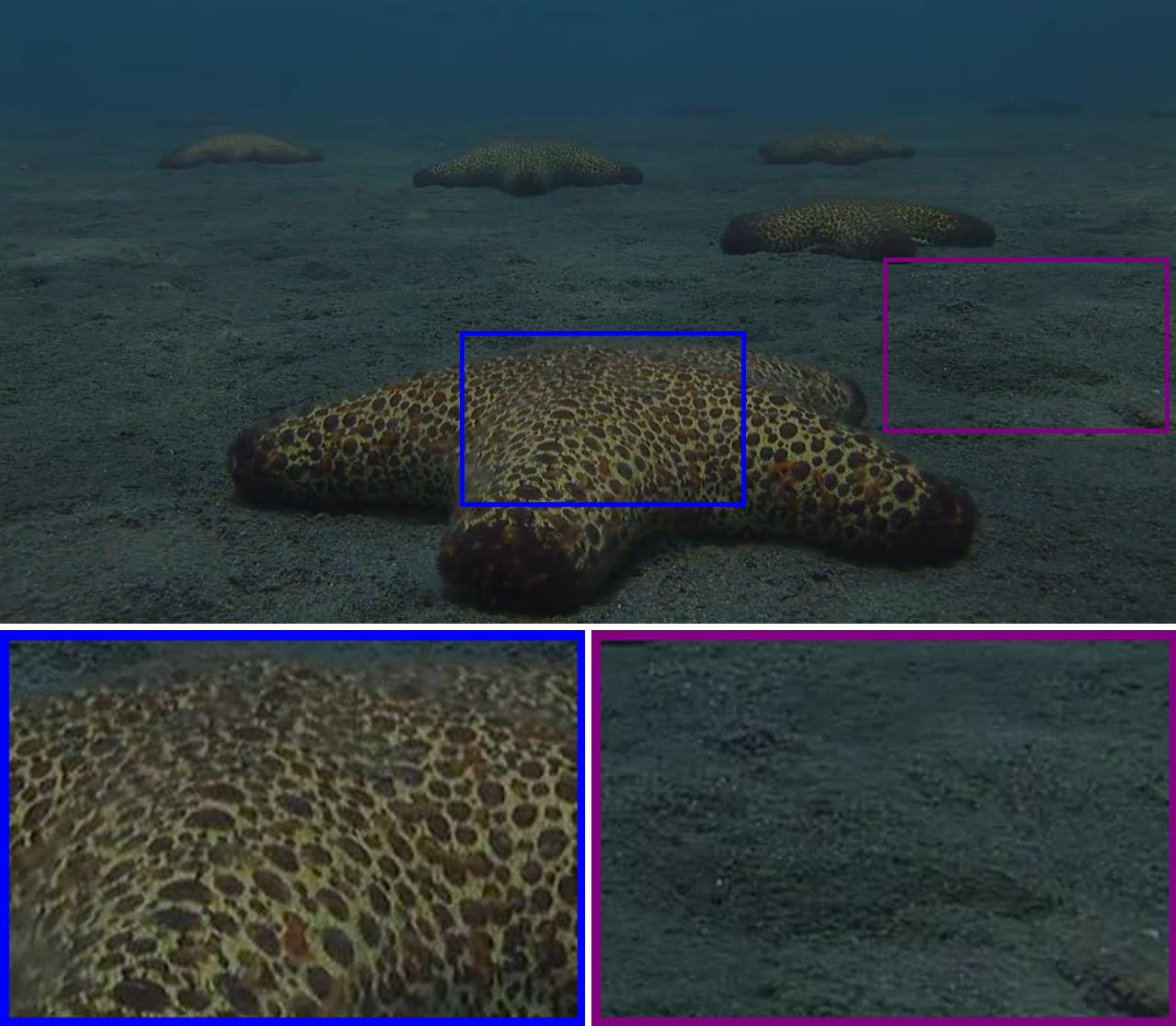} 
		\includegraphics[width=\linewidth,  height=\nuidheight]{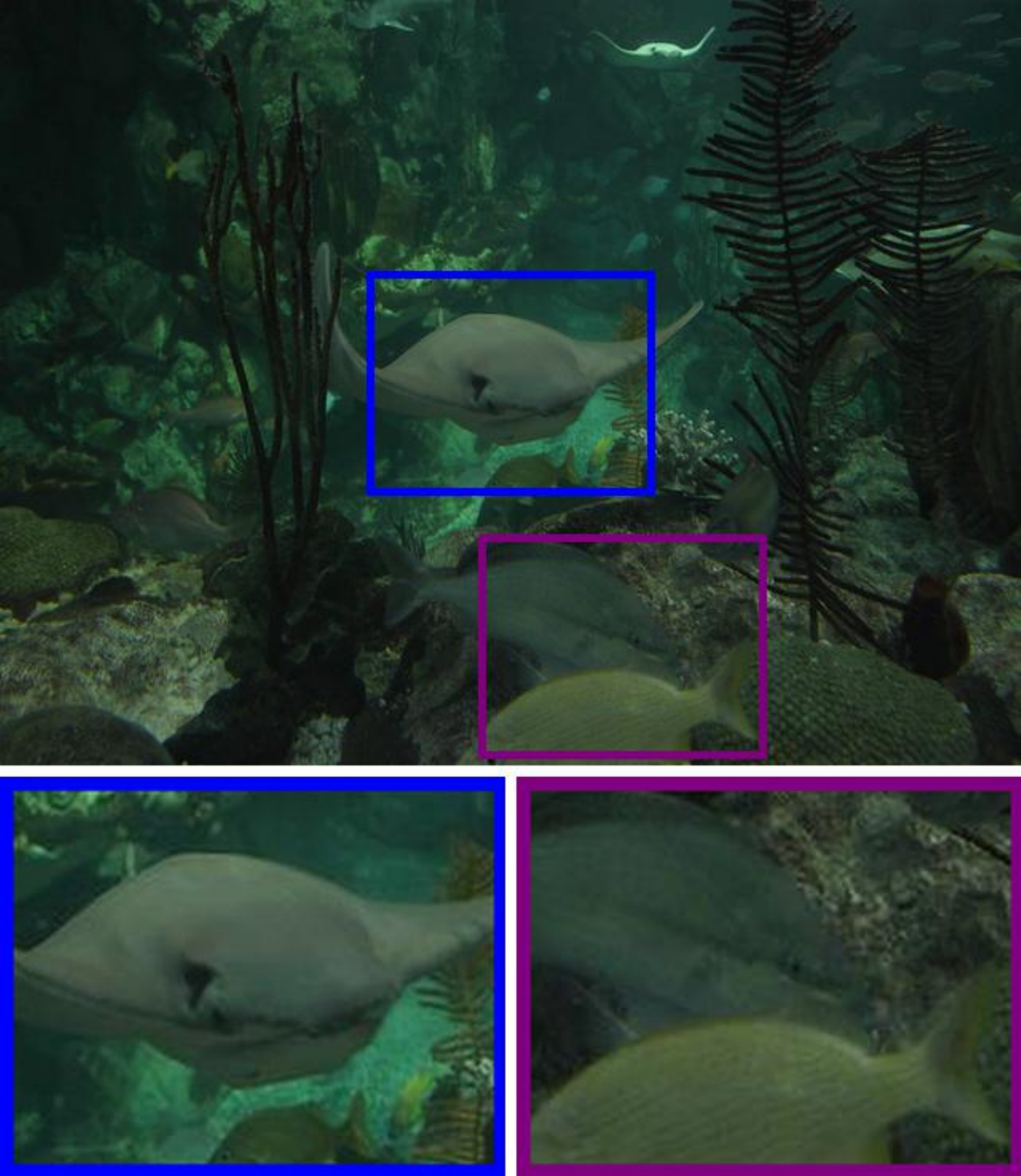} 
        \caption{\footnotesize NUI Image}
	\end{subfigure}
	\begin{subfigure}{0.105\linewidth}
		\centering
		\includegraphics[width=\linewidth,  height=\nuidheight]{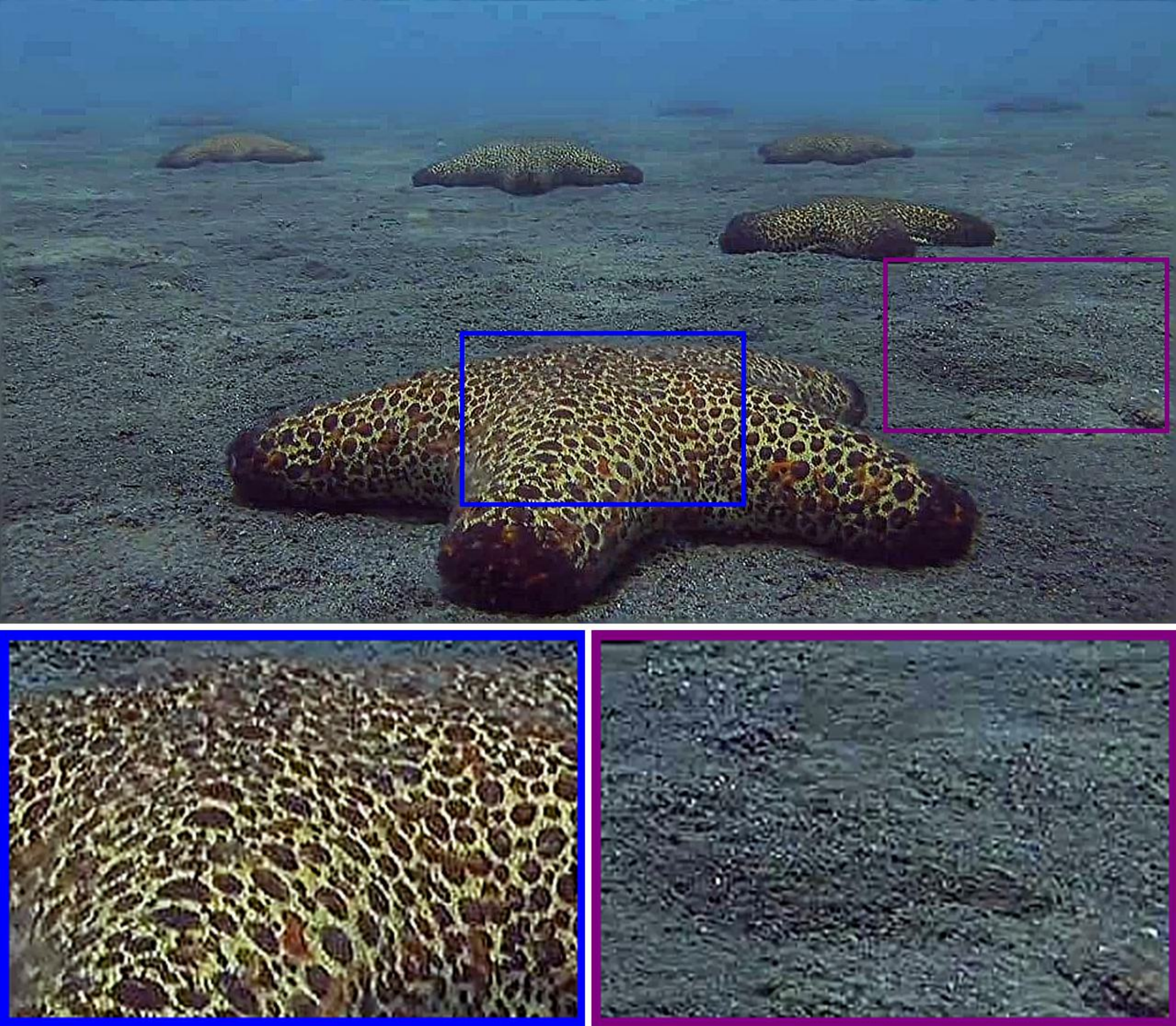} 
        \includegraphics[width=\linewidth,  height=\nuidheight]{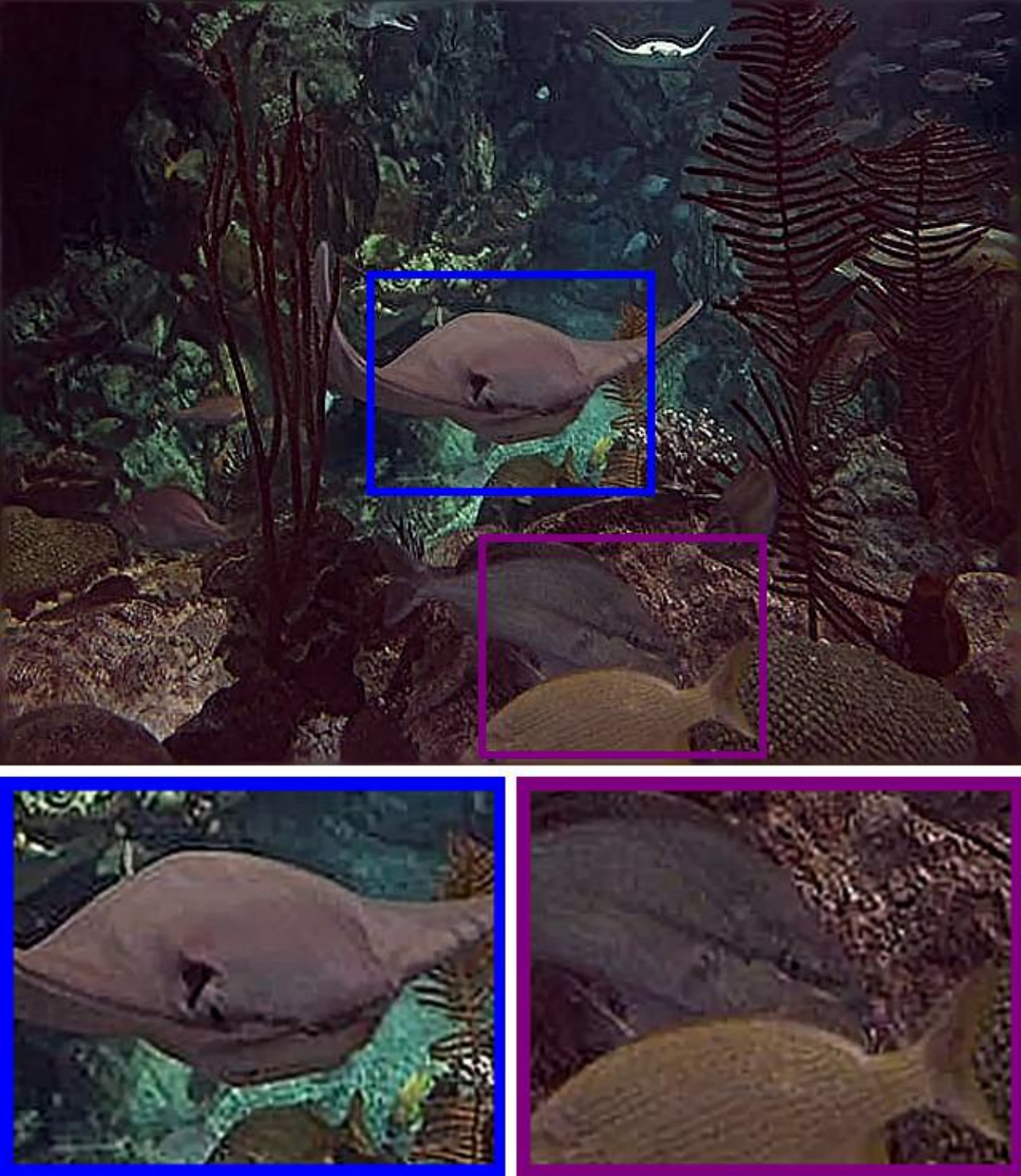}
		\caption{\footnotesize UNTV}
	\end{subfigure}
	\begin{subfigure}{0.105\linewidth}
		\centering
		\includegraphics[width=\linewidth,  height=\nuidheight]{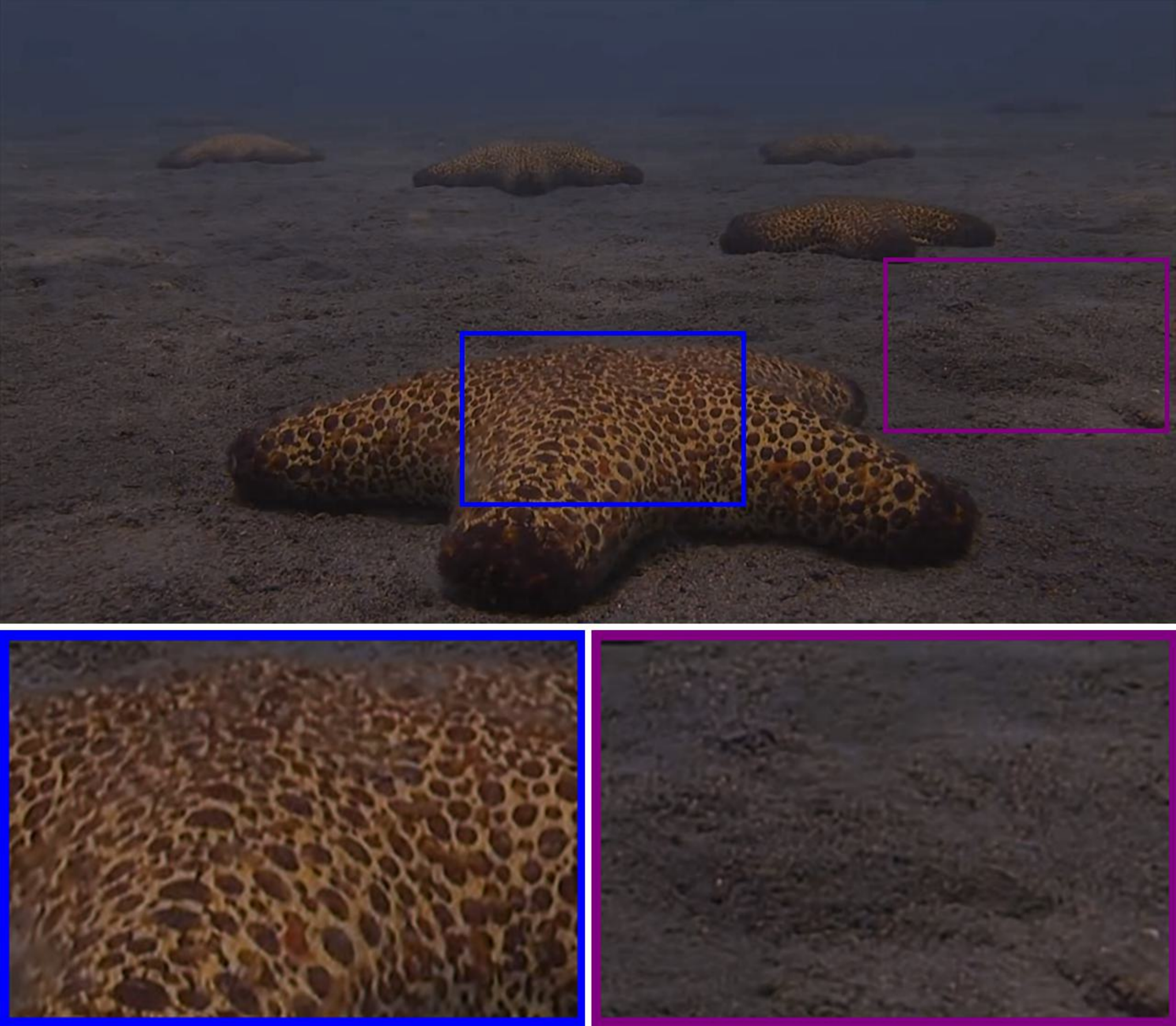}
        \includegraphics[width=\linewidth,  height=\nuidheight]{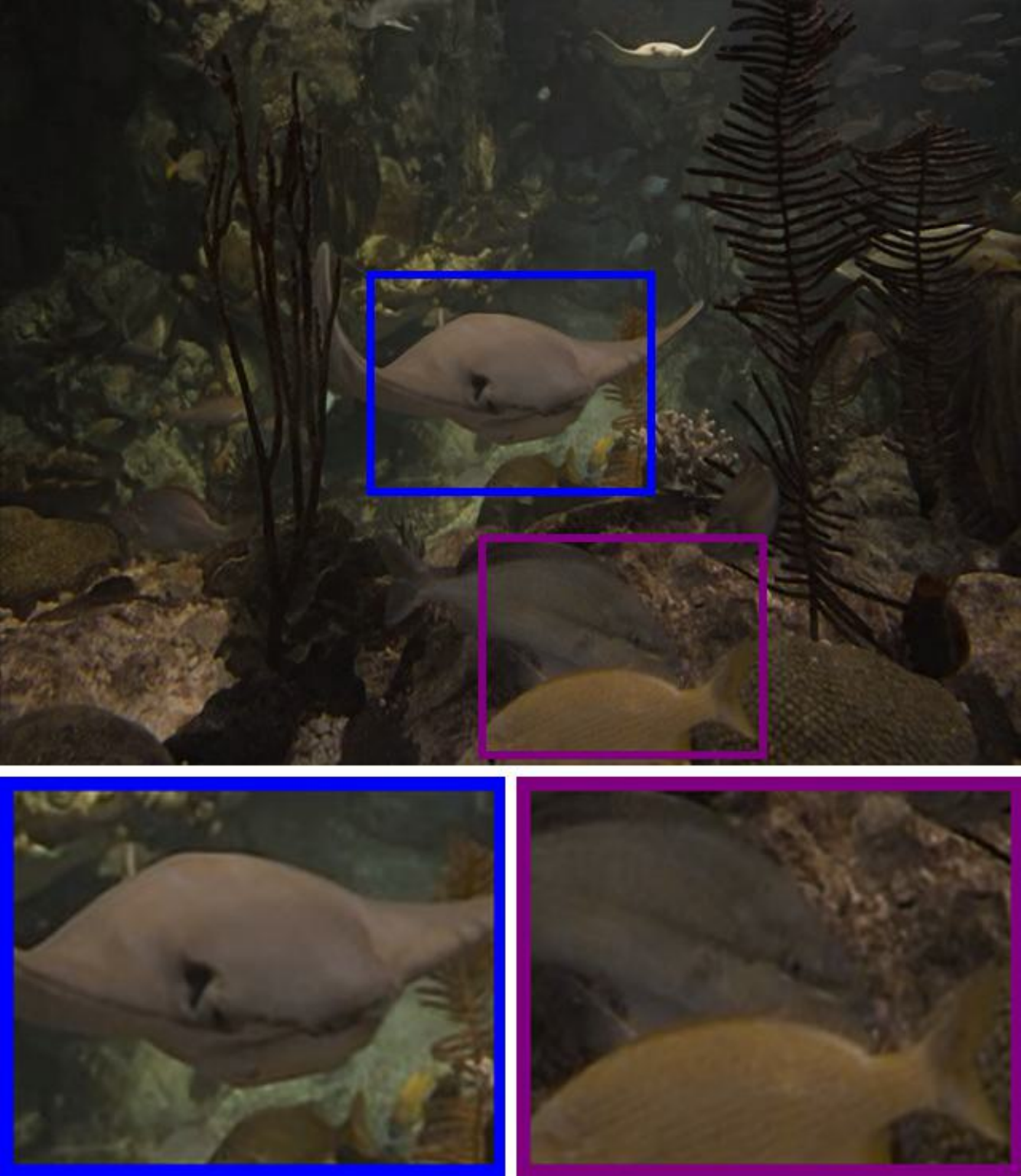}
		\caption{\footnotesize UWNet}
	\end{subfigure}
    \begin{subfigure}{0.105\linewidth}
		\centering
		\includegraphics[width=\linewidth,  height=\nuidheight]{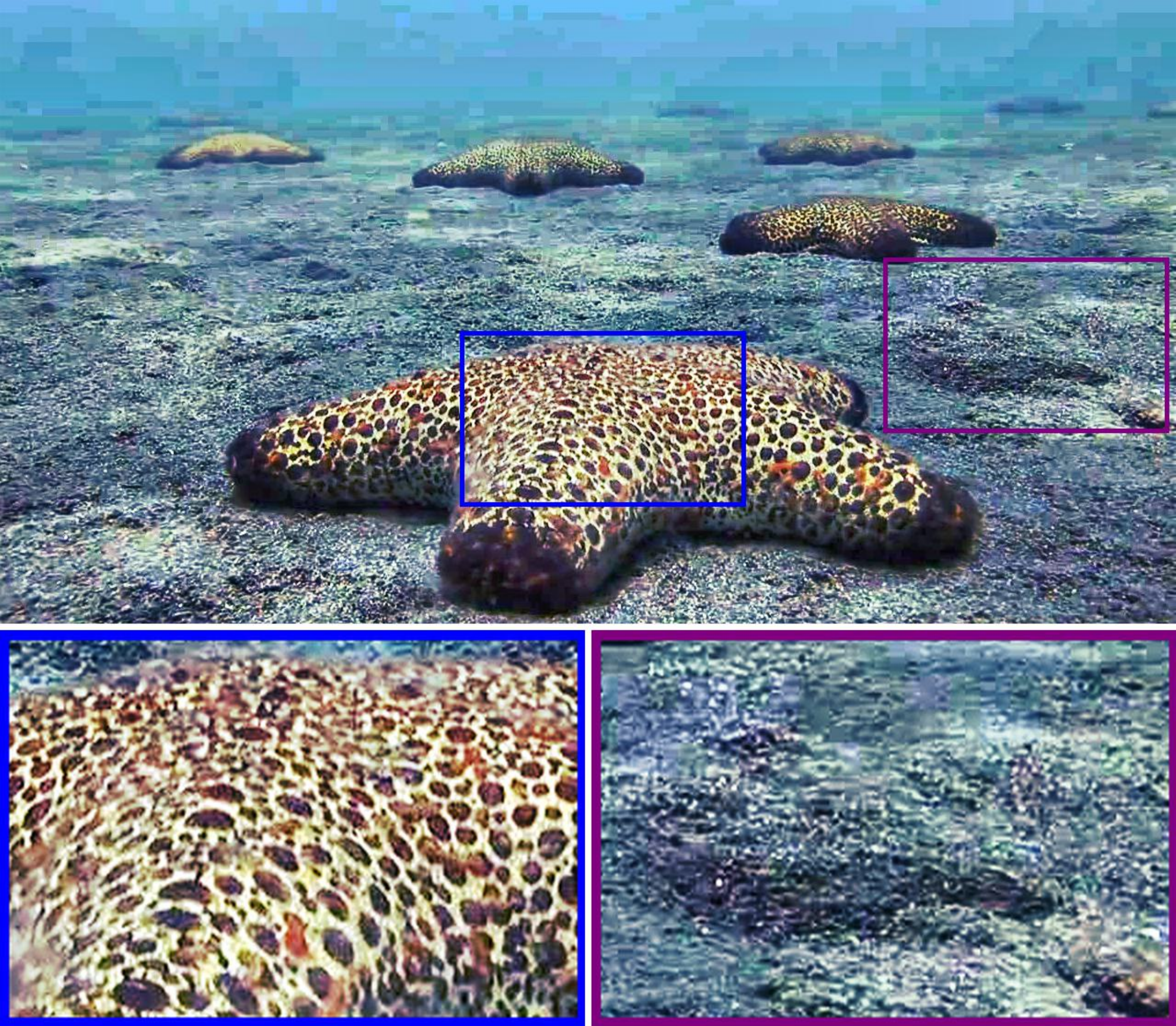} 
        \includegraphics[width=\linewidth,  height=\nuidheight]{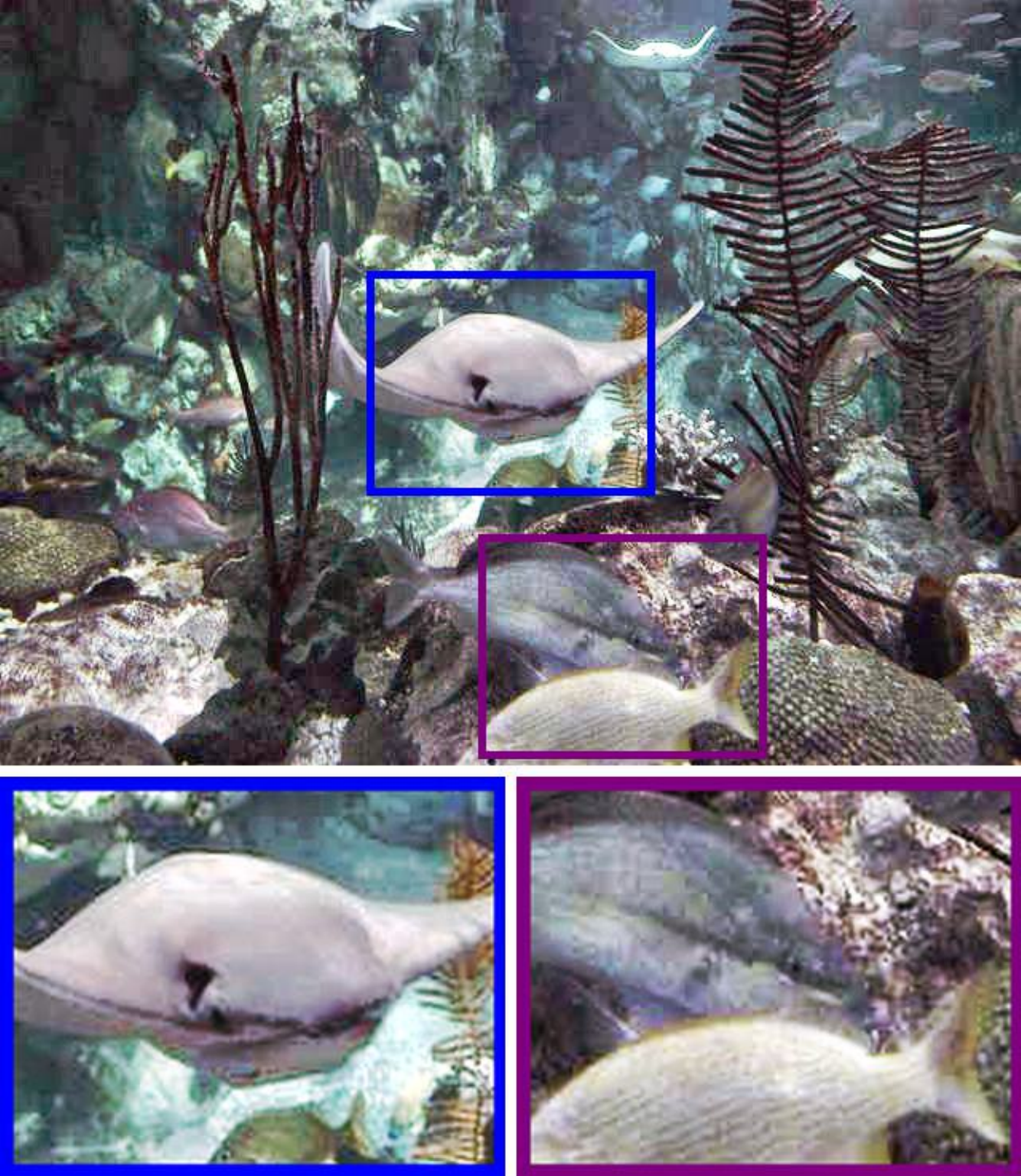} 
		\caption{\footnotesize ACDC}
	\end{subfigure}
	\begin{subfigure}{0.105\linewidth}
		\centering
		\includegraphics[width=\linewidth,  height=\nuidheight]{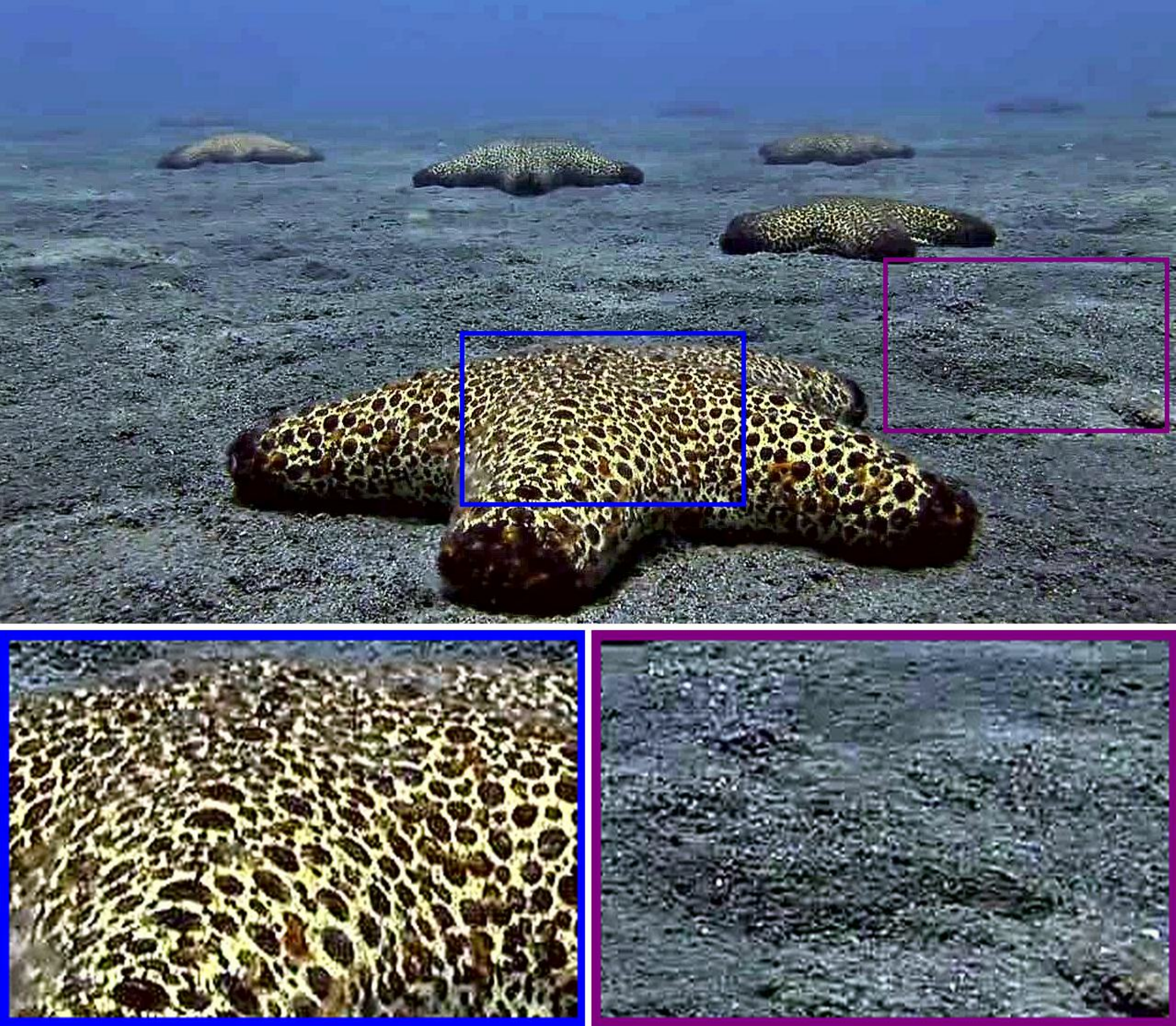} 
        \includegraphics[width=\linewidth,  height=\nuidheight]{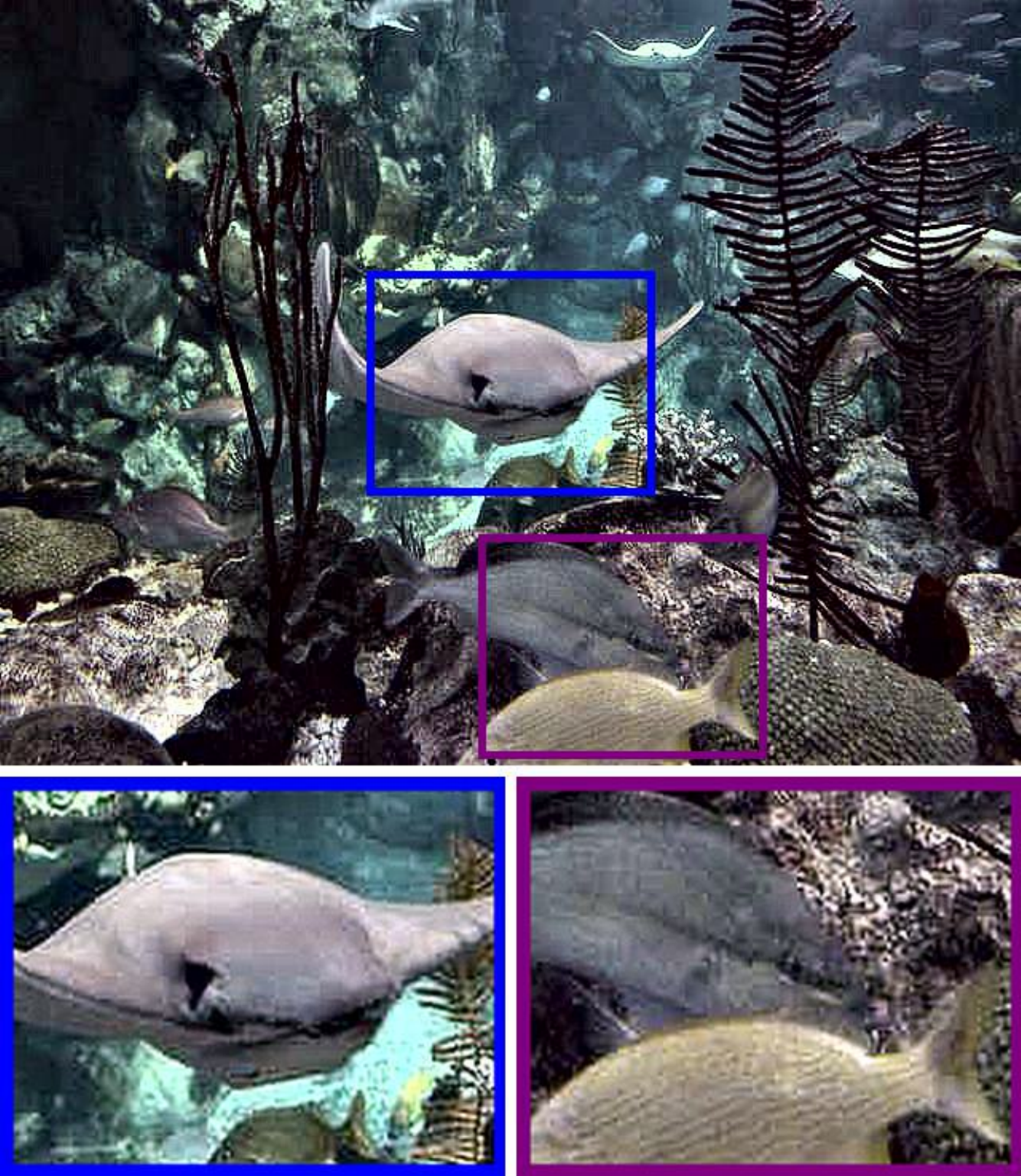} 
		\caption{\footnotesize MMLE}
	\end{subfigure}
    \begin{subfigure}{0.105\linewidth}
		\centering
		\includegraphics[width=\linewidth,  height=\nuidheight]{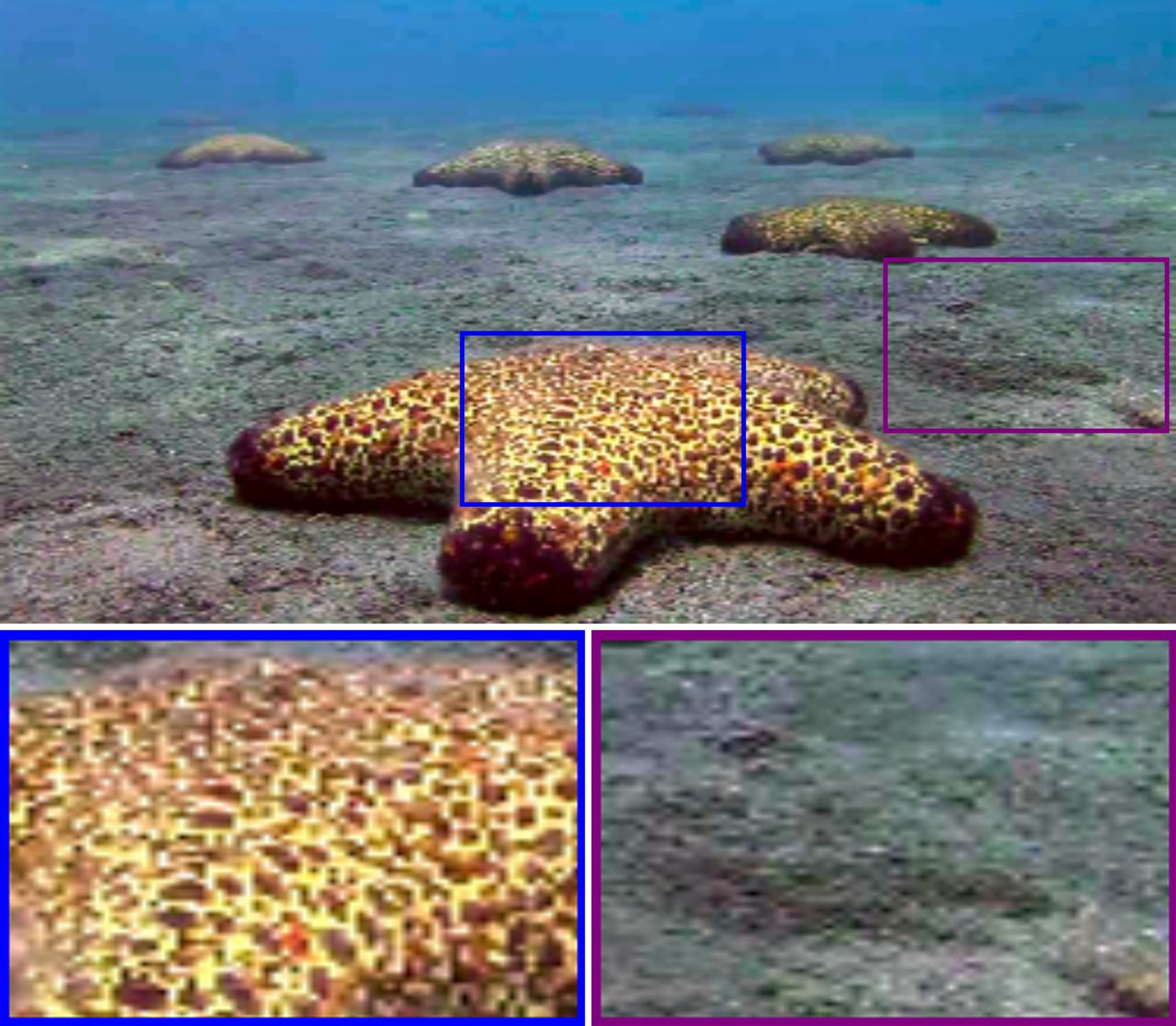} 
        \includegraphics[width=\linewidth,  height=\nuidheight]{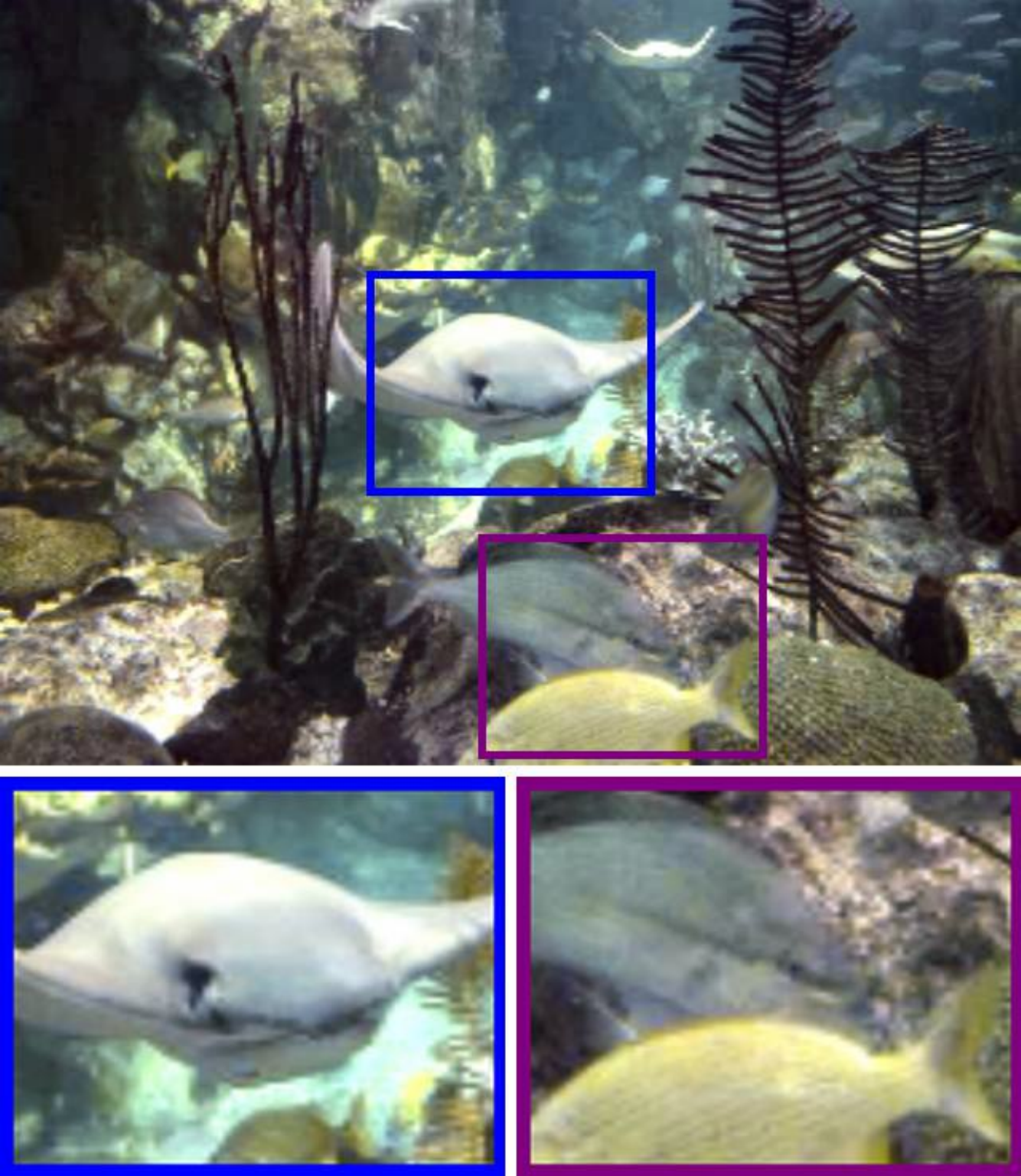} 
		\caption{\footnotesize TCTL-Net}
	\end{subfigure}
    \begin{subfigure}{0.105\linewidth}
		\centering
		\includegraphics[width=\linewidth,  height=\nuidheight]{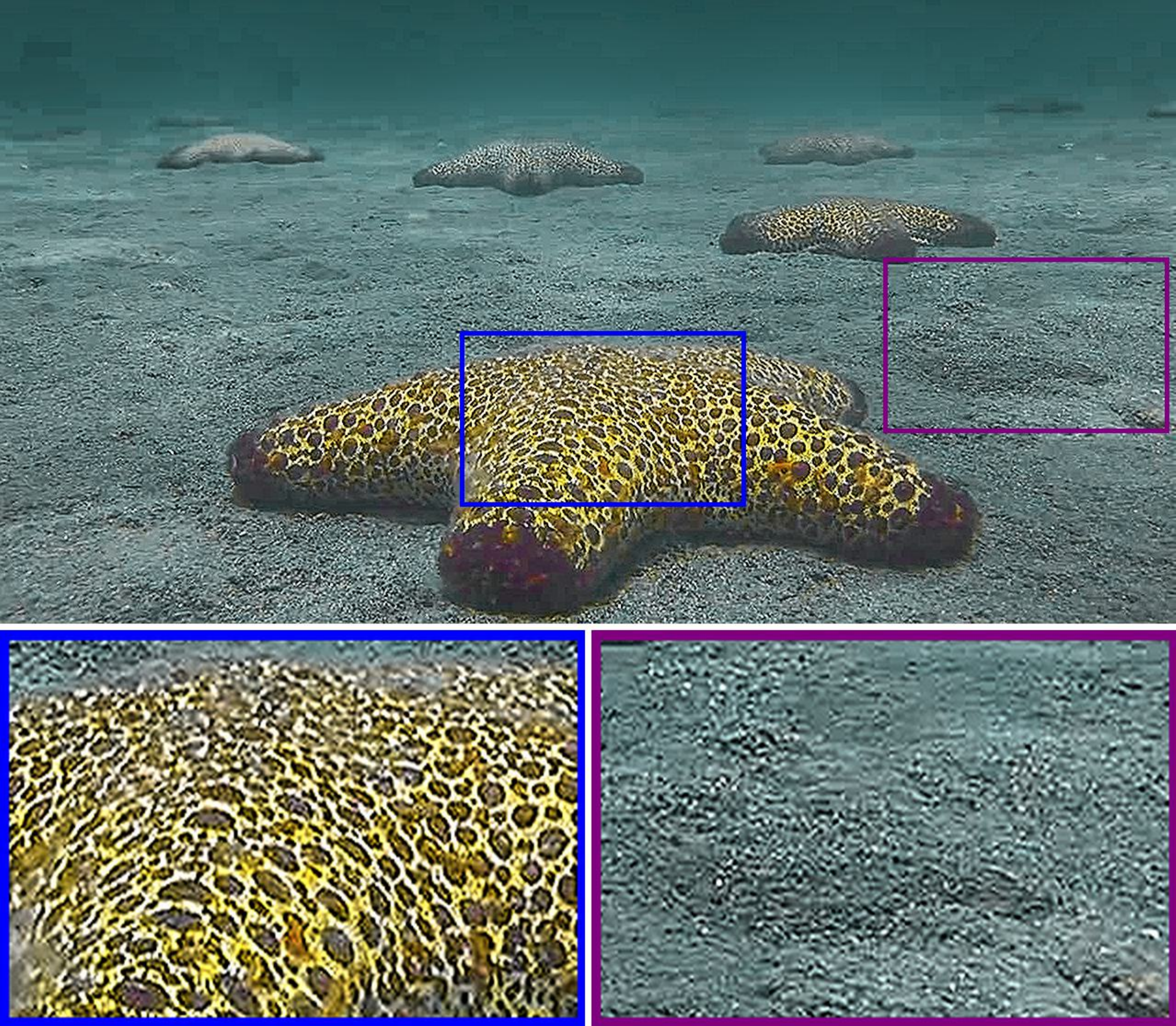} 
        \includegraphics[width=\linewidth,  height=\nuidheight]{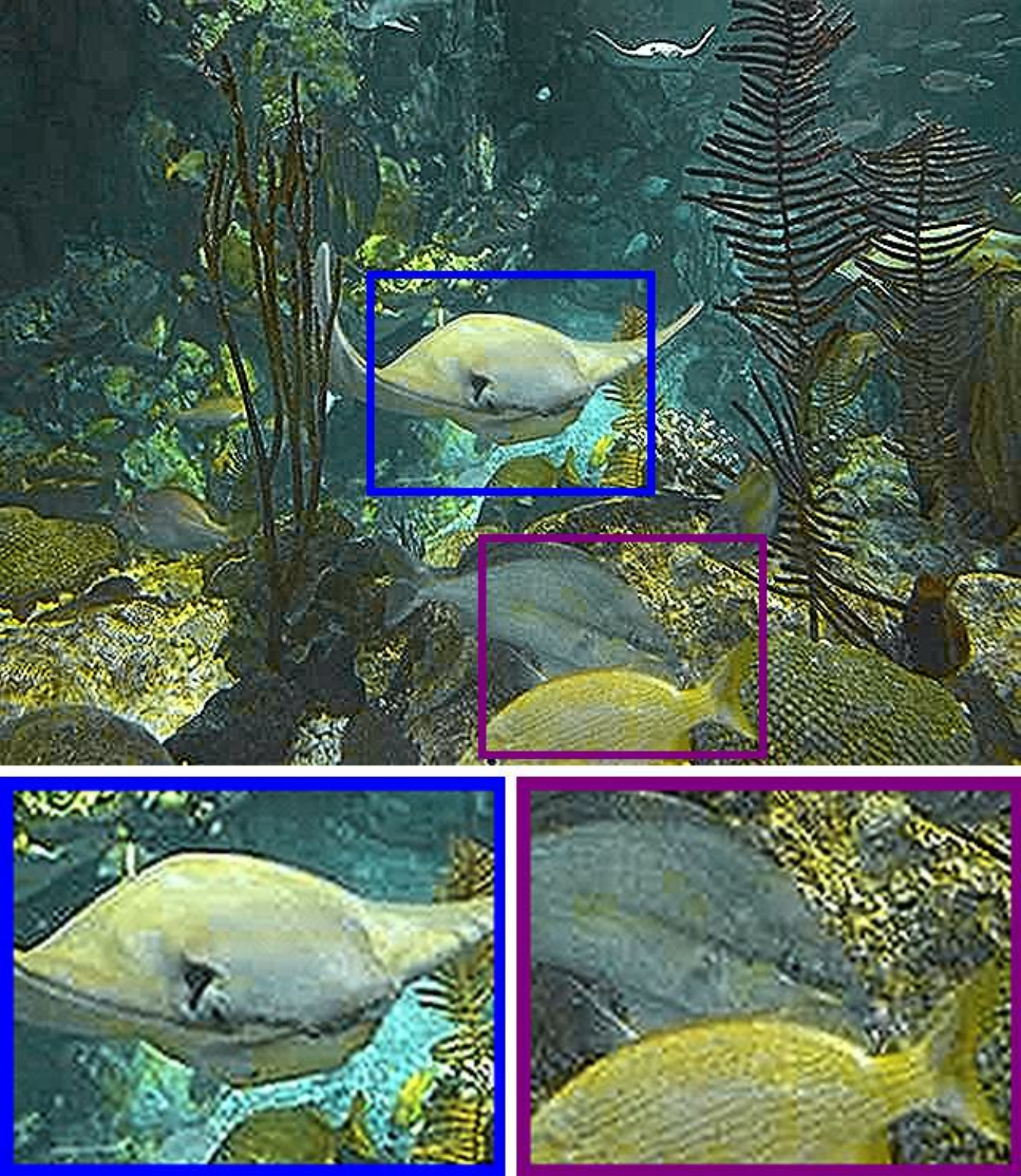} 
        \caption{\footnotesize ICSP}
	\end{subfigure}
	\begin{subfigure}{0.105\linewidth}
		\centering
		\includegraphics[width=\linewidth,  height=\nuidheight]{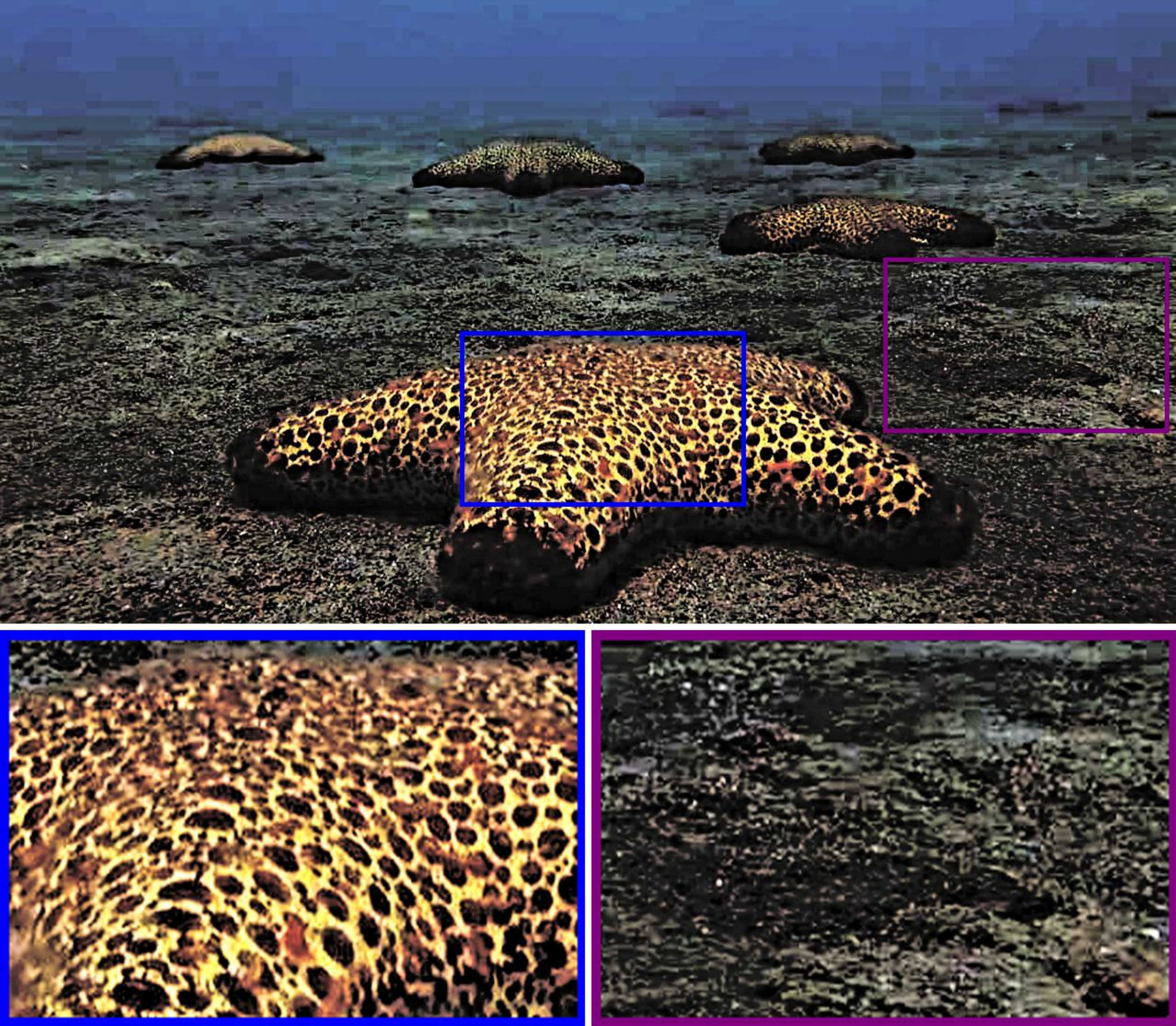} 
        \includegraphics[width=\linewidth,  height=\nuidheight]{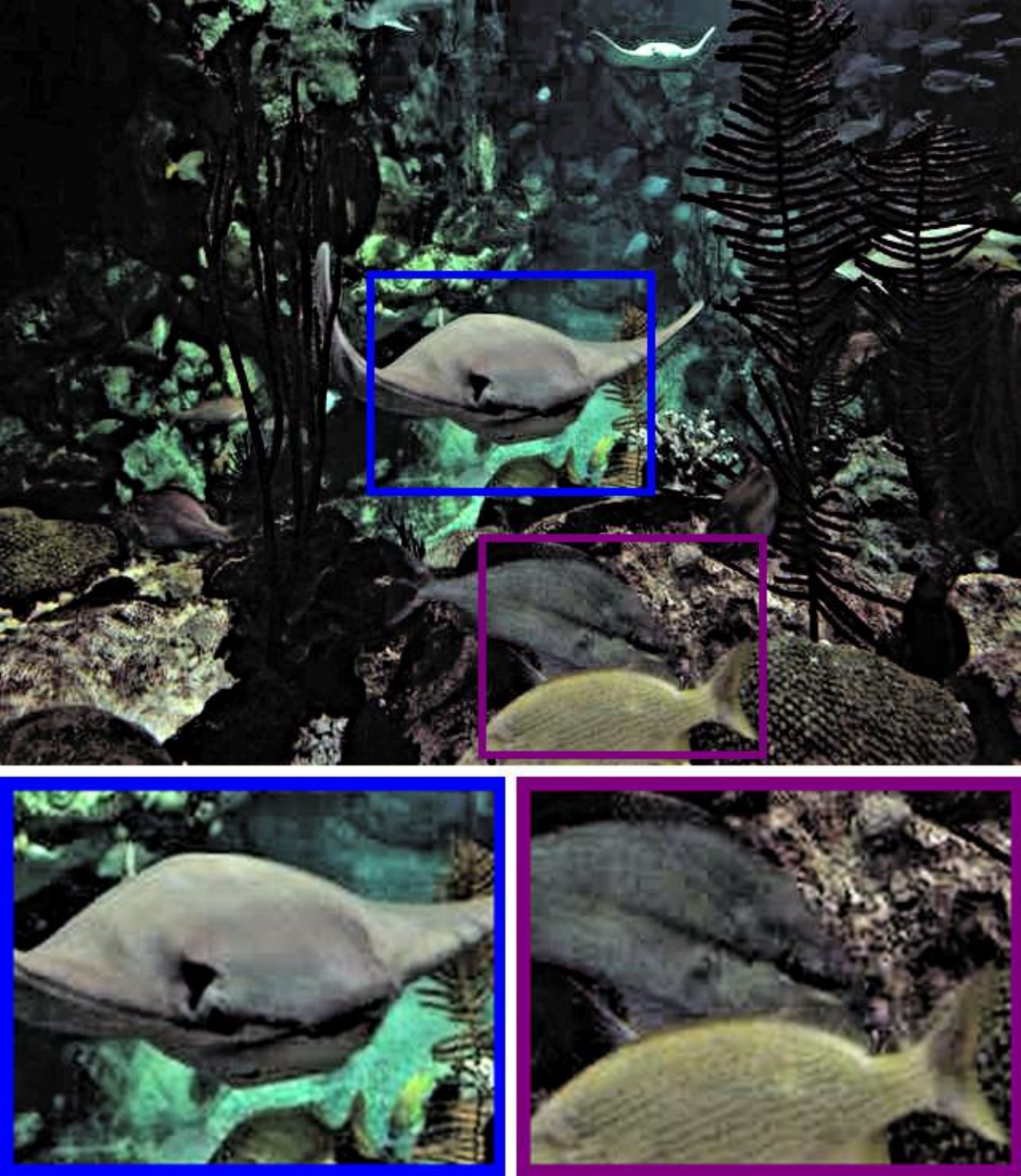} 
        \caption{\footnotesize PCDE}
	\end{subfigure}
	\begin{subfigure}{0.105\linewidth}
		\centering
		\includegraphics[width=\linewidth,  height=\nuidheight]{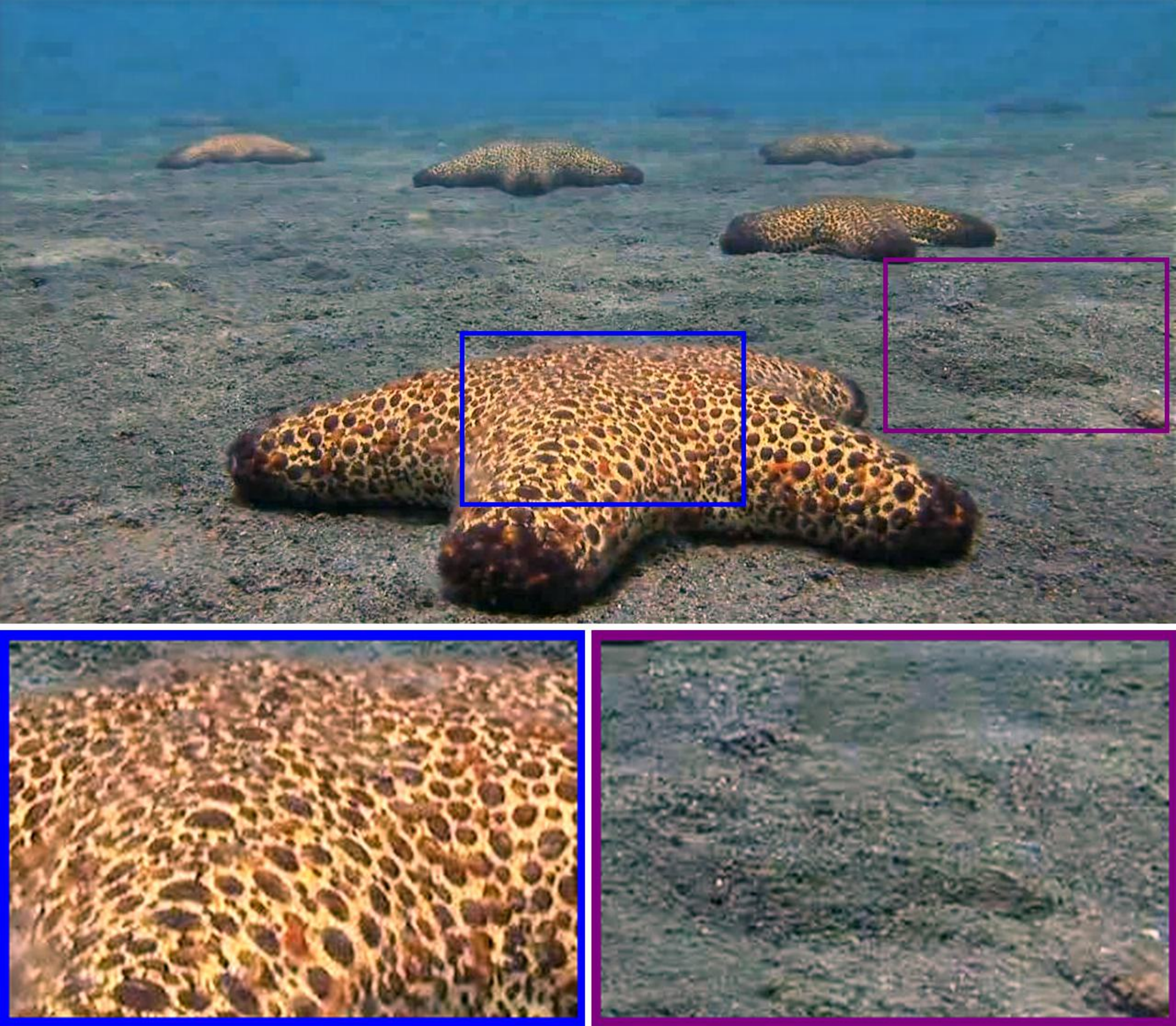}
        \includegraphics[width=\linewidth,  height=\nuidheight]{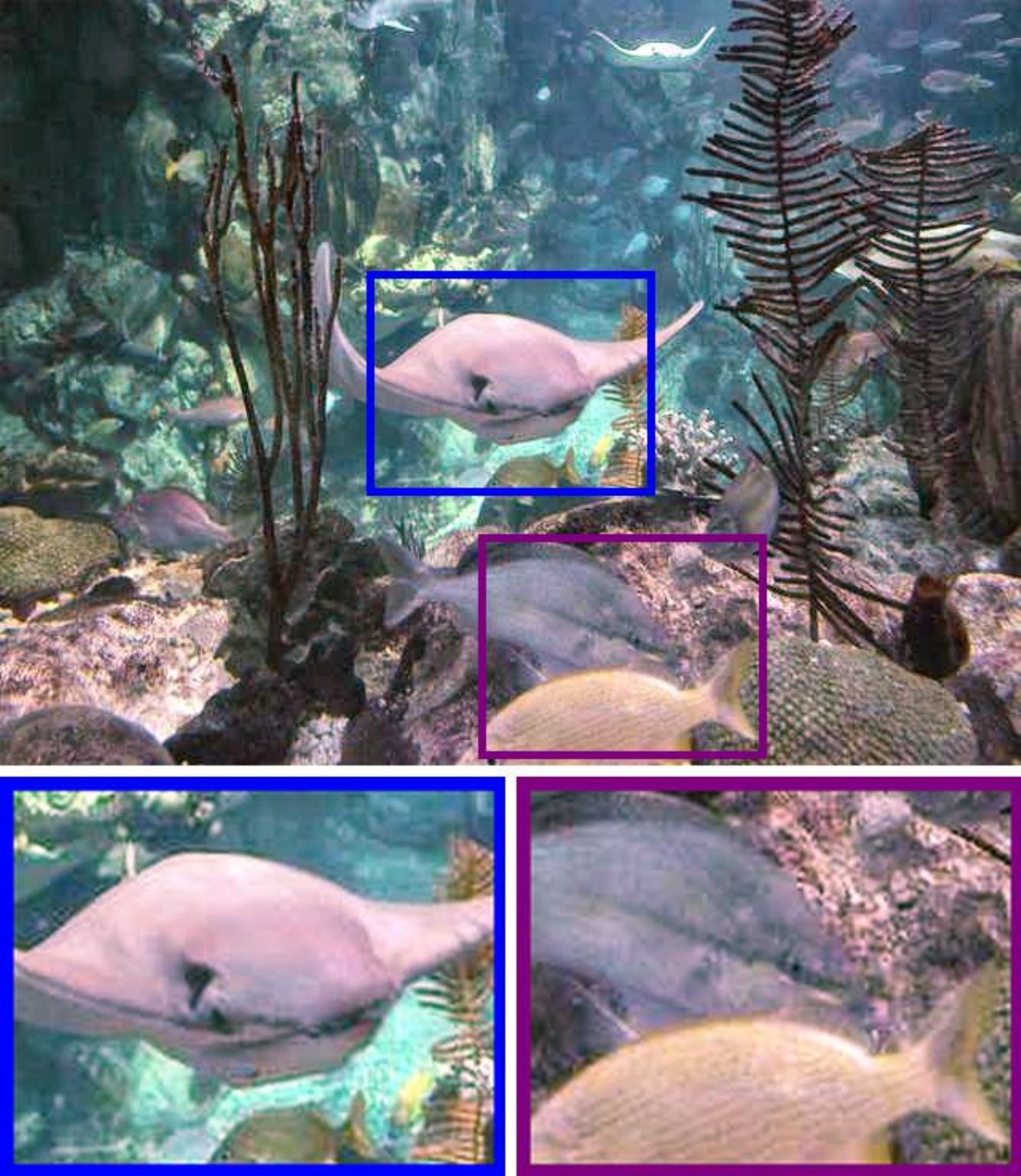}
        \caption{\footnotesize UDAformer}
	\end{subfigure}

    %%%%%%%%%%%%%%%%%%%%%%%%%%%%%

	\begin{subfigure}{0.105\linewidth}
		\centering
		\includegraphics[width=\linewidth,  height=\nuidheight]{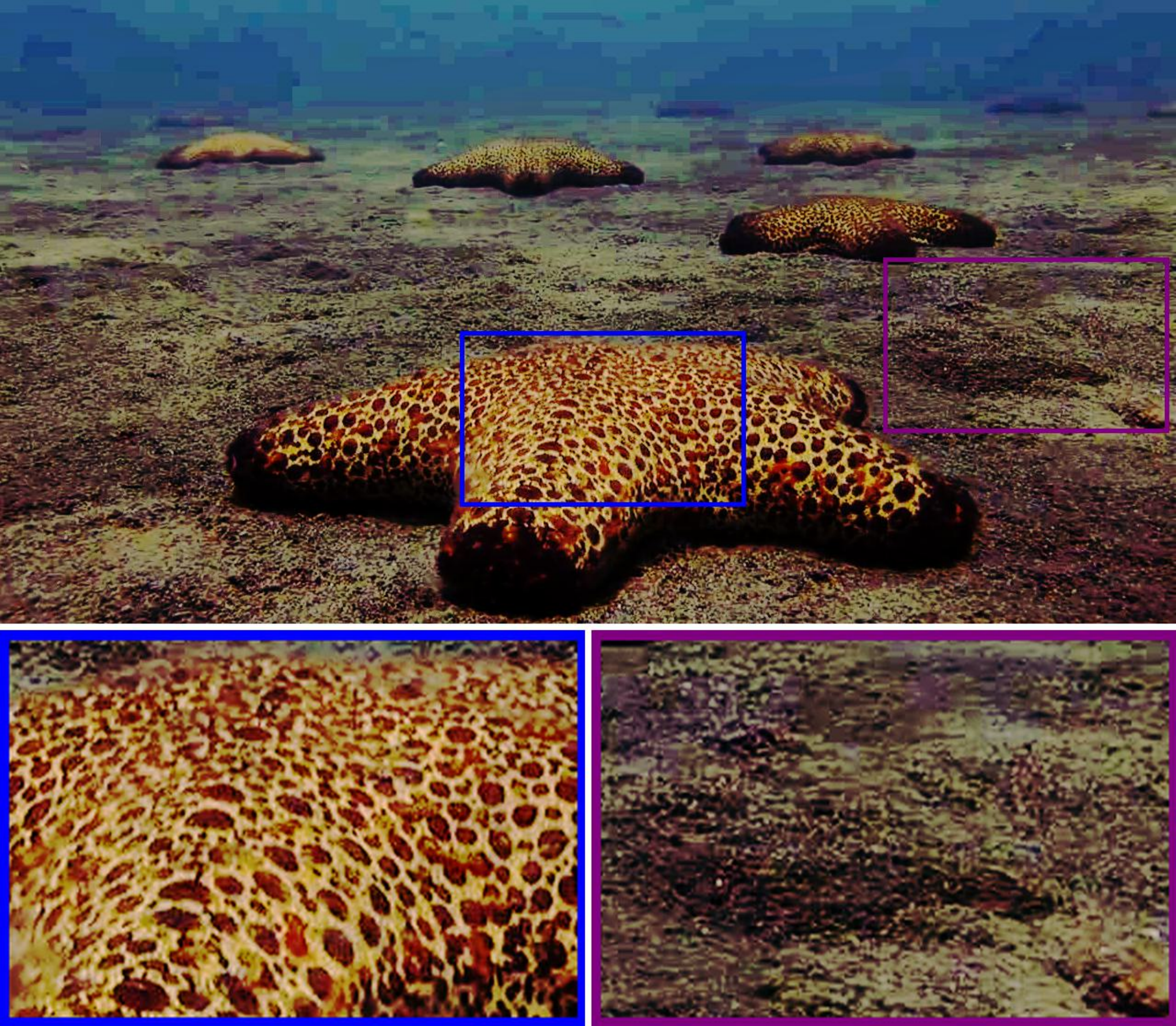} 
        \includegraphics[width=\linewidth,  height=\nuidheight]{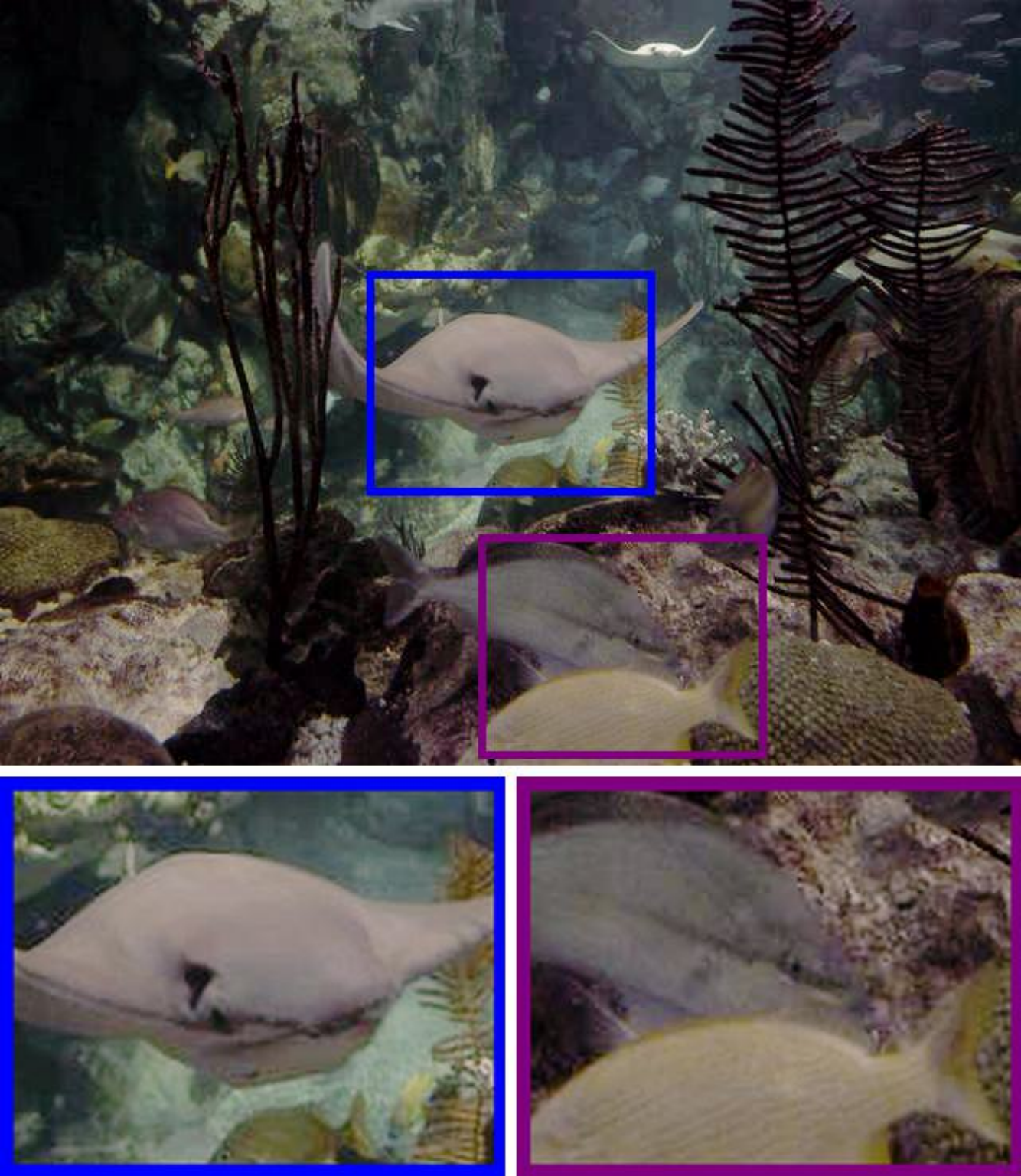} 
		\caption{\footnotesize HFM}
	\end{subfigure}
	\begin{subfigure}{0.105\linewidth}
		\centering
		\includegraphics[width=\linewidth,  height=\nuidheight]{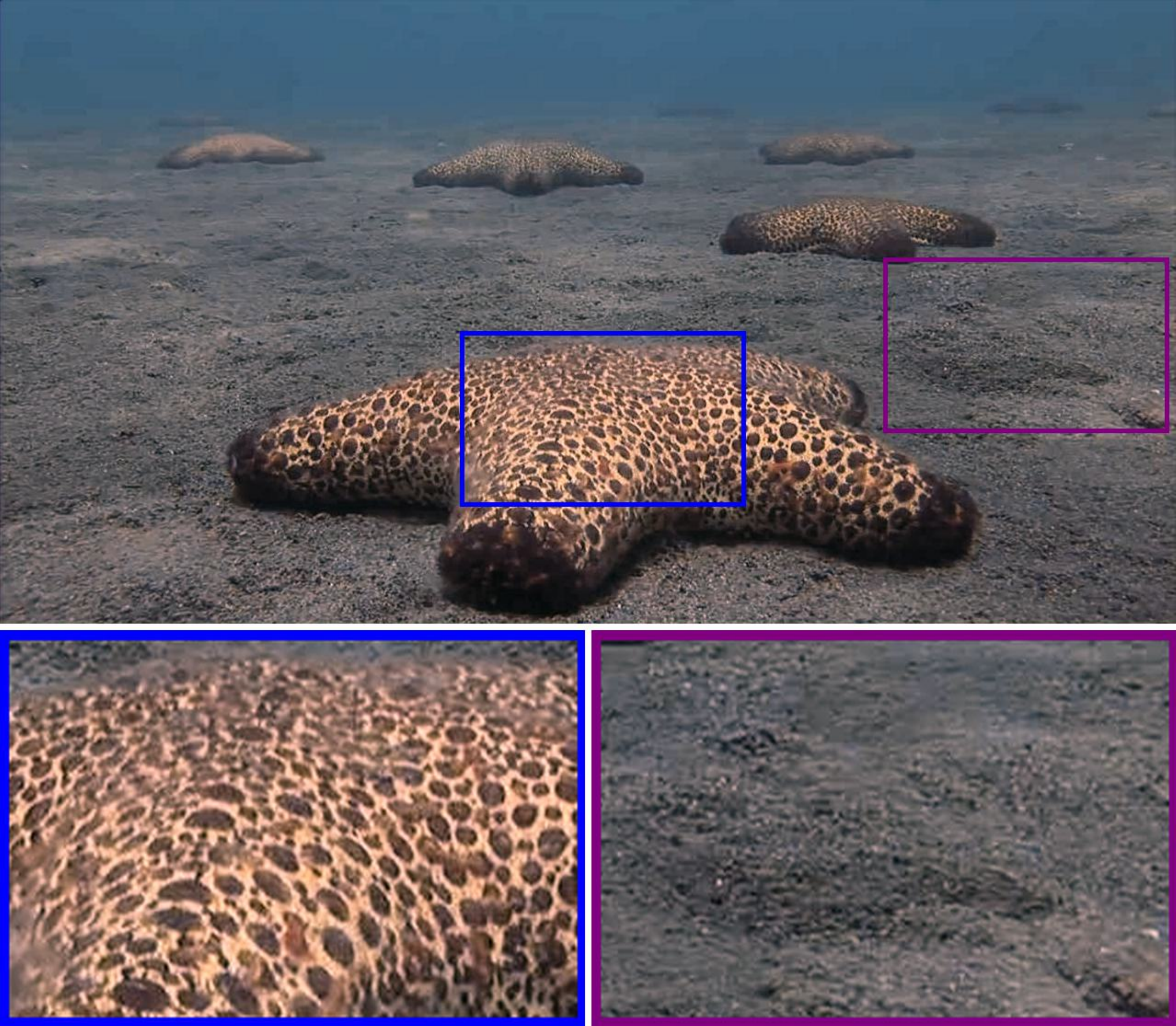} 
        \includegraphics[width=\linewidth,  height=\nuidheight]{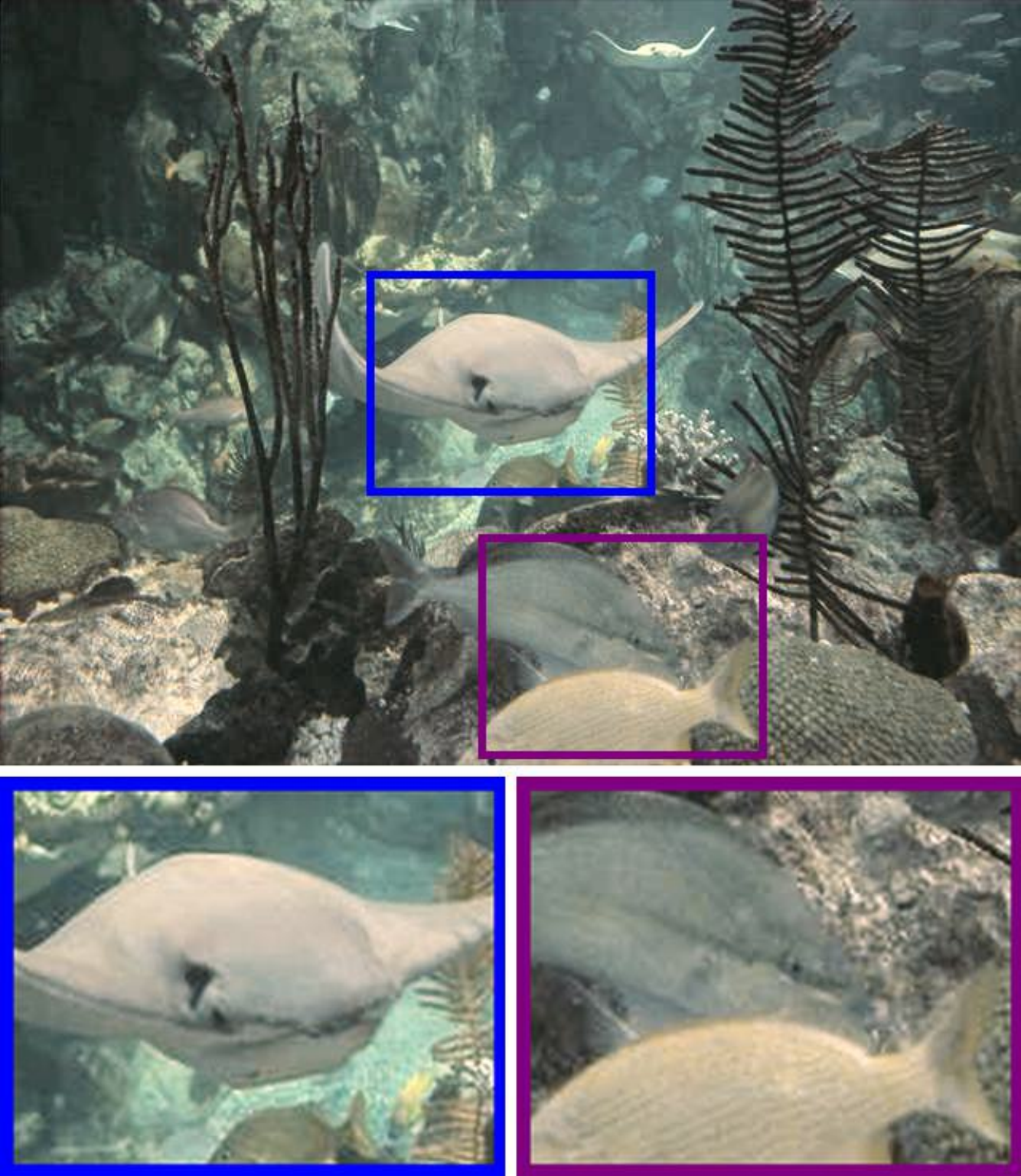} 
		\caption{\footnotesize LENet}
	\end{subfigure}
	\begin{subfigure}{0.105\linewidth}
		\centering
		\includegraphics[width=\linewidth,  height=\nuidheight]{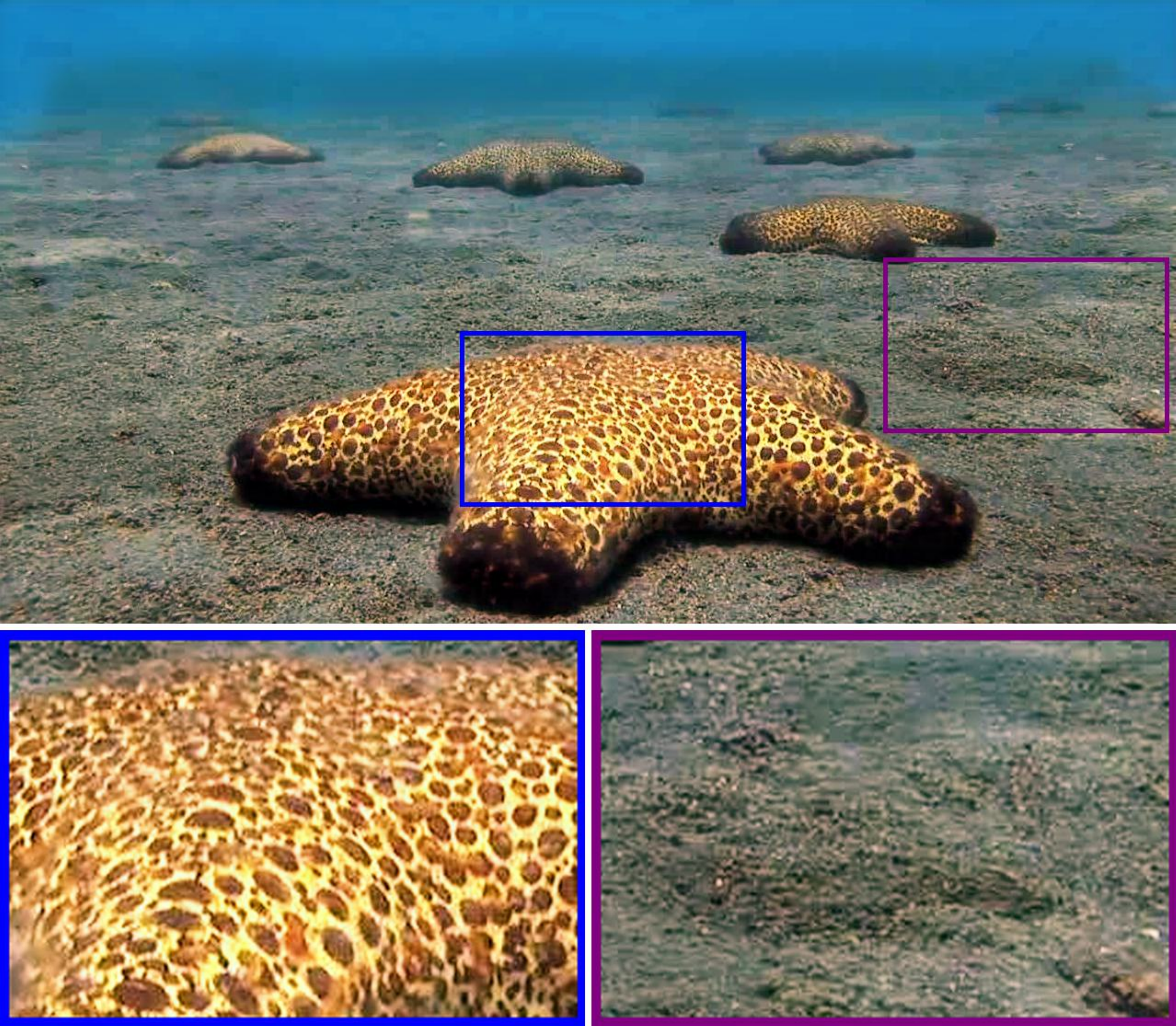}
        \includegraphics[width=\linewidth,  height=\nuidheight]{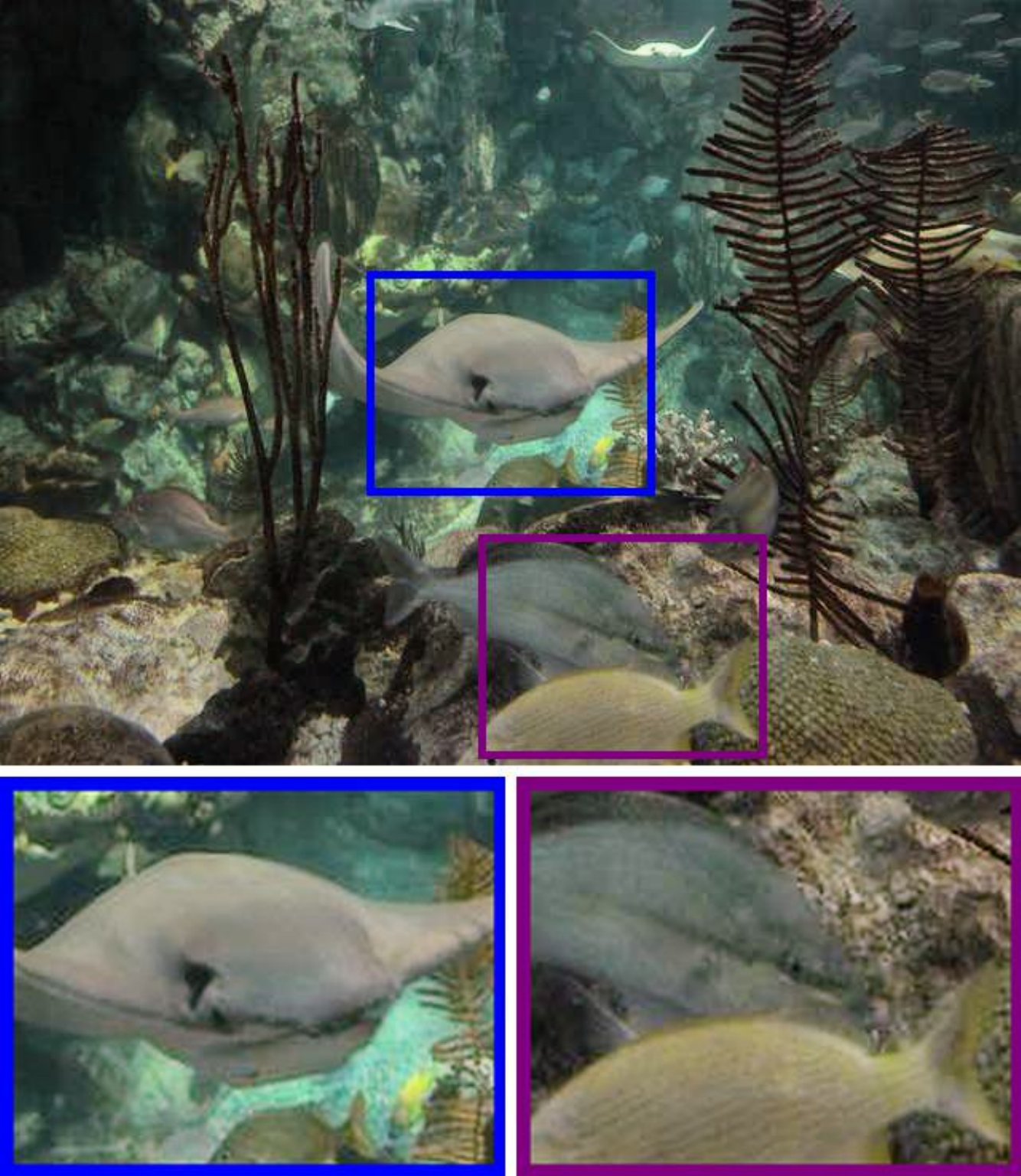}
		\caption{\footnotesize SMDR-IS}
	\end{subfigure}
	\begin{subfigure}{0.105\linewidth}
		\centering
		\includegraphics[width=\linewidth,  height=\nuidheight]{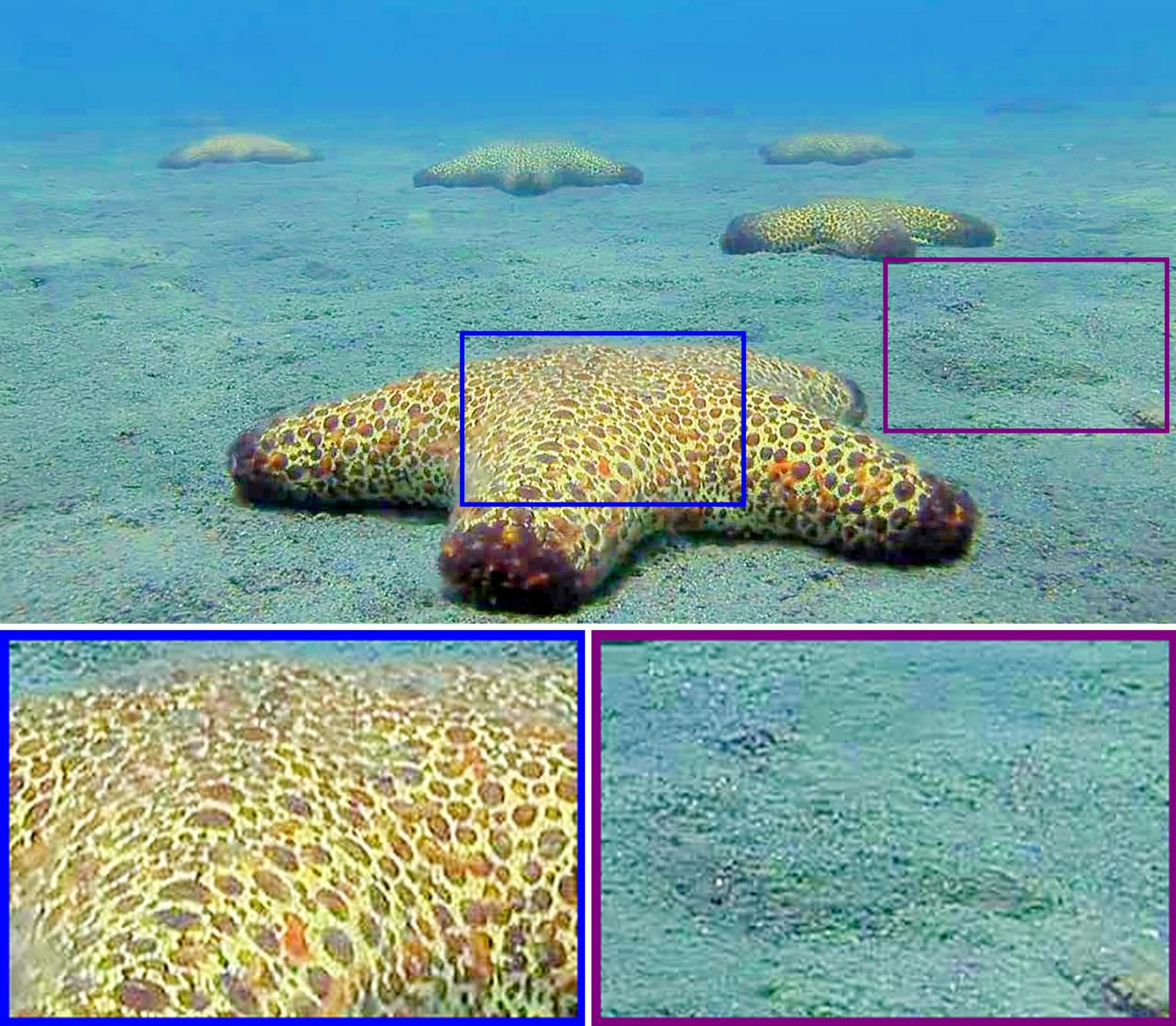} 
        \includegraphics[width=\linewidth,  height=\nuidheight]{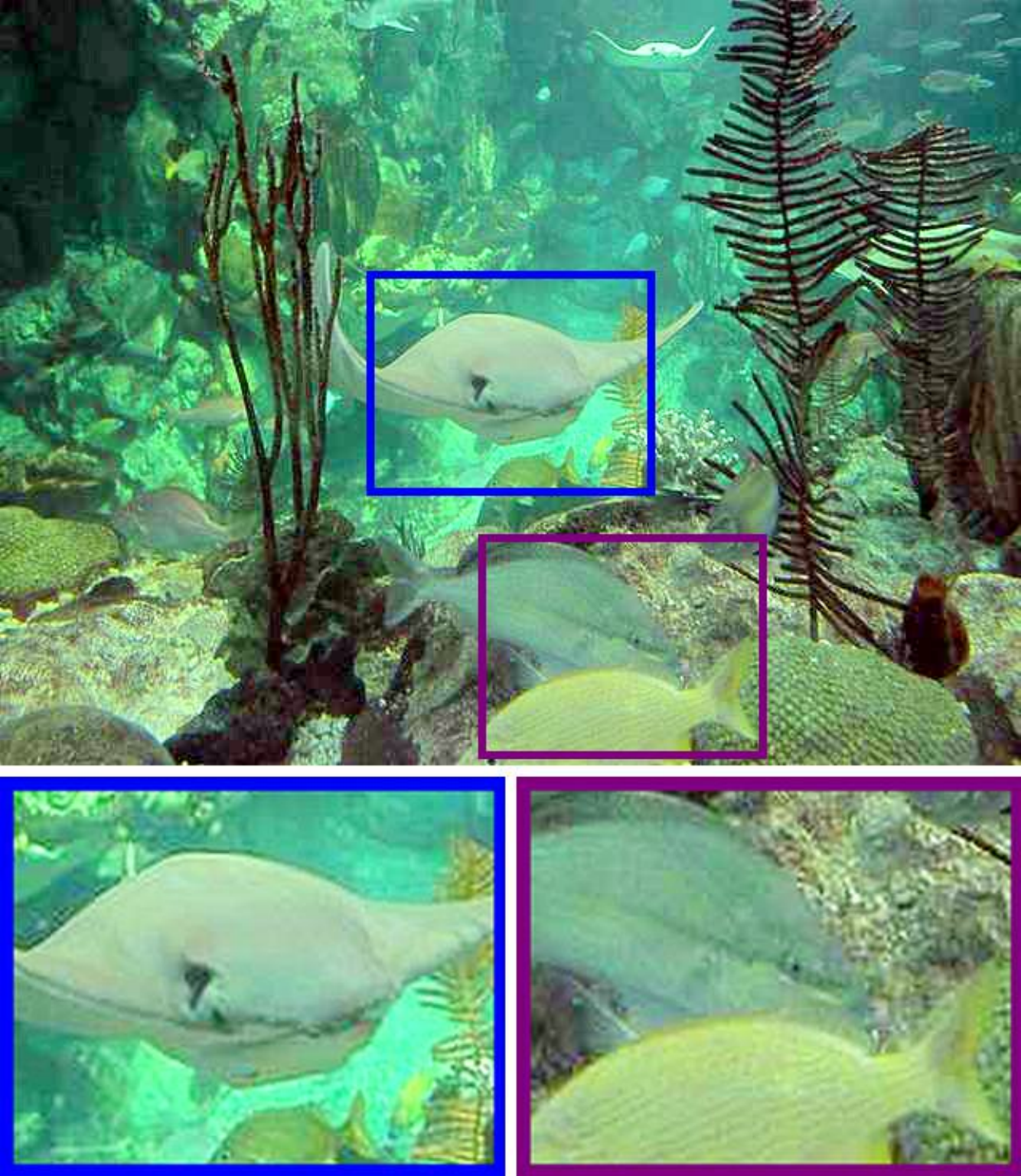} 
		\caption{\footnotesize GCP}
	\end{subfigure}
    \begin{subfigure}{0.105\linewidth}
		\centering
		\includegraphics[width=\linewidth,  height=\nuidheight]{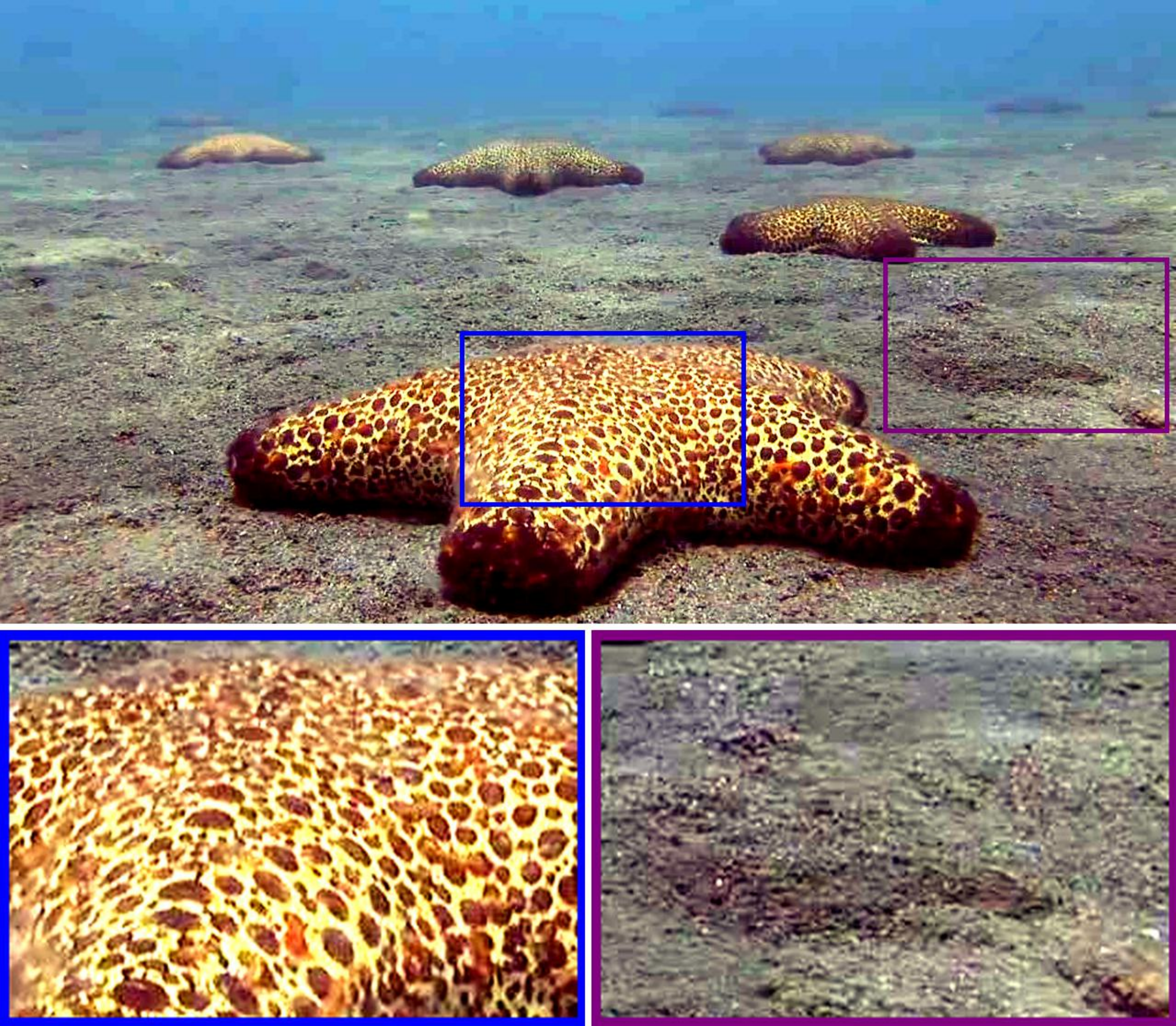} 
        \includegraphics[width=\linewidth,  height=\nuidheight]{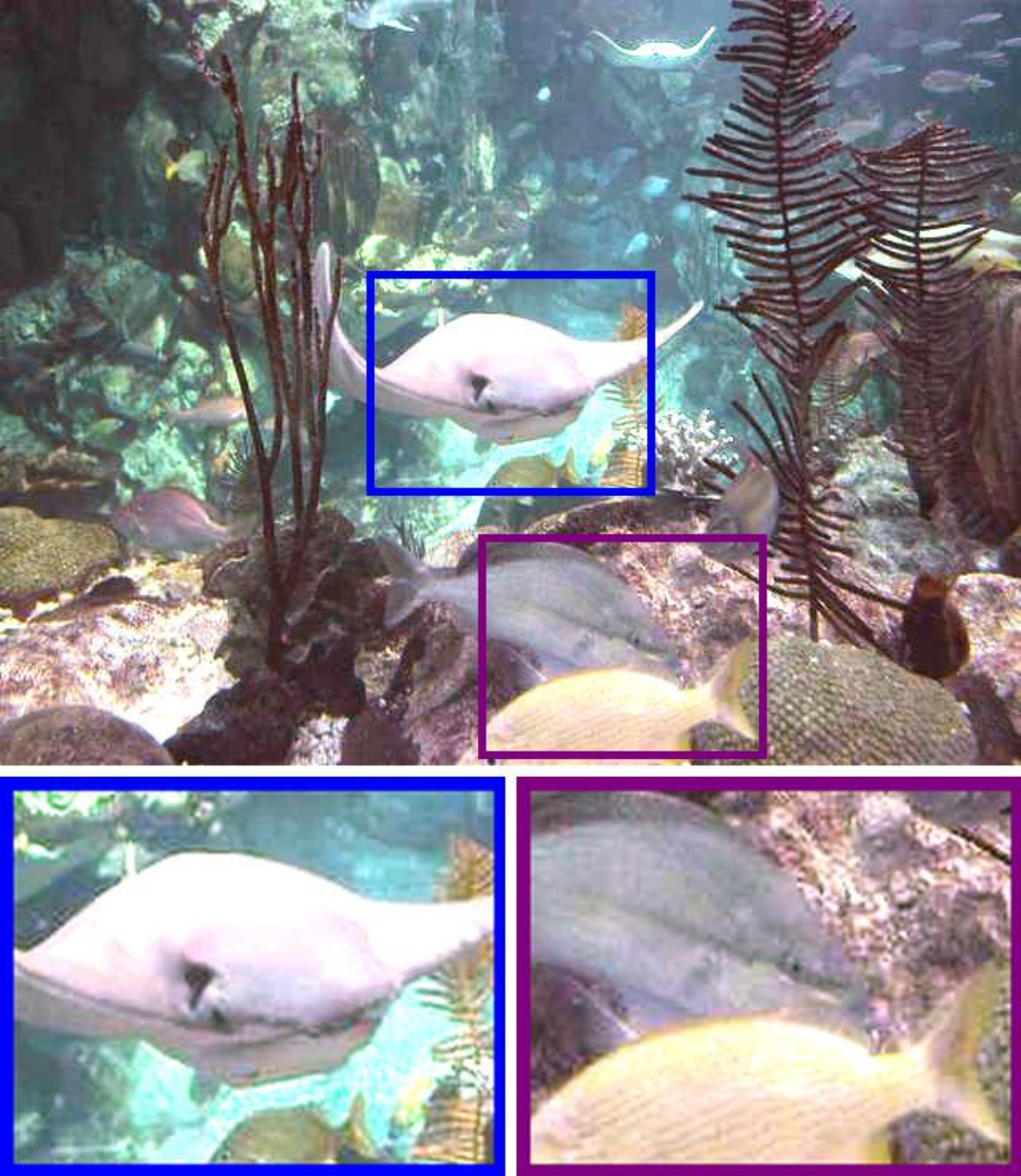} 
		\caption{\footnotesize MACT}
	\end{subfigure}
        \begin{subfigure}{0.105\linewidth}
		\centering
		\includegraphics[width=\linewidth,  height=\nuidheight]{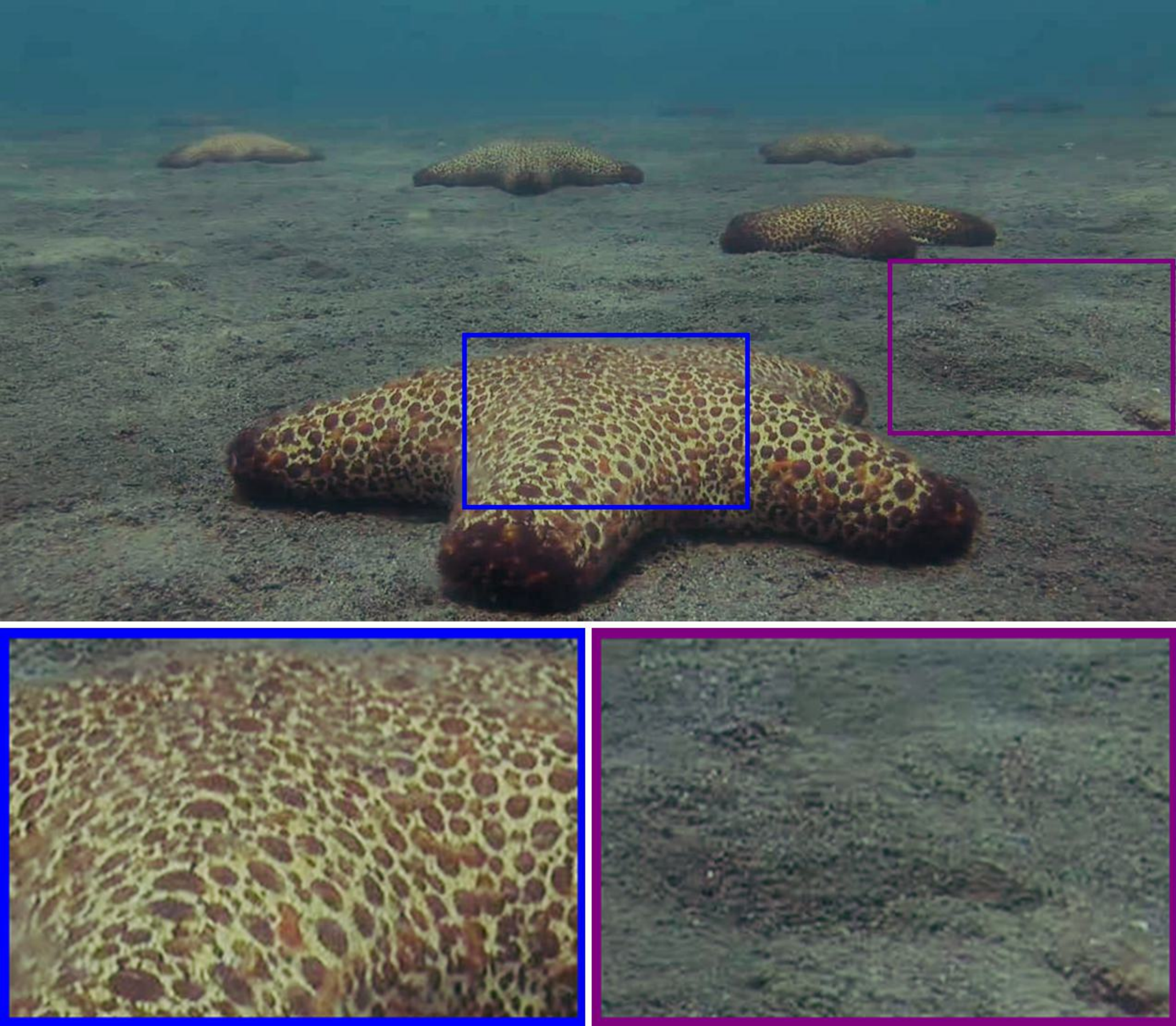} 
        \includegraphics[width=\linewidth,  height=\nuidheight]{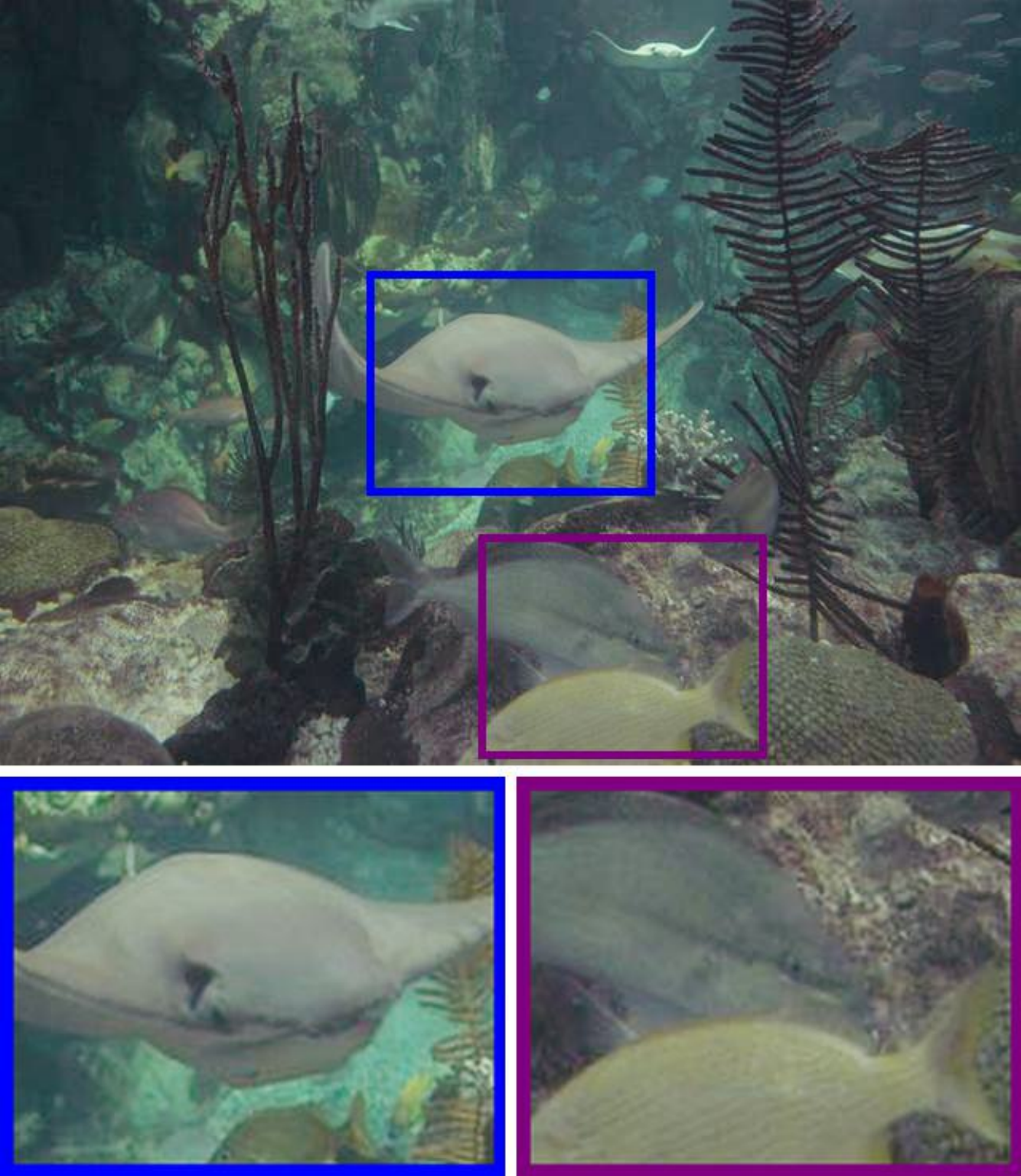} 
		\caption{\footnotesize UDnet}
	\end{subfigure}
        \begin{subfigure}{0.105\linewidth}
		\centering
		\includegraphics[width=\linewidth,  height=\nuidheight]{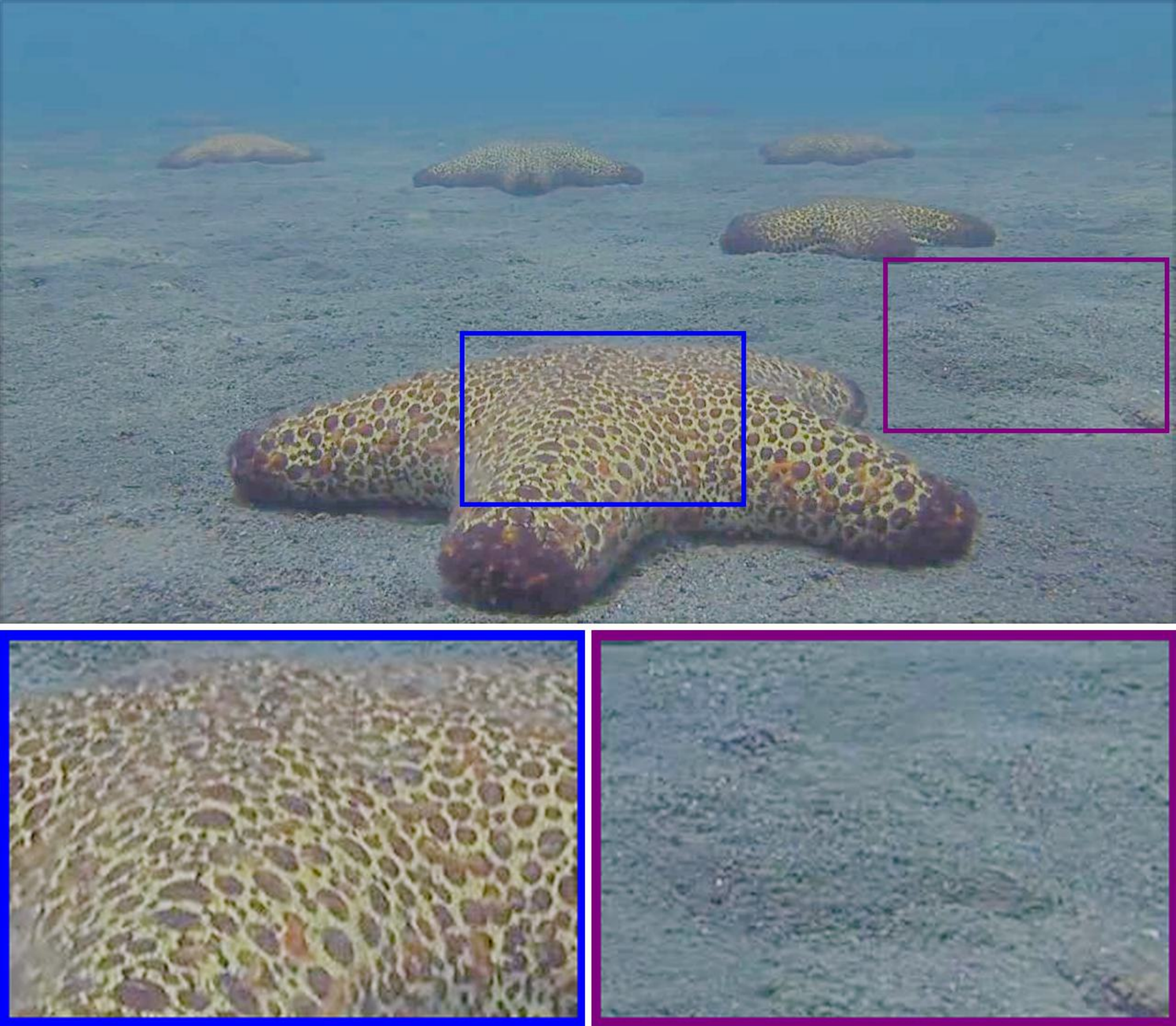} 
        \includegraphics[width=\linewidth,  height=\nuidheight]{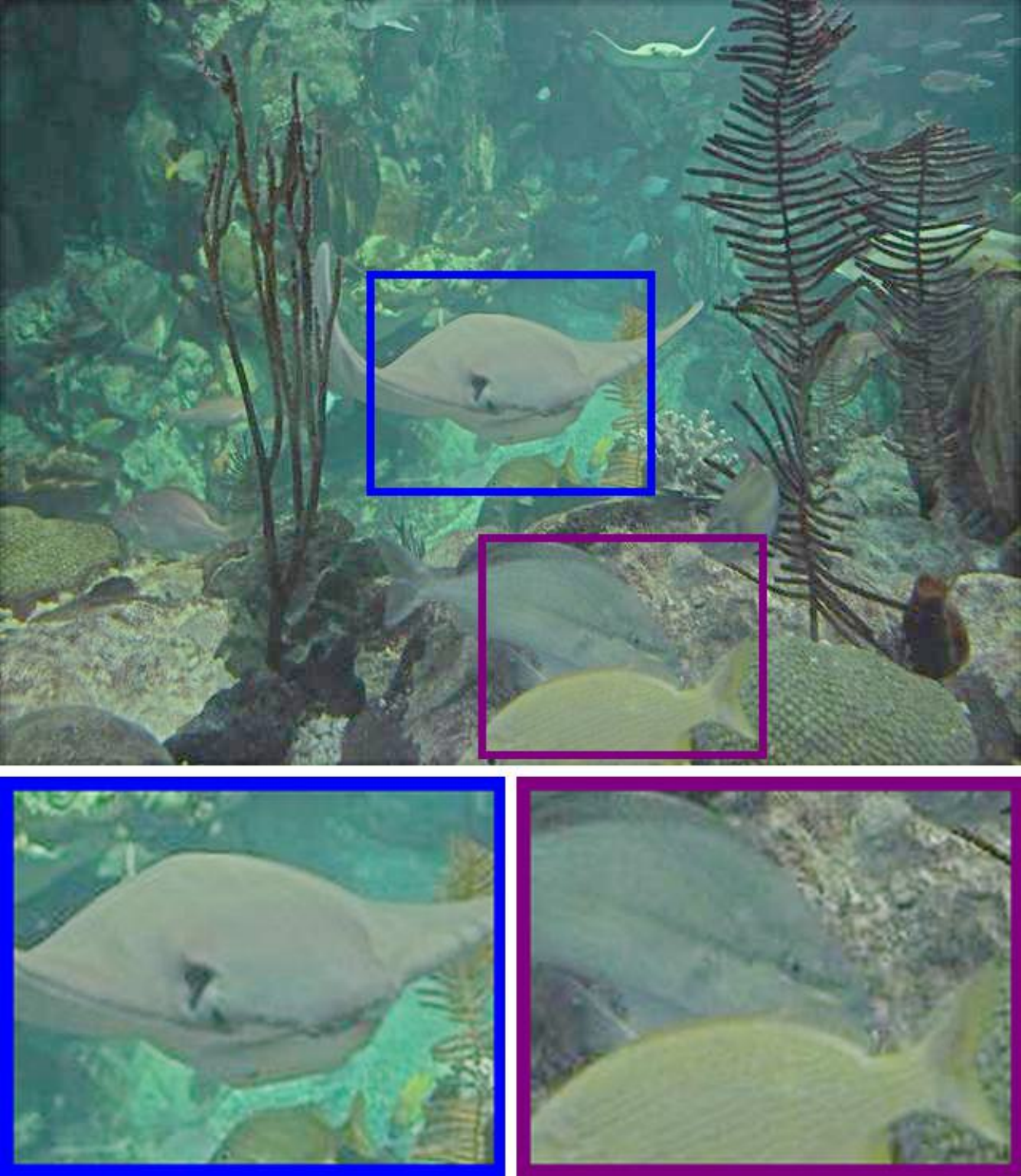} 
		\caption{\footnotesize EIB-FNDL}
	\end{subfigure}
    \begin{subfigure}{0.105\linewidth}
		\centering
		\includegraphics[width=\linewidth,  height=\nuidheight]{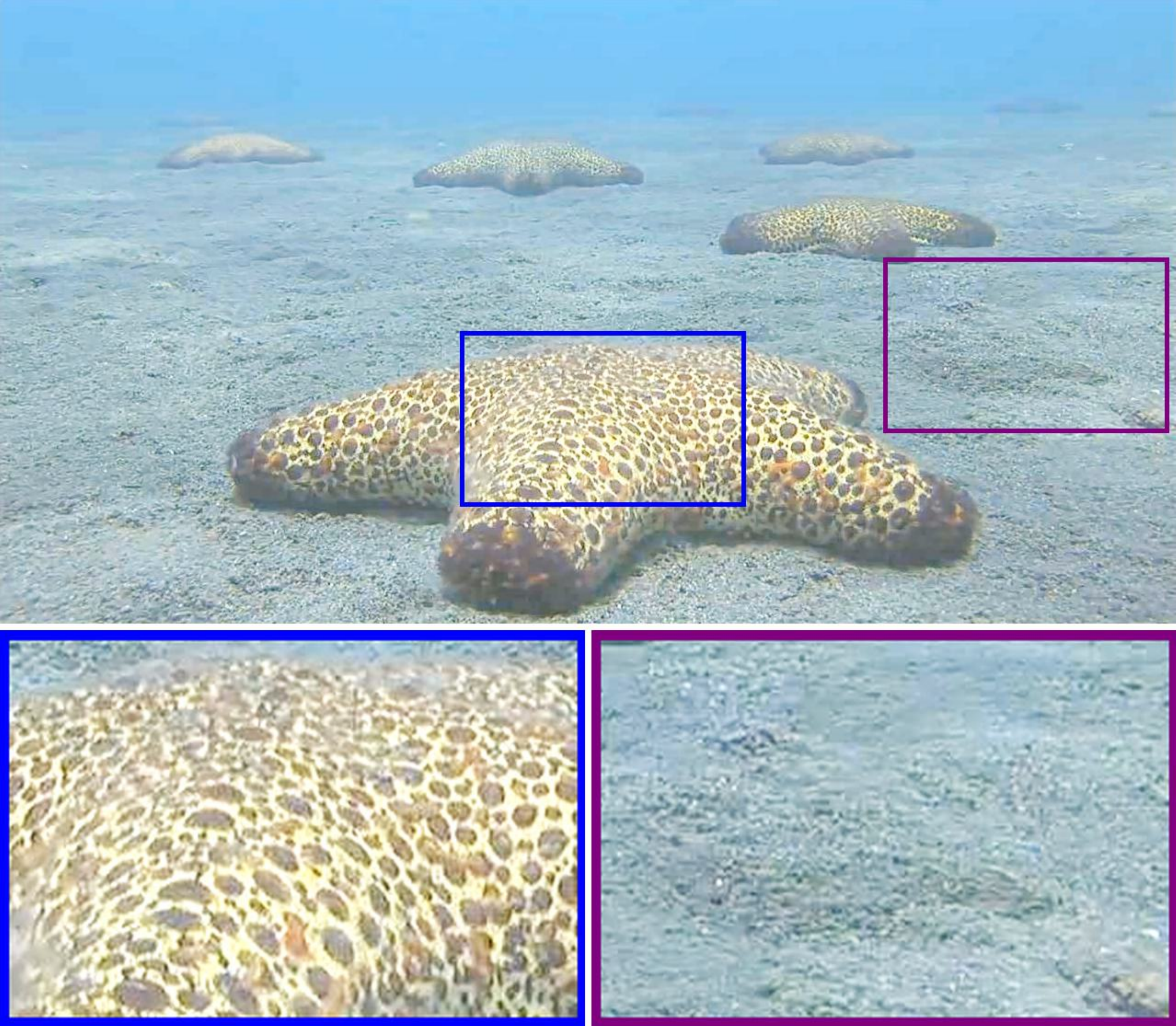} 
        \includegraphics[width=\linewidth,  height=\nuidheight]{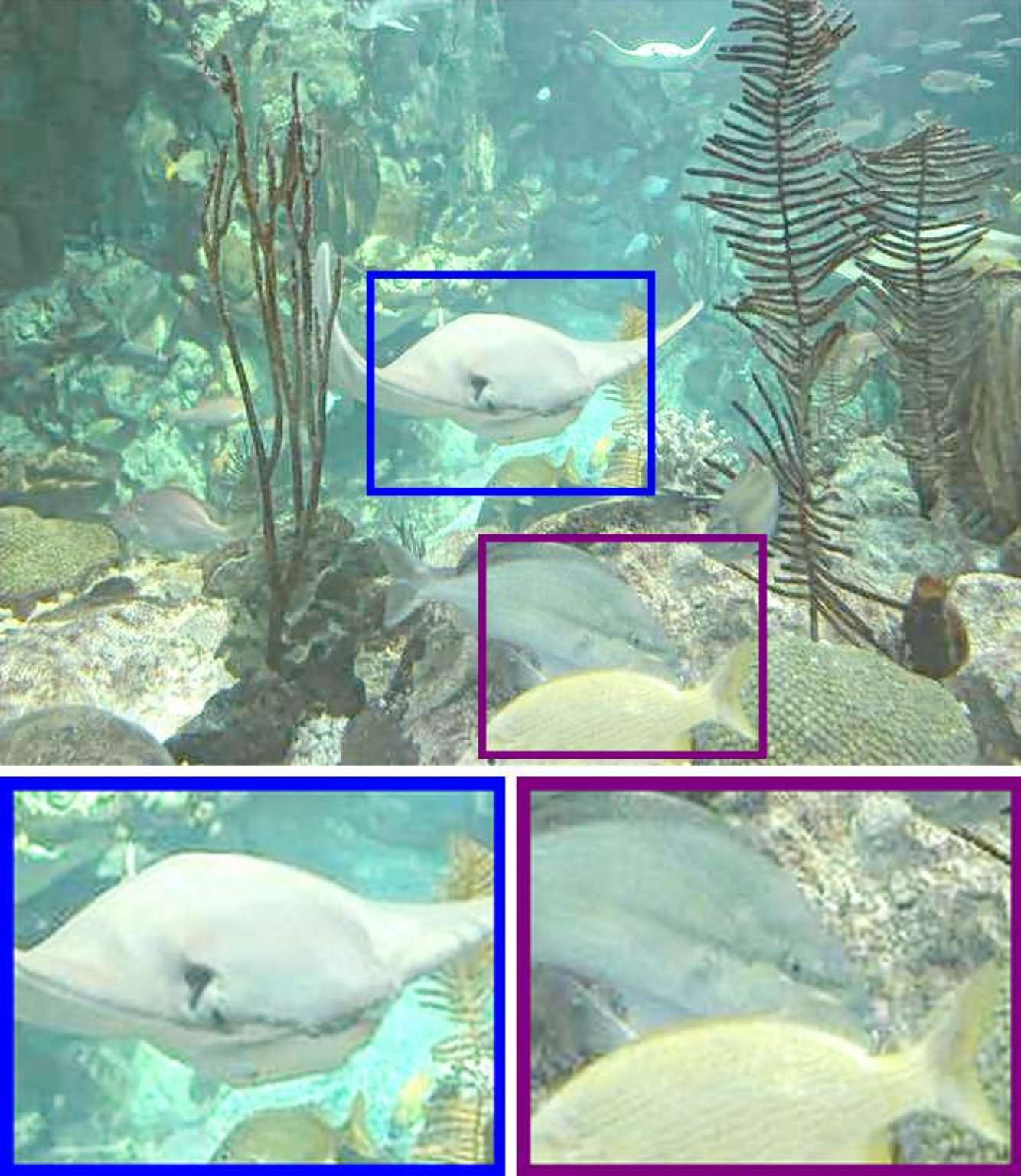} 
		\caption{\footnotesize ALEN}
	\end{subfigure}
    \begin{subfigure}{0.105\linewidth}
		\centering
		\includegraphics[width=\linewidth,  height=\nuidheight]{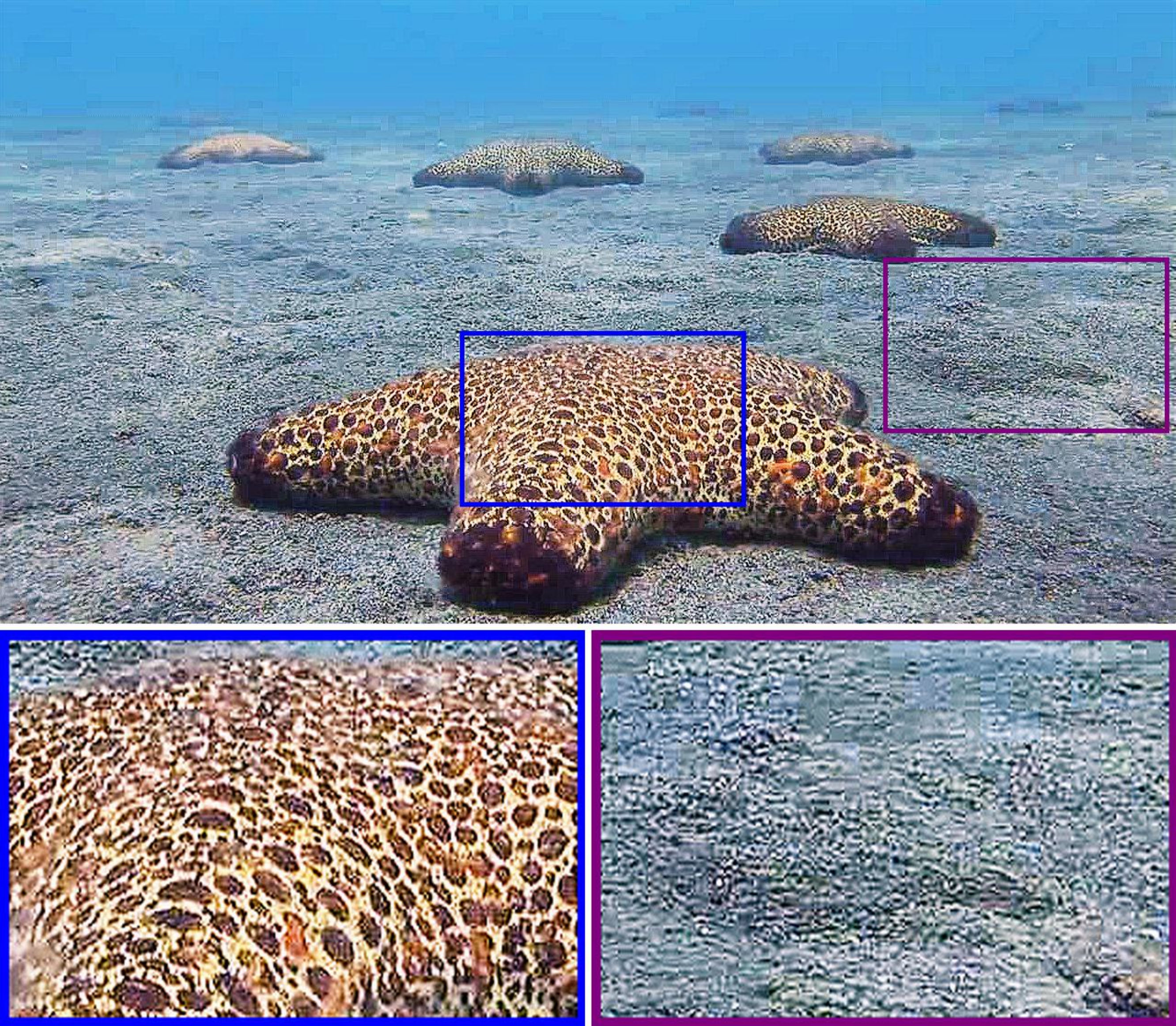} 
        \includegraphics[width=\linewidth,  height=\nuidheight]{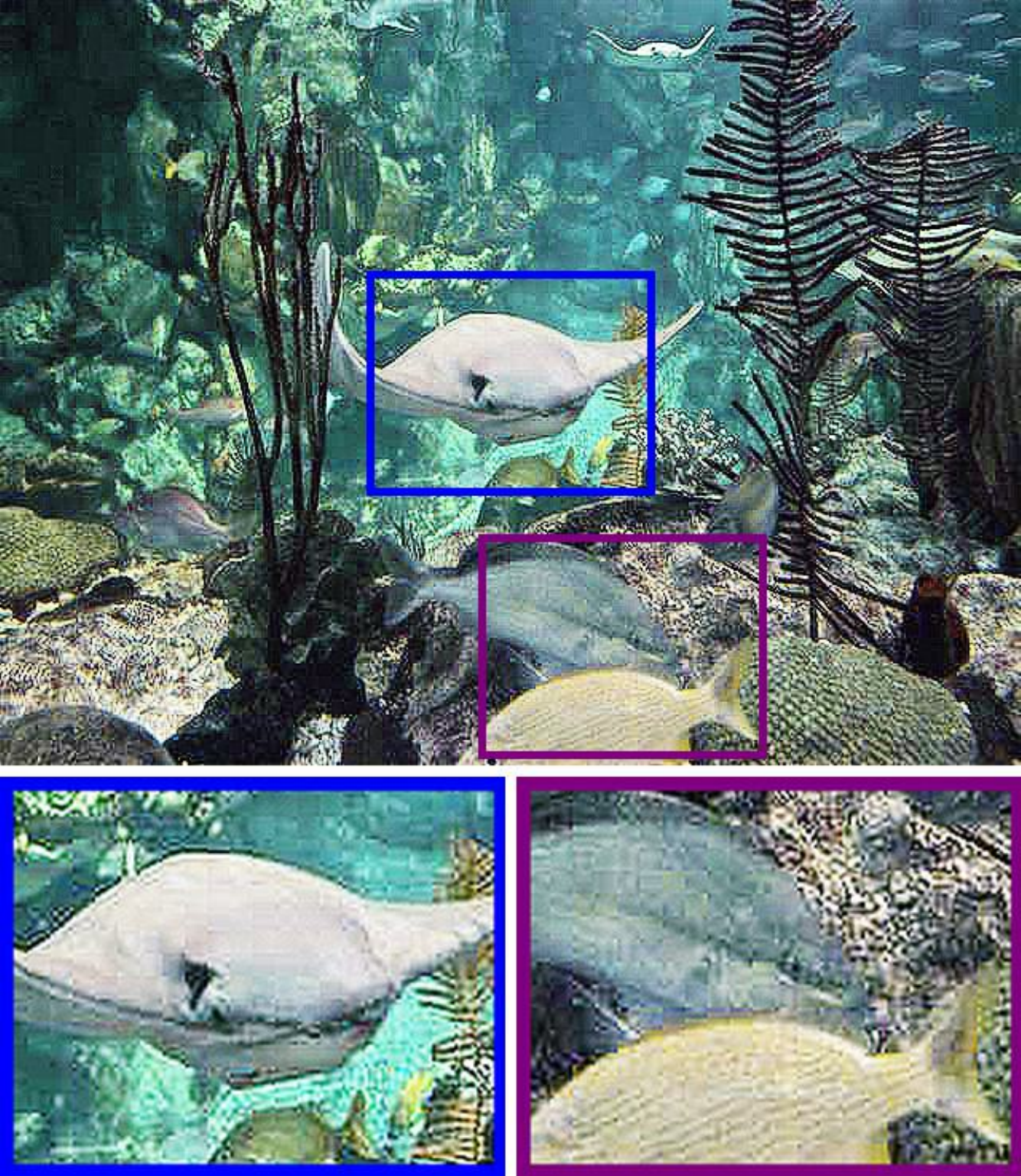} 
		\caption{\footnotesize UNIR-Net}
	\end{subfigure}
 
	\caption{Visual comparison of enhancement results on real-world underwater images from the NUID dataset, captured under low illumination during daytime conditions. Subfigure (a) shows the original image affected by poor lighting. Subfigures (b) to (r) present the outputs of various enhancement methods applied to the same input. The figure highlights the visual improvements in brightness, contrast, and overall perceptual quality achieved by each approach.}
	\label{Q2}
\end{figure*}

\begin{figure*}[!ht]
	\newlength{\nuidheighttwo}
	\setlength{\nuidheighttwo}{1.85cm}
	\centering
	\begin{subfigure}{0.105\linewidth}
		\centering
        \includegraphics[width=\linewidth,  height=\nuidheighttwo]{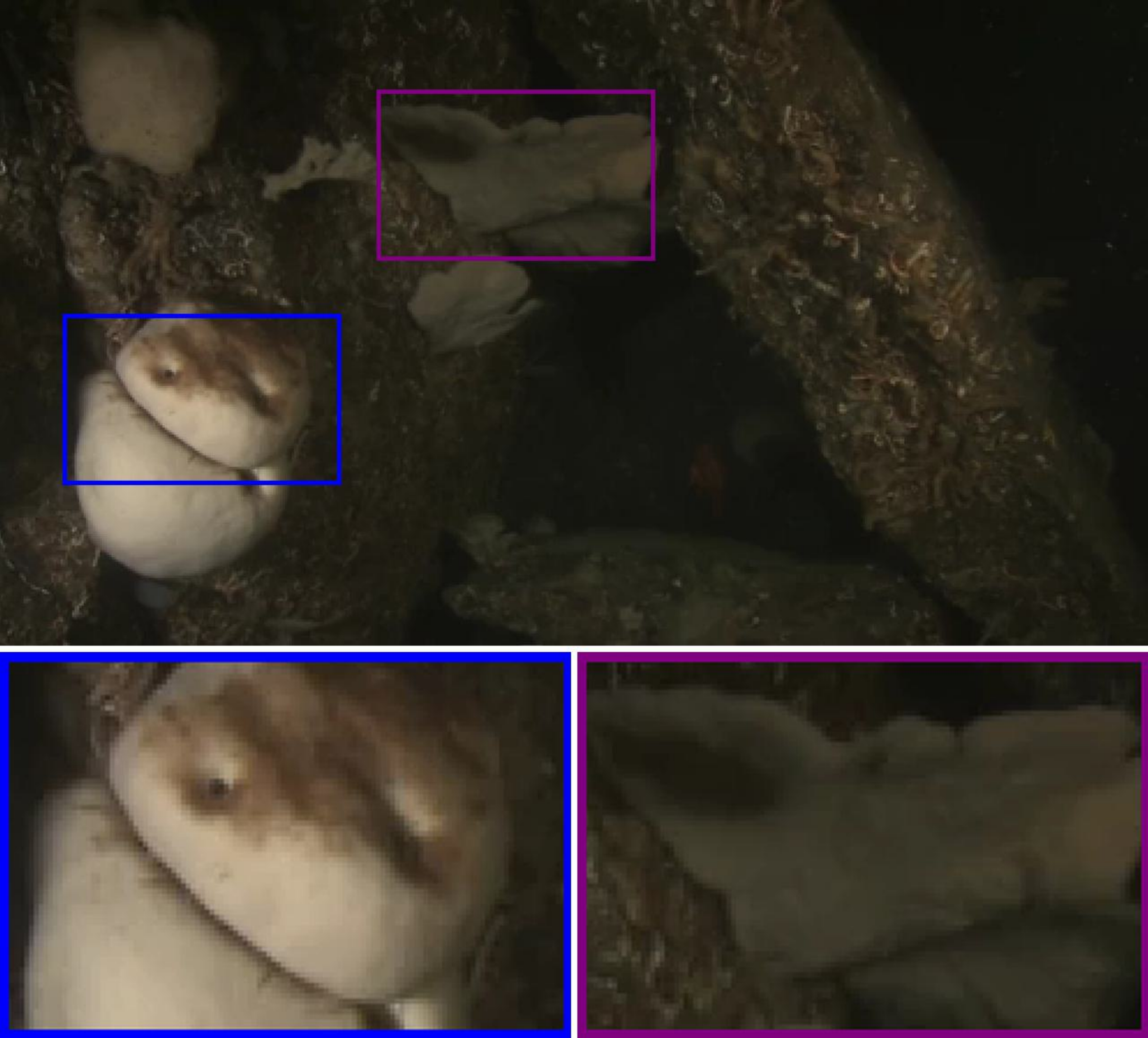} 
		\includegraphics[width=\linewidth,  height=\nuidheighttwo]{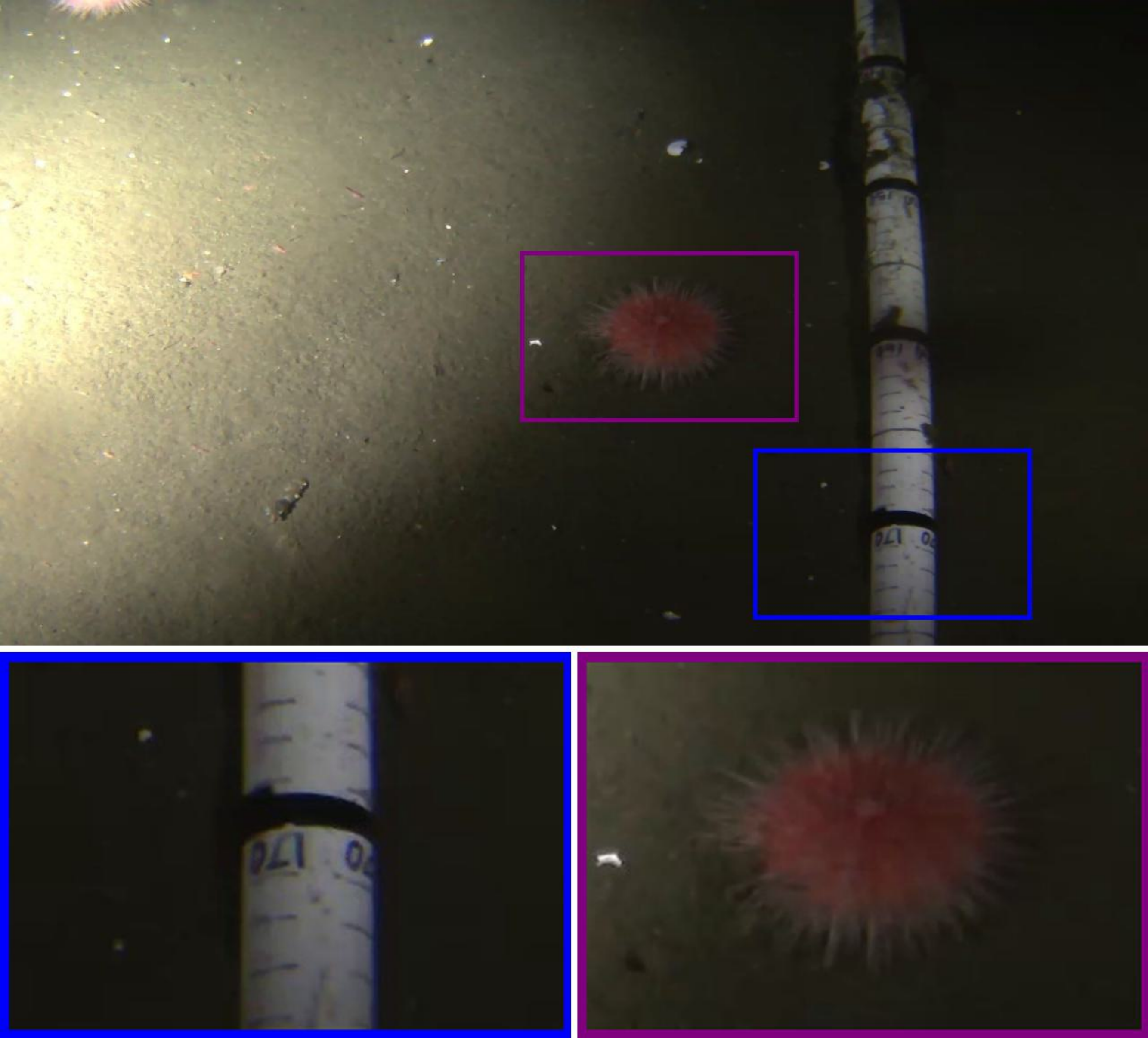} 
        \caption{\footnotesize NUI Image}
	\end{subfigure}
	\begin{subfigure}{0.105\linewidth}
		\centering
		\includegraphics[width=\linewidth,  height=\nuidheighttwo]{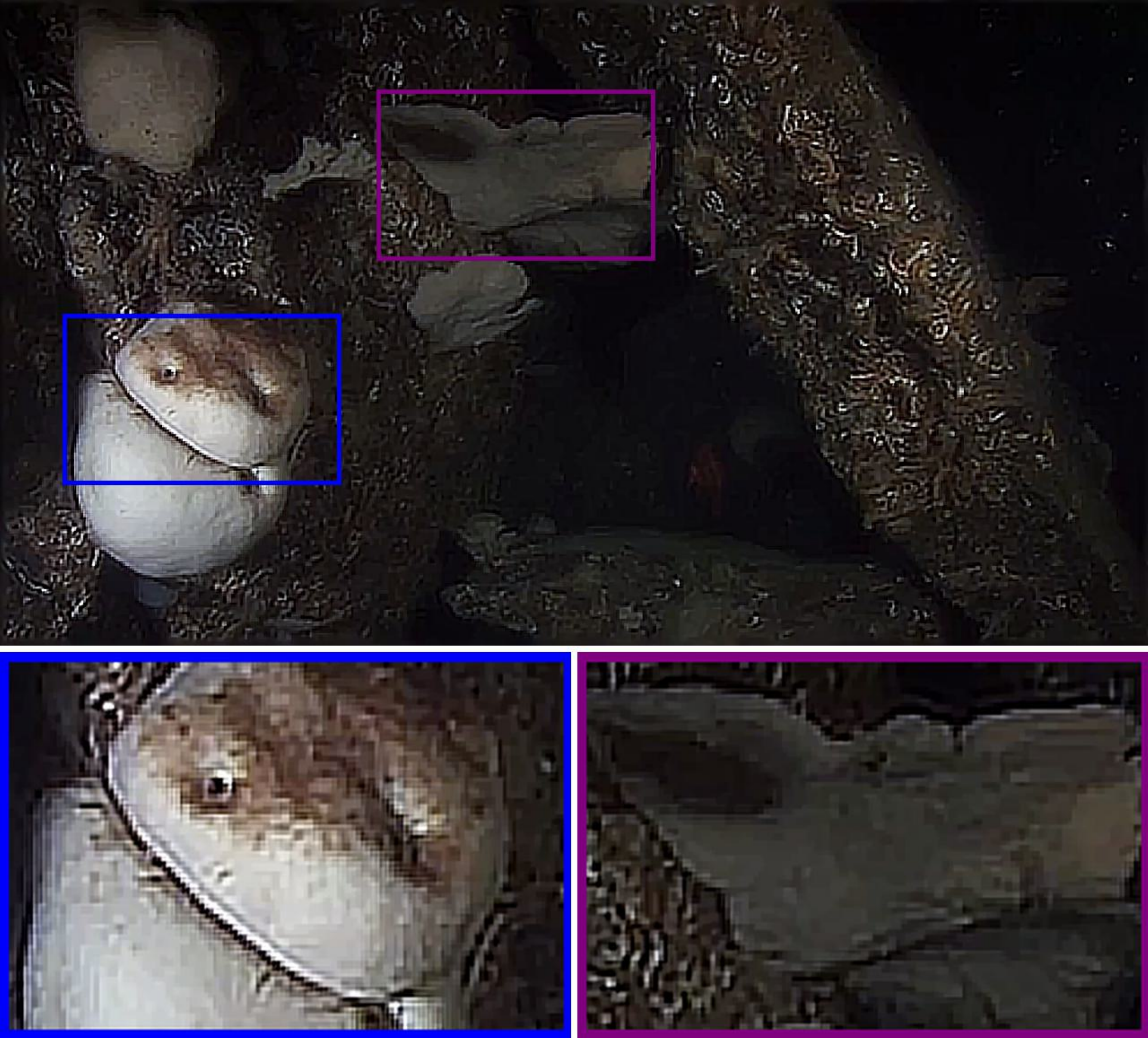} 
        \includegraphics[width=\linewidth,  height=\nuidheighttwo]{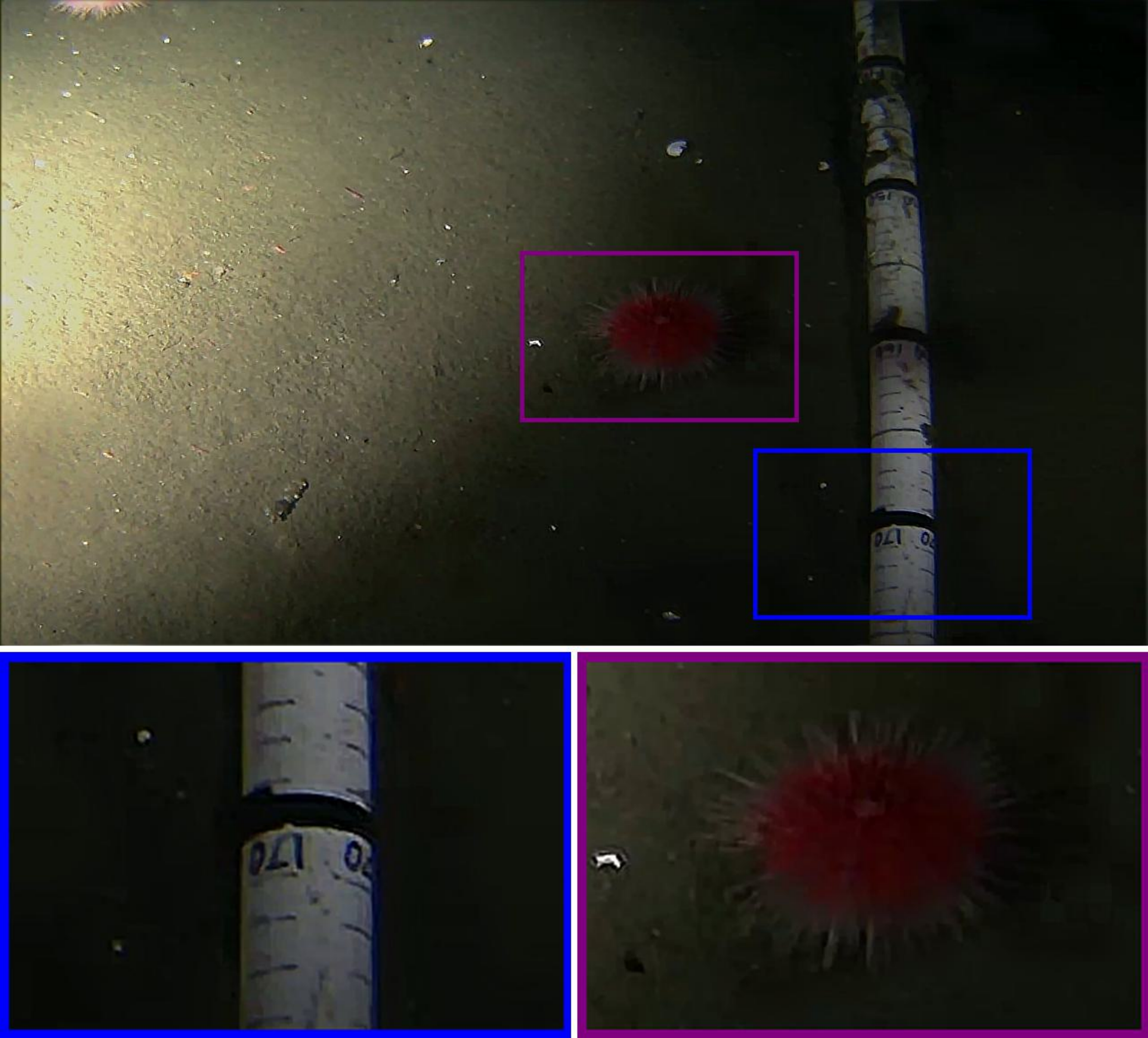}
		\caption{\footnotesize UNTV}
	\end{subfigure}
	\begin{subfigure}{0.105\linewidth}
		\centering
		\includegraphics[width=\linewidth,  height=\nuidheighttwo]{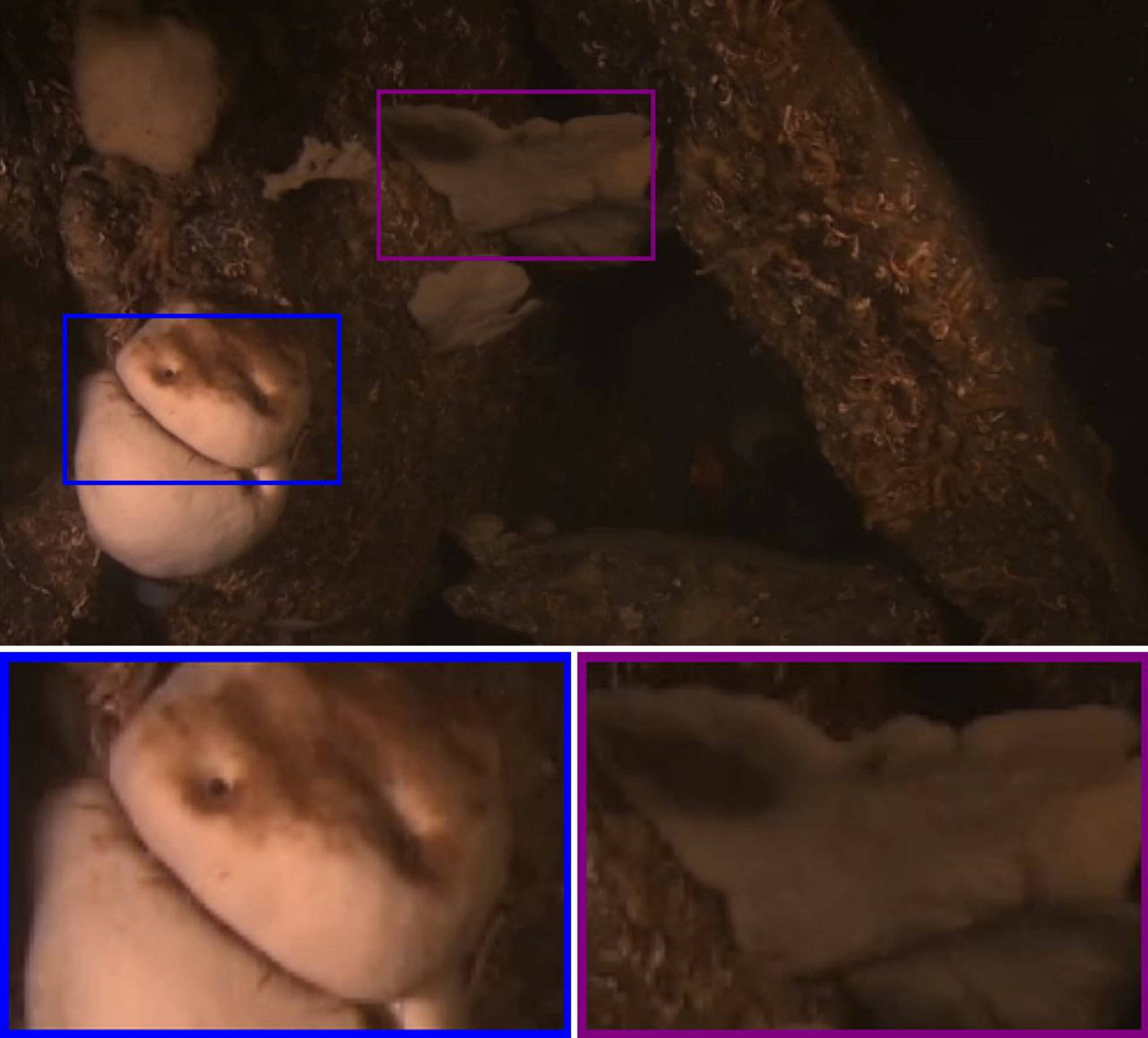}
        \includegraphics[width=\linewidth,  height=\nuidheighttwo]{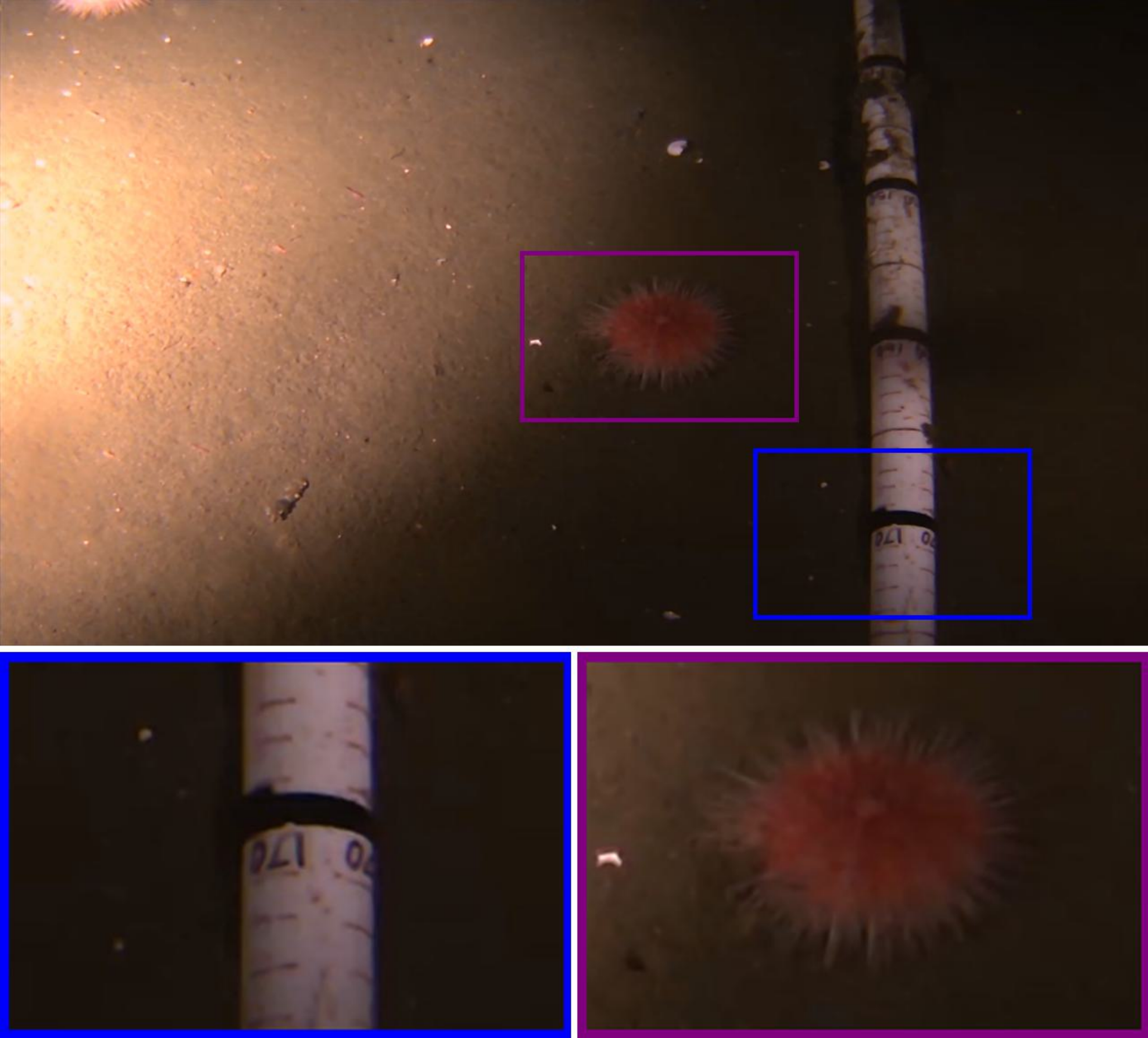}
		\caption{\footnotesize UWNet}
	\end{subfigure}
    \begin{subfigure}{0.105\linewidth}
		\centering
		\includegraphics[width=\linewidth,  height=\nuidheighttwo]{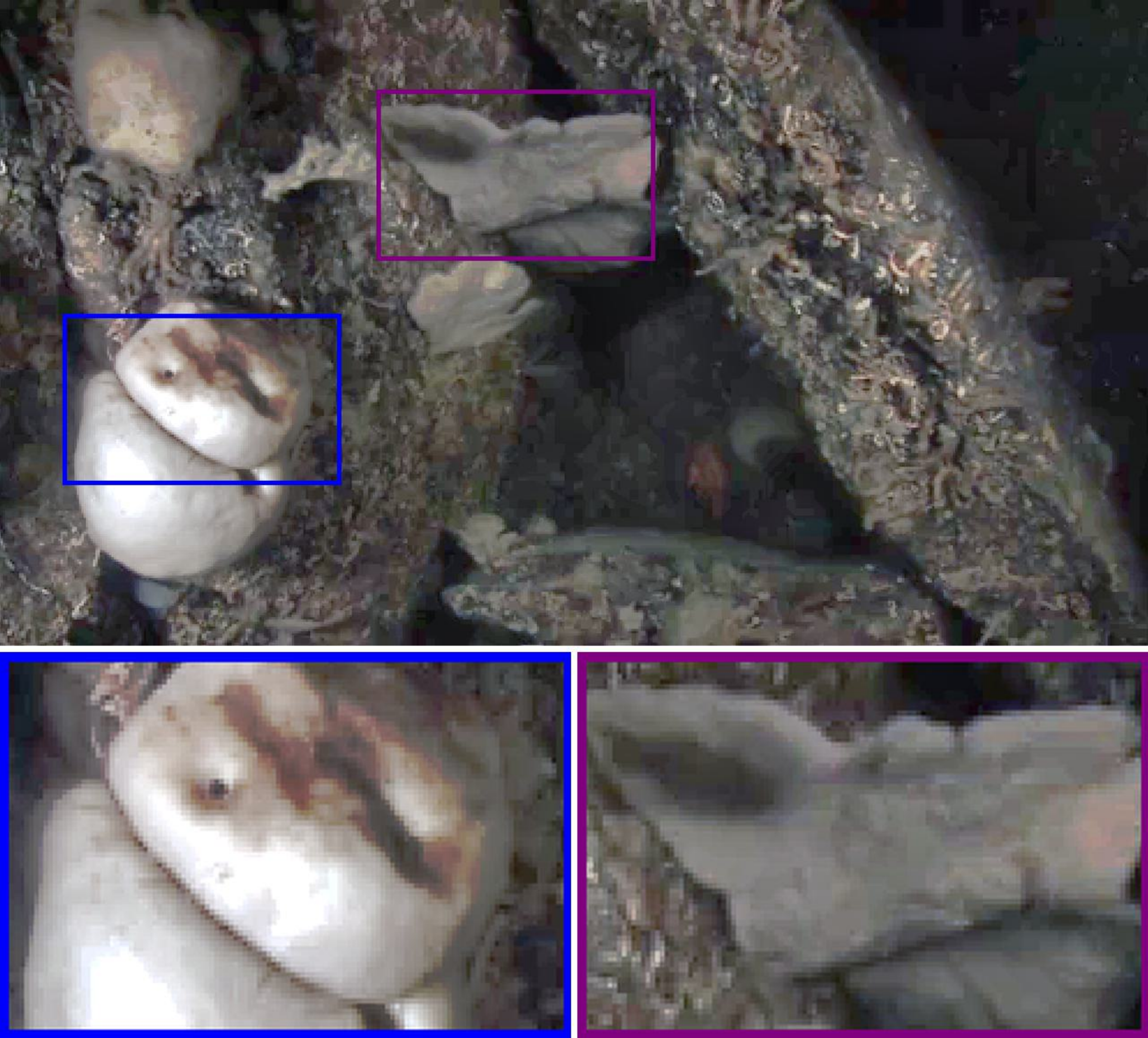} 
        \includegraphics[width=\linewidth,  height=\nuidheighttwo]{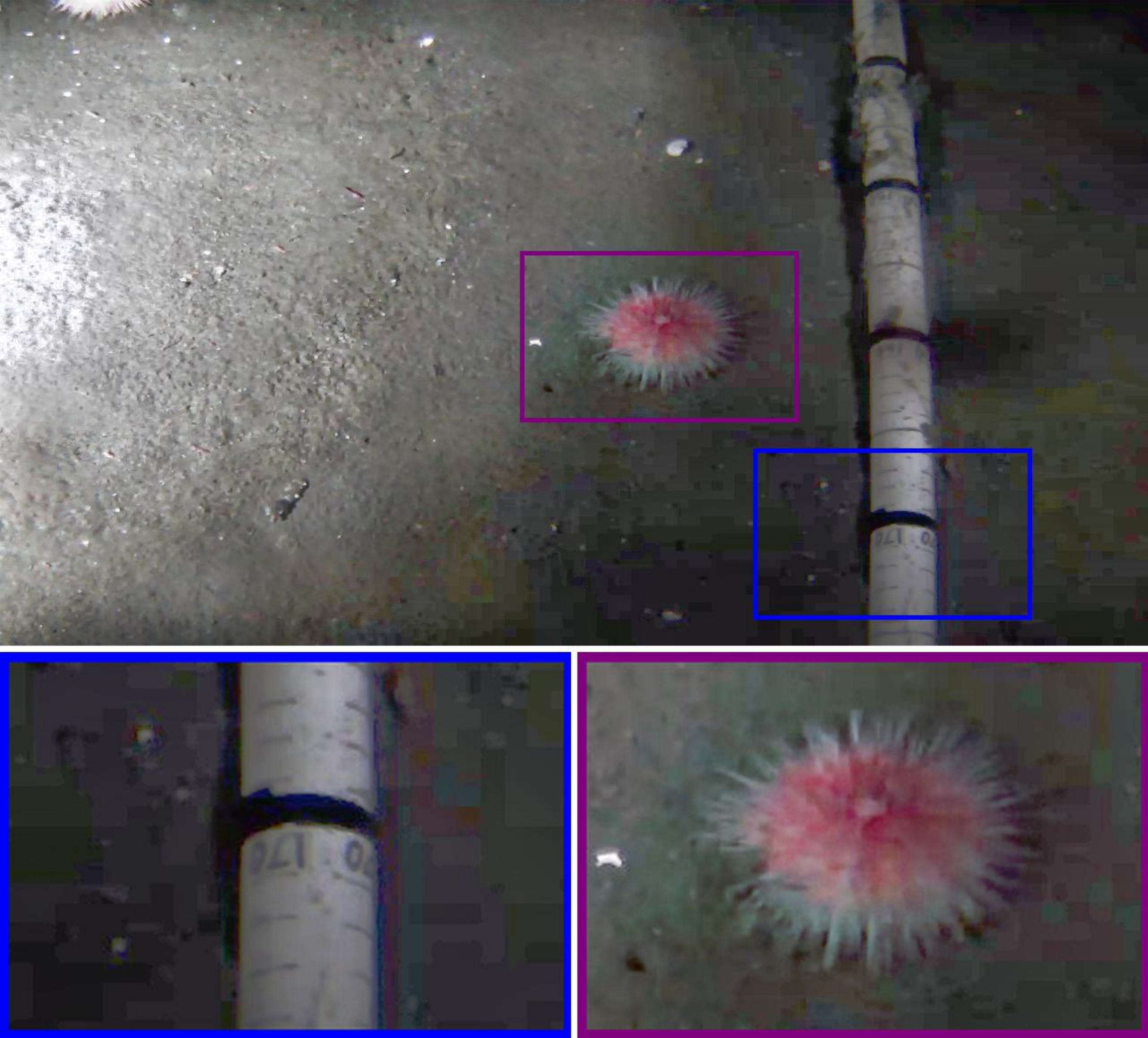} 
		\caption{\footnotesize ACDC}
	\end{subfigure}
	\begin{subfigure}{0.105\linewidth}
		\centering
		\includegraphics[width=\linewidth,  height=\nuidheighttwo]{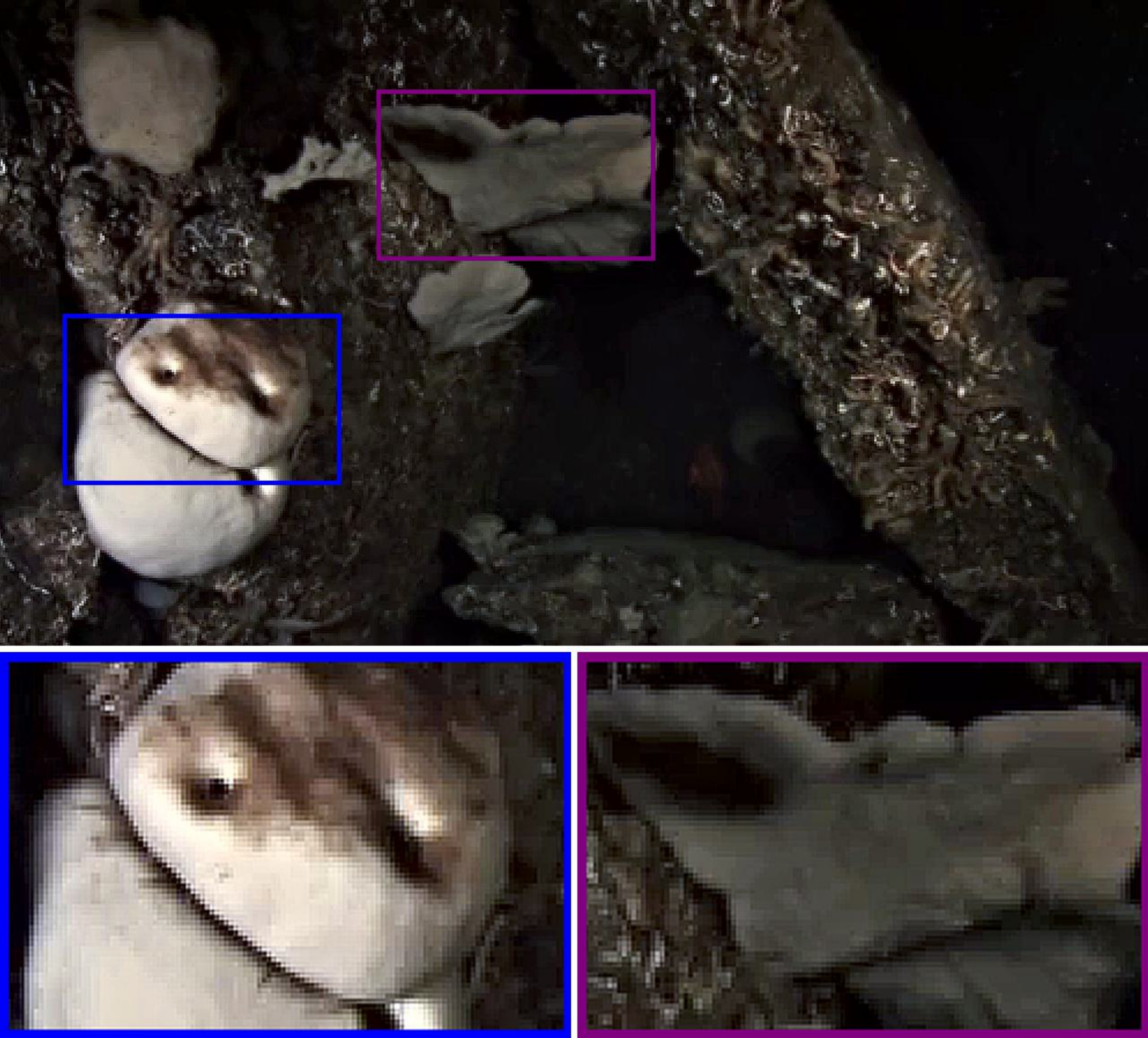} 
        \includegraphics[width=\linewidth,  height=\nuidheighttwo]{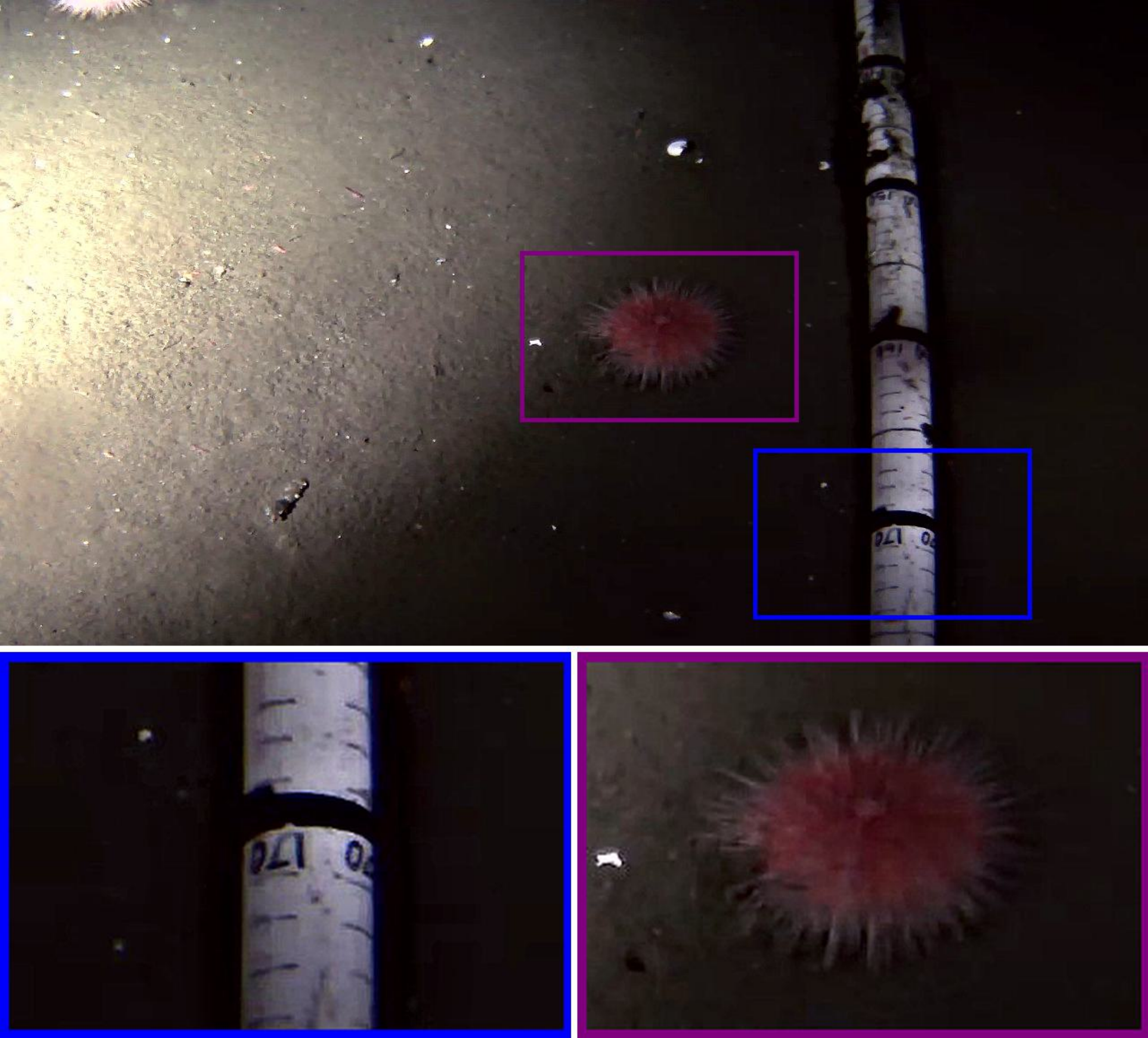} 
		\caption{\footnotesize MMLE}
	\end{subfigure}
    \begin{subfigure}{0.105\linewidth}
		\centering
		\includegraphics[width=\linewidth,  height=\nuidheighttwo]{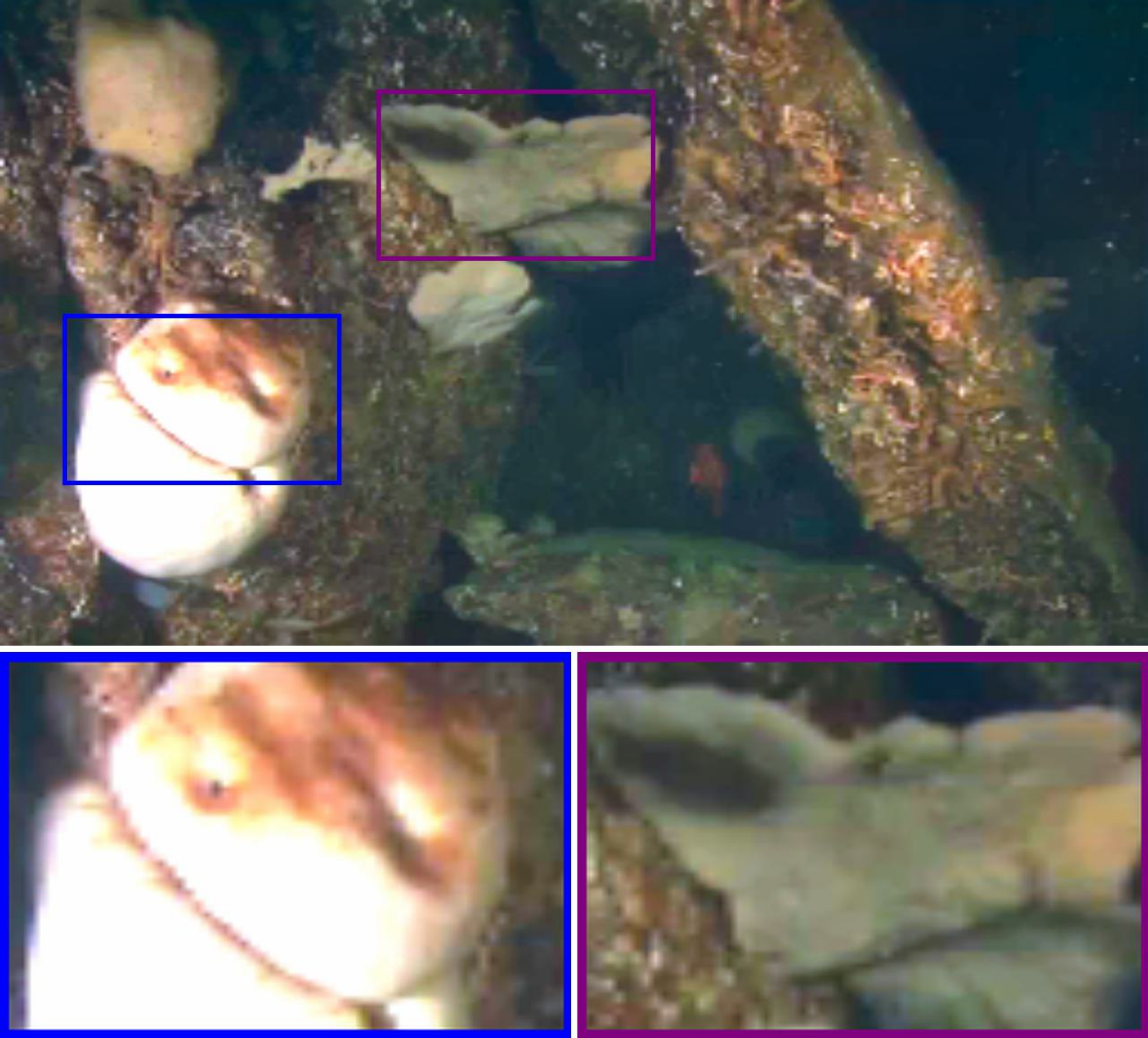} 
        \includegraphics[width=\linewidth,  height=\nuidheighttwo]{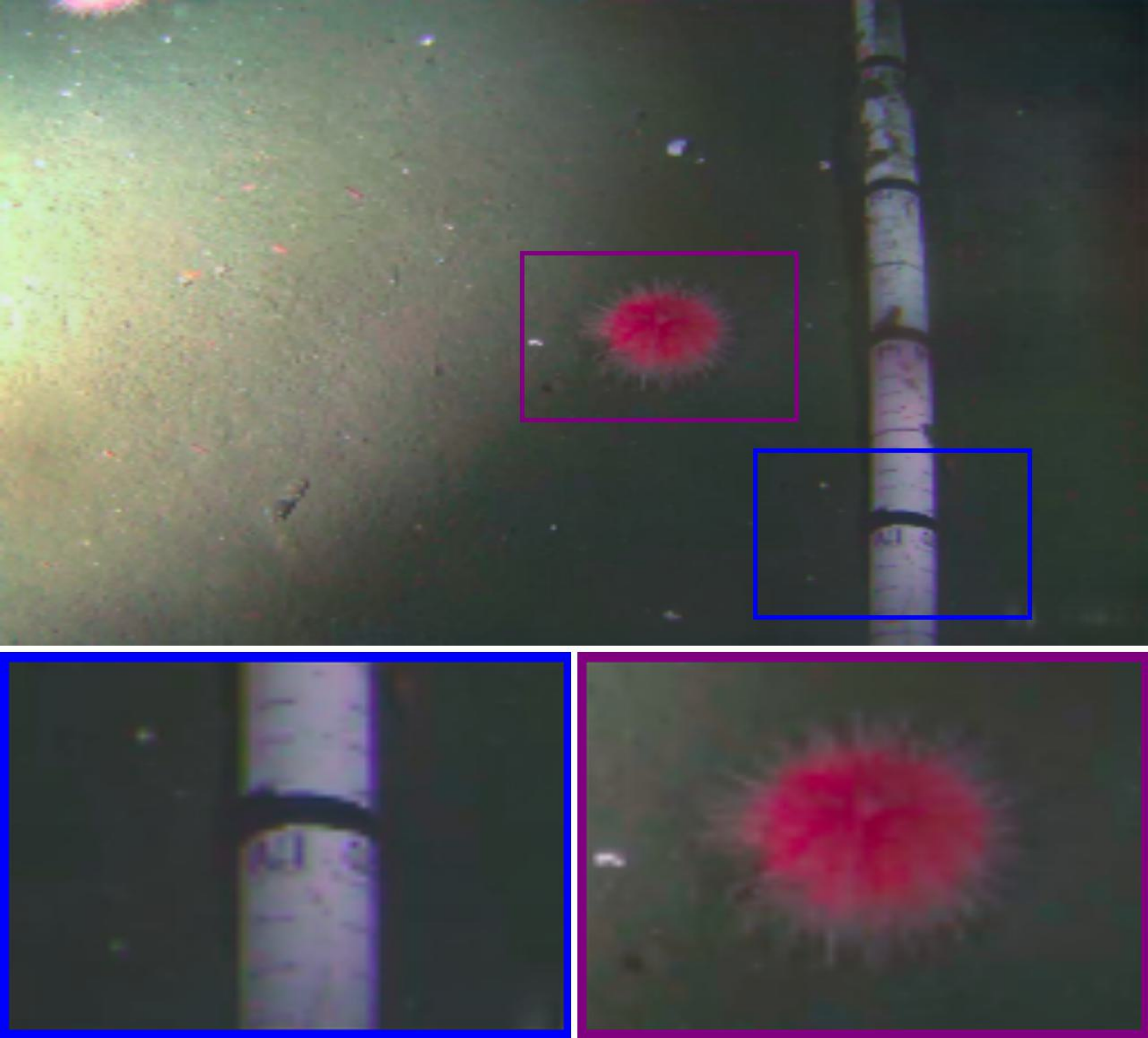} 
		\caption{\footnotesize TCTL-Net}
	\end{subfigure}
    \begin{subfigure}{0.105\linewidth}
		\centering
		\includegraphics[width=\linewidth,  height=\nuidheighttwo]{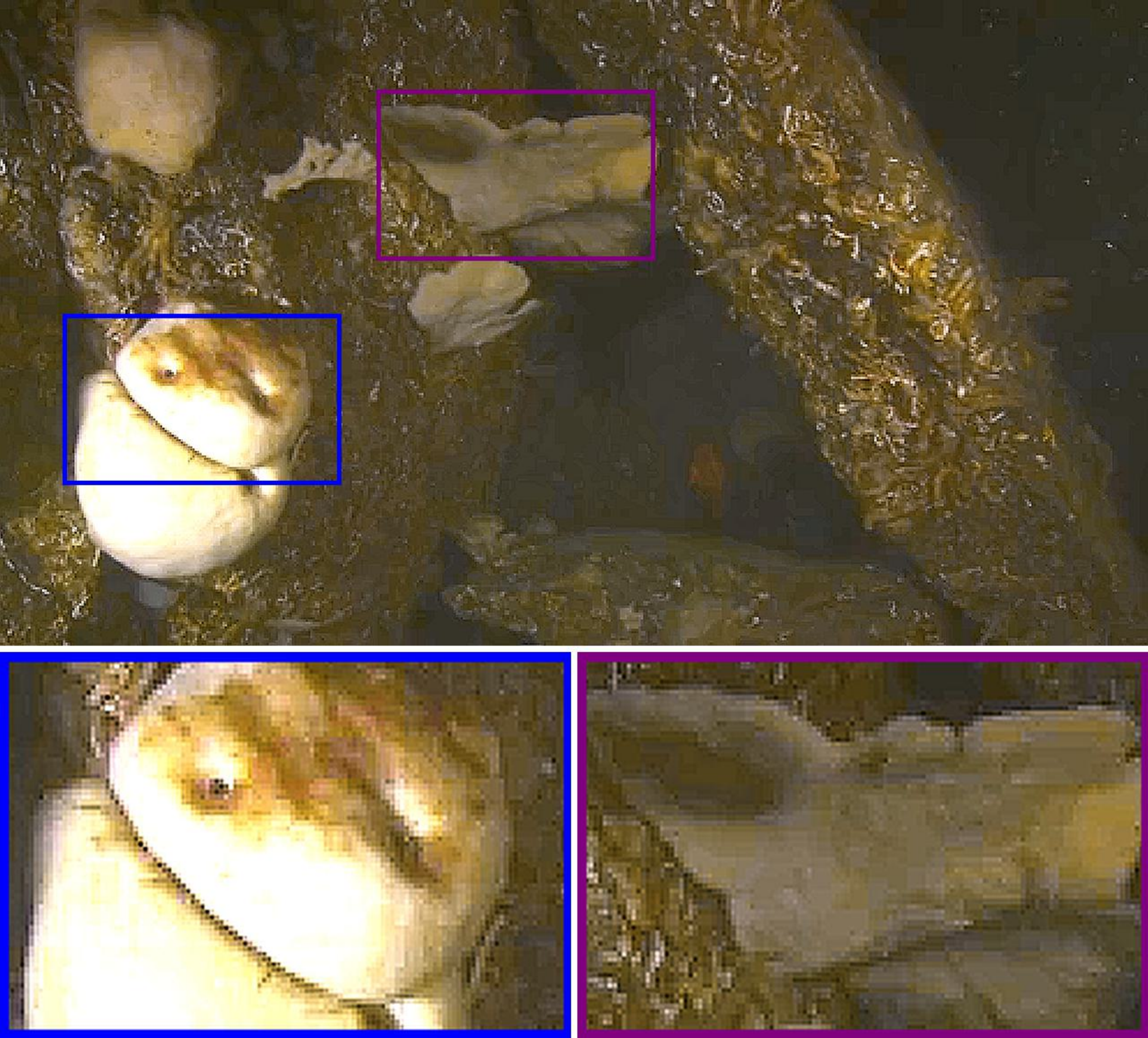} 
        \includegraphics[width=\linewidth,  height=\nuidheighttwo]{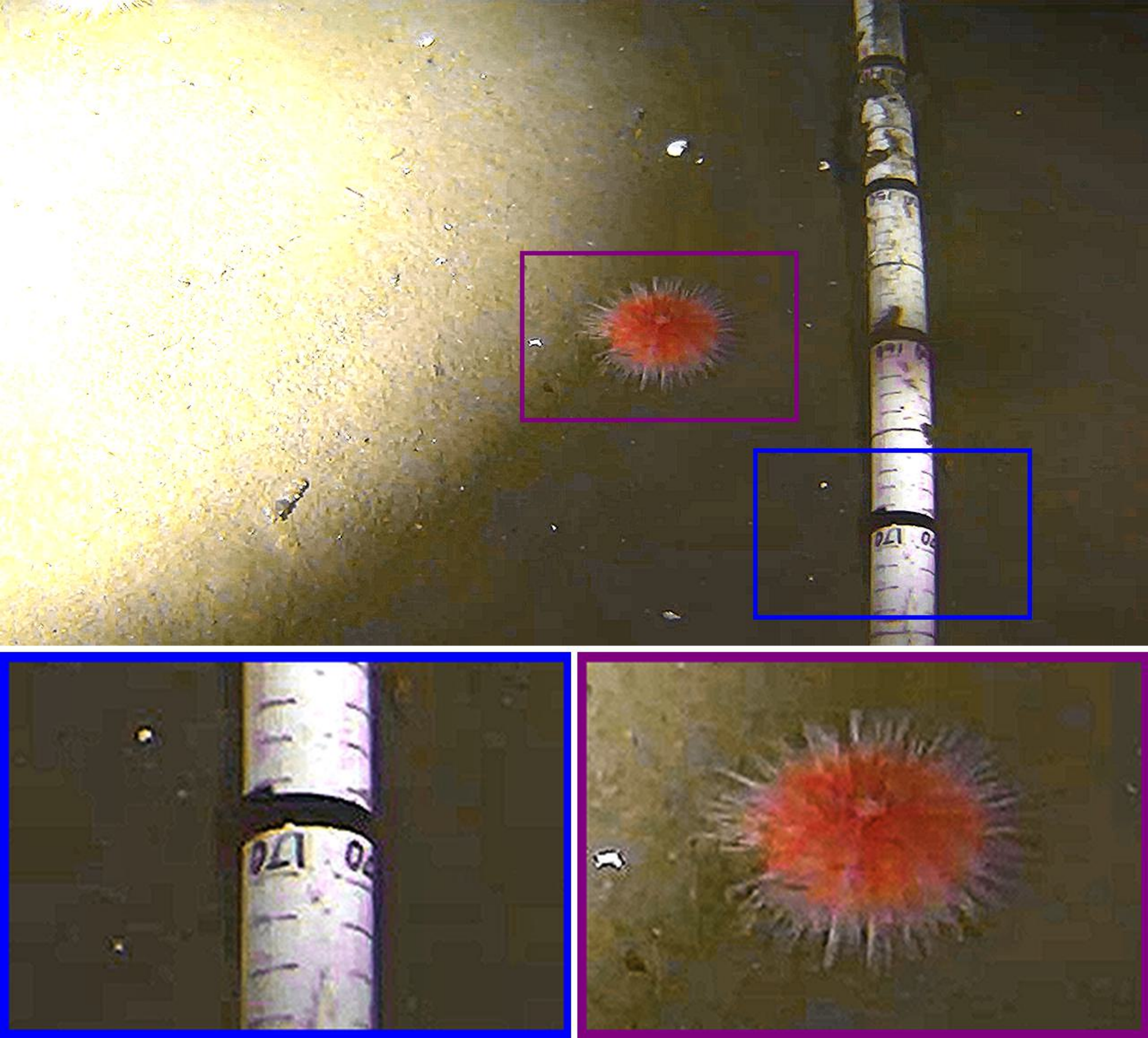} 
        \caption{\footnotesize ICSP}
	\end{subfigure}
	\begin{subfigure}{0.105\linewidth}
		\centering
		\includegraphics[width=\linewidth,  height=\nuidheighttwo]{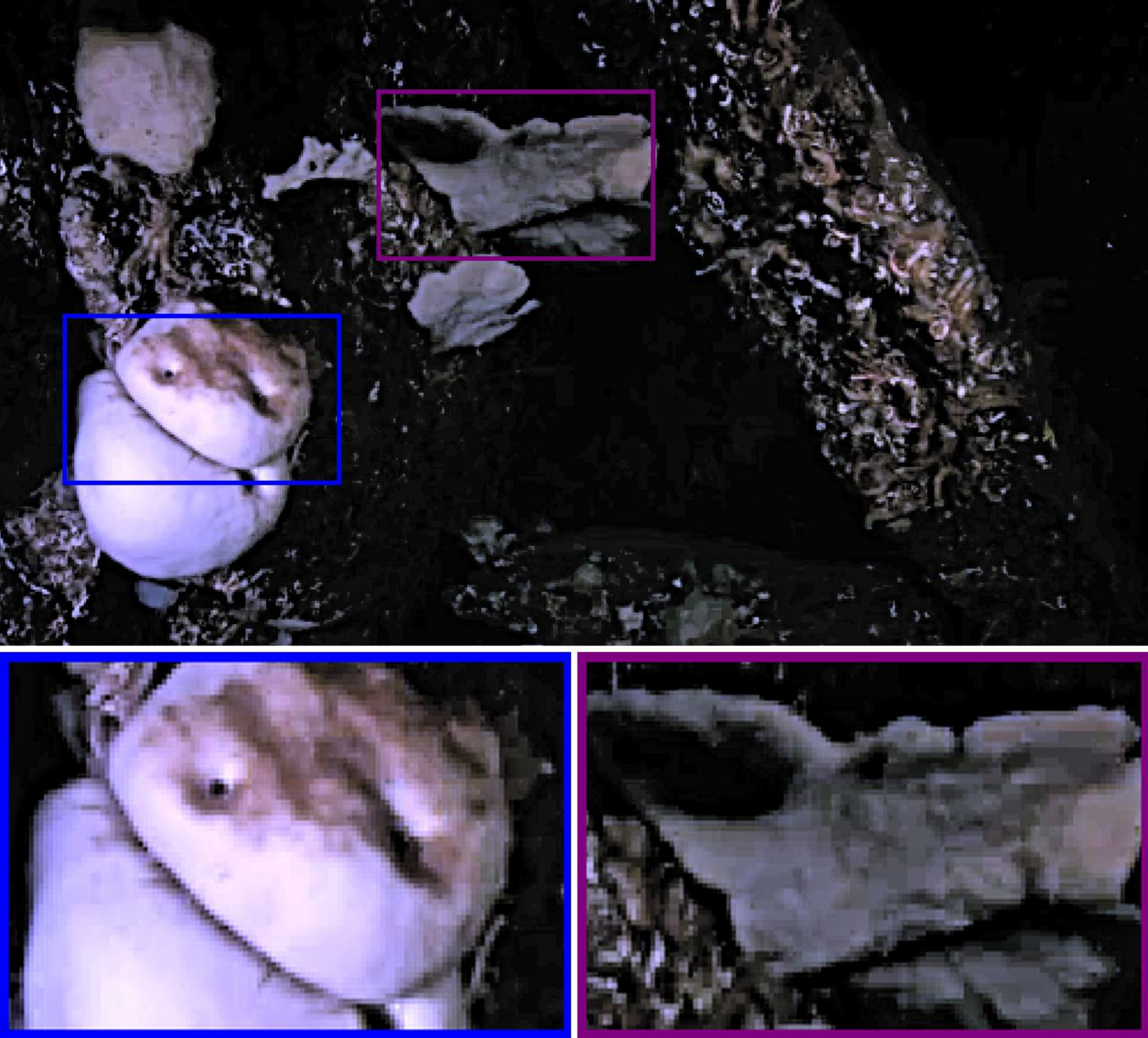} 
        \includegraphics[width=\linewidth,  height=\nuidheighttwo]{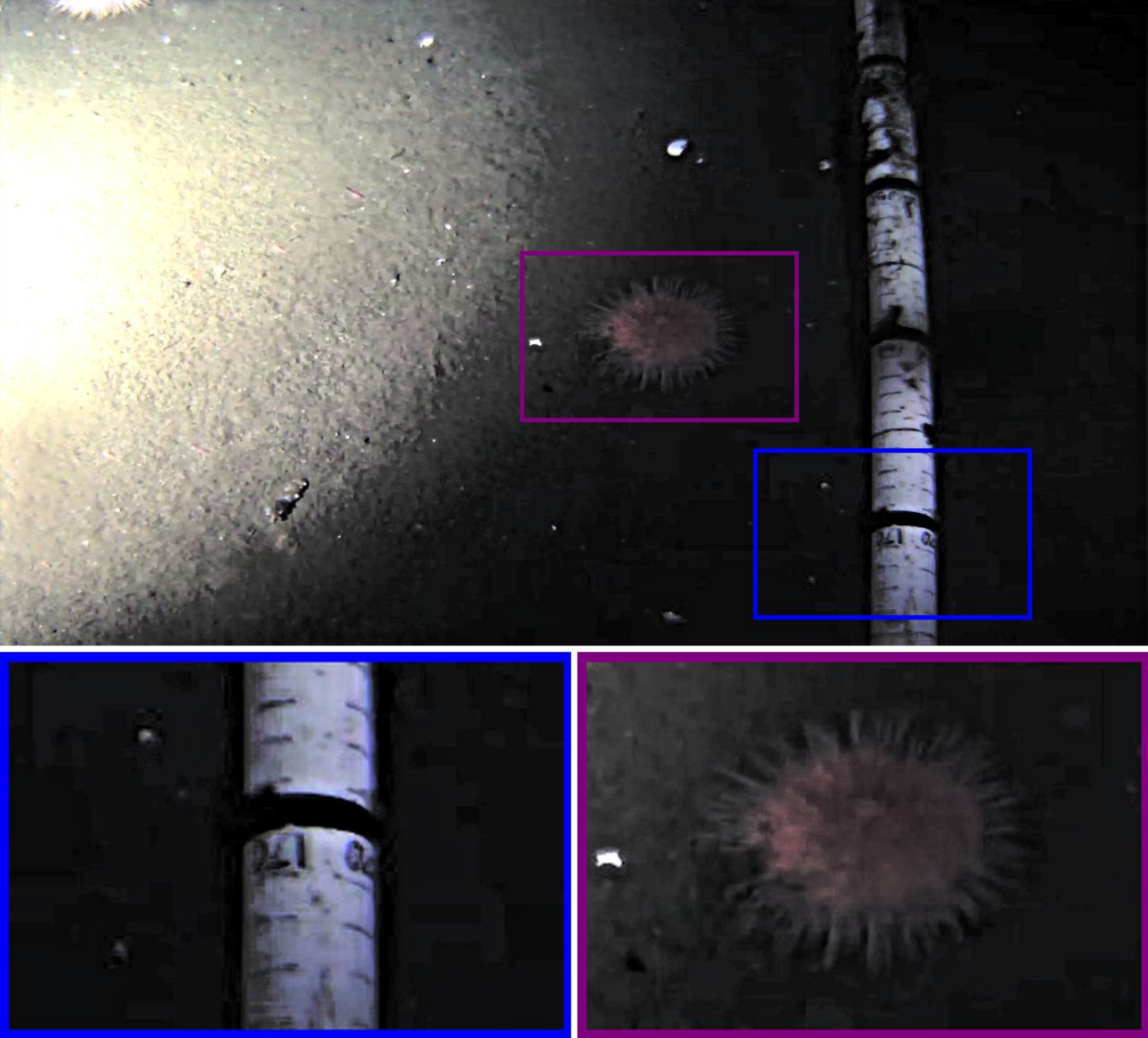} 
        \caption{\footnotesize PCDE}
	\end{subfigure}
	\begin{subfigure}{0.105\linewidth}
		\centering
		\includegraphics[width=\linewidth,  height=\nuidheighttwo]{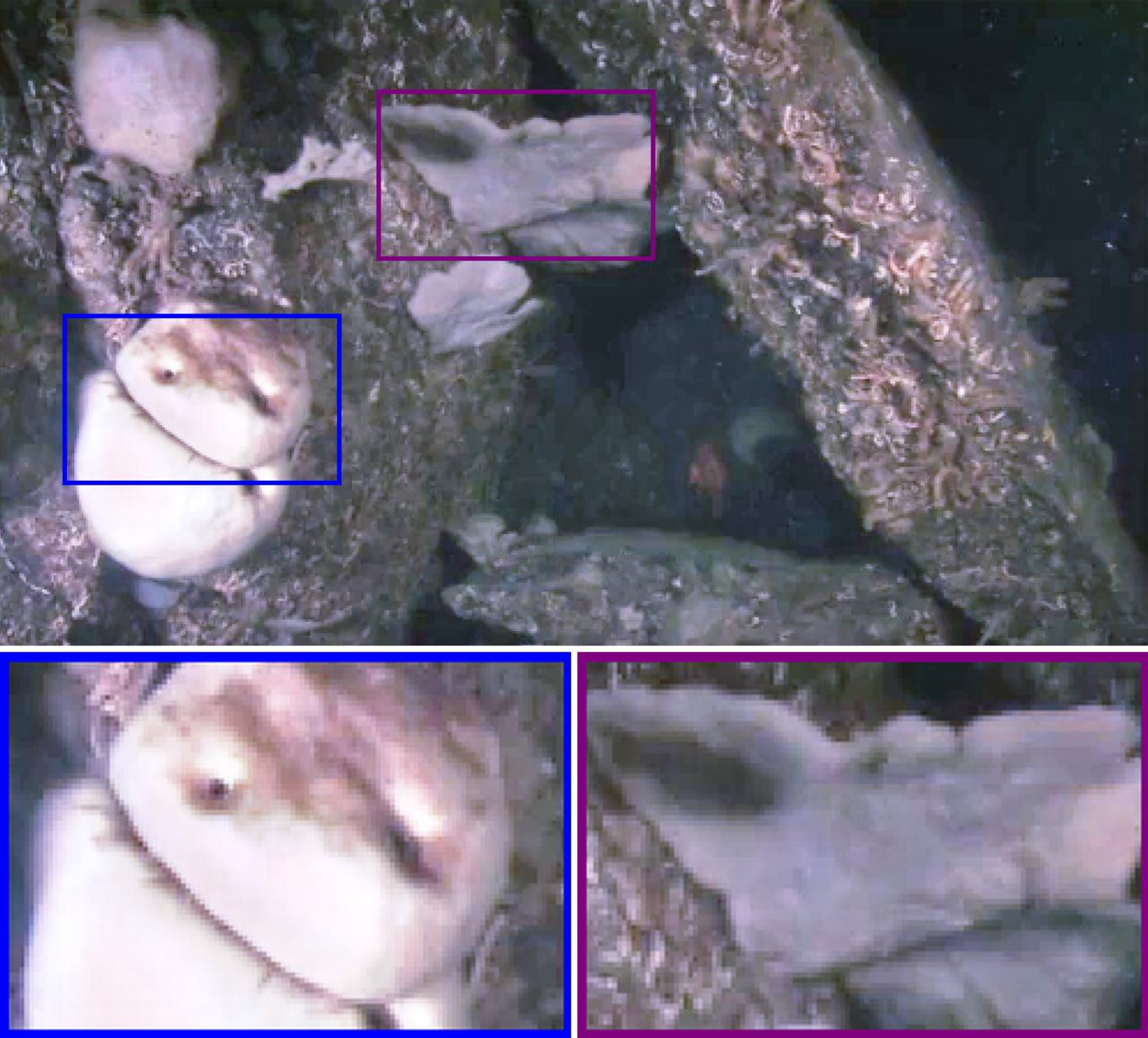}
        \includegraphics[width=\linewidth,  height=\nuidheighttwo]{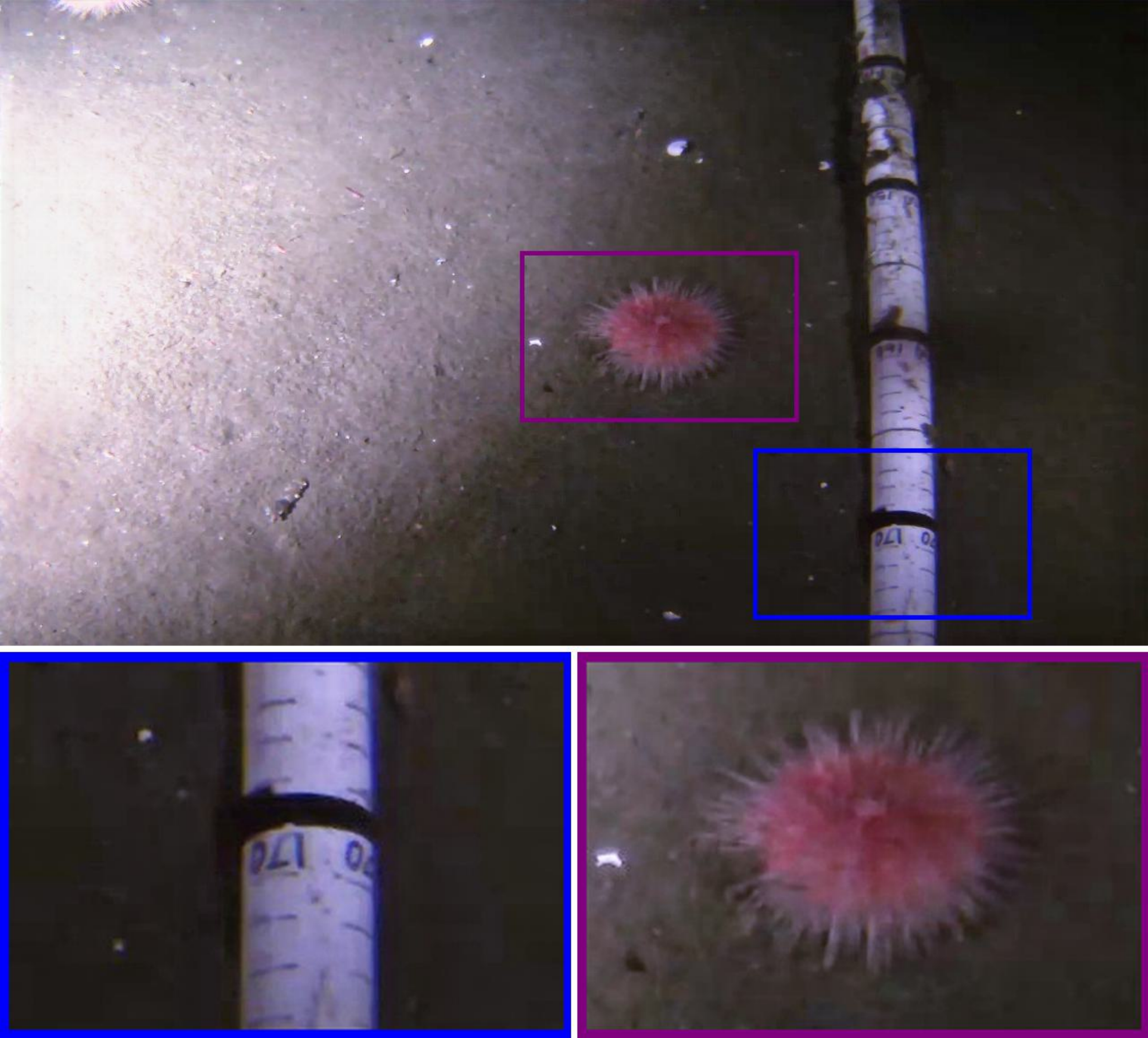}
        \caption{\footnotesize UDAformer}
	\end{subfigure}

    %%%%%%%%%%%%%%%%%%%%%%%%%%%%%
	\begin{subfigure}{0.105\linewidth}
		\centering
		\includegraphics[width=\linewidth,  height=\nuidheighttwo]{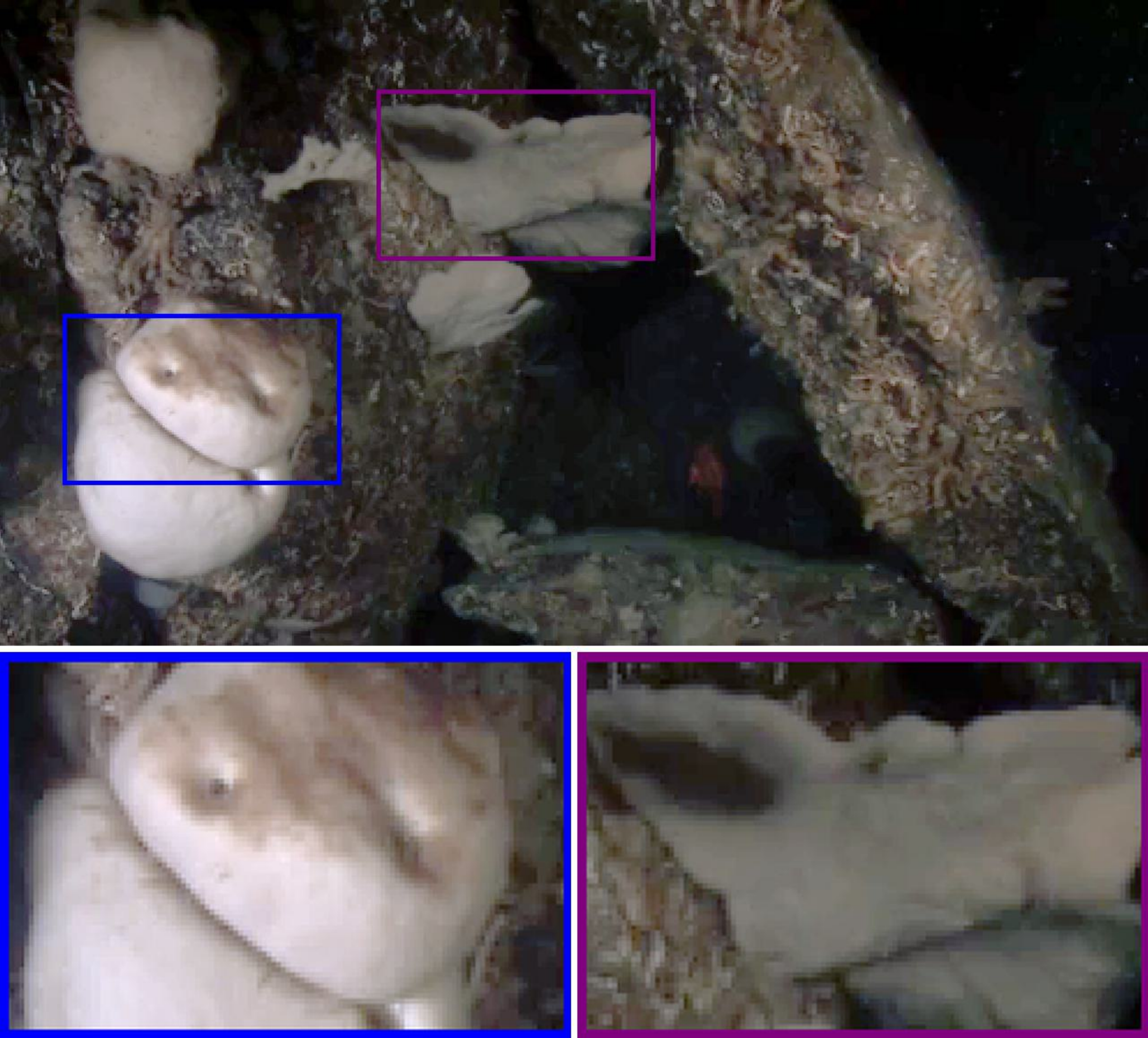} 
        \includegraphics[width=\linewidth,  height=\nuidheighttwo]{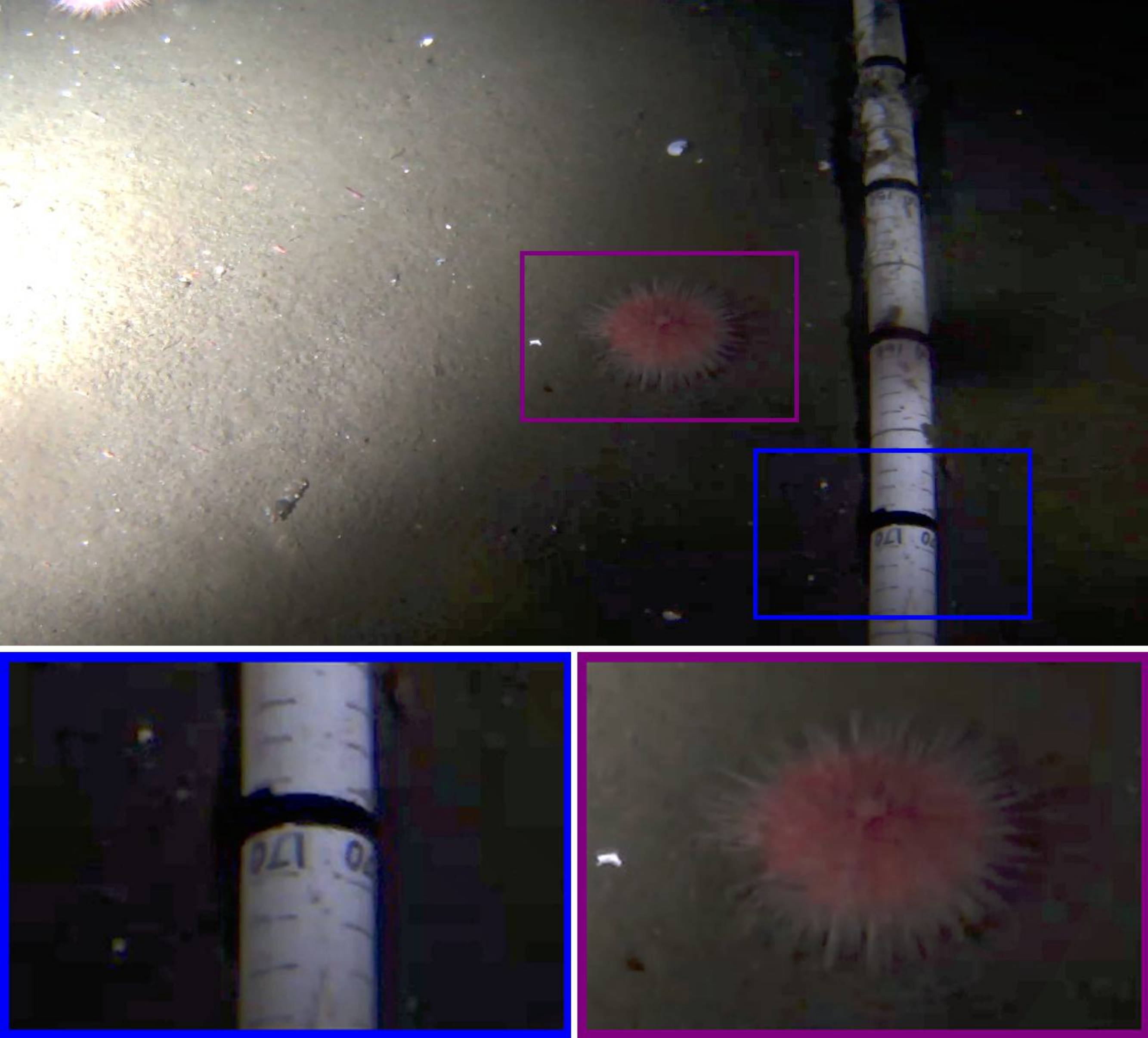} 
		\caption{\footnotesize HFM}
	\end{subfigure}
	\begin{subfigure}{0.105\linewidth}
		\centering
		\includegraphics[width=\linewidth,  height=\nuidheighttwo]{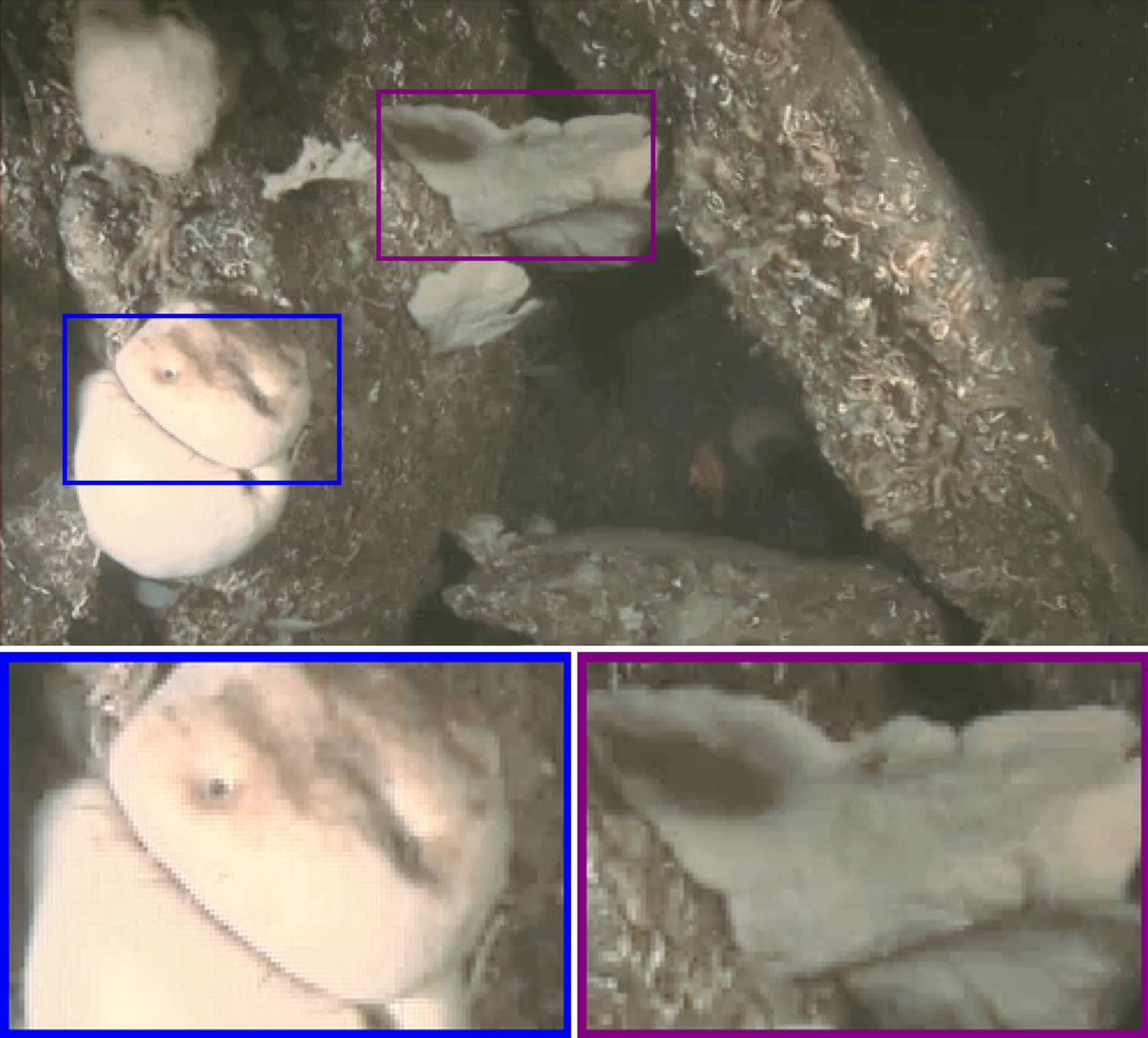} 
        \includegraphics[width=\linewidth,  height=\nuidheighttwo]{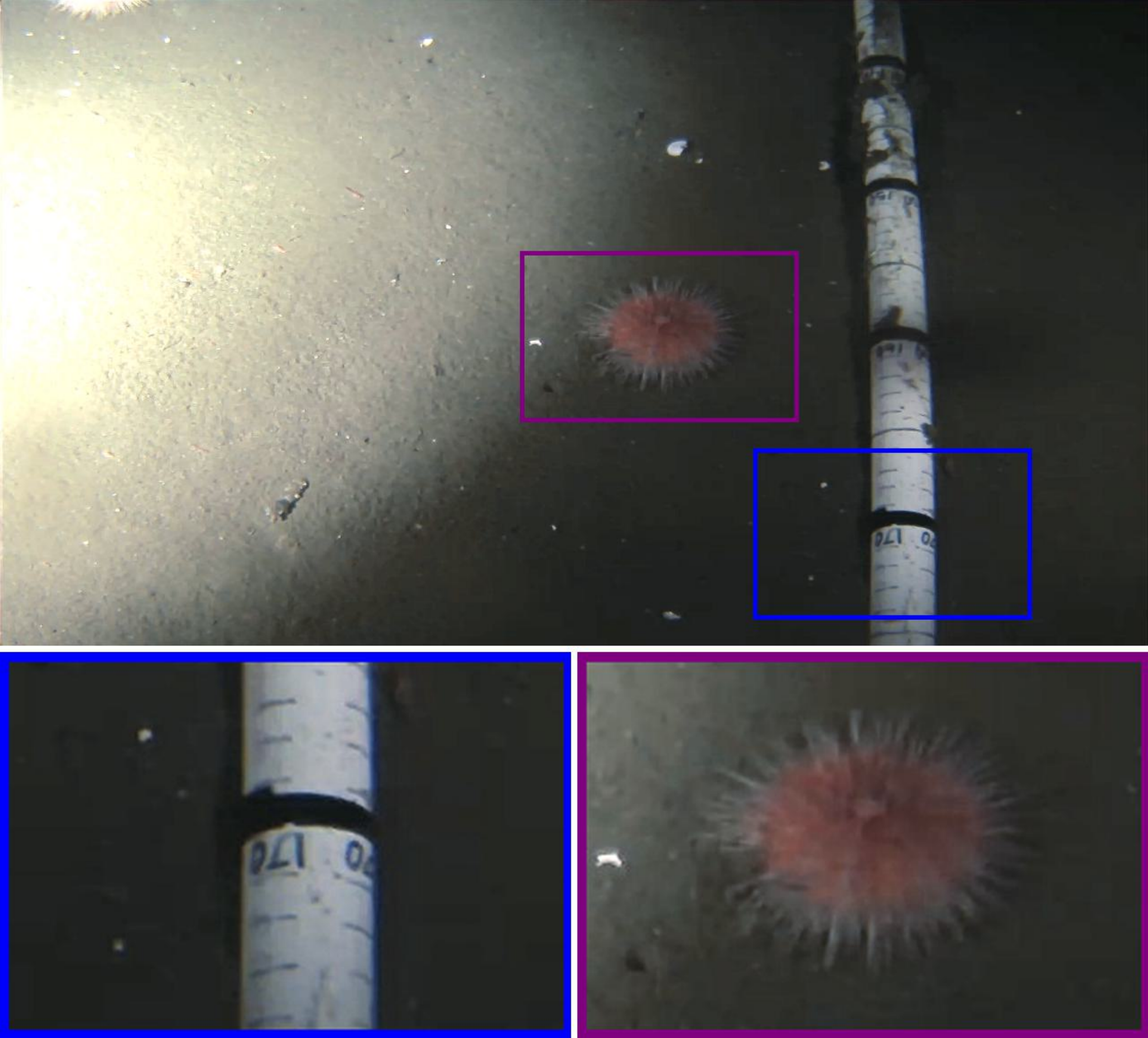} 
		\caption{\footnotesize LENet}
	\end{subfigure}
	\begin{subfigure}{0.105\linewidth}
		\centering
		\includegraphics[width=\linewidth,  height=\nuidheighttwo]{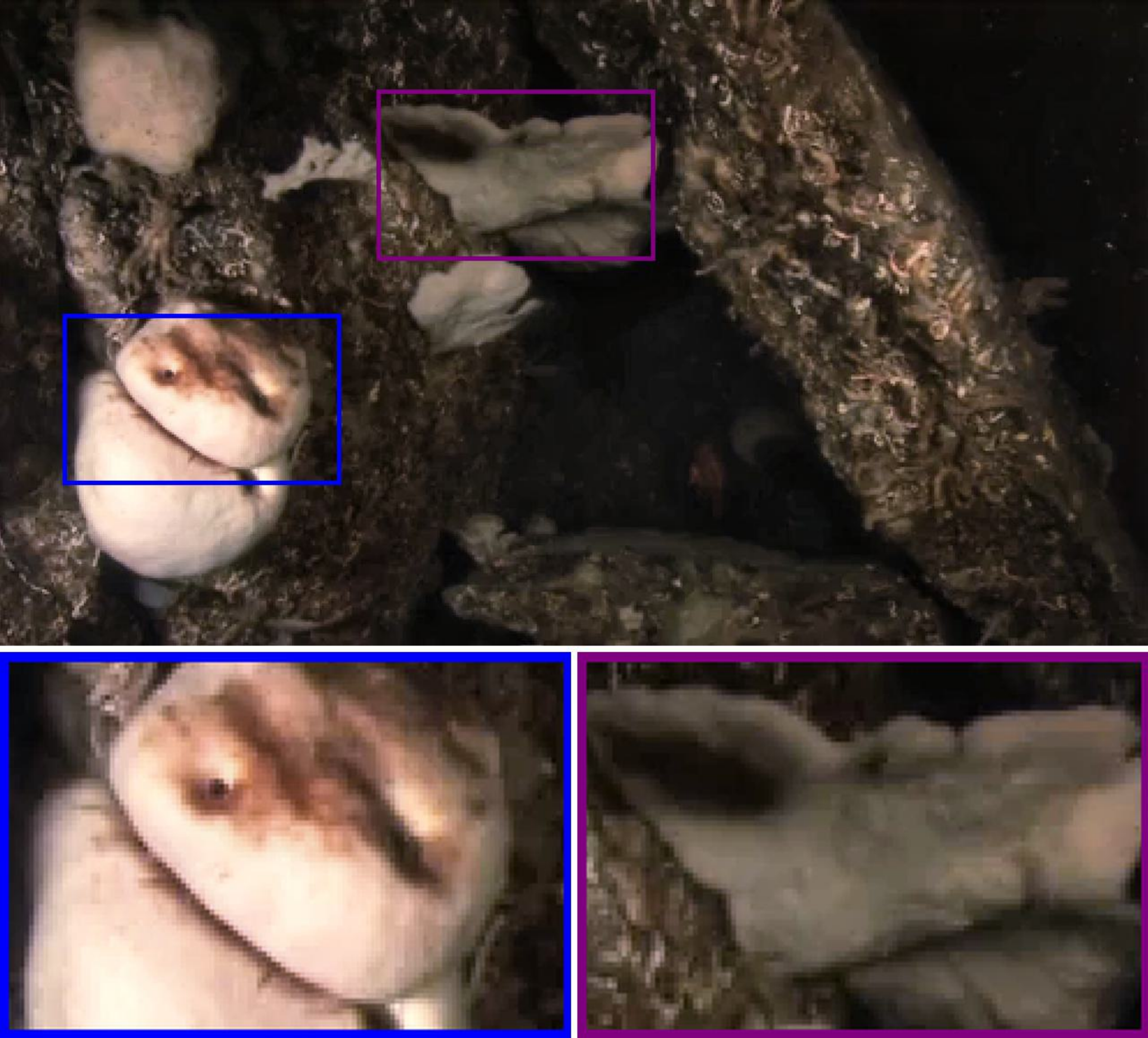}
        \includegraphics[width=\linewidth,  height=\nuidheighttwo]{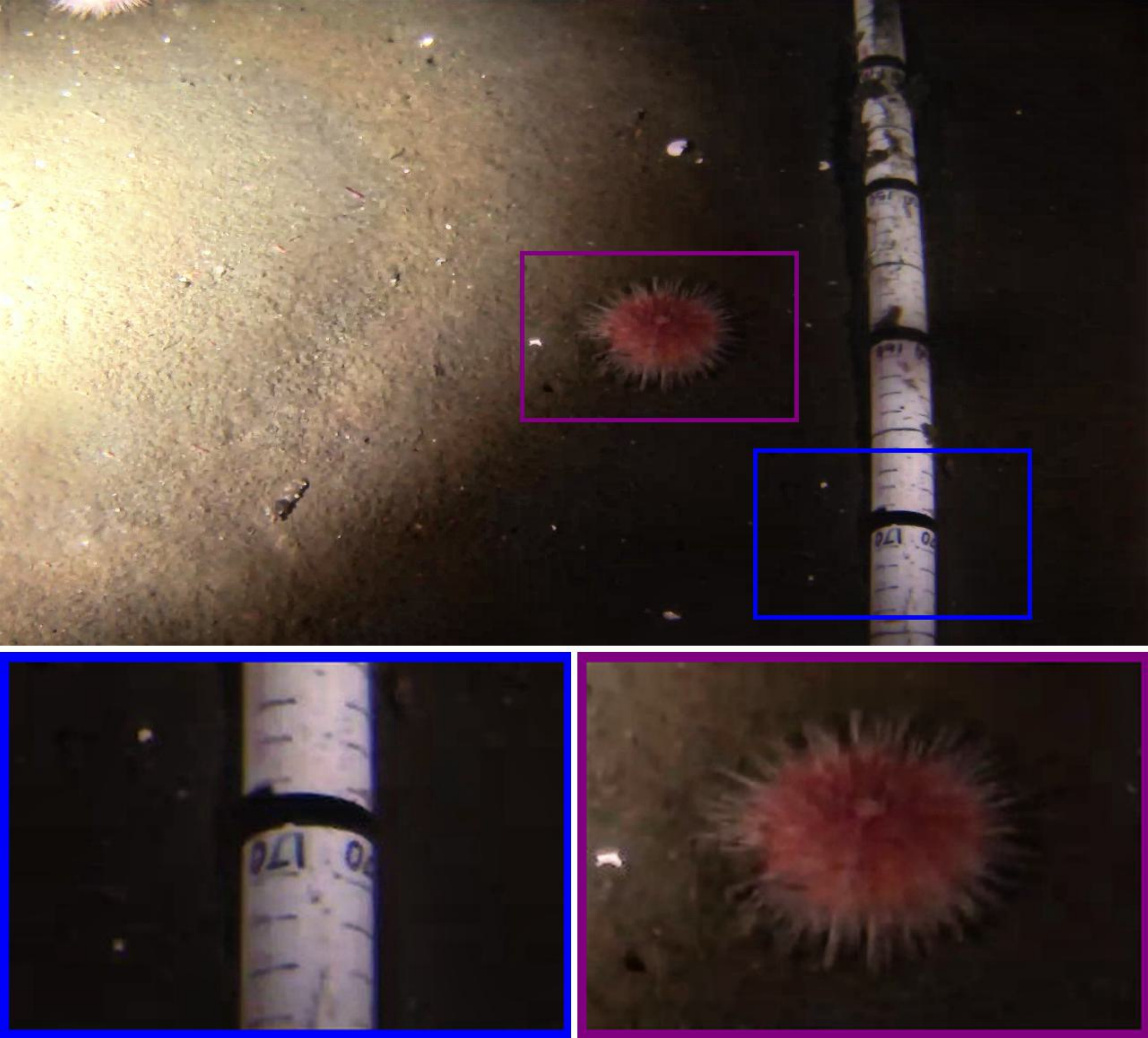}
		\caption{\footnotesize SMDR-IS}
	\end{subfigure}
	\begin{subfigure}{0.105\linewidth}
		\centering
		\includegraphics[width=\linewidth,  height=\nuidheighttwo]{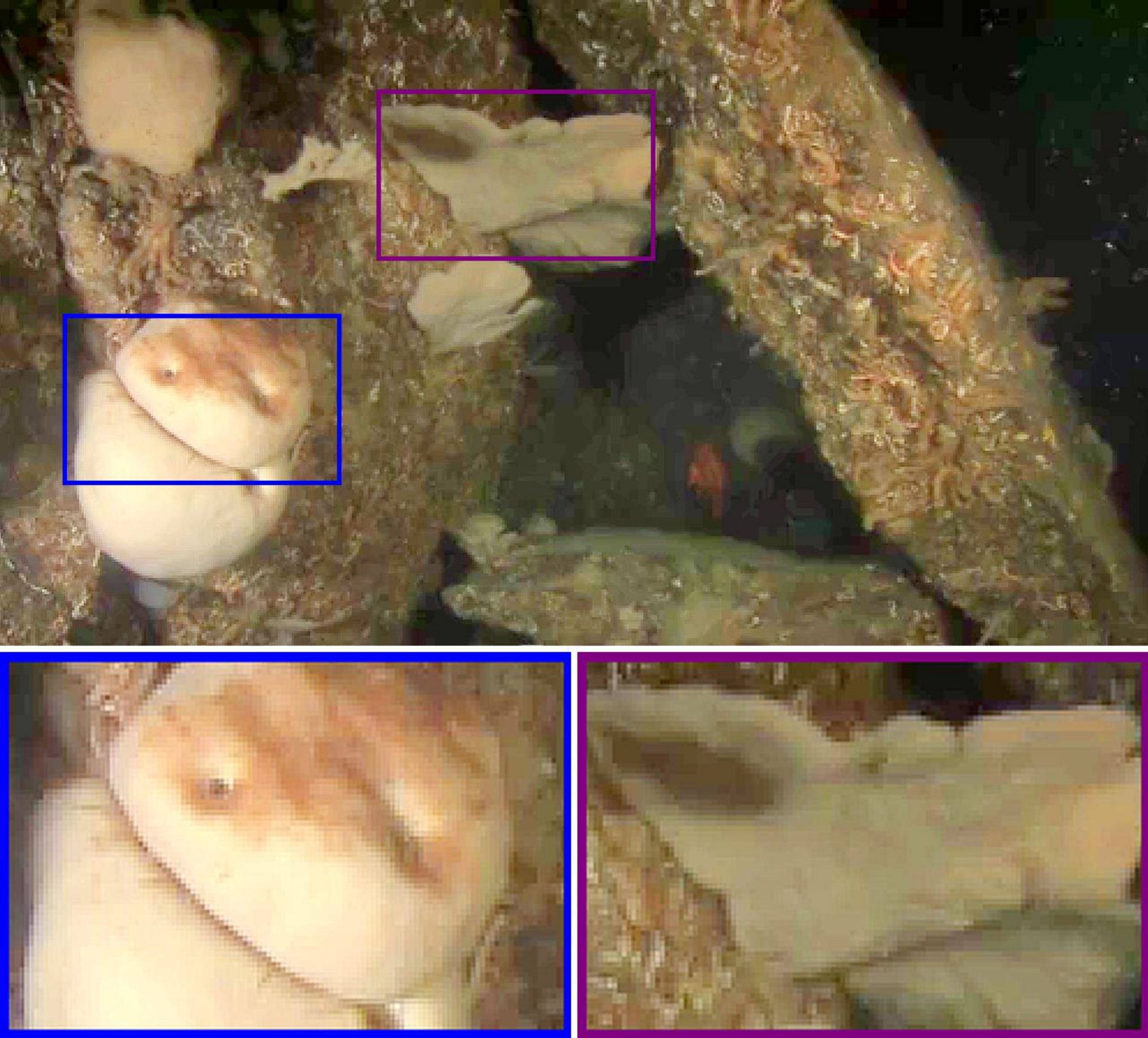} 
        \includegraphics[width=\linewidth,  height=\nuidheighttwo]{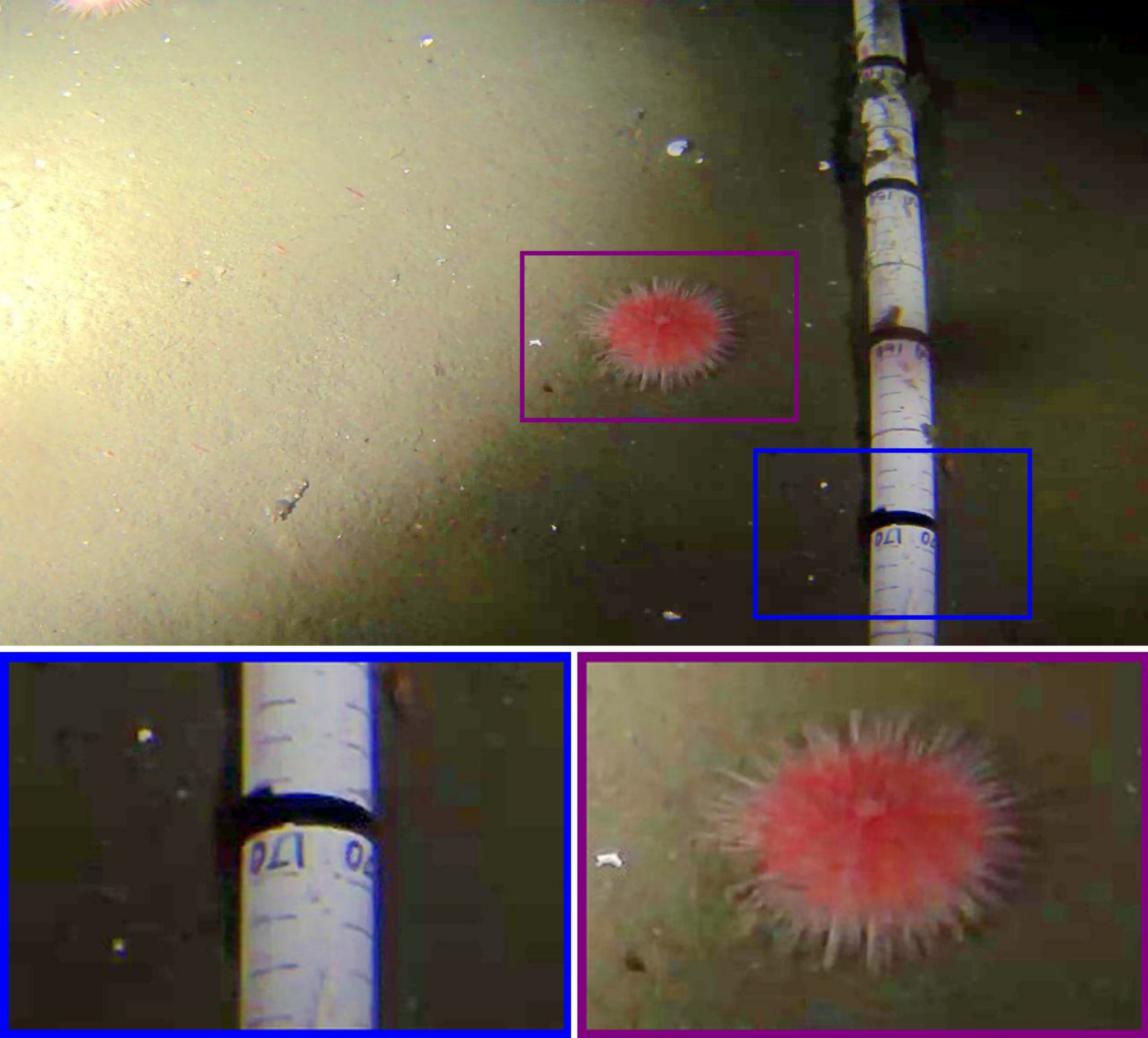} 
		\caption{\footnotesize GCP}
	\end{subfigure}
        \begin{subfigure}{0.105\linewidth}
		\centering
		\includegraphics[width=\linewidth,  height=\nuidheighttwo]{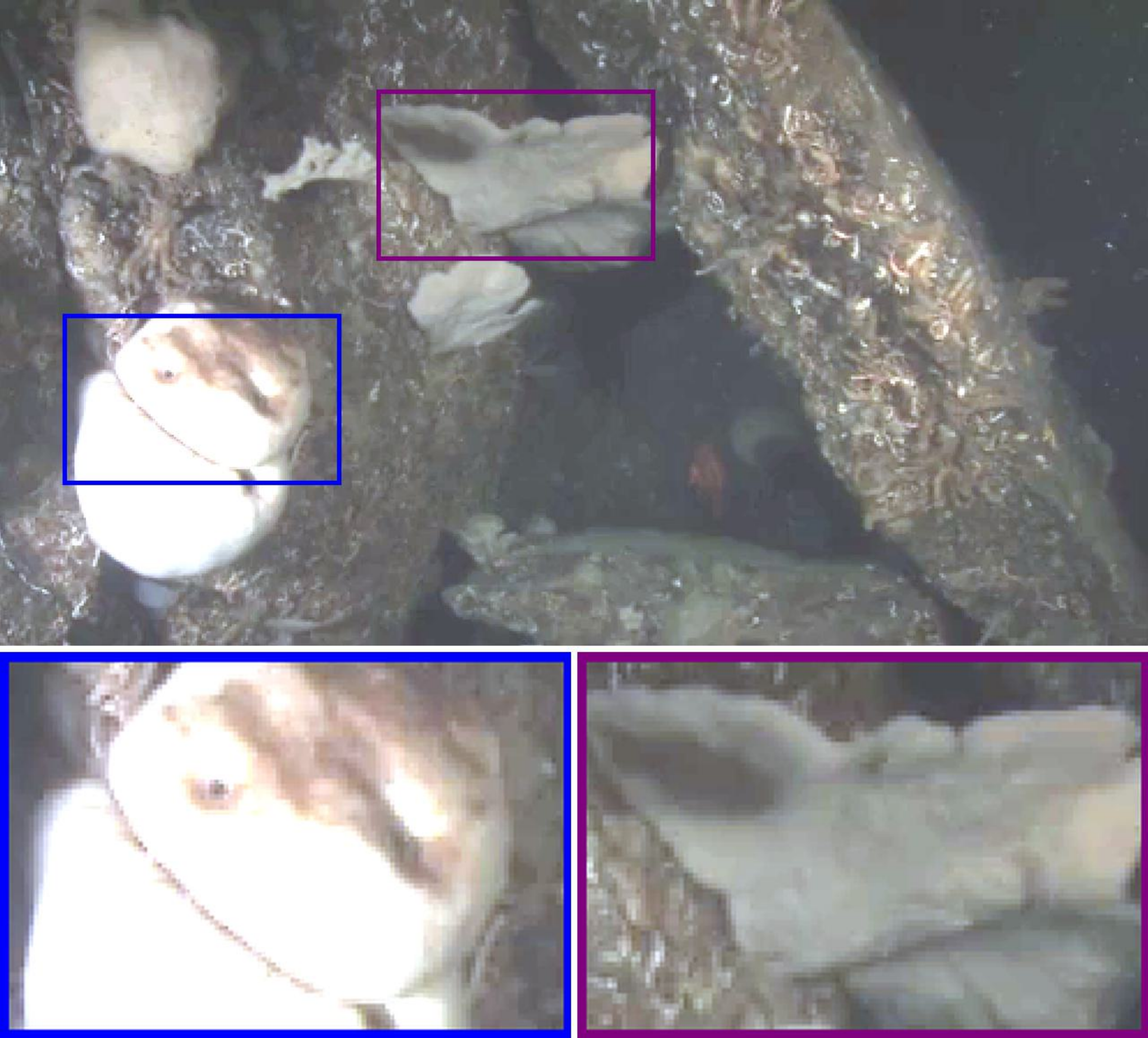} 
        \includegraphics[width=\linewidth,  height=\nuidheighttwo]{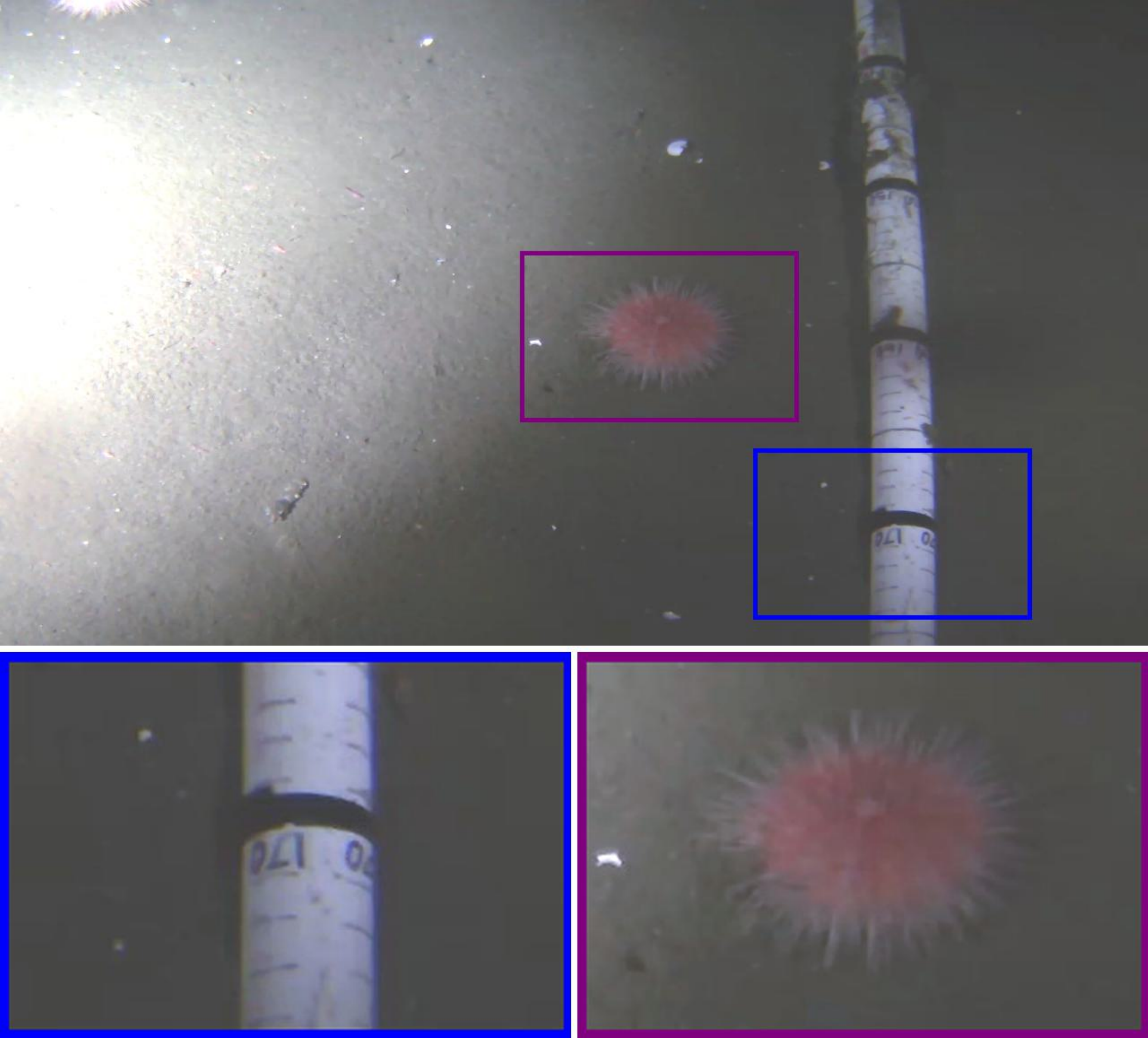} 
		\caption{\footnotesize MACT}
	\end{subfigure}
        \begin{subfigure}{0.105\linewidth}
		\centering
		\includegraphics[width=\linewidth,  height=\nuidheighttwo]{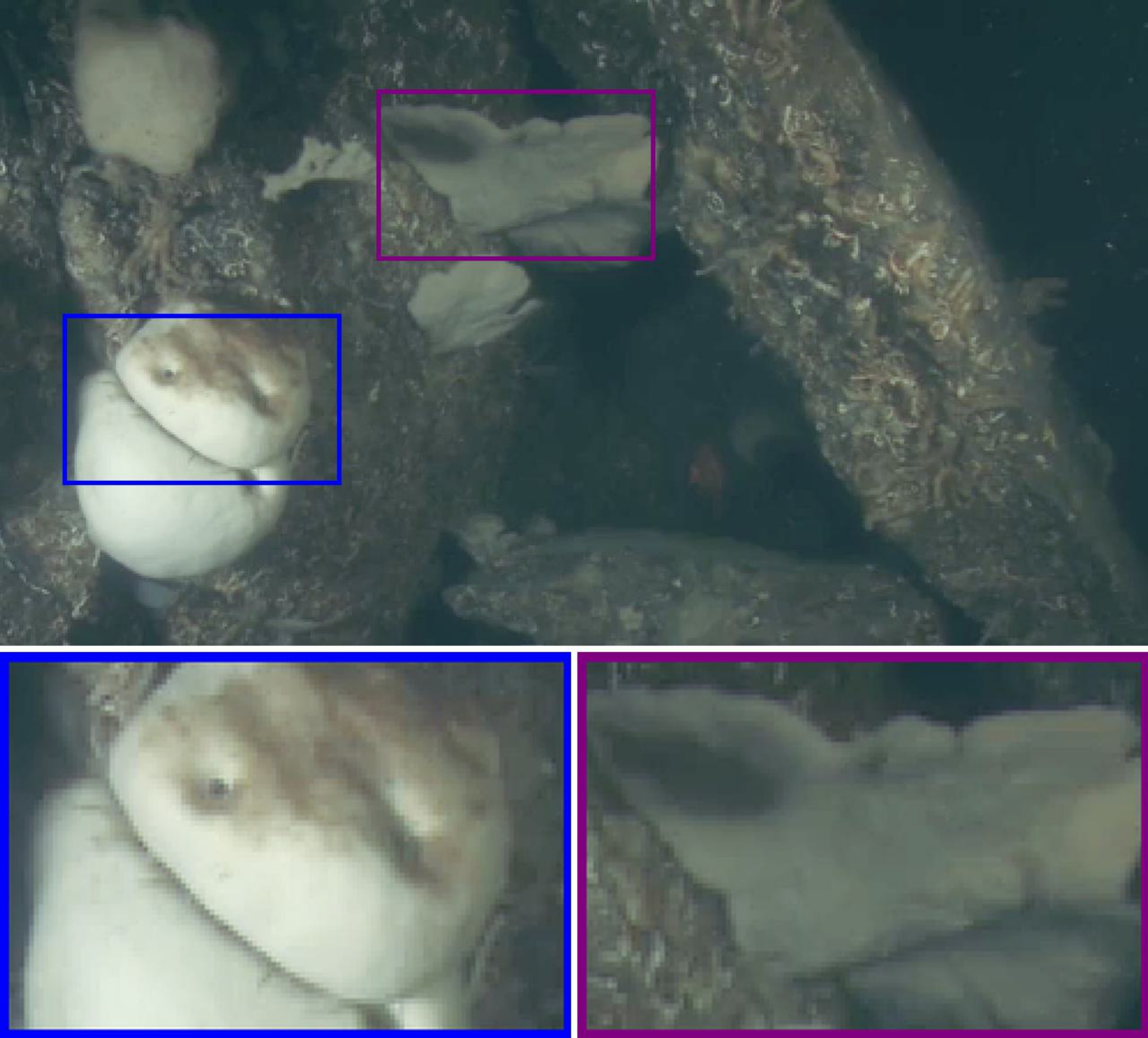} 
        \includegraphics[width=\linewidth,  height=\nuidheighttwo]{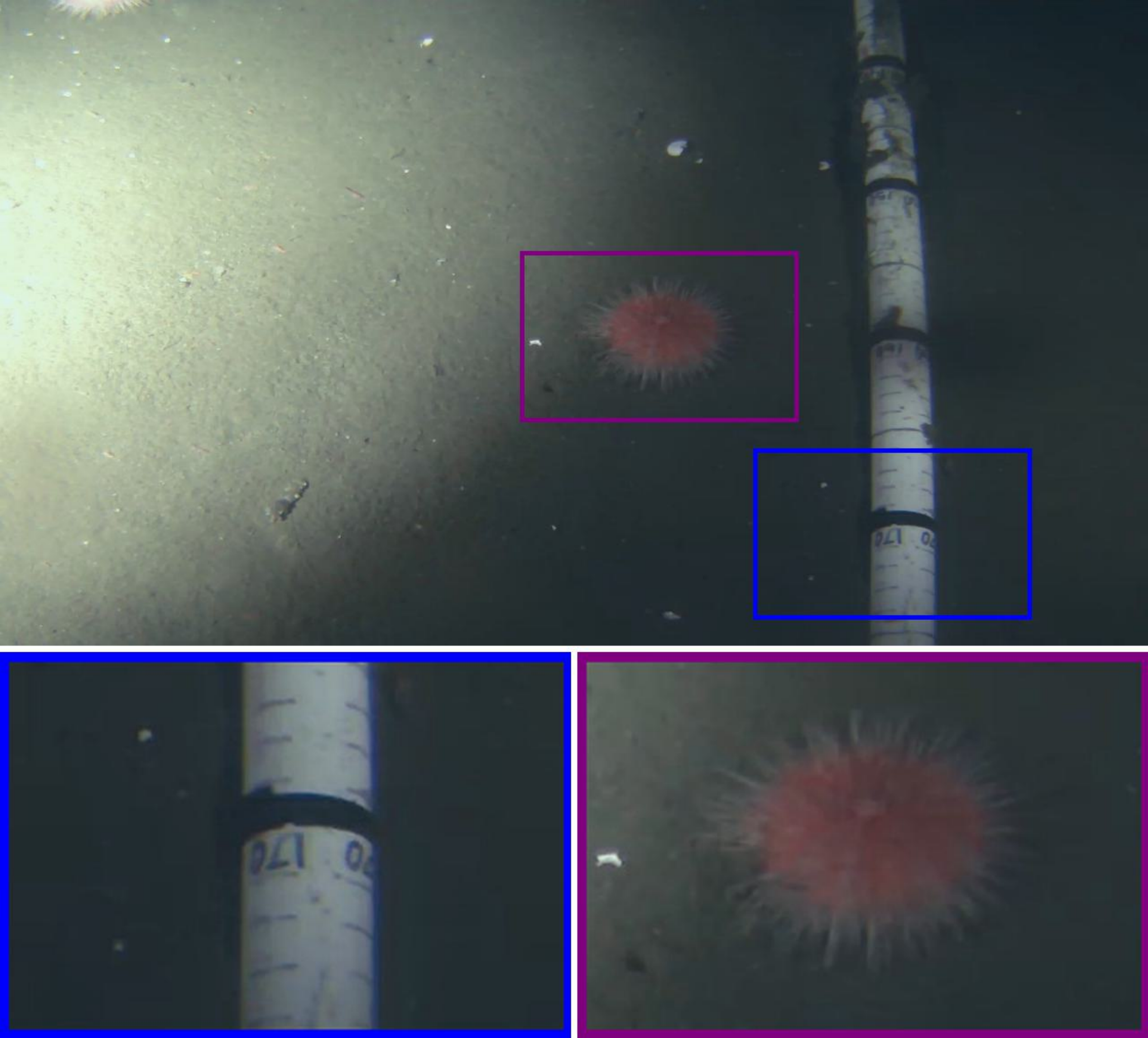} 
		\caption{\footnotesize UDnet}
	\end{subfigure}
        \begin{subfigure}{0.105\linewidth}
		\centering
		\includegraphics[width=\linewidth,  height=\nuidheighttwo]{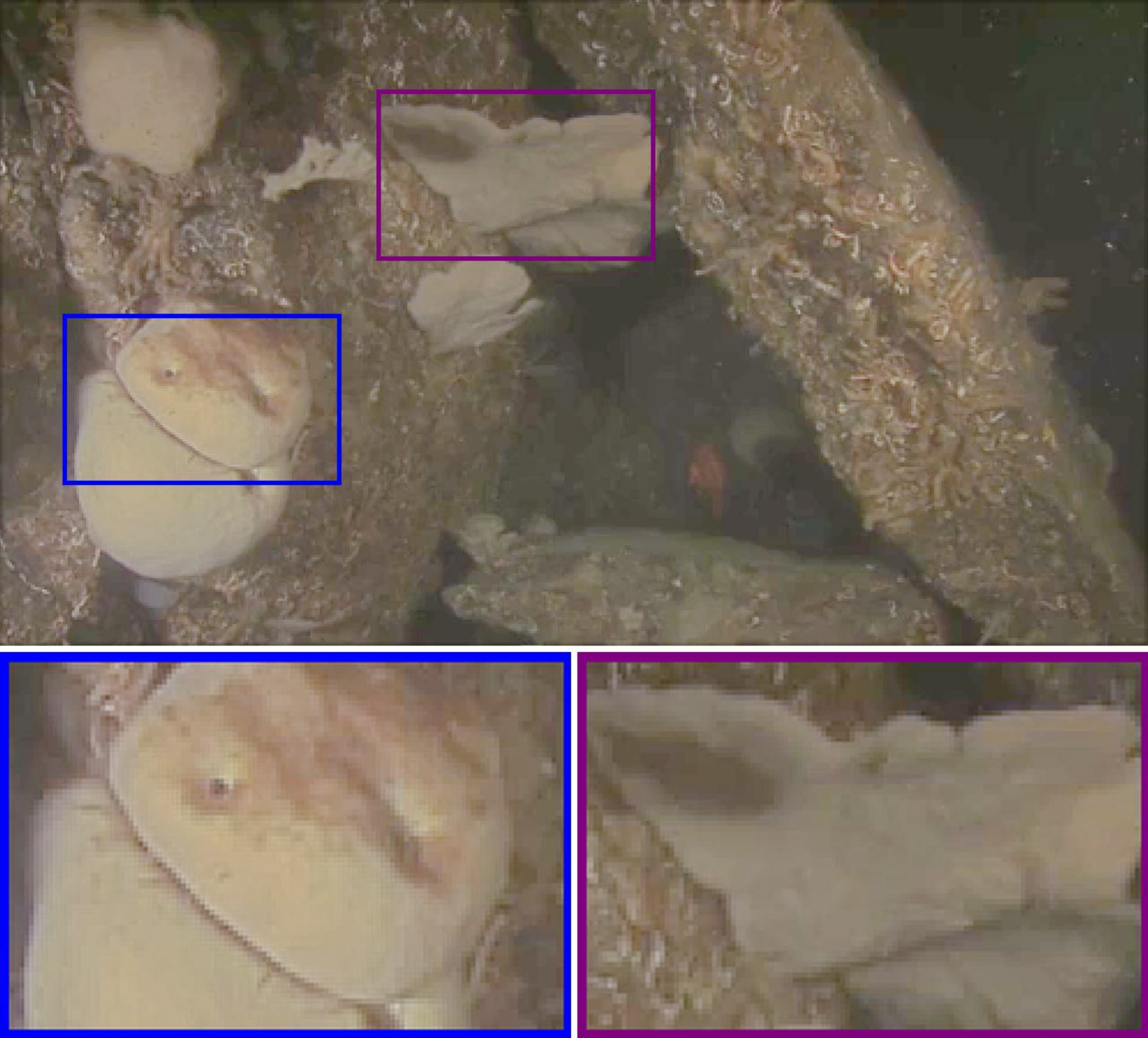} 
        \includegraphics[width=\linewidth,  height=\nuidheighttwo]{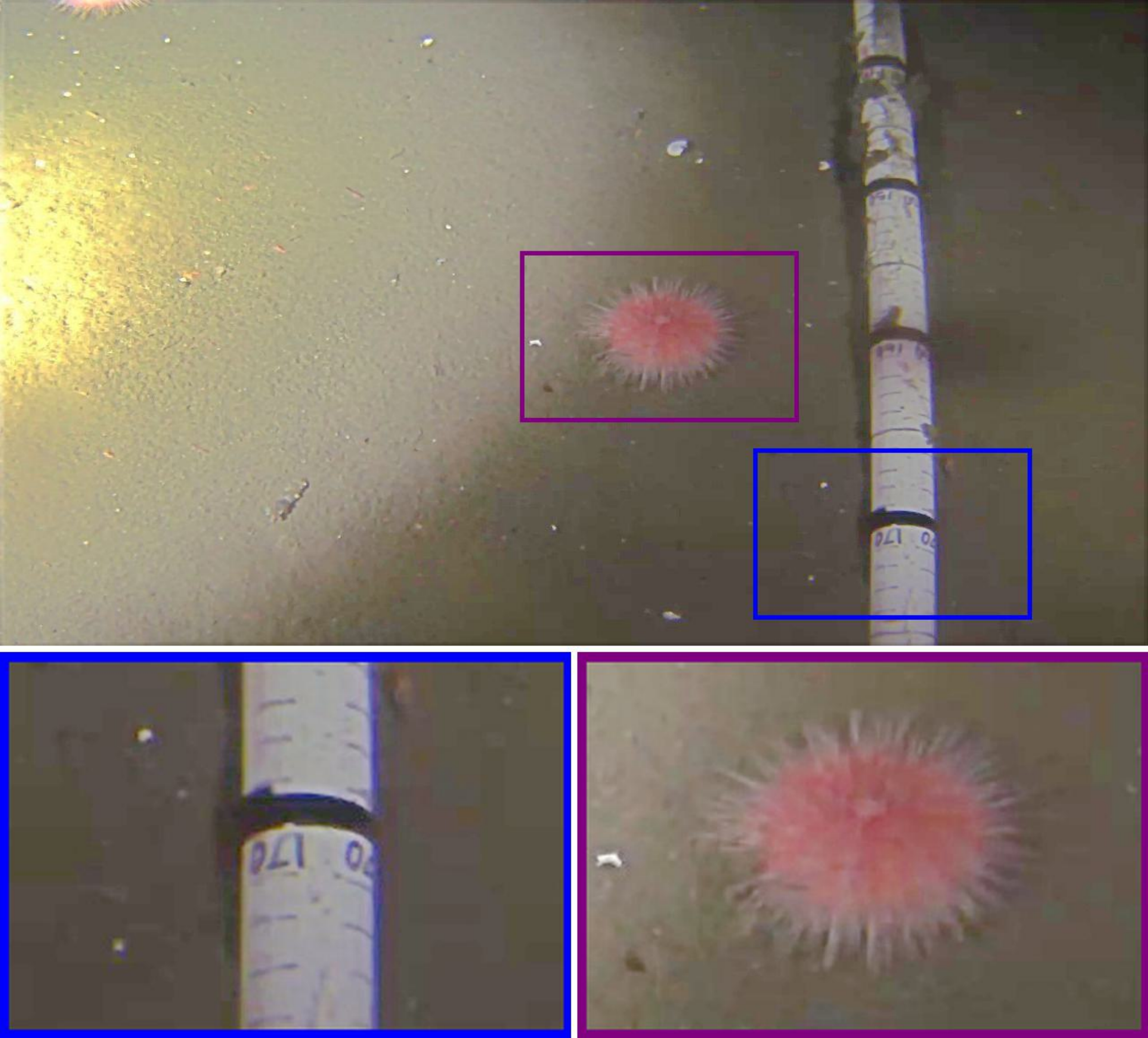} 
		\caption{\footnotesize EIB-FNDL}
	\end{subfigure}
    \begin{subfigure}{0.105\linewidth}
		\centering
		\includegraphics[width=\linewidth,  height=\nuidheighttwo]{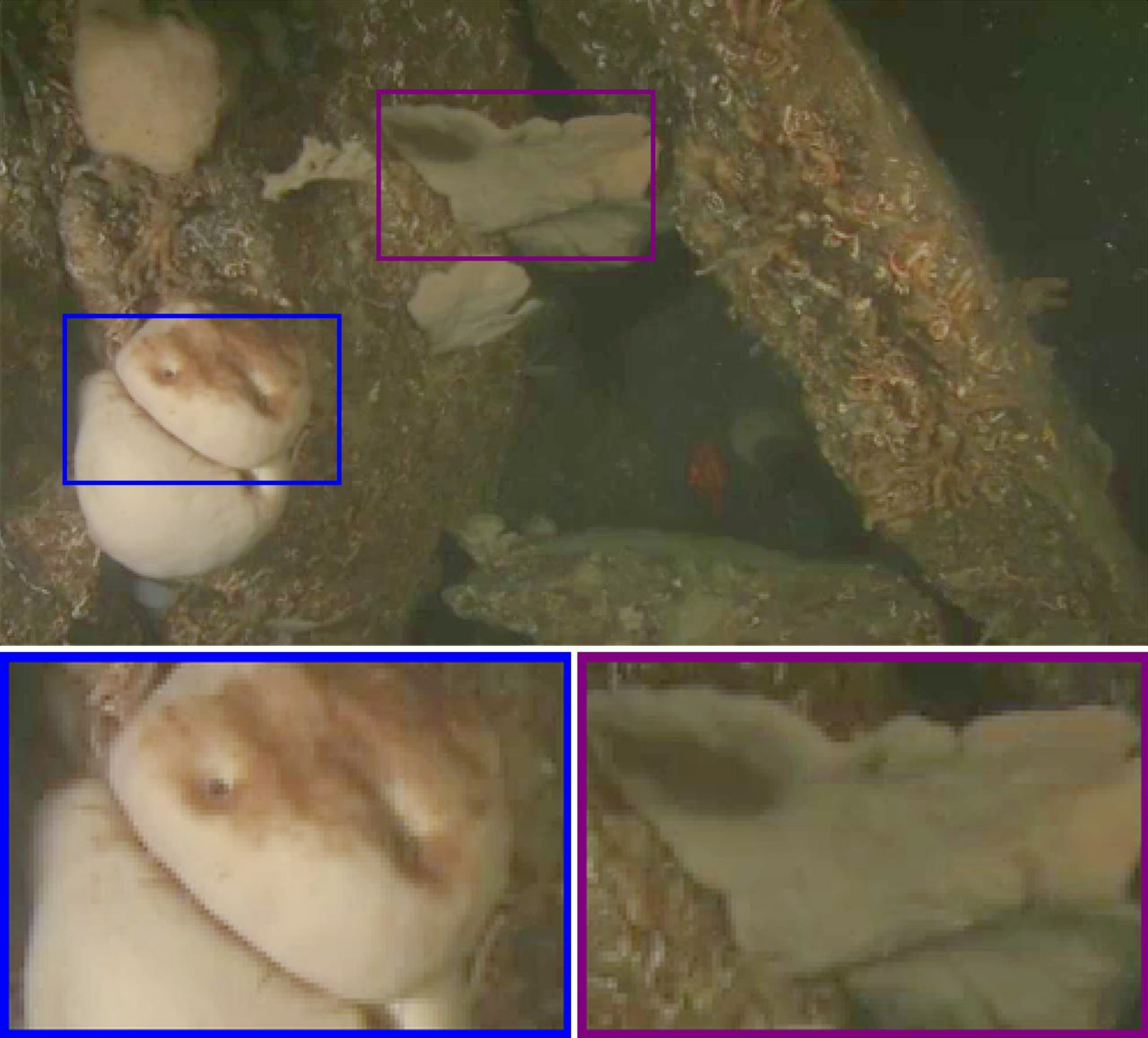} 
        \includegraphics[width=\linewidth,  height=\nuidheighttwo]{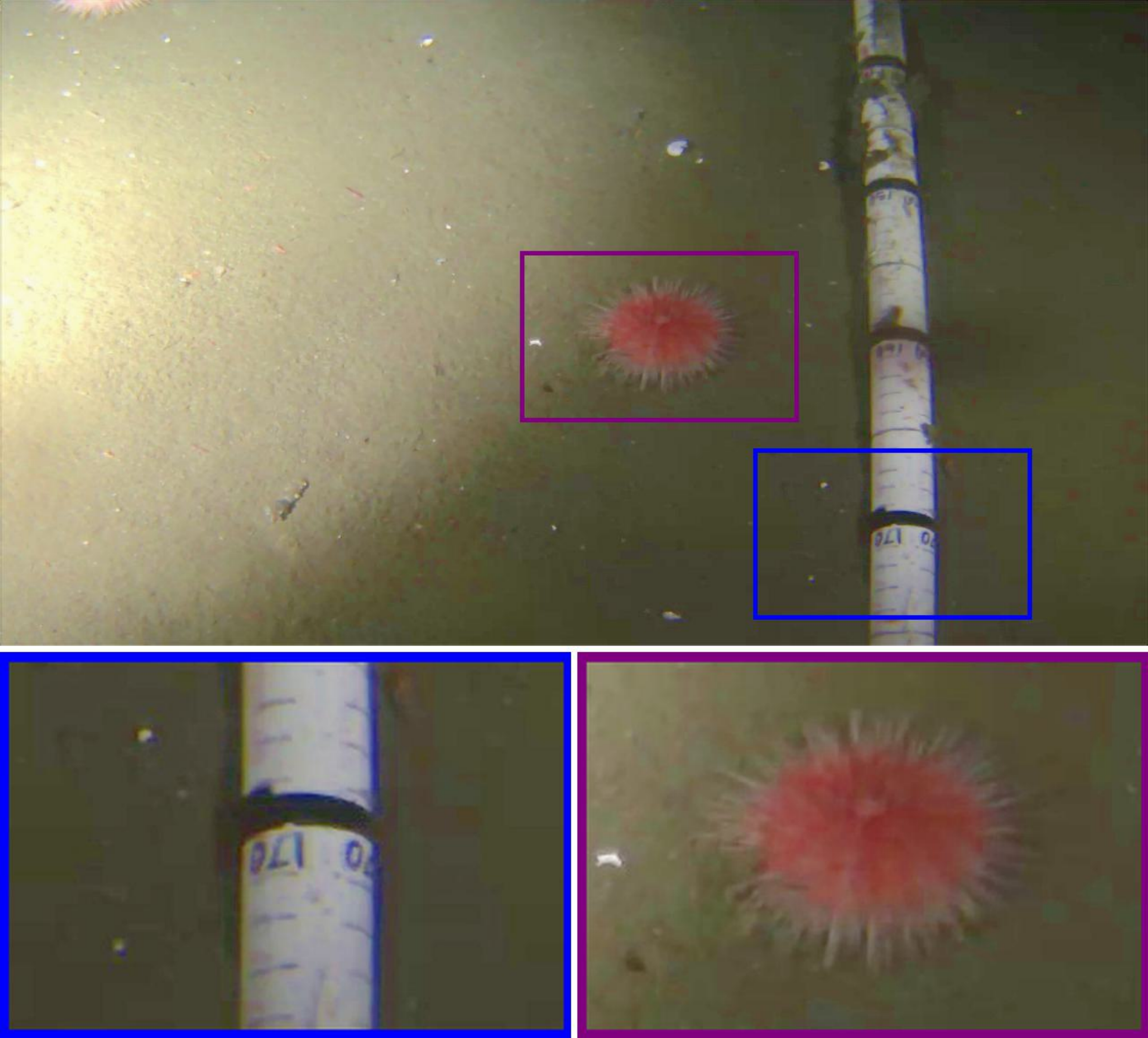} 
		\caption{\footnotesize ALEN}
	\end{subfigure}
    \begin{subfigure}{0.105\linewidth}
		\centering
		\includegraphics[width=\linewidth,  height=\nuidheighttwo]{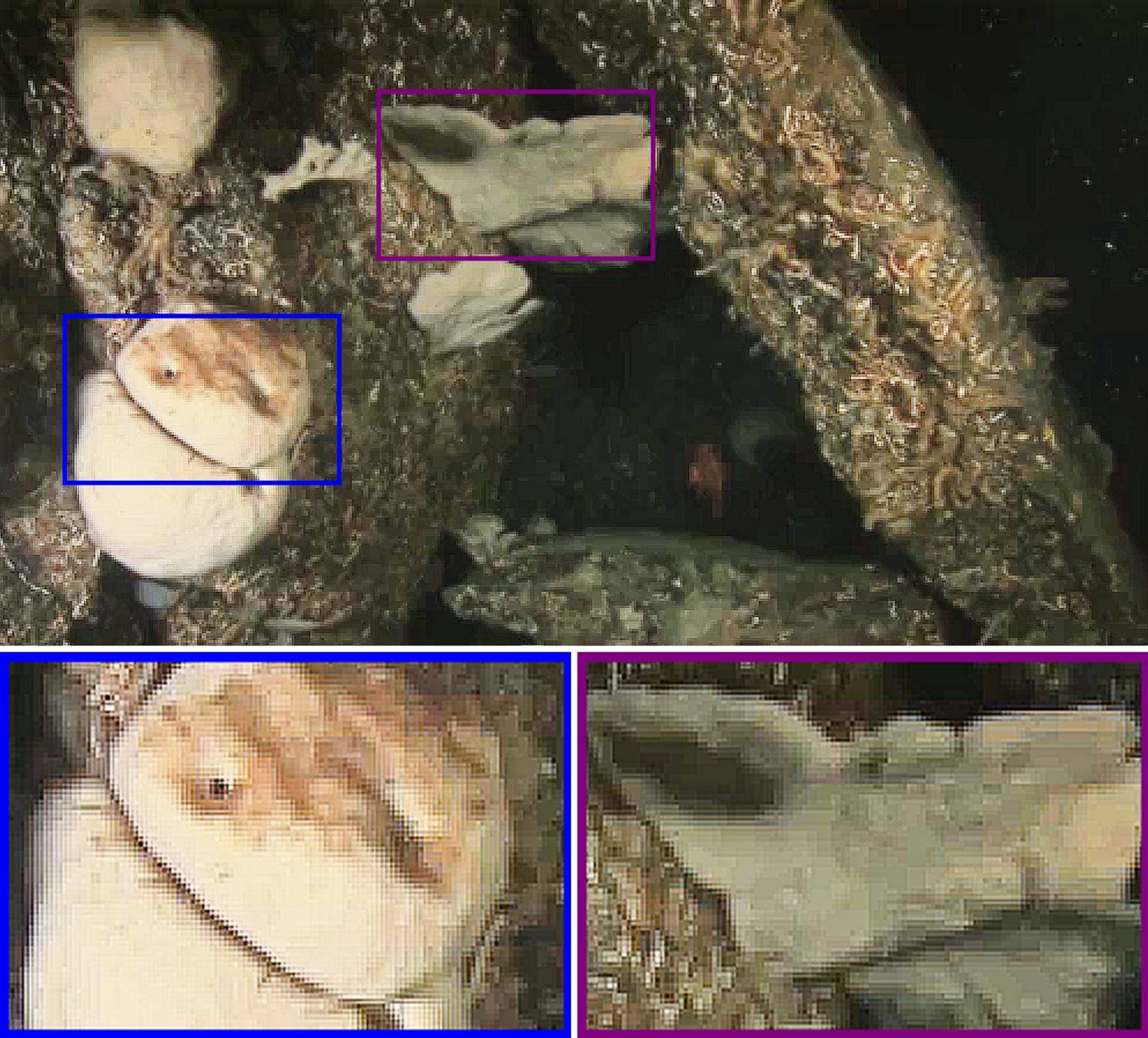} 
        \includegraphics[width=\linewidth,  height=\nuidheighttwo]{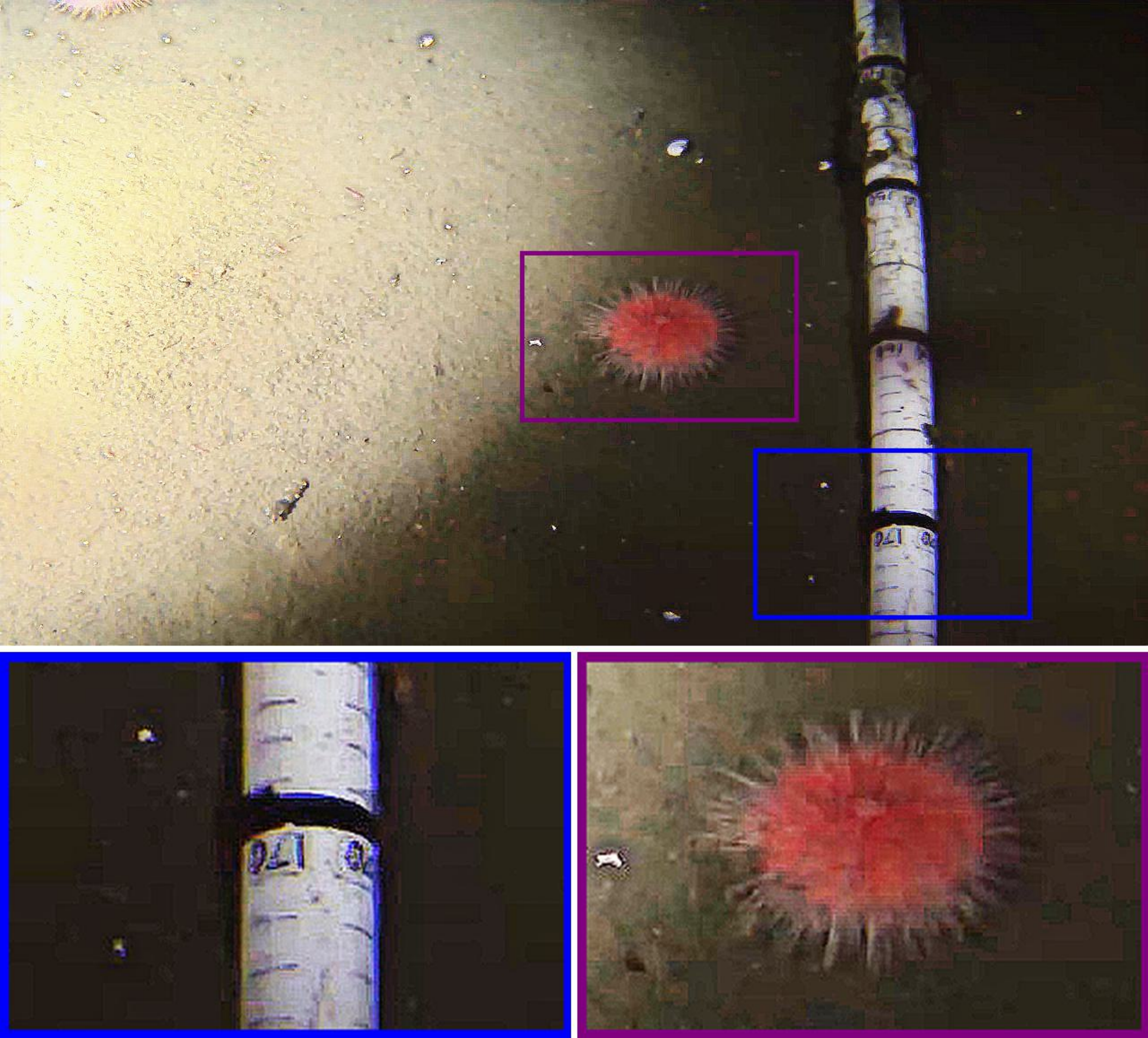} 
		\caption{\footnotesize UNIR-Net}
	\end{subfigure}
 
	\caption{Visual comparison of enhancement results on real-world underwater images from the NUID dataset, captured under night-time low-light conditions. Subfigure (a) shows the original image affected by limited illumination. Subfigures (b) to (r) present enhancement results from various methods applied to the same input image. The figure illustrates the improvements in illumination, color fidelity, and overall visual clarity achieved by each method under challenging night-time settings.}
	\label{Q3}
\end{figure*}

Figures \ref{Q2} and \ref{Q3}, corresponding to day-time and night-time underwater scenes from the NUID dataset respectively, illustrate the performance of various methods under different lighting conditions. In daytime scenarios (Fig. \ref{Q2}), methods such as ICSP and GCP achieve significant improvements in illumination; however, they tend to oversaturate green tones. Similarly, UDnet, EIB-FNDL, and ALEN enhance illumination but at the cost of a noticeable loss in contrast. UNTV, UWNet, and HFM offer greater accuracy in color reproduction, although their illumination enhancement remains limited. Methods like PCDE, MMLE, and SMDR-IS produce images with persistent dark regions, compromising overall visibility. LENet, TCTL-Net, MACT, and ACDC improve illumination, but the resulting colors tend to appear dull. UDAformer strikes a reasonable balance between illumination and color accuracy, though this comes at the expense of edge sharpness. In this context, UNIR-Net stands out by consistently producing well-illuminated images with vibrant colors and clearly defined edges.

In night-time scenes (Fig. \ref{Q3}), methods such as LENet, MACT, EIB-FNDL, and ALEN improve overall illumination, yet introduce a noticeable haze effect across the entire image. In contrast, UWNet, TCTL-Net, and UDnet enhance illumination but significantly compromise color fidelity. UDAformer and ACDC yield moderate improvements in illumination, although contrast remains subdued. UNTV, MMLE, PCDE, and SMDR-IS show limited capability to enhance illumination, leaving dark areas in the processed images. ICSP, in particular, tends to over-illuminate already well-lit regions, leading to a loss of visual detail. By contrast, HFM, GCP, and UNIR-Net deliver the most favorable results in nighttime conditions, especially regarding illumination and detail preservation. Notably, UNIR-Net excels due to its ability to produce vivid colors, consistent lighting, and sharply defined edges, outperforming other evaluated methods under both lighting scenarios.

The qualitative analysis confirms that UNIR-Net delivers outstanding performance across diverse underwater scenarios characterized by challenging lighting conditions. Its key strengths include effective illumination enhancement, preservation of vibrant colors, and precise edge definition, which are critical attributes for the visual restoration of underwater images affected by non-uniform illumination.

\subsubsection{Quantitative Results}
Table \ref{PUNI_METRICS} presents a quantitative evaluation of the PUNI dataset using various image quality metrics. UNIR-Net demonstrates superior performance by achieving the highest values in PSNR, SSIM, and UQI, along with the lowest values in LPIPS and DeltaE. This indicates that UNIR-Net produces images with greater fidelity to the original data and enhanced visual perception. In the NIQA metrics, UNIR-Net also excels, achieving the highest values in UCIQE and NIQMC, as well as the lowest value in FADE. This highlights its ability to generate more natural images with better visual balance. However, in the MUSIQ metric, the GCP method ranks first.

\begin{table*}[!ht]
	\centering
	\caption{Quantitative Comparison on the PUNI Dataset.}
	\label{PUNI_METRICS}
	\resizebox{1\textwidth}{!}{
		\begin{tabular}{l c c c c c c c c c c}
			\hline
            \multirow{2}{*}{Method} & \multirow{2}{*}{Year} & \multicolumn{5}{c}{FIQA} & \multicolumn{4}{c}{NIQA}\\ 
			\cmidrule(lr){3-7}  \cmidrule(lr){8-11}
			&  & PSNR$\uparrow$ & SSIM$\uparrow$ & UQI$\uparrow$ & LPIPS$\downarrow$ & DeltaE$\downarrow$ & UCIQE$\uparrow$ & FADE$\downarrow$ & NIQMC$\uparrow$ & MUSIQ$\uparrow$ \\ \hline

                UNTV~\cite{xie2021variational} & 2021 & 10.4148	& 0.4045 & 0.5619 &	0.3176 & 33.8359 & 0.6503 & 0.1572 & 4.9717 & 62.5657\\
                UWNet~\cite{naik2021shallow} & 2021 & 9.8603 & 0.3162 & 0.4581 & 0.5920 & 37.1908 & 0.5680 & 0.2945 & 4.2854 & 58.2181\\
                
                ACDC~\cite{zhang2022underwater} & 2022 & 13.6405 & 0.5785 & 0.7954 & 0.4190	& 22.5252 & 0.6098 & 0.2773	& 5.4007 & 66.1372\\
                MMLE~\cite{zhang2022underwaterMMLE}  & 2022 & 13.9459 & 0.6322 & 0.7019	&	0.2588	&	22.1338 & 0.6612 & 0.1868 & 5.2876 & 66.0242\\

                TCTL-Net~\cite{li2023tctl} & 2023 & 12.6514	& 0.4649 & 0.7516 &	0.5790 & 24.9678 & 0.6188 & 0.3868 & 5.2356 & 56.5916\\
                ICSP~\cite{hou2023non} & 2023 & 13.6869	&	0.7042 & 0.7723	&	0.2418	&	21.1913 & 0.6820 &	0.1534 & 5.3652 & 65.6839\\
                PCDE~\cite{zhang2023PCDE} & 2023 & 12.1812 & 0.4795 & 0.5977 & 0.4273 &	28.4354 & 0.6556 & 0.1550 & 5.0770 & 62.7529\\
                UDAformer~\cite{shen2023udaformer} & 2023 & 14.2730	& 0.6531 & 0.8211 & 0.3907 & 21.1077 & 0.6519 & 0.2362 & 5.4164 & 66.6605\\
                
                HFM~\cite{an2024hfm} & 2024 & 11.8410 & 0.4530 & 0.6564	& 0.4353 & 26.9411 & 0.6227 & 0.2590 & 5.1111 & 64.1482\\
                LENet~\cite{zhang2024liteenhancenet} & 2024 & 13.6257 & 0.5669 & 0.7913	&	0.4283	&	24.1685 & 0.6028	&	0.3749	&	5.1782 & 65.5047\\
                SMDR-IS~\cite{zhang2024synergistic} & 2024 & 11.7681 & 0.5037 & 0.6333	&	0.4327	&	28.8656 & 0.6148 & 0.2523 & 4.8751 & 65.3956\\
                GCP~\cite{jeon2024low} & 2024 & 14.1003	& 0.6833 & 0.7786	&	0.3137	&	22.3167 & 0.6765	&	0.2212	&	5.3022 & \textbf{68.2599}\\

                MACT~\cite{zhang2025mact}  & 2025 & 12.9443	&	0.5208 & 0.7494	& 0.4143 & 27.5462 & 0.6051 & 0.5072 & 5.1924 & 64.7613\\
                UDnet~\cite{saleh2025adaptive}  & 2025 & 12.0883 & 0.4296 & 0.7332 & 0.4926 & 27.7778 & 0.5604	&	0.4747	& 4.5613 & 65.7620\\
                EIB-FNDL~\cite{khajehvandi2025enhancing} & 2025 & 13.5161 & 0.6349 & 0.7743	& 0.3954 & 23.7709 & 0.6171	&	0.2616	&	4.9389 & 67.3659\\
                ALEN~\cite{perez2025alen} & 2025 & 13.3469 & 0.6235 & 0.7637 & 0.4162 & 24.2324 & 0.6372 &	0.2696	&	5.0879 & 68.2030\\

                UNIR-Net & 2025 & \textbf{19.7054} & \textbf{0.8771} & \textbf{0.8903}	&	\textbf{0.0667}	&	\textbf{11.0012} & \textbf{0.7082} &	\textbf{0.1466} & \textbf{5.5852} & 67.6252\\ \hline

		\end{tabular}
	}
\end{table*}

Meanwhile, Tables \ref{NUI_METRICS_I} and \ref{NUI_METRICS_II} detail the results obtained on the NUID dataset, evaluated using UCIQE, FADE, NIQMC, and MUSIQ metrics. The metrics were calculated for different subsets of the dataset, including Enhanced Underwater Visual Perception (EUVP), Google Image (GI), Nature Footage (NF), OceanDark (OD), and Underwater Image Enhancement Benchmark (UIEB), as well as the overall average. In this evaluation, UNIR-Net consistently shows the best average performance, standing out with the highest average values in UCIQE, NIQMC, and MUSIQ and the lowest average value in FADE. This confirms that UNIR-Net not only enhances the visual quality of real-world underwater images but also preserves a natural appearance, significantly outperforming current state-of-the-art methods.

\begin{table*}[!ht]
	\centering
	\caption{Comparison results on NUID underwater image dataset regarding UCIQE and FADE. }
	\label{NUI_METRICS_I}
	\resizebox{1\textwidth}{!}{
		\begin{tabular}{l c c c c c |c| c c c c c |c|}
			\hline
			\multirow{2}{*}{Method}  & \multicolumn{6}{c}{UCIQE$\uparrow$} & \multicolumn{6}{c}{FADE$\downarrow$}\\
                \cmidrule(lr){2-7}  \cmidrule(lr){8-13}  
                & EUVP & GI & NF & OD & UIEB & Average & EUVP & GI & NF & OD & UIEB & Average \\ \hline

                UNTV~\cite{xie2021variational} & 0.6270	&	0.6015	&	0.5581	&	0.5264	&	0.5947	&	0.5815 &	0.1941	&	0.4732	&	0.6159	&	0.8184	&	0.4816	&	0.5166\\
                UWNet~\cite{naik2021shallow}  & 0.5801	&	0.5484	&	0.5084	&	0.5715	&	0.5572	&	0.5531 &	0.3520	&	0.8651	&	1.1000	&	1.2336	&	0.8571	&	0.8816\\
                
                ACDC~\cite{zhang2022underwater}  & 0.5769	&	0.5604	&	0.5651	&	0.5315	&	0.5713	&	0.5610 &	0.3061	&	0.6949	&	1.3274	&	1.1725	&	0.7273	&	0.8456\\
                MMLE~\cite{zhang2022underwaterMMLE} & 0.6224	&	0.6049	&	0.5833	&	0.5660	&	0.6095	&	0.5972 &	0.1978	&	0.5162	&	0.7549	&	0.8515	&	0.4535	&	0.5548\\
                
                TCTL-Net~\cite{li2023tctl} & 0.5944	&	0.5915	&	0.5798	&	0.5437	&	0.6128	&	0.5844 &	0.4635	&	0.8790	&	1.4949	&	1.7122	&	0.8383	&	1.0776\\
                ICSP~\cite{hou2023non}  & 0.6471	&	0.6035	&	0.5815	&	0.5857	&	0.6205	&	0.6077 &	0.1883	&	0.4263	&	0.7367	&	0.9540	&	0.3949	&	0.5400\\
                PCDE~\cite{zhang2023PCDE}  & 0.6288	&	0.6127	& 0.5891	&	0.5670	&	0.6201	&	0.6035 &	0.1593	&	0.5398	&	\textbf{0.5552}	&	0.9016	&	0.3817	&	0.5075\\
                UDAformer~\cite{shen2023udaformer}  & 0.6225	&	0.6059	&	0.5765	&	0.5767	&	0.6271	&	0.6017  &	0.2888	&	0.6084	&	1.0143	&	1.1073	&	0.6145	&	0.7267\\
                
                HFM~\cite{an2024hfm}  & 0.6106	&	0.5942	&	0.5640	&	0.5686	&	0.5970	&	0.5869 &	0.3073	&	0.7005	&	1.1409	&	1.3086	&	0.7484	&	0.8411\\
                LENet~\cite{zhang2024liteenhancenet}  & 0.5949	&	0.5818	&	0.5304	&	0.5659	&	0.5937	&	0.5733 &	0.3850	&	0.7831	&	1.7562	&	1.4539	&	0.8541	&	1.0465\\
                SMDR-IS~\cite{zhang2024synergistic}  & 0.6130	&	0.5926	&	0.5628	&	0.5696	&	0.6083	&	0.5893 &	0.2910	&	0.5457	&	0.9054	&	0.9198	&	0.7146	&	0.6753\\
                GCP~\cite{jeon2024low} & 0.6272	&	0.5964	&	0.6019	&	0.5762	&	0.6137	&	0.6031 &	0.2567	&	0.6198	&	0.8505	&	1.1737	&	0.6053	&	0.7012\\

                MACT~\cite{zhang2025mact}  & 0.5406 & 0.5404	& 0.5090 & 0.5139 &	0.5515 & 0.5311 & 0.7384 & 1.5471 &	2.7188 & 2.3283	& 1.5328 & 1.7731\\
                UDnet~\cite{saleh2025adaptive}  & 0.5505	& 0.5373 & 0.4963 &	0.5235 & 0.5496	& 0.5314 & 0.5177 &	1.0008 & 1.6491	& 1.6886 & 0.9921 &	1.1697\\
                EIB-FNDL~\cite{khajehvandi2025enhancing}  & 0.5551 & 0.5080	& 0.5206 & 0.5161 &	0.5306 & 0.5261 & 0.3129 & 0.8620 &	1.1849 & 1.4882	& 0.7775 & 0.9251\\
                ALEN~\cite{perez2025alen}  & 0.5652 & 0.5285	& 0.5437 & 0.5357 &	0.5504 & 0.5447 & 0.3097 & 0.7523 &	1.0952 & 1.3102	& 0.7296 & 0.8394\\
                
                UNIR-Net & \textbf{0.6533}	&	\textbf{0.6232}	&	\textbf{0.6039}	&	\textbf{0.6001}	&	\textbf{0.6323}	&	\textbf{0.6225} &  \textbf{0.1572}	&	\textbf{0.3085}	&	0.6656	&	\textbf{0.6363}	&	\textbf{0.3245}	&	\textbf{0.4184} \\ \hline

		\end{tabular}
	}
\end{table*}

\begin{table*}[!ht]
	\centering
	\caption{Comparison results on the NUID underwater image dataset regarding NIQMC and MUSIQ. }
	\label{NUI_METRICS_II}
	\resizebox{1\textwidth}{!}{
		\begin{tabular}{l c c c c c |c| c c c c c |c|}
			\hline
			\multirow{2}{*}{Method}  & \multicolumn{6}{c}{NIQMC$\uparrow$} & \multicolumn{6}{c}{MUSIQ$\uparrow$}\\
                \cmidrule(lr){2-7}  \cmidrule(lr){8-13}  
                & EUVP & GI & NF & OD & UIEB & Average & EUVP & GI & NF & OD & UIEB & Average \\ \hline

                UNTV~\cite{xie2021variational} & 4.9137	&	4.8215	&	4.3360	&	4.8374	&	4.6429	&	4.7103 & 57.7946	&	52.7507	&	\textbf{37.9192}	&	48.8997	&	51.8534	&	49.8435\\
                UWNet~\cite{naik2021shallow}   & 4.6532	&	4.3660	&	3.7366	&	4.9869	&	4.4004	&	4.4286 & 55.6179	&	48.3772	&	32.6250	&	47.1962	&	47.4309	&	46.2494\\
                
                ACDC~\cite{zhang2022underwater}   & 5.2697	&	5.2990	&	4.9161	&	5.3163	&	5.3108	&	5.2224 & 57.7228	&	50.5946	&	31.5843	&	45.6607	&	47.0116	&	46.5148 \\
                MMLE~\cite{zhang2022underwaterMMLE} & 5.1769	&	\textbf{5.3519}	&	4.8767	&	5.3660	&	5.3106	&	5.2164 & 59.8614	&	54.0019	&	34.9565	&	48.6503	&	50.7224	&	49.6385\\

                TCTL-Net~\cite{li2023tctl} & 5.1747	&	5.1455	&	4.8806	&	5.1260	&	5.2888	&	5.1231 & 52.0720	&	41.1017	&	34.8615	&	36.0238	&	40.2606	&	40.8639\\
                ICSP~\cite{hou2023non}  & 5.2242	&	5.0117	&	4.5138	&	5.1488	&	5.2406	&	5.0278 & 59.8850	&	52.3104	&	35.4520	&	45.3621	&	51.7898	&	48.9599\\
                PCDE~\cite{zhang2023PCDE}  & 4.9899	&	5.0996	&	4.7900	&	5.4514	&	5.1401	&	5.0942 & 56.6199	&	52.0433	&	35.4325	&	44.8043	&	49.7388	&	47.7278\\
                UDAformer~\cite{shen2023udaformer}  & 5.3526	&	5.2446	&	4.9392	&	\textbf{5.5604}	&	5.3755	&	5.2945 & 58.4523	&	54.3923	&	34.9949	&	44.1347	&	50.6314	&	48.5211\\
                
                HFM~\cite{an2024hfm}  & 5.3131	&	5.2834	&	4.7723	&	5.4502	&	5.2289	&	5.2096 & 56.1166	&	50.0688	&	33.4501	&	44.2613	&	47.9797	&	46.3753\\
                LENet~\cite{zhang2024liteenhancenet}  & 5.2623	&	5.1139	&	4.6004	&	5.4952	&	5.2439	&	5.1431 & 58.4278	&	54.4434	&	35.8024	&	42.2083	&	49.7863	&	48.1336\\
                SMDR-IS~\cite{zhang2024synergistic}   & 5.1165	&	5.0517	&	4.6384	&	5.3784	&	5.2061	&	5.0782 & 57.5129 &	52.9491	&	33.3865	&	43.0757	&	48.2745	&	47.0398
\\

                GCP~\cite{jeon2024low} & 5.3321	&	5.1731	&	4.9009	&	5.4067	&	5.2463	&	5.2118 & 59.2625	&	53.0948	&	34.0865	&	42.7211	&	46.8959	&	47.2122\\
            
                MACT~\cite{zhang2025mact}  & 4.9123 & 5.1294 & 4.5730 & 5.2496 & 5.1586 & 5.0046 & 56.3356 &	48.7811	& 32.1339 &	41.2445	& 46.5650 &	45.0120 \\
                UDnet~\cite{saleh2025adaptive}  & 4.6344	& 4.5922 & 4.1616 & 5.0928 & 4.6384 & 4.6239 & 58.2981 & 53.0276 & 35.5988 & 43.3529 & 47.5595 & 47.5674\\
                EIB-FNDL~\cite{khajehvandi2025enhancing}  & 4.8105 & 4.3400 & 4.2503 & 4.6846 & 4.5493 & 4.5269 & 60.5743 & 55.3410 & 35.9677 & 47.0044 & \textbf{51.9323} & 50.1639\\
                ALEN~\cite{perez2025alen}  & 4.8038 & 4.6117 & 4.4593 & 5.0579 & 4.7064 & 4.7278 & 58.5969 &	52.2588	& 35.5355 &	43.2737	& 48.0356 &	47.5401\\
                
                UNIR-Net &  \textbf{5.4463}	&	5.3039	&	\textbf{5.0367}	&	5.5034	&	\textbf{5.4504}	&	\textbf{5.3481}	&	 \textbf{61.2277}	&	\textbf{58.4932}	&	35.8605	&	\textbf{52.3809}	&	49.8281	&	\textbf{51.5581} \\ \hline

		\end{tabular}
	}
\end{table*}

These quantitative results underscore the effectiveness of UNIR-Net, both in terms of fidelity to the original information and the naturalness of the enhanced images, establishing it as a robust solution for improving underwater images with non-uniform illumination.

\subsubsection{Comparison of training datasets in UNIR-Net}

To highlight the significance of the PUNI dataset in training, UNIR-Net is trained on various datasets. For terrestrial low-light images, the LOw-Light (LOL)~\cite{wei2018deep} and MIT-Adobe FiveK~\cite{bychkovsky2011learning} datasets are used. For underwater low-light images, the Low-light Underwater Image Enhancement (LUIE)~\cite{xie2022lighting} dataset is utilized, along with the synthetic Non-Uniform Illumination Paired Underwater
Image (NUIPUI)~\cite{ma2025pqgal} dataset, which specifically targets underwater images with non-uniform illumination.

LOL dataset includes 500 paired images with varying exposure times and ISO settings, illustrating both low and normal lighting conditions. Similarly, the MIT-Adobe FiveK dataset presents 5000 DSLR images, offering diverse scenarios of low-light environments. In the underwater domain, the LUIE dataset generates 2524 paired images derived from 362 underwater scenes, targeting the specific challenges posed by low-light conditions in underwater environments. Complementing this, the NUIPUI dataset provides 1200 real underwater images and 1200 synthesized ones, specifically addressing the complexities introduced by non-uniform illumination in underwater scenes.

\begin{figure*}[ht]
	\centering
	\begin{subfigure}{0.15\linewidth}
		\centering
        \includegraphics[width=\linewidth]{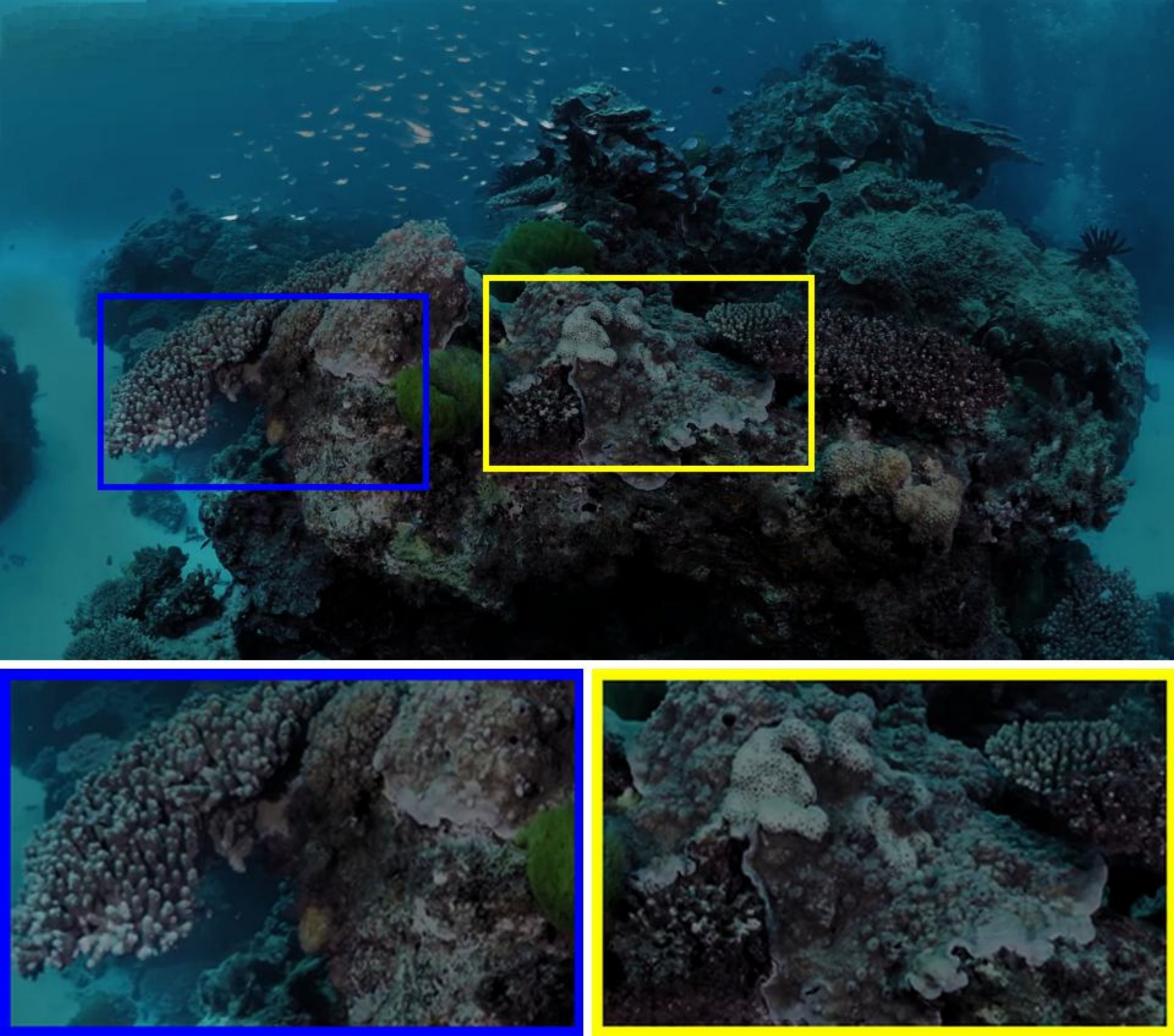} 
		\includegraphics[width=\linewidth]{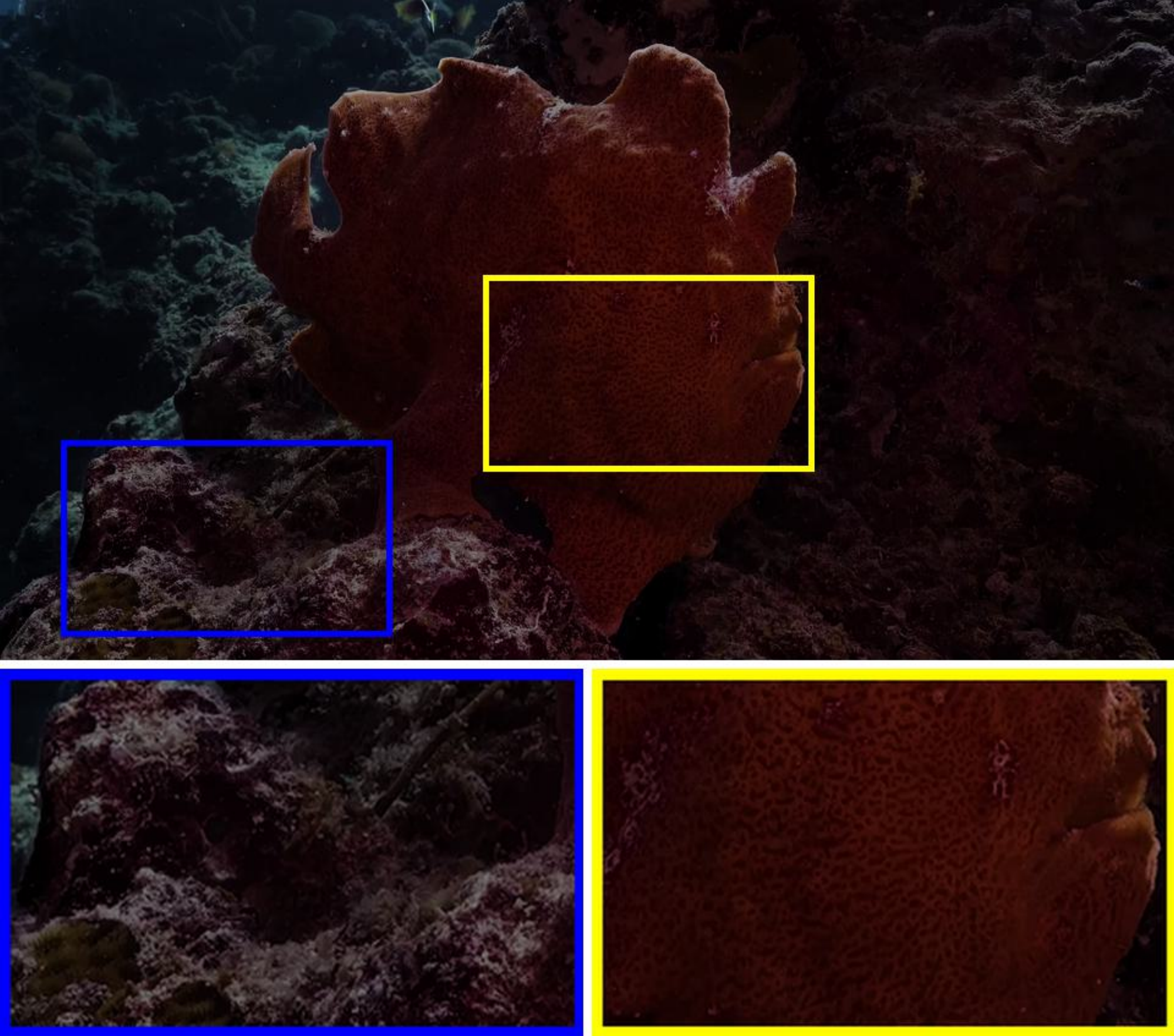} 
        \includegraphics[width=\linewidth]{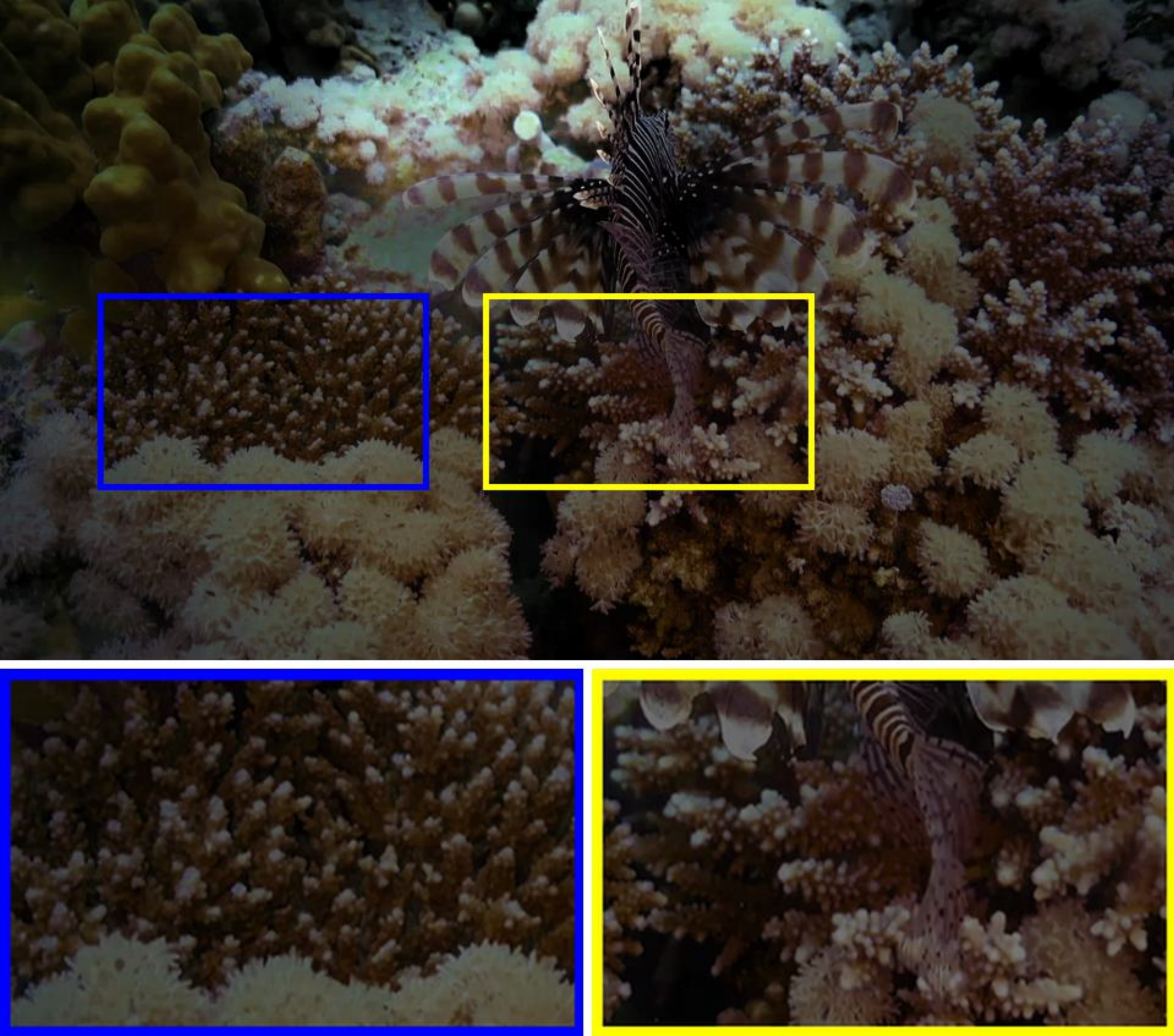} 
        \caption{\footnotesize Input}
	\end{subfigure}
	\begin{subfigure}{0.15\linewidth}
		\centering
		\includegraphics[width=\linewidth]{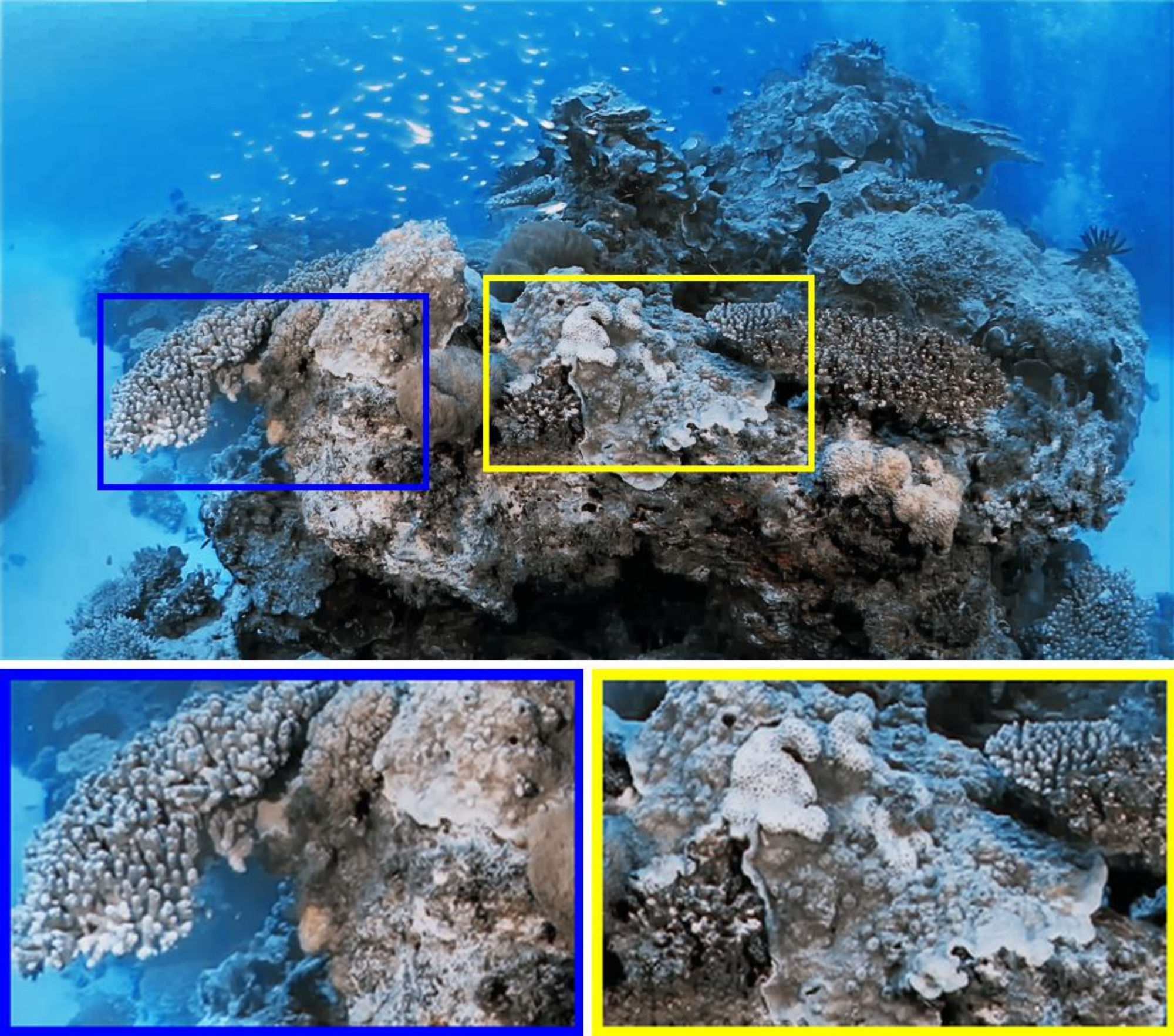}
        \includegraphics[width=\linewidth]{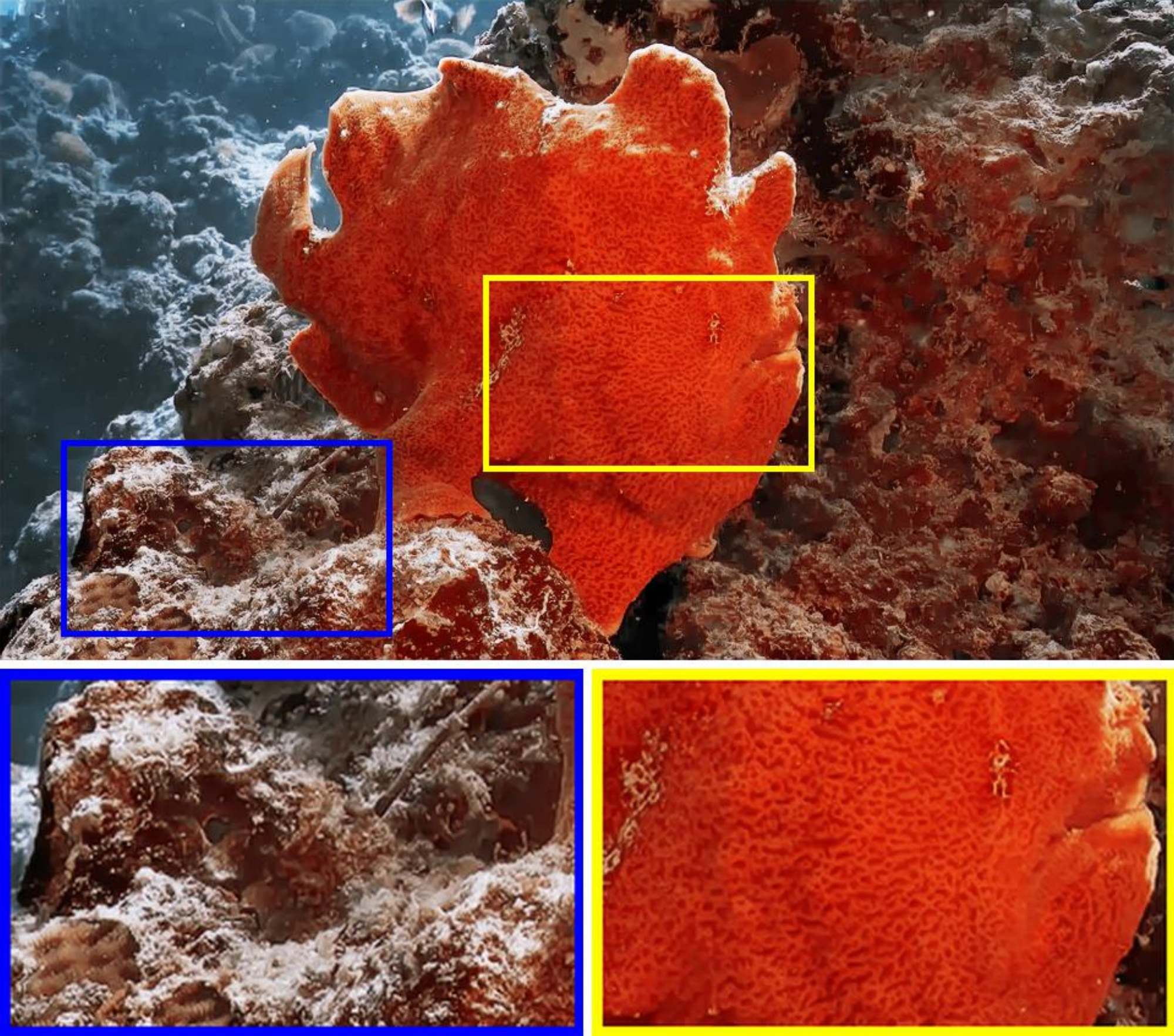} 
        \includegraphics[width=\linewidth]{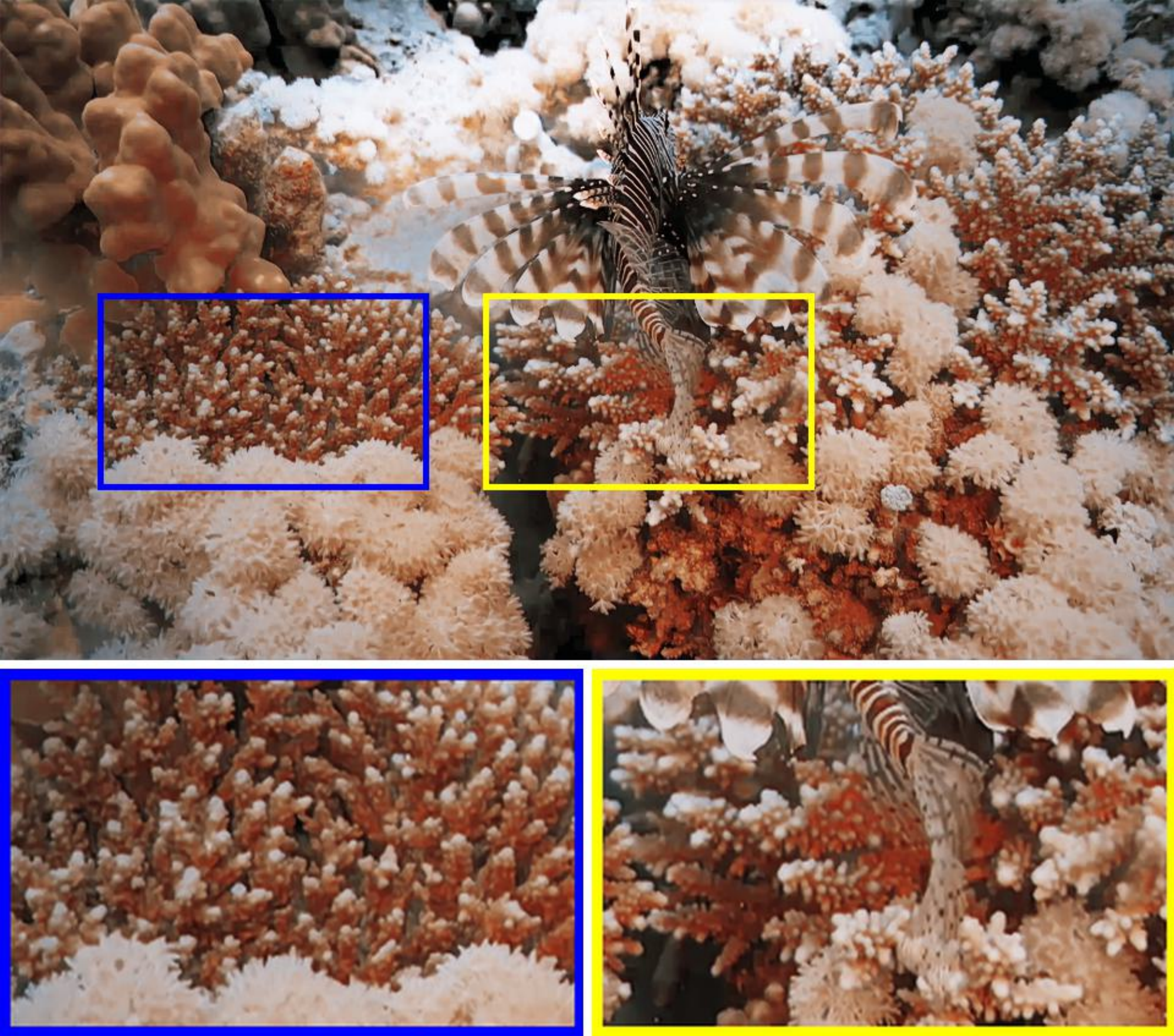} 
        \caption{\footnotesize LOL Dataset }
	\end{subfigure}
	\begin{subfigure}{0.15\linewidth}
		\centering
		\includegraphics[width=\linewidth]{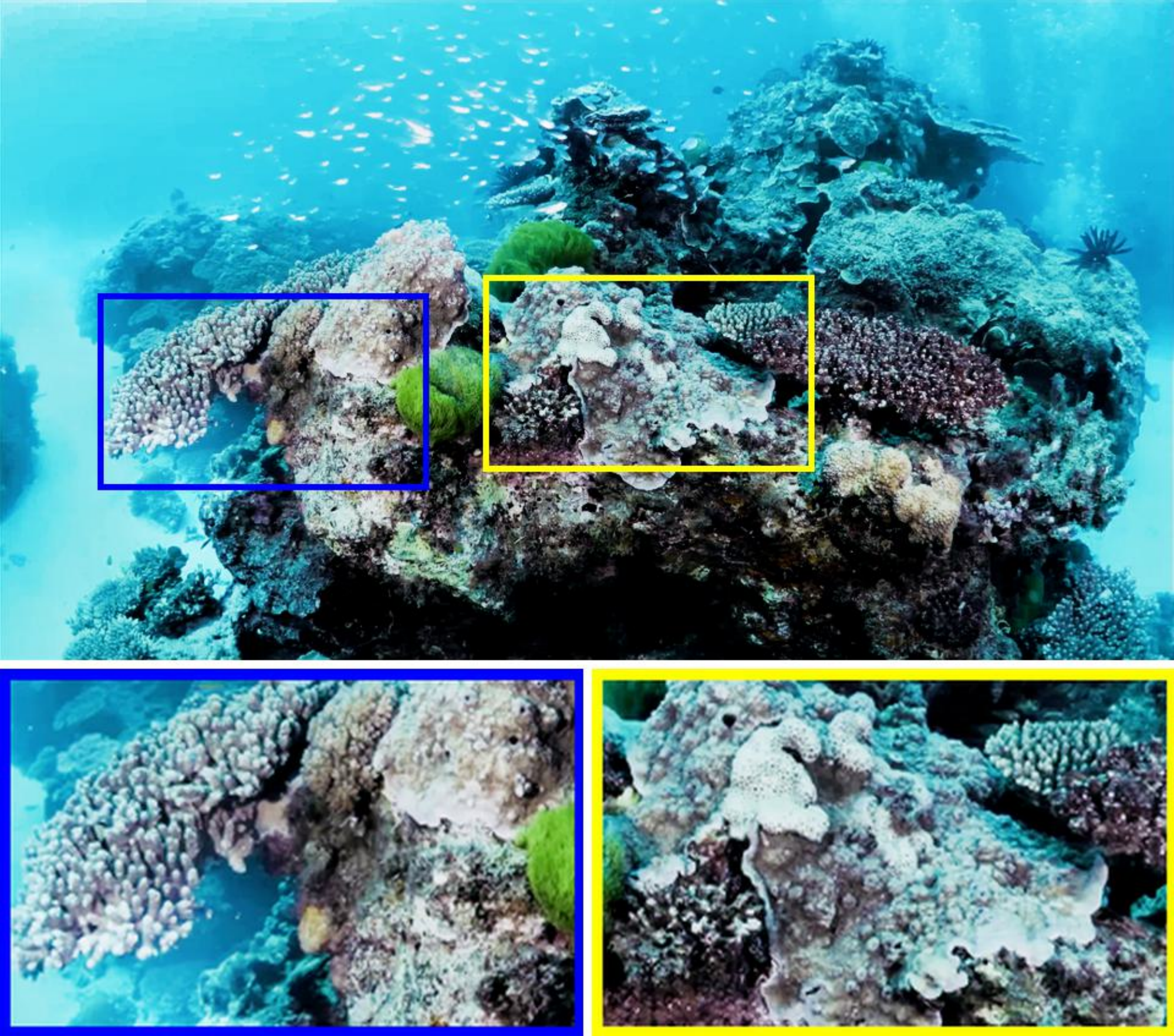}
        \includegraphics[width=\linewidth]{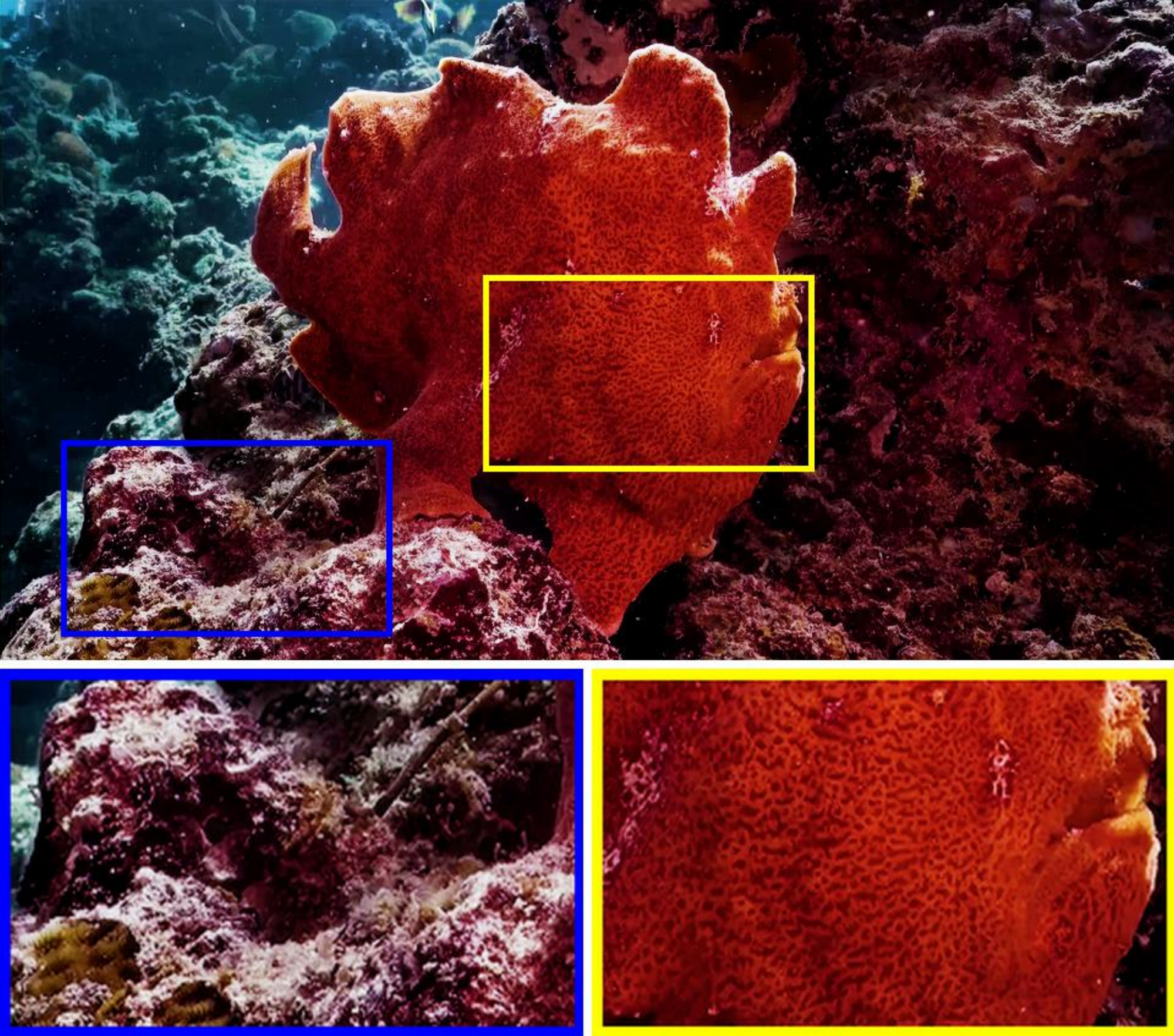}
        \includegraphics[width=\linewidth]{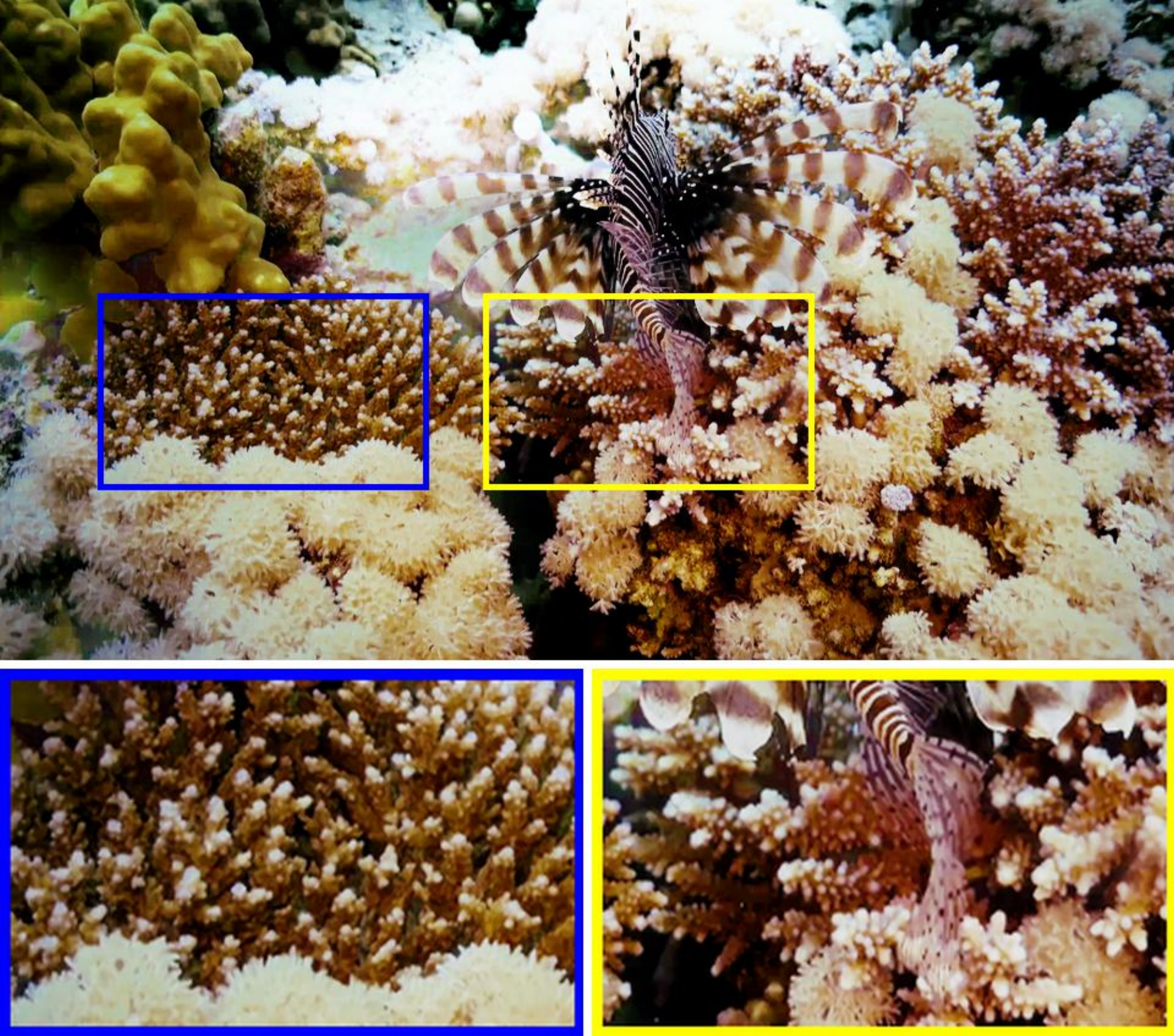} 
        \caption{\footnotesize MIT Dataset }
	\end{subfigure}
	\begin{subfigure}{0.15\linewidth}
		\centering
		\includegraphics[width=\linewidth]{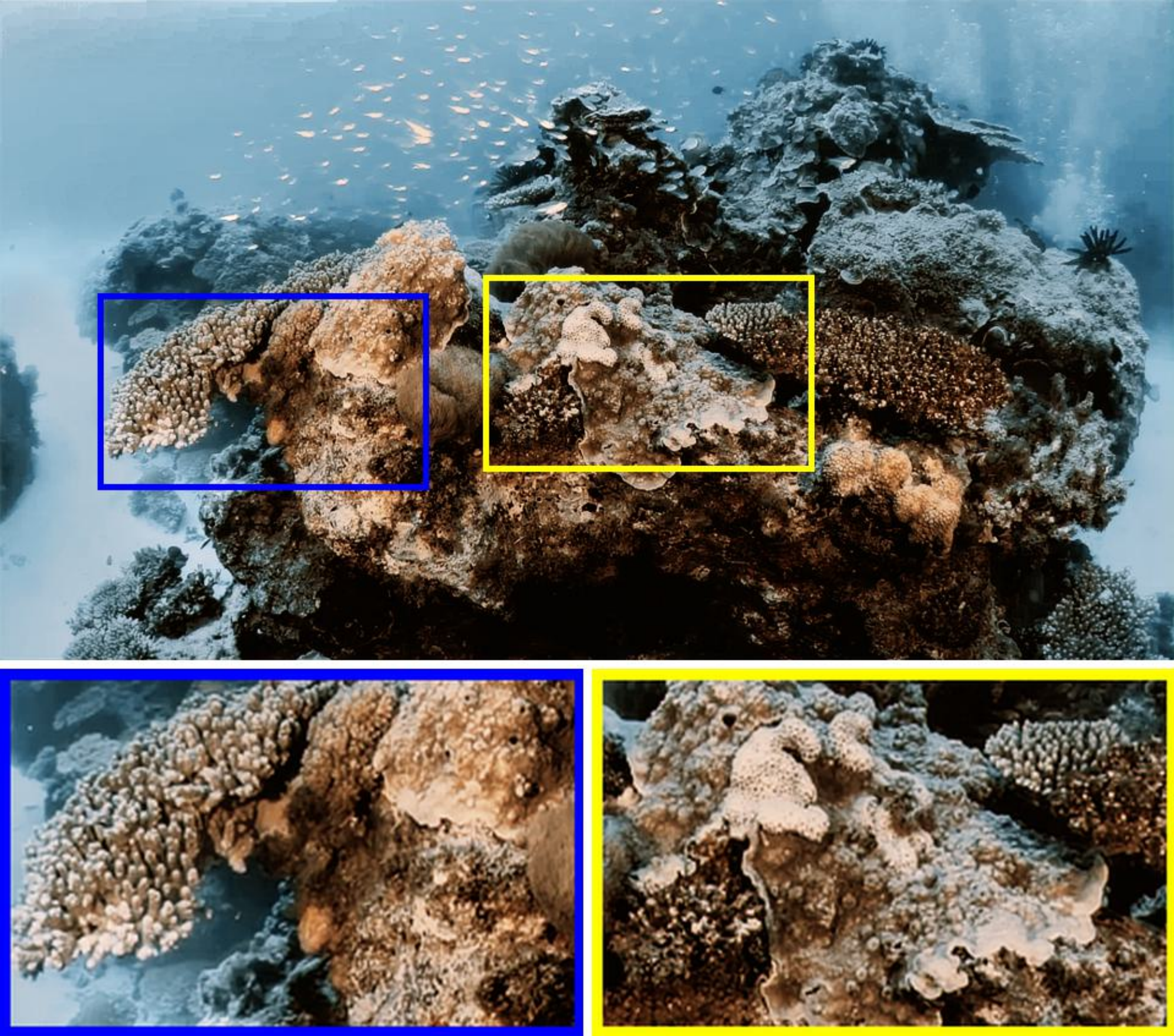} 
        \includegraphics[width=\linewidth]{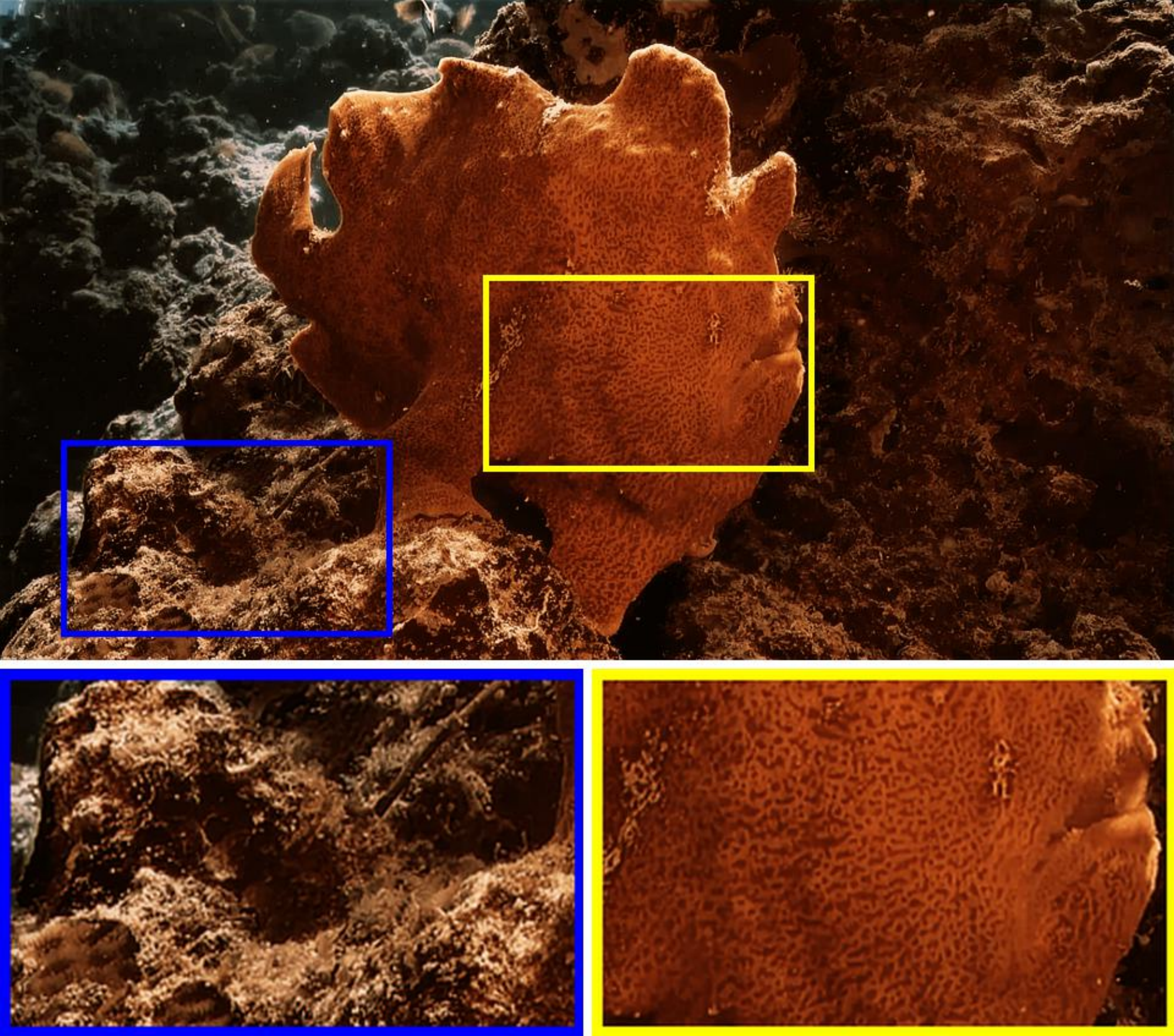} 
        \includegraphics[width=\linewidth]{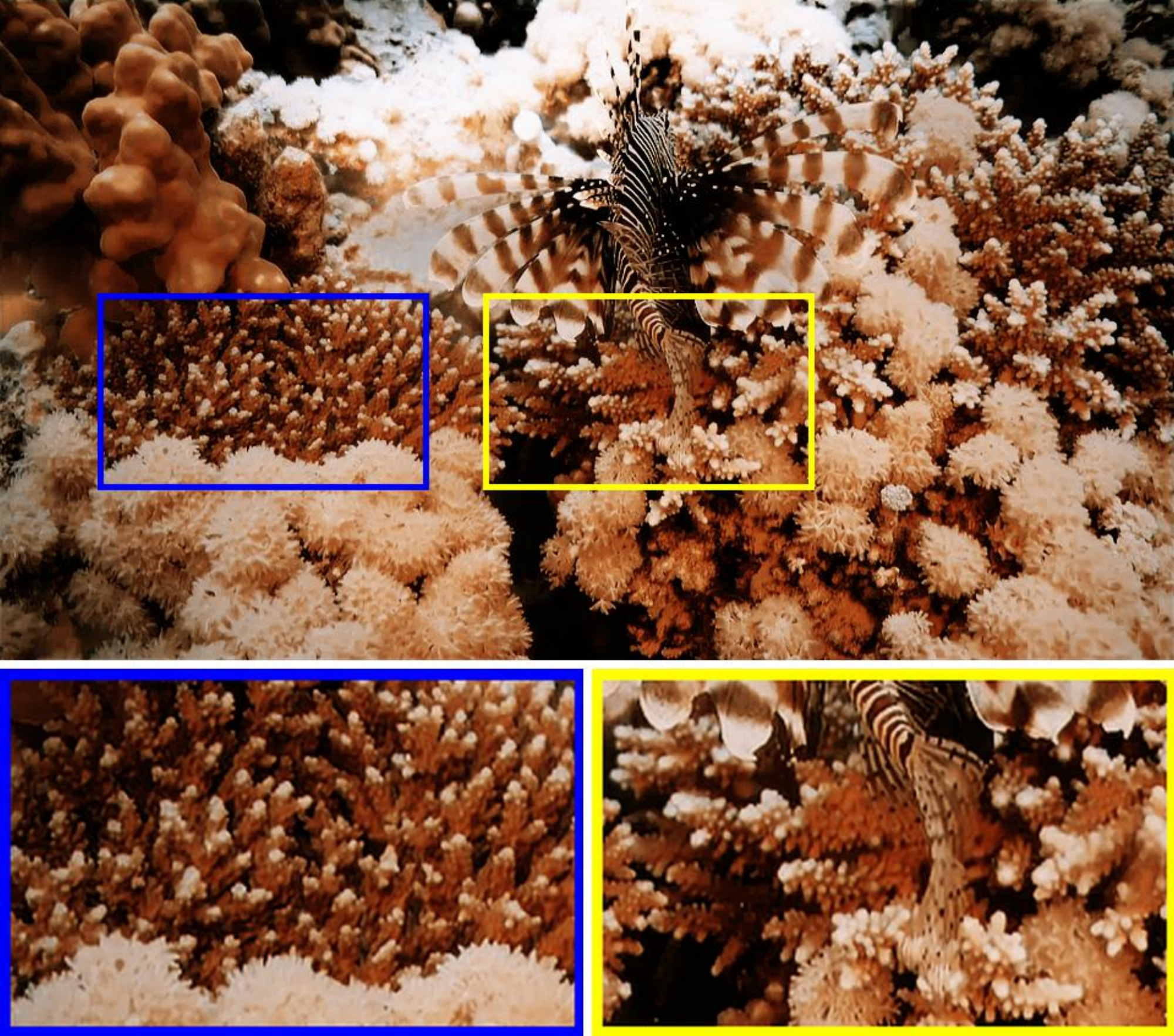} 
		\caption{\footnotesize LUIE Dataset }
	\end{subfigure}
    \begin{subfigure}{0.15\linewidth}
		\centering
		\includegraphics[width=\linewidth]{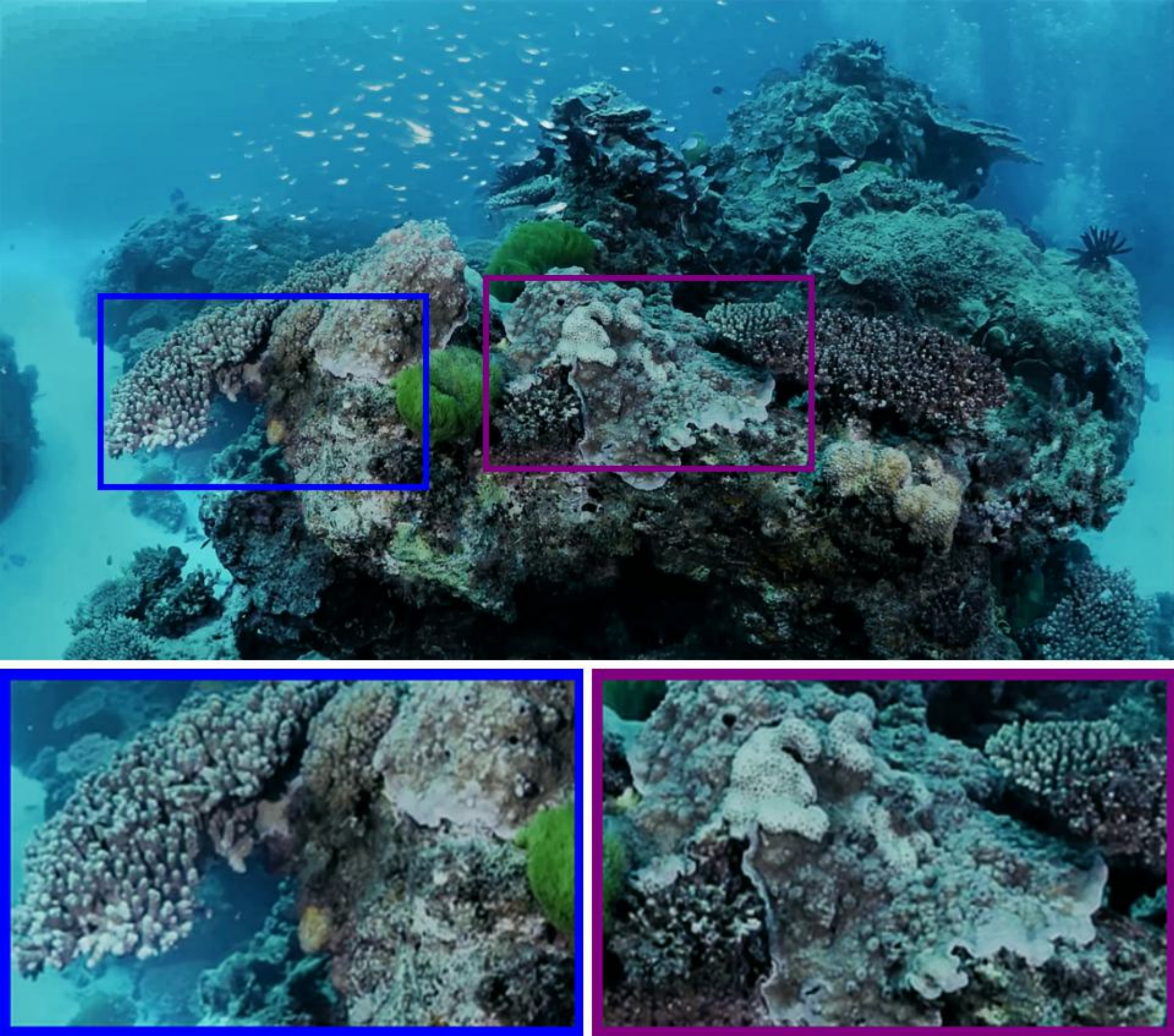} 
        \includegraphics[width=\linewidth]{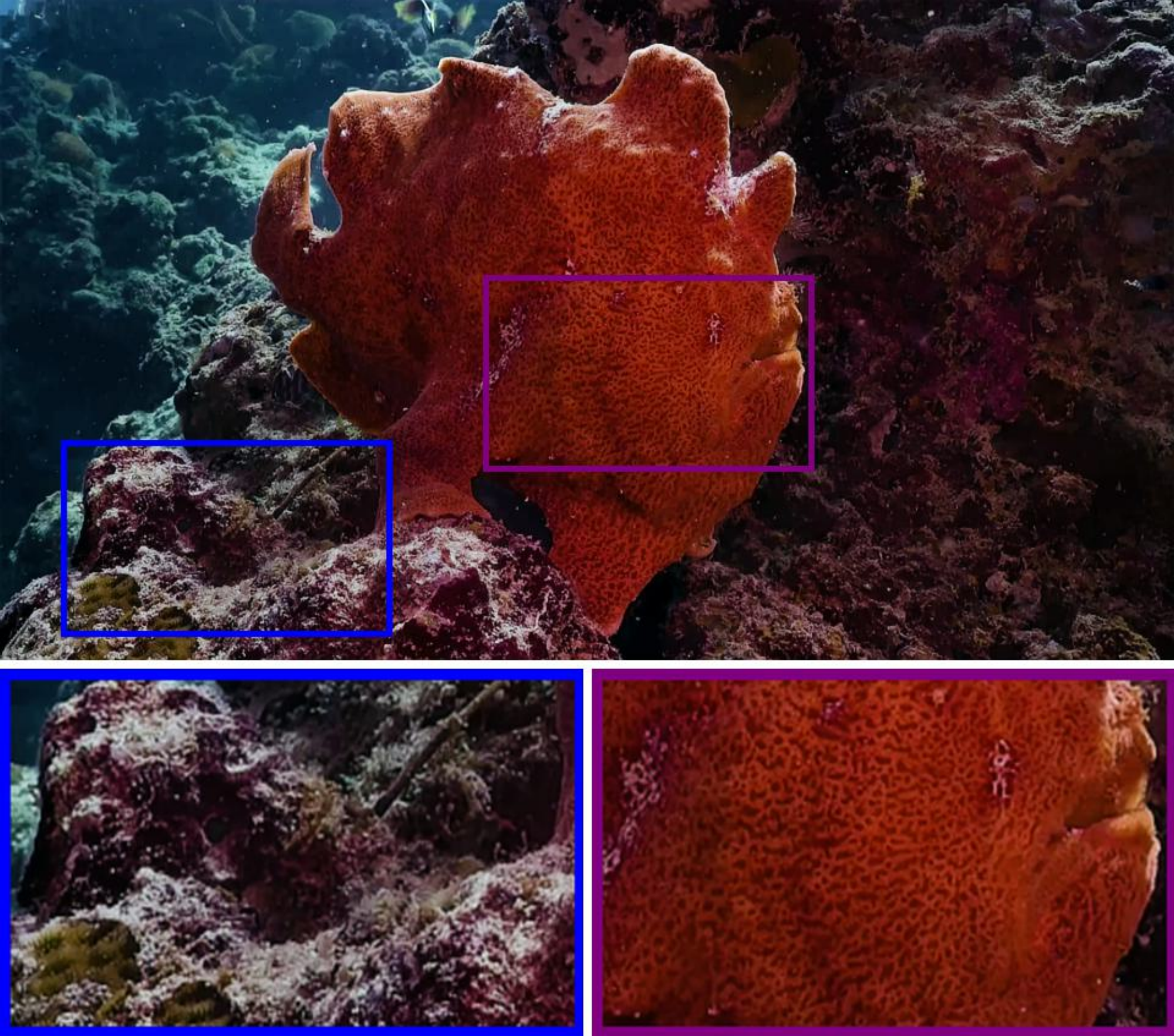}
        \includegraphics[width=\linewidth]{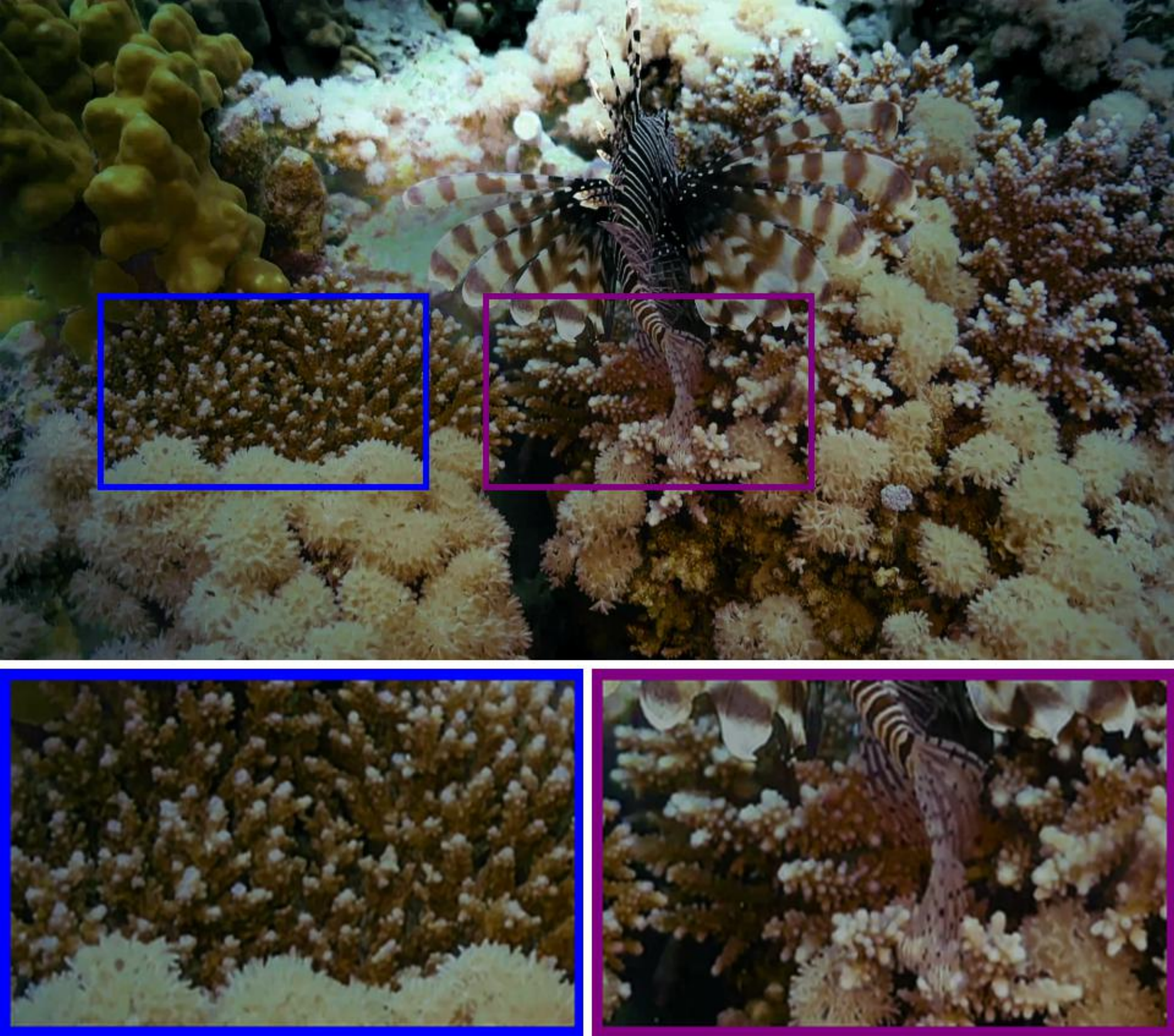} 
		\caption{\footnotesize NUIPUI Dataset}
	\end{subfigure}
	\begin{subfigure}{0.15\linewidth}
		\centering
		\includegraphics[width=\linewidth]{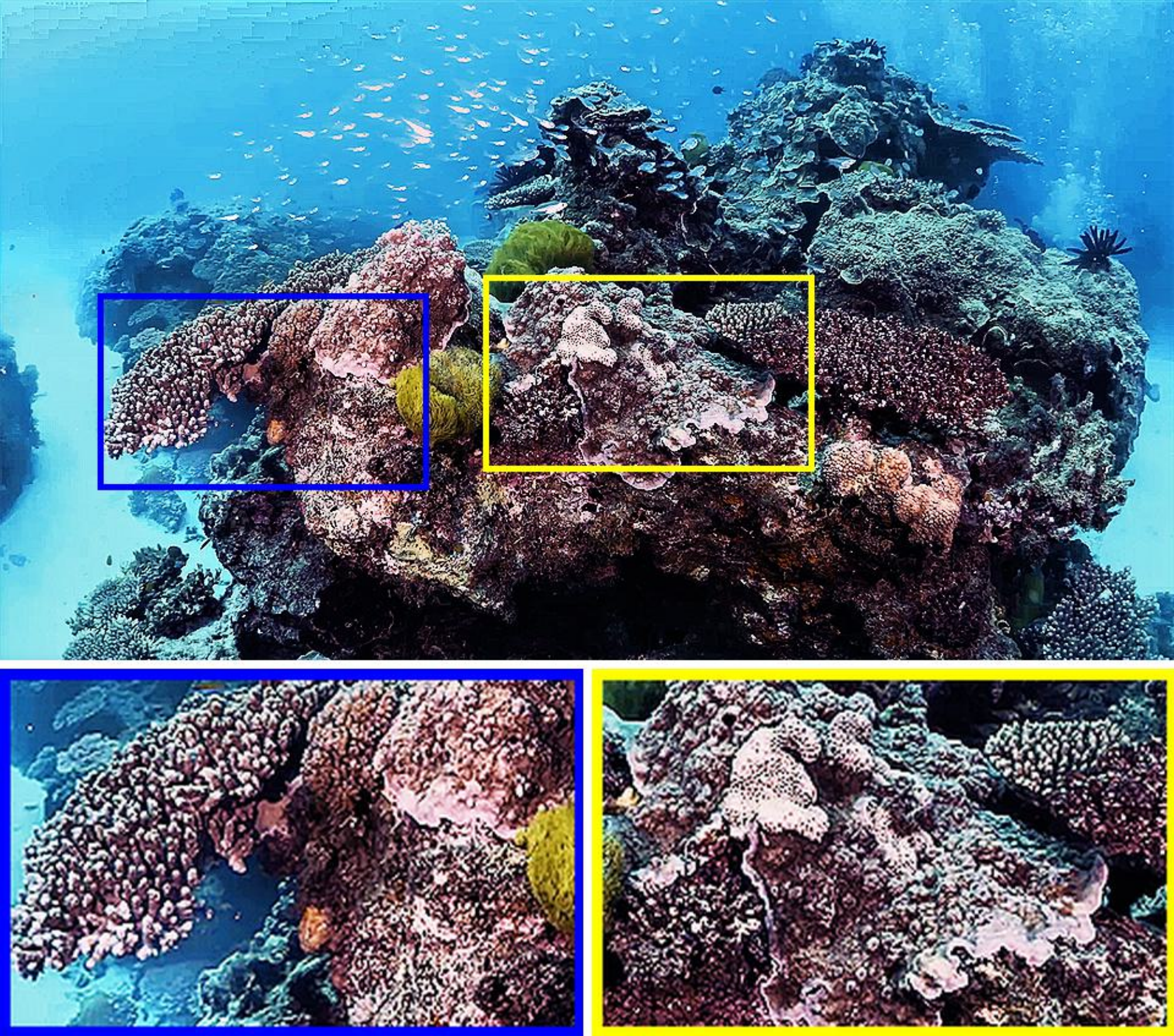} 
        \includegraphics[width=\linewidth]{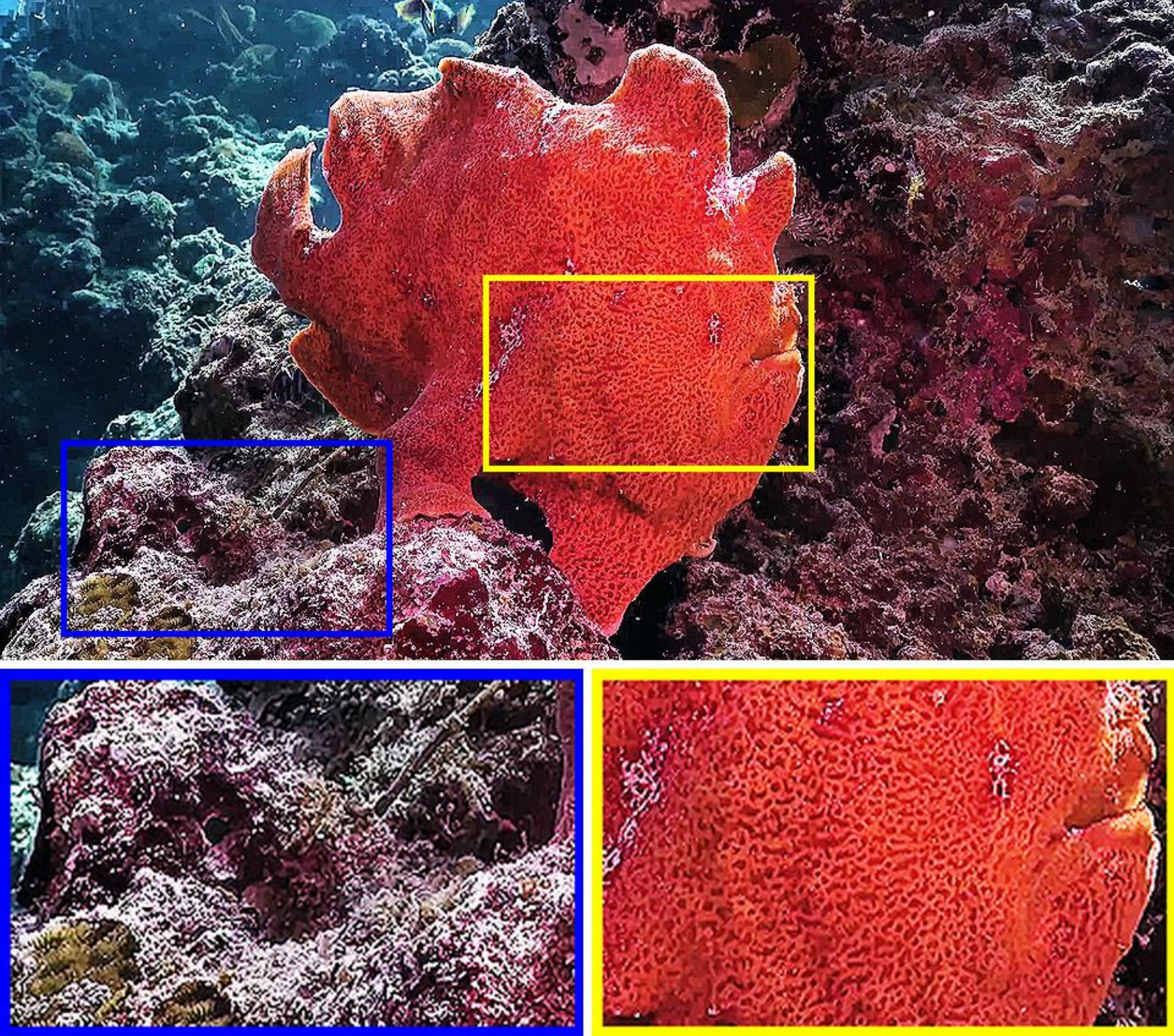}
        \includegraphics[width=\linewidth]{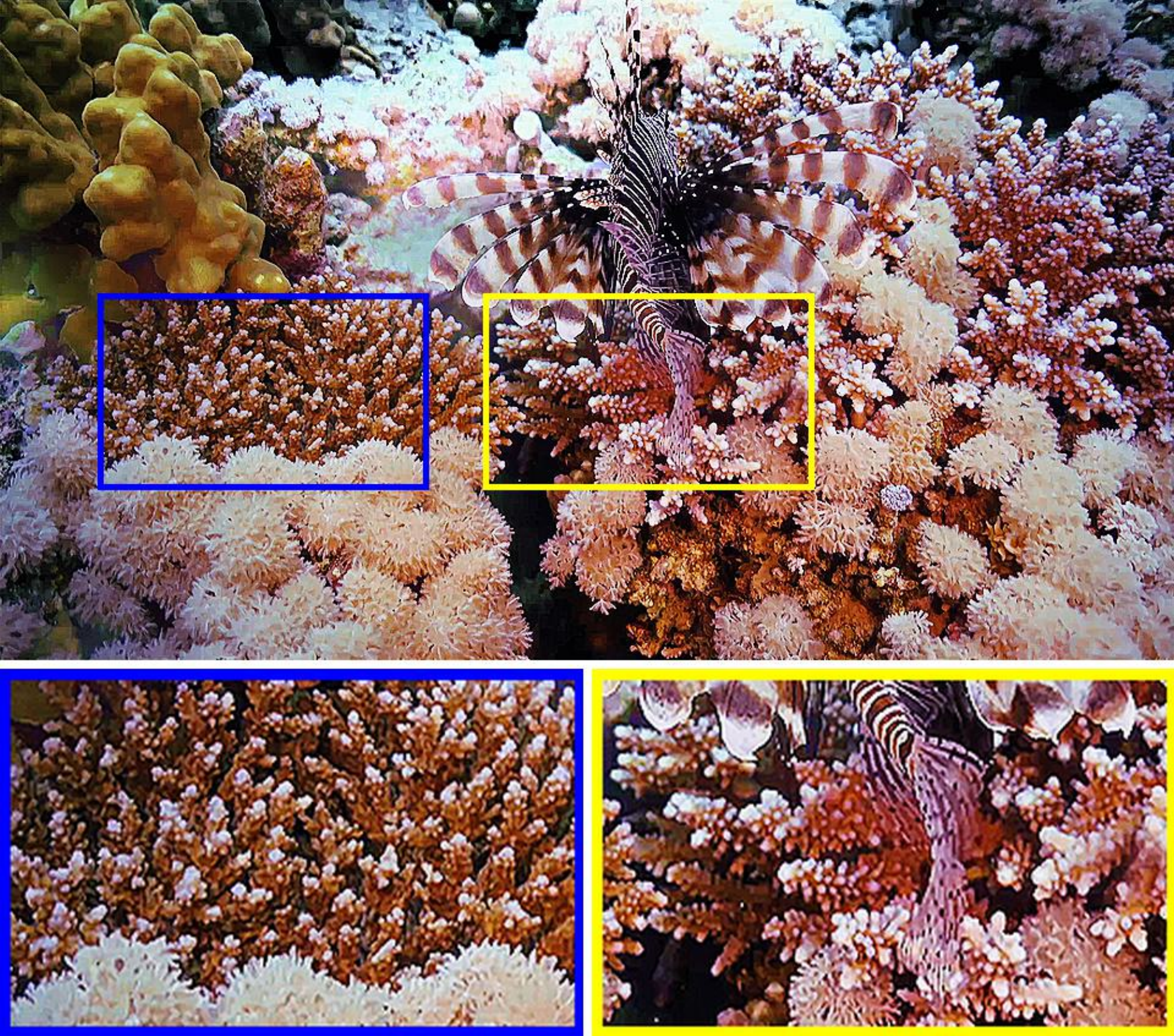} 
		\caption{\footnotesize PUNI Dataset}
	\end{subfigure}
    	\caption{Comparison of training datasets in UNIR-Net.}
	\label{DS}
\end{figure*}

Figure~\ref{DS} visually compares the results obtained by training UNIR-Net with the different datasets mentioned. The comparison shows that the LOL dataset significantly improves lighting and color; however, it does not produce notable enhancement in edge details and introduces a slight haze effect. The MIT dataset also improves lighting but leaves certain regions noticeably dark. The LUIE dataset results in color distortions, generating enhanced images with unnatural hues. In the case of the synthetic NUIPUI dataset, the illumination is not fully corrected across the entire image. Although NUIPUI is designed to address non-uniform underwater illumination, the enhanced outputs remain slightly dark overall, limiting its effectiveness for this specific challenge. In contrast, the PUNI dataset yields the best visual results, offering substantial improvements in lighting, preserving natural colors, and enhancing edge definition more effectively than the other datasets.

These visual observations are further supported by the quantitative results presented in Table~\ref{CC_DS}, which shows the performance of UNIR-Net trained on each dataset and evaluated on the NUID dataset using four non-reference image quality metrics. The model trained with the PUNI dataset leads in three metrics: UCIQE (0.6225), FADE (0.4184), and MUSIQ (51.5581), although it does not achieve the highest score in NIQMC.

\begin{table*}[ht]
	\centering
	\caption{Performance evaluation of UNIR-Net trained on different datasets and evaluated on the NUID dataset}
	\label{CC_DS}
	\resizebox{0.62\textwidth}{!}{
		\begin{tabular}{l c c c c}
			\hline
			Dataset & UCIQE $\uparrow$ & FADE $\downarrow$ & NIQMC $\uparrow$ & MUSIQ $\uparrow$  \\ \hline
			LOL~\cite{wei2018deep}	 & 0.5642 & 0.8559 & 5.1246 & 50.4725 \\
            MIT-Adobe FiveK~\cite{bychkovsky2011learning} & 0.6045 & 0.5375 & 5.1288 & 45.5405 \\
            LUIE~\cite{xie2022lighting} & 0.6014 & 0.9069 & \textbf{5.5207} & 42.3370 \\
            NUIPUI~\cite{ma2025pqgal}& 0.5733& 0.6253& 4.9371& 48.7758 \\
			PUNI & \textbf{0.6225} & \textbf{0.4184} & 5.3481 & \textbf{51.5581} \\ \hline
		\end{tabular}
	}
\end{table*}

\subsection{Additional Analyses}

\subsubsection{Ablation Study}
In this section, the contribution of the different components of the model is evaluated through an ablation study. This quantitative evaluation is conducted on the NUID dataset, which serves as the basis for assessing the effects of each architectural component in underwater image synthesis under non-uniform illumination. Metrics such as UCIQE and NIQMC were employed to comprehensively evaluate color fidelity and image quality under varying illumination conditions. The results for the PUNI model are presented in Table \ref{Ablation_1}, while the outcomes for the UNIR-Net model, after removing specific components, are detailed in Table \ref{Ablation_2}.

\begin{figure*}[ht]
	\centering
	\begin{subfigure}{0.24\linewidth}
		\centering
        \includegraphics[width=\linewidth]{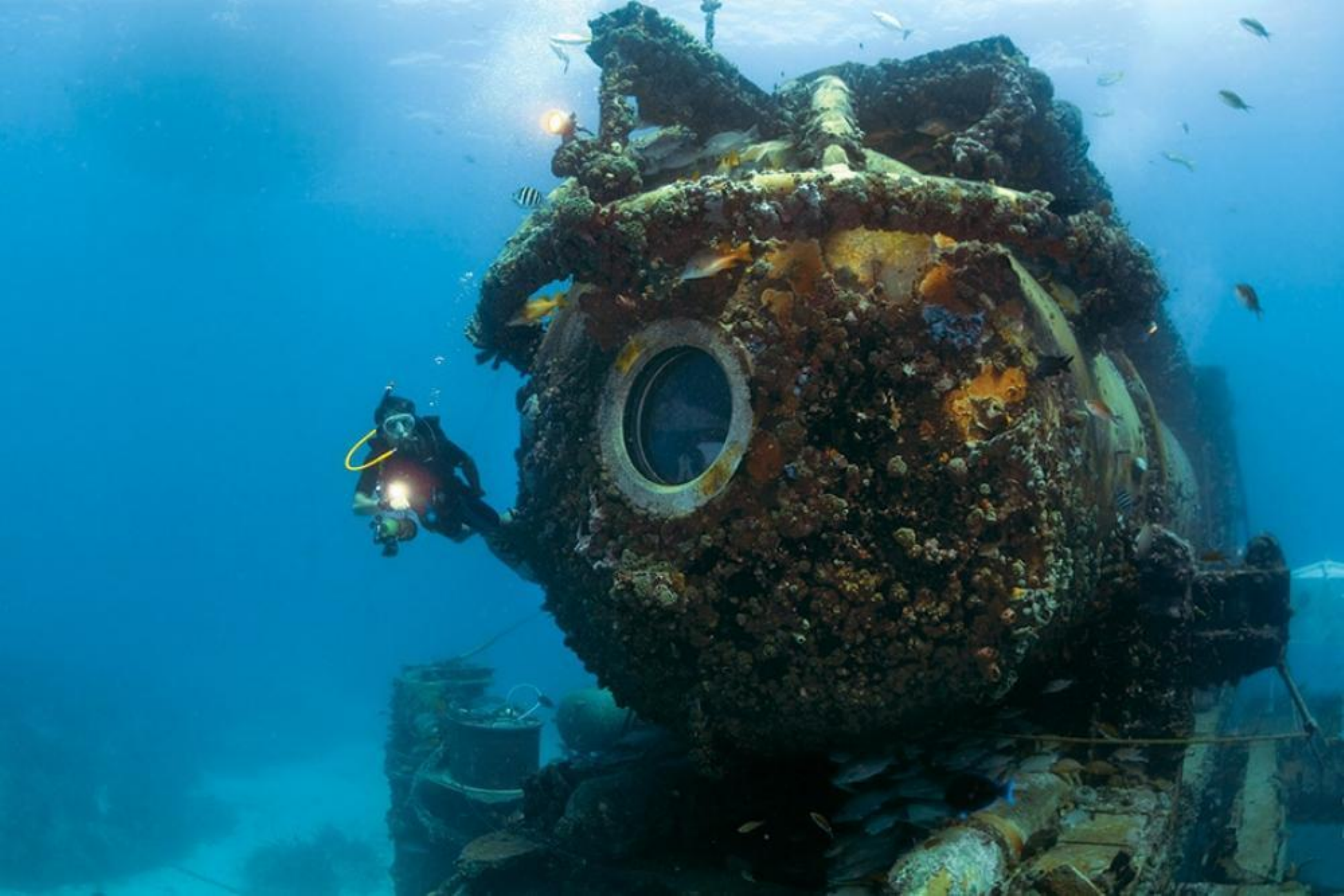} 
		\includegraphics[width=\linewidth]{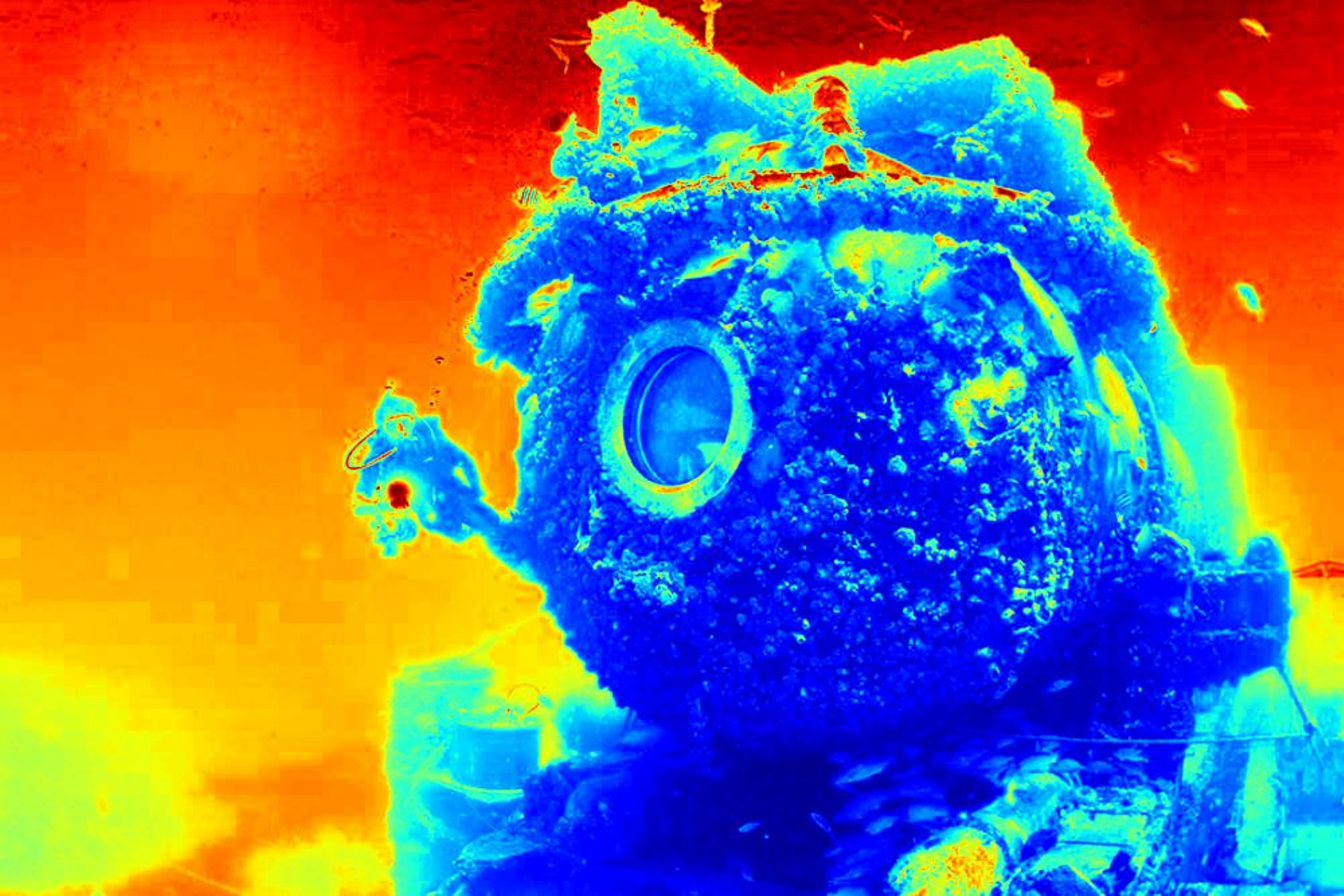} 
        \caption{\footnotesize NUI image}
	\end{subfigure}
	\begin{subfigure}{0.24\linewidth}
		\centering
		\includegraphics[width=\linewidth]{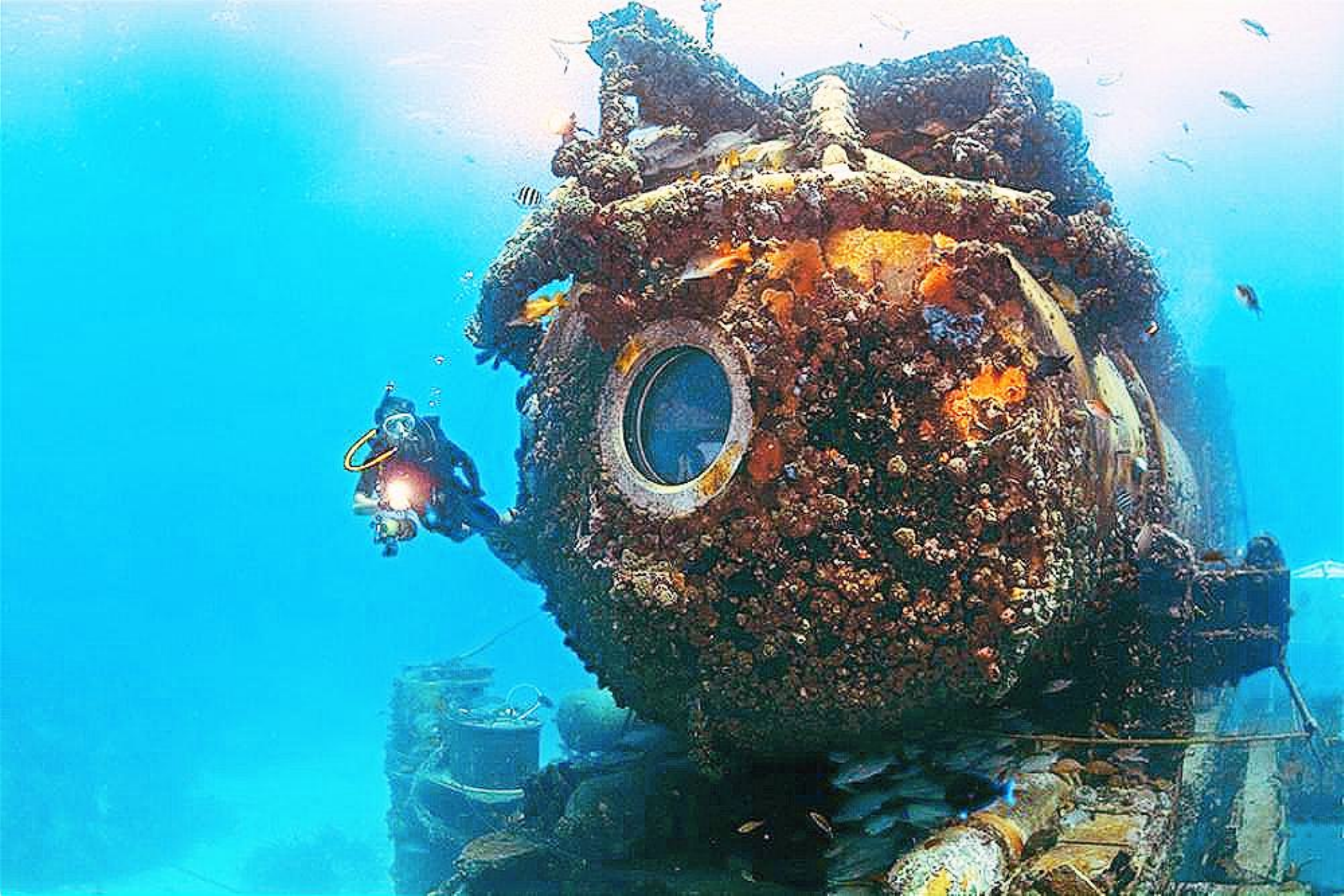}
        \includegraphics[width=\linewidth]{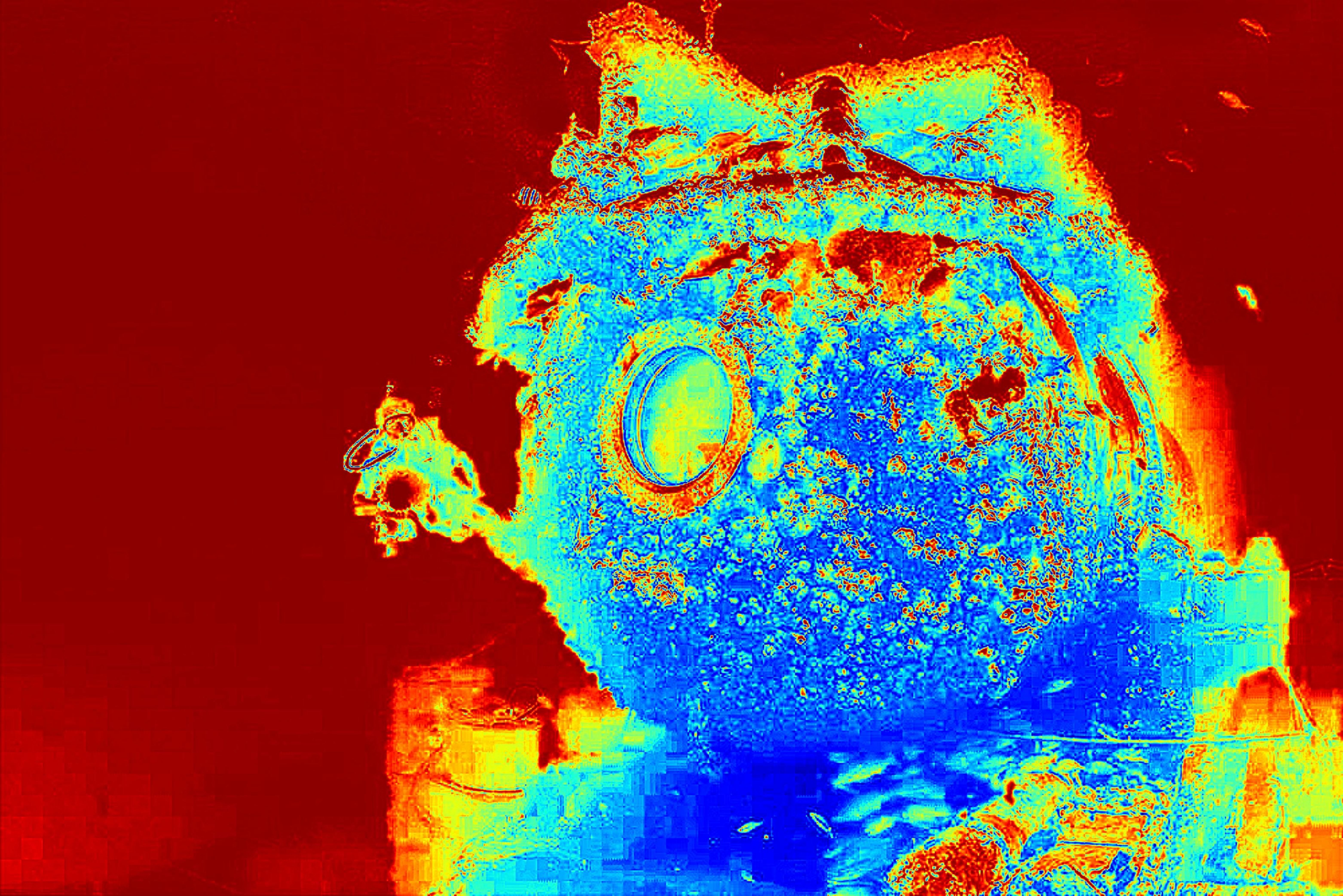} 
        \caption{\footnotesize w/o Mask}
	\end{subfigure}
	\begin{subfigure}{0.24\linewidth}
		\centering
		\includegraphics[width=\linewidth]{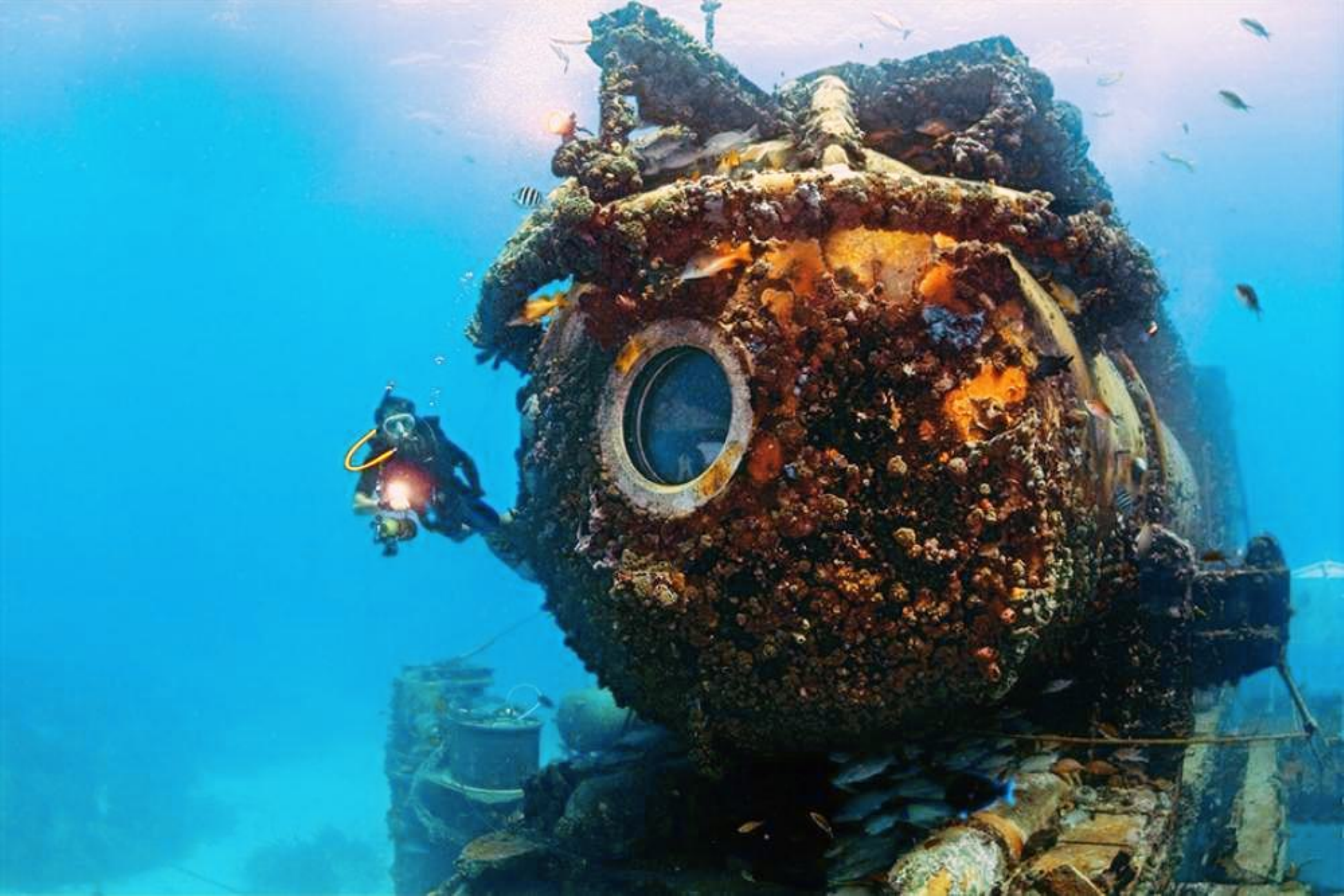}
        \includegraphics[width=\linewidth]{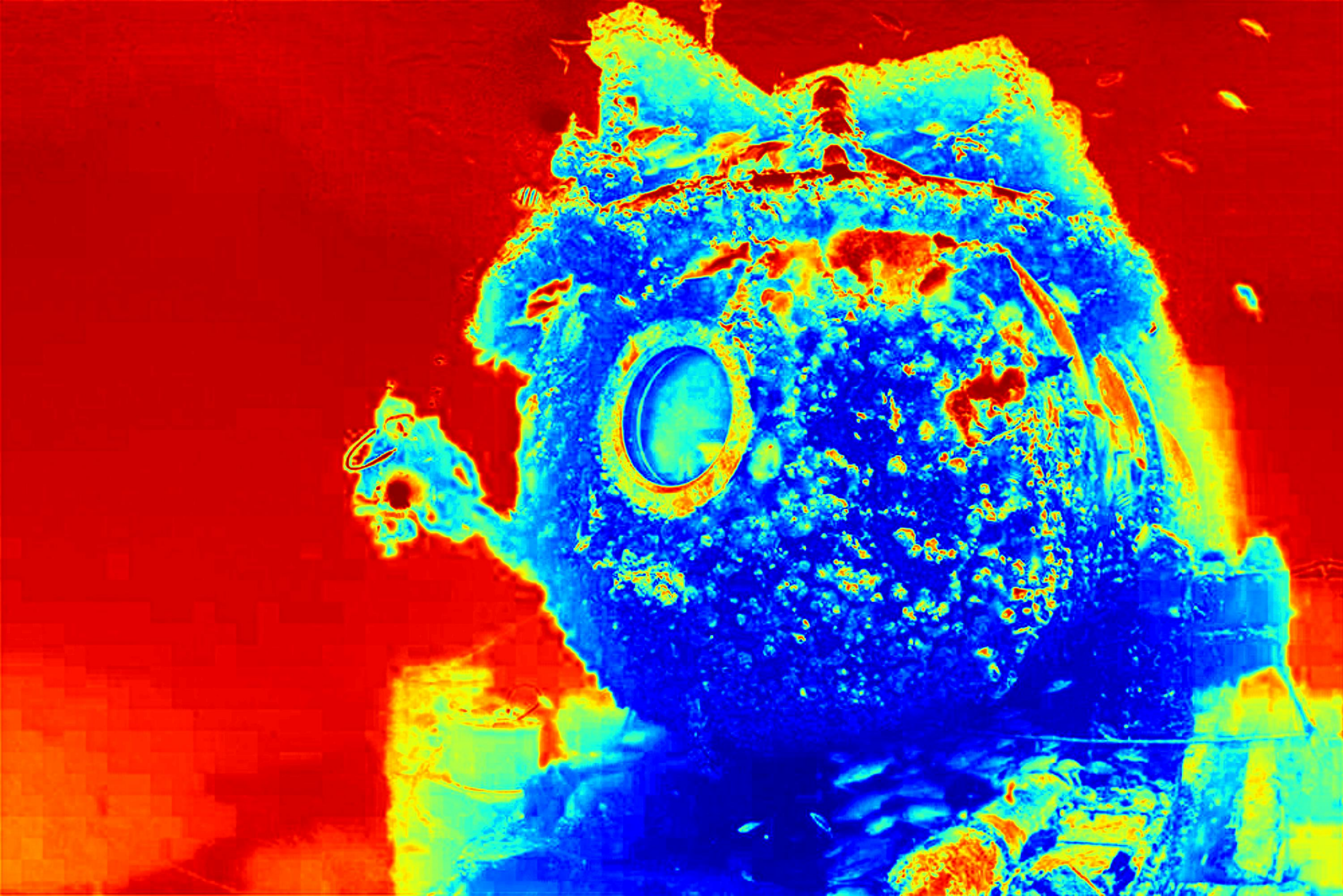}
        \caption{\footnotesize w/o Sharpening}
	\end{subfigure}
	\begin{subfigure}{0.24\linewidth}
		\centering
		\includegraphics[width=\linewidth]{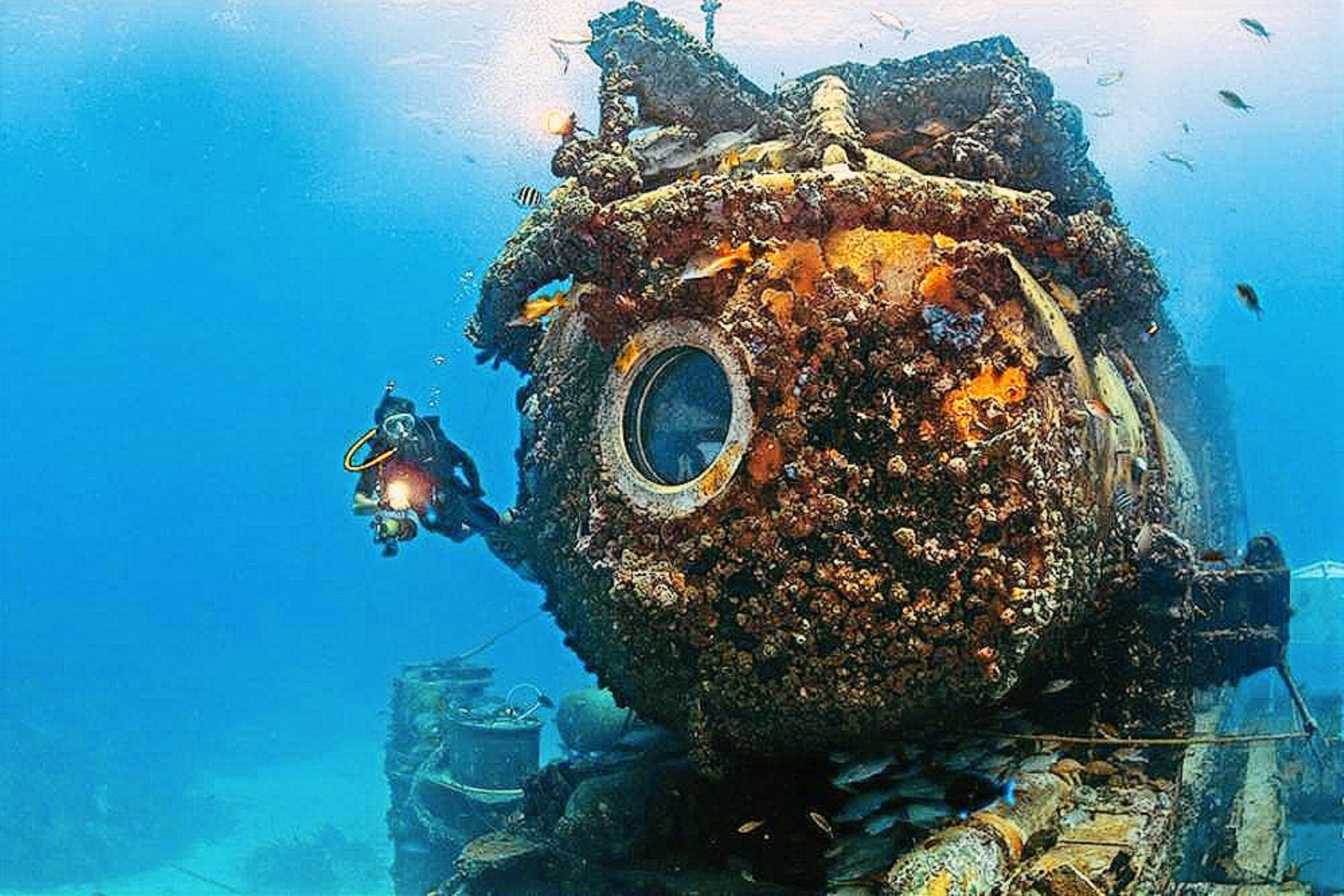} 
        \includegraphics[width=\linewidth]{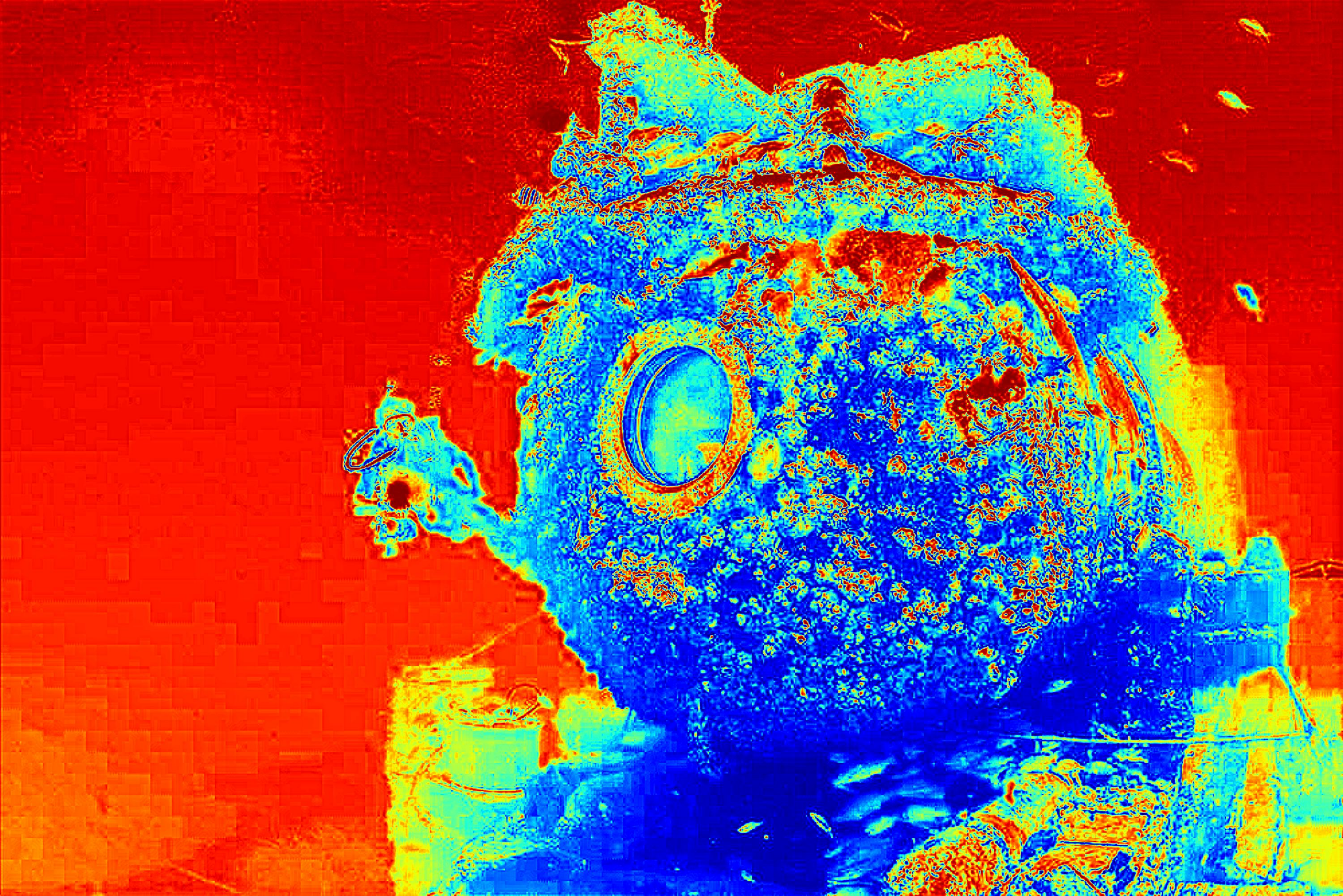} 
		\caption{\footnotesize Full Model}
	\end{subfigure}
    	\caption{Visual comparison of enhancement results using datasets synthesized with different configurations. The first column shows a real image from the NUID dataset. The second and third columns show UNIR-Net results trained on synthetic datasets generated without segmentation masks and without border sharpening, respectively. The fourth column shows results using the full configuration with both components. Each image is shown with its corresponding illumination map.}
	\label{DS_ABLATION}
\end{figure*}

Table \ref{Ablation_1} details the impact of removing individual components, such as the mask and sharpening synthesis. When the mask is removed, the UCIQE and NIQMC values significantly decrease, reaching 0.6112 and 5.2866, respectively. Similarly, suppressing the sharpening also negatively impacts performance, with UCIQE and NIQMC values dropping to 0.6064 and 5.2649, respectively. In contrast, the full configuration, which integrates both the mask and the sharpening synthesis, achieves the best results with a UCIQE of 0.6225 and a NIQMC of 5.3481. Figure \ref{DS_ABLATION} illustrates visual results showing that including all model components produces the most visually enhanced images.

\begin{table}[ht]
	\centering
	\caption{Quantitative results of the PUNI dataset synthesis evaluated under different ablation scenarios.}
	\label{Ablation_1}
    \resizebox{0.5\textwidth}{!}{
    \begin{tabular}{lccccc}
		\toprule
		\multicolumn{1}{c}{\multirow{2}{*}{Model}} & \multicolumn{2}{c}{Components} & \multicolumn{2}{c}{Metrics} \\
		\cmidrule(lr){2-3} \cmidrule(lr){4-5}
	    & Mask & Sharpening & UCIQE $\uparrow$ & NIQMC $\uparrow$ \\
		\midrule
        Model w/o Mask & $\times$ & $\checkmark$  & 0.6112 & 5.2866  \\
		Model w/o Sharpening & $\checkmark$ & $\times$   & 0.6064  & 5.2649  \\
		Full Configuration & $\checkmark$ & $\checkmark$  & \textbf{0.6225}  & \textbf{5.3481}  \\
		\bottomrule
	\end{tabular}
 }
\end{table}

On the other hand, Table \ref{Ablation_2} evaluates the impact of removing additional components, such as the Core, VRM, and CCM. Jointly removing the Core and VRM severely affects performance, resulting in UCIQE and NIQMC scores of 0.5611 and 4.5416, respectively. Furthermore, removing only the VRM leads to even lower performance, highlighting the importance of this module. Finally, excluding the CCM reduces the metrics to a UCIQE of 0.6123 and an NIQMC of 5.2927. The Full Model demonstrates superior performance in all cases, achieving a UCIQE of 0.6225 and an NIQMC of 5.3481.

These results highlight the significance of each component within the UNIR-Net and the data synthesis process. Moreover, they validate the effectiveness of the UNIR-Net model in improving visual quality and quantitative metrics for underwater images with non-uniform illumination. Although the impact of the CCM was evaluated through the ablation study (Table~\ref{Ablation_2}), the AB module could not be individually analyzed due to its deep integration into the network architecture. Future work will explore more modular designs to facilitate a detailed evaluation of such components.

\begin{table}[ht]
	\centering
	\caption{Quantitative results of the UNIR-Net model under different ablation scenarios.}
	\label{Ablation_2}
    \resizebox{0.5\textwidth}{!}{
    \begin{tabular}{lcccccccc}
		\toprule
		\multicolumn{1}{c}{\multirow{2}{*}{Model}} & \multicolumn{3}{c}{Components} & \multicolumn{2}{c}{Metrics} \\
		\cmidrule(lr){2-4} \cmidrule(lr){5-6}
		& Core & VRM & CCM & UCIQE $\uparrow$ & NIQMC $\uparrow$ \\
		\midrule
        Model w/o Core and VRM & $\times$ & $\times$ & $\checkmark$ & 0.5611 & 4.5416 \\
		Model w/o VRM & $\checkmark$ & $\times$ & $\checkmark$ & 0.5270 & 4.2766 \\
        Model w/o CCM & $\checkmark$ & $\checkmark$ & $\times$ & 0.6123 & 5.2927 \\
		Full Model & $\checkmark$ & $\checkmark$ & $\checkmark$ & \textbf{0.6225} & \textbf{5.3481} \\
		\bottomrule
	\end{tabular}
 }
\end{table}

\subsubsection{Limitations of UNIR-Net in challenging scenarios}
\begin{figure*}[ht]
	\centering
	\begin{subfigure}{0.105\linewidth}
		\centering
        \includegraphics[width=\linewidth]{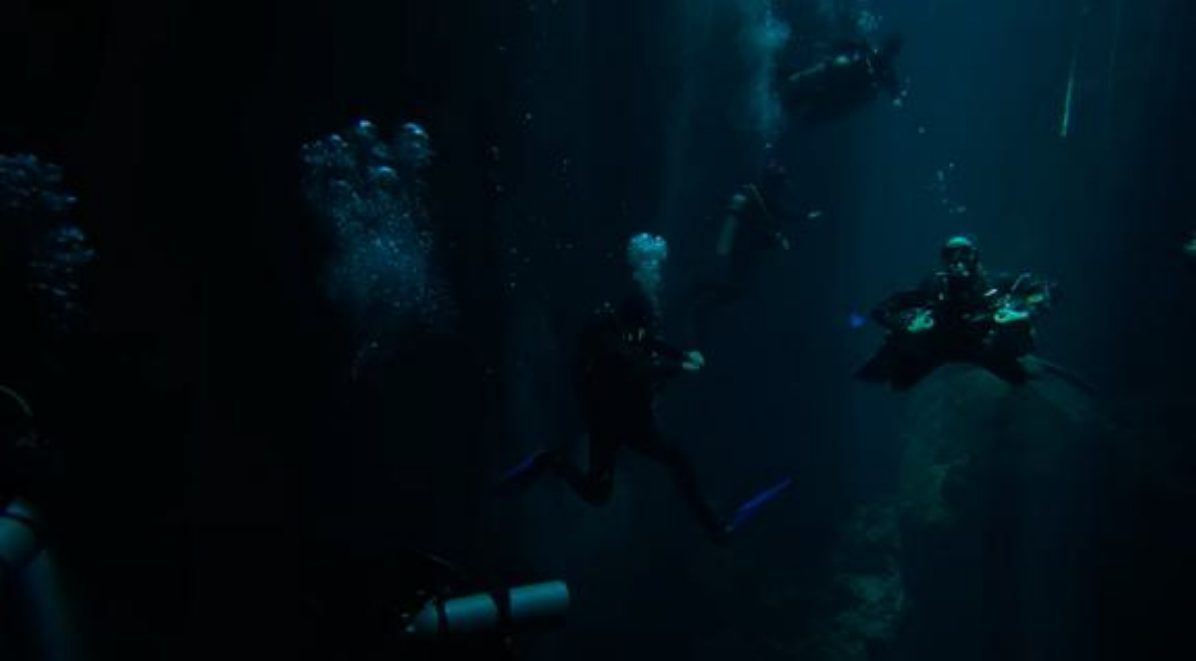} 
		\includegraphics[width=\linewidth]{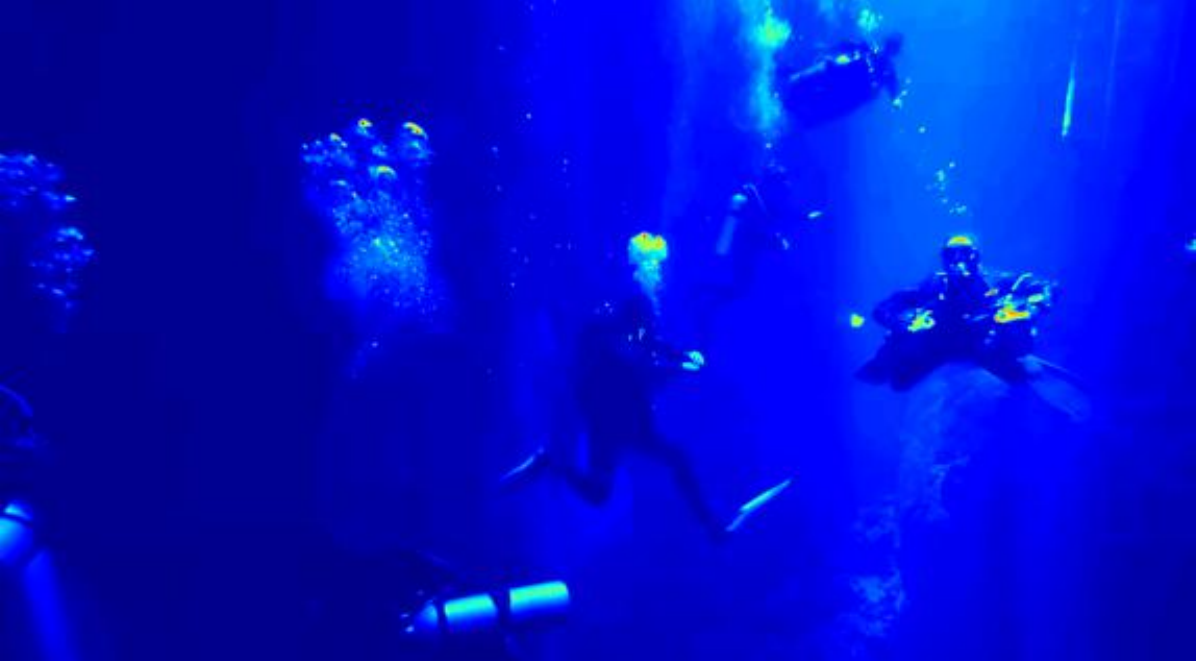} 
        \includegraphics[width=\linewidth]{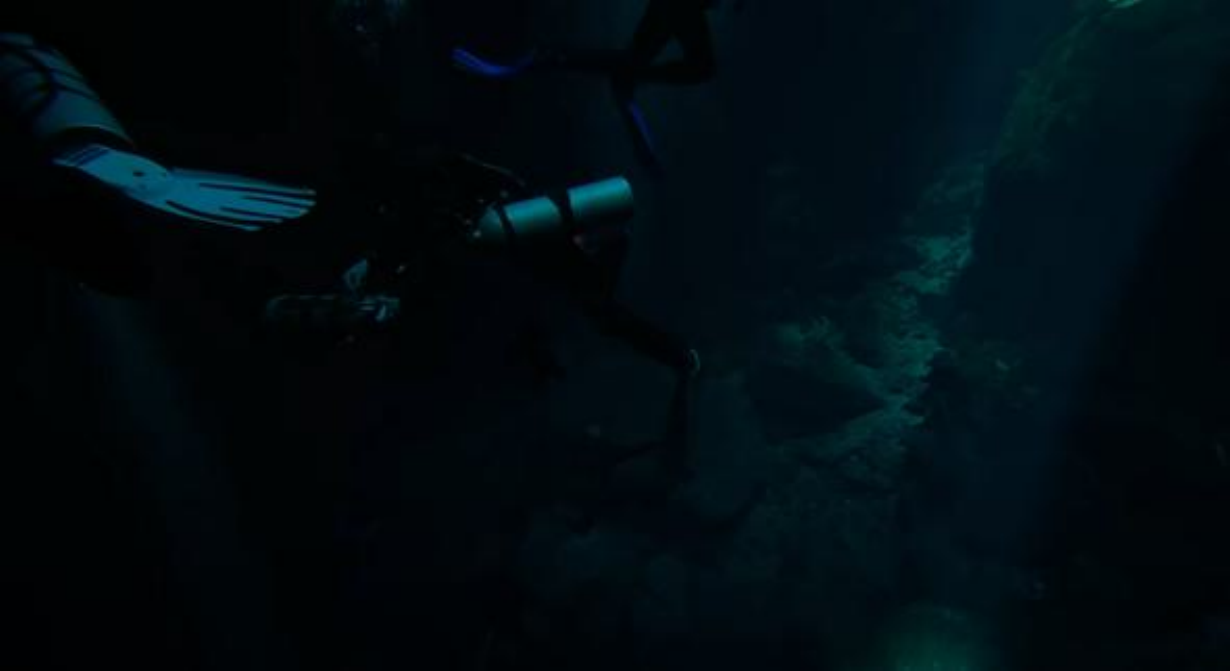} 
        \includegraphics[width=\linewidth]{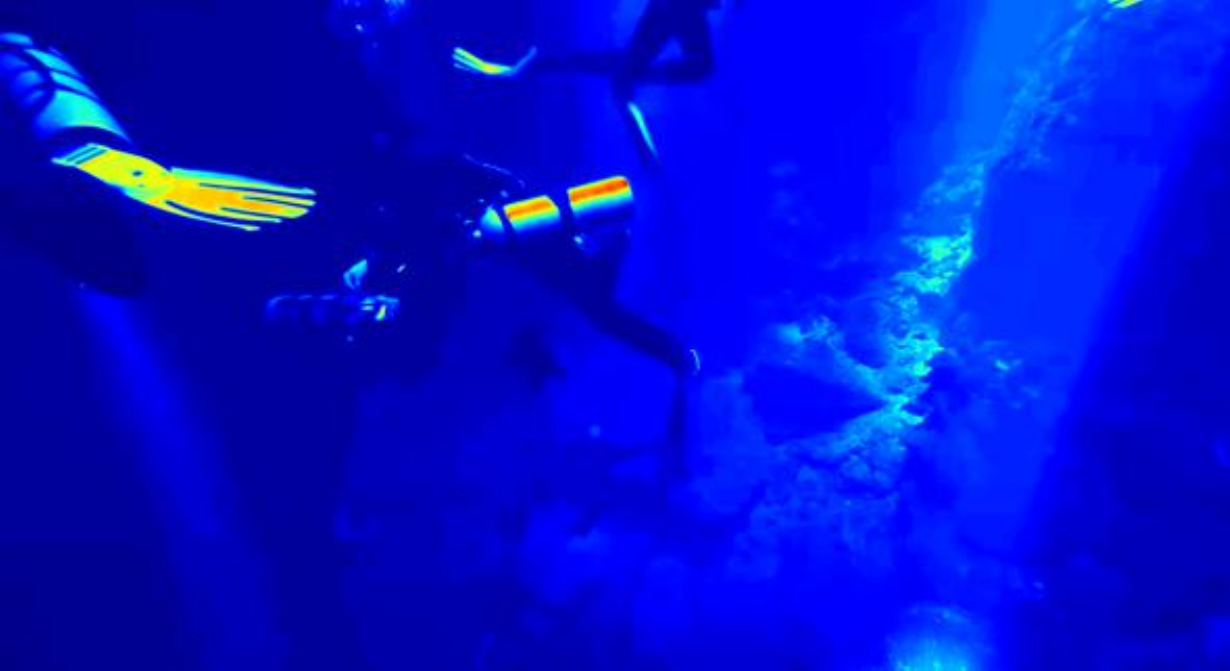} 
        \includegraphics[width=\linewidth]{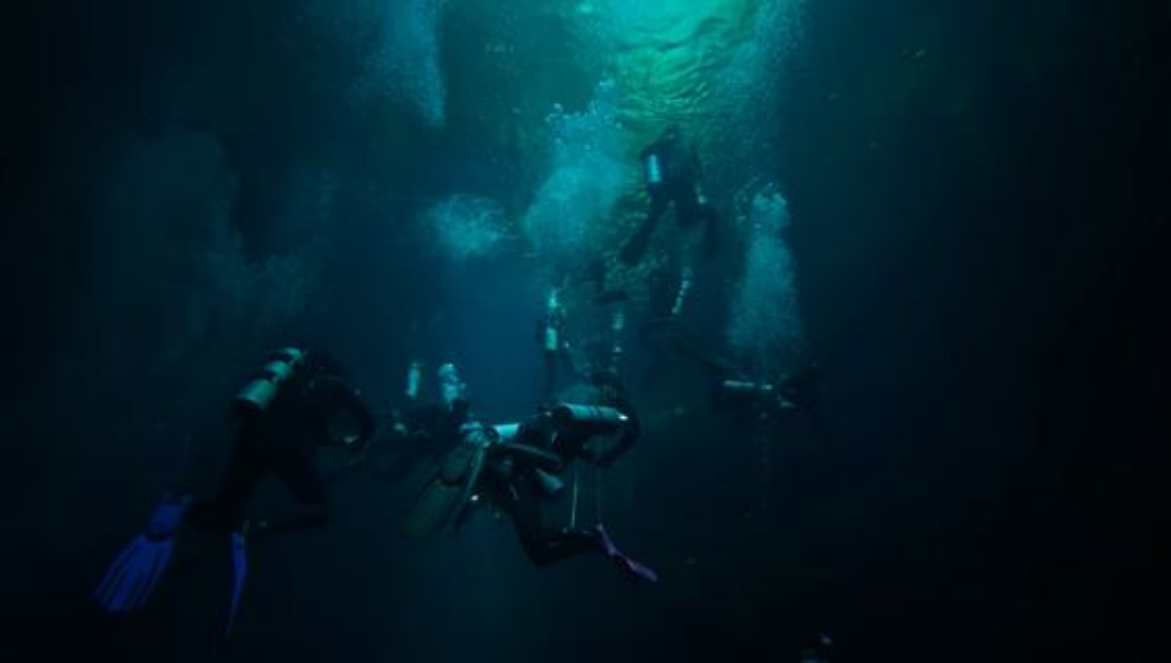} 
        \includegraphics[width=\linewidth]{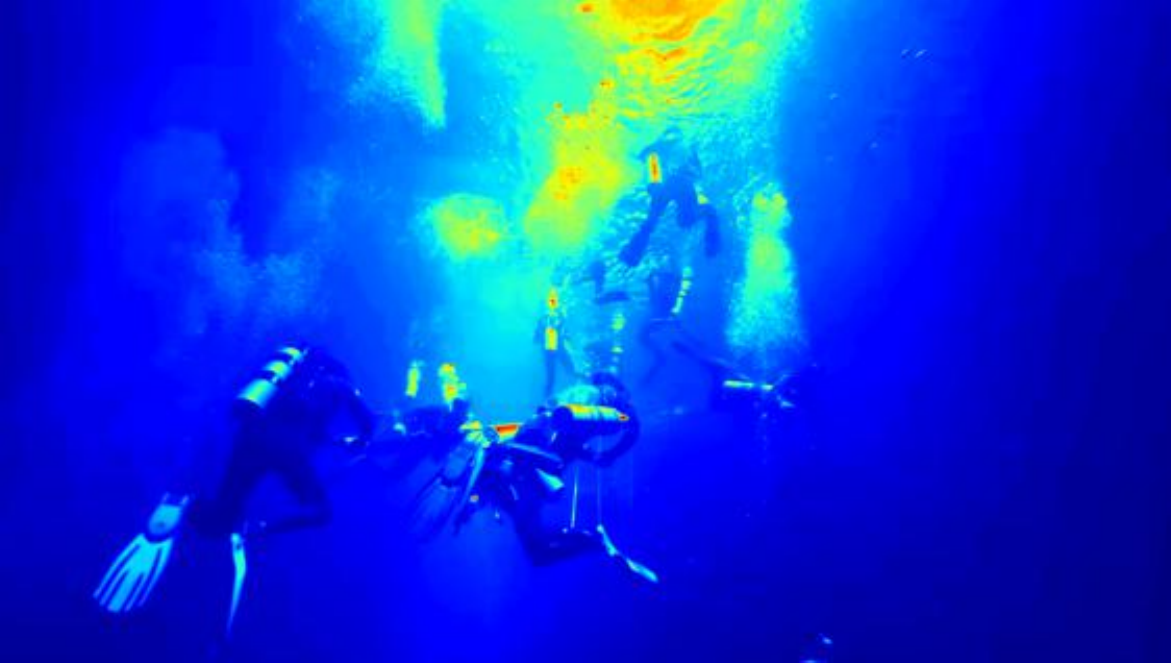} 
        \includegraphics[width=\linewidth]{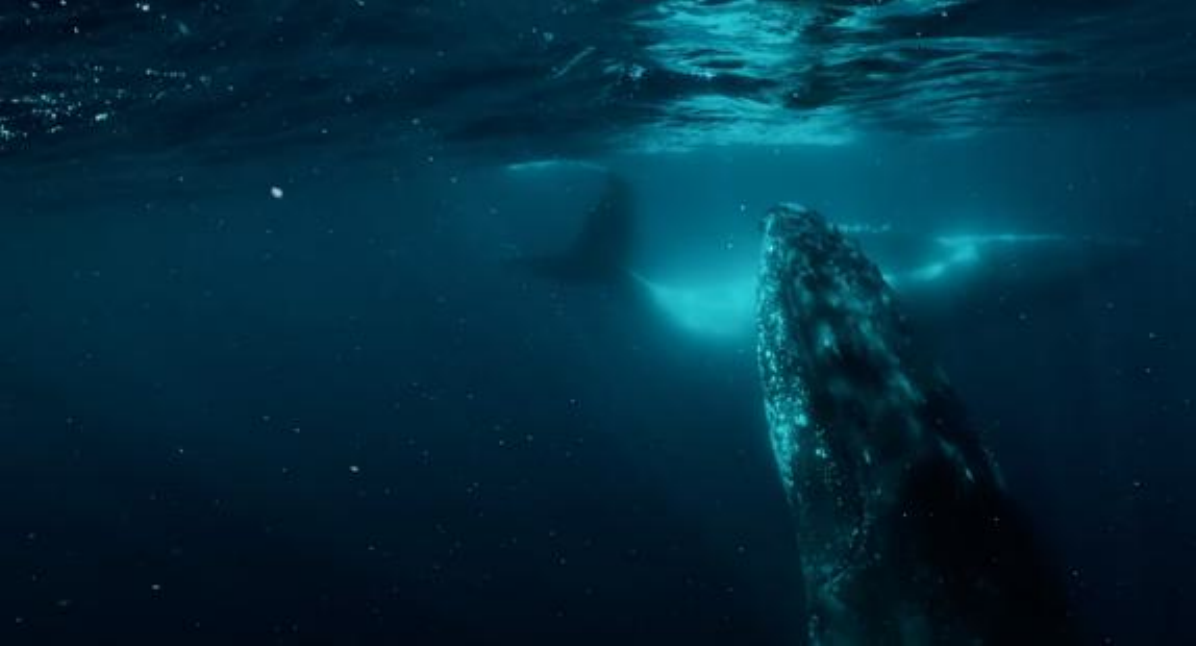} 
        \includegraphics[width=\linewidth]{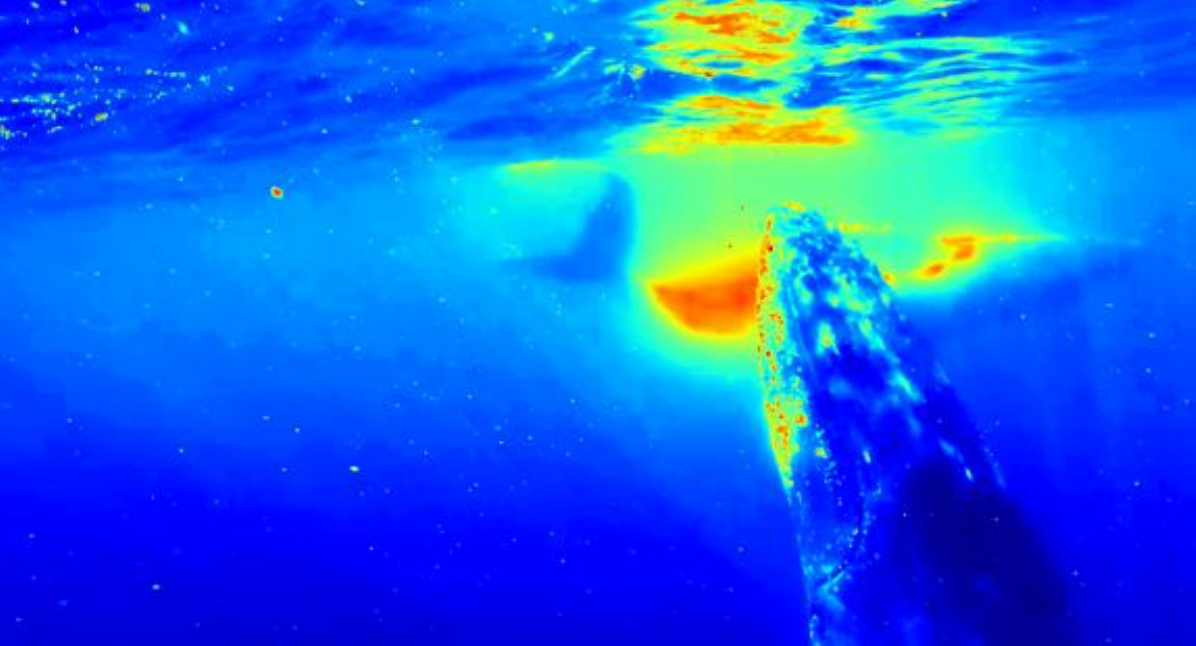} 
        \includegraphics[width=\linewidth]{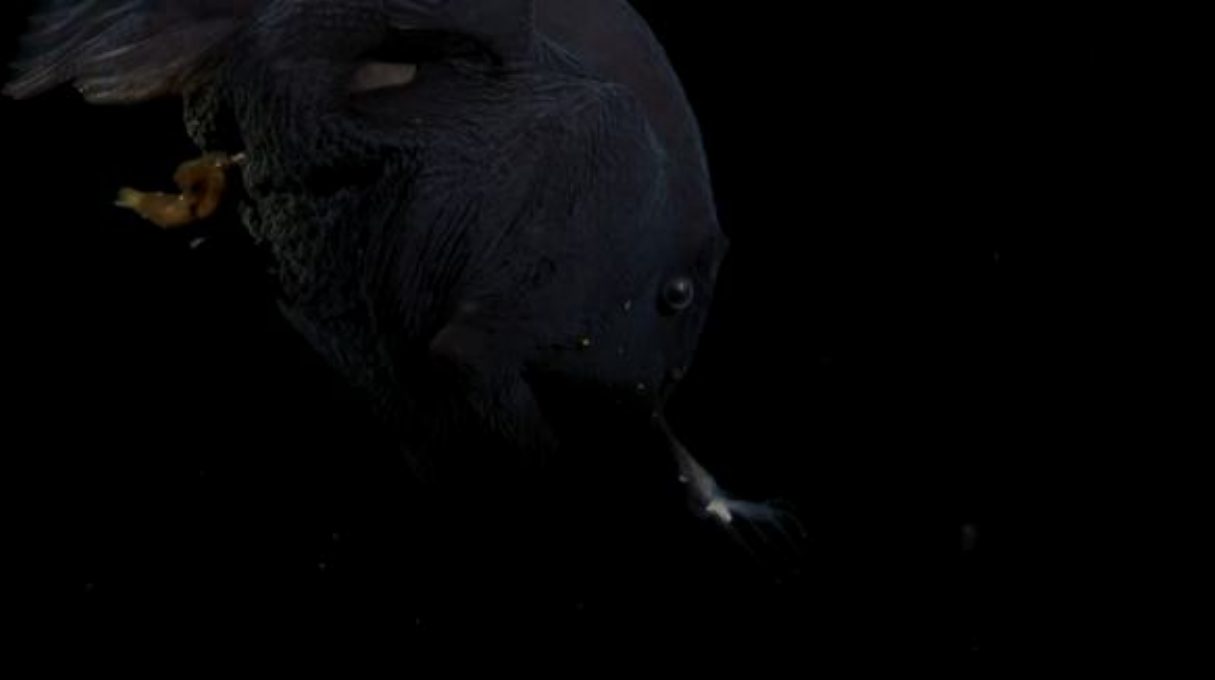} 
        \includegraphics[width=\linewidth]{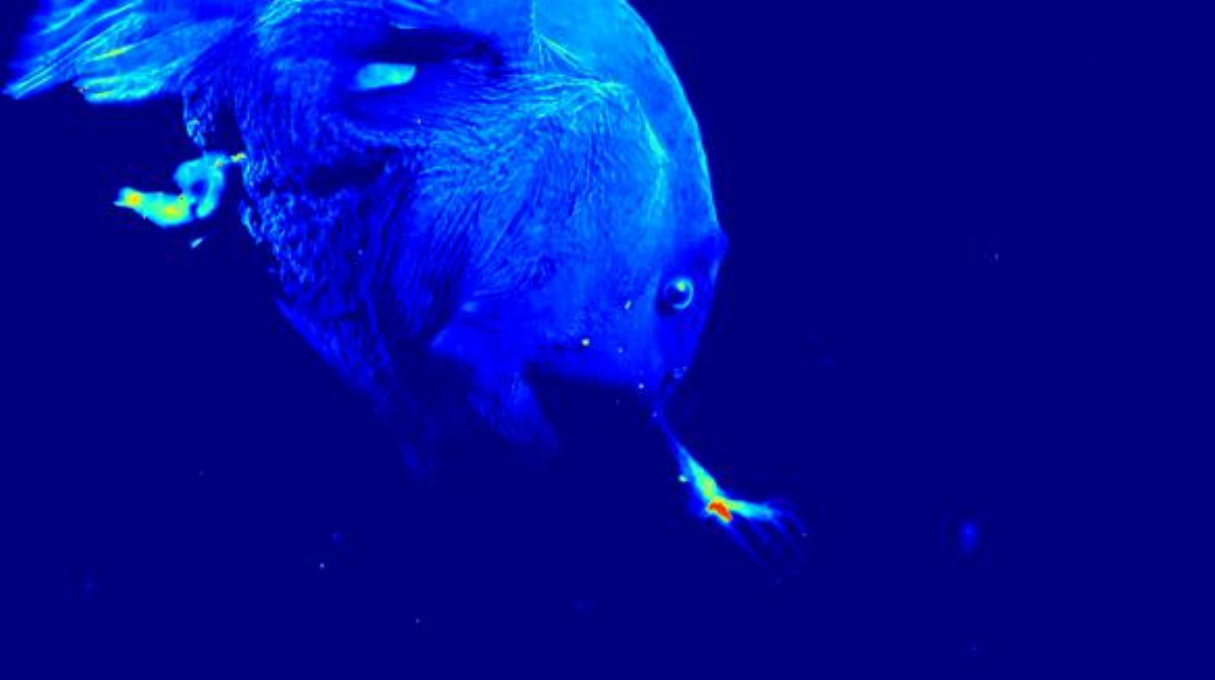} 
        \caption{\footnotesize CI Image}
	\end{subfigure}
    \begin{subfigure}{0.105\linewidth}
		\centering
		\includegraphics[width=\linewidth]{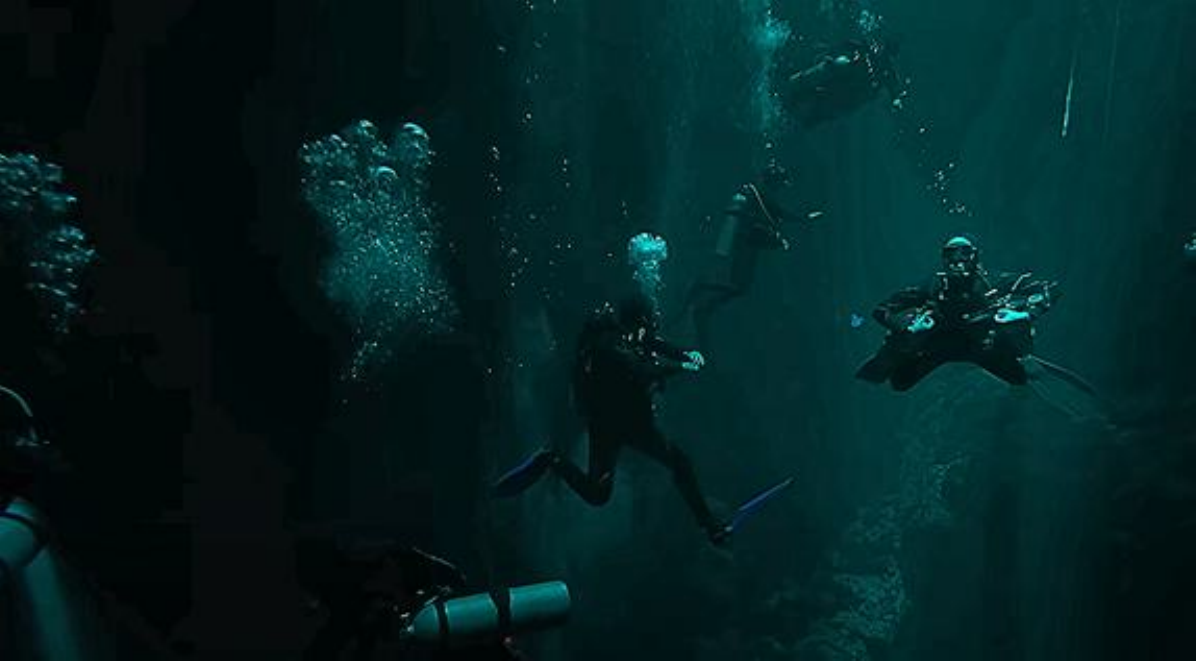} 
        \includegraphics[width=\linewidth]{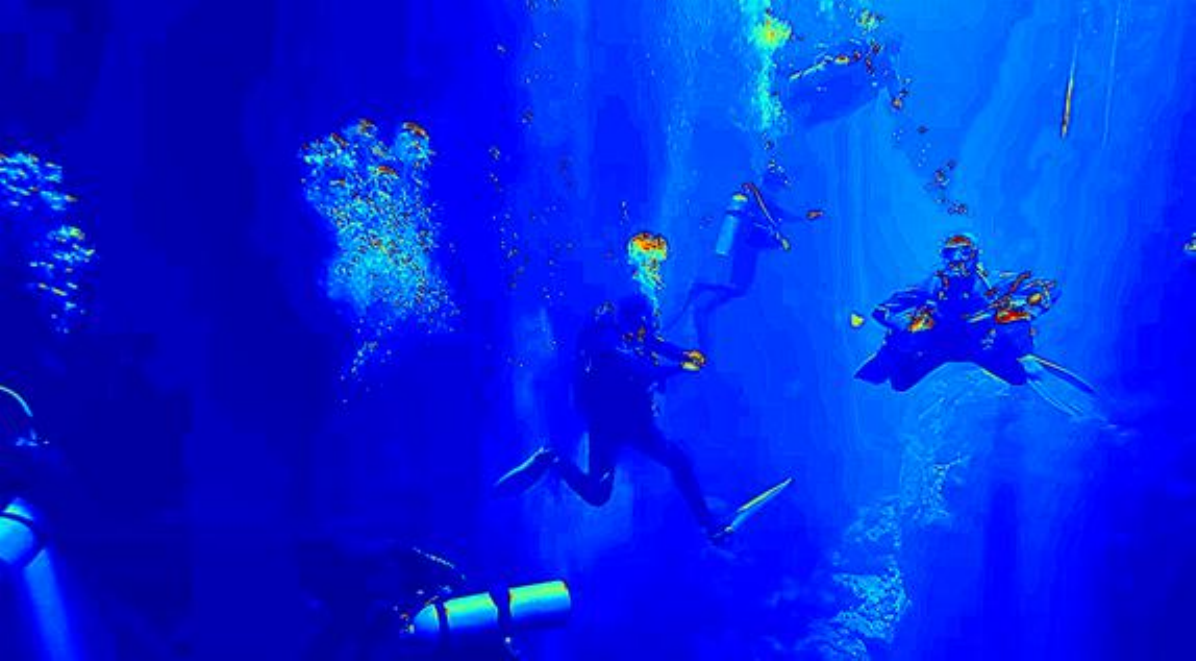} 
		\includegraphics[width=\linewidth]{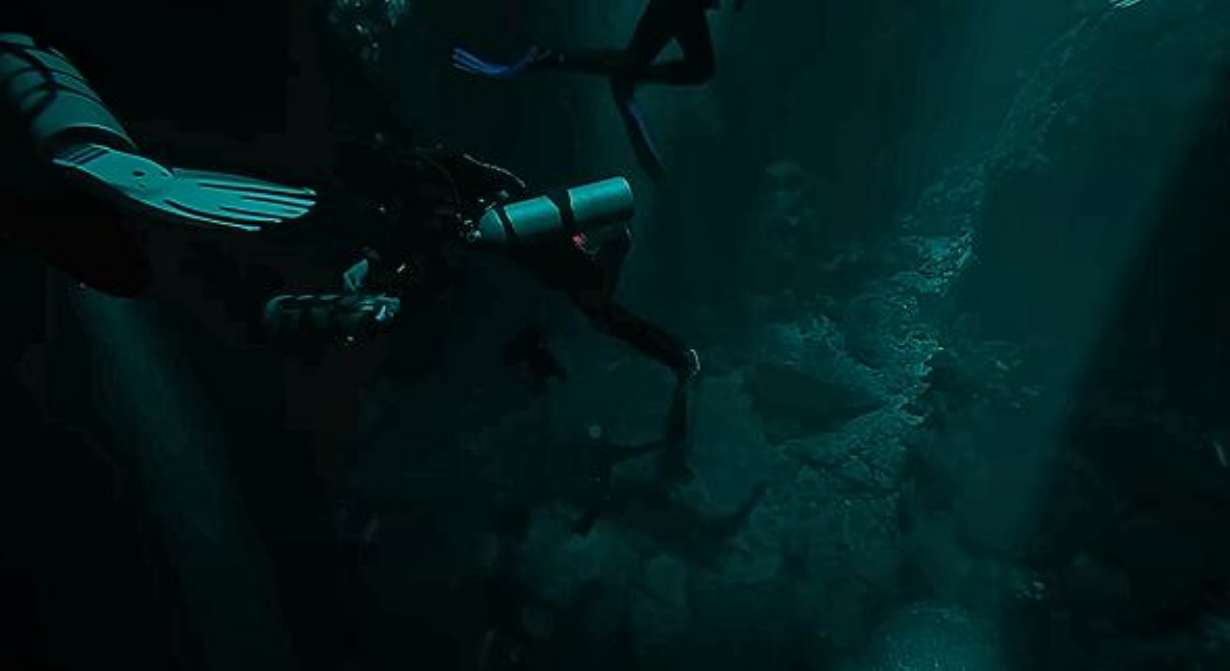} 
        \includegraphics[width=\linewidth]{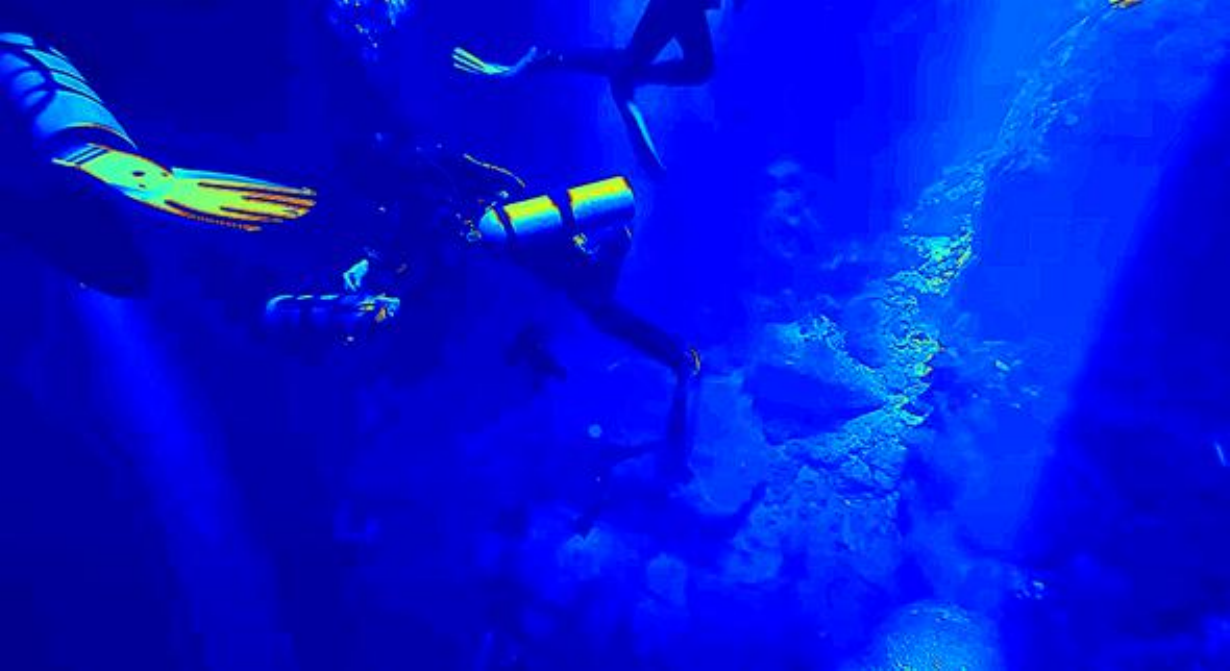} 
		\includegraphics[width=\linewidth]{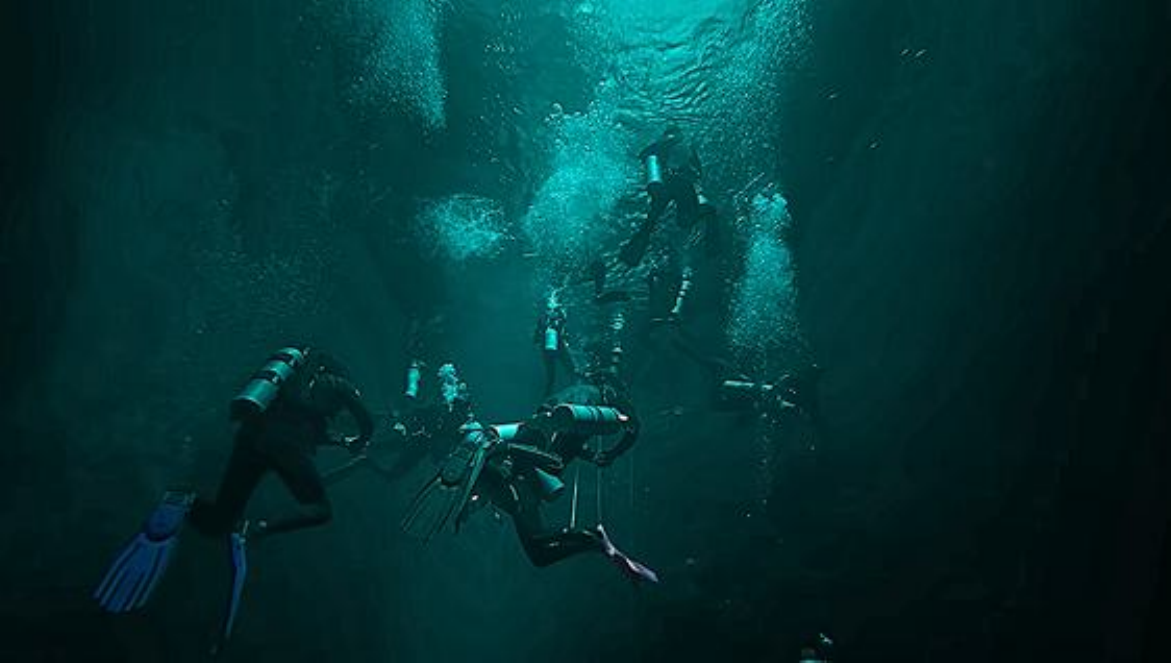} 
        \includegraphics[width=\linewidth]{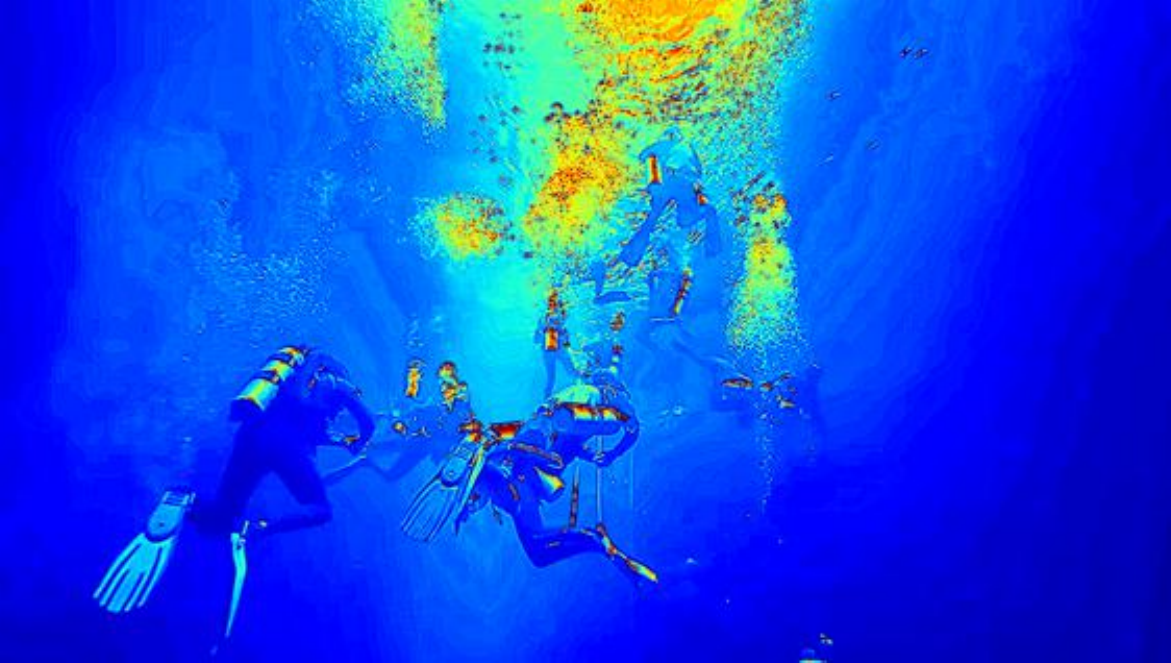} 
		\includegraphics[width=\linewidth]{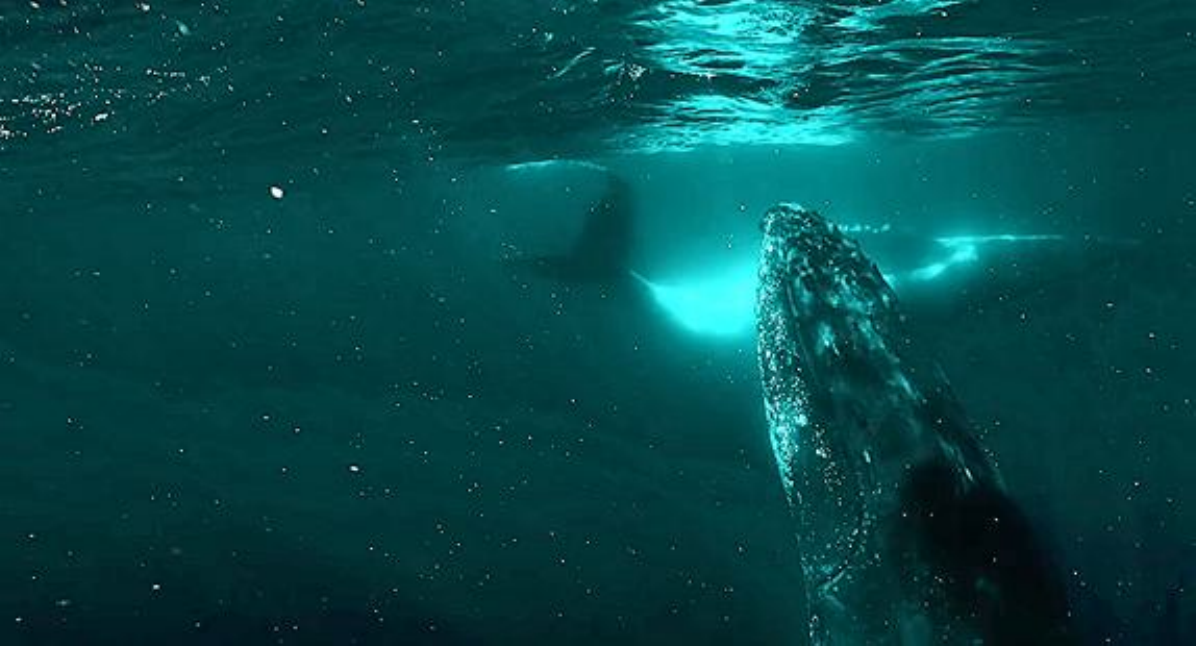} 
        \includegraphics[width=\linewidth]{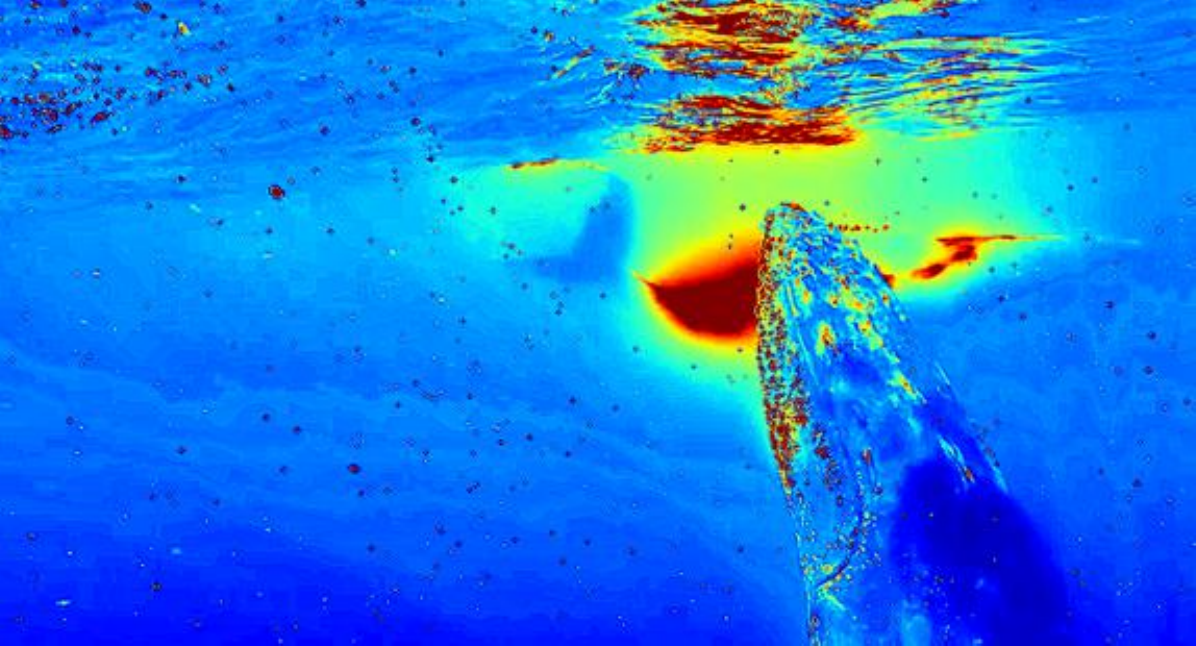} 
		\includegraphics[width=\linewidth]{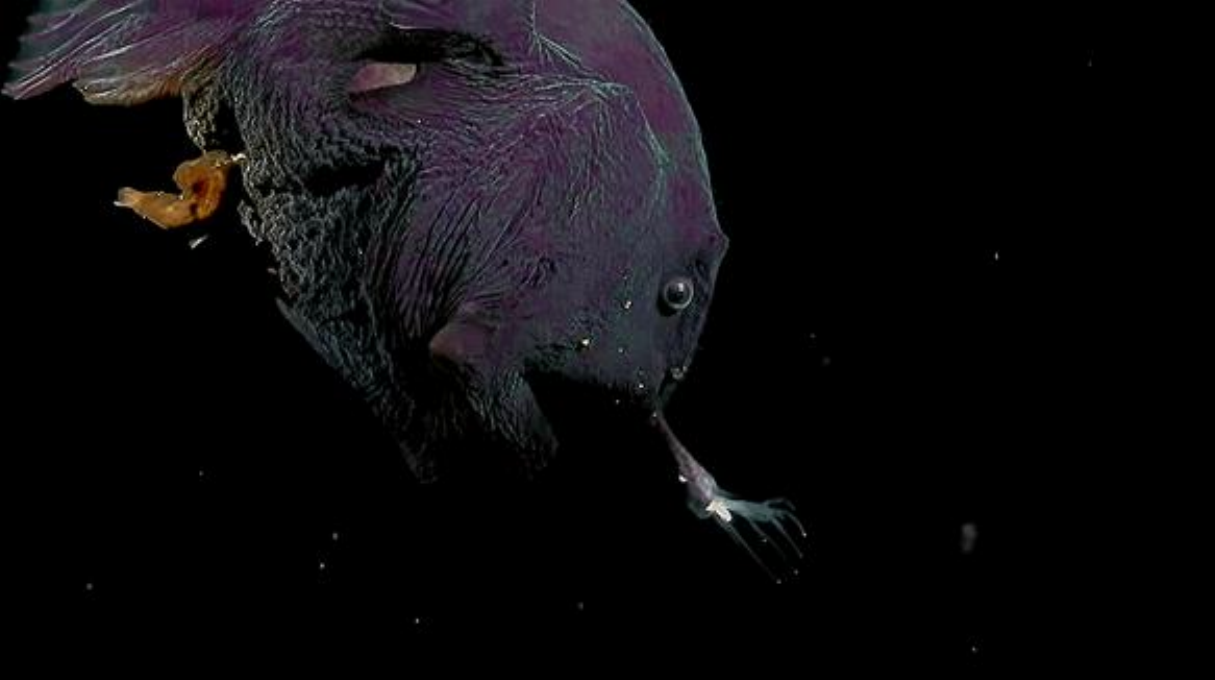} 
        \includegraphics[width=\linewidth]{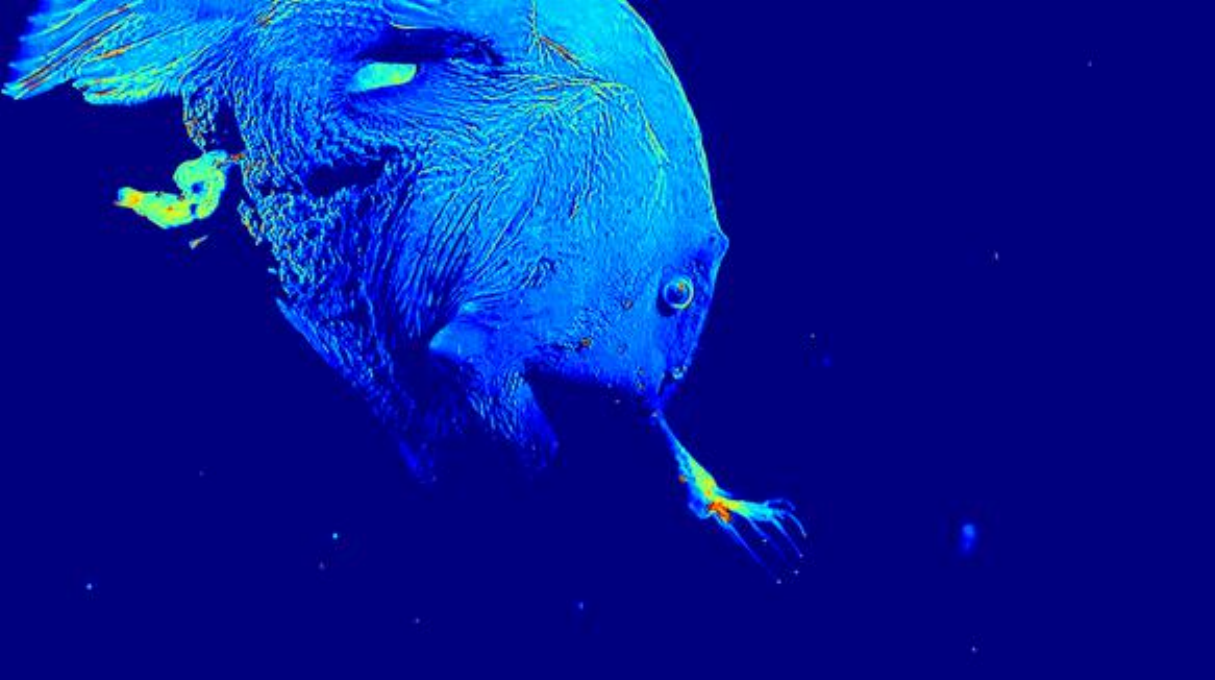} 
		\caption{\footnotesize ICSP }
	\end{subfigure}
    \begin{subfigure}{0.105\linewidth}
		\centering
		\includegraphics[width=\linewidth]{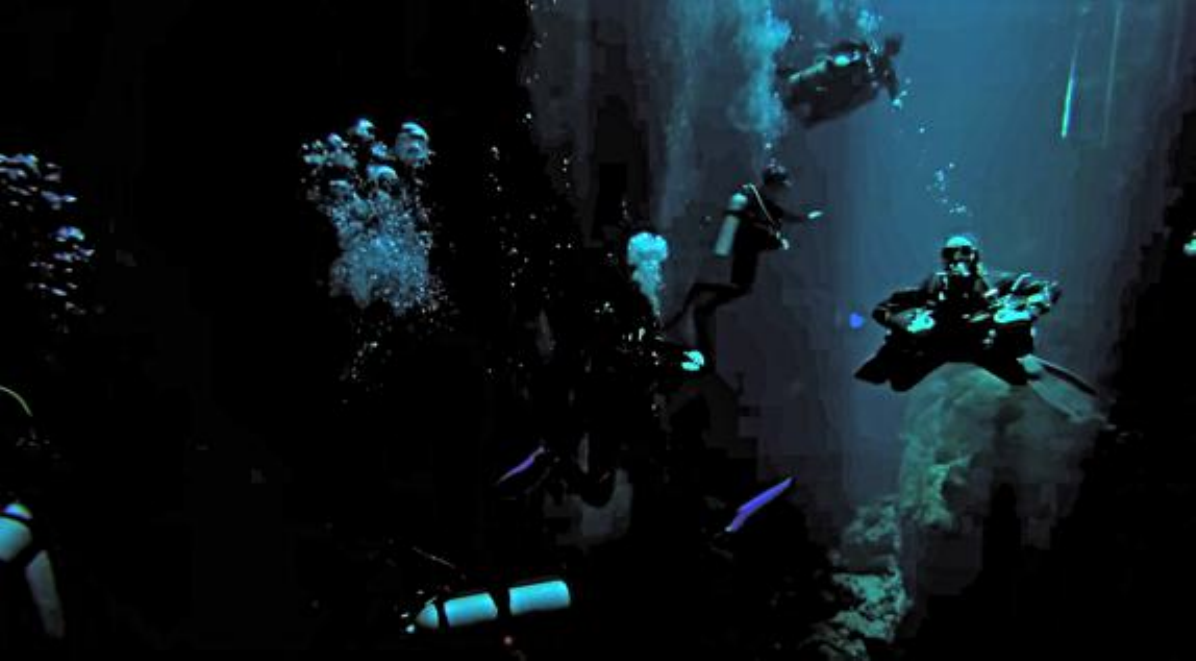} 
        \includegraphics[width=\linewidth]{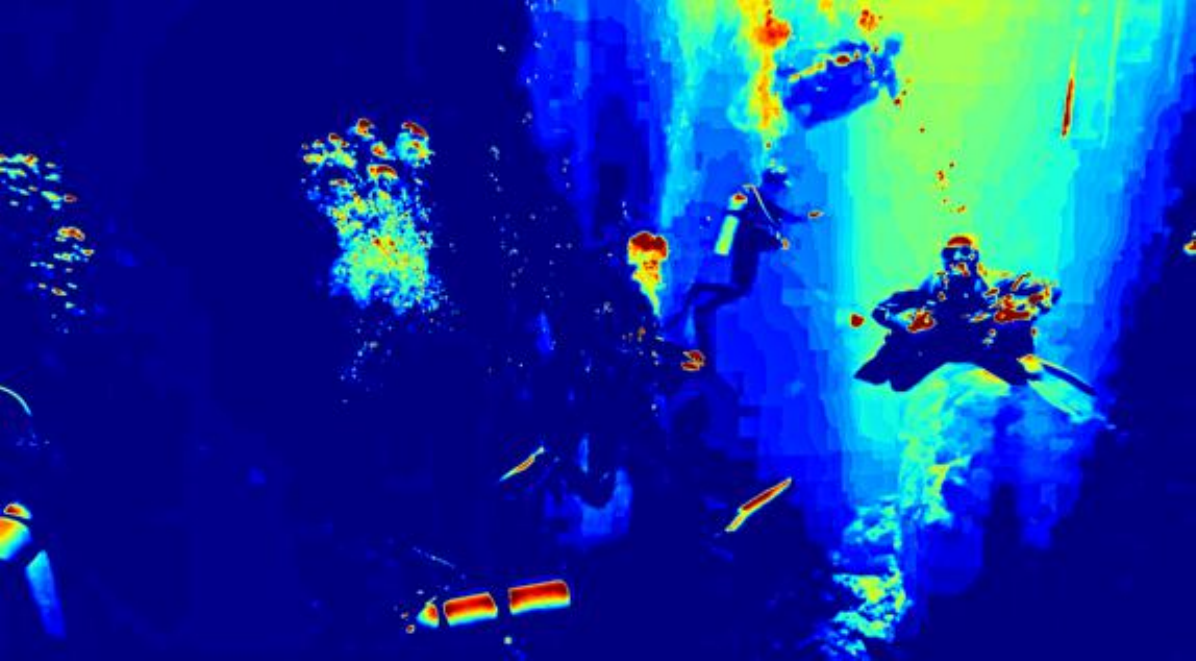} 
		\includegraphics[width=\linewidth]{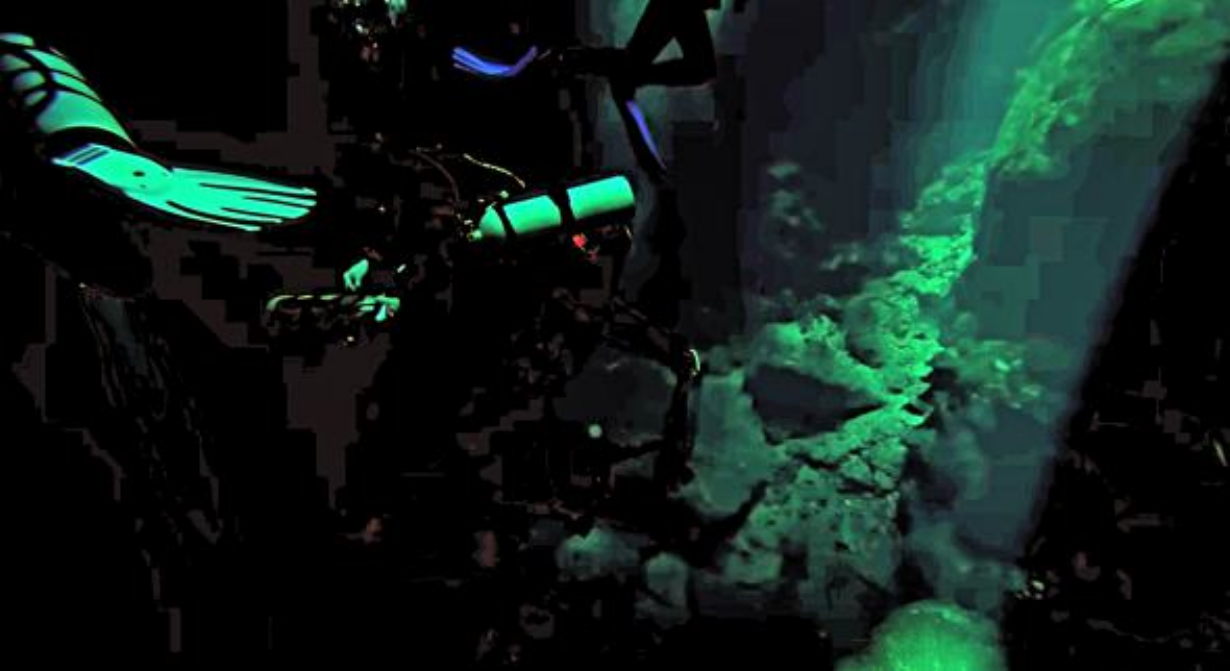} 
        \includegraphics[width=\linewidth]{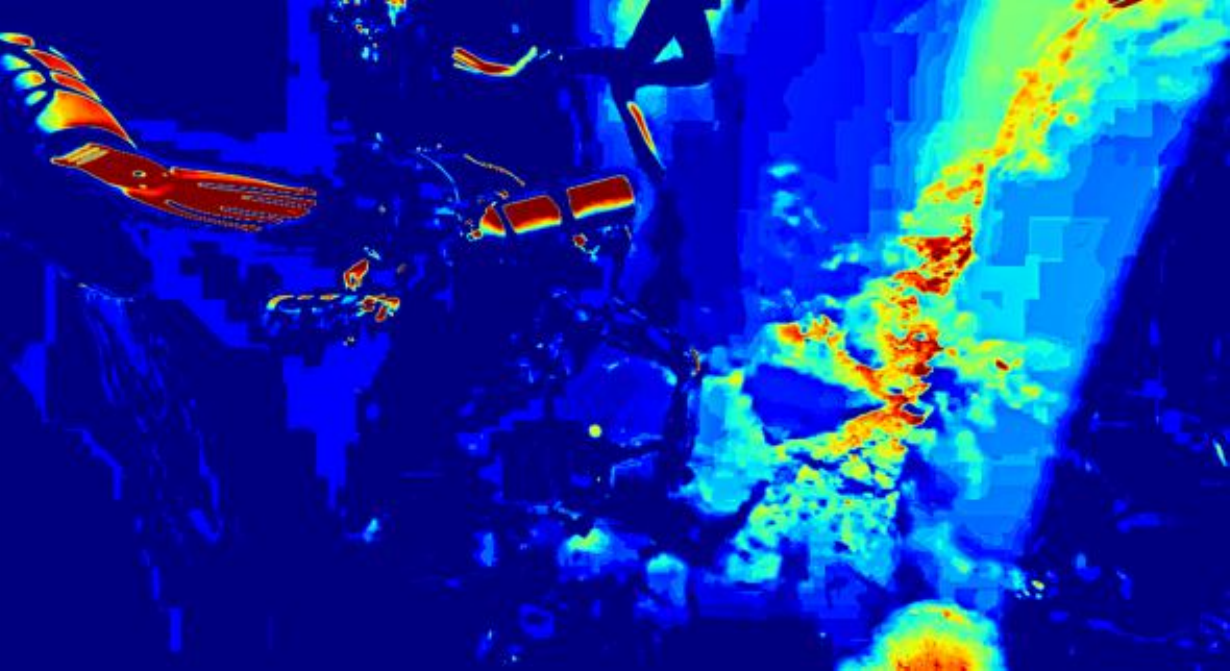} 
		\includegraphics[width=\linewidth]{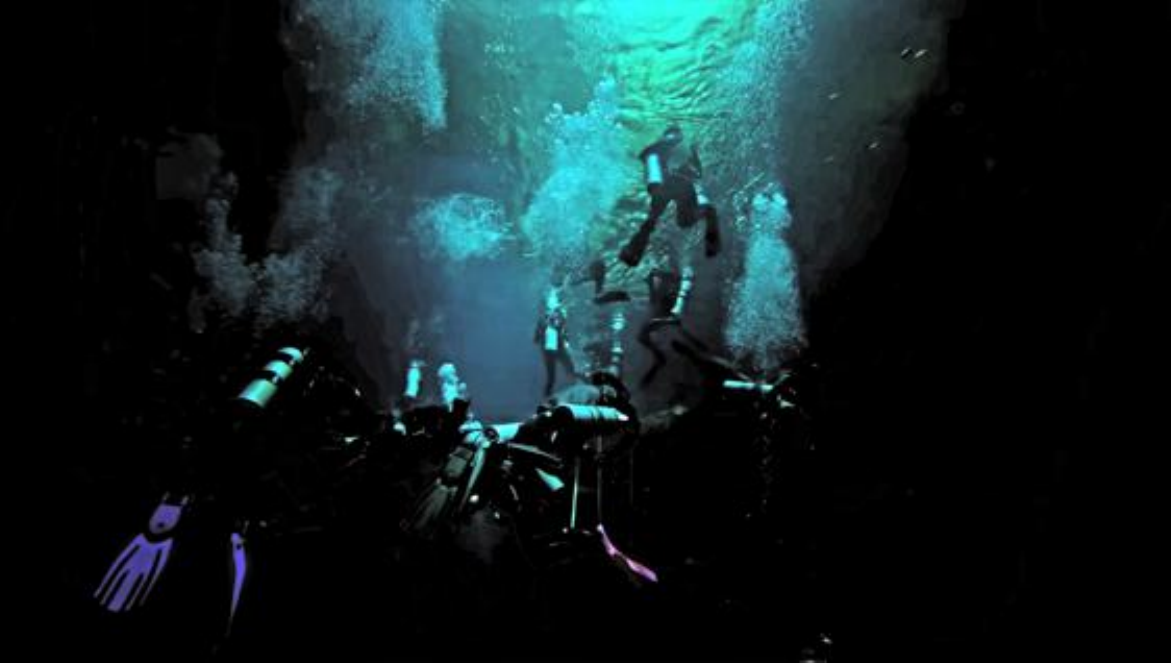} 
        \includegraphics[width=\linewidth]{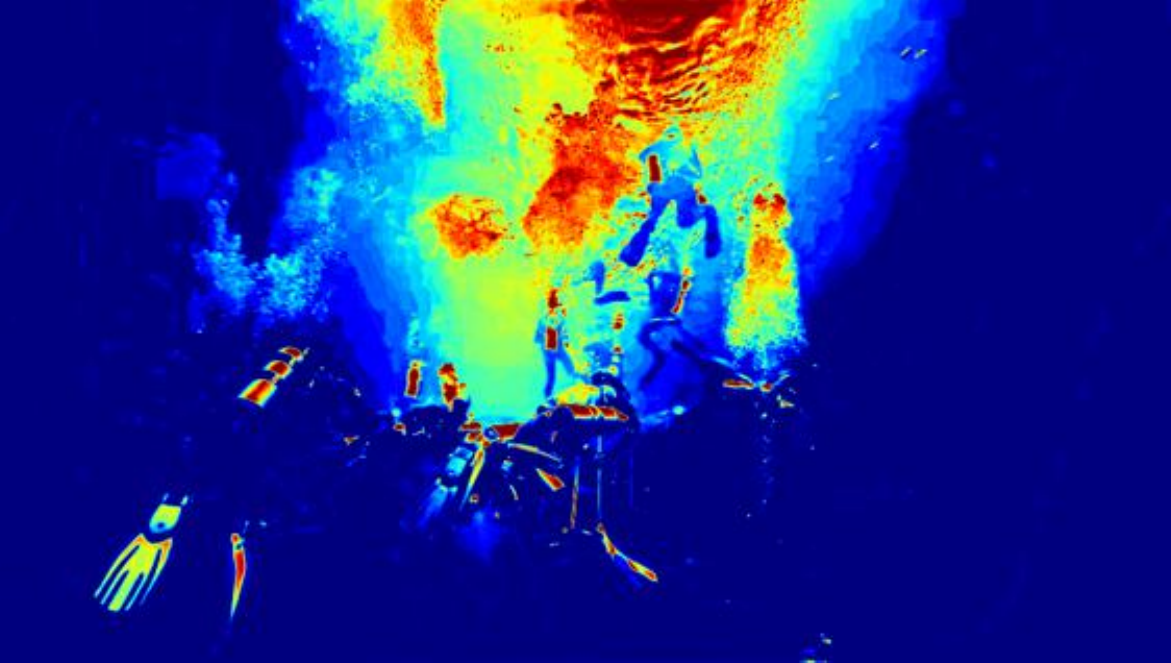} 
		\includegraphics[width=\linewidth]{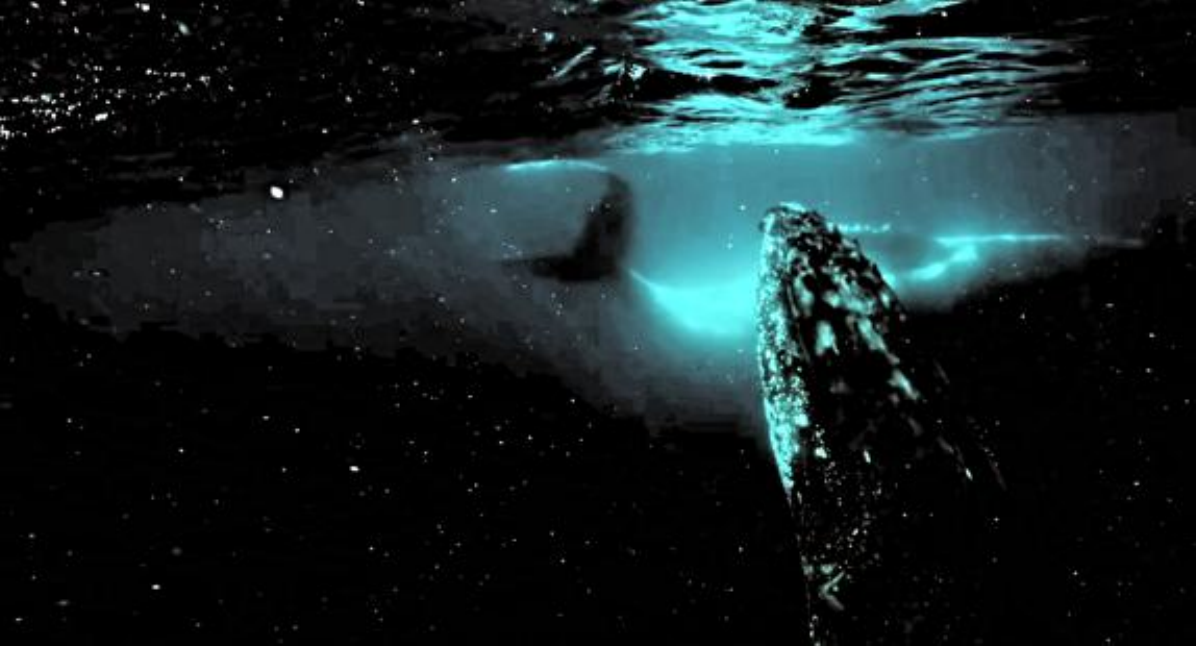} 
        \includegraphics[width=\linewidth]{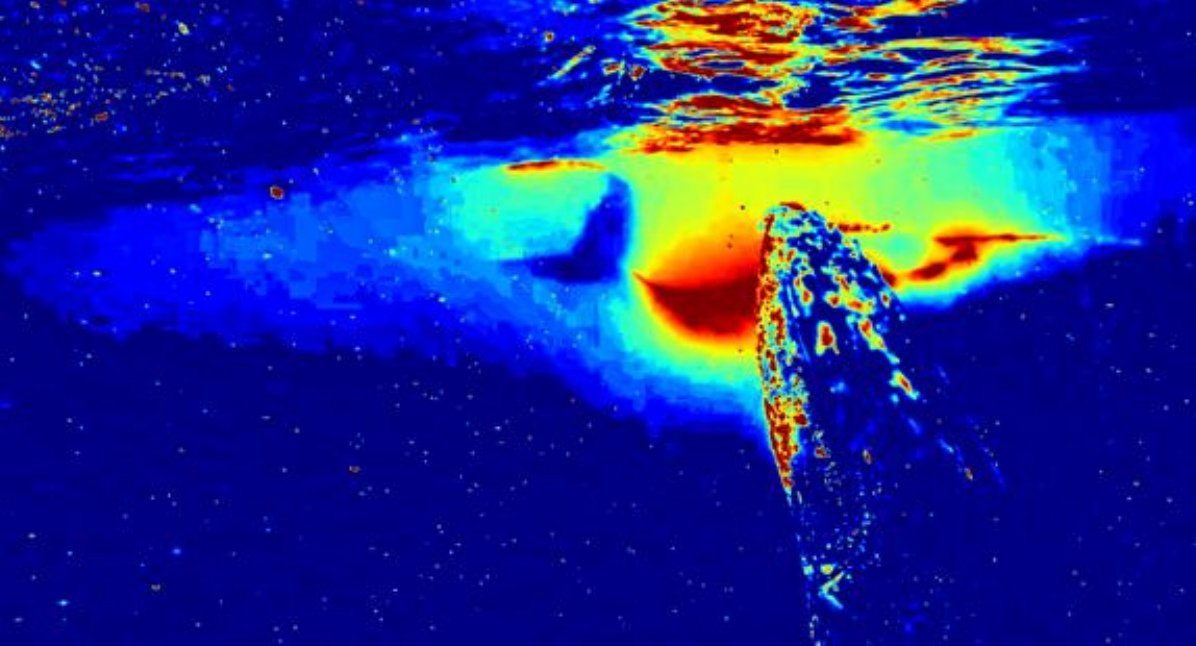} 
		\includegraphics[width=\linewidth]{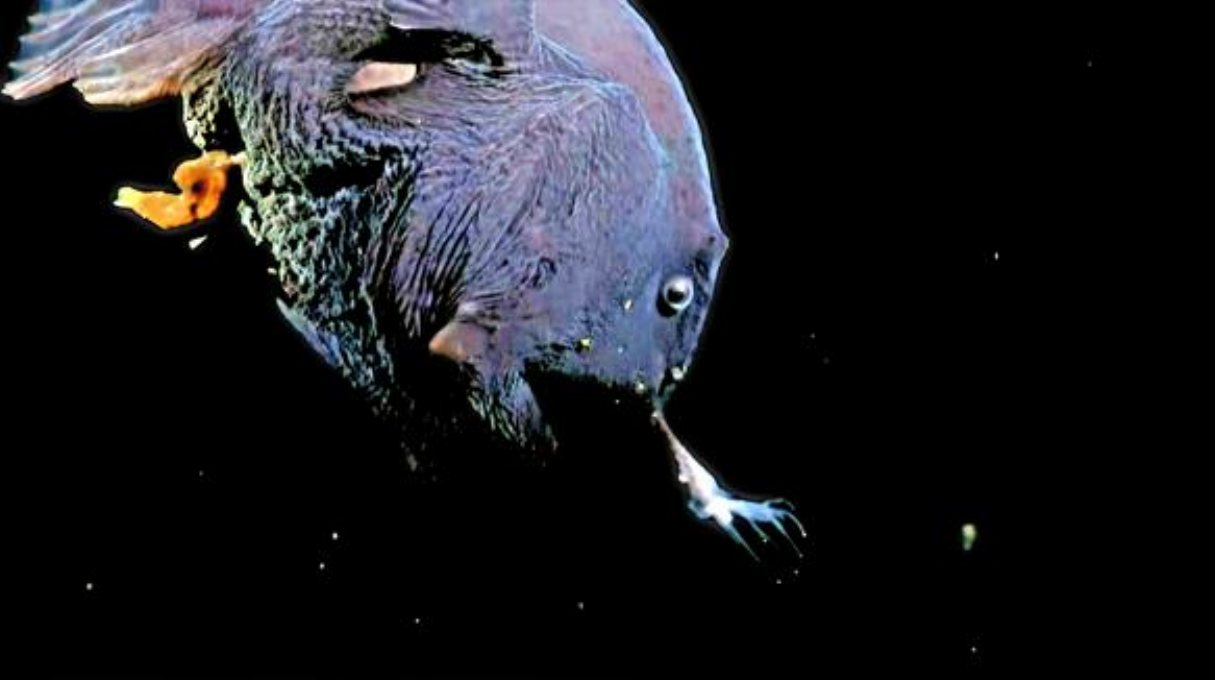} 
        \includegraphics[width=\linewidth]{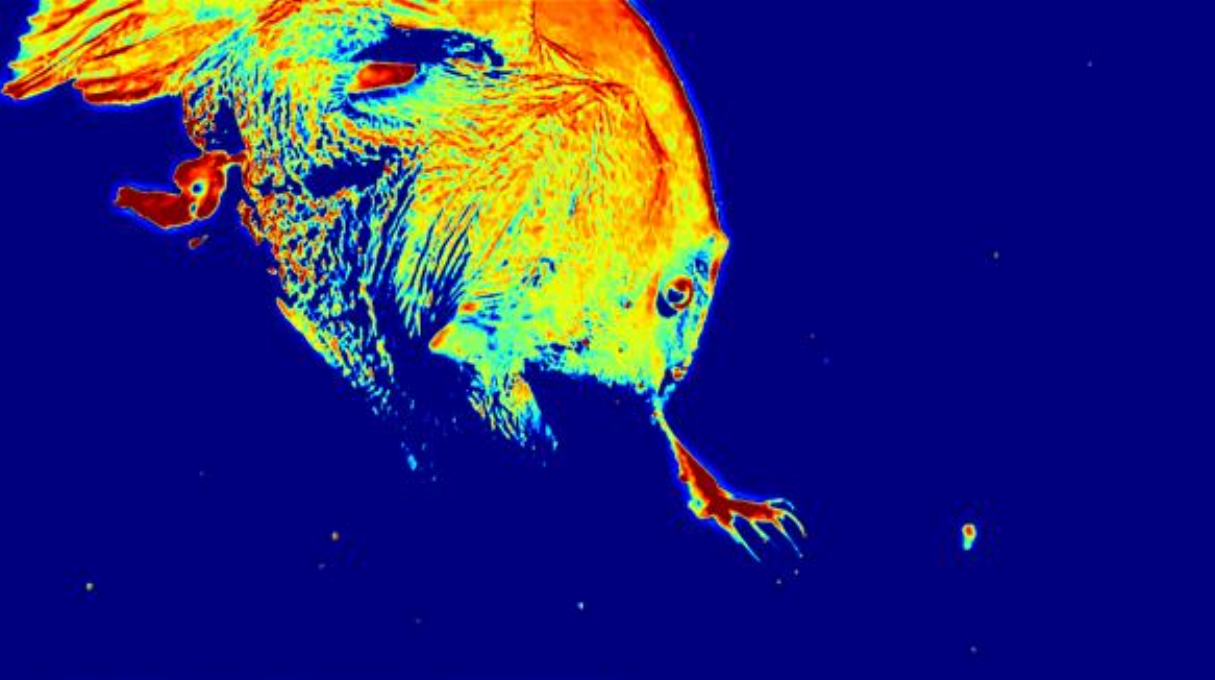} 
		\caption{\footnotesize PCDE}
	\end{subfigure}
	\begin{subfigure}{0.105\linewidth}
		\centering
		\includegraphics[width=\linewidth]{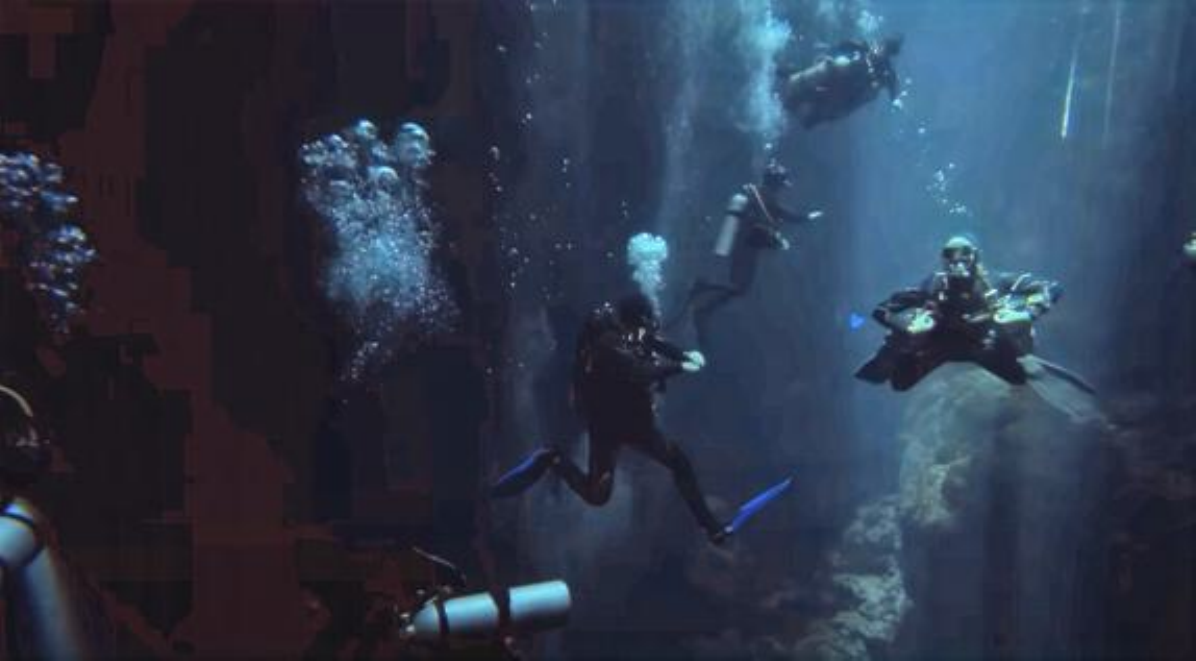}
        \includegraphics[width=\linewidth]{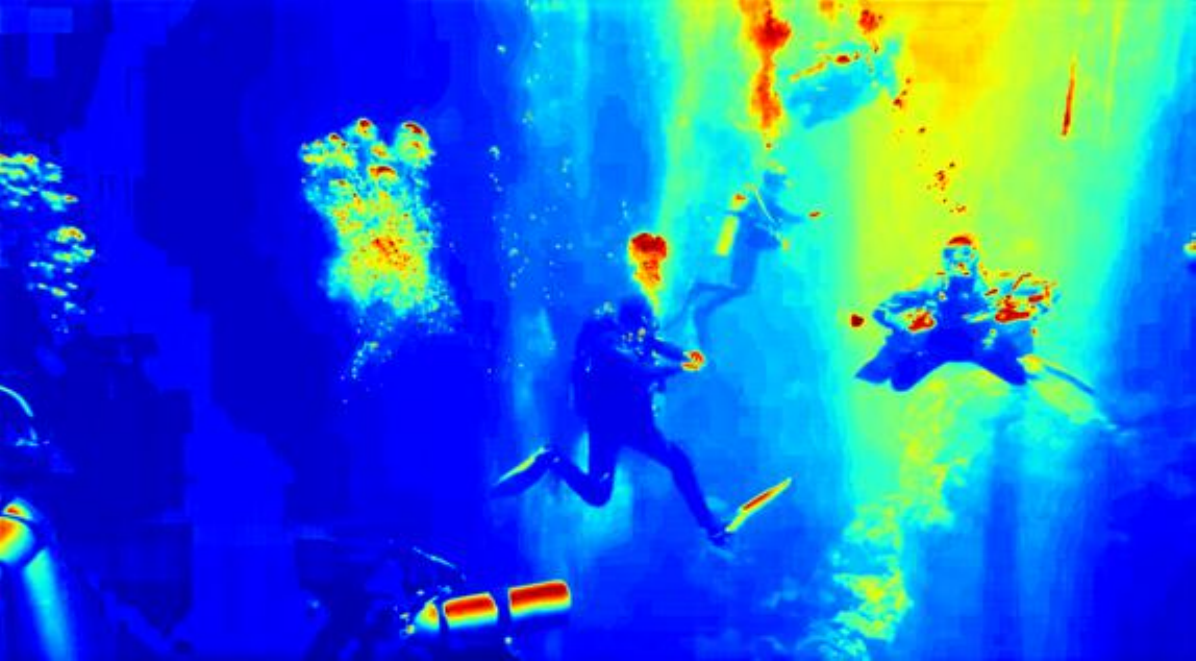}
		\includegraphics[width=\linewidth]{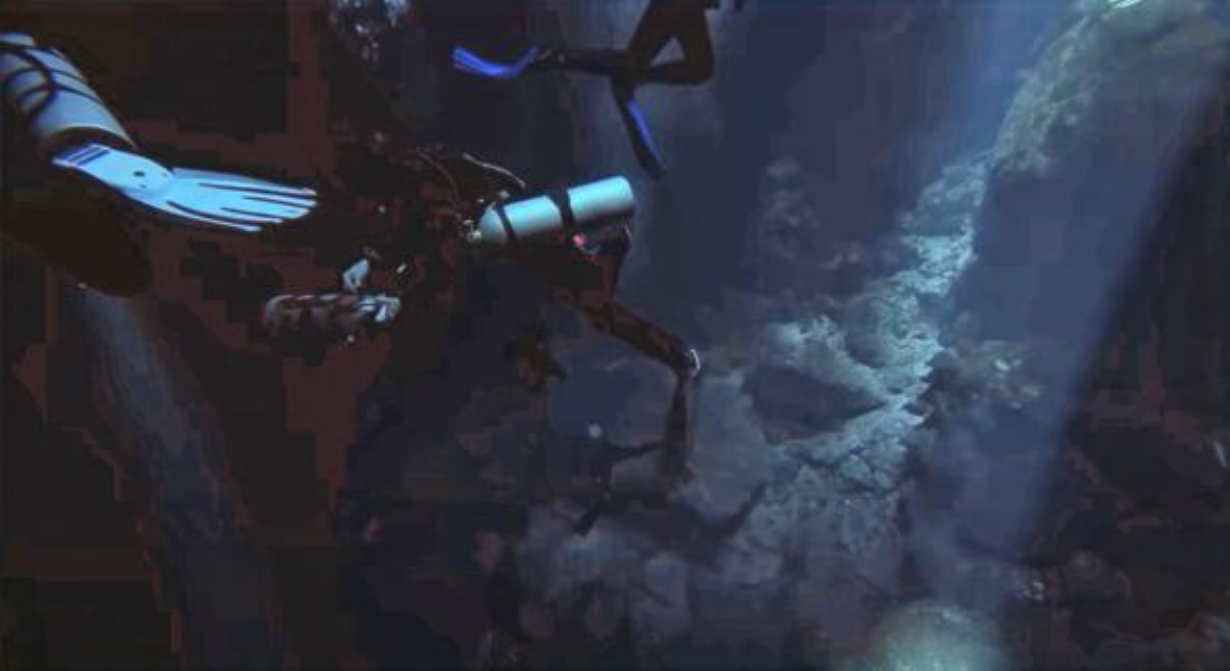}
        \includegraphics[width=\linewidth]{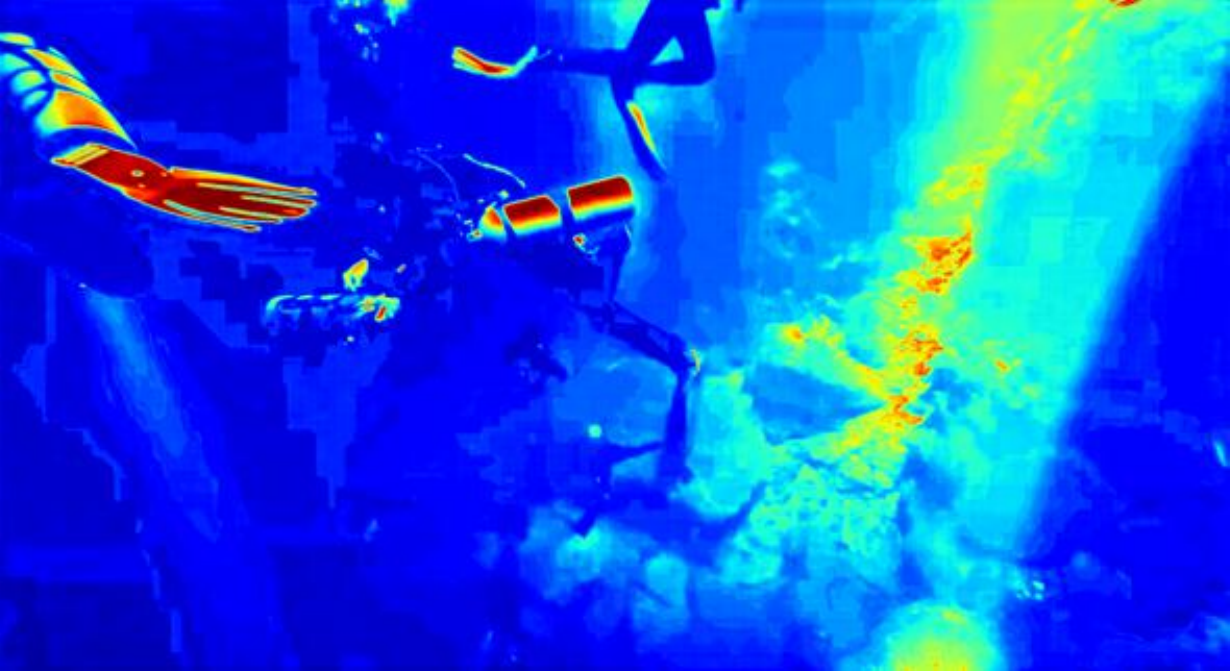}
        \includegraphics[width=\linewidth]{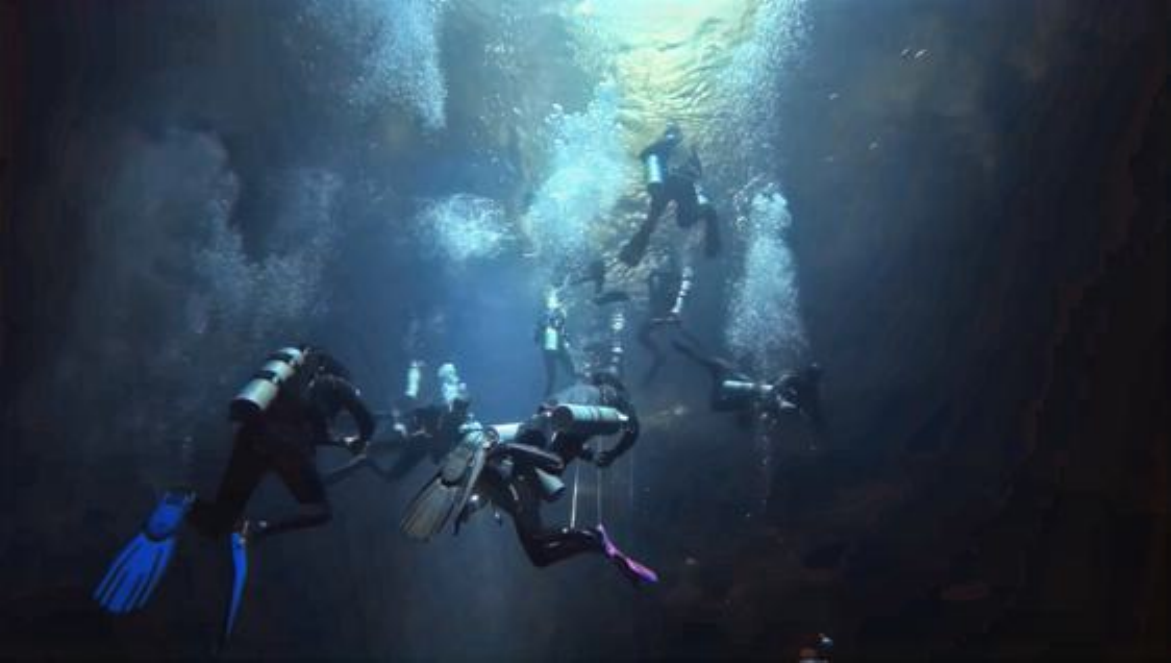}
        \includegraphics[width=\linewidth]{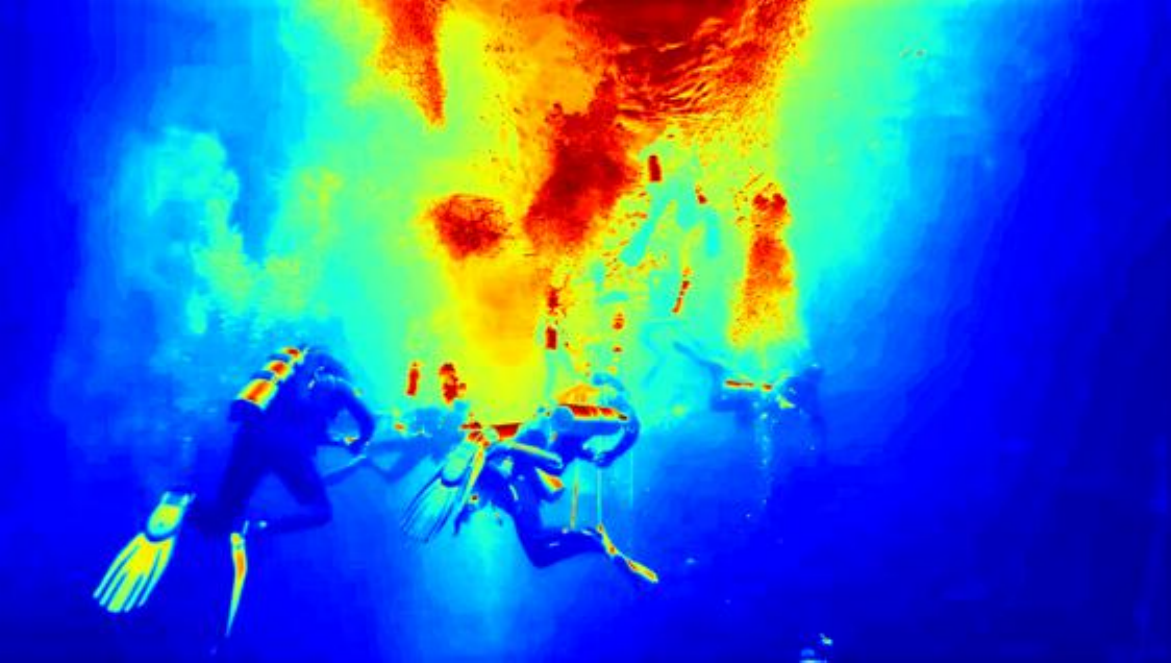}
        \includegraphics[width=\linewidth]{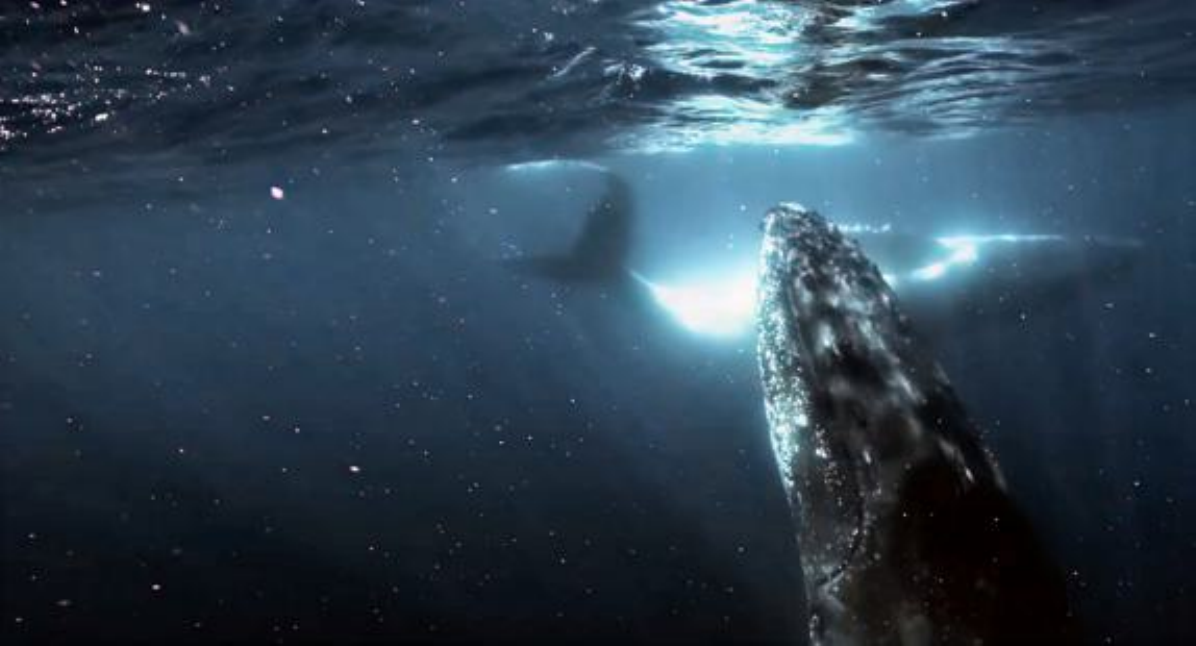}
        \includegraphics[width=\linewidth]{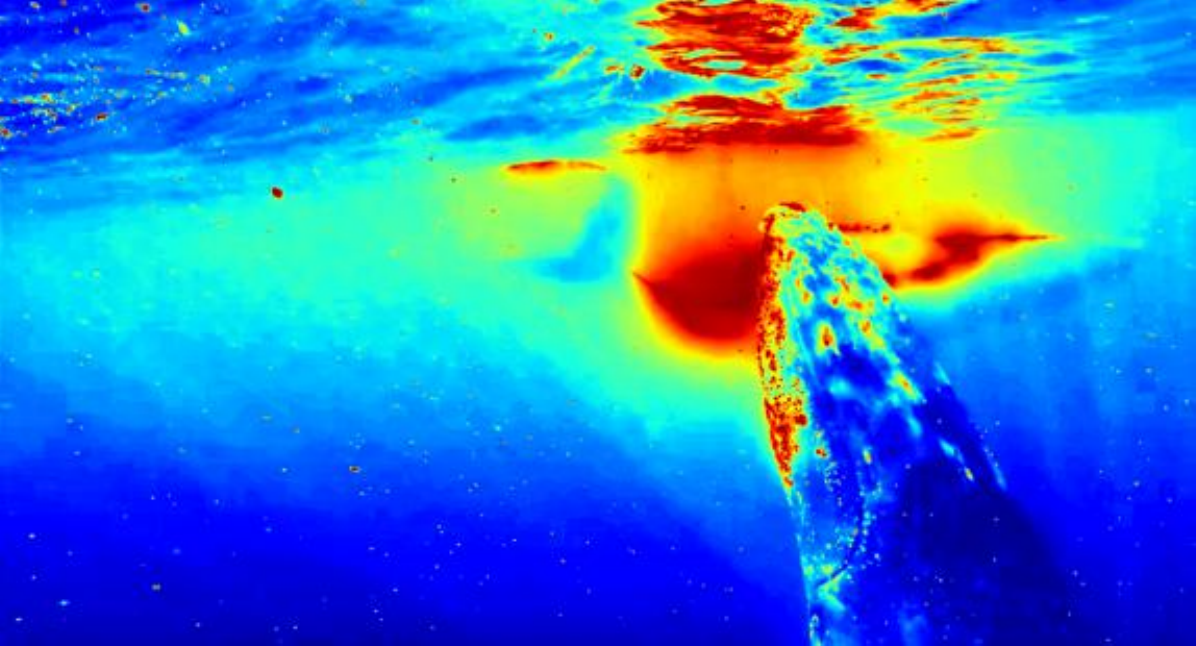}
        \includegraphics[width=\linewidth]{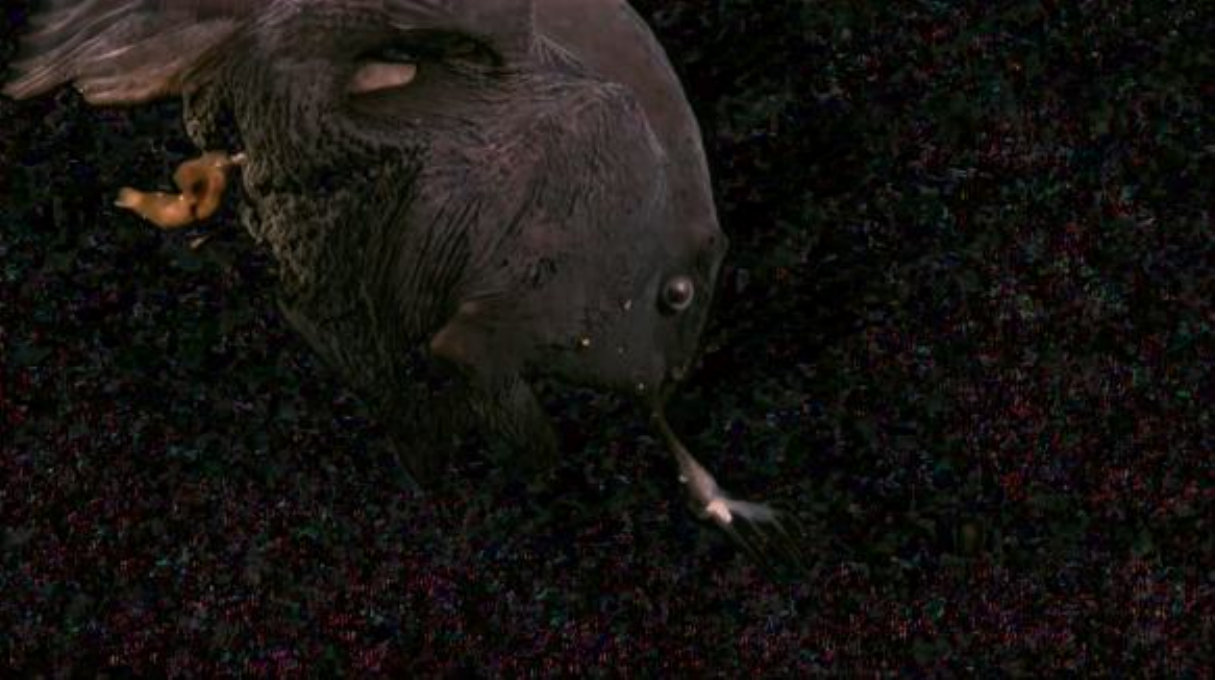}
        \includegraphics[width=\linewidth]{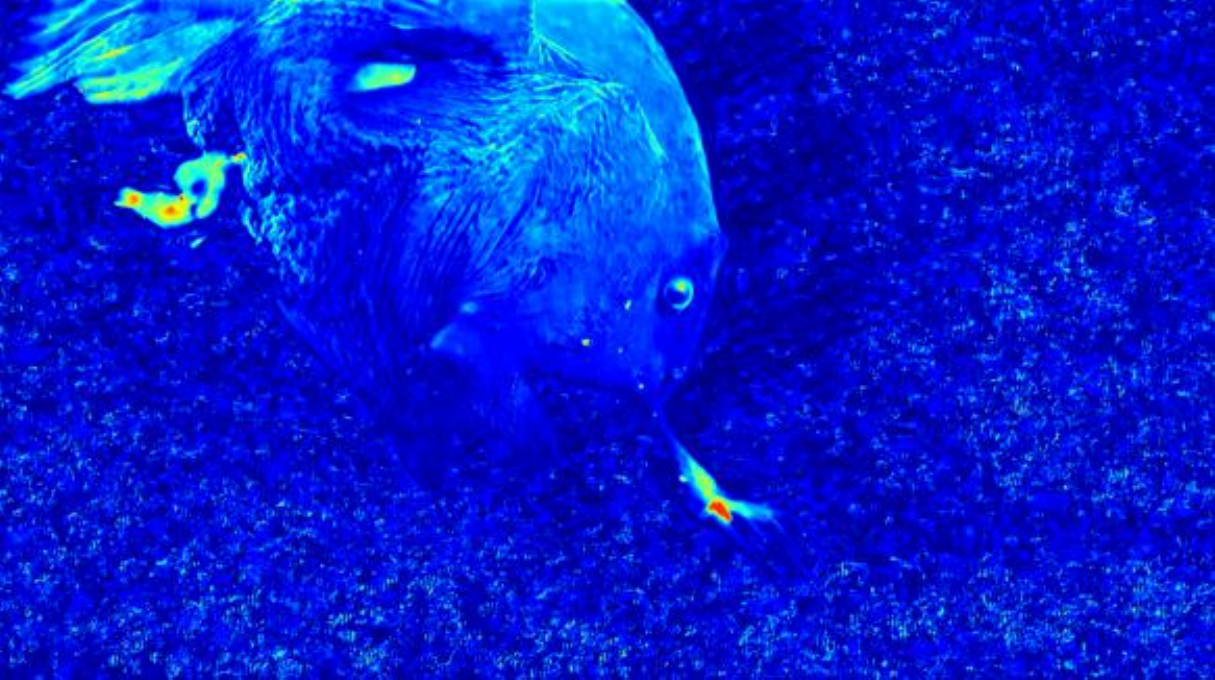}
        \caption{\footnotesize UDAformer }
	\end{subfigure}
	\begin{subfigure}{0.105\linewidth}
		\centering
		\includegraphics[width=\linewidth]{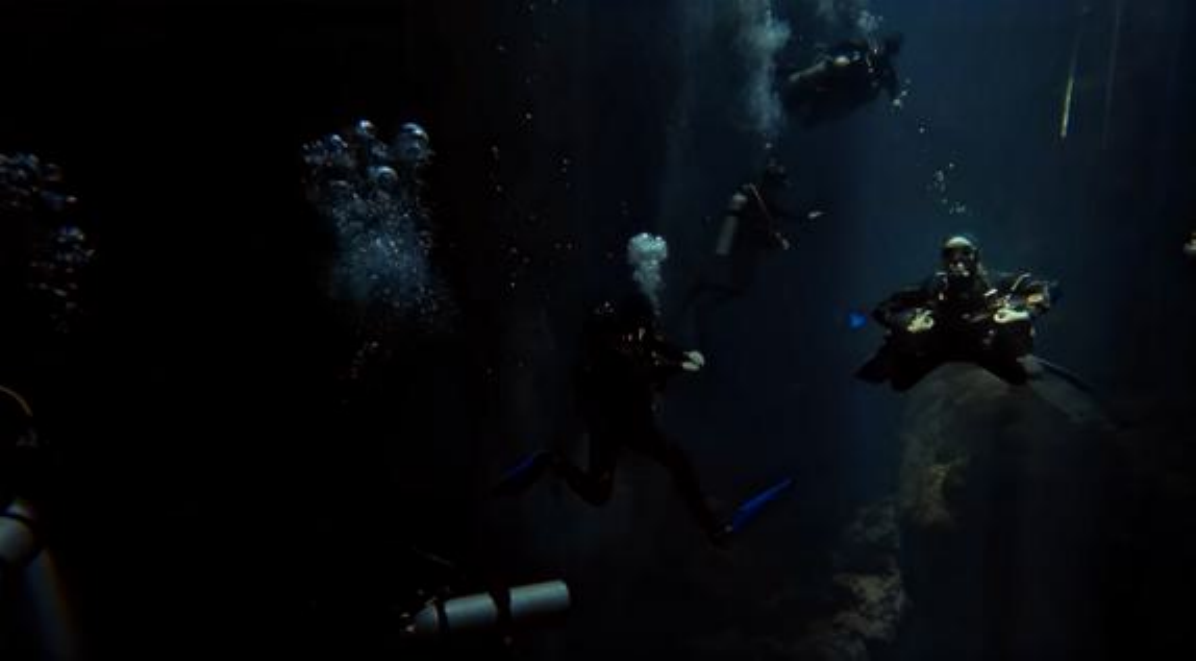}
        \includegraphics[width=\linewidth]{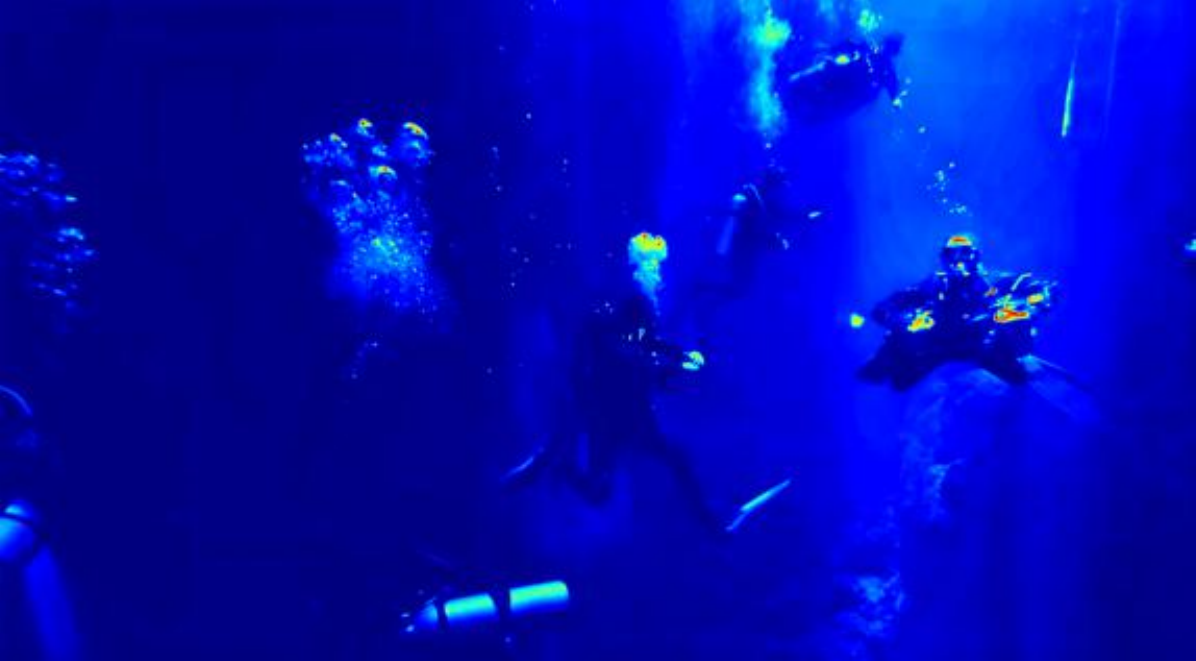}
        \includegraphics[width=\linewidth]{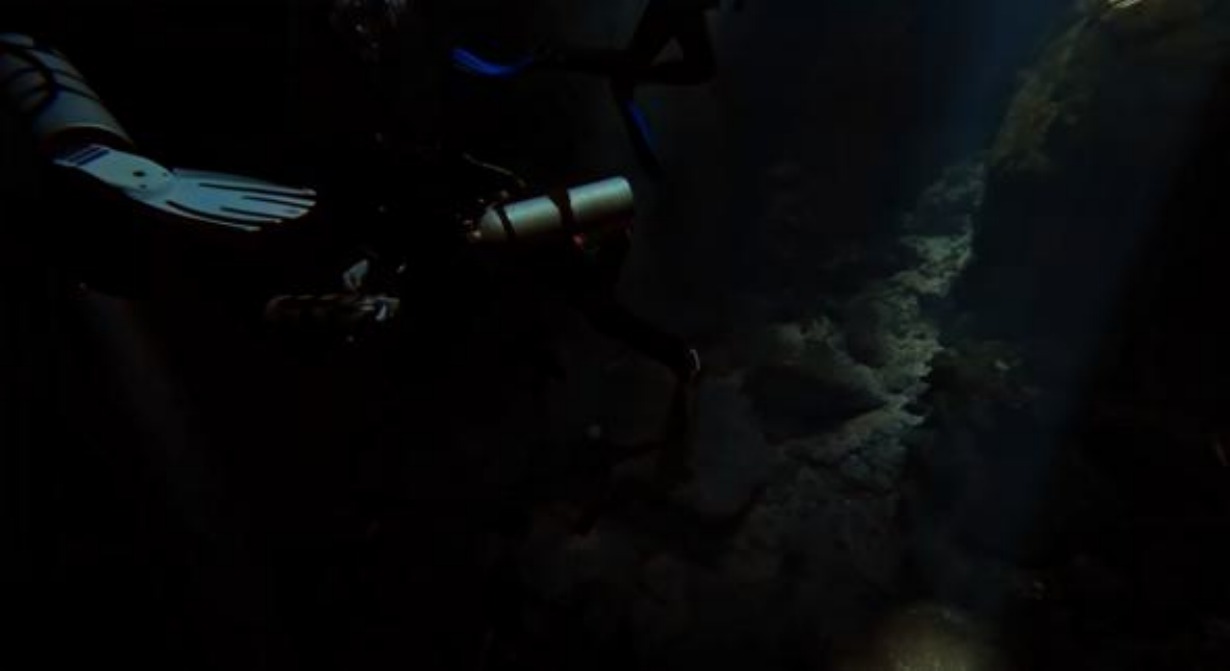}
        \includegraphics[width=\linewidth]{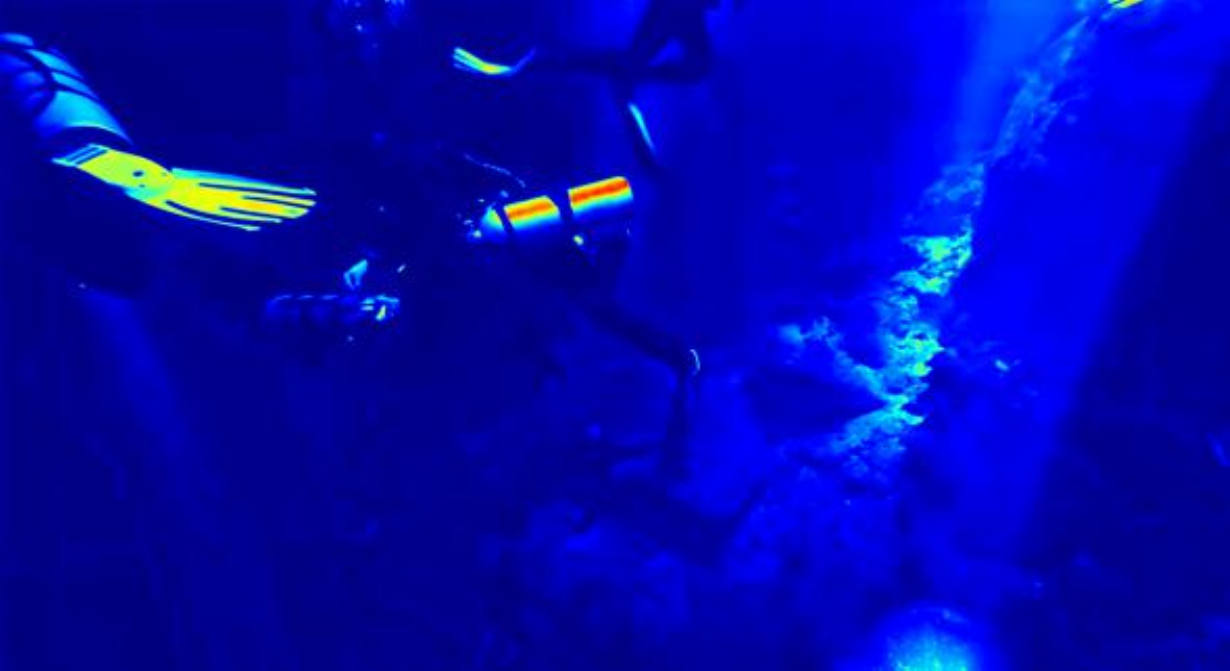}
        \includegraphics[width=\linewidth]{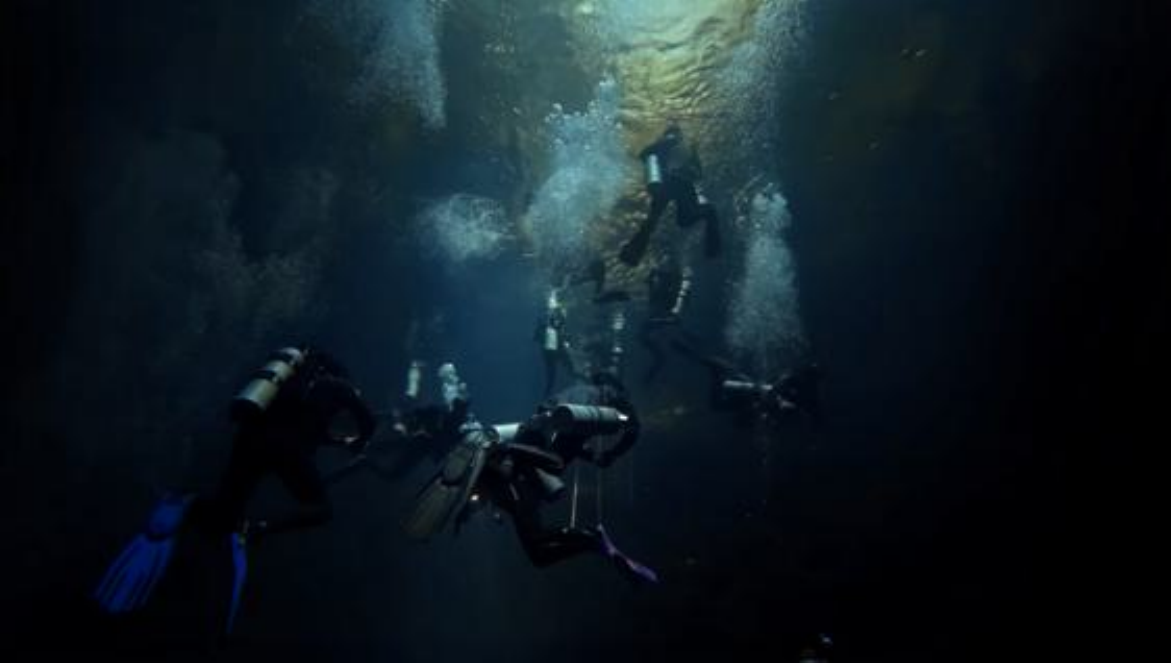}
        \includegraphics[width=\linewidth]{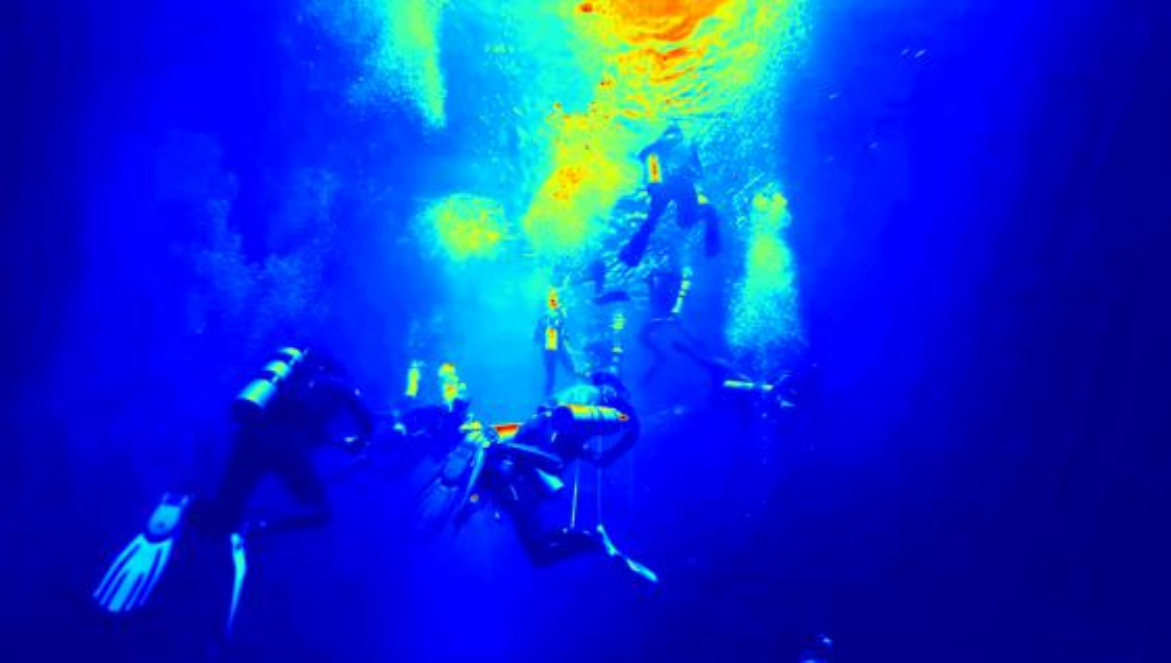}
        \includegraphics[width=\linewidth]{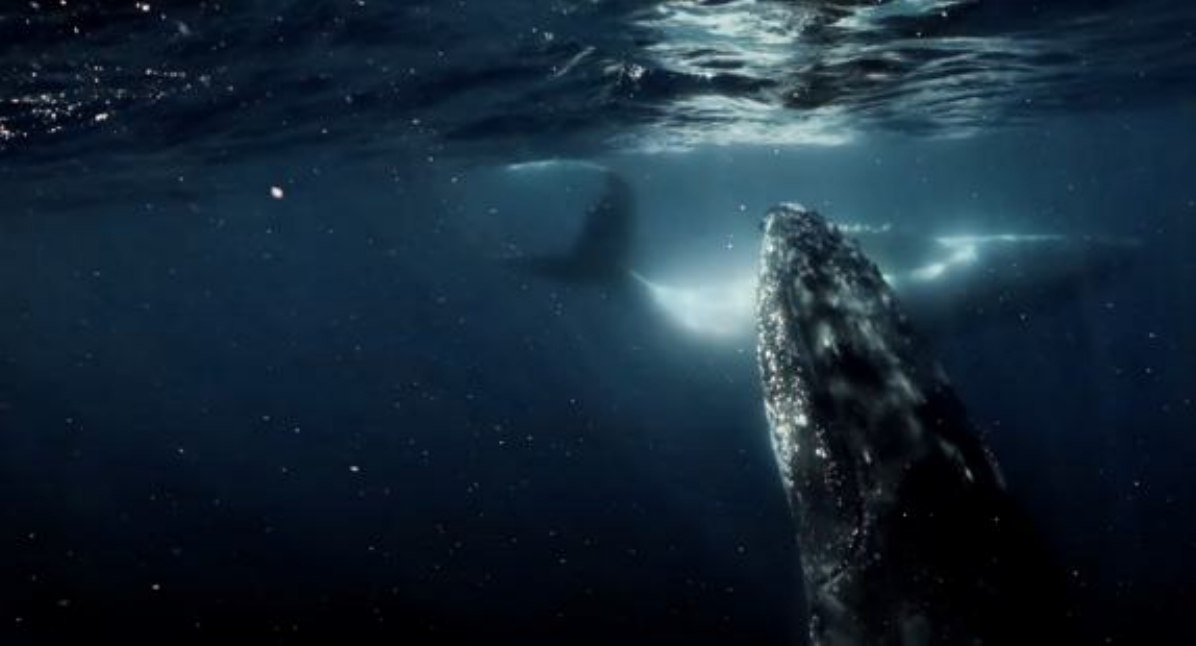}
        \includegraphics[width=\linewidth]{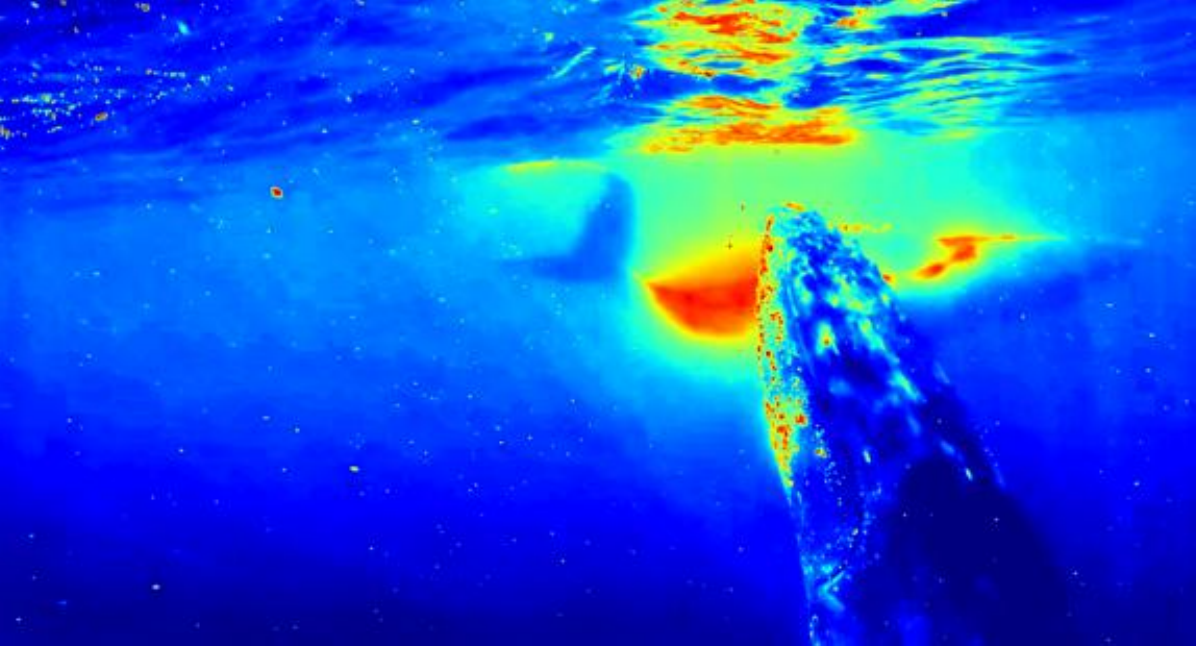}
        \includegraphics[width=\linewidth]{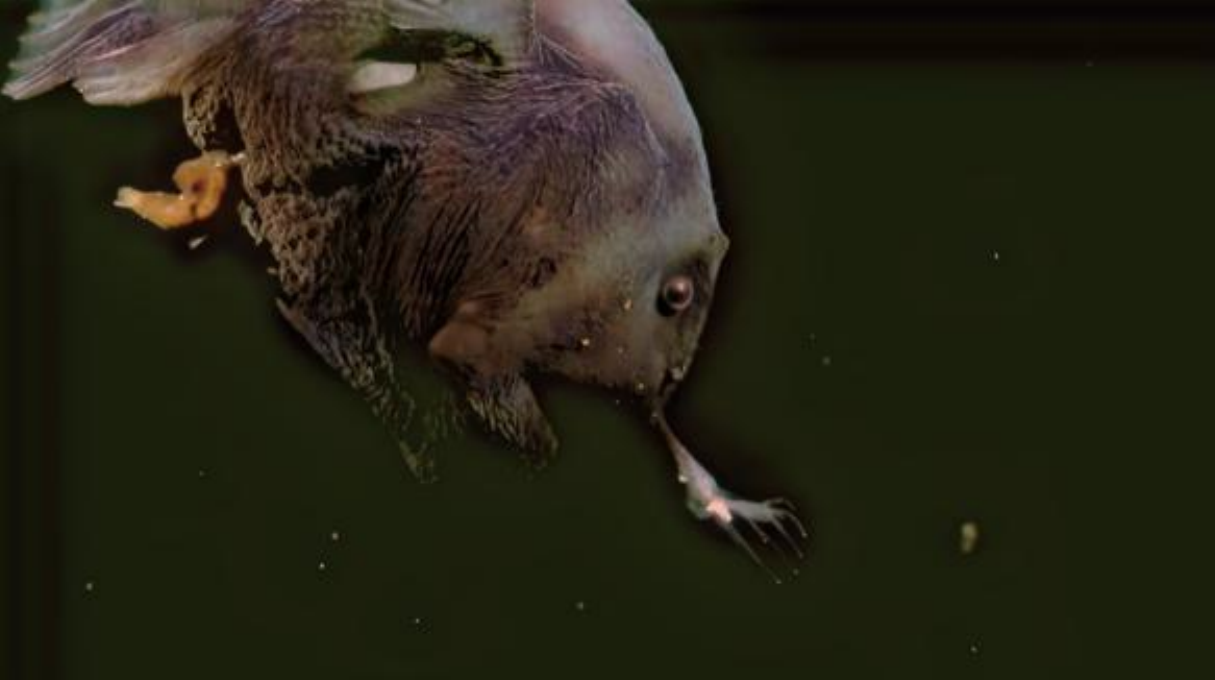}
        \includegraphics[width=\linewidth]{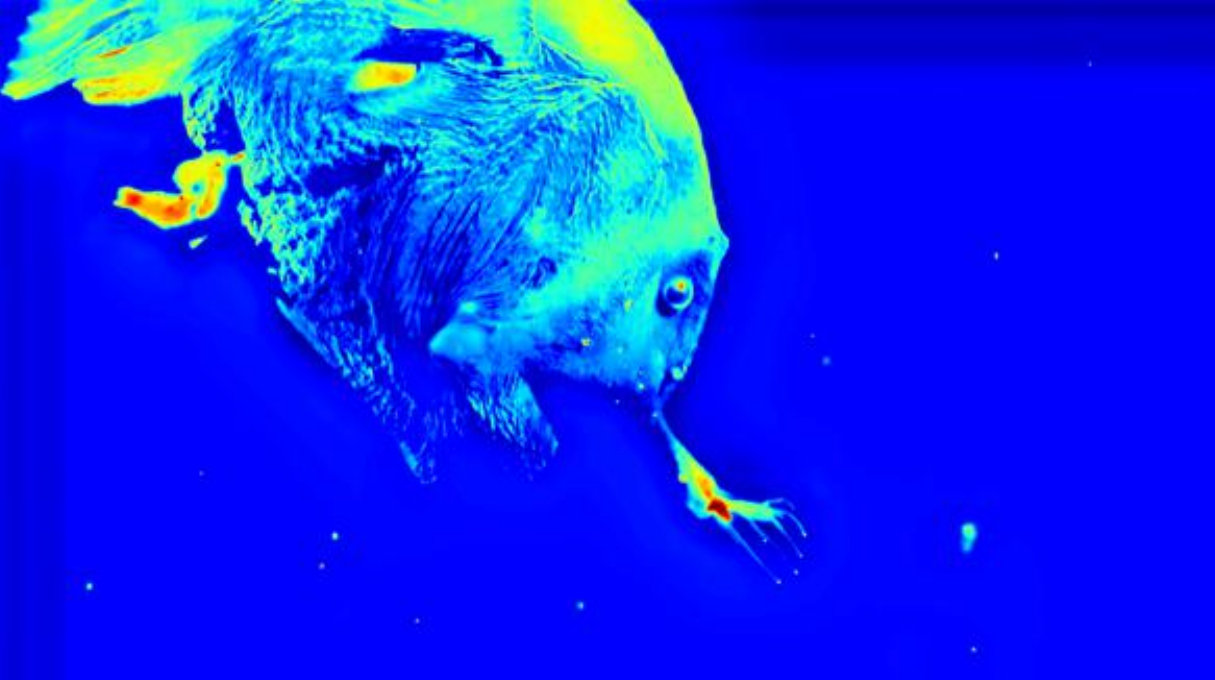}
		\caption{\footnotesize SMDR-IS}
	\end{subfigure}
    \begin{subfigure}{0.105\linewidth}
		\centering
		\includegraphics[width=\linewidth]{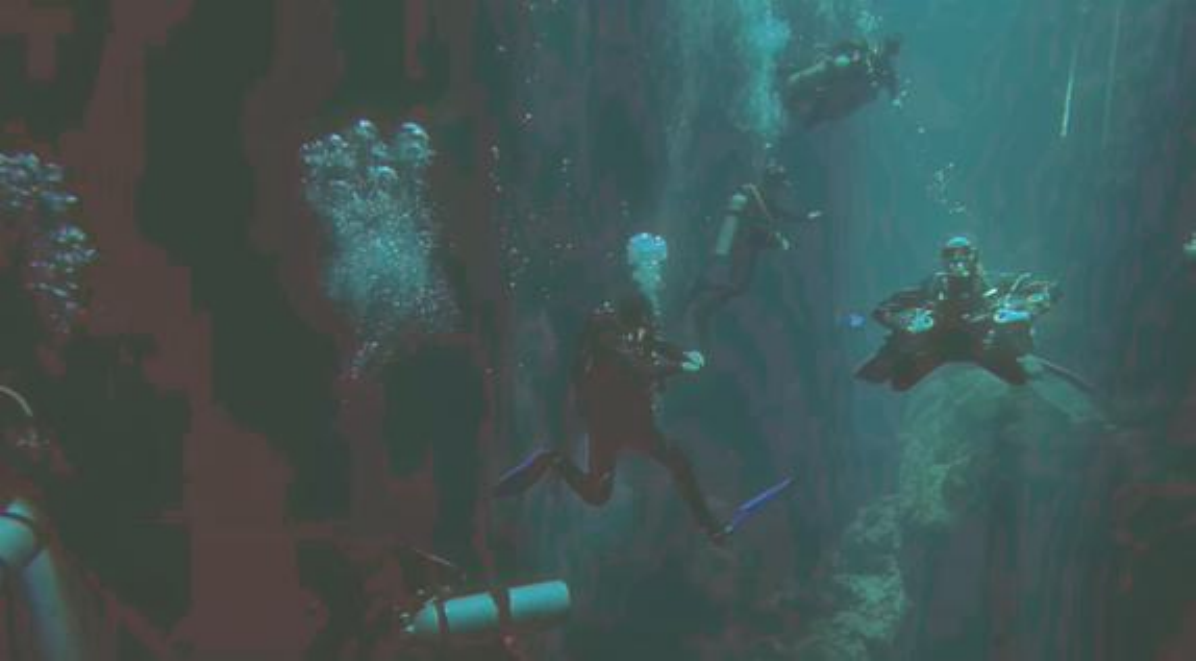} 
        \includegraphics[width=\linewidth]{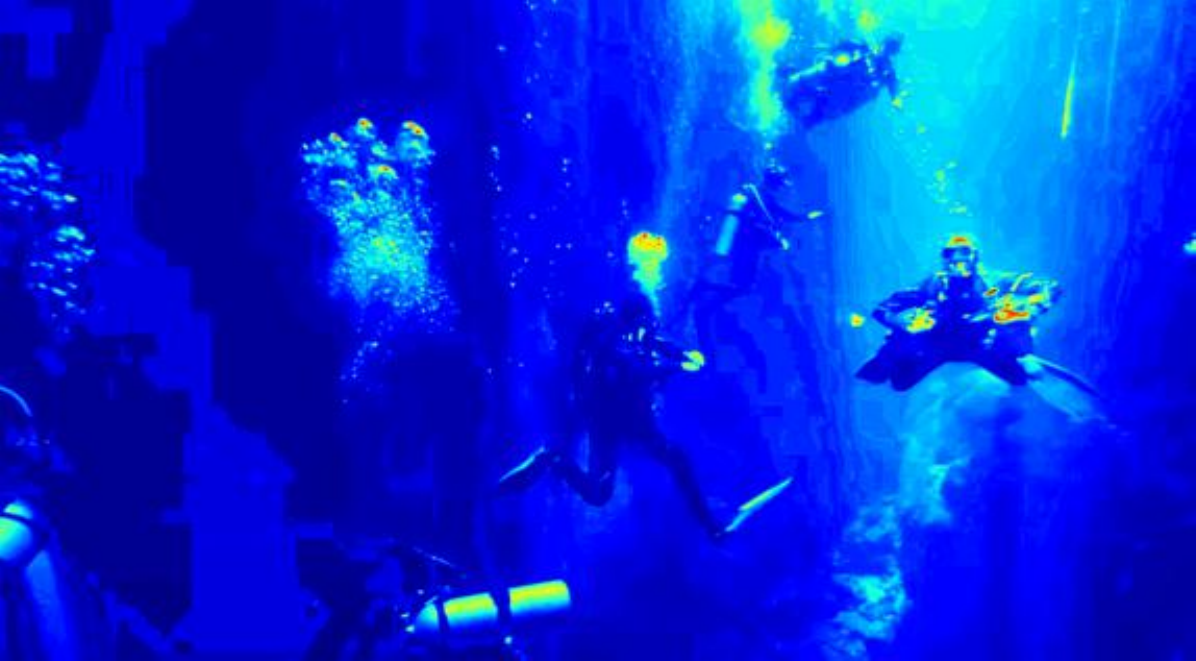}
        \includegraphics[width=\linewidth]{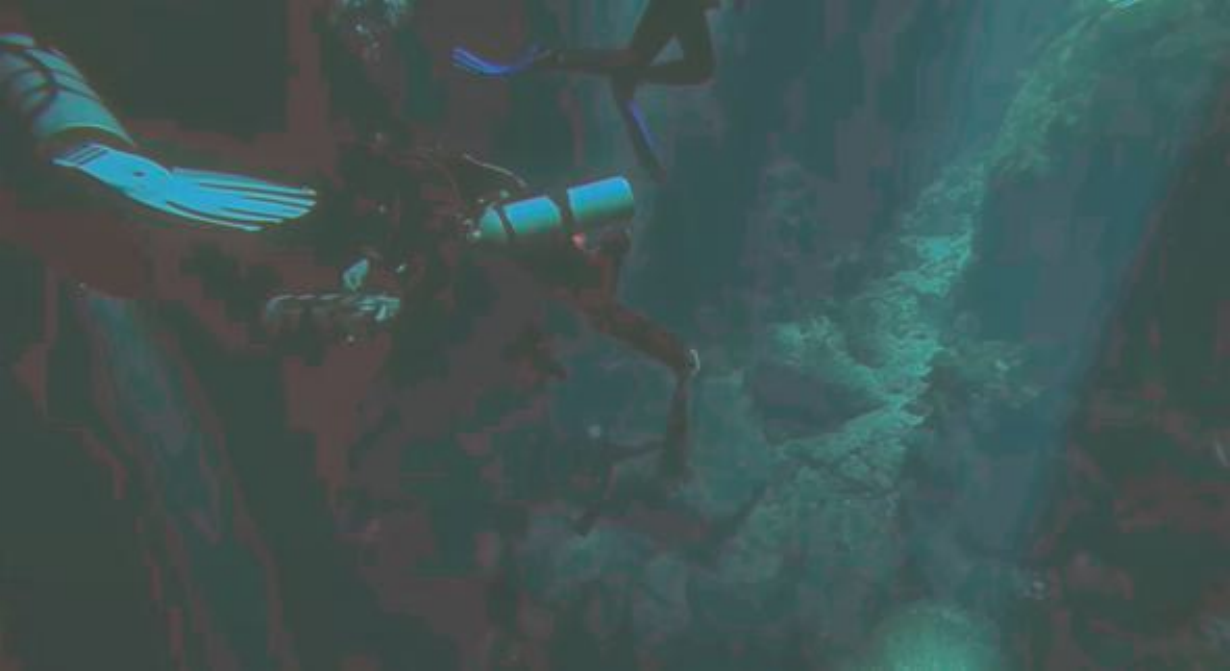} 
        \includegraphics[width=\linewidth]{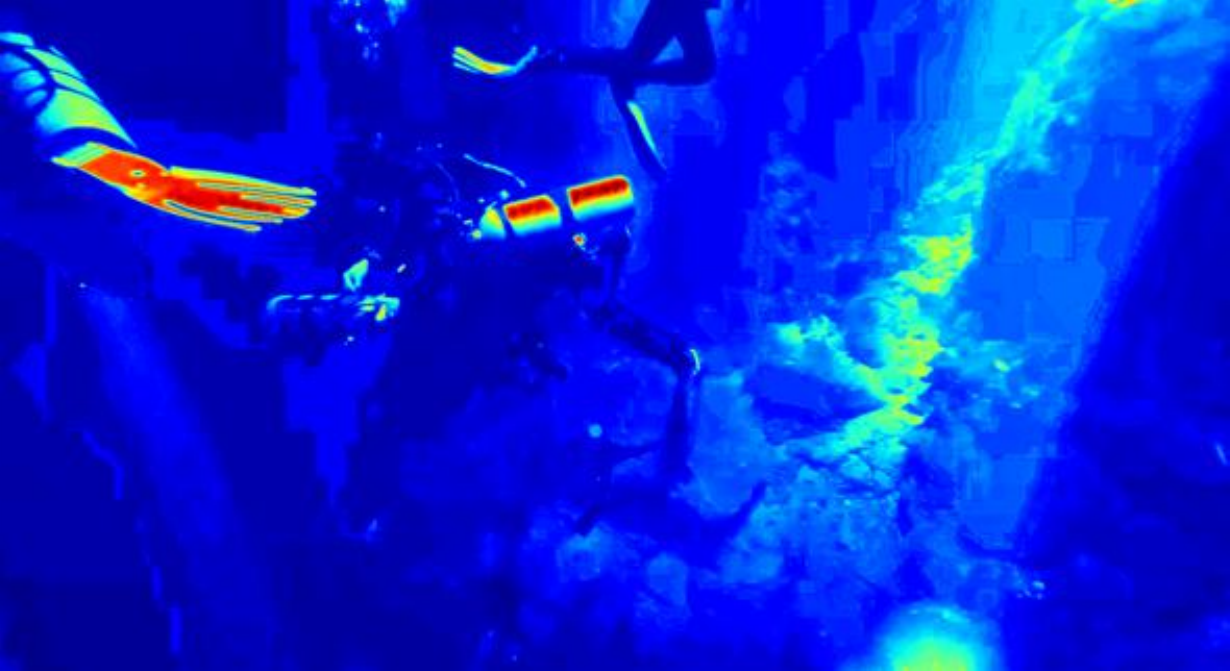}
        \includegraphics[width=\linewidth]{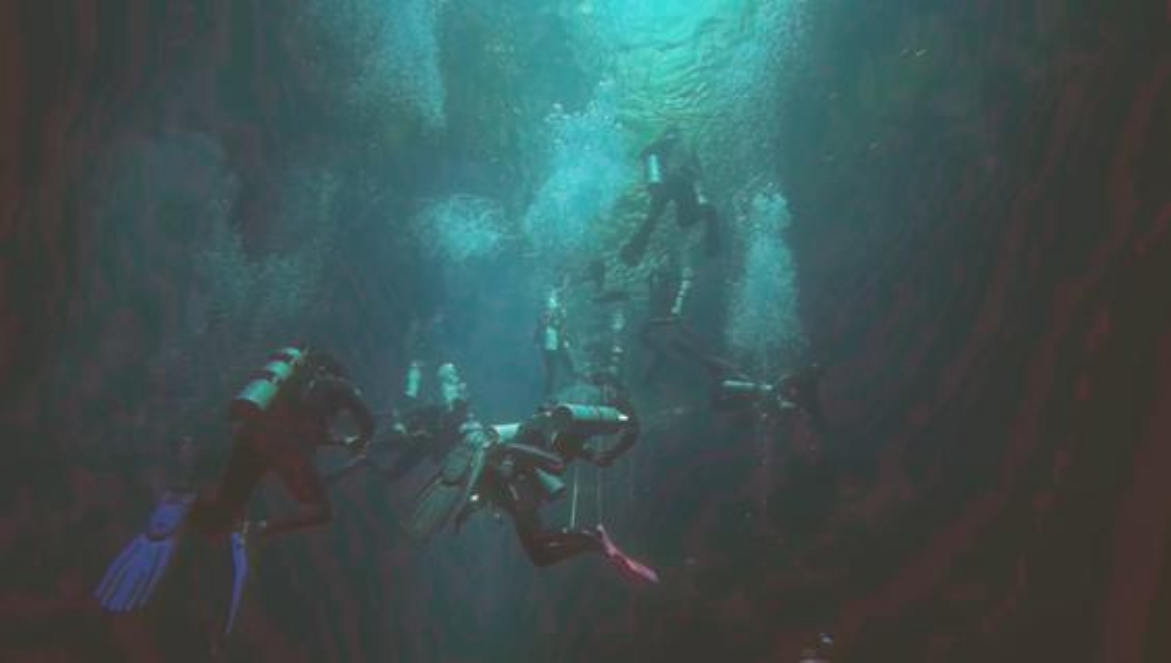} 
        \includegraphics[width=\linewidth]{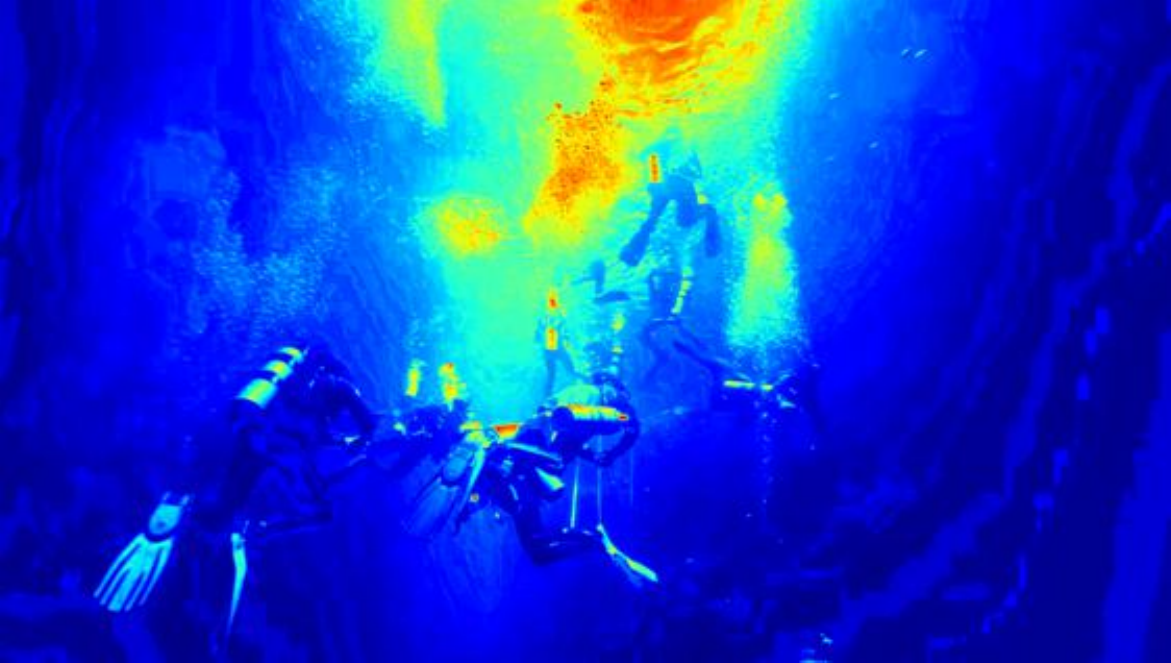}
        \includegraphics[width=\linewidth]{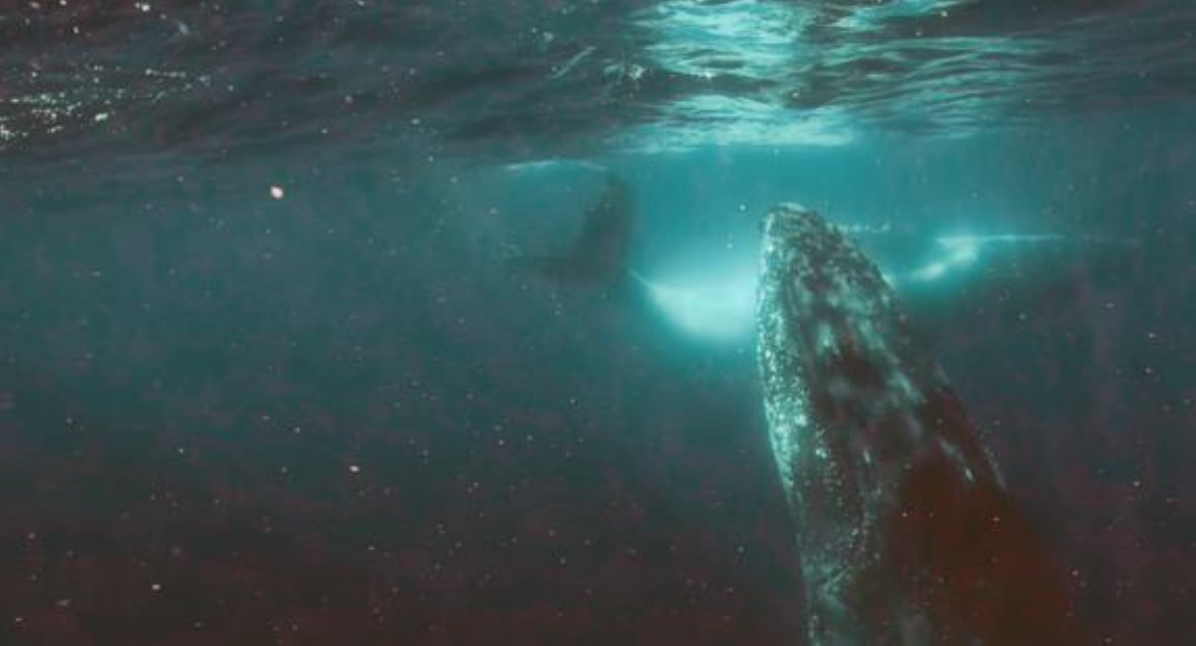} 
        \includegraphics[width=\linewidth]{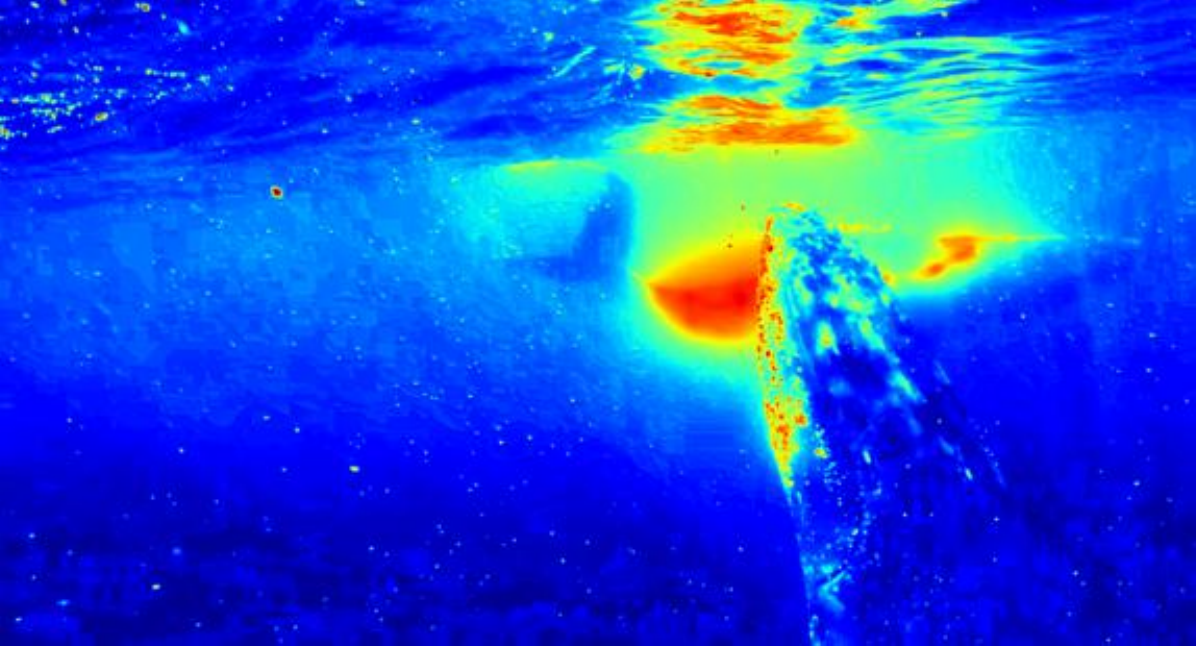}
        \includegraphics[width=\linewidth]{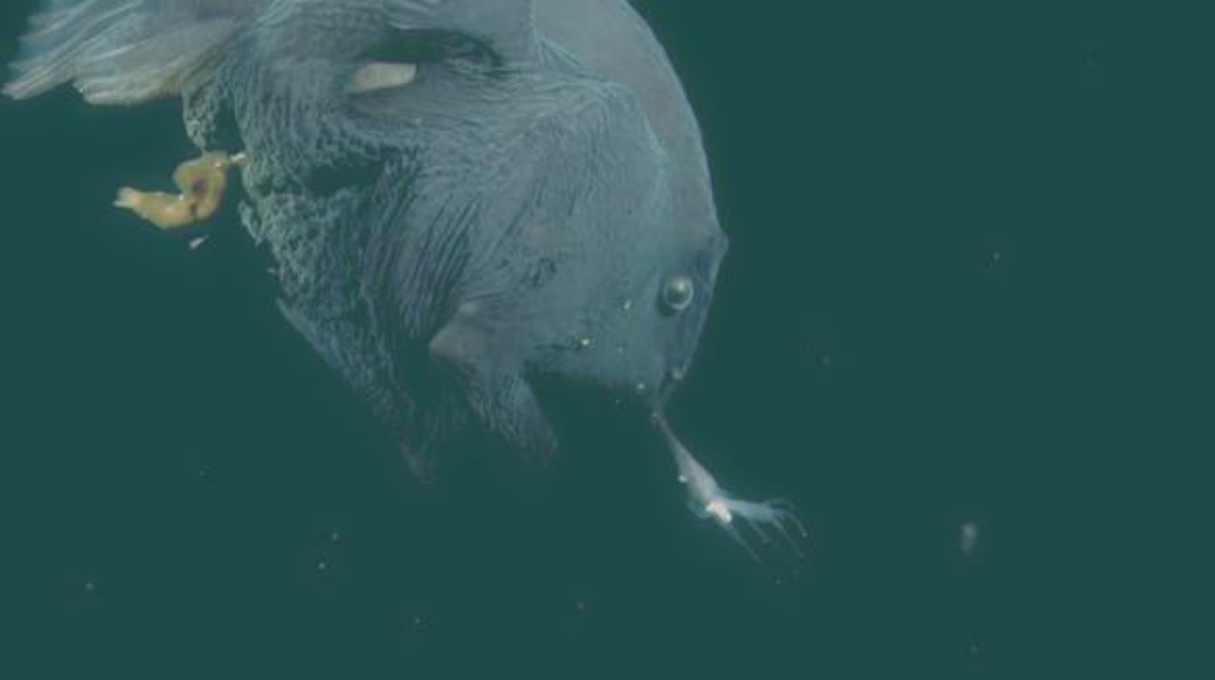} 
        \includegraphics[width=\linewidth]{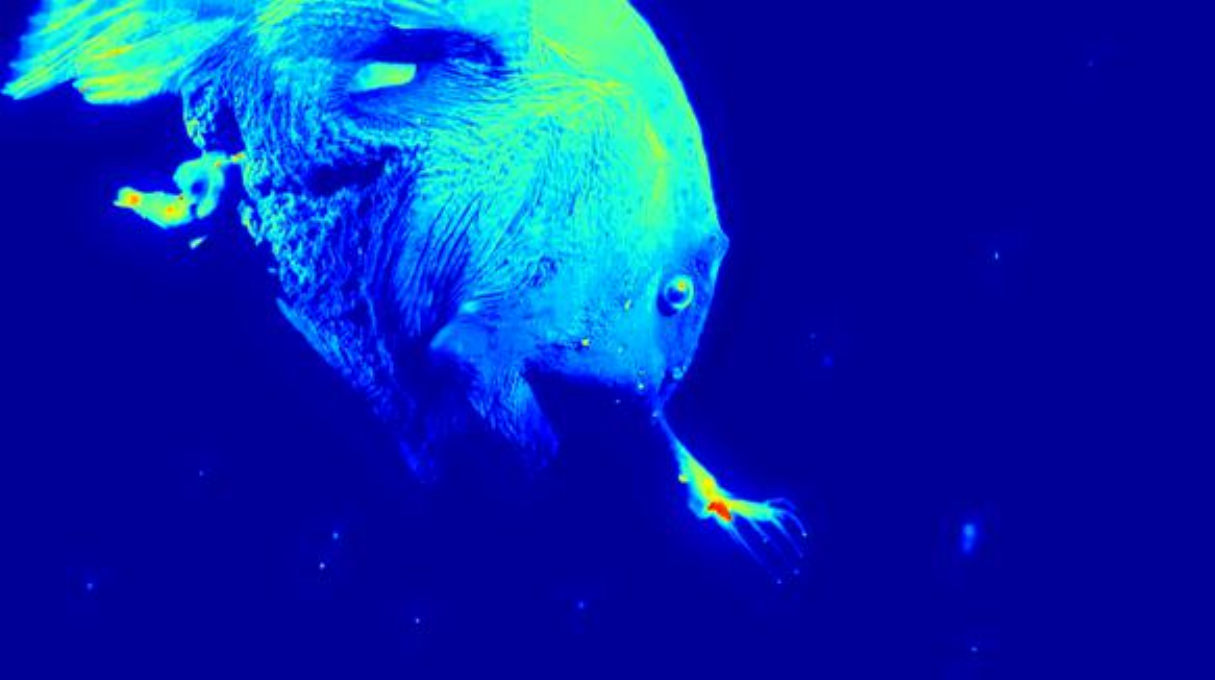}
		\caption{\footnotesize UDnet}
	\end{subfigure}
    	\begin{subfigure}{0.105\linewidth}
		\centering
		\includegraphics[width=\linewidth]{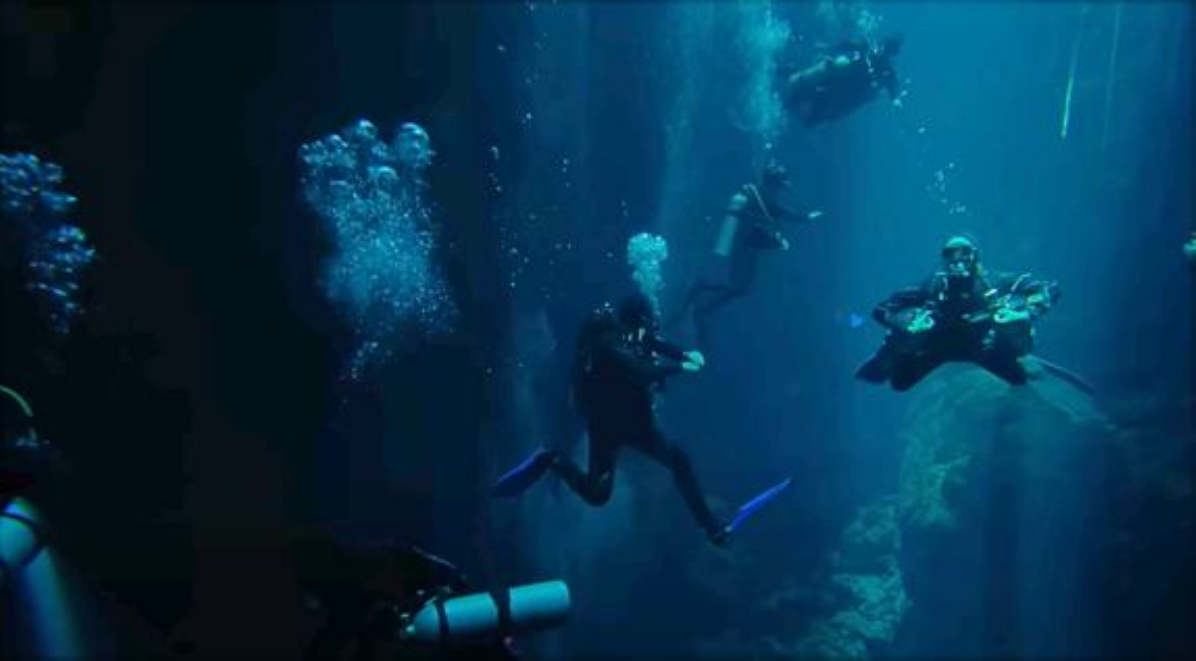}
        \includegraphics[width=\linewidth]{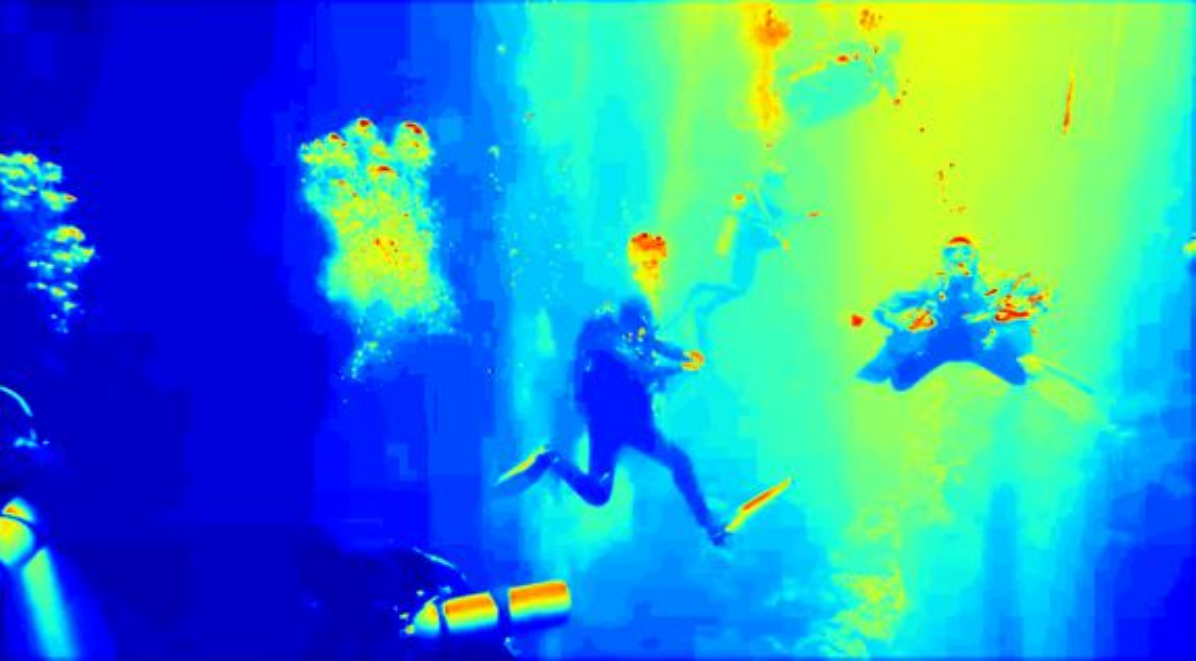} 
        \includegraphics[width=\linewidth]{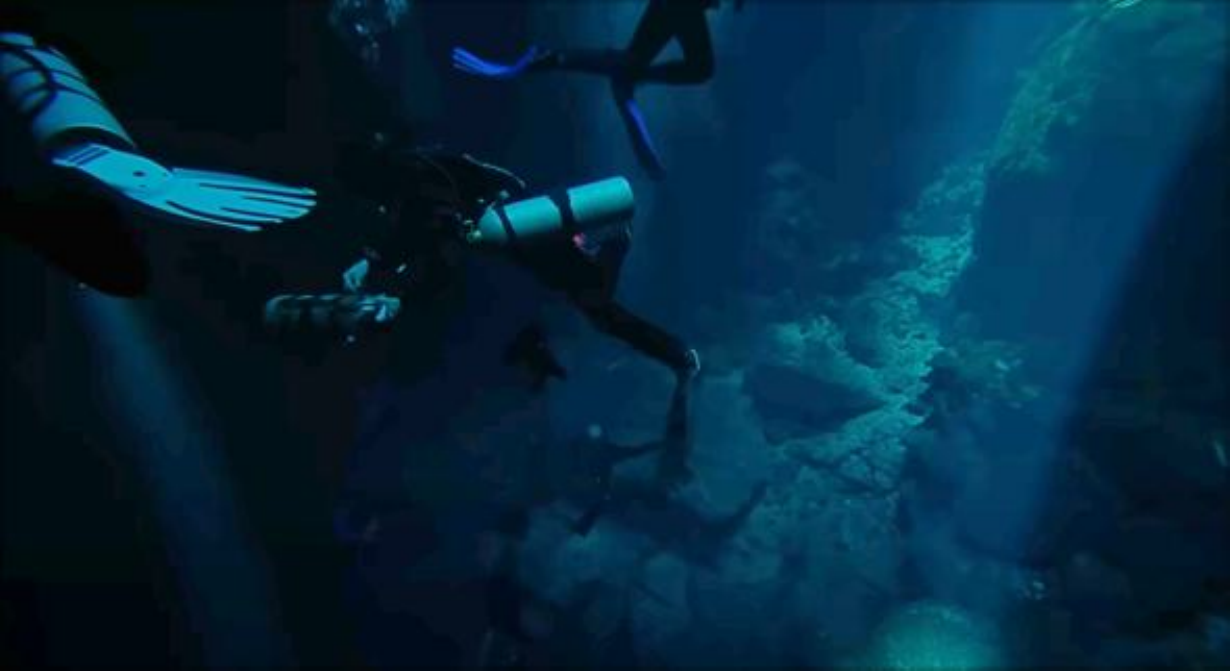}
        \includegraphics[width=\linewidth]{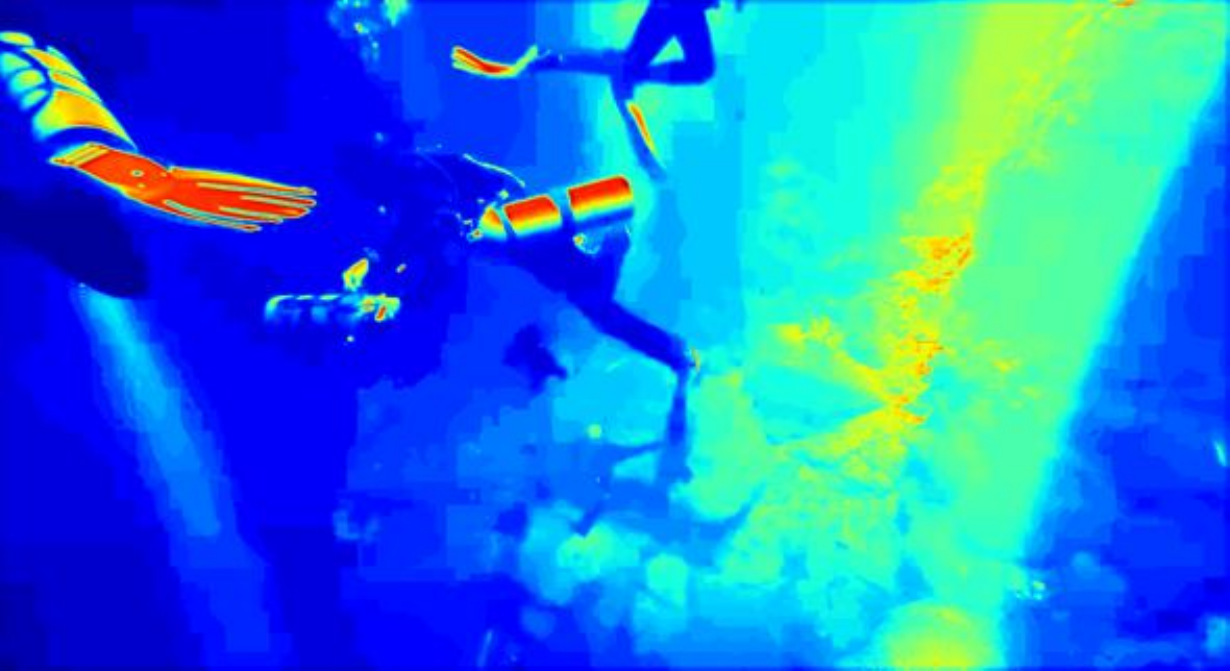} 
        \includegraphics[width=\linewidth]{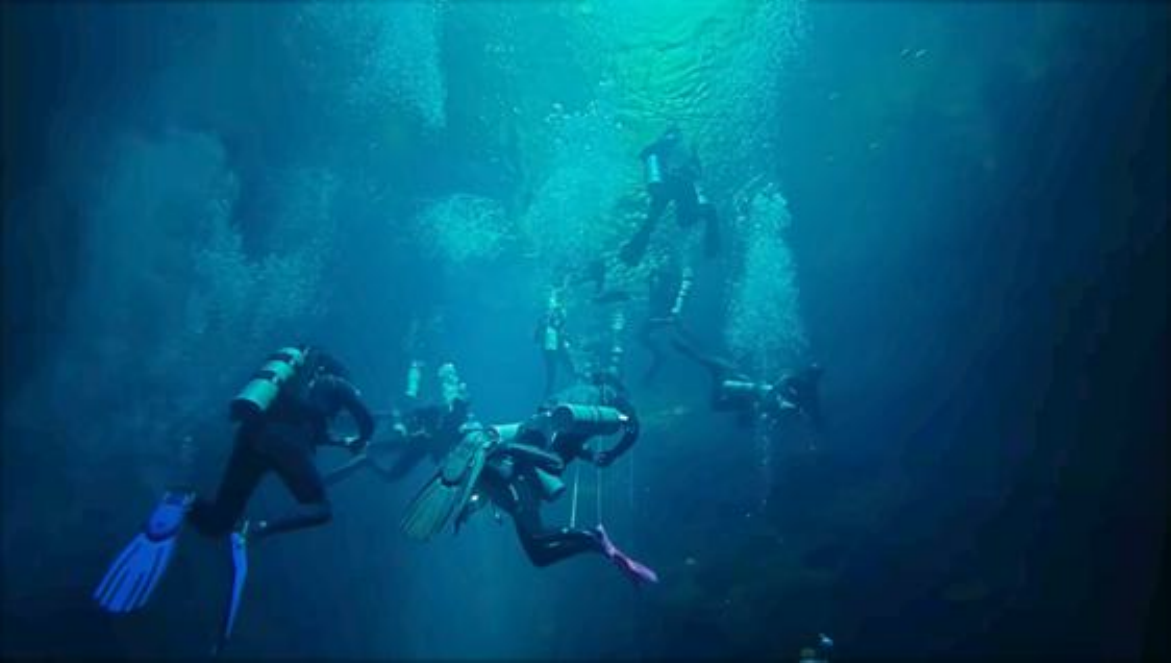}
        \includegraphics[width=\linewidth]{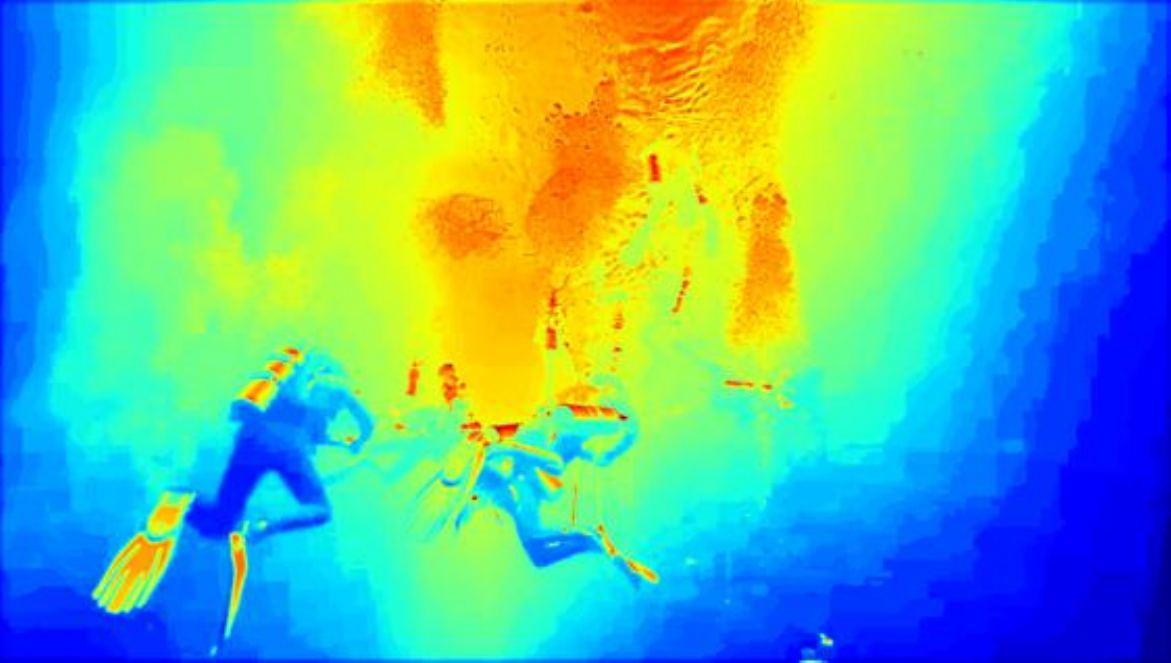} 
        \includegraphics[width=\linewidth]{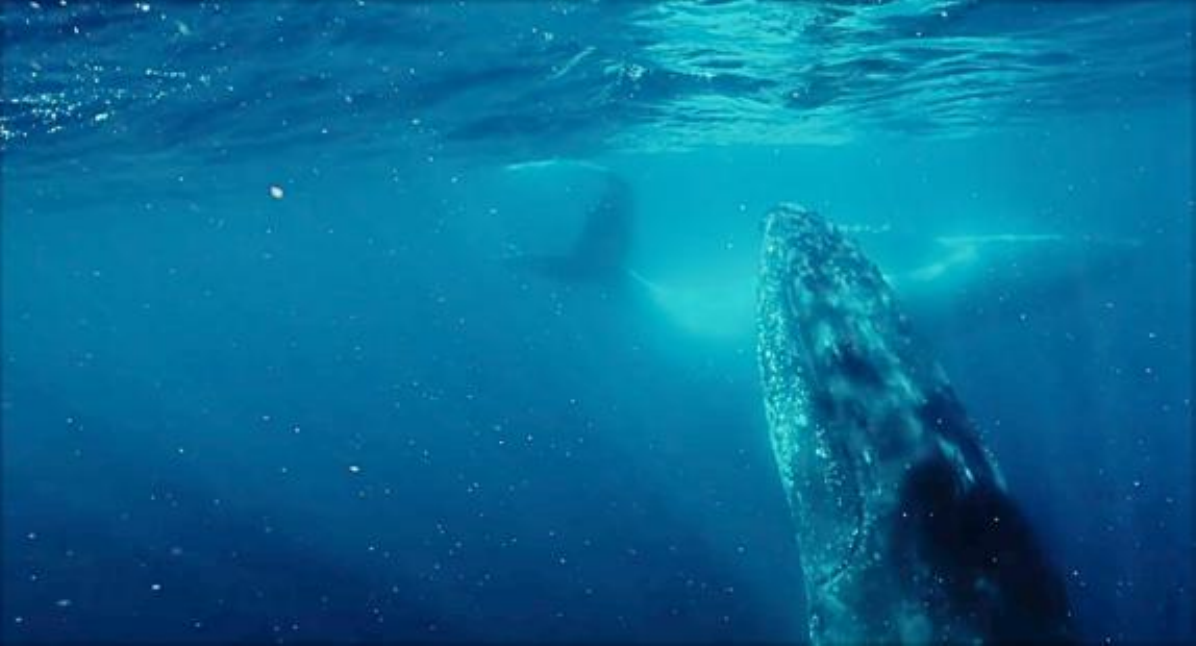}
        \includegraphics[width=\linewidth]{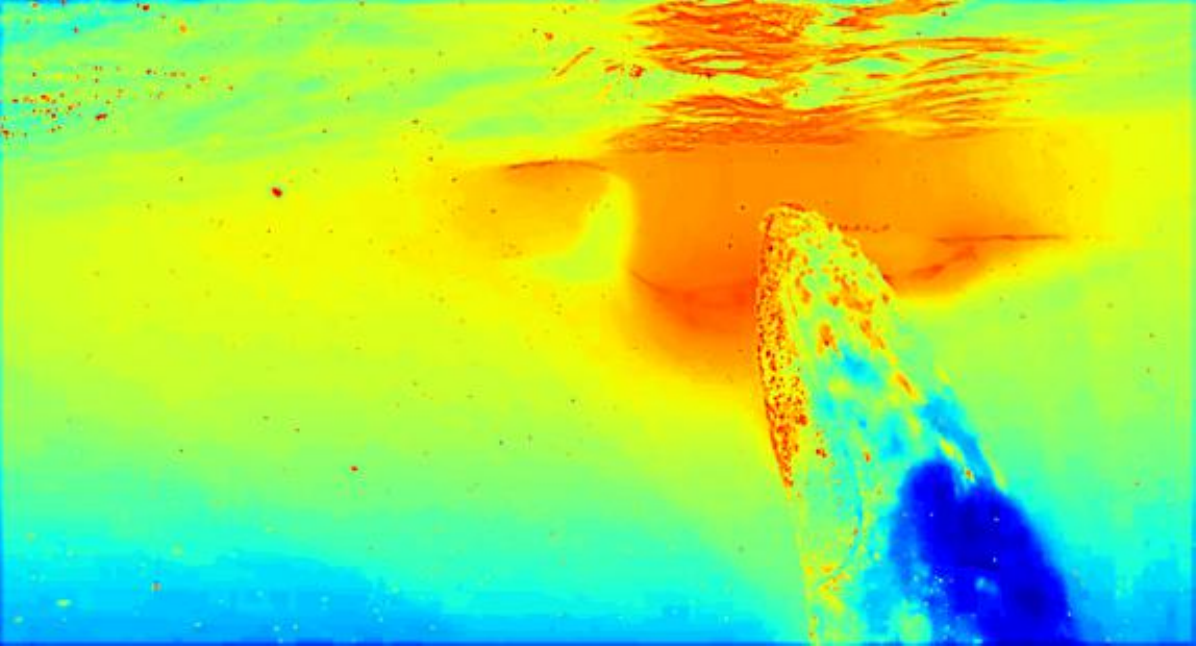} 
        \includegraphics[width=\linewidth]{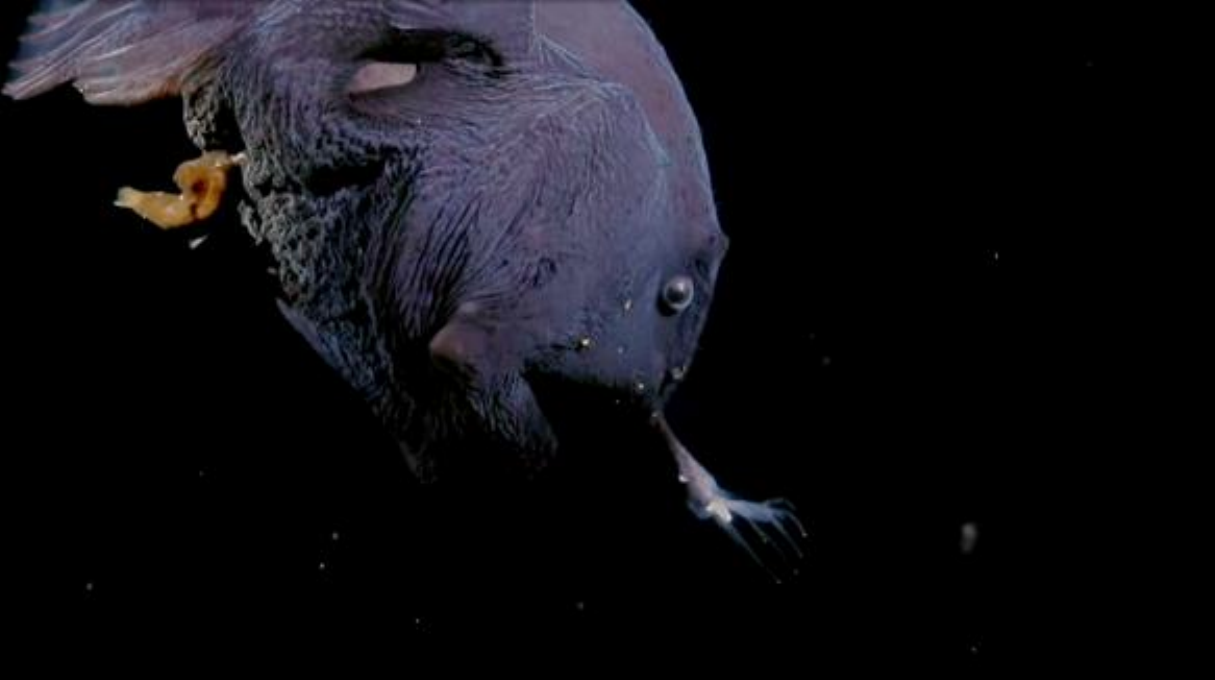}
        \includegraphics[width=\linewidth]{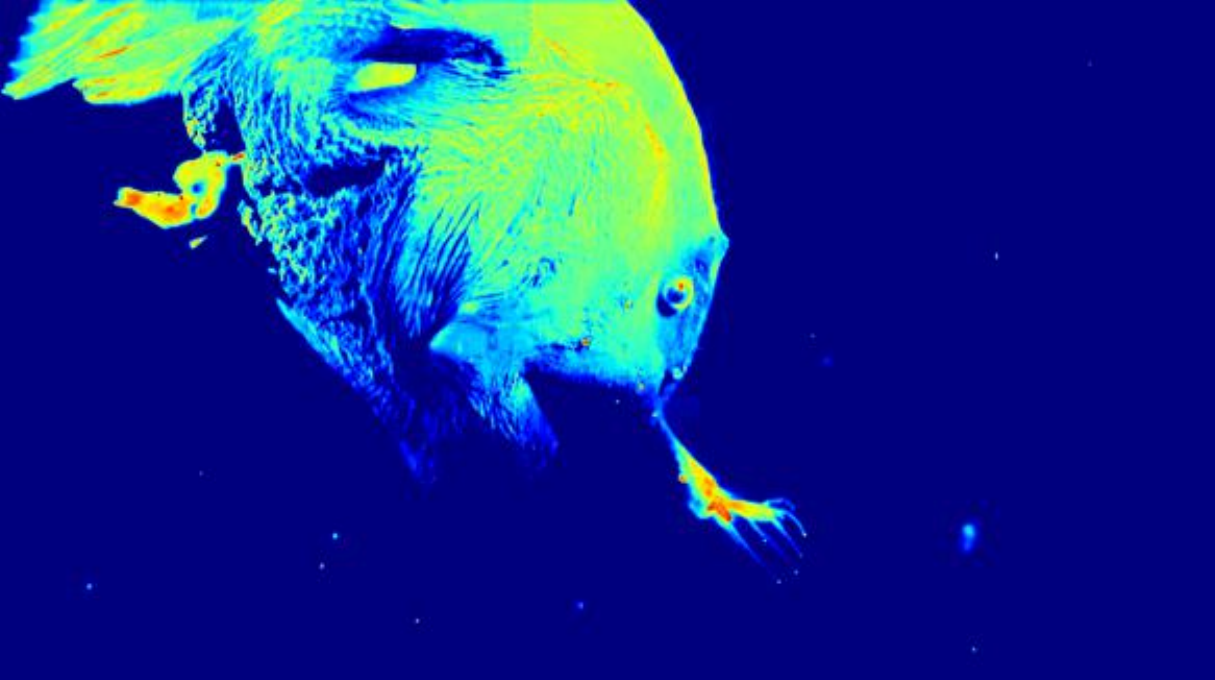} 
        \caption{\footnotesize EIB-FNDL }
	\end{subfigure}
    \begin{subfigure}{0.105\linewidth}
		\centering
		\includegraphics[width=\linewidth]{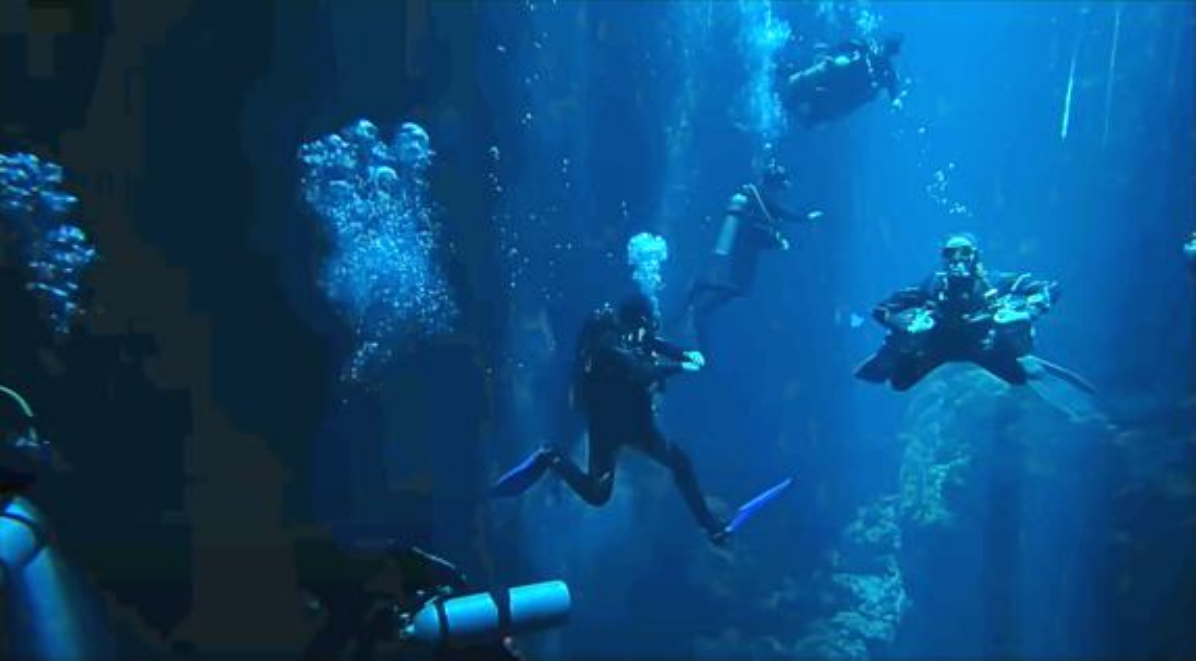} 
        \includegraphics[width=\linewidth]{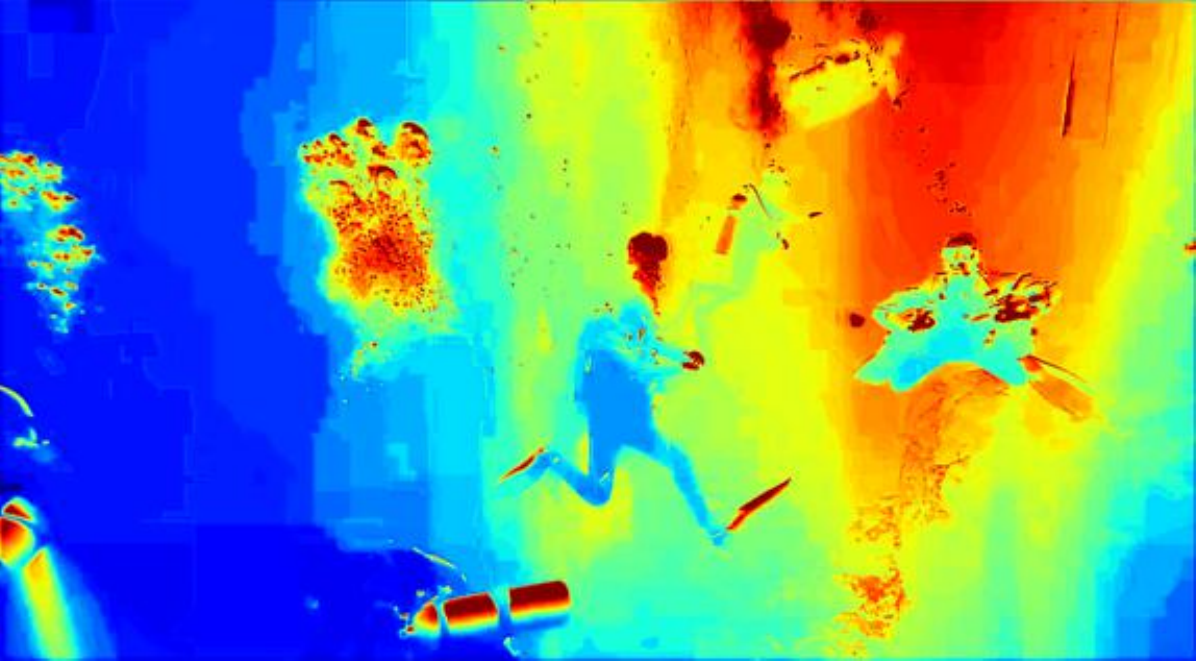} 
		\includegraphics[width=\linewidth]{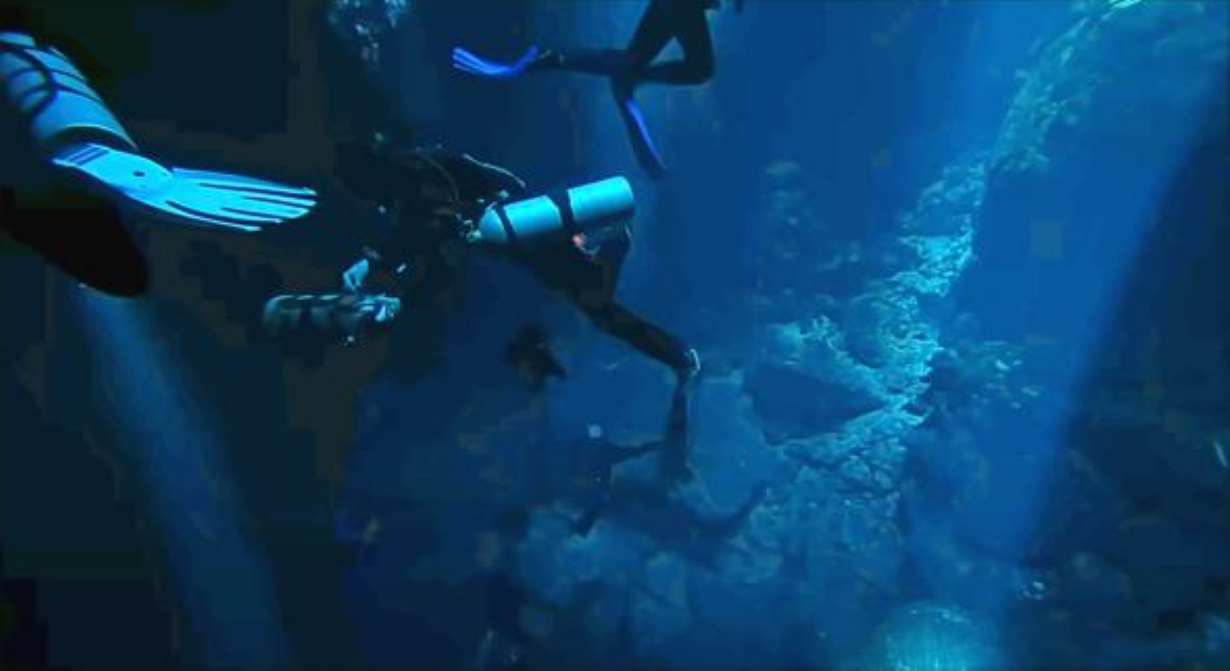} 
        \includegraphics[width=\linewidth]{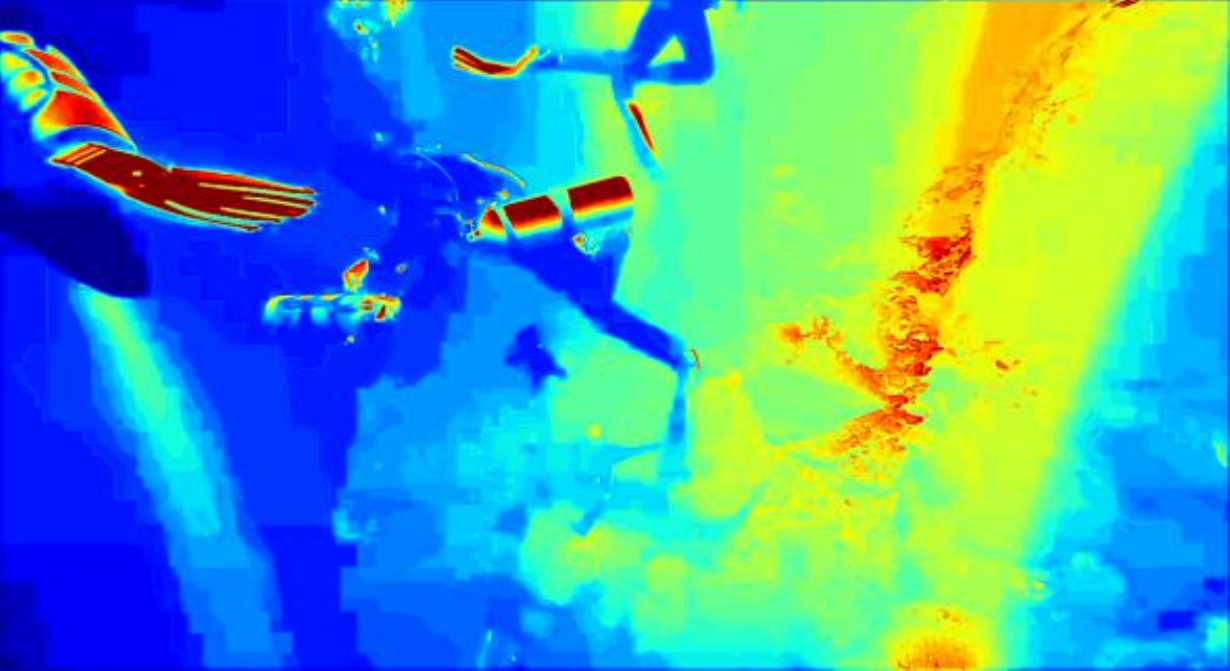} 
		\includegraphics[width=\linewidth]{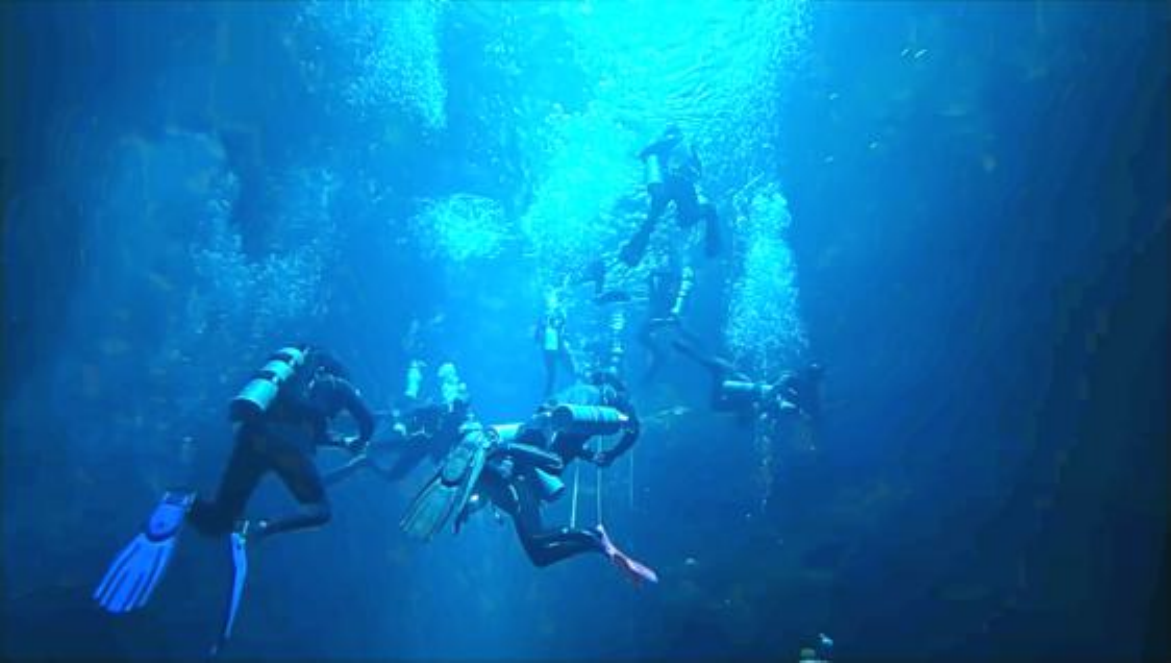} 
        \includegraphics[width=\linewidth]{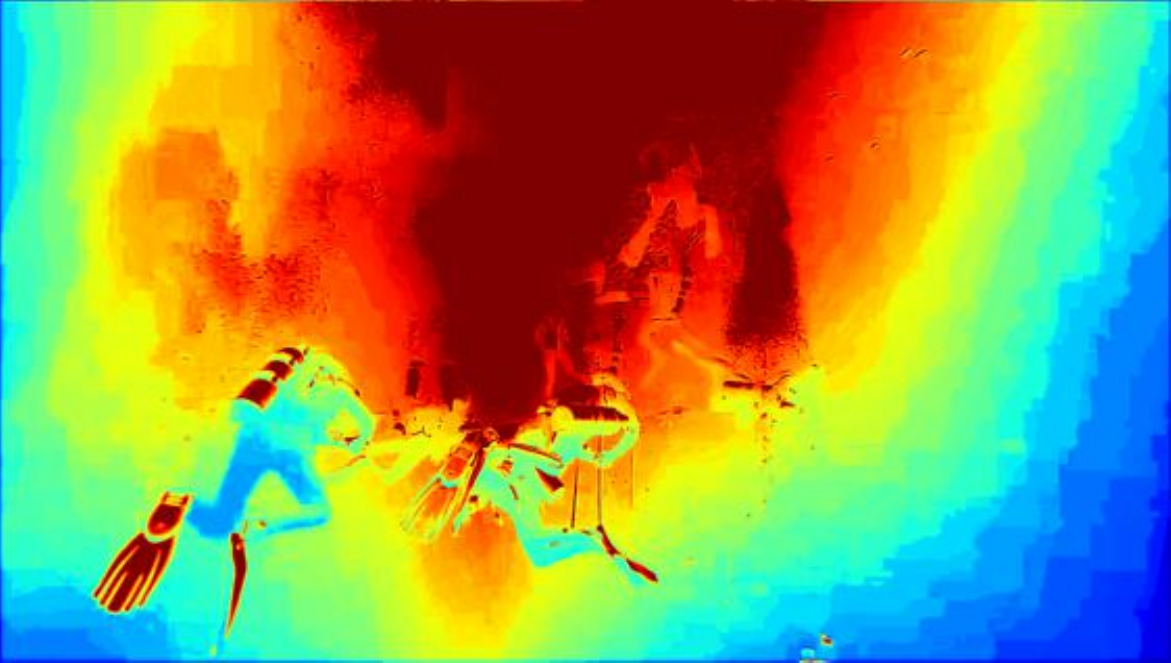} 
		\includegraphics[width=\linewidth]{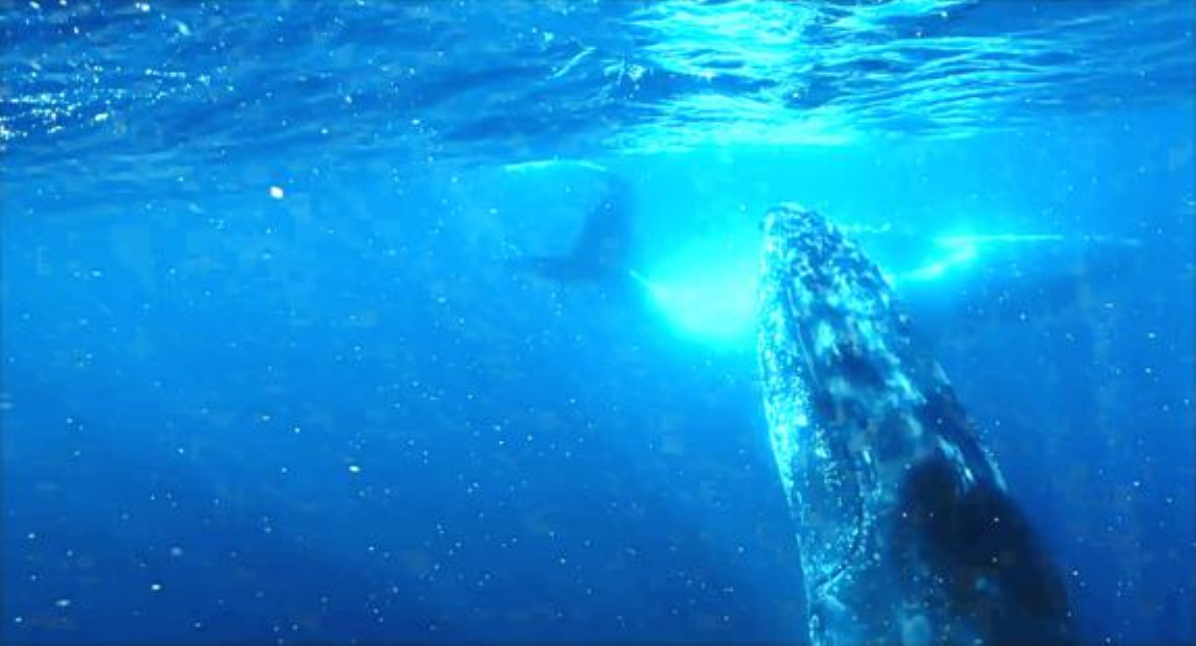} 
        \includegraphics[width=\linewidth]{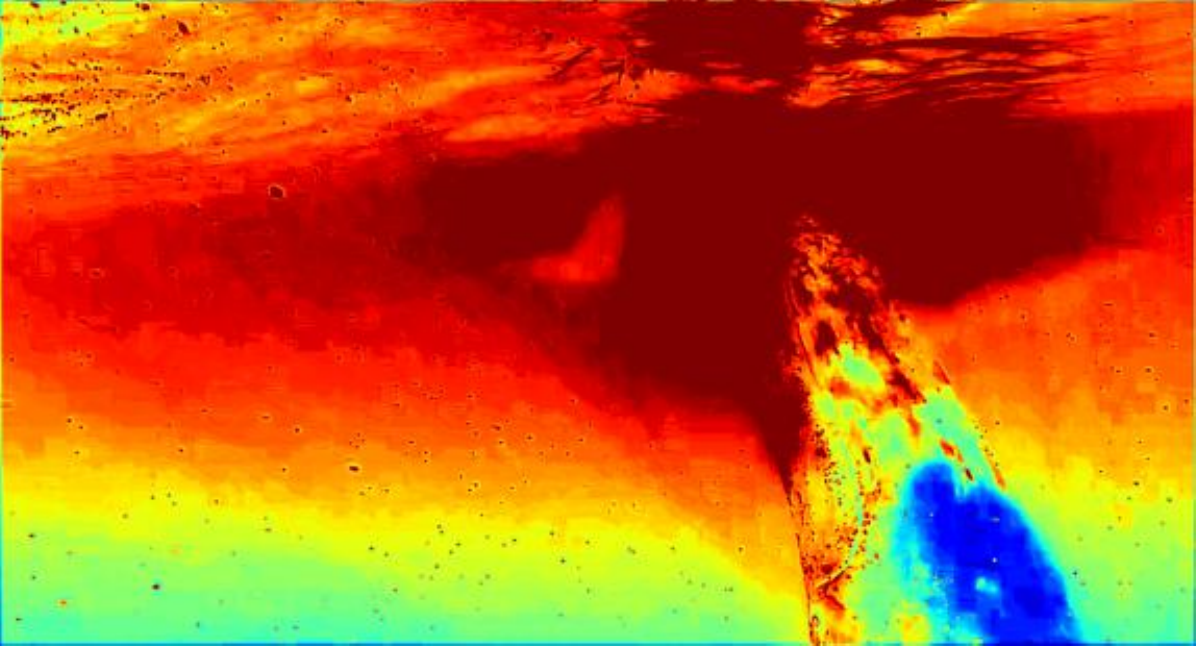} 
		\includegraphics[width=\linewidth]{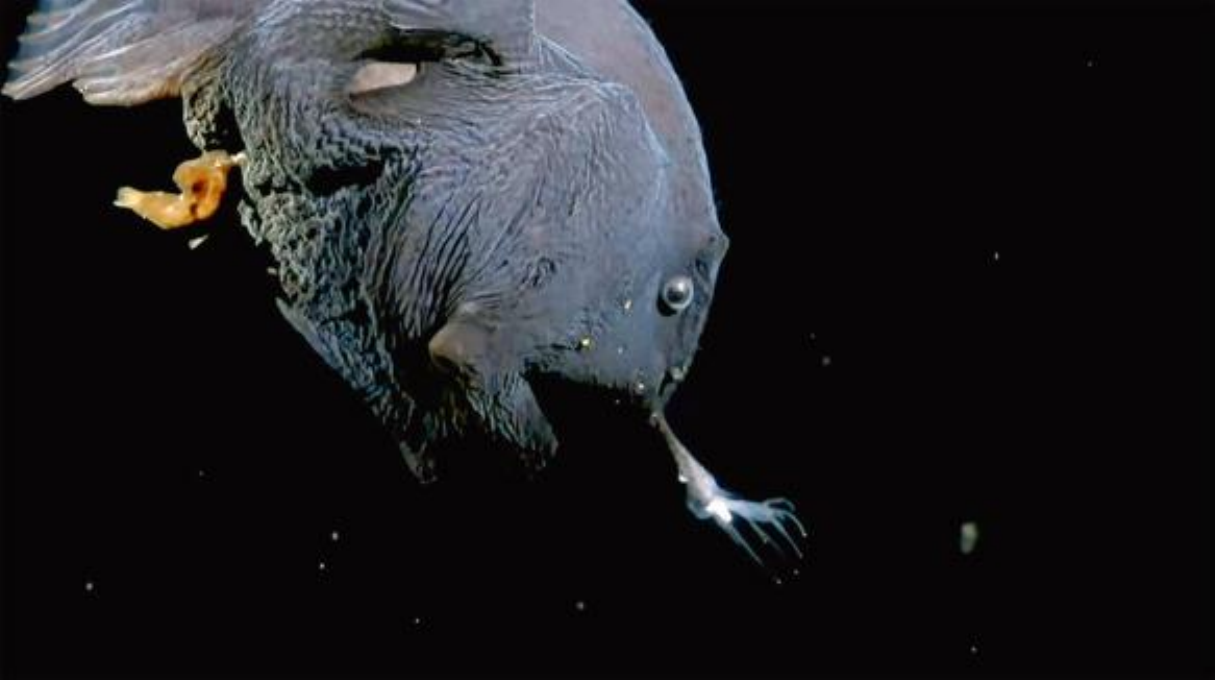} 
        \includegraphics[width=\linewidth]{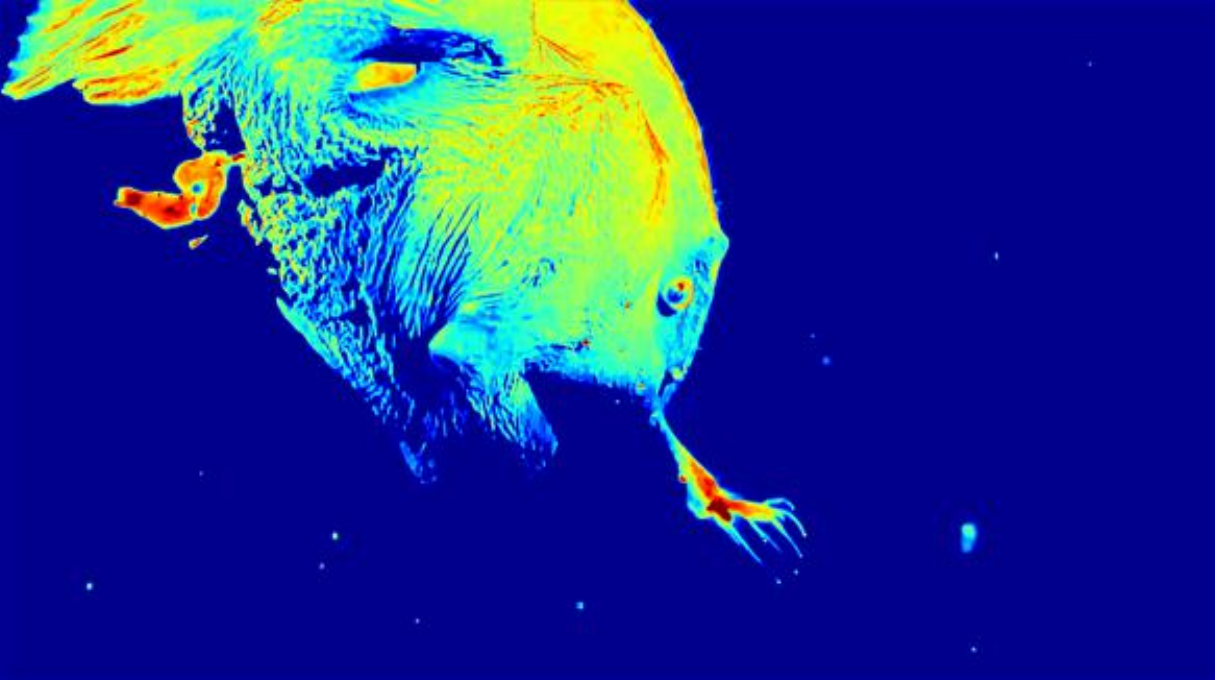} 
		\caption{\footnotesize ALEN}
	\end{subfigure}
    \begin{subfigure}{0.105\linewidth}
		\centering
		\includegraphics[width=\linewidth]{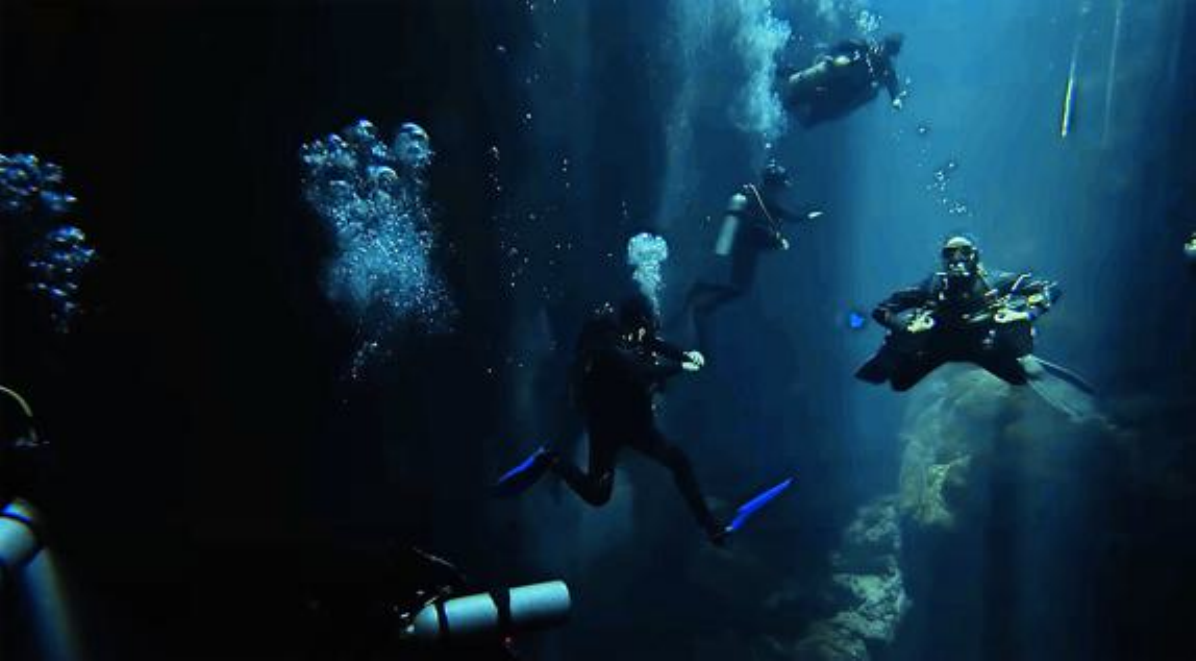} 
        \includegraphics[width=\linewidth]{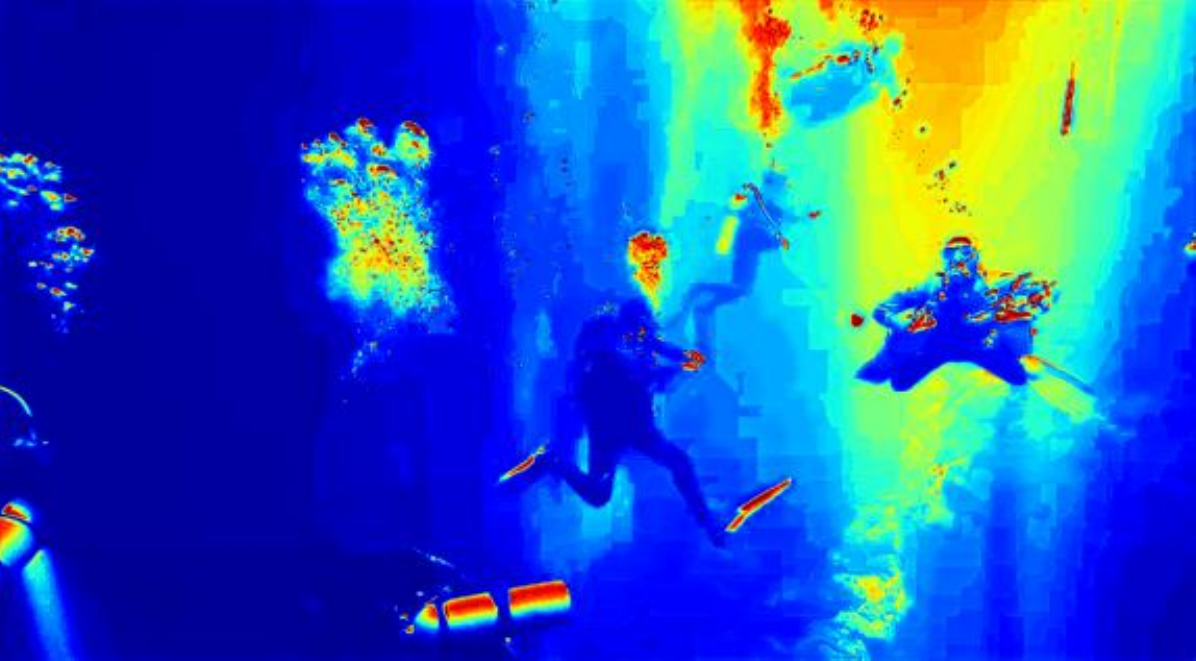} 
		\includegraphics[width=\linewidth]{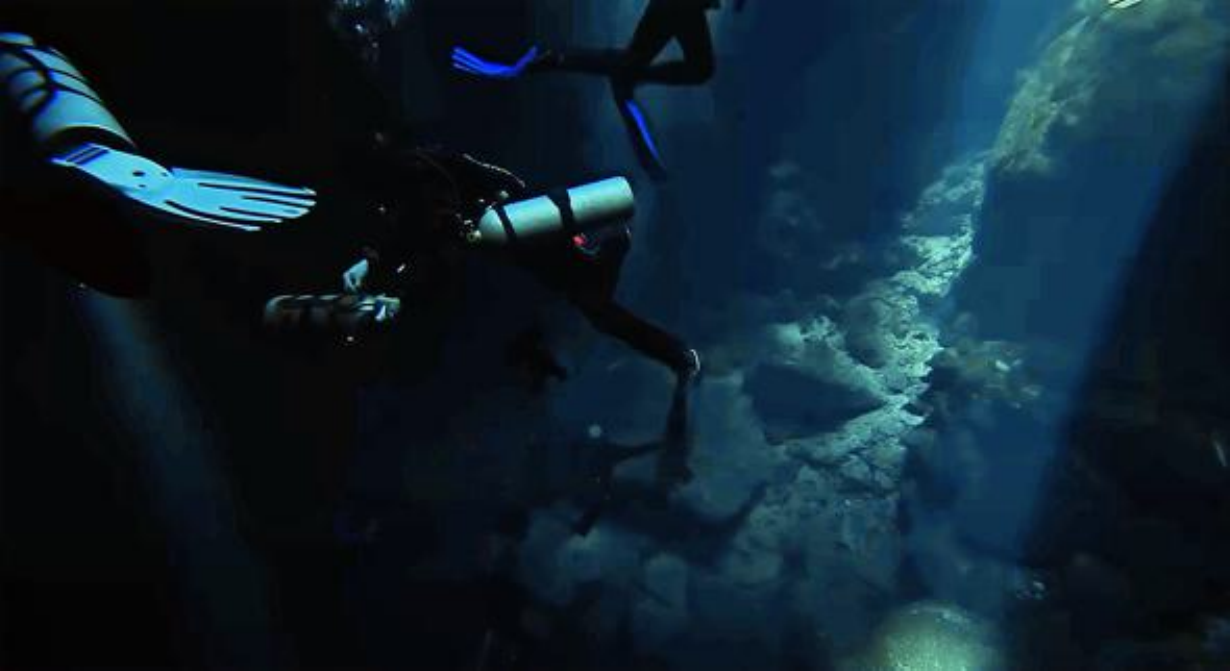} 
        \includegraphics[width=\linewidth]{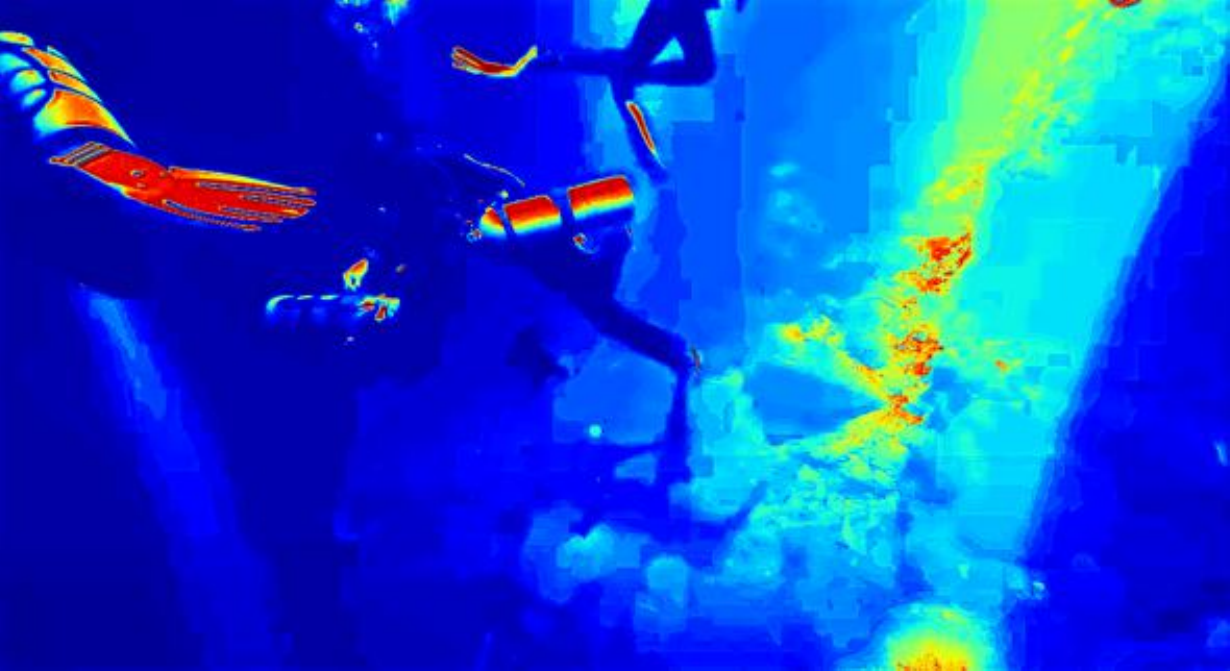} 
		\includegraphics[width=\linewidth]{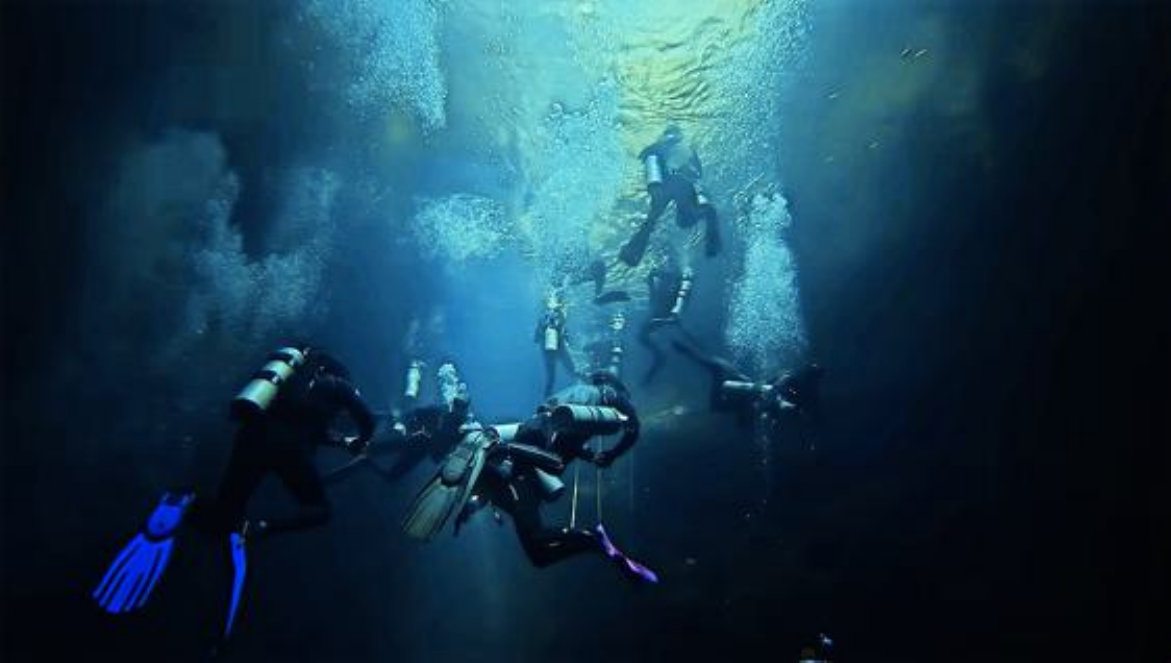} 
        \includegraphics[width=\linewidth]{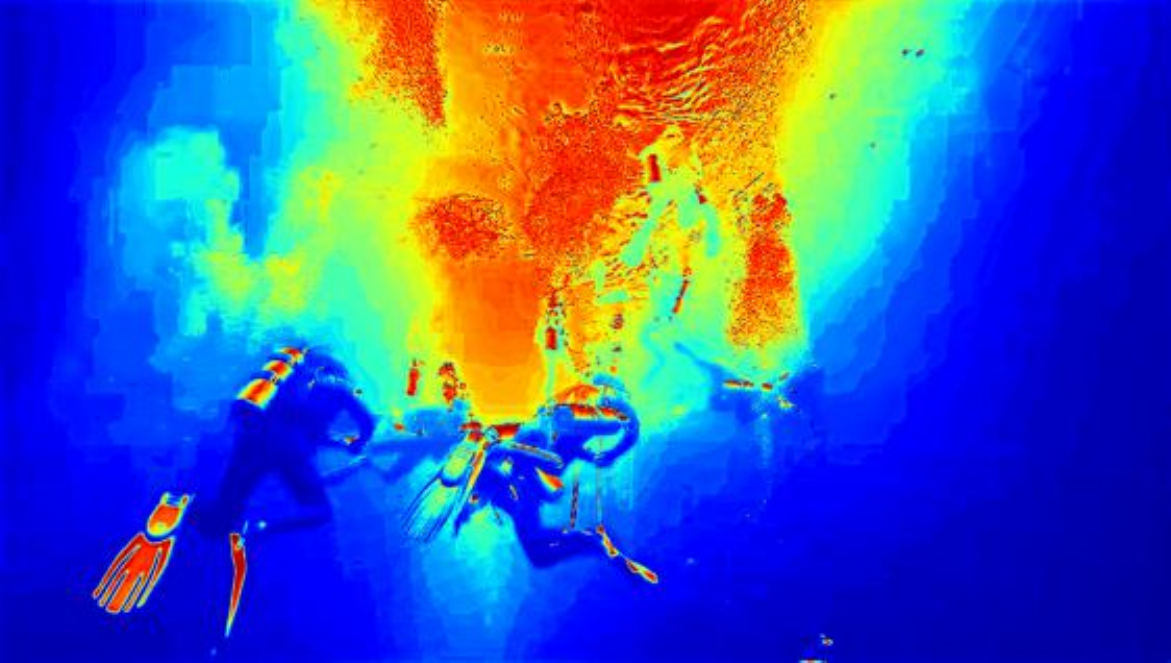} 
		\includegraphics[width=\linewidth]{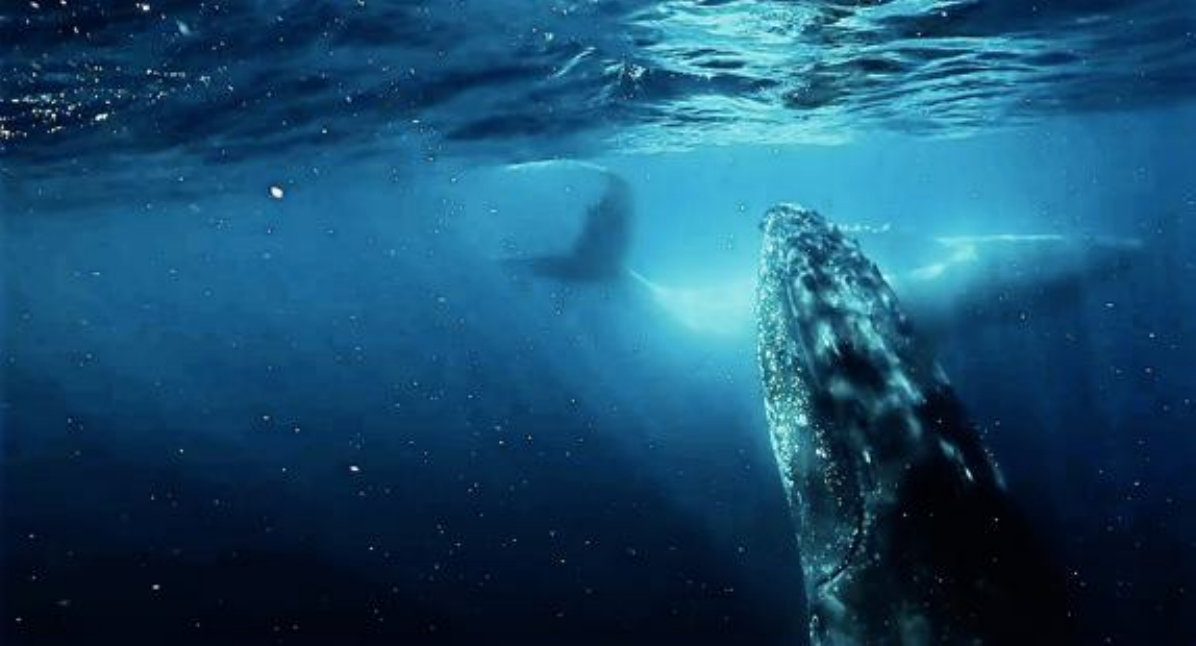} 
        \includegraphics[width=\linewidth]{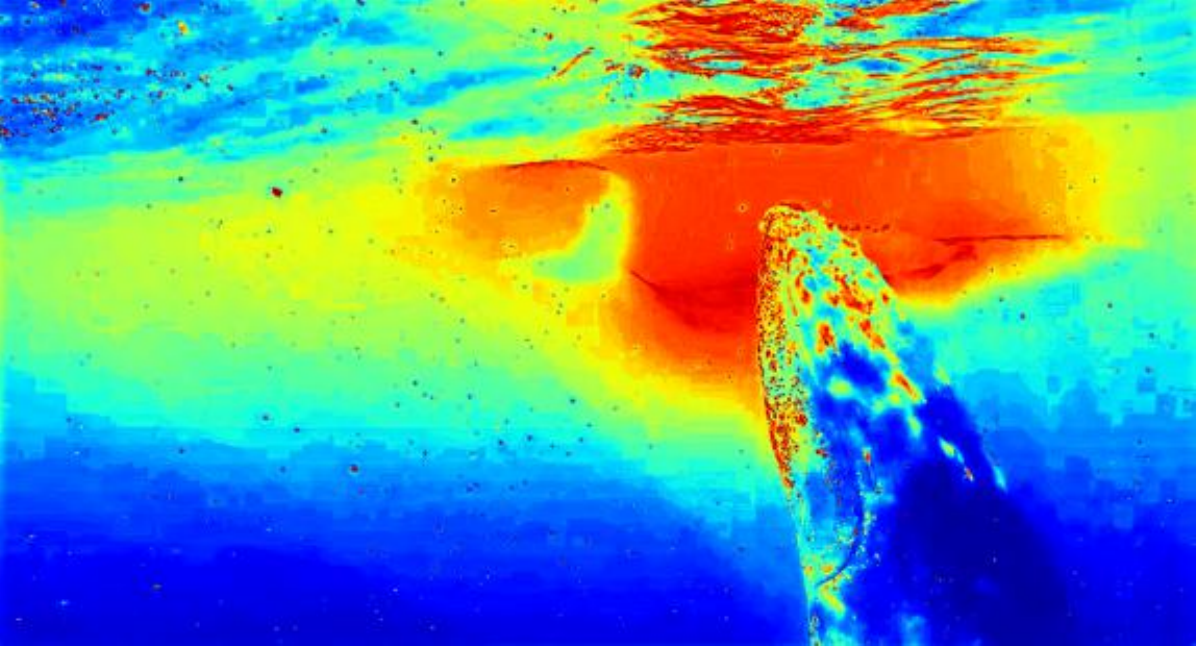} 
		\includegraphics[width=\linewidth]{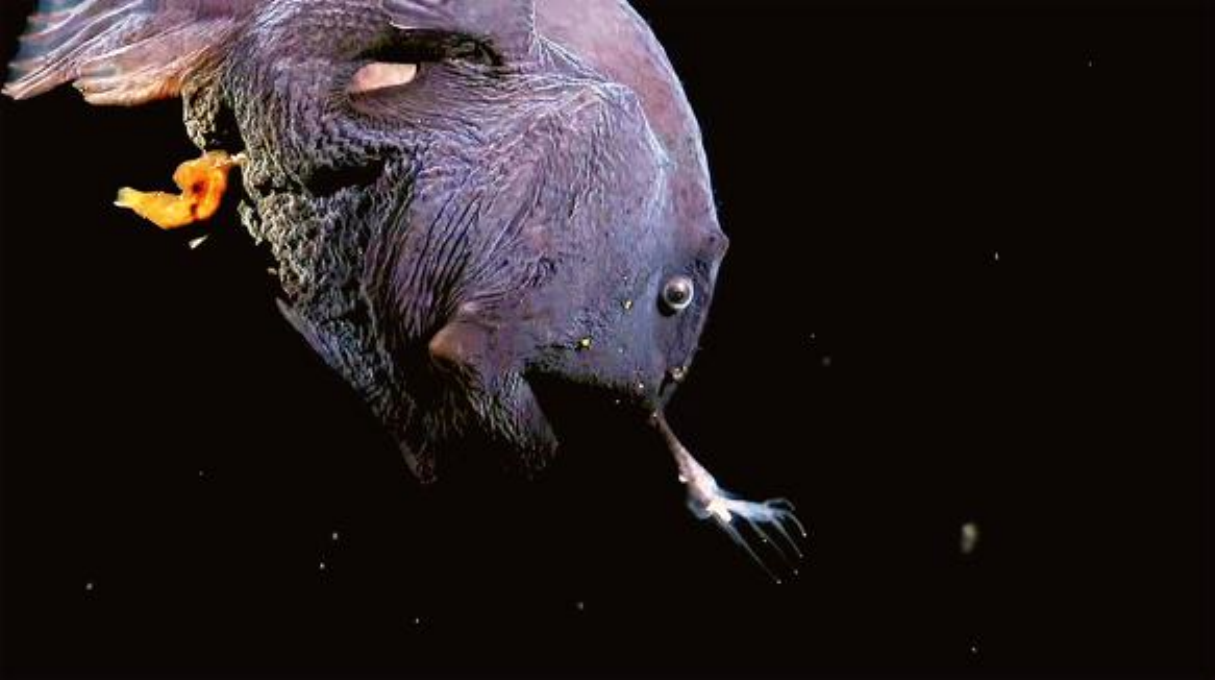} 
        \includegraphics[width=\linewidth]{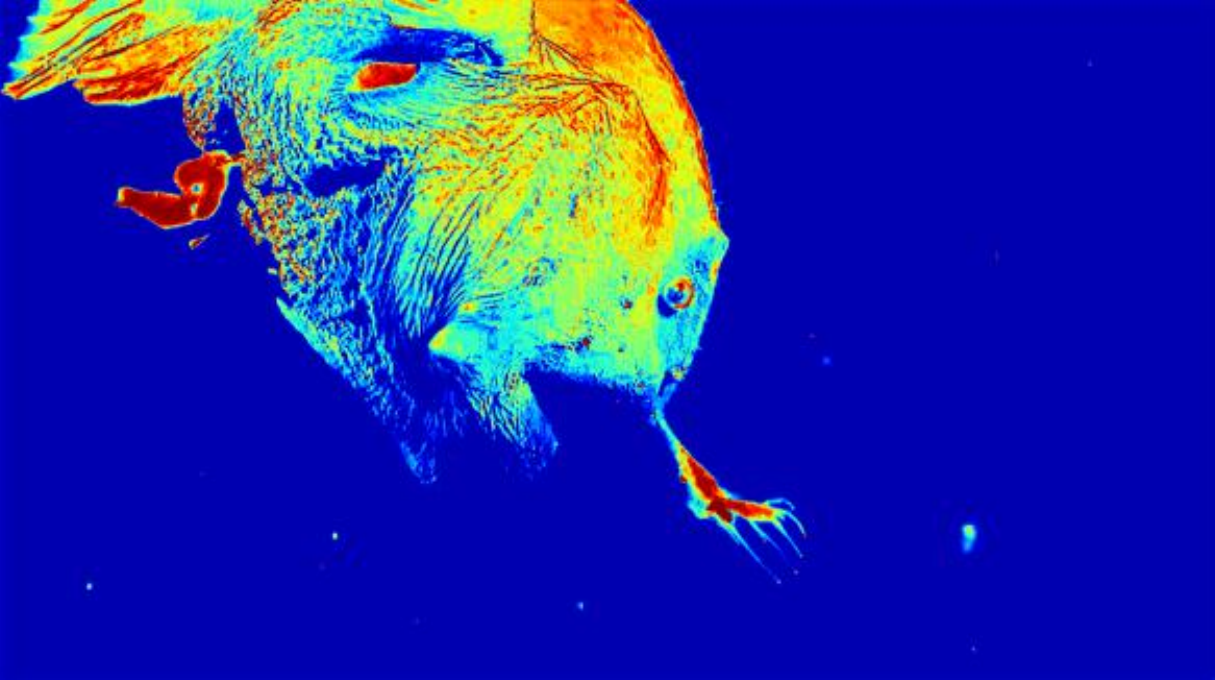} 
		\caption{\footnotesize UNIR-Net}
	\end{subfigure}
 
	\caption{Comparison of enhancement results and illumination maps for underwater images with complex illumination. From left to right: input image with complex illumination followed by various enhancement outputs. Each output includes the enhanced image and its estimated illumination map, showing differences in lighting.}
	\label{CI_STUDY}
\end{figure*}

To analyze the limitations of UNIR-Net, five representative examples were selected from Unsplash. These cases illustrate scenarios where the model performs poorly, particularly in scenes with complex illumination (CI) patterns or extremely low-contrast regions. Figure~\ref{CI_STUDY} presents these examples, along with a visual comparison between UNIR-Net and seven state-of-the-art methods: ICSP, PDE, UDAformer, SMDR-IS, UDnet, EIB-FNDL, and ALEN.

The first four images in the figure show that ICSP, PDE, UDAformer, SMDR-IS, UDnet, and UNIR-Net fail to achieve complete lighting restoration. Despite some localized enhancements, large dark regions persist in the images. In contrast, EIB-FNDL and ALEN yield better results in these scenes; however, ALEN introduces visible artifacts that reduce the perceptual quality of the enhancement. The fifth image, which shows a fish, has milder lighting conditions. In this case, most methods are able to render the shape of the fish more clearly, suggesting that model performance may improve under less challenging lighting scenarios, as illustrated by this specific case.

The limitations seen in the first four images stem primarily from the model’s limited generalization capacity. This is especially evident under extreme lighting variations not present in the training data. In addition, the current illumination estimation module struggles to accurately capture abrupt or non-uniform lighting transitions, resulting in incomplete corrections. These findings highlight the need for future improvements, such as developing a more robust illumination model and expanding the training dataset with more complex scenes that reflect real-world challenges.

\subsubsection{Refining high-level vision processing}
\begin{figure*}[!ht]
	\centering
	\begin{subfigure}{0.105\linewidth}
		\centering
        \includegraphics[width=\linewidth]{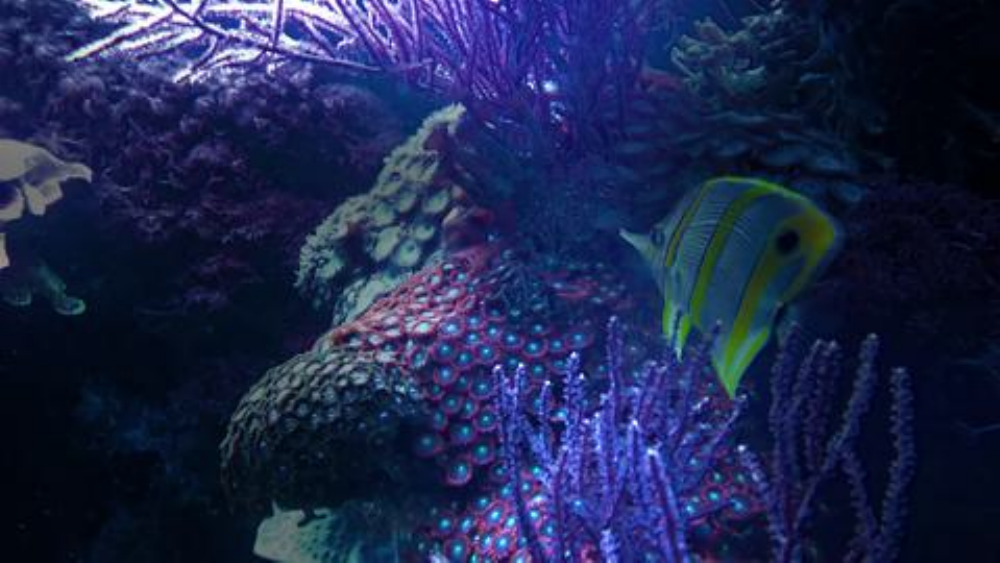} 
		\includegraphics[width=\linewidth]{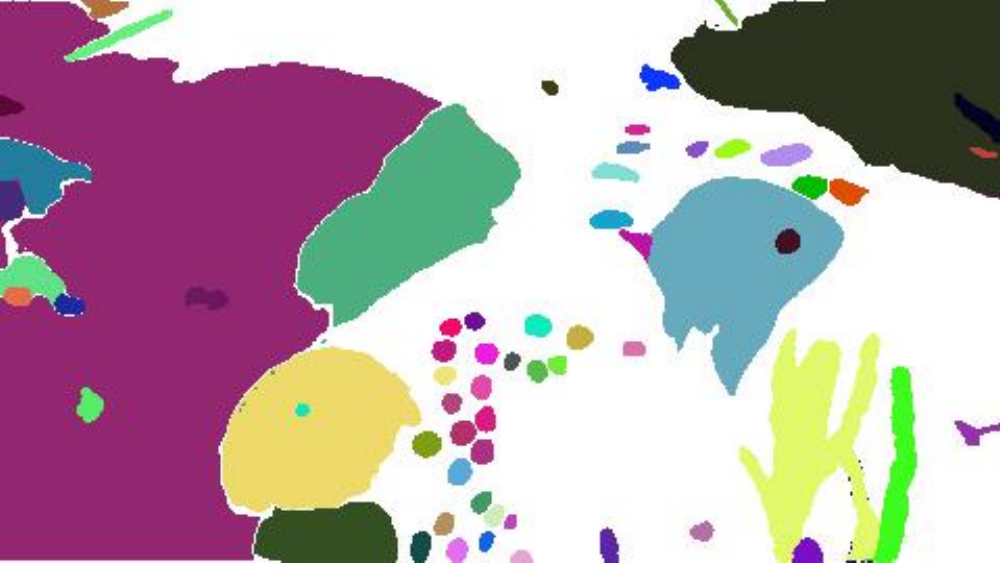} 
        \includegraphics[width=\linewidth]{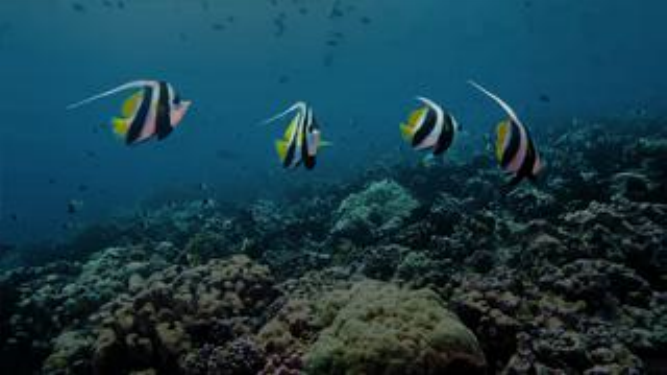} 
        \includegraphics[width=\linewidth]{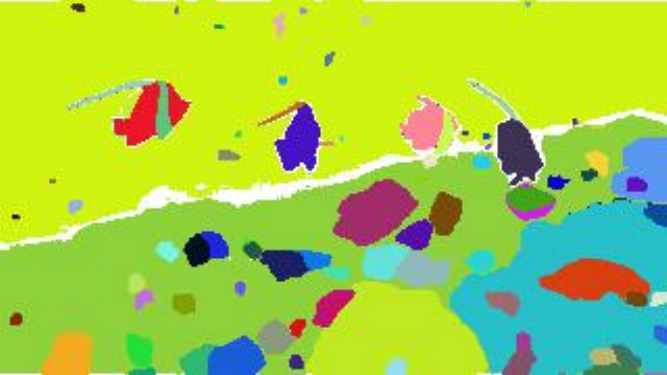} 
        \includegraphics[width=\linewidth]{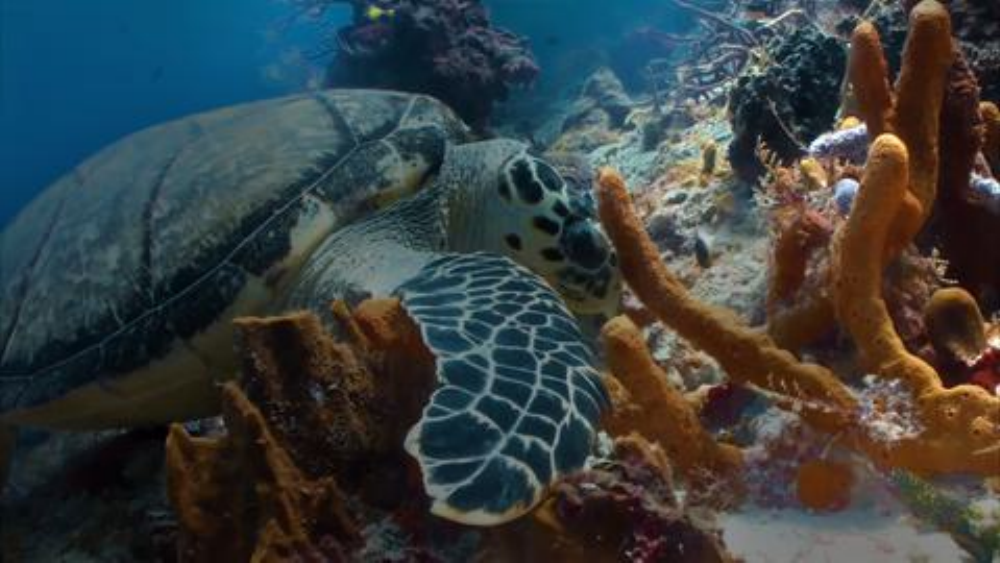} 
        \includegraphics[width=\linewidth]{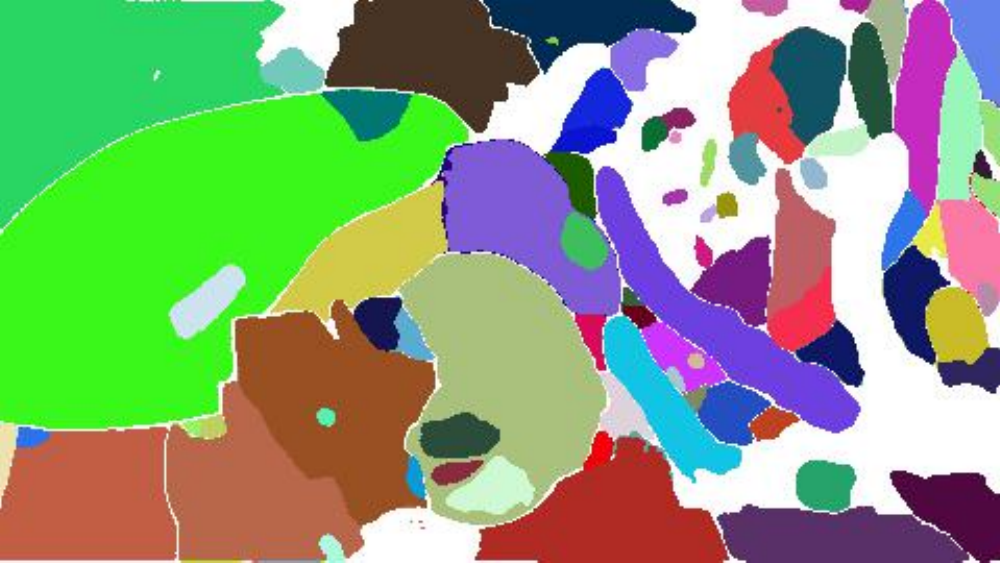} 
        \includegraphics[width=\linewidth]{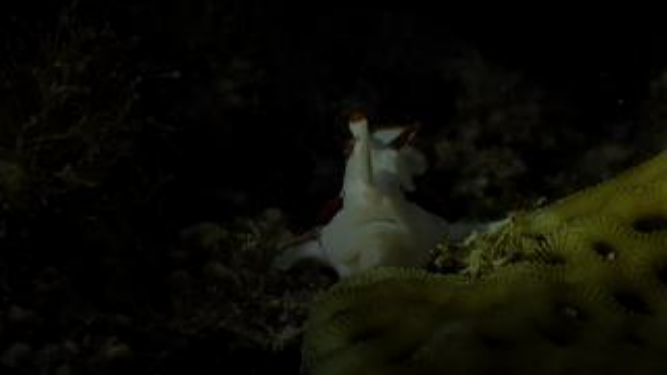} 
        \includegraphics[width=\linewidth]{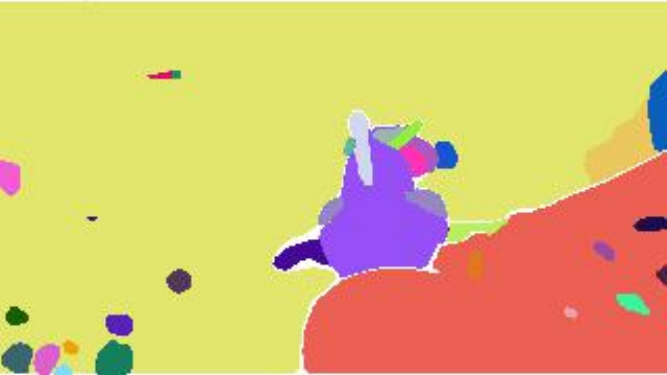} 
        \includegraphics[width=\linewidth]{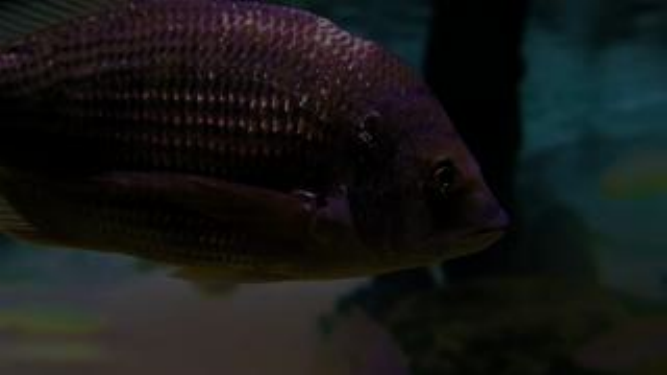} 
        \includegraphics[width=\linewidth]{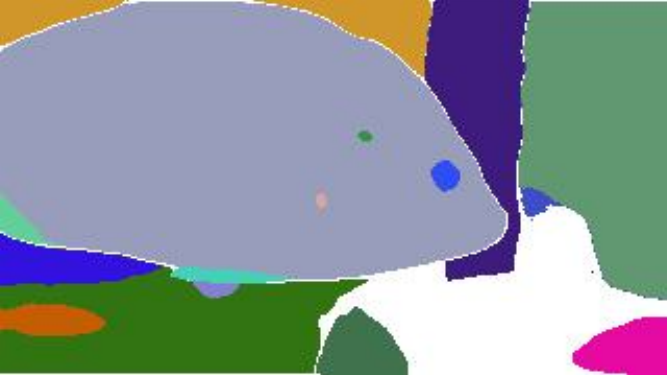} 
        \caption{\footnotesize NUI Image}
	\end{subfigure}
    \begin{subfigure}{0.105\linewidth}
		\centering
        \includegraphics[width=\linewidth]{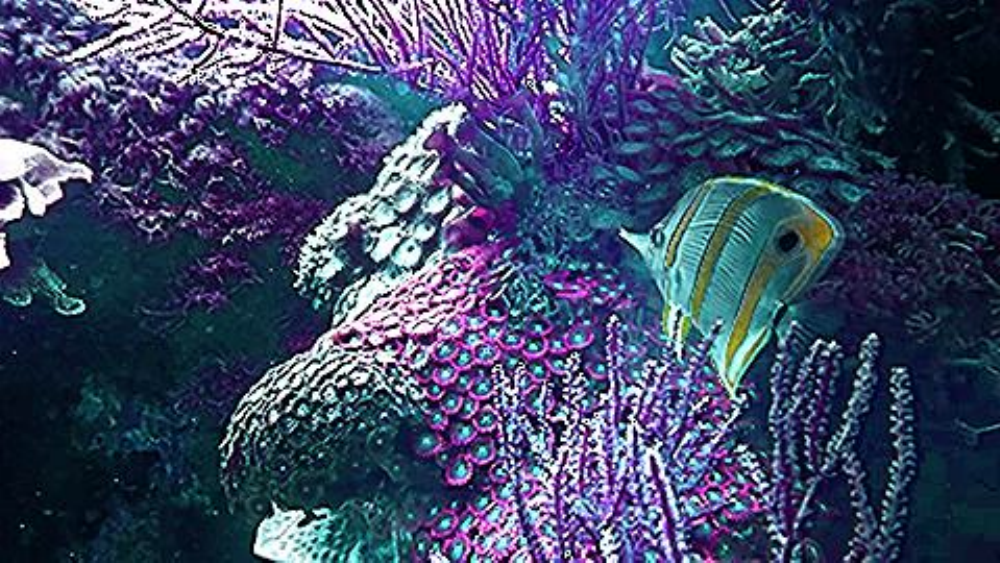} 
		\includegraphics[width=\linewidth]{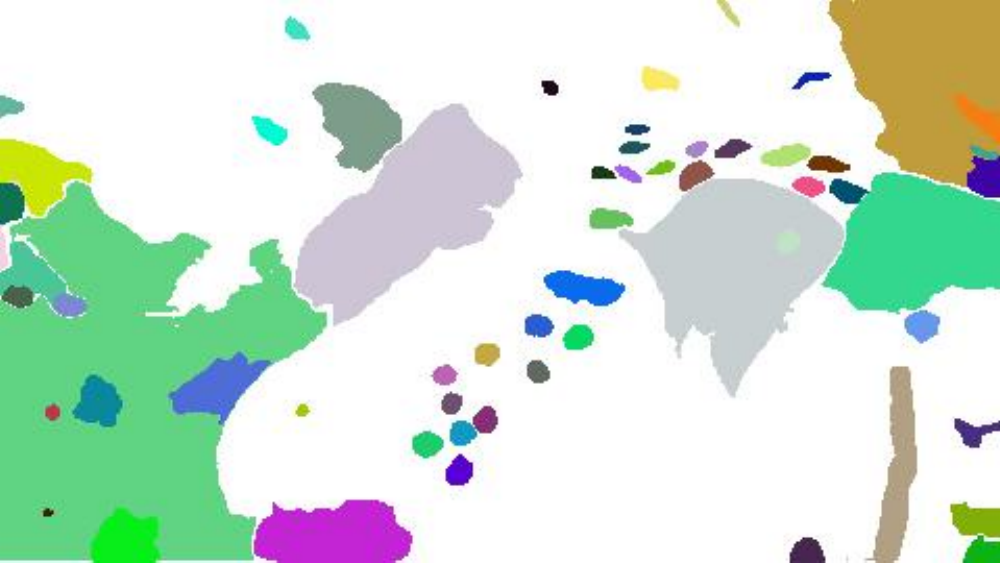} 
        \includegraphics[width=\linewidth]{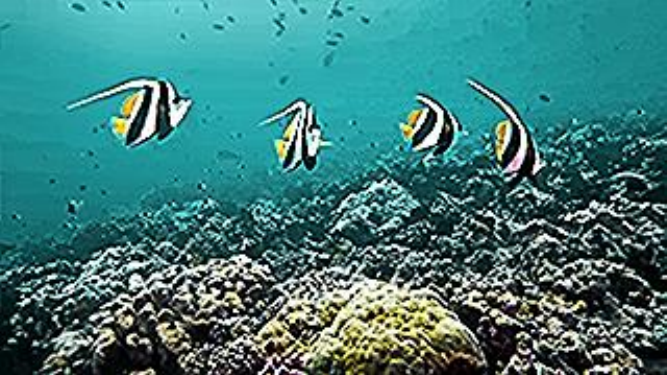} 
        \includegraphics[width=\linewidth]{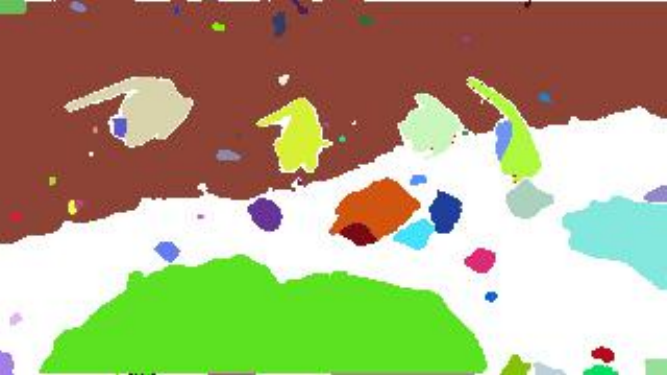} 
        \includegraphics[width=\linewidth]{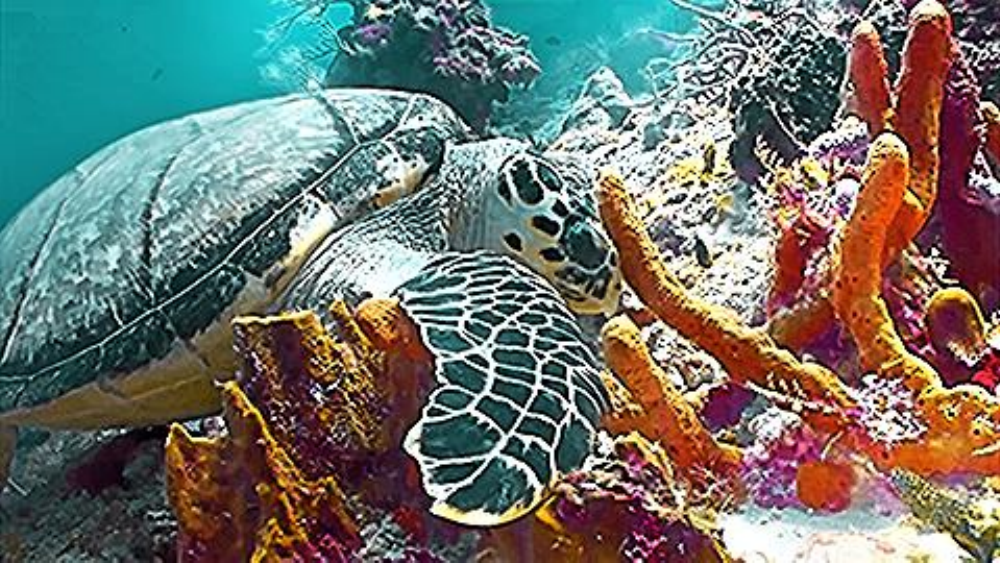} 
        \includegraphics[width=\linewidth]{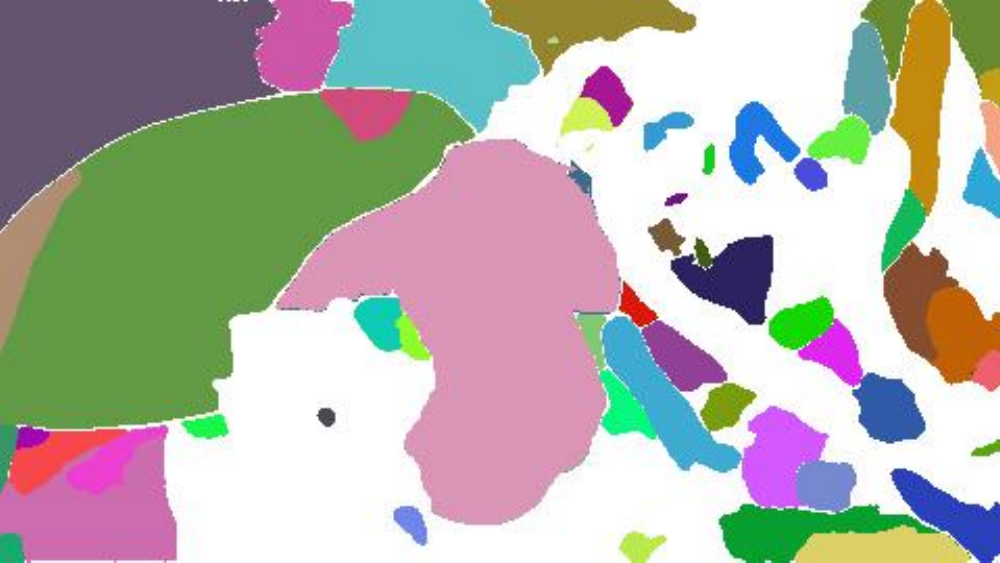} 
        \includegraphics[width=\linewidth]{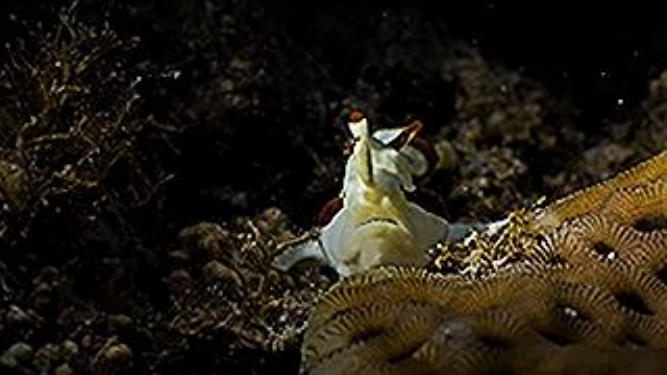} 
        \includegraphics[width=\linewidth]{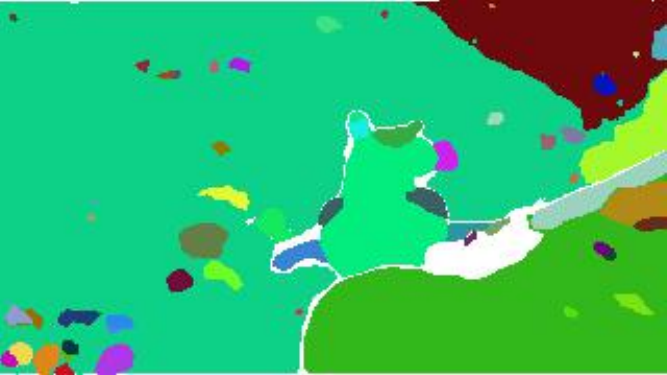} 
        \includegraphics[width=\linewidth]{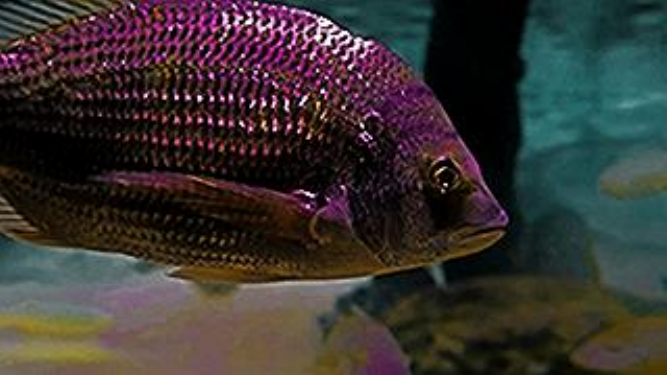} 
        \includegraphics[width=\linewidth]{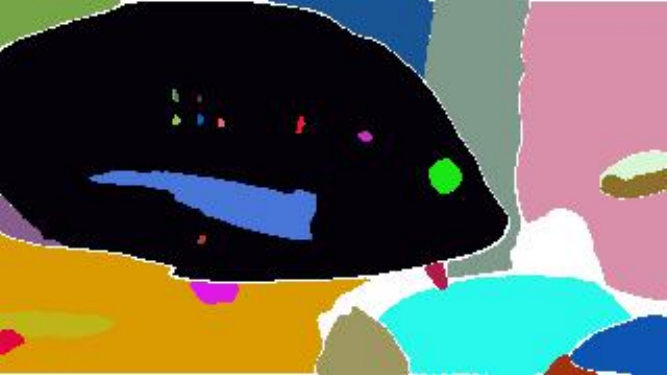} 
		\caption{\footnotesize ICSP }
	\end{subfigure}
    \begin{subfigure}{0.105\linewidth}
		\centering
        \includegraphics[width=\linewidth]{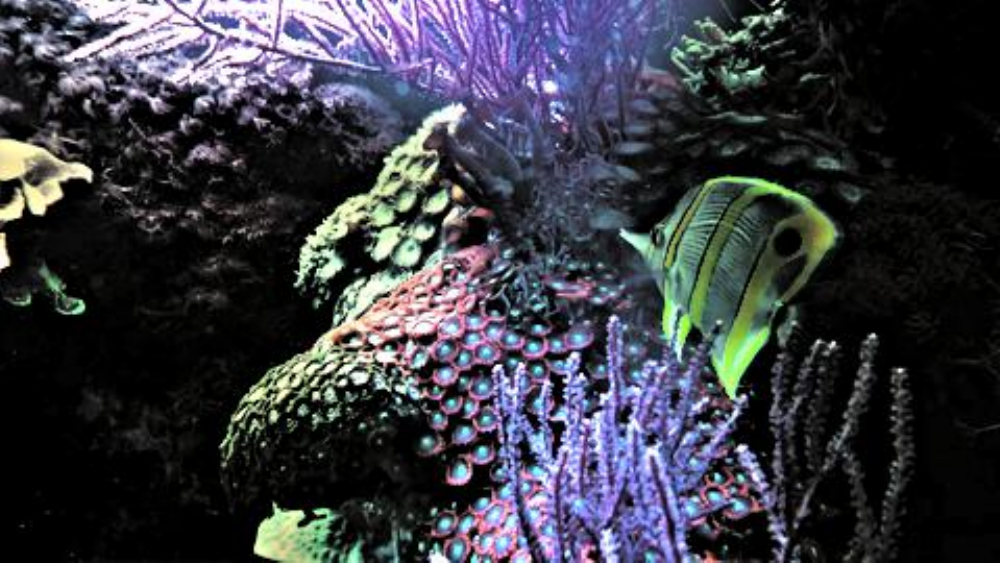} 
		\includegraphics[width=\linewidth]{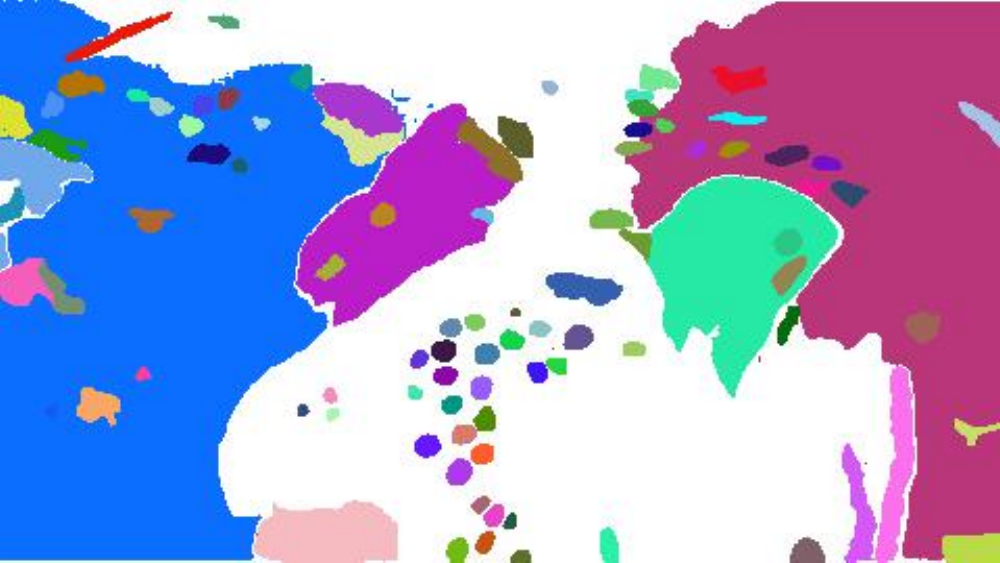} 
        \includegraphics[width=\linewidth]{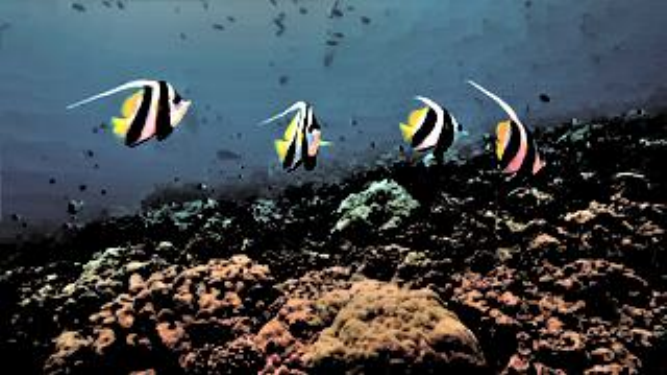} 
        \includegraphics[width=\linewidth]{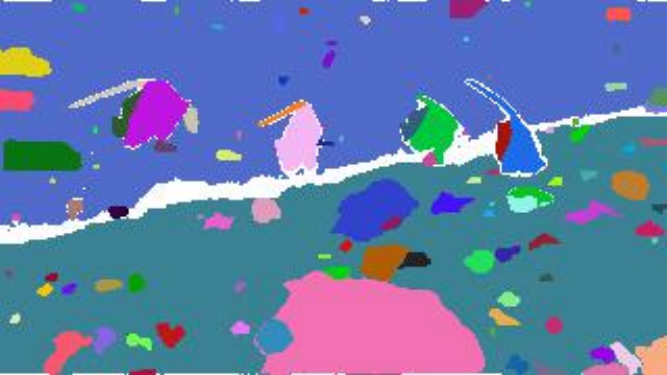} 
        \includegraphics[width=\linewidth]{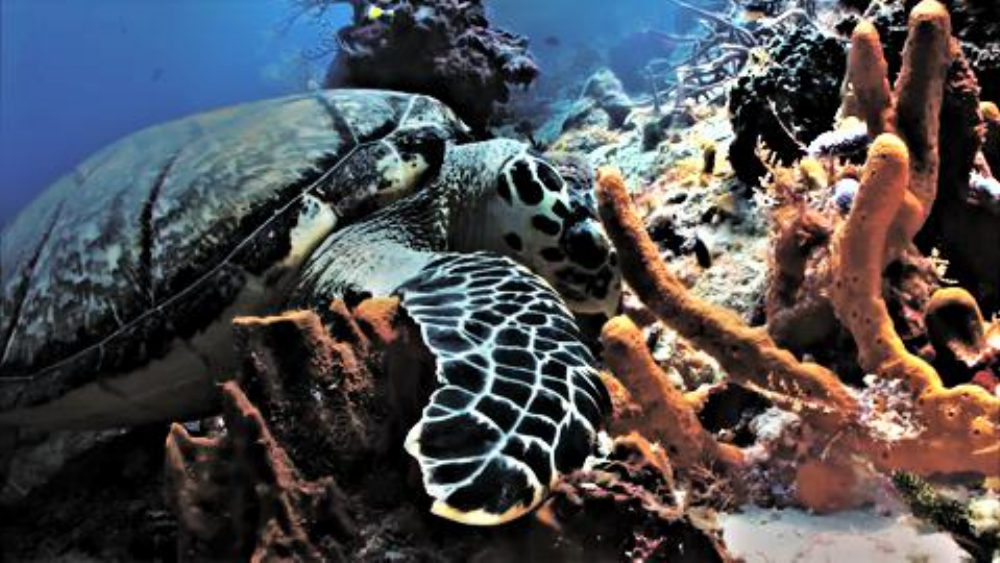} 
        \includegraphics[width=\linewidth]{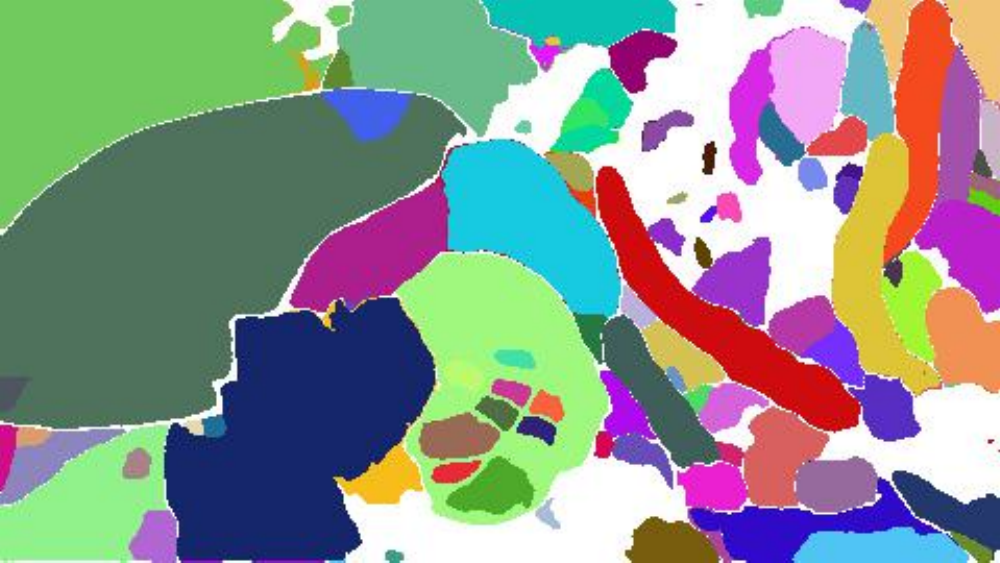} 
        \includegraphics[width=\linewidth]{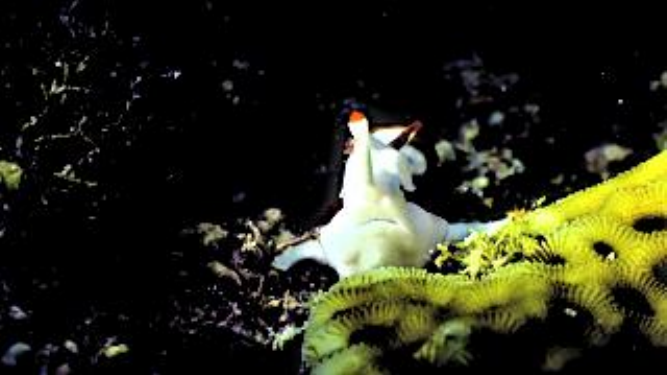} 
        \includegraphics[width=\linewidth]{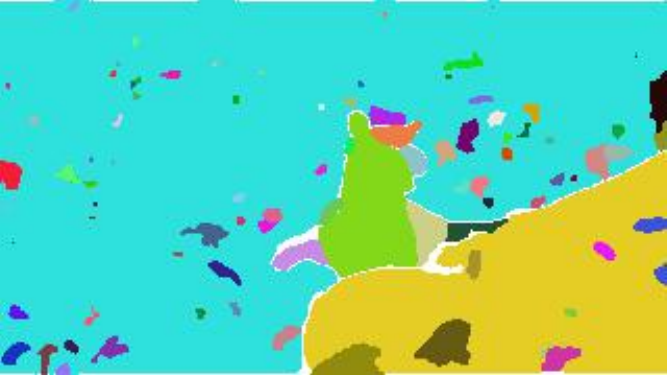} 
        \includegraphics[width=\linewidth]{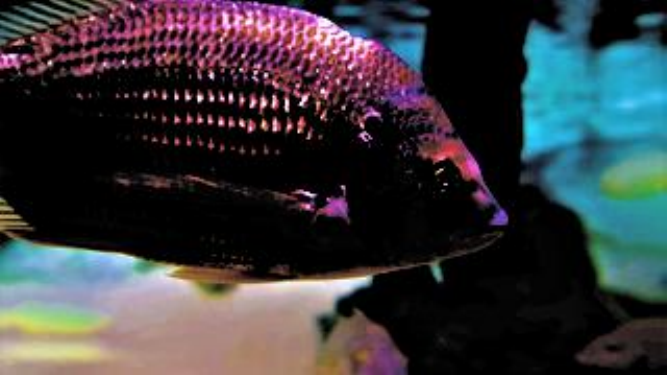} 
        \includegraphics[width=\linewidth]{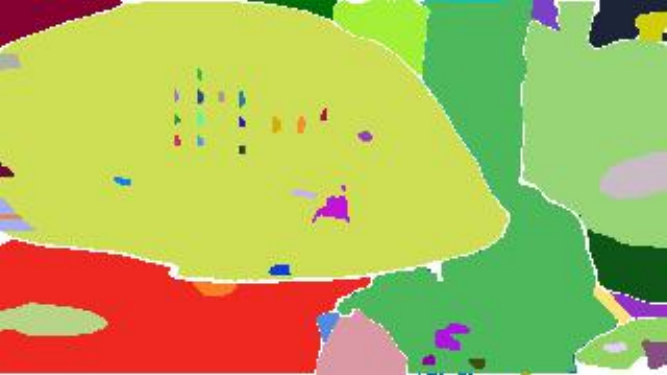} 
		\caption{\footnotesize PCDE}
	\end{subfigure}
	\begin{subfigure}{0.105\linewidth}
		\centering
        \includegraphics[width=\linewidth]{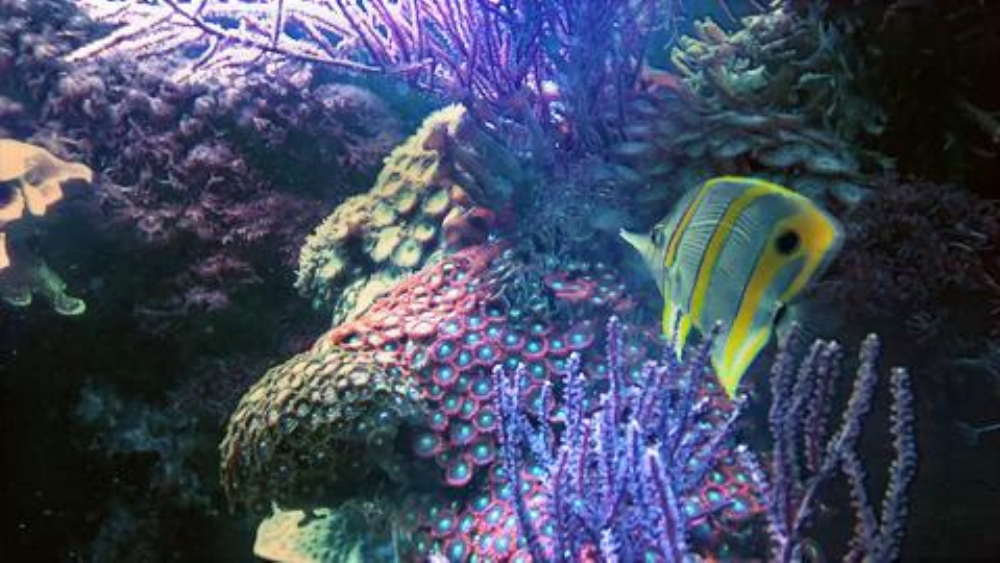} 
		\includegraphics[width=\linewidth]{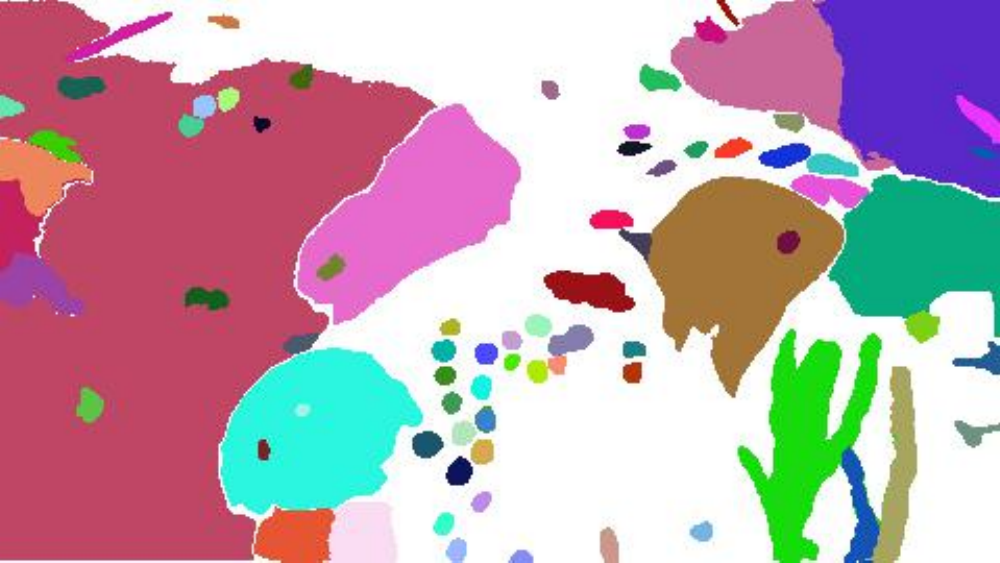} 
        \includegraphics[width=\linewidth]{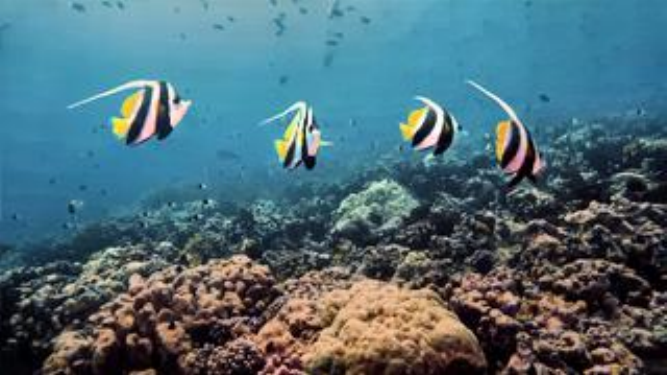} 
        \includegraphics[width=\linewidth]{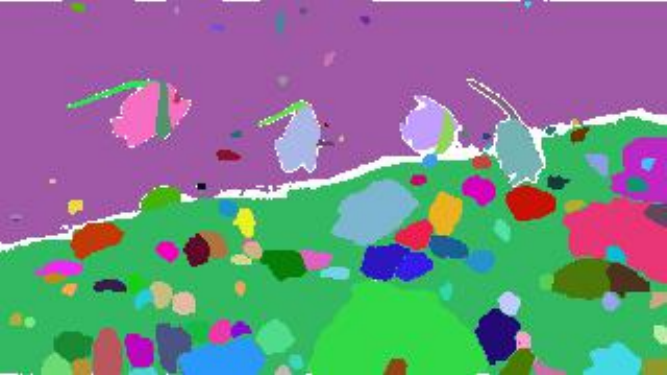} 
        \includegraphics[width=\linewidth]{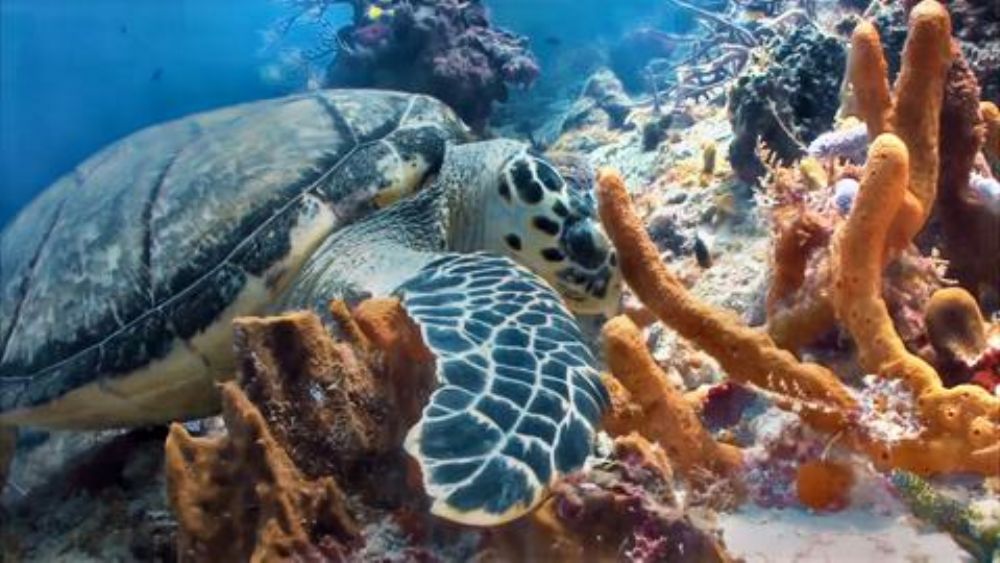} 
        \includegraphics[width=\linewidth]{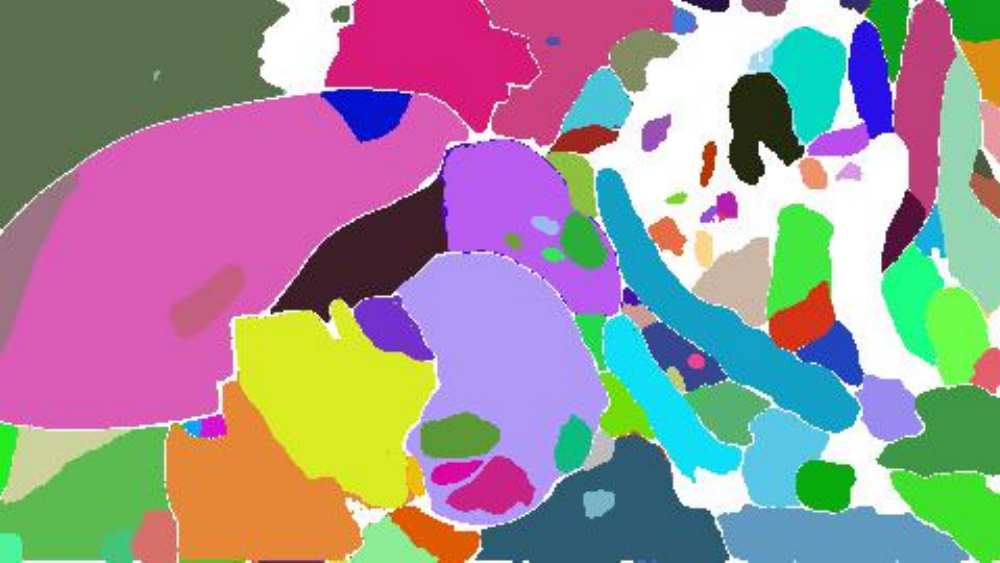} 
        \includegraphics[width=\linewidth]{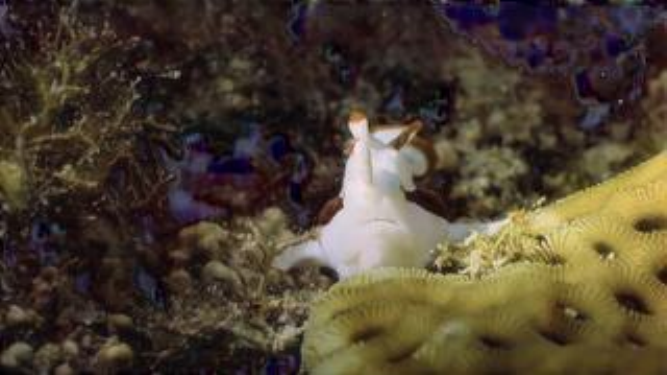} 
        \includegraphics[width=\linewidth]{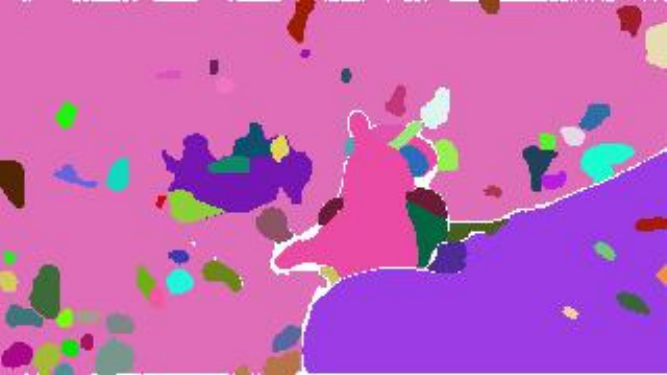} 
        \includegraphics[width=\linewidth]{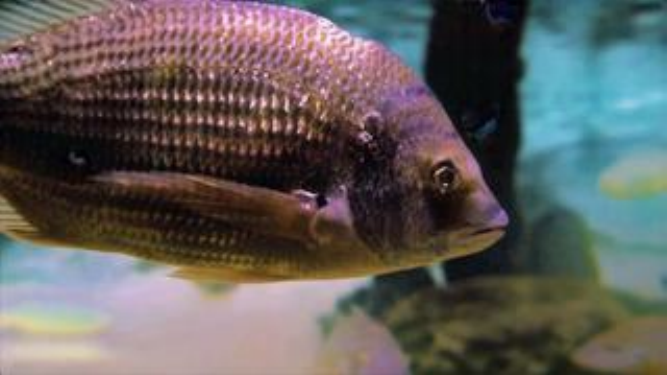} 
        \includegraphics[width=\linewidth]{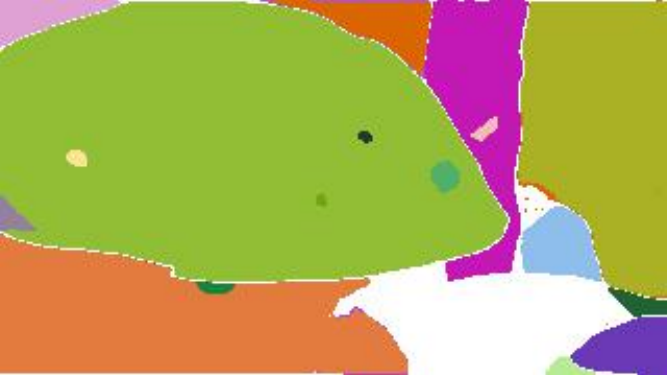} 
        \caption{\footnotesize UDAformer }
	\end{subfigure}
	\begin{subfigure}{0.105\linewidth}
		\centering
        \includegraphics[width=\linewidth]{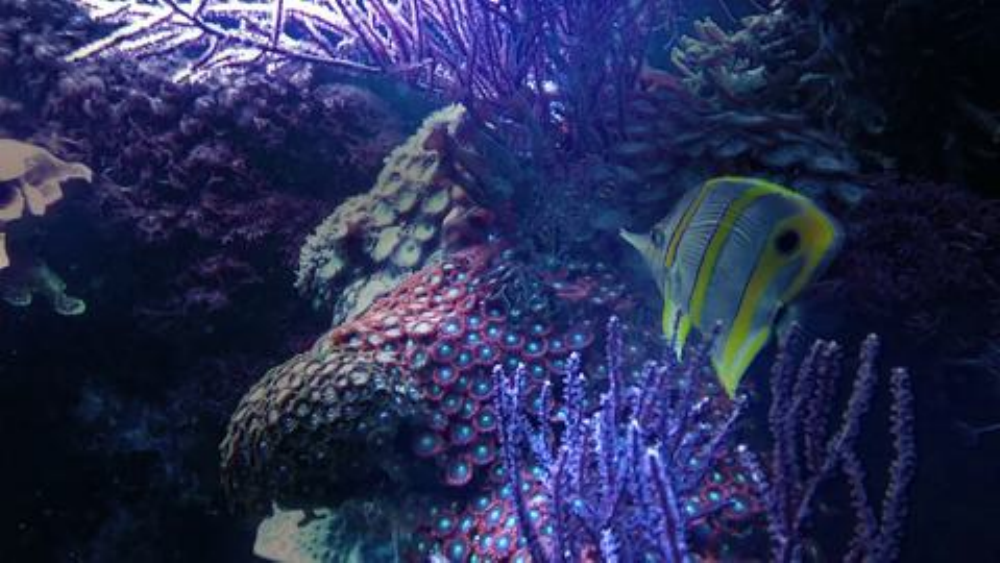} 
		\includegraphics[width=\linewidth]{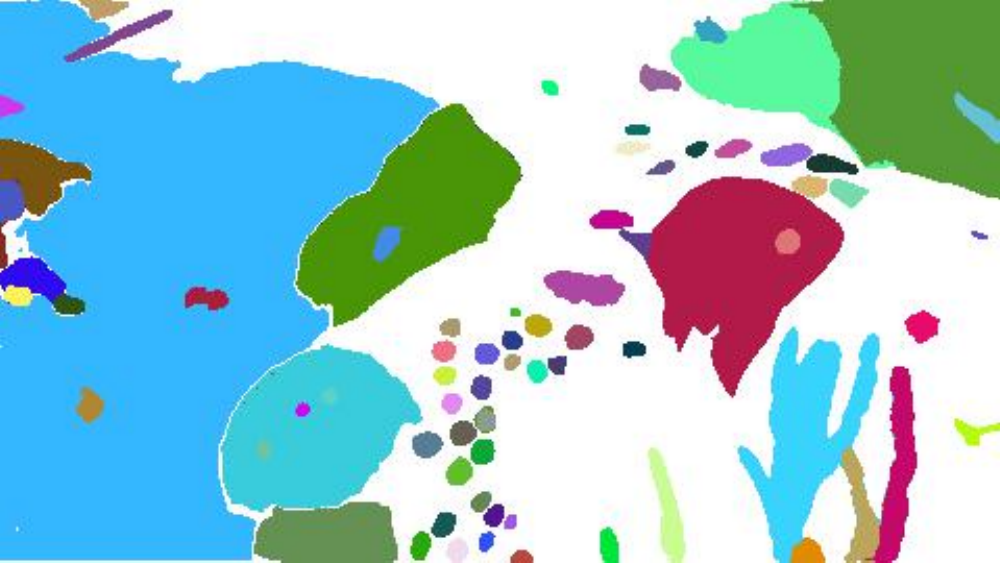} 
        \includegraphics[width=\linewidth]{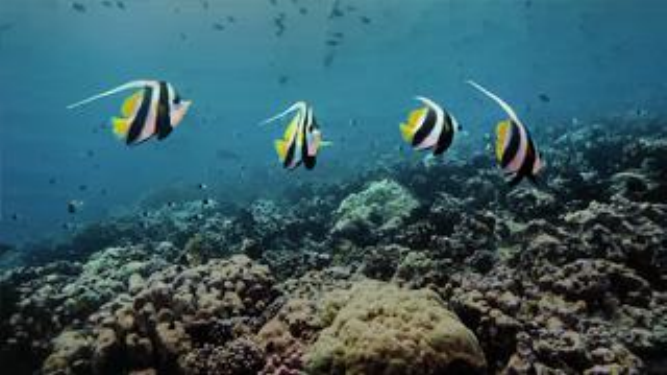} 
        \includegraphics[width=\linewidth]{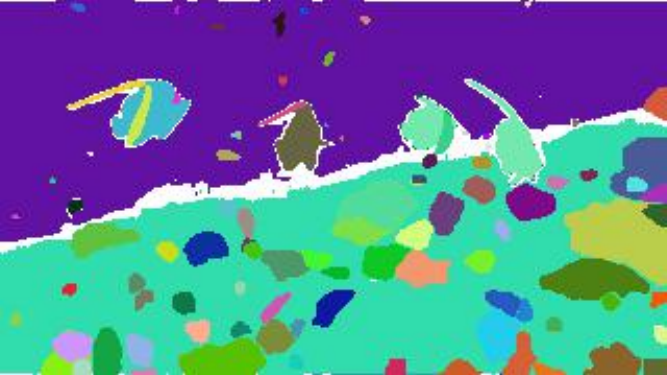} 
        \includegraphics[width=\linewidth]{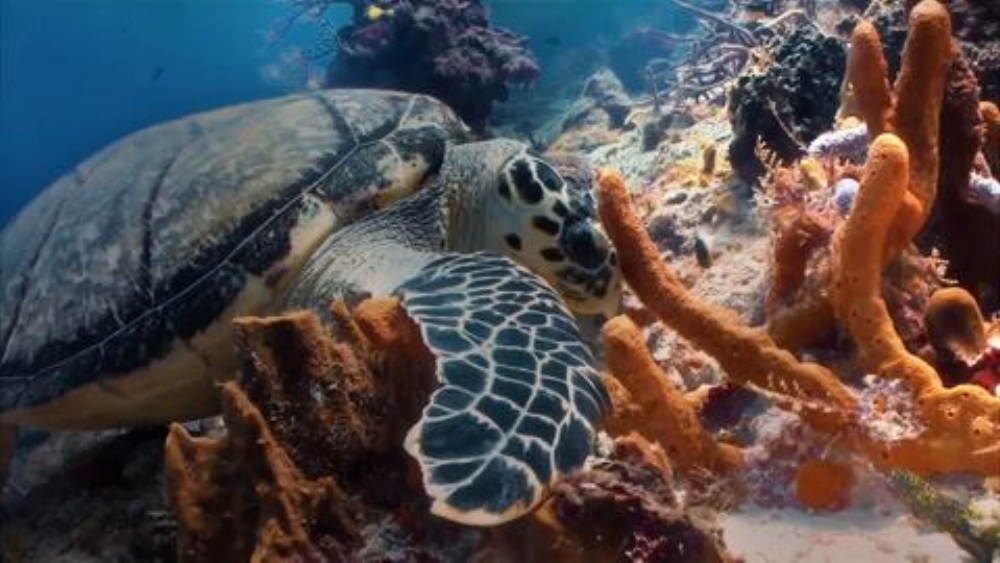} 
        \includegraphics[width=\linewidth]{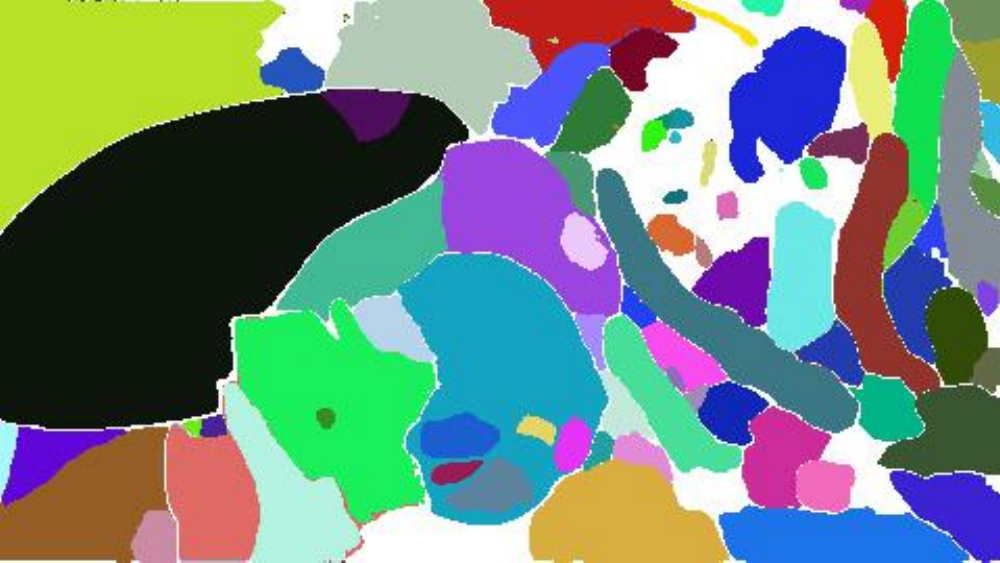} 
        \includegraphics[width=\linewidth]{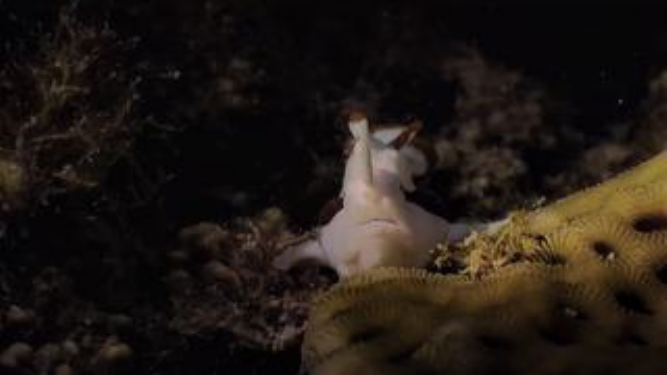} 
        \includegraphics[width=\linewidth]{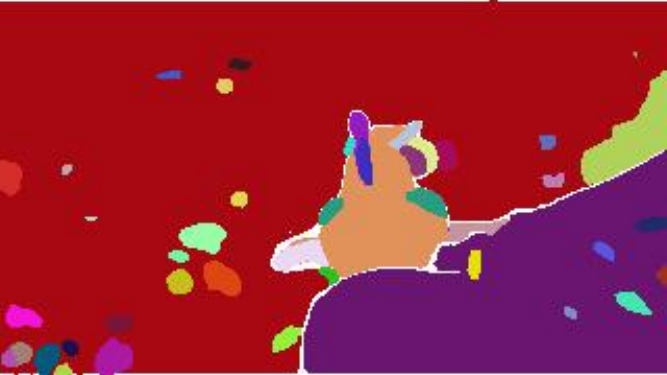} 
        \includegraphics[width=\linewidth]{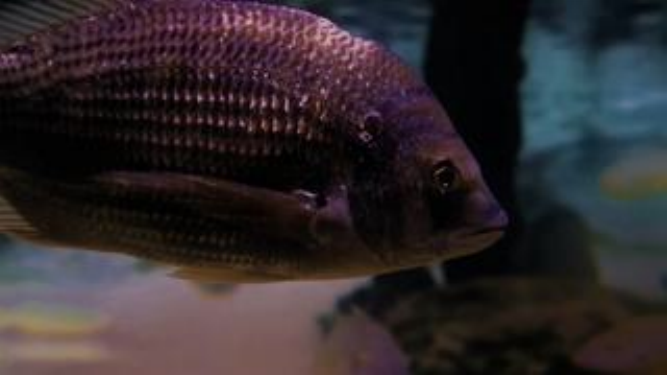} 
        \includegraphics[width=\linewidth]{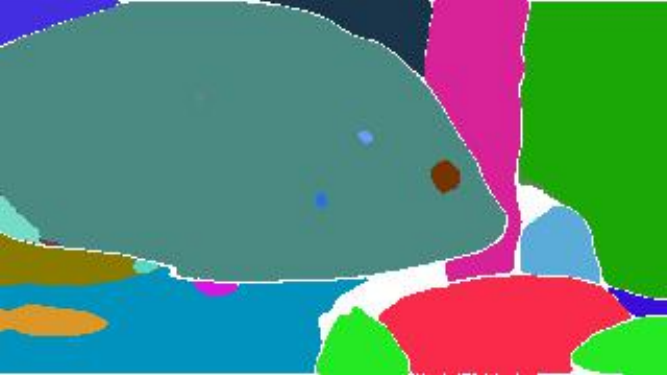} 
		\caption{\footnotesize SMDR-IS}
	\end{subfigure}
    \begin{subfigure}{0.105\linewidth}
		\centering
        \includegraphics[width=\linewidth]{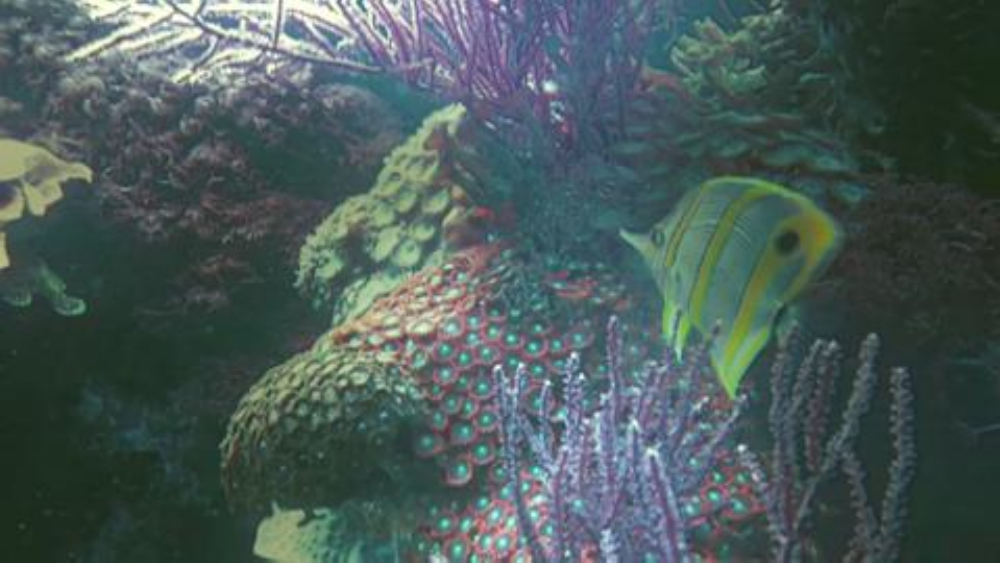} 
		\includegraphics[width=\linewidth]{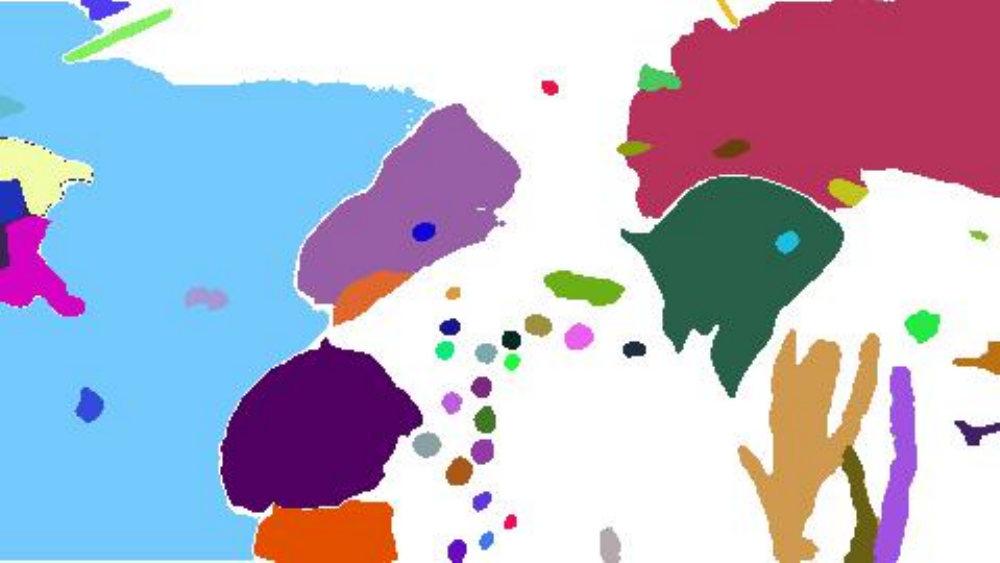} 
        \includegraphics[width=\linewidth]{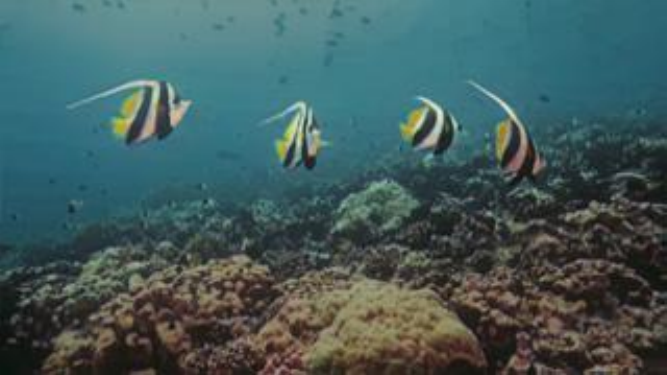} 
        \includegraphics[width=\linewidth]{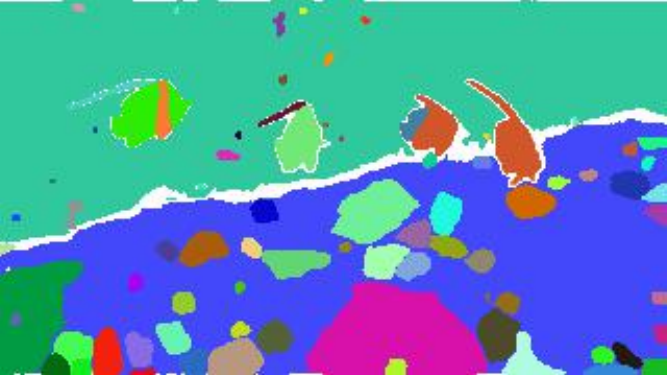} 
        \includegraphics[width=\linewidth]{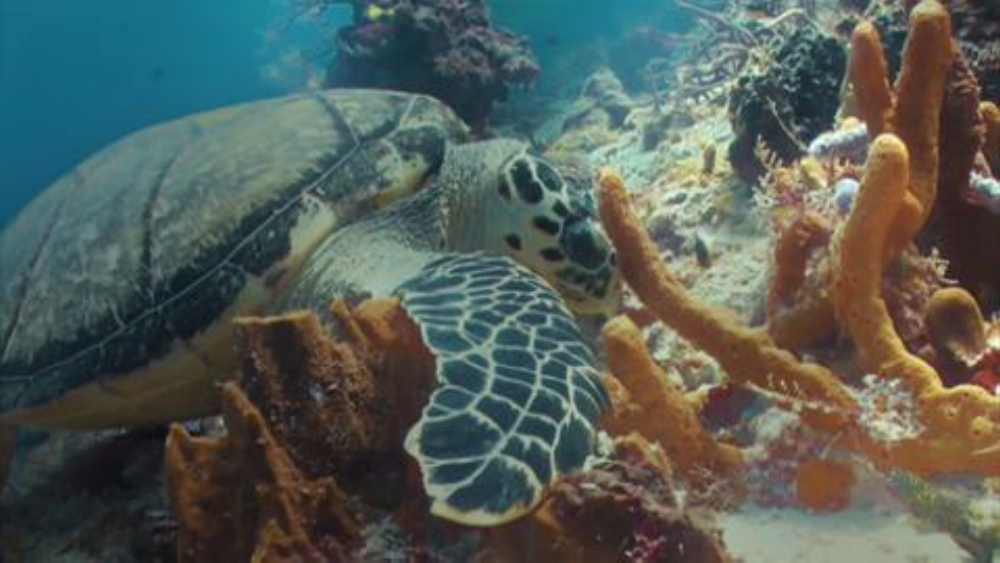} 
        \includegraphics[width=\linewidth]{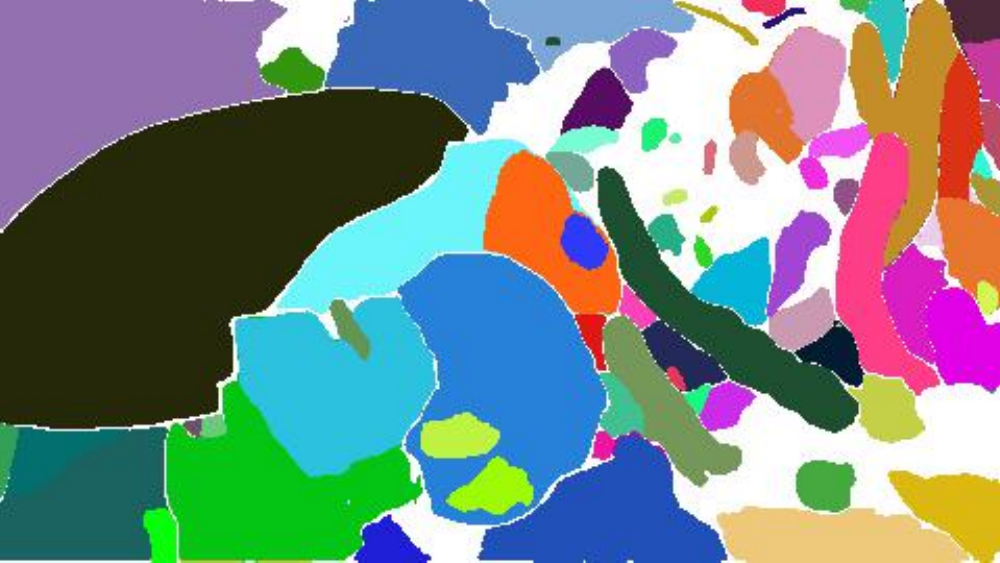} 
        \includegraphics[width=\linewidth]{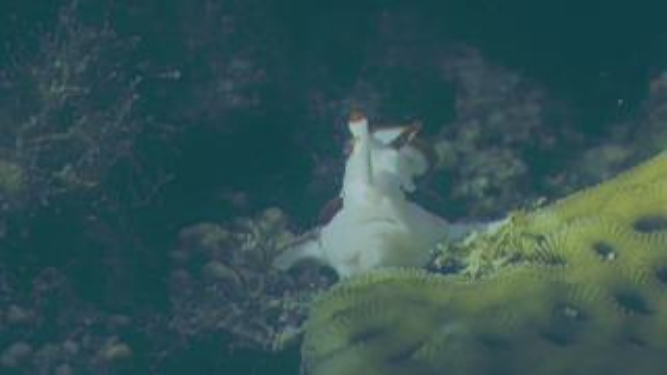} 
        \includegraphics[width=\linewidth]{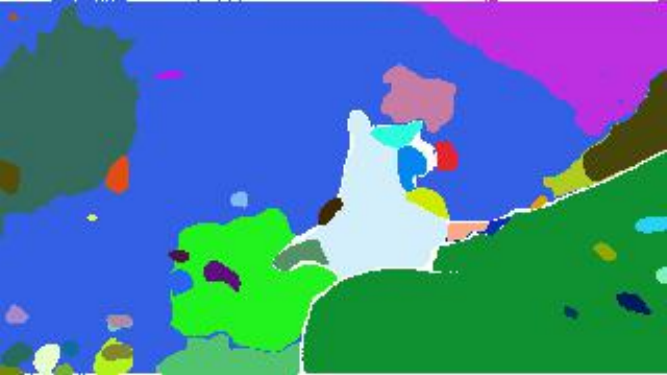} 
        \includegraphics[width=\linewidth]{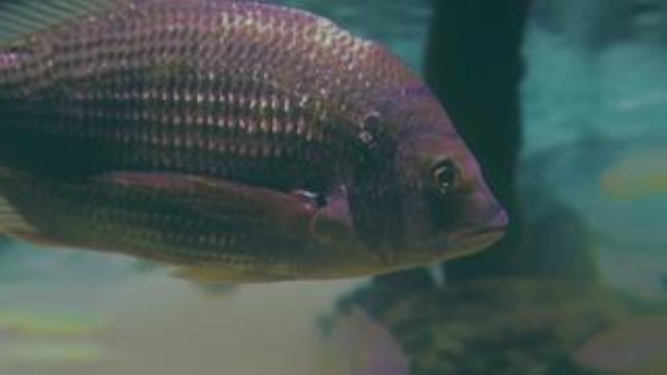} 
        \includegraphics[width=\linewidth]{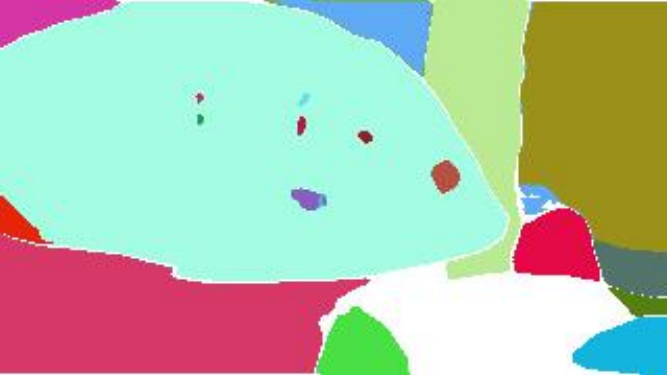} 
		\caption{\footnotesize UDnet}
	\end{subfigure}
    	\begin{subfigure}{0.105\linewidth}
		\centering
        \includegraphics[width=\linewidth]{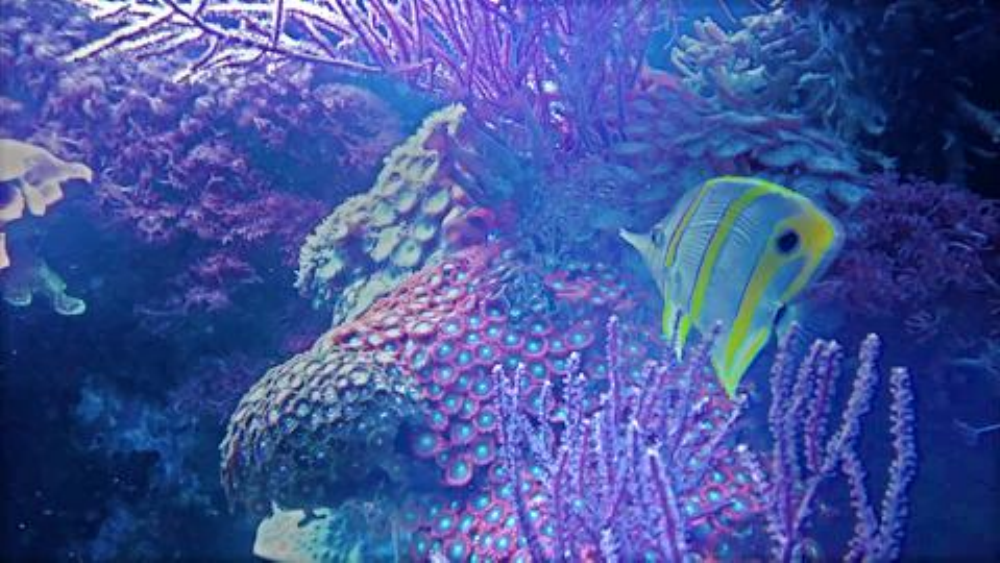} 
		\includegraphics[width=\linewidth]{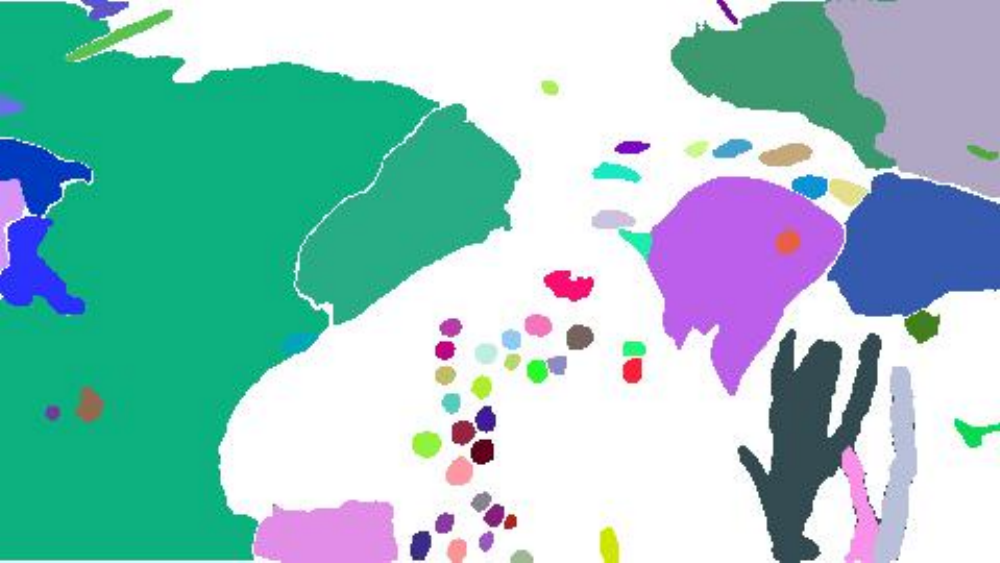} 
        \includegraphics[width=\linewidth]{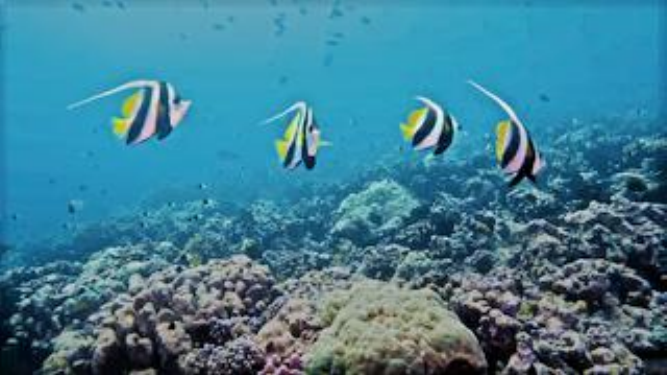} 
        \includegraphics[width=\linewidth]{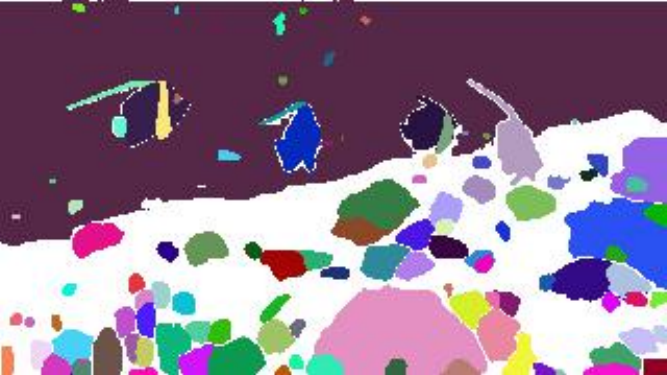} 
        \includegraphics[width=\linewidth]{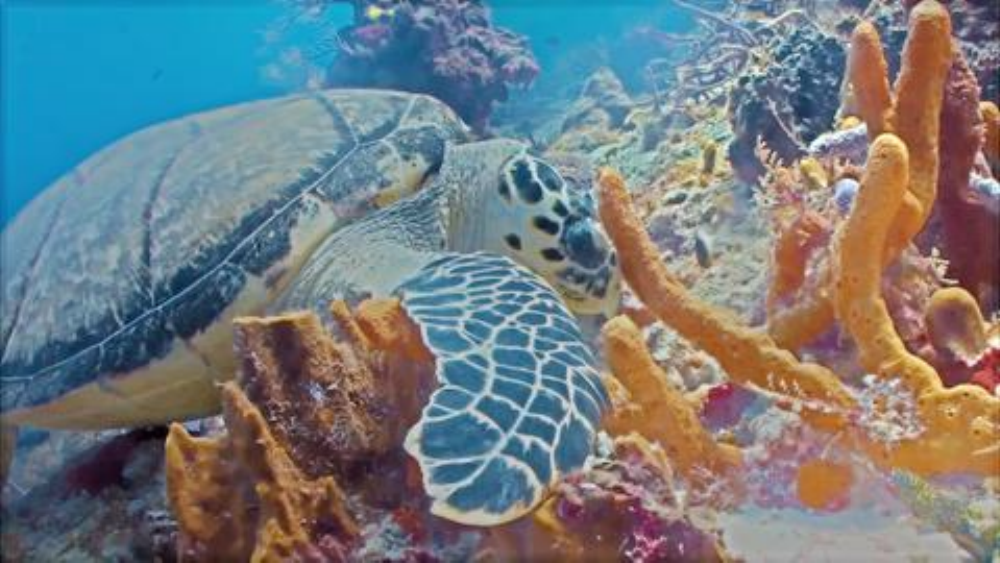} 
        \includegraphics[width=\linewidth]{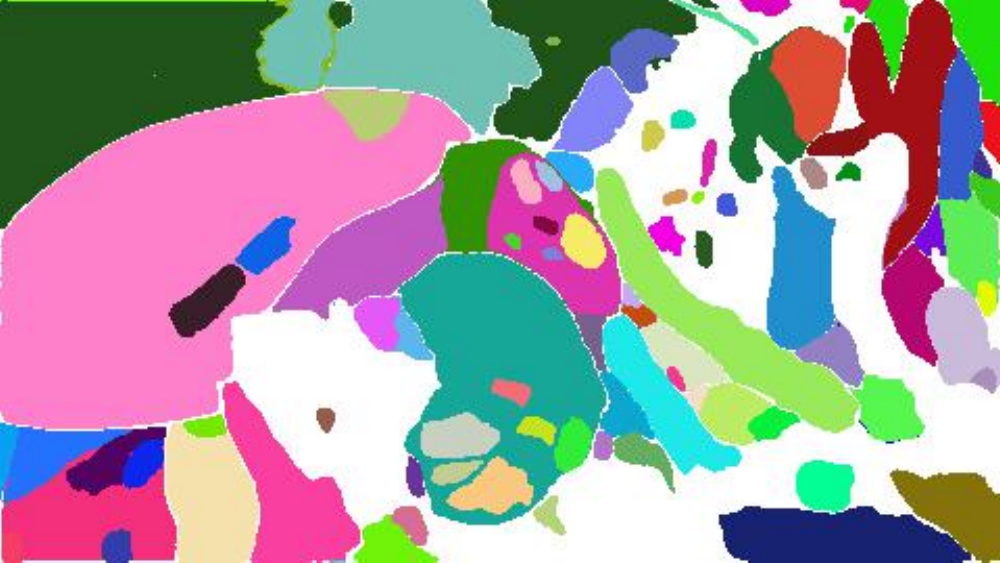} 
        \includegraphics[width=\linewidth]{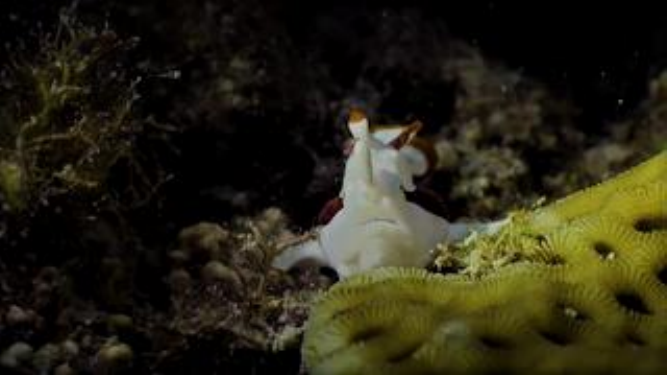} 
        \includegraphics[width=\linewidth]{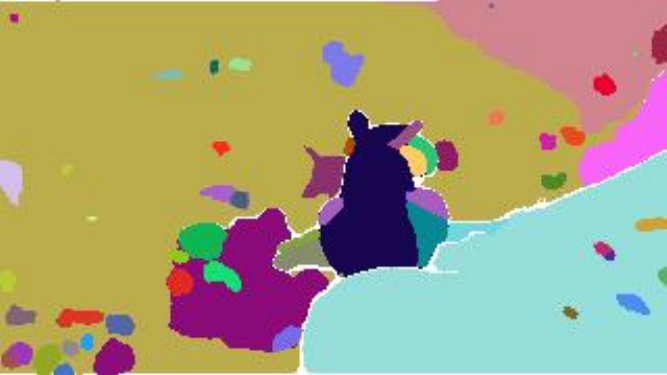} 
        \includegraphics[width=\linewidth]{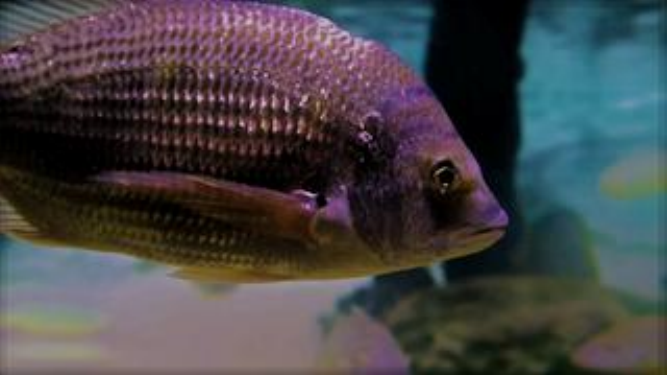} 
        \includegraphics[width=\linewidth]{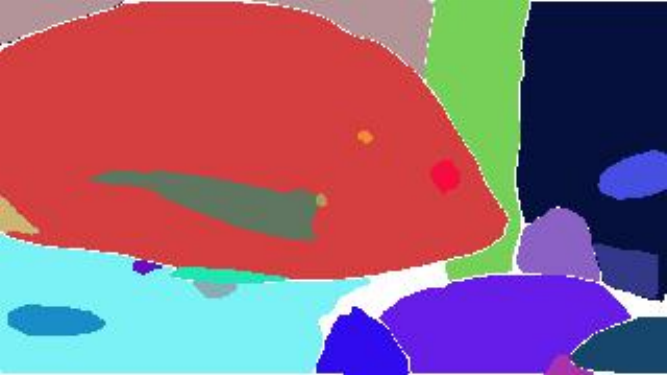} 
        \caption{\footnotesize EIB-FNDL }
	\end{subfigure}
    \begin{subfigure}{0.105\linewidth}
		\centering
        \includegraphics[width=\linewidth]{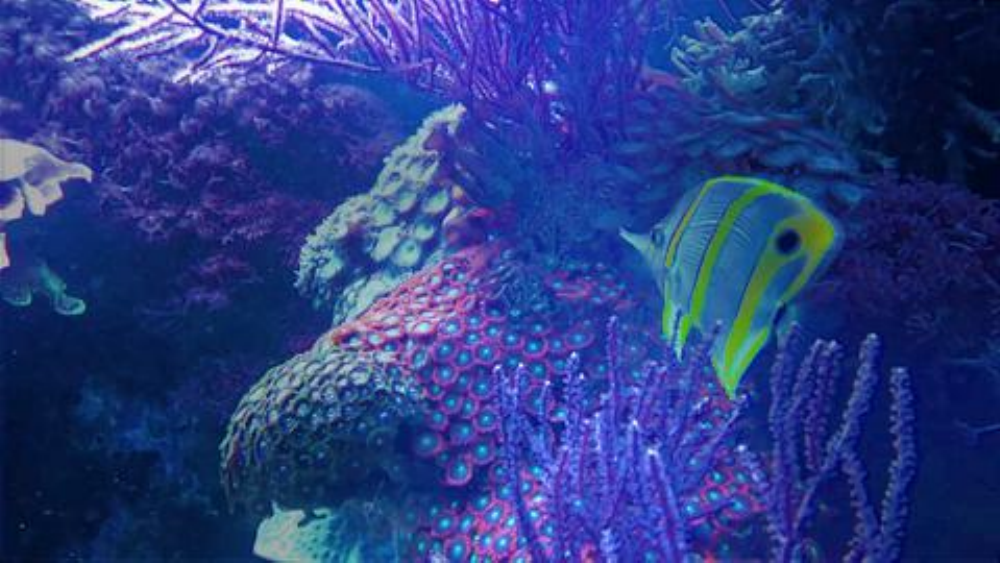} 
		\includegraphics[width=\linewidth]{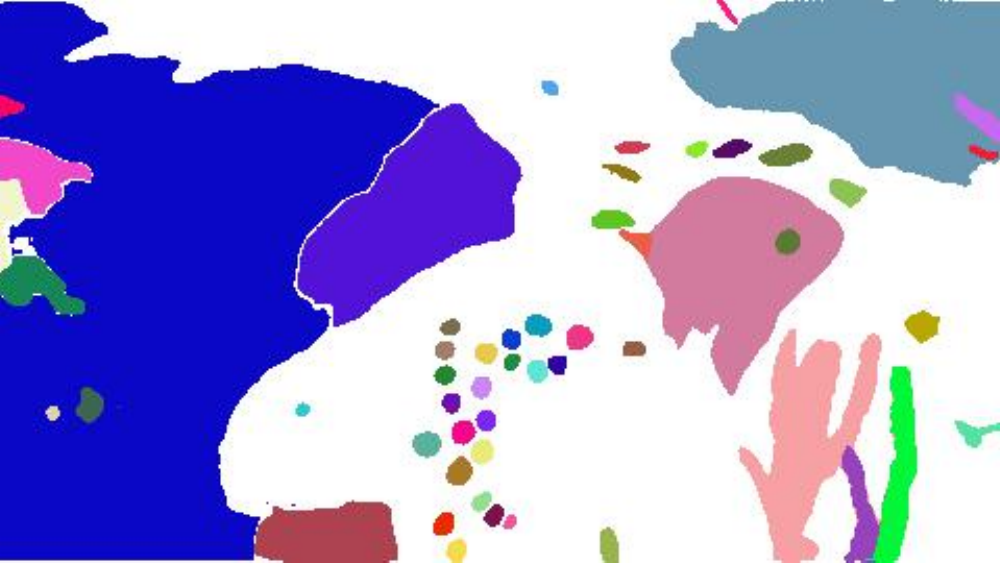} 
        \includegraphics[width=\linewidth]{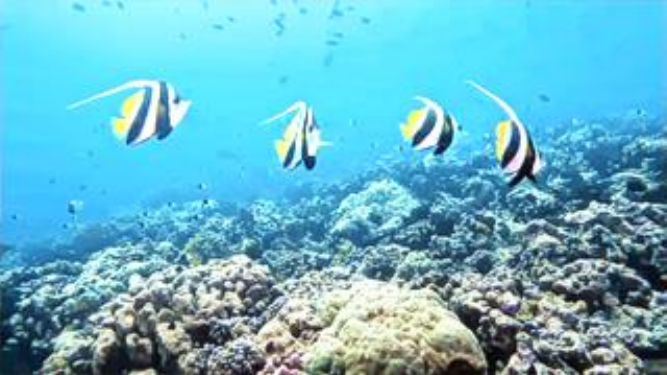} 
        \includegraphics[width=\linewidth]{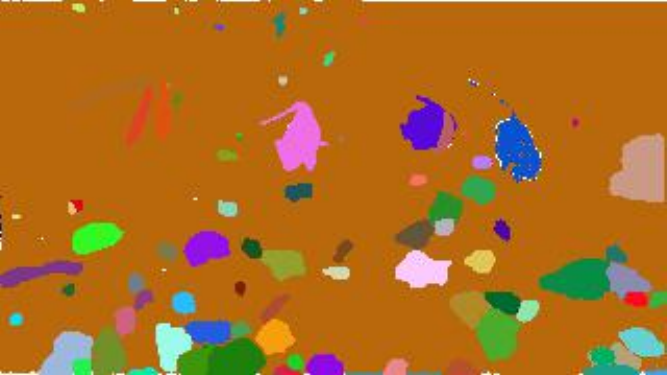} 
        \includegraphics[width=\linewidth]{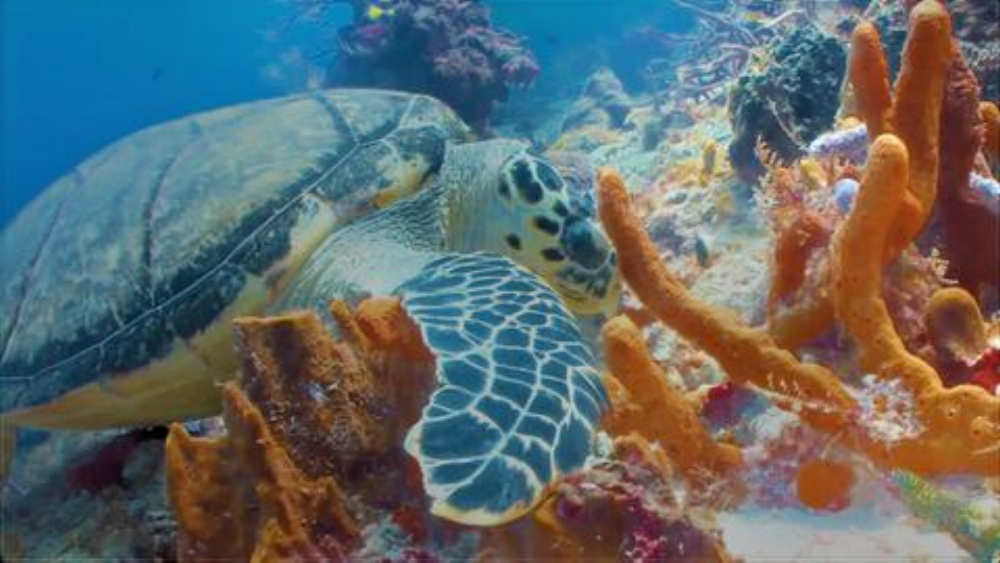} 
        \includegraphics[width=\linewidth]{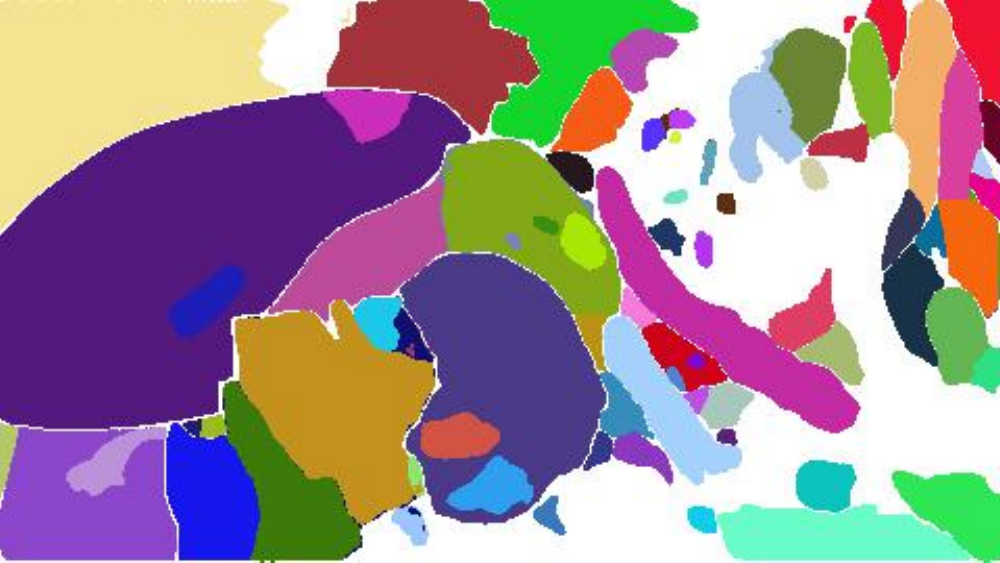} 
        \includegraphics[width=\linewidth]{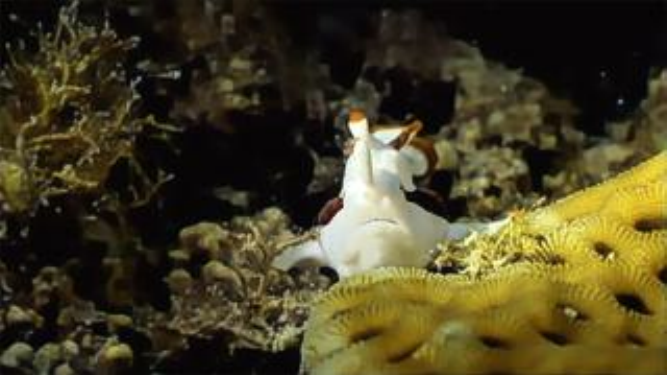} 
        \includegraphics[width=\linewidth]{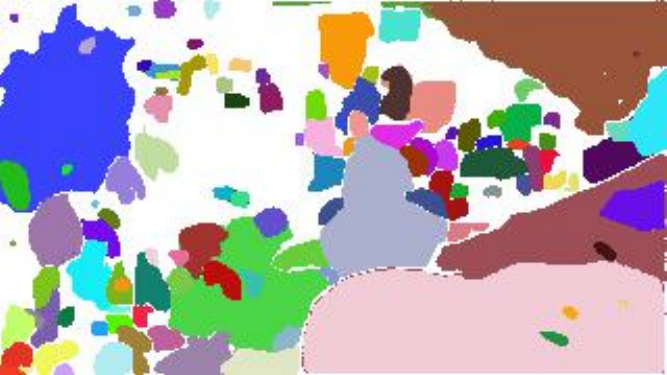} 
        \includegraphics[width=\linewidth]{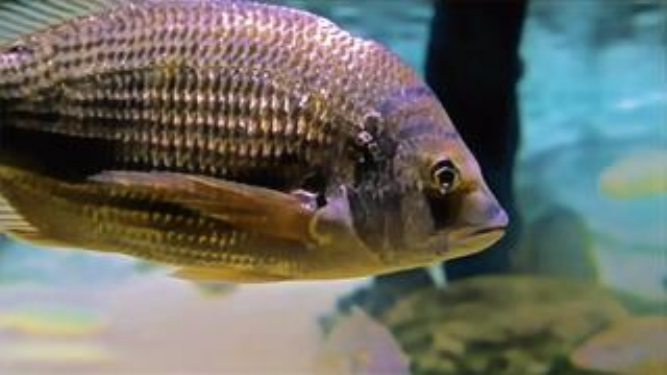} 
        \includegraphics[width=\linewidth]{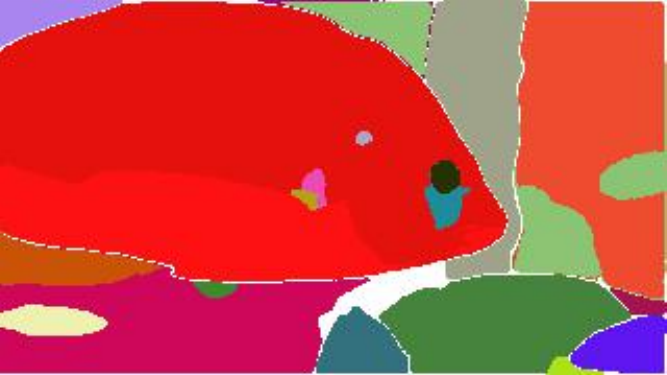} 
		\caption{\footnotesize ALEN}
	\end{subfigure}
    \begin{subfigure}{0.105\linewidth}
		\centering
        \includegraphics[width=\linewidth]{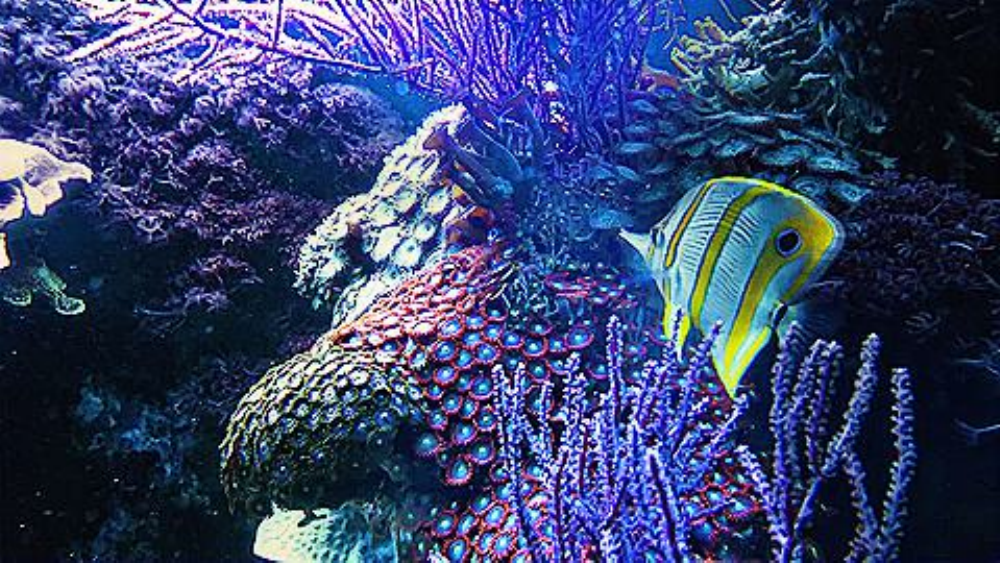} 
		\includegraphics[width=\linewidth]{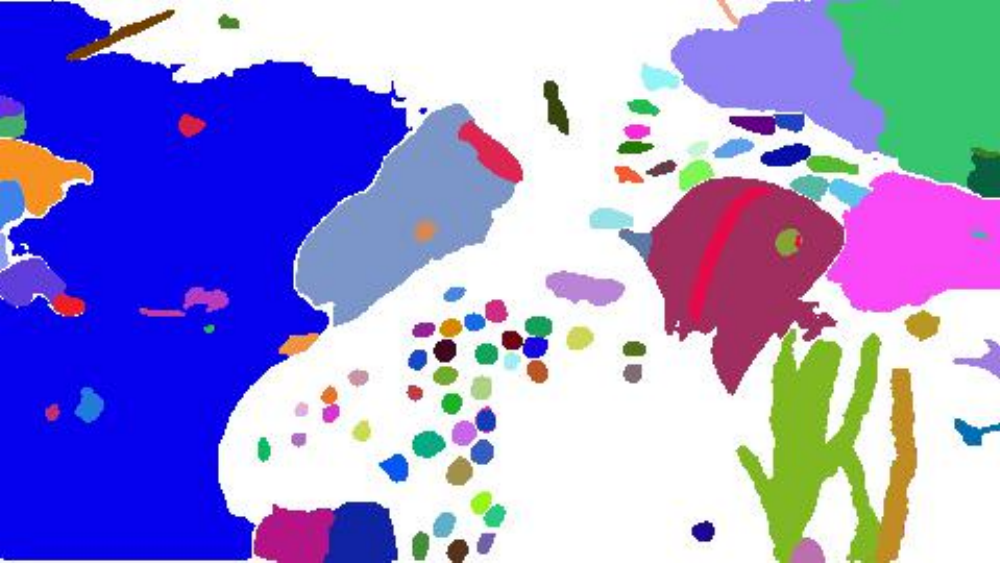}  
        \includegraphics[width=\linewidth]{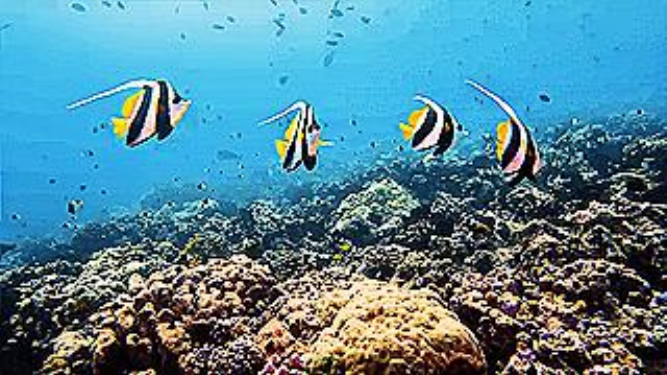} 
        \includegraphics[width=\linewidth]{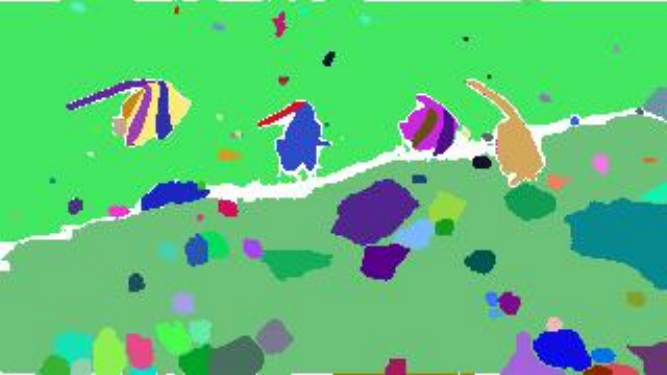} 
        \includegraphics[width=\linewidth]{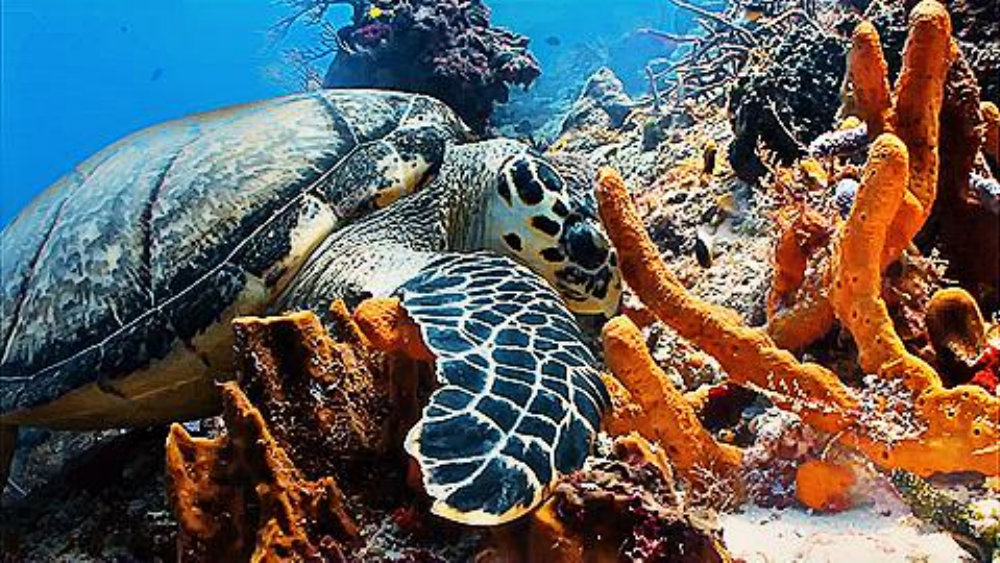} 
        \includegraphics[width=\linewidth]{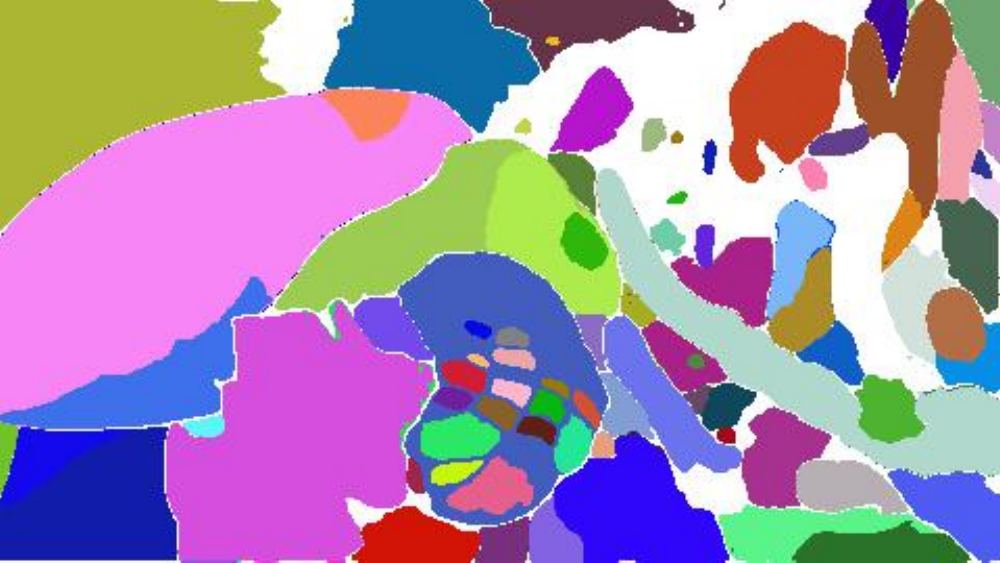} 
        \includegraphics[width=\linewidth]{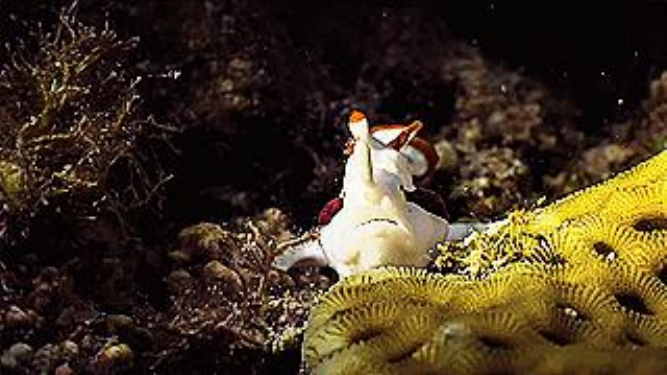} 
        \includegraphics[width=\linewidth]{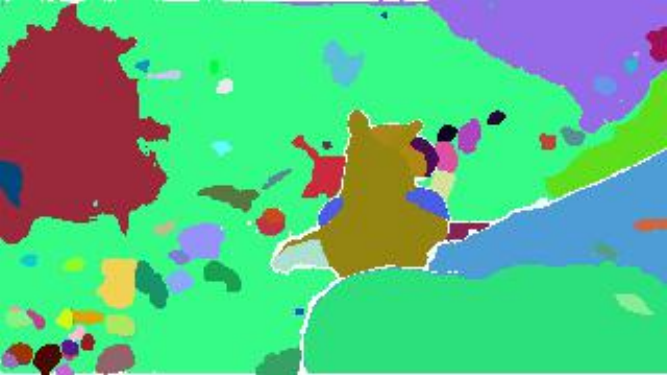} 
        \includegraphics[width=\linewidth]{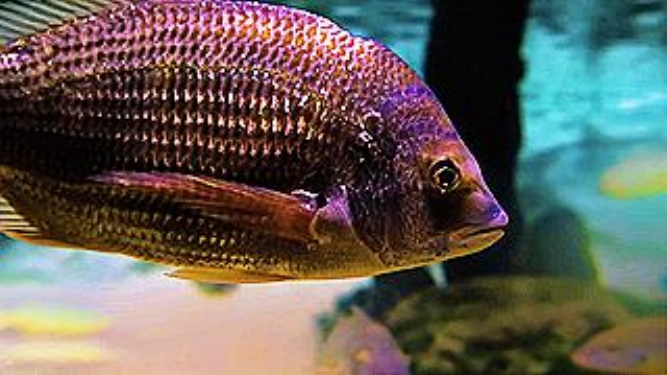} 
        \includegraphics[width=\linewidth]{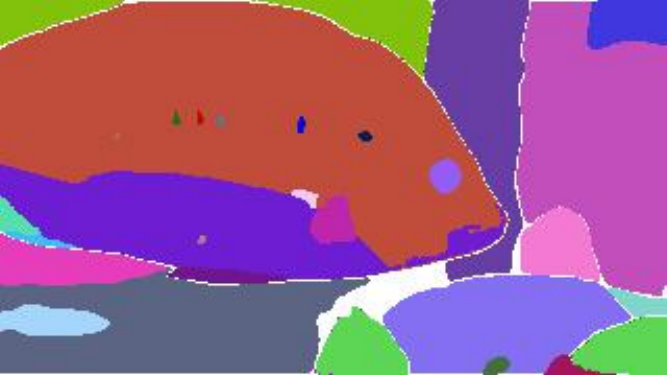} 
		\caption{\footnotesize UNIR-Net}
	\end{subfigure}
 
	\caption{Visual comparison of enhancement results and their corresponding segmentation outputs in images with non-uniform illumination. The figure demonstrates the influence of enhancement quality on segmentation performance.}
	\label{SEGMENT}
\end{figure*}

To demonstrate the practical utility of UNIR-Net, the model was applied to real underwater scenes featuring marine fauna. These included fish, nudibranchs, and sea turtles, which are commonly observed during coral reef inspections or biodiversity surveys. Figure \ref{SEGMENT} presents NUI images from these scenes, illustrating typical conditions encountered by Remotely Operated Vehicles (ROVs) and Autonomous Underwater Vehicles (AUVs), such as low-light environments and other visibility challenges.

UNIR-Net was compared against seven state-of-the-art methods: ICSP, PCDE, UDAformer, SMDR-IS, UDnet, EIB-FNDL, and ALEN. The visual comparisons, as shown in Fig.~\ref{SEGMENT}, demonstrate that most of these methods significantly enhance visibility under challenging conditions. This improvement reveals fine details and key structural features that are essential for ecological monitoring.

In addition, integrating UNIR-Net with a segmentation model such as SAM \cite{kirillov2023segment} enables more precise identification and localization of marine species. This, in turn, facilitates downstream tasks such as population assessment and habitat mapping. This setup simulates a realistic workflow in which enhancement and segmentation are jointly applied to underwater survey data. The results highlight the practical relevance of UNIR-Net for real-world deployments in marine exploration and biodiversity assessment pipelines.

\subsubsection{Computational cost comparison}

The computational cost of UNIR-Net is assessed and compared with state-of-the-art methods under two scenarios: GPU-based and CPU-based implementations. For the evaluation, 30 images with a resolution of 1280 × 720 were used. Tables \ref{DL} and \ref{NON_DL} provide detailed comparisons in terms of inference speed, parameter count, FLOPs, memory usage, and platform implementation.

Table \ref{DL} highlights the efficiency of GPU-based deep learning methods. Among these, UDnet demonstrates the fastest inference speed at 0.0618 seconds. EIB-FNDL has the smallest number of parameters, with only 0.0013 million, making it highly efficient for lightweight applications. It also requires the fewest floating-point operations, at 1.10 GFLOPs, which contributes to its overall computational efficiency, despite a slightly slower inference speed due to other factors. UNIR-Net offers a balanced trade-off between computational efficiency and model complexity, with an inference time of 0.3297 seconds, 0.34 million parameters, and 311.56 GFLOPs. However, it consumes slightly more memory than some of the lighter models, using 2,085.57 MB. ALEN stands out for its competitive performance, although it incurs a high computational cost. It has an inference time of 1.0031 seconds and requires 1,600.23 GFLOPs, highlighting the trade-off between accuracy and efficiency in transformer-based approaches.

\begin{table*}[ht]
	\centering
	\caption{Comparison of computational cost for GPU-based methods.}
	\label{DL}
	\resizebox{0.9\textwidth}{!}{
		\begin{tabular}{l c c c c c c}
			\hline
			Method & Year & Inference Speed (s) $\downarrow$ & \# Parameters $\downarrow$ & FLOPs (G) $\downarrow$ & Memory Usage (MB) $\downarrow$ & Platform \\ \hline
			%Zero-DCE~\cite{guo2020zero} & 2020 	& \textbf{0.2287} &	\textbf{0.08} (M)	& 72.99   &	1,222.23  &	PyTorch	\\
            UWNet~\cite{naik2021shallow} & 2021 & 0.2860 &	0.22 (M)	& 304.17   &	1,364.53  &	PyTorch	\\
            TCTL-Net~\cite{li2023tctl} & 2023 & 0.2371 &	99.72 (M)	& 56.62   &	765.45  &	PyTorch	\\
			UDAformer~\cite{shen2023udaformer} & 2023 &  1.7358 &	9.59 (M)	& 584.90   &	3,880.11 & PyTorch\\
            LENet~\cite{zhang2024liteenhancenet} & 2024 & 0.2596 &	0.01 (M)	& 9.70   &	1,543.99 & PyTorch\\
            SMDR-IS~\cite{zhang2024synergistic} & 2024 &  0.5575 &	12.25 (M)	& 724.78   & 2,161.63  & PyTorch\\
            %GACA~\cite{yao2024gaca} & 2024 & 0.6168 &	5.74 (M) & 652.26  &	1,488.04 & PyTorch\\
            UDnet~\cite{saleh2025adaptive}  &  2025 &  \textbf{0.0618} &	1.40 (M)	& 30.15  & \textbf{31.88}  & PyTorch\\
            EIB-FNDL~\cite{khajehvandi2025enhancing}  &  2025 &  0.1741 &	\textbf{0.0013 (M)}	& \textbf{1.10}  & 302.36  & PyTorch\\
            ALEN~\cite{perez2025alen}  &  2025 &  1.0031 &	 5.67 (M)	& 1,600.23  & 4,149.94  & PyTorch\\
            UNIR-Net & 2025  & 0.3297 &	0.34 (M)	& 311.56   &	2,085.57 &	PyTorch	\\ \hline
		\end{tabular}
	}
\end{table*}

Table \ref{NON_DL} presents a comparison of CPU-based methods, including traditional algorithms and other optimization techniques. In this category, MACT outperforms other methods with an impressive inference speed of 0.1116 seconds, making it well-suited for real-time or resource-constrained environments. UNIR-Net shows a moderate CPU inference time of 8.7085 seconds. Although it is not the fastest, UNIR-Net remains competitive among recent CPU-based learning models. In contrast, methods like UNTV show the slowest inference speed, taking 21.4268 seconds.

\begin{table}[ht]
	\centering
	\caption{Comparison of computational cost for CPU-based methods.}
	\label{NON_DL}
	\resizebox{0.5\textwidth}{!}{
		\begin{tabular}{l c c c c}
			\hline
			Method & Year & Inference Speed (s) $\downarrow$ & Platform \\ \hline
                %LIME~\cite{guo2016lime} & 2016 & 11.0941 & Python\\
                %DUAL~\cite{zhang2019dual} & 2019 & 22.3193 & Python \\
                UNTV~\cite{xie2021variational} & 2021 & 21.4268 & Matlab \\
                ACDC~\cite{zhang2022underwater} & 2022 & 2.1339 & Matlab \\
                MMLE~\cite{zhang2022underwaterMMLE} & 2022 & 0.9006 & Matlab \\
                ICSP~\cite{hou2023non} & 2023 & 1.4530 & Matlab\\
                PCDE~\cite{zhang2023PCDE} & 2023 & 3.0262 & Matlab\\
                HFM~\cite{an2024hfm} & 2024 & 3.2908 & Matlab\\
                GCP~\cite{jeon2024low}  & 2024 & 0.2375 & Python\\
                MACT~\cite{zhang2025mact} & 2025 & \textbf{0.1116} & Matlab\\
                UNIR-Net & 2025  & 8.7085 &	Pytorch\\ \hline
		\end{tabular}
	}
\end{table}

The results indicate that UNIR-Net is optimized for GPU acceleration, striking a balance between inference speed, computational resources, and memory usage. While the CPU-based implementation could be further optimized, the focus remains on leveraging GPU efficiency to handle the complex task of underwater image enhancement under non-uniform illumination.

\section{Conclusion}
\label{conlusions}
This article introduces UNIR-Net, an architecture specifically designed to address the challenges of limited visibility in underwater environments with non-uniform illumination. This method combines illumination enhancement and attention blocks with visual refinement and contrast correction modules, effectively preserving both the visual quality of images and maintaining appropriate illumination and edge detail.

Additionally, PUNI, a synthetic dataset, is presented and developed specifically to study and resolve issues related to uneven lighting conditions in underwater environments. Experimental results, both qualitative and quantitative, demonstrate that UNIR-Net outperforms state-of-the-art methods across various metrics, excelling in FIQA metrics while achieving competitive performance in the MUSIQ metric. UNIR-Net shows superior performance on the real-world NUID dataset, consistently achieving high average values across all evaluated subsets. Although the PUNI dataset is synthetically generated, it demonstrates strong performance when applied to real underwater images such as those in the NUID dataset. Nevertheless, the absence of real paired ground-truth data hinders a comprehensive evaluation of its effectiveness in real-world scenarios. As such, some domain gap may exist when deploying models trained on PUNI to real-world underwater imagery. Nonetheless, the model’s robust performance on the real, unpaired NUID dataset suggests promising generalization capabilities beyond synthetic data.

This limitation is acknowledged, and future research would benefit from the development or collection of real paired datasets captured under diverse illumination conditions. Furthermore, in practical applications, UNIR-Net significantly impacts high-level tasks such as marine fauna segmentation, thanks to its ability to enhance visual details and edges. In terms of computational cost, although UNIR-Net is not the most efficient method in resource usage, its balance between visual quality and computational consumption positions it favorably compared to lighter alternatives that produce inferior visual results.

As future research directions, expanding the PUNI dataset is proposed to incorporate a greater diversity of characteristics typical of underwater environments with non-uniform illumination. Additionally, optimizing the UNIR-Net architecture aims to reduce its computational cost further without compromising performance.

\section{Acknowledgments}
This work is supported by the Zhejiang Provincial Natural Science Foundation of China (No.LY24F020004, No.LZ23F020004), the Zhejiang Gongshang University "Digital+" Disciplinary Construction Management Project (No.SZJ2022B016) and the University of Guadalajara.

 \bibliographystyle{elsarticle-num} 
 \bibliography{bibfile}

\vspace{5\baselineskip}

\noindent
\begin{minipage}{0.15\textwidth}
    \includegraphics[width=\linewidth]{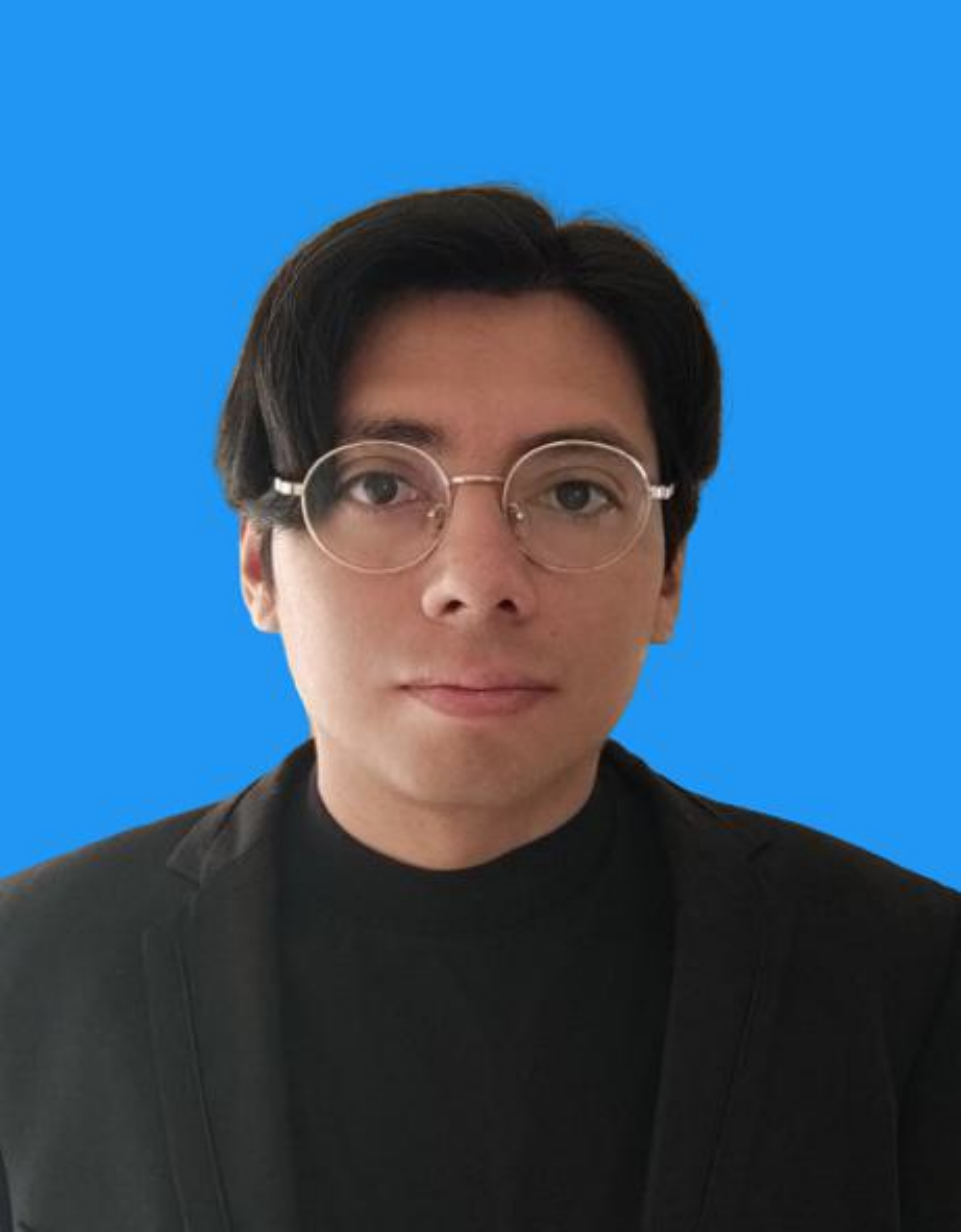}
\end{minipage}%
\hspace{0.02\textwidth}%
\begin{minipage}{0.82\textwidth}
    \vspace{-1cm}
    \hspace{1cm} \textbf{Ezequiel Perez-Zarate} received the B.S. degree in Communications and Electronics Engineering from the University of Guadalajara, Mexico, in 2018, and the M.E. degree in Computer Science and Technology from Zhejiang Gongshang University, China, in 2025. His research interests include computer vision, computer graphics, machine and deep learning, digital image processing, and real-time image processing.
\end{minipage}

\vspace{3\baselineskip}

\noindent
\begin{minipage}{0.15\textwidth}
    \includegraphics[width=\linewidth]{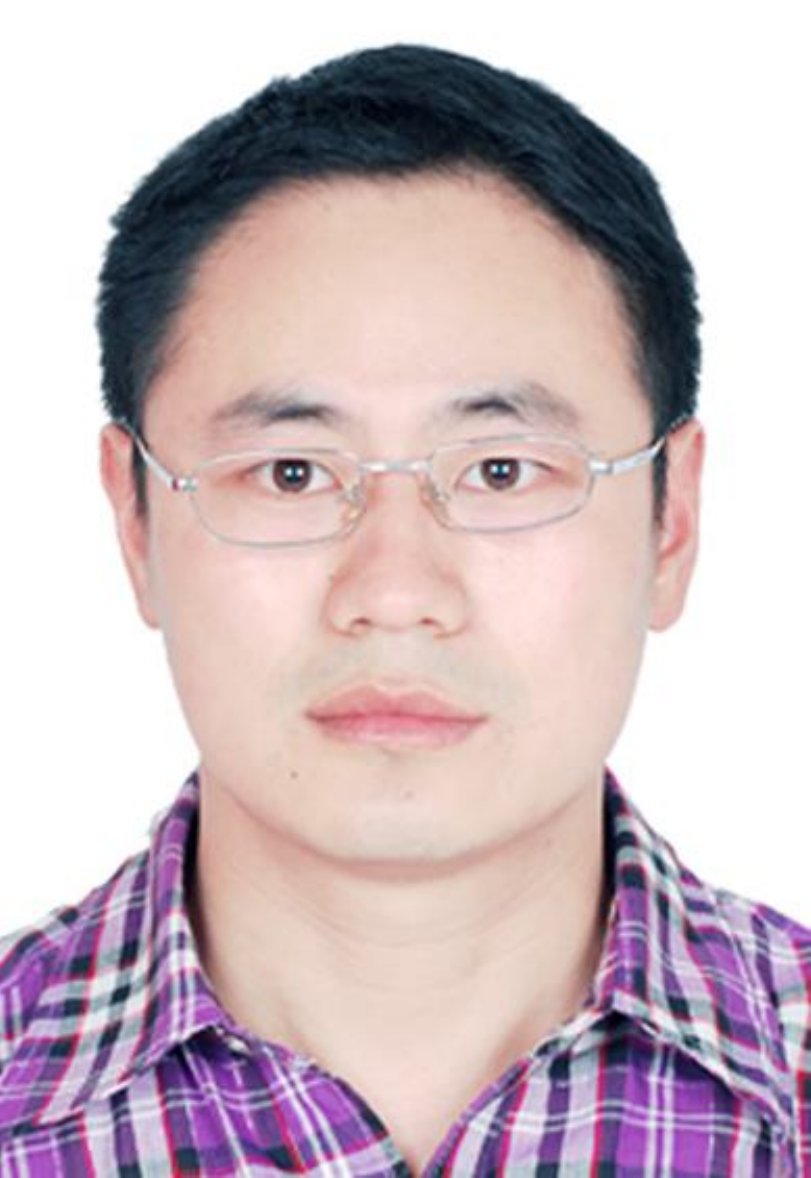}
\end{minipage}%
\hspace{0.02\textwidth}%
\begin{minipage}{0.82\textwidth}
    \hspace{1cm} \textbf{Chunxiao Liu} is an associate professor and master supervisor in Computer Science and Technology at the School of Computer Science and Technology, Zhejiang Gongshang University. He got his Ph.D in Mathematics from the State Key Lab of CAD\&CG, Zhejiang University, Hangzhou, China, in 2019. His current research interests include image and video processing, computer vision, computer graphics, machine learning, intelligent systems.
\end{minipage}

\vspace{2\baselineskip}

\noindent
\begin{minipage}{0.15\textwidth}
    \includegraphics[width=\linewidth]{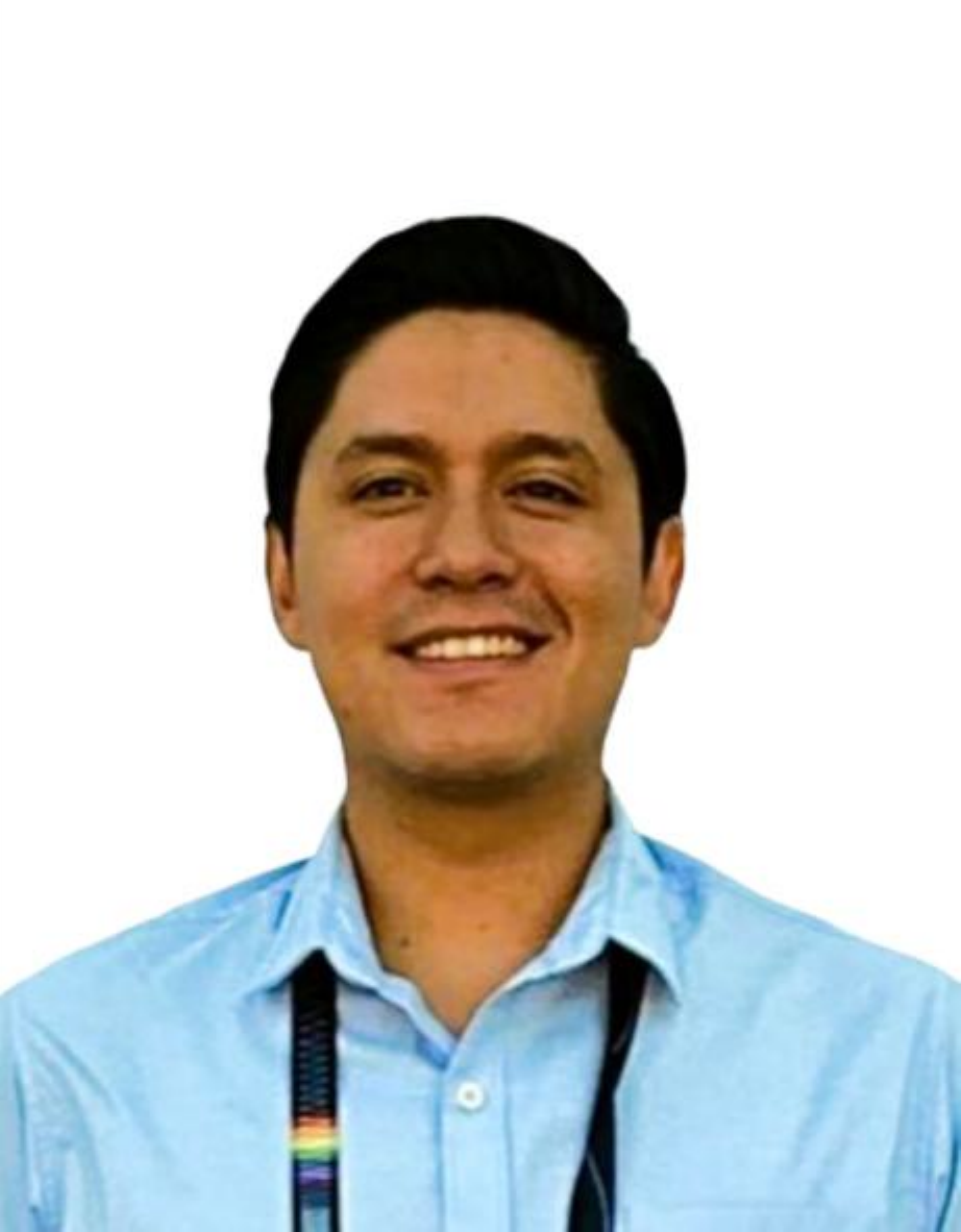}
\end{minipage}%
\hspace{0.02\textwidth}%
\begin{minipage}{0.82\textwidth}
    \hspace{1cm} \textbf{Oscar Ramos-Soto} received the B.S. degree in Communications and Electronics Engineering from the National Polytechnic Institute, Mexico, in 2019, and the M.Sc. degree in Electronic Engineering and Computer Sciences from University of Guadalajara, Mexico, in 2021. Currently, he is pursuing a Ph.D. degree in Electronics and Computer Sciences at the University of Guadalajara, Mexico. His research interests primarily involve areas such as computer vision, digital image processing, machine and deep learning, biomedical engineering, and vision science.
\end{minipage}

\vspace{3\baselineskip}

\noindent
\begin{minipage}{0.15\textwidth}
    \includegraphics[width=\linewidth]{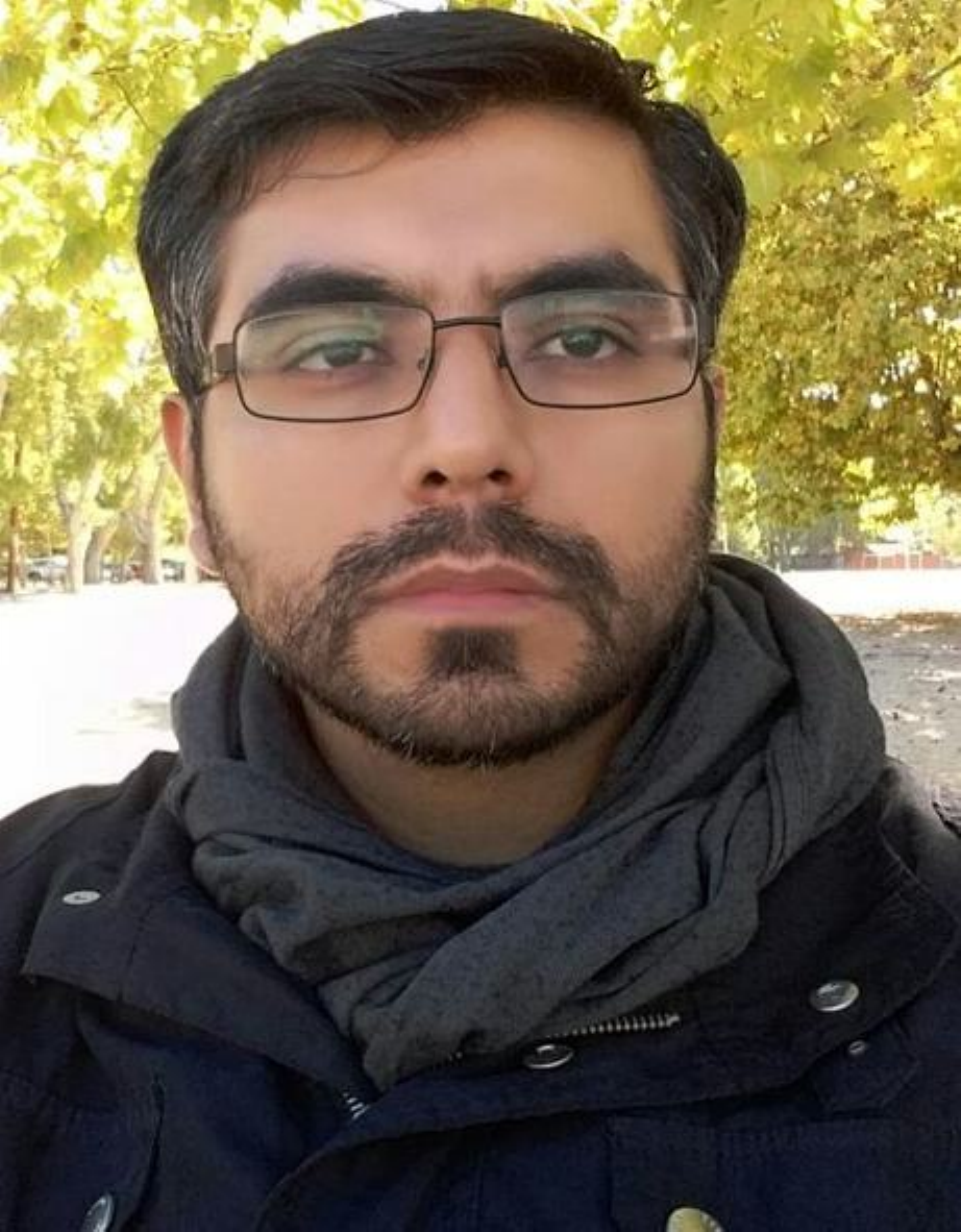}
\end{minipage}%
\hspace{0.02\textwidth}%
\begin{minipage}{0.82\textwidth}
    \hspace{1cm} \textbf{Prof. Diego Oliva} received a B.S. degree in Electronics and Computer Engineering from the Industrial Technical Education Center (CETI) of Guadalajara, Mexico, in 2007 and an M.Sc. degree in Electronic Engineering and Computer Sciences from the University of Guadalajara, Mexico, in 2010. He obtained a Ph. D. in Informatics in 2015 from the Universidad Complutense de Madrid. Currently, he is an Associate Professor at the University of Guadalajara in Mexico. He is a member of the Mexican National Research System (SNII), a Senior member of the IEEE, a member of the Association for Computing Machinery (ACM), and a member of the Mexican Academy of Computer Sciences (AMEXCOMP). His research interests include evolutionary and swarm algorithms, hybridization of evolutionary and swarm algorithms, and computational intelligence.
\end{minipage}

\vspace{3\baselineskip}

\noindent
\begin{minipage}{0.15\textwidth}
    \includegraphics[width=\linewidth]{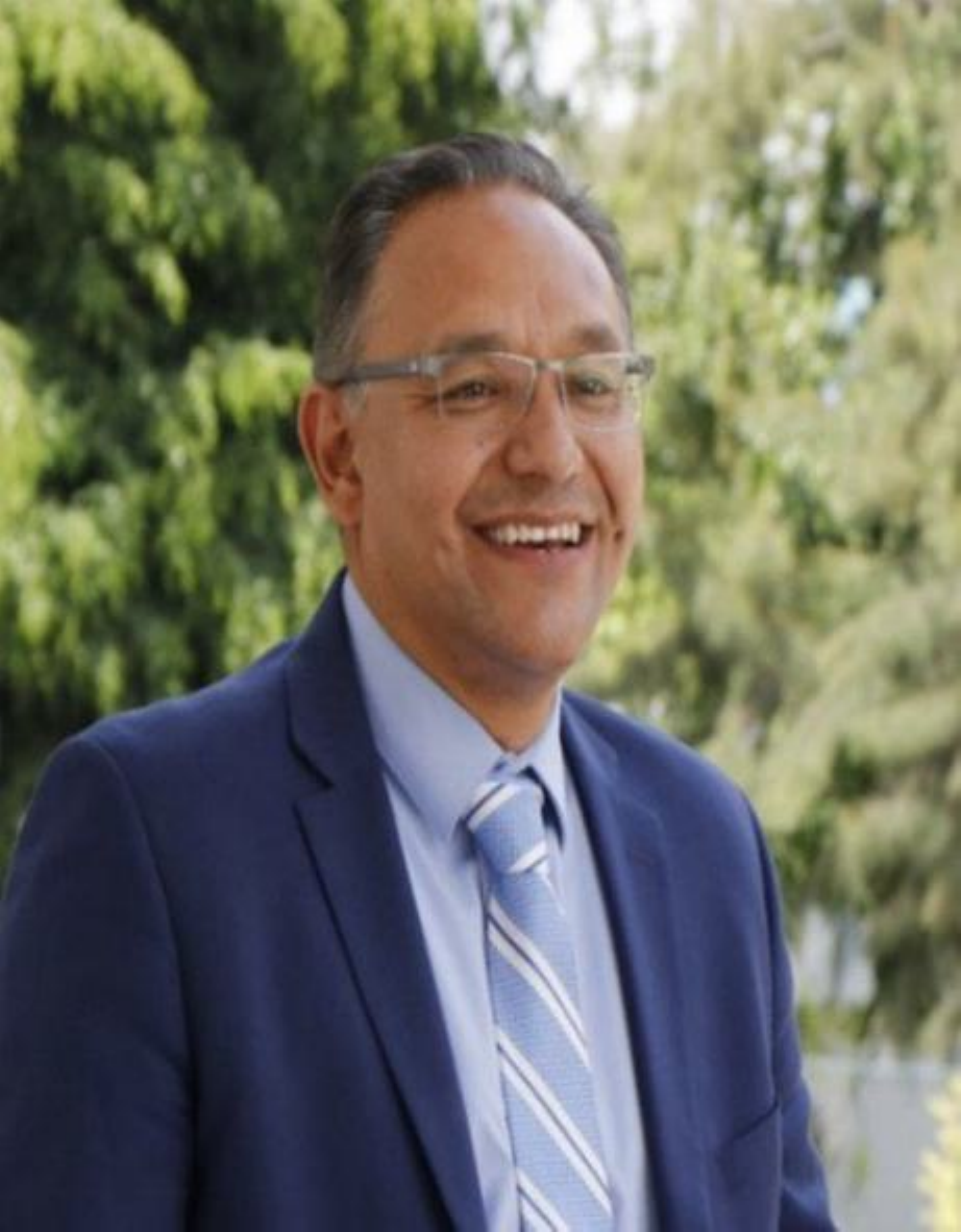}
\end{minipage}%
\hspace{0.02\textwidth}%
\begin{minipage}{0.82\textwidth}
    \hspace{1cm} \textbf{Marco Pérez-Cisneros} received the B.S. degree (Hons.) in electronics and communications engineering from Universidad de Guadalajara, Mexico, in 1995, the M.Sc. degree in industrial electronics from ITESO University, Guadalajara, in 2000, and the Ph.D. degree from the Institute of Science and Technology, UMIST, The University of Manchester, U.K., in 2004. Since 2005, he has been with CUCEI, Universidad de Guadalajara, where he is currently a Professor and the Rector of CUCEI. He has published over 75 indexed papers, currently holding an H-index of 17. He is the author of seven textbooks about his research interests. His current research interests include computational intelligence and evolutionary algorithms and their applications to robotics, computational vision, and automatic control. He is a member of Mexican Science Academy, Mexican National Research System (SNII), and a Senior Member of IET.
\end{minipage}

\end{document}